\documentclass[11pt]{book}

  \usepackage[english]{babel}
  \usepackage[utf8]{inputenc}

  \usepackage[T1]{fontenc}          %

   \usepackage[a4paper]{geometry}
  \usepackage{url}        
  \usepackage{graphicx}   
  \usepackage{mathrsfs}   
  \usepackage{array}      

    \usepackage{pifont}     
    \usepackage{enumitem}   

  \usepackage{amsmath}    
  \usepackage{amssymb}    
  \usepackage{mathtools}  
  \usepackage{mathdots}   
  \usepackage{wasysym}    
  \usepackage{cancel}     
  \usepackage{stackrel}   
  \usepackage{dsfont}     

  \numberwithin{table}{section}
  \numberwithin{equation}{section}
  \numberwithin{figure}{section}

  \usepackage{xcolor}
    \definecolor{colNumberTitle}{RGB}{220,220,220}
    \definecolor{greyforboxes}{RGB}{235,235,235}

  \usepackage{fancyhdr}
  \usepackage{titletoc}
  \usepackage[toctitles, raggedright]{titlesec}
  
    \titlespacing*{\chapter}{0pt}{-30pt}{20pt}
    \newcommand\auxtitlenumformat[1]{{\fontsize{70pt}{70pt}\selectfont\textcolor{colNumberTitle}{#1}}}
    \titleformat{\chapter}[hang]{\bfseries}{\auxtitlenumformat{\thechapter}\hspace{3mm}}{0pt}{\huge\raggedright}

  \allowdisplaybreaks[1]            

    \PassOptionsToPackage{normalem}{ulem}
    \usepackage{ulem}

  \usepackage{float}  

\usepackage[margin=1cm]{caption}
  \usepackage{amsthm}

  \theoremstyle{plain}
    \ifx\thechapter\undefined
      \newtheorem{thm}{\protect\theoremname}   
    \else
      \newtheorem{thm}{\protect\theoremname}[chapter]
    \fi

    \newtheorem{prop}[thm]{\protect\propositionname}  
    \newtheorem{lem}[thm]{\protect\lemmaname}
    \newtheorem{cor}[thm]{\protect\corollaryname}

  \theoremstyle{definition}
    \newtheorem{defn}[thm]{\protect\definitionname}
    \newtheorem{example}[thm]{\protect\examplename}

  \theoremstyle{remark}

  \providecommand{\theoremname}{Theorem}
  \providecommand{\propositionname}{Proposition}
  \providecommand{\lemmaname}{Lemma}
  \providecommand{\corollaryname}{Corollary}
  \providecommand{\definitionname}{Definition}
  \providecommand{\examplename}{Example}

  \usepackage{tikz}                            
  \usepackage[framemethod=tikz]{mdframed}      
  \mdfsetup{splittopskip=\topskip, 
            linewidth=0.5}

  \newcommand\boxsimple [1]{\begin{mdframed} #1 \end{mdframed} }

  \newcommand\boxquote[2]{\begin{mdframed}[topline=false, leftline=false, bottomline=false, 
                                           userdefinedwidth=.8\linewidth,align=right]
                                          {\small\textit{#1} \begin{flushright}--- #2 \par\end{flushright}}\end{mdframed}}

  \newcommand\boxmath[1]
    {\begin{mdframed}[topline=false,rightline=false, bottomline=false,
                      linewidth=2mm, linecolor=black, backgroundcolor=greyforboxes, innertopmargin = 0pt, innerbottommargin = 1pt] #1 \end{mdframed} }
  \newcommand\boxmathclear[1]
    {\begin{mdframed}[topline=false,rightline=false, bottomline=false,
                      linewidth=2mm, linecolor=greyforboxes] #1 \end{mdframed} }

  \newcommand\boxtheorem     [1]{\boxmath{#1}}
  \newcommand\boxproposition [1]{\boxmath{#1}}
  \newcommand\boxlemma       [1]{\boxmath{#1}}
  \newcommand\boxcorollary   [1]{\boxmath{#1}}
  \newcommand\boxproof       [1]{\boxmathclear{#1}}
  \newcommand\boxexample     [1]{\boxmathclear{#1}}
  \newcommand\boxdefinition  [1]{\begin{mdframed}
                                 [leftline=false, rightline=false, backgroundcolor=greyforboxes, skipbelow=0cm] #1 
                                 \end{mdframed} }

  \usepackage[unicode=true, 
              pdfusetitle,            
              bookmarks=true,         
              bookmarksnumbered=false,
              bookmarksopen=false,
              breaklinks=true,        
              backref=false,          
              colorlinks=true,        
              linktocpage             
              ]{hyperref}
  
  \hypersetup{linkcolor = blue,
              citecolor = orange,
              urlcolor  = magenta}

\usepackage[backend=bibtex8,
            sorting=none,               
            citestyle=numeric-comp,     
            bibstyle=PhDThesisBibStyle, 
            giveninits=true,            
            isbn=false,                 
            maxbibnames=4               
           ]{biblatex}

  \newcommand{\iN}{\mathrm{i}}  
  \newcommand{\eN}{\mathrm{e}}  
  \newcommand{\mpl}{M_{\rm Pl}} 

  \newcommand{\dform}[1]{\boldsymbol{#1}}        
  \newcommand{\bvec} [1]{\boldsymbol{#1}}        
  \newcommand{\ten}  [1]{\boldsymbol{#1}}        

  \newcommand{\dfal}{\dform{\alpha}}
  \newcommand{\dfbe}{\dform{\beta}}
  \newcommand{\dfga}{\dform{\gamma}}
  \newcommand{\dfGa}{\dform{\Gamma}}
  \newcommand{\dfDe}{\dform{\Delta}}
  \newcommand{\dfvarDe}{\dform{\varDelta}}
  
  \newcommand{\dfThe}{\dform{\Theta}}
  \newcommand{\dfLa}{\dform{\Lambda}}
  \newcommand{\dfvarLa}{\dform{\varLambda}}
  \newcommand{\dfmu}{\dform{\mu}}
  \newcommand{\dfnu}{\dform{\nu}}
  \newcommand{\dfom}{\dform{\omega}}
  \newcommand{\dfOm}{\dform{\Omega}}
  \newcommand{\dfsi}{\dform{\sigma}}
  \newcommand{\dfSi}{\dform{\Sigma}}

  \newcommand{\dfPhi}{\dform{\Phi}}
  
  \newcommand{\dfPsi}{\dform{\Psi}}
  \newcommand{\dfvarPsi}{\dform{\varPsi}}
  \newcommand{\dfvarUps}{\dform{\varUpsilon}}
  \newcommand{\dfxi}{\dform{\xi}}
  \newcommand{\dfXi}{\dform{\Xi}}

  \newcommand{\dfA}{\dform{A}}
  \newcommand{\dfB}{\dform{B}}

  \newcommand{\dfE}{\dform{E}}
  \newcommand{\dfF}{\dform{F}}
  
  \newcommand{\dfH}{\dform{H}}
  \newcommand{\dfI}{\dform{I}}

  \newcommand{\dfL}{\dform{L}}

  \newcommand{\dfP}{\dform{P}}
  \newcommand{\dfQ}{\dform{Q}}
  \newcommand{\dfR}{\dform{R}}
  
  \newcommand{\dfT}{\dform{T}}
  \newcommand{\dfU}{\dform{U}}
  
  \newcommand{\dfW}{\dform{W}}
  \newcommand{\dfX}{\dform{X}}
  \newcommand{\dfY}{\dform{Y}}
  \newcommand{\dfZ}{\dform{Z}}

  \newcommand{\dfa}{\dform{a}}
  
  \newcommand{\dfh}{\dform{h}}
  \newcommand{\dfk}{\dform{k}}
  \newcommand{\dfl}{\dform{l}}
  \newcommand{\dfm}{\dform{m}}
  \newcommand{\dfq}{\dform{q}}
  \newcommand{\dfu}{\dform{u}}

  \newcommand{\vecv}{\bvec{v}}

  \newcommand{\vecE}{\bvec{E}}
  \newcommand{\vecT}{\bvec{T}}
  \newcommand{\vecU}{\bvec{U}}
  \newcommand{\vecV}{\bvec{V}}
  \newcommand{\vecW}{\bvec{W}}
  \newcommand{\vecX}{\bvec{X}}
  
  \newcommand{\vecZ}{\bvec{Z}}
	
  \newcommand{\vecxi}{\bvec{\xi}}

  \newcommand{\teng}{{\ten{g}}}
  \newcommand{\tenS}{{\ten{S}}}
  \newcommand{\tenT}{{\ten{T}}}
  \newcommand{\tent}{{\ten{t}}}

  \newcommand{\identity}{\mathds{1}}             

  \DeclareMathOperator{\sign}{sgn}         

  \newcommand{\der}{\mathrm{d}}            

  \newcommand\dex{\mathrm{d}}            
  \newcommand\Dex{\mathbf{D}}            
  \newcommand\dint[1]{#1\lrcorner}       
  \newcommand\dLie{\mathfrak{L}}         
  \newcommand\DLie{\mathbb{L}}           
  \newcommand\anabla{\vec{\nabla}}     

\newcommand{\dfQNoTr}{\nearrow\!\!\!\!\!\!\!\!\dfQ{~}}
\newcommand{\QNoTr}  {\nearrow\!\!\!\!\!\!\!Q}
\newcommand{\dfZNoTr}{\nearrow\!\!\!\!\!\!\!\dfZ}
\newcommand{\dfPNoTr}{\nearrow\!\!\!\!\!\!\!\dfP}
\newcommand{\rmLNoTr}{\nearrow\!\!\!\!\!\!\!{\rm L}}

  \newcommand{\vpartial}{\bvec{\partial}}
  \newcommand{\cofr}{\dform{\vartheta}}
  \newcommand{\vfre}{\bvec{e}}
  \newcommand{\volf}{\dform{\omega}_\mathrm{vol}} 
  \newcommand{\volfg}{\mathbf{vol}_{g}} 
  \newcommand{\sgng}{\sign(g)}                    
  \newcommand{\dimM}{\mathtt{D}}                           
  \newcommand{\LCten}{\mathcal{E}}                

  \newcommand{\irrQ}[1]  {{}^{\scriptscriptstyle(#1)\!}Q{}}
  \newcommand{\irrT}[1]  {{}^{\scriptscriptstyle(#1)\!}T{}}
  \newcommand{\irrW}[1]  {{}^{\scriptscriptstyle(#1)\!}W{}}
  \newcommand{\irrZ}[1]  {{}^{\scriptscriptstyle(#1)\!}Z{}}

  \newcommand{\irrdfQ}[1]{{}^{\scriptscriptstyle(#1)\!}\dfQ}
  \newcommand{\irrdfT}[1]{{}^{\scriptscriptstyle(#1)\!}\dfT}
  \newcommand{\irrdfW}[1]{{}^{\scriptscriptstyle(#1)\!}\dfW}
  \newcommand{\irrdfZ}[1]{{}^{\scriptscriptstyle(#1)\!}\dfZ}
  \newcommand{\irrdfLCW}[1]{{}^{\scriptscriptstyle(#1)\!}\mathring{\dfW}}

\newcommand*{\TakeFourierOrnament}[1]{{\fontencoding{U}\fontfamily{futs}\selectfont\char#1}}
\newcommand*{\danger}{\text{\TakeFourierOrnament{66}}}       
\newcommand{\Danger}{\textcolor{red}{\danger}}

\newcommand{\dfomtL}{{}^{\scriptscriptstyle \mathrm{L}}\!\tilde{\dfom}{}}
\newcommand{\dfomtT}{{}^{\scriptscriptstyle \mathrm{T}}\!\tilde{\dfom}{}}

\newcommand{\dfAL}{{}^{\scriptscriptstyle \mathrm{L}}\!\dfA{}}
\newcommand{\dfAT}{{}^{\scriptscriptstyle \mathrm{T}}\!\dfA{}}

\newcommand{\dfBL}{{}^{\scriptscriptstyle \mathrm{L}}\!\dfB{}}
\newcommand{\dfBT}{{}^{\scriptscriptstyle \mathrm{T}}\!\dfB{}}

\newcommand{\dfGaL}{{}^{\scriptscriptstyle \mathrm{L}}\dfGa{}}
\newcommand{\dfGaT}{{}^{\scriptscriptstyle \mathrm{T}}\dfGa{}}

\newcommand{\sobj}[1]{\underline{#1}}
\newcommand{\sdex}{\sobj{\dex}}
\newcommand{\spartial}{\sobj{\partial}{}}
\newcommand{\spatcofr}{\sobj{\cofr}{}}
\newcommand{\sstar}{\sobj{\star}}
\newcommand{\svolf}{\sobj{\mathbf{vol}}_\delta}
\newcommand{\sDelta}{\sobj{\Delta}}

\newcommand{\metriccompatible}[1]{\tilde{#1}}
\newcommand{\mcdfom}{\metriccompatible{\dfom}}

\newcommand{\mcdfT}{\metriccompatible{\dfT}}
\newcommand{\mcdfR}{\metriccompatible{\dfR}}
\newcommand{\mcDex}{\metriccompatible{\Dex}}
\newcommand{\mcdfL}{\metriccompatible{\dfL}}
\newcommand{\mcS}{\metriccompatible{S}}

\newcommand{\overeq}[1]{\overset {\scriptstyle #1} = {}}
\newcommand{\eqonshell}[1]{\overset{{\scriptscriptstyle #1}}{\approx}}

\newcommand{\lrho}{\ell_\rho^2}

\newcommand{\kA}{\hat{A}}
\newcommand{\sA}{{\sobj{A}}{}}
\newcommand{\kB}{\hat{B}}
\newcommand{\sB}{{\sobj{B}}{}}
\newcommand{\kC}{\hat{C}}
\newcommand{\sC}{{\sobj{C}}{}}
\newcommand{\kP}{\hat{P}}
\newcommand{\sP}{{\sobj{P}}{}}
\newcommand{\calF}{\mathcal{F}}
\newcommand{\VdHC}{{\sobj{N}}{}}

\newcommand{\pW}{\overset {\scriptscriptstyle (+)}W {}} 
\newcommand{\pmW}{\overset {\scriptscriptstyle (\pm)}W {}}
\newcommand{\pOm}{\overset {\scriptscriptstyle (+)}\dfOm {}}
\newcommand{\mOm}{\overset {\scriptscriptstyle (-)}\dfOm {}}
\newcommand{\pmOm}{\overset {\scriptscriptstyle (\pm)}\dfOm {}}
\newcommand{\irrpOm}[1]  {{}^{\scriptscriptstyle(#1)}\!\pOm }
\newcommand{\irrmOm}[1]  {{}^{\scriptscriptstyle(#1)}\!\mOm }
\newcommand{\irrpmOm}[1] {{}^{\scriptscriptstyle(#1)}\!\pmOm}
\newcommand{\irrpmOmc}[1]{{}^{\scriptscriptstyle(#1)}\!{\overset {\scriptscriptstyle (\pm)}\Omega} {}}
\newcommand{\wUps}{\overset {\scriptscriptstyle (w)}\dfvarUps {}}
\newcommand{\zUps}{\overset {\scriptscriptstyle (z)}\dfvarUps {}}
\newcommand{\oUps}{ \dfvarUps {}}
\newcommand{\wphi}{\overset{\scriptscriptstyle (w)}{\varphi}}
\newcommand{\zphi}{\overset{\scriptscriptstyle (z)}{\varphi}}

\newcommand{\uW}[1]{{W_{\sobj{#1}}}}
\newcommand{\uV}[1]{{V_{\sobj{#1}}}}
\newcommand{\uU}[1]{{U^{\sobj{#1}}}}
\newcommand{\upartial}[1]{{\partial^{\sobj{#1}}}}

\newcommand{\cU}{\mathcal{U}}
\newcommand{\cV}{\mathcal{V}}
\newcommand{\cW}{\mathcal{W}}
\newcommand{\cX}{\mathcal{X}}
\newcommand{\cUb}{\oddvar{\mathcal{U}}}
\newcommand{\cVb}{\oddvar{\mathcal{V}}}
\newcommand{\cWb}{\oddvar{\mathcal{W}}}
\newcommand{\cXb}{\oddvar{\mathcal{X}}}

\newcommand{\mm}{{\hat a}}
\newcommand{\nn}{{\hat b}}

\newcommand\KMaxSym{K}

\newcommand{\pECG}{\beta}
\newcommand{\eanis}{\epsilon_\sigma}
\newcommand{\HubH}{\mathrm{H}}
\newcommand{\Qb}{\check{Q}{}}
\newcommand{\cc}{\tilde{\gamma}{}}

\newcommand{\bgobject}[1]{\widehat{#1}{}}
\newcommand{\bgDex}{\bgobject{\Dex}}
\newcommand{\bgg}{\bgobject{g}}
\newcommand{\bgnabla}{\bgobject{\nabla}}

\newcommand{\bgcofr}{\bgobject{\cofr}}

\newcommand{\bgdfcon}{\bgobject{\dfom}}
\newcommand{\bgdfR}{\bgobject{\dfR}}

\newcommand{\perg}{\mu}
\newcommand{\percofr}{\dform{\chi}}
\newcommand{\perdfcon}{\dform{\gamma}}

\newcommand{\oddvar}[1]{\overline{#1}{}}
\newcommand{\odda}{\oddvar{a}} 
\newcommand{\oddA}{\oddvar{A}} 
\newcommand{\oddb}{\oddvar{b}}
\newcommand{\oddc}{\oddvar{c}}
\newcommand{\oddv}{\oddvar{v}}
\newcommand{\oddw}{\oddvar{w}}
\newcommand{\oddz}{\oddvar{z}}
\newcommand{\oddT}{\oddvar{T}}
\newcommand{\oddX}{\oddvar{X}}
\newcommand{\odddfT}{\oddvar{\dfT}}
\newcommand{\odddfPsi}{\oddvar{\dfPsi}}
\newcommand{\odddfPhi}{\oddvar{\dfPhi}}
\newcommand{\odddfvarLa}{\oddvar{\dfvarLa}}
\newcommand{\odddfX}{\oddvar{\dfX}}
\newcommand{\odddfm}{\oddvar{\dfm}}
\newcommand{\odddfh}{\oddvar{\dfh}}
\newcommand{\odddfP}{\oddvar{\dfP}}
\newcommand{\odddfA}{\oddvar{\dfA}}
\newcommand{\odddfY}{\oddvar{\dfY}}

\begin{document}


\pagestyle{empty} 

\begin{center}

\vspace{10mm} \includegraphics[height=0.15\paperheight]{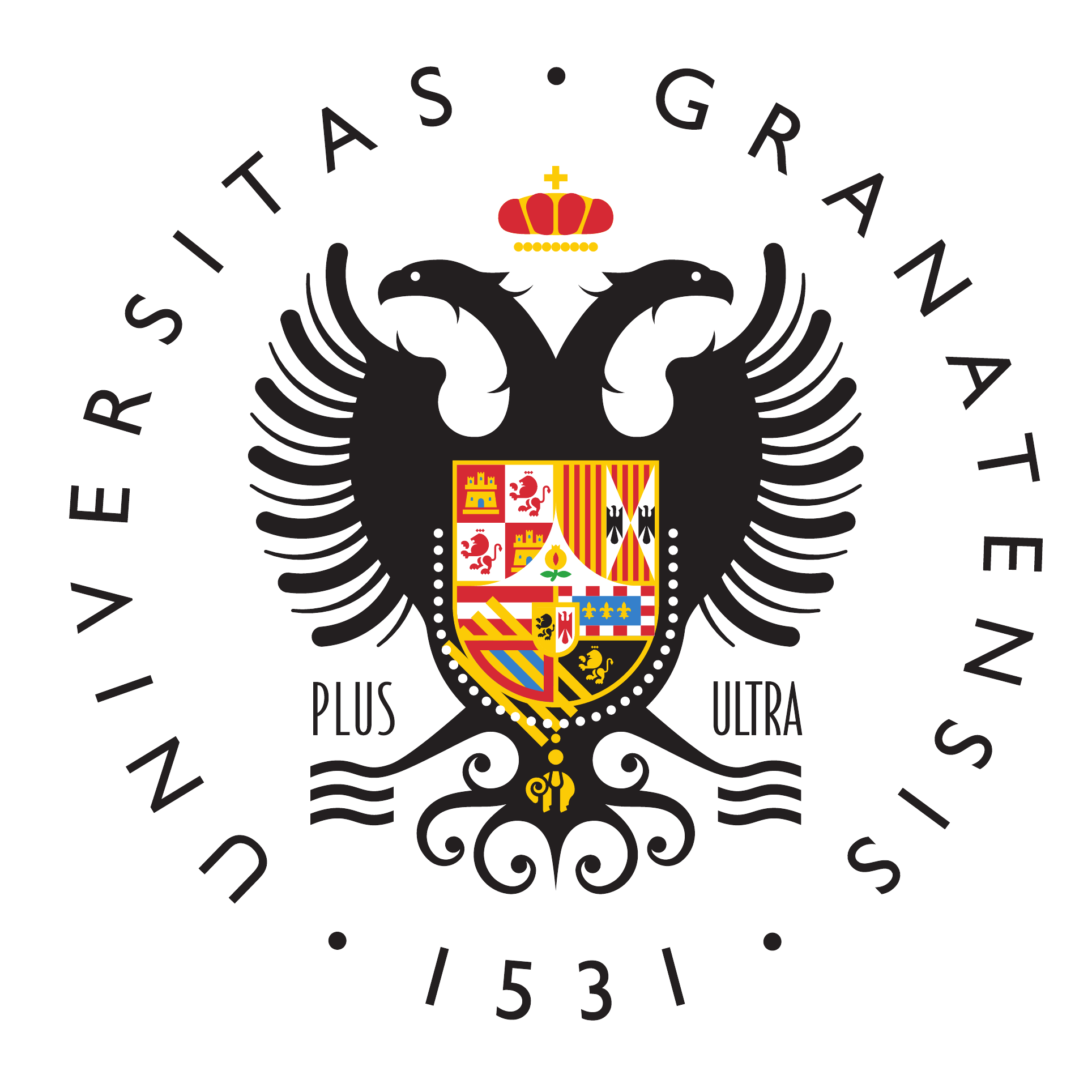}

\vspace{30mm} \textbf{\LARGE METRIC-AFFINE GAUGE THEORIES OF GRAVITY}

\vspace{4mm}  \textbf{\LARGE Foundations and new insights}

\vspace{16mm} {\Large A}{\large LEJANDRO} {\Large J}{\large IM\'ENEZ} {\Large C}{\large ANO}

\vspace{30mm} {\large \textit{Thesis submitted for the degree of}}

\vspace{2mm}  {\textbf{DOCTOR OF PHILOSOPHY}}

\vspace{5mm}  {\Large  \textbf{Programa de Doctorado en F\'isica y Matem\'aticas}}

\vspace{8mm}  {May 2021}

\vspace{8mm}  {Defended on 07 September 2021}

{\color{red}\small
\vspace{2mm}   \texttt{[Revised version: 30 January 2022 (see last page)]}}

\end{center}

\vfill{}      {\large Supervisor: Bert Janssen}

\vspace{2mm}  {\large Departamento de F\'isica Te\'orica y del Cosmos}

\vspace{2mm}  {\large Universidad de Granada}

\newpage \mbox{} 

\newpage
\makeatletter\@openrightfalse\makeatother  
\pagenumbering{Roman}                      

\begin{center}
\textbf{\LARGE List of publications}
\end{center}

The following list collects the articles and publications in which the candidate has participated during the realization of this thesis (reverse chronological order).

{\small

\begin{enumerate}
  \item 
     A. Jim\'enez-Cano, Y. N. Obukhov.\\
     \textit{Gravitational waves in metric-affine gravity theory}\\
     Physical Review D \textbf{103}, 024018 (2021). \\
     \texttt{DOI: \href{https://doi.org/10.1103/PhysRevD.103.024018}{10.1103/PhysRevD.103.024018}  \hfill{}
                      arXiv: \href{https://arxiv.org/abs/2010.14528}{2010.14528 [gr-qc]}}\qquad{}\qquad{}\cite{JCAObukhov2021a}

  \item 
     J. Beltr\'an Jim\'enez, A. Jim\'enez-Cano.\\
     \textit{On the strong coupling of Einsteinian Cubic Gravity and its generalisations}\\
     Journal of Cosmology and Astroparticle Physics  \textbf{01}, 069 (2021).  \\
     \texttt{DOI: \href{https://doi.org/10.1088/1475-7516/2021/01/069}{10.1088/1475-7516/2021/01/069} \hfill{}
                      arXiv: \href{https://arxiv.org/abs/2009.08197}{2009.08197 [gr-qc]}}\qquad{}\qquad{}\cite{BeltranJCA2021}

  \item 
     J. Arrechea, A. Delhom, A. Jim\'enez-Cano.\\
     \textit{Inconsistencies in four-dimensional Einstein-Gauss-Bonnet gravity} \\
     Chinese Physics C \textbf{45}, 013107 (2021).  \\
     \texttt{DOI: \href{https://doi.org/10.1088/1674-1137/abc1d4}{10.1088/1674-1137/abc1d4} \hfill{}
                      arXiv: \href{https://arxiv.org/abs/2004.12998}{2004.12998 [gr-qc]}}\qquad{}\qquad{}\cite{ArrecheaDelhomJCA2021}

  \item 
     J. Arrechea, A. Delhom, A. Jim\'enez-Cano.\\
     \textit{Comment on ``Einstein-Gauss-Bonnet Gravity in four-dimensional space-time''}\\
     Physical Review Letters \textbf{125}, 149002 (2020).  \\
     \texttt{DOI: \href{https://doi.org/10.1103/PhysRevLett.125.149002}{10.1103/PhysRevLett.125.149002} \hfill{}
                      arXiv: \href{https://arxiv.org/abs/2009.10715}{2009.10715 [gr-qc]}}\qquad{}\qquad{}\cite{ArrecheaDelhomJCA2020}

  \item 
     A. Jim\'enez-Cano.\\
     \textit{New metric-affine generalizations of gravitational wave geometries} \\
     The European Physical Journal C \textbf{80}, 672 (2020).  \\
     \texttt{DOI: \href{https://doi.org/10.1140/epjc/s10052-020-8239-5}{10.1140/epjc/s10052-020-8239-5} \hfill{}
                      arXiv: \href{https://arxiv.org/abs/2005.02014}{2005.02014 [gr-qc]}}\qquad{}\qquad{}\cite{JCA2020}

  \item 
     J. Beltr\'an Jim\'enez, L. Heisenberg, D. Iosifidis, A. Jim\'enez-Cano, T. S. Koivisto.\\
     \textit{General Teleparallel Quadratic Gravity} \\
     Physics Letters B \textbf{805}, 135422 (2020).  \\
     \texttt{DOI: \href{https://doi.org/10.1016/j.physletb.2020.135422}{10.1016/j.physletb.2020.135422} \hfill{}
                      arXiv: \href{https://arxiv.org/abs/1909.09045}{1909.09045 [gr-qc]}}\qquad{}\qquad{}\cite{BeltranHeisenbergJCA2020}

  \item 
    C. Bejarano, A. Delhom, A. Jim\'enez-Cano, G. J. Olmo, D. Rubiera-Garcia.\\
    \textit{Geometric inequivalence of metric and Palatini formulations of General Relativity} \\
    Physics Letters B \textbf{802}, 135275 (2020).  \\
    \texttt{DOI: \href{https://doi.org/10.1016/j.physletb.2020.135275}{10.1016/j.physletb.2020.135275} \hfill{}
                      arXiv: \href{https://arxiv.org/abs/1907.04137}{1907.04137 [gr-qc]}}\qquad{}\qquad{}\cite{BejaranoDelhomJCA2019}

  \item 
    B. Janssen, A. Jim\'enez-Cano.\\
    \textit{On the topological character of metric-affine Lovelock Lagrangians in critical dimensions} \\
    Physics Letters B \textbf{798}, 134996 (2019).  \\
    \texttt{DOI: \href{https://doi.org/10.1016/j.physletb.2019.134996}{10.1016/j.physletb.2019.134996} \hfill{}
                      arXiv: \href{https://arxiv.org/abs/1907.12100}{1907.12100 [gr-qc]}}\qquad{}\qquad{}\cite{JanssenJCA2019b}

  \item B. Janssen, A. Jim\'enez-Cano, J. A. Orejuela.\\
    \textit{A non-trivial connection for the metric-affine Gauss-Bonnet theory in $D=4$} \\
    Physics Letters B \textbf{795}, 42--48 (2019).  \\
    \texttt{DOI: \href{https://doi.org/10.1016/j.physletb.2019.06.002}{10.1016/j.physletb.2019.06.002} \hfill{}
                      arXiv: \href{https://arxiv.org/abs/1903.00280}{1903.00280 [gr-qc]}}\qquad{}\qquad{}\cite{JanssenJCA2019a}

  \item 
    B. Janssen, A. Jim\'enez-Cano.  \\
    \textit{Projective symmetries and induced electromagnetism in metric-affine gravity.} \\
    Physics Letters B \textbf{786}, 462--465 (2018).  \\
    \texttt{DOI: \href{https://doi.org/10.1016/j.physletb.2018.10.032}{10.1016/j.physletb.2018.10.032} \hfill{}
                      arXiv: \href{https://arxiv.org/abs/1807.10168}{1807.10168 [gr-qc]}}\qquad{}\qquad{}\cite{JanssenJCA2018}

  \item
    A. N. Bernal, B. Janssen, A. Jim\'enez-Cano, J. A. Orejuela, M. S\'anchez, P. S\'anchez-Moreno.\\
    \textit{On the (non-)uniqueness of the Levi-Civita connection in the Einstein-Hilbert-Palatini formalism.} \\
    Physics Letters B \textbf{768}, 280--287 (2017).  \\
    \texttt{DOI: \href{https://doi.org/10.1016/j.physletb.2017.03.001}{10.1016/j.physletb.2017.03.001} \hfill{}
                      arXiv: \href{https://arxiv.org/abs/1606.08756}{1606.08756 [gr-qc]}}\qquad{}\qquad{}\cite{BernalJCA2017}

\end{enumerate}
}

\newpage
The results of 1, 2, 5, 6 and 8, as well as part of the results of 3, 4 and 9, are presented in this thesis. The style, definitions and conventions have been adapted and unified (and any exception in this regard is explicitly indicated). In some cases, even the notation has been changed to avoid confusion between different chapters. Some parts have been directly extracted from those articles, and others have been rewritten and complemented with intermediate steps to make the derivations crystal clear.

~

The following list shows the papers on which each chapter is based:

~

{\noindent}{\bf\small  Chapter \ref{ch:introduction}} \dotfill No original results

\subsubsection*{Part \ref{part:foundations}. Geometry and metric-affine theories}

{\noindent}{\bf\small  Chapter \ref{ch:mathtools}} \dotfill No original results

{\noindent}{\bf\small Chapter \ref{ch:MAGfoundations}} \dotfill No original results

{\noindent}{\bf\small Chapter \ref{ch:Lovelock}} \dotfill Based on \cite{JanssenJCA2019b} (also contains results from \cite{JanssenJCA2019a})

\subsubsection*{Part \ref{part:GW}. Gravitational wave exact solutions in MAG}

{\noindent}{\bf\small Chapter \ref{ch:GWgen}} \dotfill Based on \cite{JCA2020}

{\noindent}{\bf\small Chapter \ref{ch:GWsolutions}} \dotfill Based on \cite{JCAObukhov2021a}

\subsubsection*{Part \ref{part:viability}. Viability of MAG and other modified theories of gravity}

{\noindent}{\bf\small Chapter \ref{ch:gravityviability}} \dotfill Includes part of the results of \cite{ArrecheaDelhomJCA2020, ArrecheaDelhomJCA2021}

{\noindent}{\bf\small Chapter \ref{ch:ECG}} \dotfill Based on \cite{BeltranJCA2021}

{\noindent}{\bf\small Chapter \ref{ch:QTG}} \dotfill Based on \cite{BeltranHeisenbergJCA2020}

{\noindent}{\bf\small Chapter \ref{ch:MAGspectrum}} \dotfill Based on \cite{JCAObukhov2021b} (\textit{work in progress})

\newpage
\begin{center}
\vspace{20mm} \textbf{\LARGE Agradecimientos / Acknowledgements}
\end{center}

Si he de recalcar algo que he aprendido durante estos años de doctorado es la importancia de abrirse a otros grupos, gente e ideas, por encima de la física en sí. Hacer una tesis en física fundamental es algo realmente  apasionante, pero no ha sido un camino fácil. Aparte de la investigación y los constantes callejones sin salida hay que lidiar con los sentimientos de superación y autocrítica, la constante lucha contra el síndrome del impostor y el afrontar un futuro postdoctoral que se muestra desafiante como poco. Antes de comenzar con la tesis en sí, siento necesario detenerme un momento a agredecer a todas las personas que han contribuido a este proyecto.

A los que han pasado por el despacho y me habéis dejado un poco de vosotros: Fran, Pablo, Carlos, Juanmi, Antonio, Pablo Guerrero, Carmen, Maria, Guilhaume... Y, en especial, a Chema, Dani y Juansi por aguantar mis tonterías y entrar en resonancia con ellas, por recordarme cada día que la física que he estado trabajando estos años es mentira y por hacerme perder el tiempo con eternas discusiones matemáticas y filosóficas. A Juan Carlos y Javi Lizana por vuestras observaciones y críticas a mi trabajo y por vuestra ayuda siempre desinteresada. Y también al resto de compañeros del departamento, Jose Alberto, Javier Olmedo, Manel, Mar, Manolo, José y otros tantos que me dejo. Mención especial a José Ignacio, por el trabajo en equipo con el curso de cuántica (del que aprendí mucho) y por esa fascinación y entusiasmo que desprendes. Gracias.

A mis amigos (Aitor, María, Álvaro, Edu Pi, Fer, Sara...) y también a mis compañeros de promoción (Kike, Manu, Jesús, Edu, Alejandro...) por estar siempre ahí y por tantos buenos ratos. A Charlie Flin, por tus visitas sorpresa al despacho y por tantas y tantas risas. A Edu M, por nuestro contacto constante y por animarme desde la otra punta del mundo. A Cris, por tu enorme apoyo, por aguantarme en esta época tan dura y por haberme robado tantísimo tiempo; de verdad que lo necesitaba. Gracias.

{\it To those who hosted me during my doctoral stay in Tartu. Tomi, Laur, Manuel, Christian, Sebastian and Daniel, and all the others I met in Estonia, for giving me the opportunity to learn from you and for making me feel as if I were at home. To my all collaborators (Gonzalo O., Diego R., Lavinia H., Pablo S. and the rest) for teaching me so much; to M. Sánchez, R. Tresguerres, J. Zanelli, J. Maldonado and P. Cano, for useful discussions and comments. Special mention to José Beltrán and Yuri Obukhov, for your invaluable emails, full of useful information, and for your great contribution to my training and research}; y, finalmente, a Adrià y Julio, por esos ratos compartiendo ``movidas'' y ``geometrical mandanga'' y por ese equipazo que hacemos (¿paellita en Valencia?). Gracias.

A los rostros interesados y caras boquiabiertas de mis talleres y charlas divulgativas de física y Relatividad. 
A Esther y a mis estudiantes de prácticas y problemas por ayudarme a esculpir mi figura de profesor. Por soportarme (con lo minucioso y riguroso que puedo llegar a ser, ¡que me doy asco a mí mismo a veces!) y por responder siempre tan bien. Me alegro de haberos visto evolucionar y (hasta que el virus lo permitió) de seguir saludándoos por los pasillos. Gracias.

Hombre, no podía faltar: Bert. Que has sido más que un director de tesis; has sido colaborador y amigo. Por esos cafés en los que me enseñaste física, pero también filosofía, música y mucho más; porque has sido una referencia a la hora de formarme como divulgador y docente; y por la sonrisilla con que salía de tu despacho cuando me decías ``buen trabajo''. Gracias, de verdad.

Terminar esta parte dando gracias a mi familia, mis abuelos, titos y primos, y dirigiéndome en especial a mis padres y a mi hermano, a los que agradezco la constante preocupación, y vuestro impulso y ánimo para seguir en esta difícil carrera. Porque gran parte de lo que soy os lo debo a vosotros. Muchas gracias.

\newpage

Finalmente agradecer a los principales organismos que han financiado mi investigación y que han hecho posible que establezca la multitud de colaboraciones que han enriquecido este proyecto; sin las cuales no podría haber llegado tan lejos. Al Gobierno de España por financiarme a través del contrato doctoral FPU (ref. FPU15/02864) y los proyectos FIS2016-78198-P y PID2019-105943GB-I00; al Plan Propio de la Universidad de Granada, en especial por cubrir mi estancia doctoral mediante el programa ``Estancias breves en centros de investigación nacionales y extranjeros''; y a los diferentes grupos de la Universidad Complutense de Madrid, la Universidad de Salamanca y la Universidad de Valencia que han financiado visitas cortas. Gracias.

\newpage 
\vspace*{0.25\paperheight}  \begin{flushright}{\Large To: Pablo, Jos\'e and Paqui} \end{flushright}

\newpage \mbox{} 

\newpage

\begin{center}
 
\vspace*{20mm}{\Large METRIC-AFFINE GAUGE THEORIES OF GRAVITY}

\vspace{2mm}  {\Large Foundations and new insights}

\vspace{6mm}  {\large Alejandro Jim\'enez Cano}
  
\vspace{20mm} \textbf{Summary (in English)}

\end{center}

\vspace{5mm}

General Relativity (GR) is the geometric theory we currently use as standard framework for gravitation. In GR, gravity is understood as a manifestation of the deformations of the spacetime caused by the energy-momentum content. The basic mathematical object in this theory is the metric, which defines the notion of distance. In addition, this metric canonically induces a connection (essentially a notion of parallelism), which has a curvature associated with it. It is this curvature what explains the gravitational effects at large scales.

This thesis is about developments performed in the so-called metric-affine framework. This alternative framework extends GR by considering a connection more general than the one induced by the metric. Interestingly, the resulting structure can be integrated within a gauge procedure, similarly as we do with the other interactions of Nature. The resulting theory is known as Metric-Affine Gauge (MAG) gravity. Contributing to the development of this theory, as well as to other modified theories of gravity, is the central goal of this thesis.

In Chapter \ref{ch:introduction} we revise the open problems that are present in GR with an extensive bibliographic revision and use this to motivate modifications of GR. After this introduction, the contents of the thesis are divided into three main parts, covering different aspects of metric-affine theory and other modified theories of gravity.

In the  {\bf first part} we revise more formal aspects of the metric-affine framework. It contains three chapters:
\begin{itemize}
\item Chapter \ref{ch:mathtools}. This chapter is a compendium of mathematical definitions, results and formulae that we will need to properly work in metric-affine theories.
\item In Chapter \ref{ch:MAGfoundations}, we introduce the gauge approach that leads to metric-affine gravity and collect some general properties of these theories. We also construct the most general action up to quadratic order in curvature, torsion and nonmetricity. 
\item In Chapter \ref{ch:Lovelock}, we discuss the topological nature of the Lovelock terms when formulated in a metric-affine framework in their critical dimensions. At the end we also comment on other topological invariants.
\end{itemize}

\newpage
In the two chapters that constitute the {\bf second part}, we study gravitational wave geometries:
\begin{itemize}
\item In Chapter \ref{ch:GWgen}, we analyze (kinematically) different ways to generalize the geometry of a gravitational wave metric to the metric-affine framework. We apply some of the results and criteria obtained to particular geometries.
\item In Chapter \ref{ch:GWsolutions}, we perform, for a particular geometry (Ansatz), the analysis of the dynamical equations for quadratic metric-affine gravity (only with even parity invariants), and search for exact solutions of that type.
\end{itemize}

The {\bf third part} is much more physical. The aim here is to analyze the viability of different extensions of GR by guaranteeing the stability of their degrees of freedom:
\begin{itemize}
\item In Chapter \ref{ch:gravityviability}, we present and discuss some of the most important field-theoretical pathologies that we can find in general theories and, in particular, in gravity. We see some particular cases and analyze other types of problems in the particular case of 4-dimensional Einstein-Gauss-Bonnet gravity.
\item In Chapter \ref{ch:ECG}, we collect the results of an analysis of the stability of cosmological backgrounds in Einsteinian Cubic gravity and other extensions of this theory. The idea is to explicitly show the consequences of doing physics in a strongly coupled background. This chapter is not directly related to metric-affine gravity, but it is a nice example of a modified theory of gravity with problems that could also be expected in the metric-affine case.
\item In Chapter \ref{ch:QTG}, we concentrate on the  null-curvature restriction of the even quadratic metric-affine action. We express the well known teleparallel equivalents of GR as particular gauge-fixed versions of it. In addition, we show that extra symmetries are needed to avoid the presence of ghosts.
\item In Chapter \ref{ch:MAGspectrum}, we present  some preliminary results on the particle spectrum of the full quadratic metric-affine Lagrangian (with even and odd invariants) in four dimensions around Minkowski space.
\end{itemize}

The chapters based on published works contain at the end their own conclusions (together with a list of limitations of that work). In Chapter \ref{ch:conclusions}, we just briefly revise the most important ones, as well as some general ideas and lessons that one could extract from the whole thesis. This chapter is followed by several appendices that collect complementary contents, expressions and proofs.

\newpage

\begin{center}
 
\vspace*{20mm}{\Large METRIC-AFFINE GAUGE THEORIES OF GRAVITY}

\vspace{2mm}  {\Large Foundations and new insights}

\vspace{6mm}  {\large Alejandro Jim\'enez Cano}
  
\vspace{20mm} \textbf{Resumen (en español)}

\end{center}

\vspace{5mm}

Relatividad General (GR) es la teoría geométrica que actualmente constituye el marco estándar de trabajo en gravitación. En ella, la gravedad se entiende como una manifestación de las deformaciones provocadas en el espaciotiempo por el contenido de energía-momento. El objeto matemático básico de esta teoría es la métrica, la cual define la noción de distancia. Esta métrica, además, induce canónicamente una conexión (esencialmente, una noción de paralelismo) que tiene asociada una curvatura. Y es esta curvatura la que explica a gran escala los efectos gravitatorios. 

Esta tesis trata sobre desarrollos llevados a cabo en el llamado marco métrico-afín de gravedad. Se trata de un marco alternativo que extiende Relatividad General al considerar una conexión más general que la que induce la métrica. Curiosamente, la estructura resultante puede enmarcarse dentro de un procedimiento gauge, tal y como hacemos para el resto de interacciones de la Naturaleza. A la teoría resultante se la conoce como {\it Metric-Affine Gauge} (MAG) {\it gravity}. Contribuir al desarrollo de esta y, en general, al campo de la gravedad modificada, es el objetivo central de esta tesis.

En el Capítulo \ref{ch:introduction} revisamos los problemas abiertos de GR con una extensa batería bibliográfica y los usaremos para motivar modificaciones de GR. Tras esta introducción, los contenidos de la tesis se agrupan en tres grandes bloques, cubriendo diferentes aspectos de la teoría métrico-afín y de otras teorías modificadas de gravedad.

En la  {\bf primera parte} abarcamos aspectos más formales del marco métrico-afín. Contiene tres capítulos:
\begin{itemize}
\item Capítulo \ref{ch:mathtools}. Este capítulo es un compendio de definiciones, resultados y fórmulas matemáticas que necesitaremos para trabajar en teorías métrico-afines.
\item En el Capítulo \ref{ch:MAGfoundations} introducimos el procedimiento gauge que conduce a las teorías métrico-afines y recopilamos algunas propiedades generales de estas. Construimos también la acción más general hasta orden cuadrático en la curvatura, la torsión y la no-metricidad.

\item En el Capítulo \ref{ch:Lovelock} discutimos la naturaleza topológica de los términos de Lovelock formulados en el marco métrico-afín en sus dimensiones críticas. Al final, comentamos también sobre otros invariantes topológicos.
\end{itemize}

\newpage
En los dos capítulos que constituyen la {\bf segunda parte} estudiamos geometrías de onda gravitacional:
\begin{itemize}
\item En el Capítulo \ref{ch:GWgen} analizamos (cinemáticamente) diferentes modos de generalizar métricas de onda gravitacional al marco métrico-afín. Aplicaremos también algunos de los resultados a geometrías particulares.
\item En el Capítulo \ref{ch:GWsolutions} llevamos a cabo, para una geometría particular (Ansatz), el análisis de las ecuaciones dinámicas de la teoría cuadrática métrico-afín (solo con invariantes pares bajo paridad), y exploramos soluciones exactas de aquel tipo.
\end{itemize}

La {\bf tercera parte} es mucho más física. La idea ahora es analizar la viabilidad de diferentes extensiones de GR, usando como criterio la estabilidad de sus grados de libertad:
\begin{itemize}
\item En el Capítulo \ref{ch:gravityviability} presentamos y discutimos algunas de las patologías más importantes que pueden aparecer en teorías de campos generales y, en particuar, en gravedad. Veremos algunos casos particulares y analizaremos otros tipos de problemas en el caso particular de {\it 4-dimensional  Einstein-Gauss-Bonnet gravity}.

\item En el Capítulo \ref{ch:ECG} recopilamos los resultados de un análisis sobre la estabilidad de fondos cosmológicos en {\it Einsteinian Cubic gravity} y extensiones de esta. La idea es mostrar explícitamente las consecuencias de hacer física en un fondo fuertemente acoplado. Este capítulo no está directamente relacionado con gravedad métrico-afín, pero es un buen ejemplo de teoría modificada de gravedad con problemas que podrían esperarse también en el caso métrico-afín.

\item En el Capítulo \ref{ch:QTG} nos concentramos en la restricción de curvatura nula de la parte par de la acción cuadrática métrico-afín. Expresamos los bien conocidos equivalentes de GR como versiones de dicha teoría bajo {\it gauge-fixings} particulares. Además, demostramos que son necesarias simetrías extra para evitar la presencia de modos fantasma.

\item En el Capítulo \ref{ch:MAGspectrum} presentamos algunos resultados preliminares sobre el espectro de partículas del lagrangiano cuadrático métrico-afín general (incluyendo invariantes pares e impares), en cuatro dimensiones y alrededor del espacio de Minkowski.
\end{itemize}

Los capítulos basados en trabajos publicados contienen al final sus propias conclusiones (junto a una lista de limitaciones de dicho estudio). En el Capítulo \ref{ch:conclusions}, brevemente revisaremos las más importantes, así como algunas ideas generales y lecciones que podemos extraer de toda la tesis. A este capítulo le siguen varios apéndices que recopilan contenidos, expresiones y demostraciones complementarias.

\pagestyle{fancy}  

\tableofcontents

\newpage\phantomsection 
\markboth{List of Figures}{List of Figures}
\addcontentsline{toc}{chapter}{\listfigurename}
\listoffigures

\newpage\phantomsection 
\addcontentsline{toc}{chapter}{\listtablename}
\listoftables

\chapter*{List of Abbreviations}

\addcontentsline{toc}{chapter}{List of Abbreviations}
\markboth{List of Abbreviations}{List of Abbreviations}

\noindent \begin{center}
\begin{tabular}{>{\raggedright}p{0.2\textwidth}>{\raggedright}p{0.75\textwidth}}
  l(r).h.s. & Left(right) hand side                    \tabularnewline
  dof(s)    & Degree(s) of freedom                     \tabularnewline
  GR        & General Relativity                       \tabularnewline
  PG        & Poincaré Gauge (gravity)                 \tabularnewline 
  MAG       & Metric-Affine Gauge (gravity)            \tabularnewline 
  4DEGB     & 4-dimensional Einstein-Gauss-Bonnet      \tabularnewline
  TEGR      & (Standard) Teleparallel Equivalent of GR \tabularnewline
  STEGR     & Symmetric Teleparallel Equivalent of GR  \tabularnewline
  GTEGR     & General Teleparallel Equivalent of GR    \tabularnewline
  ECG       & (Cosmological extension of) Einsteinian Cubic Gravity   \tabularnewline
  EFT       & Effective Field Theory                   \tabularnewline
  GQTG      & Generalized Quasi-Topological Gravity    \tabularnewline
  FLRW      & Friedmann-Lema\^itre-Robertson-Walker    \tabularnewline
  GW        & Gravitational Wave                       \tabularnewline
  g.c.t.    & General coordinate transformation(s)     \tabularnewline
  Diff(s) or diff(s)   & Diffeomorphism(s)             \tabularnewline
  TDiff(s)  & Transversal Diffeomorphism(s)            \tabularnewline
  WTDiff(s) & Weyl + Transversal Diffeomorphism(s)     \tabularnewline
\end{tabular}
\par\end{center}

\chapter*{Conventions}\label{chap:Conventions}

\addcontentsline{toc}{chapter}{Conventions}
\markboth{Conventions}{Conventions}

\begin{itemize}
\item We take the speed of light in vacuum to be the unit $c=1$ (natural units).
\item {\bf Mostly minus signature} for the metric: we assign $+$ to the timelike directions and $-$ to the spacelike ones. For example, for a Lorentzian metric in four dimensions we would have $(+---)$.
\begin{itemize}
\item[\Danger] Only in Chapter \ref{ch:ECG} we will turn to mostly plus $(-+++)$.
\end{itemize}

\item $\dimM$ is the dimension of the spacetime.

\item {\bf Indices}

We will constantly apply the \emph{Einstein summation convention}: repeated indices up and down are supposed to be contracted, which means summation over all their possible values (in general, the dimension of the space in which the involved objects live). For tensors, upper indices correspond to the contravariant part (vector) and the lower ones to the covariant part (covector). 

\end{itemize}

\begin{table}[H]
\begin{centering}
\renewcommand{\arraystretch}{1.5}
{\small
\begin{tabular}{|>{\centering}p{2.8cm}|>{\centering}p{11cm}|}
  \hline 
  Indices     & Meaning\tabularnewline
  \hline 
  \hline $\mathfrak{a},\mathfrak{b},\mathfrak{c}$
         &\raggedright{} Indices in some abstract Lie algebra (used in Chapter \ref{ch:MAGfoundations}). \tabularnewline
  \hline $\mathtt{M},\mathtt{N}$
         &\raggedright{} Generic internal indices in a vector space. Used for generic vector-valued forms (Chapter \ref{ch:mathtools}) and for the matter fields (Chapter \ref{ch:MAGfoundations}). \tabularnewline
  \hline $\mu,\nu,\rho,\lambda,\sigma,\tau,\alpha,...$
         &\raggedright{} Components in a holonomic (coordinate) basis of the manifold. \tabularnewline
  \hline $a,b,c,d,e,f...$
         &\raggedright{} Components in a anholonomic frame of the manifold. Sometimes they are used as generic indices (both anholonomic and holonomic). \tabularnewline
  \hline $i,j,k...$
         &\raggedright{} Spatial coordinate indices. In cosmological situations they will cover all the $\dimM-1$ spatial directions. In gravitational wave situations they refer to the transversal $(\dimM-2)$-dimensional space. \tabularnewline
  \hline $A,B,C,D,E,...$  $\sobj{A},\sobj{B},\sobj{C}...$
         &\raggedright{} Same as the previous ones but with respect to an arbitrary frame. The underlined version indicates that they have been raised/lowered with the $\delta_{AB}$ metric (only in Chapter \ref{ch:GWsolutions}). \tabularnewline
  \hline $\mm,\nn$
         &\raggedright{}They cover the two non-transversal directions in a GW anholonomic basis (used in Chapter \ref{ch:GWsolutions}). \tabularnewline
  \hline 
\end{tabular}
}
\renewcommand{\arraystretch}{1}
\par\end{centering}
\caption{Different types of indices used in this thesis.}
\end{table}

\begin{itemize}

\item Convention for the covariant derivative (index order in the connection):
  \begin{equation}
    \nabla_{\mu}H_{\nu}{}^{\rho}\coloneqq\partial_{\mu}H_{\nu}{}^{\rho}-\Gamma_{\mu\nu}{}^{\sigma}H_{\sigma}{}^{\rho}+\Gamma_{\mu\sigma}{}^{\rho}H_{\nu}{}^{\sigma}\,.
  \end{equation}
  \begin{equation}
    \nabla_{\mu}H_{a}{}^{b}\coloneqq\partial_{\mu}H_{a}{}^{b}-\omega_{\mu a}{}^{c}H_{c}{}^{b}+\omega_{\mu c}{}^{b}H_{a}{}^{c}\,.
  \end{equation}
\item Conventions for the Riemann tensor, the Ricci tensor, the Ricci scalar,
the torsion and the nonmetricity components, respectively:
  \begin{align}
    R_{\mu\nu\rho}{}^{\lambda} & \coloneqq2\partial_{[\mu}\Gamma_{\nu]\rho}{}^{\lambda}+2\Gamma_{[\mu|\sigma}{}^{\lambda}\Gamma_{|\nu]\rho}{}^{\sigma}\,, & R_{\mu\nu} & \coloneqq R_{\mu\rho\nu}{}^{\rho}\,, & R & \coloneqq g^{\mu\nu}R_{\mu\nu}\,,\nonumber\\
    T_{\mu\nu}{}^{\lambda} & \coloneqq2\Gamma_{[\mu\nu]}{}^{\lambda}\,,
    &Q_{\mu\nu\sigma} &\coloneqq-\nabla_{\mu}g_{\nu\sigma} \label{eq:convR}\,.
\end{align}
\begin{itemize}
\item[\Danger] In Chapter \ref{ch:ECG}, the curvature tensors $R_{\mu\nu\rho}{}^{\lambda}$ and $R_{\mu\nu}$ will be as in \eqref{eq:convR} but with a global minus sign ($R$ remains the same due to the change of signature).
\end{itemize}

\vspace{-2mm}
\item Symmetrization and antisymmetrization of indices.
  \begin{align}
    H_{(\mu_{1}...\mu_{k})} & \coloneqq\frac{1}{k!}\sum_{\sigma\,\text{perm}}H_{\sigma(\mu_{1})...\sigma(\mu_{k})}\,. & H_{[\mu_{1}...\mu_{k}]} & \coloneqq\frac{1}{k!}\sum_{\sigma\,\text{perm}}\sign(\sigma)H_{\sigma(\mu_{1})...\sigma(\mu_{k})}\,.
  \end{align}
  Every index between the parenthesis (or square-brakets) participates in the (anti-) symmetrization process:
  \begin{equation}
    V_{(\mu}{}^{\lambda}W_{a\nu)}\coloneqq\frac{1}{6}\left(V_{\mu}{}^{\lambda}W_{a\nu}+V_{\mu}{}^{\lambda}W_{\nu a}+V_{a}{}^{\lambda}W_{\mu\nu}+V_{a}{}^{\lambda}W_{\nu\mu}+V_{\nu}{}^{\lambda}W_{a\mu}+V_{\nu}{}^{\lambda}W_{\mu a}\right)\,.
  \end{equation}
  Observe that $\lambda$ is an upper index so it is not affected by the symmetrization. \\
  To exclude indices we use bars:
  \begin{equation}
    V_{(\mu}W_{|a\rho|\nu)}=\frac{1}{2}\left(V_{\mu}W_{a\rho\nu}+V_{\nu}W_{a\rho\mu}\right)\qquad(a,\,\rho\text{ are invisible for the symmetrization}).
  \end{equation}
  We should not use several of these symbols one inside the other because we can create confusion. Suppose, for instance, that we want to antisymmetrize the object $V_{[\mu}{}^{\lambda}W_{\nu]\sigma\rho}$ in $\nu\rho$. Since one of the indices is involved in two antisymmetrizations, what we will do is using an auxiliary delta to avoid confusion:
  \begin{equation}
    V_{[\mu}{}^{\lambda}W_{[\nu]|\sigma|\rho]}\,\text{(ugly/confusing notation)}\quad\longrightarrow\quad V_{[\mu}{}^{\lambda}W_{\alpha]\sigma\beta}\delta_{[\nu}^{\alpha}\delta_{\rho]}^{\beta}=V_{[\mu}{}^{\lambda}W_{\alpha]\sigma[\rho}\delta_{\nu]}^{\alpha}\,.
  \end{equation}

\item We will sometimes refer to the metric through the corresponding line element, using the identifications:
  \begin{equation}
    \dex s^{2}\equiv\teng\,,\qquad\qquad2\dex x\dex y\equiv\dex x\otimes\dex y+\dex y\otimes\dex x\,.
  \end{equation}

\item In Chapters \ref{ch:MAGfoundations} and \ref{ch:MAGspectrum}, overlined objects are odd parity ones: $\oddvar{U}$, $\oddvar{\dfT}$, $\oddvar{\dfLa}$... The overlined parameters are those of the odd parity terms of the action: $\odda_1$, $\oddb_5$, $\oddc_1$, $\oddc_2$...

\item In Chapter \ref{ch:Lovelock}, objects with tilde are associated to the metric-compatible connection $\mcdfom_a{}^b$ defined in Proposition \ref{prop:defmcconn}. Examples: $\mcdfR_a{}^b$, $\mcDex$...

\item In Chapters \ref{ch:GWgen} and \ref{ch:GWsolutions}, underlined objects are transversal in a gravitational wave scenario. For tensor-valued differential forms (of non-zero rank) the external indices can be non-transversal. Examples: $\sobj{\cofr}^a$, $\sobj{g}{}_{ij}$ $\sobj{\dex}$...

\item In Chapters \ref{ch:gravityviability} and \ref{ch:MAGspectrum}, objects with a big hat on them are the background values of the fields in perturbation theory. Examples: $\bgg_{ab}$, $\bgcofr^a$ $\bgDex$...

\end{itemize}


\newpage \mbox{} 

\newpage
\makeatletter\@openrighttrue\makeatother   
\pagenumbering{arabic}                     

\chapter{Introduction}\label{ch:introduction}

\boxquote{Autoritätsdusel ist der größte Feind der Wahrheit  \\ / Blind obedience to authority is the greatest enemy of truth.}{Albert Einstein, in a letter to Jost Winteler (July 8th, 1901)}

\section{Motivation: General Relativity and reasons to modify it}

In Newton's theory, space and time play the role of a static stage where physics takes place, in which gravity is just an interaction between masses governed by a certain law. This changed drastically at the beginning of the 20th century. Two years after his Special Relativity theory, Albert Einstein discovered that non-accelerated observers and freely falling observers are two sides of the same coin, and this fact (the \emph{equivalence principle}) guided him to relate gravity with intrinsic properties of the spacetime. Finally, in 1915, he established a new framework for the gravitational interaction, \emph{General Relativity} (which we will abbreviate sometimes as GR). In this theory, gravity is a manifestation of the nontrivial geometry of the spacetime. To properly describe this geometry, one needs to understand two mathematical structures. The first one is the \emph{metric} $g_{\mu\nu}$ that allows to measure distances (and any type of hyper-volume) and also distinguishes between space and time directions. The other structure is the \emph{connection} $\Gamma_{\mu\nu}{}^\rho$, which establishes the notion of parallel transport, i.e. a criterion to compare tensors in different points of the spacetime. This connection defines the \emph{curvature} in each point.\footnote{
    In the following chapter we will provide a precise definition of all of these objects.} 

There is a very particular connection which is uniquely defined by the metric, the \emph{Levi-Civita connection} $\mathring{\Gamma}_{\mu\nu}{}^\rho$, whose components $\{{}_{\mu\nu}{}^\rho\}$ (called Christoffel symbols) depend on the first derivatives of the metric. The curvature of this Levi-Civita connection, $\mathring{R}_{\mu\nu\rho}{}^\lambda$, is what explains the gravitational interaction in the context of GR. Indeed, the core of GR is the \emph{Einstein equation},
\begin{equation}\label{eq:EinsteinEq}
  \mathring{R}_{\mu\nu}-\frac{1}{2}g_{\mu\nu}\mathring{R} = \kappa \mathcal{T}_{\mu\nu}\,,\qquad \kappa\coloneqq 8\pi G_\mathrm{N}
\end{equation}
where $G_\mathrm{N}$ is the Newton gravitational constant, $\mathring{R}_{\mu\nu}$ and $\mathring{R}$ are the Ricci tensor and Ricci scalar of the geometry (objects derived from $\mathring{R}_{\mu\nu\rho}{}^\lambda$), and $\mathcal{T}_{\mu\nu}$ is the energy-momentum tensor. Therefore, the equation \eqref{eq:EinsteinEq} tells that the curvature of the spacetime (l.h.s.) is dynamically related to the energy and momentum of the matter and fields that live in the spacetime (r.h.s.). By following a Lagrangian approach, the equation \eqref{eq:EinsteinEq} can be derived by varying with respect to the metric a matter action coupled to one of the following purely gravitational actions (that differ in a boundary term):
\begin{equation}\label{eq:LagEINvsEH}
  S_\mathrm{Eins}=\frac{1}{2\kappa}\int \dex^4x\sqrt{|g|} \ g^{\mu\nu}(\{{}_{\lambda\sigma}{}^\sigma\}\{{}_{\mu\nu}{}^\lambda\}-\{{}_{\mu\lambda}{}^\sigma\}\{{}_{\nu\sigma}{}^\lambda\})\,,\qquad
  S_\mathrm{EH} = \frac{1}{2\kappa}\int \dex^4x\sqrt{|g|} \ \mathring{R} \,,
\end{equation}
which we will call \emph{Einstein action} and \emph{Einstein-Hilbert action}, respectively. 

General Relativity is an extraordinarily successful theory that agrees with all observations performed in a broad range of scales, from sub-milimetric to Solar System. It describes the planetary orbits together with their anomalies (e.g. the precession of the perihelion of Mercury), the light deflection in the presence of very massive objects (e.g. gravitational lenses), time delays due to the effects of curvature (experiments with atomic clocks, Shapiro effect...) among others. The theory has also been used to construct the most robust cosmological model that we have: $\Lambda$-Cold-Dark-Matter ($\Lambda$CDM). This model has survived a great deal of observational tests and has provided a notable understanding of the early Universe (post-inflation), the abundances of different elements and particles, and the origin of large scale structure. Additionally, the GR dynamics predicts the emission of perturbations of the metric from extremely violent astronomical processes. These are the so called \emph{gravitational waves} (GW), and were finally confirmed in the last years by LIGO and Virgo Collaborations, through the observations of several black hole mergers (see e.g. \cite{AbbottLIGO2016, AbbottLIGO2016b}) and a binary neutron star merger \cite{AbbottLIGO2017}.\footnote{However, before these detections, there was some evidence of their existence in the orbit decay rate of different binary systems (e.g. the Hulse-Taylor binary).}

Nevertheless, GR is not completely absent of problematic issues, which may come from both the experimental and the theoretical side. 

Experimentally, there are several phenomena whose nature is currently under debate:
\begin{itemize}
\item  \textbf{The Dark Matter problem}. There are astrophysical observations that seem to require extra matter that we do not see: galaxy rotation curves, gravitational lenses, the Bullet cluster, the observed Cosmic Microwave Background, etc. This problem is usually addressed by postulating the existence of some weakly interacting field that accounts for the missing mass (Dark Matter). These effects can also be interpreted as consequence of some unknown features of the gravitational interaction,\footnote{
    Although some observations put serious constraints on this: Bullet cluster observations,  baryon acoustic oscillation, etc.} 
or a combination of them. In the context of $\Lambda$CDM, a Dark Matter distribution with a density parameter $\Omega_\mathrm{c}=0.265(7)$ ($\sim$5 times the amount of ordinary/baryonic matter in the Universe) is required \cite{ReviewParticlePhysics2020, PlanckCollab2020VI}. However, the model does not explain the origin of this exotic matter.

\item  \textbf{The Dark Energy problem}. Dark Energy is how we call the unknown form of energy that is responsible for the accelerated expansion of the Universe. According to observations, Dark Energy behaves as a perfect barotropic fluid ($p=w\rho$) with $w=-1.03\pm 0.03$ \cite{PlanckCollab2020VI}. Although we do not know exactly the nature of this energy, this value is consistent with $\Lambda$CDM, which assumes a cosmological constant $\Lambda$ (i.e. $w=-1$). 

\item \textbf{The Hubble tension} is a disagreement between the value of the Hubble parameter (the expansion rate of the Universe) obtained by the Planck Collaboration from the Cosmic Microwave Background (CMB) and $\Lambda$CDM \cite{PlanckCollab2020VI} and the one coming from late-time cosmological observations \cite{RiessCasertanoYuan2018}, respectively:
\begin{equation}
  \HubH_0 = (67.4\pm 0.5)\ {\rm km}/{\rm s}/{\rm Mpc},\qquad\qquad  \HubH_0 = (73.52\pm 1.62)\ {\rm km}/{\rm s}/{\rm Mpc}\,.
\end{equation}
\end{itemize}

In addition to these observational problems, there are also some theoretical ones in GR, such as:
\begin{itemize}
\item \textbf{Singularities}. The theory predicts singularities, e.g. in the interior of black holes and in cosmological solutions.

\item \textbf{The cosmological constant problem}. The problem with invoking a cosmological constant $\Lambda$ to describe Dark Energy is that $\Lambda$ suffers from a naturalness problem. Basically the idea is that the quantum corrections to the vacuum energy are many orders of magnitude (at least 40) higher than the observed valued \cite{Weinberg2000, Martin2012}.

\item \textbf{Renormalizability and unitarity}. This is considered as one of the most important problems in quantum gravity. GR is known to be non-renormalizable by power counting \cite{tHooftVeltman1974}. One attempt to solve this was \emph{quadratic gravity}, which makes use of the Weyl tensor (conformally invariant) to construct a quadratic Lagrangian which is renormalizable. However, the theory contains a violation of unitarity (see e.g. \cite{Stelle1977, JulveTonin1978}) due to a massive spin-2 ghost.\footnote{
    A ghost is a type of unstable field whose modes carry negative kinetic energy. We will see more details on this in Chapter \ref{ch:gravityviability}.}

\end{itemize}

~

In the last decades there have been many attempts to modify and/or extend the framework of General Relativity. On the one hand, modified theories of gravity usually propose alternative corrections to GR that presumably will appear before reaching the Planck scale and that could contribute to solve some of these problems. On the other hand, the philosophy behind most of them is also to immerse GR within a larger family of theories, so that its actual status can be contrasted with other alternatives. In principle, modified theories of gravity are not intended to be final theories, so solving the renormalizability problem is not generally perceived as a goal. On the contrary, they search for alternative perspectives to attack e.g. the cosmological problems by enriching gravity with extra structure (mainly extra fields).\footnote{
    In this thesis we focus on bottom-up approaches, i.e. we move from low to high energies. It is however worth mentioning the outstanding status of \emph{String Theory} and \emph{Loop Quantum Gravity}, which are the most relevant formulations that regularize the behavior of gravity directly at high energies (top-down approaches). Successful theories in these two regimes should of course match appropriately.}

Nonetheless, modifying GR is not an easy task. When constructing or proposing a new theory one could start computing observables and comparing with experiments to set bounds on the parameters of the theory. However, one should not directly trust any claim derived from the theory. It is important to be careful and rigorous about how reliable are the results and under what hypothesis. Moreover, in many cases, the problems can be hidden under the bed, waiting patiently. The key point is that, even before contrasting it with experiments, there are some theoretical tests that the theory should pass; some of them are just a matter of mathematical consistency, and others are connected, at the end of the day, with observable effects. For instance, if we propose a field theory in the usual sense, it is crucial to check which are the degrees of freedom of that theory and if they exhibit any kind of pathological behavior. If so, one should analyze whether the region (in phase space or solution space) where this happens can be discarded by physical arguments or if the pathologies cannot be reached by any configuration of the system (in a reasonable time scale), etc. The structure of GR is robust from a field-theoretical point of view and it is not easy to modify it without introducing pathologies. 

\section{The landscape of modified theories of gravity}\label{sec:landscapeMTG}

There are many approaches to modified gravity. In this section we will present and comment on some of them. First, let us briefly characterize GR. From the very fundamental level, gravity is described as the 4-dimensional relativistic theory of a massless spin-2 particle.\footnote{
    Let us recall the reasons for this (we follow the Feynman Lectures \cite{Feynman2018GravLec}). The particle must be massless because gravity is a long-range interaction. The particle must have integer spin to produce a static force and not just scattering. Since it is universally attractive the spin must be even ($s=0,2,4...$). Spin-0 particles do not produce light deflection, whereas spins higher than 2 are constrained by several no-go theorems (e.g., Weinberg-Witten theorem). This leads to $s=2$ as unique choice.} 
The massless spin-2 irreducible representation of the Poincaré group correspond to a particle with two degrees of freedom (helicities). From a the field-theoretical point of view, the dynamics of the graviton can be described by a symmetric tensor field $h_{\mu\nu}$ subjected to the dynamics of the (massless) \emph{Fierz-Pauli Lagrangian},
\begin{equation}
  \mathcal{L}_\mathrm{FP} =\tfrac{1}{2}\partial_{\mu}h_{\nu\rho}\partial^{\mu}h^{\nu\rho}-\partial_{\rho}h{}^{\rho\mu}\partial_{\sigma}h^{\sigma}{}_{\mu}+\partial_{\sigma}h^{\sigma}{}_{\mu}\partial^{\mu}h-\tfrac{1}{2}\partial_{\mu}h\partial^{\mu}h \,,
\end{equation}
which admits GR as non-linear extension.\footnote{See e.g. the Deser's argument in \cite{Deser1970, Ortin2004}.} 
Having these ideas in mind, one way of modifying GR is by relaxing some of its basic principles:
\begin{itemize}
\item {\bf Massive gravity}. These theories describe a gravity framework where the propagating graviton is massive. In the original formulation, the theory did not recover the GR prediction in the massless limit at linear order (vDVZ discontinuity) and, additionally, the non-linear corrections develop an instability called the Boulware-Deser ghost \cite{BoulwareDeser1972}. The first one was solved in \cite{Vainshtein1972}, whereas the ghost can be eliminated by a non-linear completion called \emph{dRGT massive gravity} \cite{deRhamGabadadze2010, deRhamGabadadze2011} (see e.g. the reviews \cite{Hinterbichler2012,Rham2014}). The mass of the graviton in this theory is highly constrained \cite{deRhamDeskins2017}.

\item {\bf Lorentz violation}. Frameworks have been formulated where Lorentz invariance is violated at a very fundamental level. Examples of this are: \emph{Einstein-Aether theory} \cite{Gasperini1987} in which, together with the metric, there exists a timelike vector field (\emph{aether}) that provides a preferred reference frame; or \emph{Hořava--Lifshitz gravity} \cite{Horava2009}, in which the time is treated in a separated way (with respect to the space), and the relativistic notion of time emerges at large distances. 

\item {\bf Higher dimensions}. For instance, one can also develop geometry and field theory in higher dimensions and study the effective theory in a particular submanifold (\emph{braneworld}) of the bulk space. Some examples are the Randall-Sundrum model \cite{RandallSundrum1999} and the DGP model \cite{DvaliGabadadzePorrati2000}.

\end{itemize}

However, probably the most common approach to extend gravity is by adding new degrees of freedom:
\begin{itemize}
\item {\bf Scalar-tensor theories}.
  In addition to the metric, these theories consider an extra propagating scalar as a basic element to construct the action. One example is \emph{Brans-Dicke theory} \cite{BransDicke1961},
    \[ S_{\text{Brans-Dicke}}[g_{\mu\nu},\,\phi]=\frac{1}{2\kappa}\int\mathrm{d}^{4}x\sqrt{|g|}\left(\phi\mathring{R}-\frac{\omega_\mathrm{BD}}{\phi}\partial_{\mu}\phi\partial^{\mu}\phi\right)+\text{matter} 
    \]
    where the presence of the scalar field gives rise to a point-dependent gravitational coupling. This theory is part of a very important family of scalar-tensor theories, \emph{Horndeski gravity}. This was formulated in \cite{Horndeski1974}, as the most general scalar-tensor theory (in 4 dimensions) with at most second derivatives of the scalar field that gives second-order equations of motion (see  \cite{DeffayetDeser2009, DeffayetEsposito2009, NicolisRattazzi2009}). The theory has many applications in cosmology, specially to tackle the inflation and the Dark Energy problems \cite{CliftonFerreira2012, Kobayashi2011}. Theories with higher derivatives contain in principle Ostrogradski ghosts (a type of instability), but they can be avoided by forcing the Lagrangian to be degenerate. This gives rise to theories \emph{beyond} Horndeski \cite{GleyzesLanglois2015, GleyzesLanglois2015b, LangloisNoui2016, MotohashiNoui2016, MotohashiSuyama2018, BenAchour2016, KleinRoest2016, MotohashiSuyama2015, Zumalacarregui2013}. Scalar-tensor theories have received important constraints from the LIGO and Virgo observations \cite{SaksteinJain2017, Ezquiaga2017, BakerBellini2017}.

\item {\bf Vector-tensor theories}. Similar to the previous ones but now an extra vector variable is considered. A massless ${\rm U}(1)$ gauge vector does not allow cosmological scenarios (homogeneous and isotropic) \cite{Heisenberg2018}. There are different ways to circumvent this, e.g. by going to non-abelian groups \cite{Maleknejad2013} or by breaking the ${\rm U}(1)$ symmetry with a mass term. The latter approach led to \emph{generalized Proca} theory \cite{Heisenberg2014,Tasinato2014,BeltranHeisenberg2016} (see also \cite{BeltranHeisenbergKoivisto2016}). The case with several vectors has also been explored \cite{BeltranHeisenberg2017}.
\item {\bf Tensor-tensor theories}. These theories are characterized for having additional dynamical metrics. For instance, one can add a kinetic term for the extra metric  $f_{\mu\nu}$ in massive gravity (dRGT), and the result is known as \emph{bigravity} \cite{HassanRosen2012}. Again, GW observations put very strong constraints on these theories.
\end{itemize}

It is also possible to think of starting from the Einstein-Hilbert action and trying to generalize it. Nevertheless, most of the resulting theories can be seen as particular cases of those mentioned above. Let us describe some examples:
\begin{itemize}
\item $f(R)$\emph{-gravity}. This theory considers as Lagrangian an arbitrary function of the Ricci scalar and was thought to tackle the Dark Matter and Dark Energy problems. If we consider a non-trivial case $f''(\mathring{R})\neq0$ (i.e. when it is different from the Einstein-Hilbert action), there is a field redefinition that turns the action into the (Jordan frame of a) Brans-Dicke theory with $\omega_\mathrm{BD}=0$,
\begin{equation}
  \frac{1}{2\kappa}\int\mathrm{d}^{4}x\sqrt{|g|}\left(\phi\mathring{R}-V(\phi)\right)+\text{matter}\,, 
\end{equation}
such that, on-shell, the scalar is given by $\phi=f'(\mathring{R})$ (see for example \cite{Chiba2003}). Note that it includes a potential for the Brans-Dicke scalar that has a purely gravitational origin \cite{DeFeliceTsujikawa2010,SotiriouFaraoni2010}, and which did not appear in the original formulation of the Brans-Dicke theory \cite{BransDicke1961}.\footnote{
    The Palatini version of this theory, i.e. treating the connection as an independent field, can also be rewritten as a Brans-Dicke theory with $\omega_\mathrm{BD}=-3/2$. Interestingly, for this particular value of the parameter the scalar turns out to be non-dynamical \cite{OlmoSanchis2011}.}
Consequently, these theories essentially propagate a graviton plus a non-minimally coupled scalar. For more information and applications see \cite{DeFeliceTsujikawa2010, Olmo2011,SotiriouFaraoni2010,Sotiriou2007e, OlmoRubieraWojnar2019, Koivisto2010}. 

\item Another theory that has been explored is $f(G)$\emph{-gravity}, whose Lagrangian is an arbitrary function of the Gauss-Bonnet invariant (see e.g. \cite{NojiriOdintsov2005, CognolaElizalde2006,DeFeliceTsujikawa2009}).

\item \emph{Ricci-based gravity}. The Lagrangian is now an arbitrary function of the symmetric part of the Ricci tensor for an arbitrary connection (they are formulated \`a la Palatini). In \cite{AfonsoBejarano2017} it is shown that if $Z^{\mu\nu}=\frac{\partial\mathcal{L}}{\partial R_{\mu\nu}}(=Z^{\nu\mu})$ admits an inverse tensor, $(Z^{-1})_{\mu\nu}$, the general solution of the connection equation is the Levi-Civita connection associated to the metric $q_{\mu\nu}\coloneqq(Z^{-1})_{\mu\nu}$ up to a projective mode that can be removed with an appropriate gauge choice. Indeed, the theory admits a field redefinition to the Einstein frame of $q_{\mu\nu}$ (i.e. an Einstein-Hilbert action with  $q_{\mu\nu}$ playing the role of the metric). In the presence of matter, this redefinition complicates the matter sector but, if the equations allow some invertibility, there exists a mapping between the solutions of Ricci-based gravity with certain matter and those of GR with another matter Lagrangian \cite{AfonsoOlmoRubiera2018, AfonsoOlmoOraziRubiera2019, AfonsoOlmoOraziRubiera2018,DelhomOlmoOrazi2019}. One relevant sub-case is \emph{Born-Infeld inspired gravity} (see \cite{BeltranHeisenbergOlmoRubiera2018, PaniDelsateCardoso2012}) that, as the name suggests, was motivated by the Born-Infeld model for non-linear electromagnetism \cite{BornInfeld1934}. In these theories it is possible to construct stable wormhole solutions that extend the interior of Schwarzschild and Reissner-Nordstr\"om black holes, i.e. avoiding the singularity problem. The stability of the theory and its generalizations is discussed in \cite{BeltranDelhom2019,BeltranDelhom2020}
\end{itemize}

Finally, let us also mention \emph{unimodular gravity} \cite{AlvarezBlasGarriga2006, AlvarezHerrero2012}. This theory is a slight modification of GR constructed as the Einstein-Hilbert action but subjected to the constraint that (the absolute value of) the determinant of the metric must be 1. One of the interesting features of it is that it solves the naturalness problem of the cosmological constant value \cite{Smolin2009}.

We have performed a general overview of gravity theories. In the next section, we will focus on the metric-affine framework and theories therein, which will be very relevant in this thesis.

\section{The metric-affine framework and MAG}

In the original Palatini formalism, the connection is used as a tool to obtain modified equations of motion for the metric. In fact, not much attention is paid to its physical meaning. On the contrary, in what we are going to call the \emph{metric-affine framework}, the connection is treated as another geometrical field with its own dynamics and physical implications. In the presence of a metric, an arbitrary connection is characterized by two tensorial quantities, the torsion and the nonmetricity, which are vanishing in the Levi-Civita case. These two tensors will be the basic blocks to construct theories in this framework.

The gauge principle is one of the cornerstones of our current understanding of fundamental physical interactions. Inspired by \cite{Weyl1929}, in which Weyl establishes the basis of the gauge procedure, as well as the subsequent works by Yang, Mills \cite{YangMills1954} and Utiyama  \cite{Utiyama1956}, one can try to extend such gauge formalism to spacetime symmetries. Kibble and Sciama \cite{Kibble1961, Sciama1962,Sciama1963} formulated the gauge theory of the Poincar\'e group, from which a connection with nontrivial torsion (but zero nonmetricity) emerges. The resulting theory is called \emph{Poincar\'e Gauge gravity} (PG) (see e.g. \cite{Hehl1979a, Obukhov2006b, Blagojevic2001}). 
At this point, a natural question is: is it possible to extend this formalism in order to get a completely general connection, with nontrivial torsion and nonmetricity? The answer is yes and the key will be to extend the Poincar\'e group to the full affine group.\footnote{
    The affine group contains the translations together with all possible basis transformations.} 
This gives rise to \emph{Metric Affine Gauge gravity} (MAG) \cite{Hehl1995}. As in any other gauge theory, there are intrinsic properties of the matter (charges) that couple to the gauge fields. In MAG, the spin density, the dilation and the shear currents enter the game as new fundamental properties of the matter associated to the dynamics of the connection \cite{Hehl1976a,Hehl1976c,ObukhovTresguerres1993}, in addition to the energy-momentum tensor. This formulation and its viability as a quantum gravity model has been discussed in \cite{Lee1992,LeeNeeman1990, PaganiPercacci2015, PercacciSezgin2020, Percacci2020}. For a revision of these and other gauge theories of gravity (involving other groups, such as the conformal group), see \cite{BlagojevicHehl2012} and references therein. 

It is worth remarking that apart from MAG and PG gravity, many theories have been formulated considering an additional connection with certain properties or restrictions. Examples are the already mentioned Ricci-Based gravity, and the teleparallel equivalents and their generalizations \cite{BeltranHeisenbergKoivisto2018a, BeltranHeisenbergKoivisto2018b, AldrovandiPereira2012, BeltranDialektopoulos2020, KoivistoTsimperis2018, KrssakHoogenPereira2018, BeltranHeisenbergJCA2020, BeltranHeisenbergKoivisto2019, HohmannJarvKrssakPfeifer2019, GolovnevKoivisto2018}.\footnote{Teleparallel means that the curvature of the connection is identically zero. These theories will be discussed in more detail in Chapter \ref{ch:QTG}.}

Finally, it is worth noticing that the formulation of theories within the metric-affine framework has a very interesting motivation. The deformations of an ideal crystal can be described by the same techniques used in GR, i.e. with a metric and its associated curvature. However, real crystals present local defects such as dislocations or point defects and, to describe them, other structures beyond the metric are needed. In particular, to deal with the previously mentioned defects, respectively torsion and nonmetricity are invoked (see e.g. \cite{Falk1981, KupfermanMaor2015, KupfermanMaor2017}; see also \cite{Hehl1976a, Hehl2007, GronwaldHehl1991} and references therein). Assuming that some of these theories turn out to be viable gravitational theories, this analogy suggests that the connection is indeed encoding information about microscopic defects in the very fabric of the spacetime that a metric tensor cannot reproduce. Of course, this should be, at the end of the day, in total agreement with some established quantum gravity framework.

\section{Last comments before starting}

This thesis contains both mathematical and physical results and is an invitation to enjoy both: 
\begin{itemize}
  \item The exploration of theories, looking for interesting mathematical properties and structures that can change our way to see the world.
  \item Being extremely critical with any proposed theory (in a constructive sense). Learning about singularities and pathologies is another way to learn about Nature and how limited is our knowledge of it.
\end{itemize}
The PhD program is called ``Physics and mathematics'', so I felt free to sail between these two currents: the first half (approx.) of the thesis is of a more mathematical nature, whereas the second one is more physically focused. It is also worth mentioning that the derivations and computations of this manuscript are essentially classical (classical solutions, analysis of stability from the classical Lagrangian...). Going beyond that can be one of the next steps, but one can imagine how difficult that could be just by looking at what we know of quantum gravity models. Also one  can explore the literature about semiclassical gravity (quantum matter in classical GR gravity) and see that most of the computations are extraordinarily difficult to perform (they require numerical analysis) and are plagued with ambiguities. For these reasons, we found it convenient to first focus on understanding these theories at the classical level.\footnote{One interesting point here could be ``why should we start with a classical Lagrangian?'' (or even further: ``do we need a Lagrangian?''), but this is of course a philosophical debate far beyond the goal of this thesis.} Finally, just mentioning that for the realization of the calculations we made use of the xAct package (Wolfram Mathematica) \cite{xAct}.

\part{Geometry and metric-affine theories \label{part:foundations}}

\chapter{Mathematical tools for metric-affine gravity\label{ch:mathtools}}

\boxquote{If I were again beginning my studies, I would follow the advice of Plato and start with mathematics.}{Galileo Galilei}

This chapter provides an extensive and complete collection of the mathematical concepts and useful tools to work in the metric-affine framework. We start from the very beginning with the definition of manifold and then we focus on deriving practical objects and expressions. The idea of this chapter is essentially to give a self-contained character to the full manuscript and also to be an adequate mathematical starting point for those interested in metric-affine gravity.\footnote{Some important references for this chapter are \cite{SanchezCaja2012,Ortin2004} (see also \cite{Nakahara2003,KobayashiNomizu1963}).}

\section{Basic differential geometry. Manifolds and tensors}

\subsection{Differentiable manifolds and bundles}

\boxdefinition{
\begin{defn}
\textbf{(Topological manifold)} A $\dimM$\emph{-dimensional topological manifold} $\mathcal{M}$ is a topological space that satisfies the following requirements: (1) Hausdorff, (2) second-countable and (3) it is locally homeomorphic to the Euclidean space $\mathbb{R}^{\dimM}$.
\end{defn}
}

The third point means that for each point $p\in\mathcal{M}$, there is an open \foreignlanguage{british}{neighbourhood} $\mathcal{U}$ and a homeomorphism\footnote{A bijective continuous map between topological spaces whose inverse is also continuous.} $\varphi$  between $\mathcal{U}$ and some open set of $\mathbb{R}^{\dimM}$. The pair $(\mathcal{U},\,\varphi)$ is called a \emph{chart around $p$}, and the component maps $x^{\mu}\,:\,\mathcal{U}\rightarrow\mathbb{R}$ ($\mu=1,...,\dimM$) of $\varphi$  are called \emph{coordinate functions}. A set of charts that covers the whole manifold is an \emph{atlas}.

If we consider two charts $(\mathcal{U},\,\varphi)$ and $(\mathcal{V},\,\psi)$ around the same point $p$, the \emph{transition functions} are the maps representing the coordinate transformation between them, i.e.
\begin{equation}
\psi\circ\varphi^{-1}\,:\,\varphi(\mathcal{U}\cap \mathcal{V})\subset\mathbb{R}^{\dimM}\longrightarrow\psi(\mathcal{U}\cap \mathcal{V})\subset\mathbb{R}^{\dimM}\,.
\end{equation}
An atlas containing all possible charts with $C^{\infty}$-differentiable transition functions between them is called a \emph{$C^{\infty}$-differentiable structure}. 

\boxdefinition{
\begin{defn}
\textbf{(Smooth manifold)} A $\dimM$\emph{-dimensional smooth manifold} is a pair $(\mathcal{M},\,\mathfrak{D})$ where $\mathcal{M}$ is a $\dimM$-dimensional topological manifold and $\mathfrak{D}$ is a $C^{\infty}$-differentiable structure on it.
\end{defn}
}

\newpage
\boxdefinition{
\begin{defn}
\textbf{(Fiber bundle)} A \emph{fiber bundle} is a set $(\mathcal{B},\mathcal{M},\pi,\mathcal{F})$ where $\mathcal{B}$, $\mathcal{M}$ and $\mathcal{F}$ are topological manifolds and $\pi$ is a continuous surjective map 
\begin{equation}
\pi\,:\,\mathcal{B}\to\text{\ensuremath{\mathcal{M}}}\,,
\end{equation}
which satisfies the following local triviality condition: for each $p\in\mathcal{M}$ there exists an open neighborhood of it $\mathcal{U}$ and a homeomorphism $\Phi\,:\,\pi^{-1}(\mathcal{U})\to\text{\ensuremath{\mathcal{U}}}\times\text{\ensuremath{\mathcal{F}}}$ such that
\begin{equation}
\pi=\mathrm{proj}_1\circ\Phi\,,
\end{equation}
where $\mathrm{proj}_{1}$ is the projection onto the first factor, i.e. $\mathrm{proj}_{1}(a,b)=a$.
\end{defn}
}
$\mathcal{B}$ is the \emph{total space}, $\mathcal{M}$ the \emph{base manifold}, $\mathcal{F}$ the \emph{(abstract) fiber} and $\pi$ the \emph{projection map} of the fiber bundle. 
For each point $p\in\mathcal{M}$, the set $\pi^{-1}(p)$ is homeomorphic to $\mathcal{F}$ and is called the \emph{fiber over $p$}. 
For a given covering $\{\mathcal{U}_{i}\}$ of $\mathcal{M}$, one can take the set $\{(\mathcal{U}_{i},\,\Phi_{i})\}$ where $\Phi_{i}\,:\,\pi^{-1}(\mathcal{U}_{i})\to\text{\ensuremath{\mathcal{U}}}_{i}\times\text{\ensuremath{\mathcal{F}}}$ are the maps we previously introduced. The pair $(\mathcal{U}_{i},\,\Phi_{i})$ is called a \emph{local trivialization over} $\mathcal{U}_{i}$.

The preimage by $\pi$ of any open set  $\mathcal{U} \subset\mathcal{M}$ is homeomorphic to a product space  $\mathcal{U} \times \mathcal{F}$, so the intuitive idea is that $\mathcal{B}$ is locally the base manifold $\mathcal{M}$ with a ``copy'' of $\mathcal{F}$ attached to each point, but not globally. We will make an abuse of language and refer to the total space as the entire bundle and use the abbreviations $(\mathcal{B},\mathcal{M},\pi,\mathcal{F})\equiv\mathcal{B}\xrightarrow{\pi}\mathcal{M}\equiv\mathcal{B}$ when the other elements are clear.

\boxdefinition{
\begin{defn}
\textbf{(Sections on fiber bundles)} Let $(\mathcal{B},\mathcal{M},\pi,\mathcal{F})$ be a fiber bundle. A \emph{local section} around the point $p\in\mathcal{M}$ (consider some coordinate neighbourhood $\mathcal{U}$ of it) is a continuous map $\sigma\,:\,\mathcal{U}\to\mathcal{B}$ such that $\pi\circ\sigma$ is the identity map.

A \emph{(global) section} of a fiber bundle is a continuous map $\sigma\,:\,\mathcal{M}\to\mathcal{B}$ such that $\pi\circ\sigma$ is the identity map\emph{.}

The set of all global sections is represented as  $\Gamma(\mathcal{B})$.
\end{defn}
}

Intuitively, a section associates to each point of its domain (within the base manifold) an element of the fiber over it, but in a continuous way. So, it is like if we were ``lifting'' part of the base manifold to the total space (keeping each point in its own fiber).\footnote{
    The name ``section'' comes from the idea that after applying the section to its whole domain the image looks as if we were ``cutting'' the total space transversely to the fibers. }

A very special case of manifolds are \emph{Lie groups}, i.e. groups which also carry a manifold structure. With them we can introduce the following concept, extremely important in gauge theories:

\boxdefinition{
\begin{defn}\label{def:princbundle}
\textbf{(Principal fiber bundle)} Let $\mathcal{G}$ be a Lie group. A \emph{principal fiber bundle with structure group} $\mathcal{G}$ is a fiber bundle $\mathcal{P}\xrightarrow{\pi}\mathcal{M}$ equipped with a right action of $\mathcal{G}$ on $\mathcal{P}$ which preserves the fibers and acts freely and transitively on them.
\end{defn}
}

Locally, a principal fiber bundle looks as the product space $\mathcal{M}\times\mathcal{G}$, but not globally. 

\newpage
\subsection{Tangent and cotangent space}

Let $(\mathcal{M},\,\mathfrak{D})$ be a $\dimM$-dimensional smooth manifold, $p$ an arbitrary point of $\mathcal{M}$ and $(U,\,\varphi=(x^{\mu}))\in\mathfrak{D}$  a chart around $p$.

\boxdefinition{
\begin{defn}
\textbf{\label{Def: diff function}(Differentiable function on a manifold)} A map $f\,:\,\mathcal{M}\rightarrow\mathbb{R}$ is a\emph{ $C^{k}$-differentiable function over $\mathcal{M}$} if $f\circ\varphi^{-1}\,:\,\mathbb{R}^{\dimM}\rightarrow\mathbb{R}$ is \emph{$C^{k}$}-differentiable. 

$C^{k}(\mathcal{M})$ is the set of all \emph{$C^{k}$}-differentiable functions over the manifold.
\end{defn}
}

We are going to focus on $C^{\infty}(\mathcal{M})$ from now on. Notice that the function $f\circ\varphi^{-1}$ admits partial derivatives on $\mathbb{R}^{\dimM}$, so we can extend the notion of partial derivative to the manifold as follows
\begin{equation}
\frac{\partial f}{\partial x^{\mu}}(p)\coloneqq\frac{\partial(f\circ\varphi^{-1})}{\partial z^{\mu}}(\varphi(p))\,.
\end{equation}
where $\partial/\partial z^{\mu}$ is the usual partial derivative in $\mathbb{R}^{\dimM}$ with respect to the $\mu$-th variable. Obviously this result is chart-dependent, but when working with a fixed set of coordinates we will use the compact notation $\partial_{\mu}f(p)$. 

For a given chart, consider the set of differentiable operators
\begin{eqnarray}
\frac{\partial}{\partial x^{\mu}}\,:\,C^{\infty}(\mathcal{M}) & \longrightarrow & C^{\infty}(\mathcal{M})\nonumber \\
f & \longmapsto & \frac{\partial f}{\partial x^{\mu}}\,,
\end{eqnarray}
so we can define:

\boxdefinition{
\begin{defn}
\textbf{(Tangent space)} For a given point $p\in\mathcal{M}$, we define the \emph{tangent space} at it, $T_{p}\mathcal{M}$ as the vector space (isomorphic to $\mathbb{R}^{\dimM}$) generated by the previous operators, i.e.
\begin{equation}
T_{p}\mathcal{M}\coloneqq\mathrm{span}\left\{ \frac{\partial}{\partial x^{\mu}}|_{p}\,:\,\mu=1,\,...,\,\dimM\right\} \,.
\end{equation}
\end{defn}
}

The vector $\frac{\partial}{\partial x^{\mu}}|_{p}\equiv\vpartial_{\mu}|_{p}$ points in the direction in which the $x^{\mu}$ coordinate grows. These derivatives depend on the chart, but the whole tangent space does not. An arbitrary vector in the tangent space $\vecv=v^{\mu}\vpartial_{\mu}|_{p}$ ($v^{\mu}\in\mathbb{R}$) acts on functions $f\in C^{\infty}(\mathcal{M})$
as $\vecv(f)=v^{\mu}\partial_{\mu}f(p)$.

If we consider the disjoint union of all tangent spaces, we get the \emph{tangent bundle} 
\begin{equation}
T\mathcal{M}\coloneqq\bigsqcup_{p\in\mathcal{M}}T_{p}\mathcal{M}\,,
\end{equation}
whose fibers are the tangent spaces $T_{p}\mathcal{M}=\pi^{-1}(p)$. The underlying bundle structure is $\left(T\mathcal{M},\mathcal{M},\mathrm{proj}_1,\mathbb{R}^{\dimM}\right)$, with total space $T\mathcal{M}$.

Once we have the notion of tangent space, i.e., vectors in one particular point $p\in\mathcal{M}$, it is natural to built, by duality, the set of \emph{1-forms} on $p$. The dual basis of $\{\vpartial_{\mu}|_{p}\}$ will be denoted as $\dex x^{\mu}|_{p}$, so the duality relation is $\dex x^{\mu}(\vpartial_{\nu})=\delta_{\nu}^{\mu}$. We can then define:\footnote{
    However, the cotangent space $T_{p}^{*}\mathcal{M}$ can be defined independently of the tangent space, as the quotient vector space $I_{p}/I_{p}^{2}$ where $I_{p}$ is the ideal of functions vanishing at $p$ and $I_{p}^{2}$ the set of functions of the form $\sum_{i}f_{i}g_{i}$ with $f_{i},\,g_{i}\in I_{p}$ (notice that both $I_{p}$ and $I_{p}^{2}$ are vector spaces). The intuition behind this definition is that the space of 1-forms on $p$ can be seen as the set of all possible first-order behaviors that the functions can have around $p$ (in the sense of the Taylor expansion).}
\boxdefinition{
\begin{defn}
\textbf{(Cotangent space)} For a given point $p\in\mathcal{M}$, we define the \emph{cotangent space} at it $T_{p}^{*}\mathcal{M}$ as the dual vector space of the tangent space $T_{p}\mathcal{M}$, i.e. 
\begin{equation}
T_{p}^{*}\mathcal{M}\coloneqq(T_{p}\mathcal{M})^{*}=\mathrm{span}\left\{ \dex x^{\mu}|_{p}\,:\,\mu=1,\,...,\,\dimM\right\} \,.
\end{equation}
\end{defn}
}

Similarly, we can also introduce the notion of \emph{cotangent bundle,}
\begin{equation}
T^{*}\mathcal{M}\coloneqq\bigsqcup_{p\in\mathcal{M}}T_{p}^{*}\mathcal{M}\,,
\end{equation}
which has the bundle structure $\left(T^{*}\mathcal{M},\mathcal{M},\mathrm{proj}_1,\mathbb{R}^{\dimM}\right)$.

\subsection{Natural tensor bundles and differential forms}

Once we have vectors and 1-forms, it is natural to build other kinds of objects in each point of the manifold. For instance, by taking the tensor product $\otimes$ of the tangent space and the cotangent space several times we can generate tensor spaces:

\boxdefinition{
\begin{defn}
\textbf{($(r,s)$-tensor space)} For a given point $p\in\mathcal{M}$, we define the $(r,s)$\emph{-tensor space} on it $T_{p}^{(r,\,s)}\mathcal{M}$ as the vector space given by
\begin{equation}
T_{p}^{(r,\,s)}\mathcal{M}\coloneqq T_{p}\mathcal{M}\otimes\overset{r\,\text{times}}{\ldots}\otimes T_{p}\mathcal{M}\otimes T_{p}^{*}\mathcal{M}\otimes\overset{s\,\text{times}}{\ldots}\otimes T_{p}^{*}\mathcal{M}\,.
\end{equation}
Or, in terms on a particular chart $x^{\mu}$,
\begin{align}
T_{p}^{(r,\,s)}\mathcal{M} & \coloneqq\mathrm{span}\Big\{\vpartial_{\mu_1}|_{p}\otimes\overset{r\,\text{times}}{\ldots}\otimes\vpartial_{\mu_{r}}|_{p}\otimes\dex x^{\nu_1}|_{p}\otimes\overset{s\,\text{times}}{\ldots}\otimes\dex x^{\nu_{s}}|_{p}\nonumber \\
 & \qquad\qquad\qquad\,:\,\,\mu_1,...,\mu_{r},\nu_1,...,\nu_{s}=1,\,...,\,\dimM\Big\}\,.
\end{align}
\end{defn}
}

The elements of the tensor space $T_{p}^{(r,\,s)}\mathcal{M}$ are called $(r,\,s)$-tensors or $r$-contraviant $s$-covariant tensors on $p$. Consider an element $\tent\in T_{p}^{(r,\,s)}\mathcal{M}$. Once we have a coordinate basis we can just work with the components of the tensor:
\begin{equation}
\tent=t^{\mu_{1}...\mu_{r}}{}_{\nu_{1}...\nu_{s}}\vpartial_{\mu_1}|_{p}\otimes\overset{r\,\text{times}}{\ldots}\otimes\vpartial_{\mu_{r}}|_{p}\otimes\dex x^{\nu_1}|_{p}\otimes\overset{s\,\text{times}}{\ldots}\otimes\dex x^{\nu_{s}}|_{p}\qquad\to\qquad t^{\mu_{1}...\mu_{r}}{}_{\nu_{1}...\nu_{s}}\in\mathbb{R}\,.
\end{equation}
We recall now a few basic concepts related to tensors:
\begin{itemize}
\item We say that a tensor is \emph{symmetric} (or \emph{antisymmetric}) \emph{in a pair of indices} $\mu\nu$ if, $t^{...\mu...\nu...}=t^{...\nu...\mu...}$ (respectively, $t^{...\mu...\nu...}=-t^{...\nu...\mu...}$). 
\item In a tensor expression, \emph{contracting} two indices means to make them equal in the Einstein summation notation, i.e. to sum in all their possible values. This decreases in 1 unit the covariance and the contravariance of the tensor
\begin{align}
T_{p}^{(3,\,2)}\mathcal{M}\ni\qquad t^{\mu_1\mu_2\mu_3}{}_{\nu_1\nu_2}\nonumber \\
\downarrow\,\,\,\,\,\,\,\nonumber \\
\text{Contraction}_{1}^{2}(t^{\mu_1\mu_2\mu_3}{}_{\nu_1\nu_2}) &= \sum_{\mu_{2}=\nu_{1}=1}^{\dimM}t^{\mu_1\mu_2\mu_3}{}_{\nu_1\nu_2}=t^{\mu_1\sigma\mu_3}{}_{\sigma\nu_2}\in T_{p}^{(2,\,1)}\mathcal{M}\,.
\end{align}
\item In this component notation the product of components is directly associated with the components of the tensor product, i.e. 
\begin{equation}
t^{\mu_1\mu_2}{}_{\nu_1\nu_2}\in T_{p}^{(2,\,2)}\mathcal{M}\,,~~ h^{\mu_1\mu_2\mu_3}{}_{\nu_1}\in T_{p}^{(3,\,1)}\mathcal{M}\quad\to\quad t^{\mu_1\mu_2}{}_{\nu_1\nu_2}h^{\mu_{3}\mu_{4}\mu_5}{}_{\nu_3}\in T_{p}^{(5,\,3)}\mathcal{M}\,.
\end{equation}
\end{itemize}
Special attention must be paid to the totally antisymmetric covariant tensors which are called \emph{differential forms}.

\boxdefinition{
\begin{defn}
\textbf{($k$-forms space)} For a given point $p\in\mathcal{M}$, we define the space of $k$-forms (or \emph{differential forms of rank $k$}) on it $\Lambda_{p}^{k}\mathcal{M}$ as the subspace of $T_{p}^{(0,\,k)}\mathcal{M}$ of totally antisymmetric tensors.
\end{defn}
}

The relevance of these objects will be clear later when we talk about integration and volume forms. As we did with the tangent and cotangent spaces, we can also introduce the corresponding bundles of tensors and $k$-forms:
\begin{equation}
T^{(r,\,s)}\mathcal{M}\coloneqq\bigsqcup_{p\in\mathcal{M}}T_{p}^{(r,\,s)}\mathcal{M}\,,\qquad\Lambda^k\mathcal{M}\coloneqq\bigsqcup_{p\in\mathcal{M}}\Lambda_p^k\mathcal{M}\,.
\end{equation}

\subsection{Diffeomorphisms, fields and transformation rules}

\boxdefinition{

Let $\mathcal{M}_{1}$ and $\mathcal{M}_{2}$ be two smooth manifolds (we omit the differentiable structure from now on) of dimensions $\dimM_{1}$ and $\dimM_{2}$, and two arbitrary open subsets of them, respectively, $\mathcal{V}_{1}$ and $\mathcal{V}_{2}$.
\begin{defn}
\textbf{(Differentiable map)} Consider a map $F\,:\,\mathcal{V}_1\rightarrow\mathcal{V}_{2}$. We say that $F$ is a\emph{ $C^{k}$-differentiable map }if $\forall p\in\mathcal{V}_{1}$ we can find some coordinate chart around it $(\mathcal{U},\,\varphi)$ and a map $\psi$ defining a chart $(F(\mathcal{U}),\psi)$ around the image point $F(p)$, such that $\psi\circ F\circ\varphi^{-1}\,:\,\mathbb{R}^{\dimM_1}\rightarrow\mathbb{R}^{\dimM_2}$ is \emph{$C^{k}$}-differentiable (in the usual sense). 

The set of differentiable maps between $\mathcal{V}_{1}$ and $\mathcal{V}_{2}$ is often denoted as $C^{\infty}(\mathcal{V}_1,\mathcal{V}_2)$.\footnotemark

In particular, $\mathcal{V}_{i}$ can be the entire manifolds $\mathcal{M}_{i}$.
\end{defn}
\begin{defn}
\textbf{(Diffeomorphism)} A homeomorphism $\Phi\,:\,\mathcal{M}_1\rightarrow\mathcal{M}_{2}$ is said to be a \emph{diffeomorphism} if $\Phi$ and $\Phi^{-1}$ are $C^{\infty}$-differentiable maps.
\end{defn}
}\footnotetext{
    Observe that smooth functions (Definition \ref{Def: diff function}) are particular cases. Actually the set of smooth functions $C^{\infty}(\mathcal{M})$ is nothing but $C^{\infty}(\mathcal{M},\mathbb{R})$. }

We will be specially interested in the diffeomorphisms $\mathcal{M}\to\mathcal{M}$, which constitute the set $\mathrm{Diff}(\mathcal{M})$ that, together with the composition, has the structure of an infinite-dimensional Lie group. Its algebra is generated by transformations that can be seen from a passive point of view as infinitesimal coordinate transformations. Hence, an arbitrary diffeomorphism can be identified with a \emph{general coordinate transformation} (g.c.t.) $x^{\mu}\rightarrow y^{\alpha}$. The so-called fields (scalar, vector, tensor, spinor...) over a given manifold $\mathcal{M}$ can be seen, from a mathematical point of view, as sections of bundles. Two important definitions are:

\boxdefinition{
\begin{defn}
\textbf{(Smooth bundle)} A bundle $(\mathcal{B},\mathcal{M},\pi,\mathcal{F})$ is called \emph{smooth} if $\mathcal{B}$, $\mathcal{M}$ and $\mathcal{F}$ are smooth manifolds and the projection $\pi$ is a $C^{\infty}$-differentiable map.
\end{defn}
}

\boxdefinition{
\begin{defn}
\textbf{(Smooth section)} Let $(\mathcal{B},\mathcal{M},\pi,\mathcal{F})$ be a smooth bundle. We say that a section $\sigma$ (either local, $\sigma\,:\,\mathcal{U}\to\mathcal{B}$, or global, $\sigma\,:\,\mathcal{M}\to\mathcal{B}$) is \emph{smooth} if it is a $C^{\infty}$-differentiable map. 
\end{defn}
}

According to how their components transform under g.c.t. we can define:
\begin{itemize}
\item \emph{Scalar fields}. These are the elements of $C^{\infty}(\mathcal{M})$, i.e. objects that associate a real number to each point of the manifold. They do not change under g.c.t..
\item \emph{Vector fields}. They are the sections of the tangent bundle. At each point, the vector field $\vecV$ picks up an element of the tangent space $\vecV(p)=V^{\mu}(p)\vpartial_{\mu}|_{p}$, where $V^{\mu}(p)\in C^{\infty}(\mathcal{M})$. If we apply a g.c.t. $x^{\mu}\rightarrow y^{\alpha}$, one can show that vectors transform as
\begin{equation}
V'^{\alpha}=\frac{\partial y^{\alpha}}{\partial x^{\mu}}V^{\mu}\,,\qquad\mbox{since the basis transforms as (chain rule):}\qquad\vpartial_{\alpha}=\frac{\partial x^{\mu}}{\partial y^{\alpha}}\vpartial_{\mu}\,.
\end{equation}
The set of smooth vector fields, $C^{\infty}(\mathcal{M},T\mathcal{M})$, is also denoted as $\mathfrak{X}(\mathcal{M})$. 
\item \emph{1-form/covector fields}. These are sections of the cotangent bundle. Analogously as the vector fields, they have a local expression
$\dfal(p)=\alpha_{\mu}(p)\dex x^{\mu}|_{p}$, where $\alpha_{\mu}(p)\in C^{\infty}(\mathcal{M})$. Under a g.c.t. $x^{\mu}\rightarrow y^{\alpha}$, 1-forms transform as \begin{equation}
\alpha'{}_{\alpha}=\frac{\partial x^{\mu}}{\partial y^{\alpha}}\alpha_{\mu}\,,\qquad\mbox{since the basis transforms as (chain rule):}\qquad\dex y^{\alpha}=\frac{\partial y^{\alpha}}{\partial x^{\mu}}\dex x^{\mu}\,.
\end{equation}
The set of smooth 1-form fields, $C^{\infty}(\mathcal{M},T^{*}\mathcal{M})\equiv C^{\infty}(\mathcal{M},\Lambda^1\mathcal{M})$,
is also denoted as $\Omega^{1}(\mathcal{M})$.
\item \emph{$(r,\,s)$- tensor field}. They are sections of the corresponding tensor bundle. Locally they can be expressed as $\tenT=T^{\mu_1\mu_{2}...\mu_{r}}{}_{\nu_1\nu_{2}...\nu_{s}}\vpartial_{\mu_1}\otimes...\otimes\vpartial_{\mu_{r}}\otimes\dex x^{\nu_1}\otimes...\otimes\dex x^{\nu_{s}}$ and the transformation rule under g.c.t. is
\begin{equation}
T'^{\alpha_1\alpha_{2}...\alpha_{r}}{}_{\beta_1\beta_{2}...\beta_{s}}=\frac{\partial y^{\alpha_1}}{\partial x^{\mu_1}}...\frac{\partial y^{\alpha_{r}}}{\partial x^{\mu_{r}}}\frac{\partial x^{\nu_1}}{\partial y^{\beta_1}}...\frac{\partial x^{\nu_{s}}}{\partial y^{\beta_{s}}}\,T^{\mu_1\mu_{2}...\mu_{r}}{}_{\nu_1\nu_{2}...\nu_{s}}\,.
\end{equation}
The set of smooth \emph{$(r,\,s)$- tensor} fields, $C^{\infty}(\mathcal{M},T^{(r,\,s)}\mathcal{M})$, is also denoted as $\mathscr{T}^{(r,\,s)}(\mathcal{M})$. 
\item \emph{$k$-form over $\mathcal{M}$}. They are sections of the corresponding exterior bundle. They transform as $(0,s)$-tensor fields. We will denote the set $C^{\infty}(\mathcal{M},\Lambda^k\mathcal{M})$ of smooth $k$-form fields as $\Omega^{k}(\mathcal{M})$, as it is usual in the literature.

\item \emph{Vector/tensor-valued $k$-form over $\mathcal{M}$}. They are sections $\Gamma(\Lambda^k\mathcal{M}\otimes W)\eqqcolon \Omega^k(\mathcal{M};W)$ where $W$ is a vector space. In particular, they can be expressed as $\dfal = \dfal^\mathtt{M}\otimes\zeta_\mathtt{M}$, where $\dfal^\mathtt{M}\in\Omega^k(\mathcal{M})$ and $\{\zeta_\mathtt{M}\}$ is a basis of $W$. Sometimes we will refer to the part $\dfal^\mathtt{M}$ as tensor-valued form. 

\end{itemize}
One can also find other kinds of objects like the following ones, which we do not introduce formally in order not to complicate these notes unnecessarily.\footnote{One has to introduce orientation bundles and other related topics to do it properly.} Tensor densities and pseudotensors are objects whose transformation rules under g.c.t. involve the Jacobian determinant
\begin{equation}
\text{Jac}_{x\to y}\coloneqq\det\left(\frac{\partial y^{\alpha}}{\partial x^{\mu}}\right)\,.
\end{equation}

\begin{itemize}
\item \emph{$(r,\,s)$-tensor density field of weight $w$}. They transform as tensors do but with an extra factor consisting on some power of the Jacobian determinant:
\begin{equation}
\mathfrak{T}'{}^{\alpha_1\alpha_{2}...\alpha_{r}}{}_{\beta_1\beta_{2}...\beta_{s}}=\left|\text{Jac}_{x\to y}\right|^{w}\frac{\partial y^{\alpha_1}}{\partial x^{\mu_1}}...\frac{\partial y^{\alpha_{r}}}{\partial x^{\mu_{r}}}\frac{\partial x^{\nu_1}}{\partial y^{\beta_1}}...\frac{\partial x^{\nu_{s}}}{\partial y^{\beta_{s}}}\,\mathfrak{T}^{\mu_1\mu_{2}...\mu_{r}}{}_{\nu_1\nu_{2}...\nu_{s}}\,.
\end{equation}
(We use mathfrak-type font for them) \item \emph{$(r,\,s)$-pseudotensor}. They transform as tensors do but with the sign of the Jacobian determinant:
\begin{equation}
T'^{\alpha_1\alpha_{2}...\alpha_{r}}{}_{\beta_1\beta_{2}...\beta_{s}}=\sign(\text{Jac}_{x\to y})\frac{\partial y^{\alpha_1}}{\partial x^{\mu_1}}...\frac{\partial y^{\alpha_{r}}}{\partial x^{\mu_{r}}}\frac{\partial x^{\nu_1}}{\partial y^{\beta_1}}...\frac{\partial x^{\nu_{s}}}{\partial y^{\beta_{s}}}\,T^{\mu_1\mu_{2}...\mu_{r}}{}_{\nu_1\nu_{2}...\nu_{s}}\,.
\end{equation}
Particular physically relevant cases are pseudoscalars and pseudovectors (also known as axial vectors).
\end{itemize}

Finally, we introduce a very important type of differentiable maps: curves. 
\vspace{-0.2cm}
\boxdefinition{
\begin{defn}
\textbf{(Curve, velocity, trajectory)} Let $\mathcal{M}$ be a smooth manifold, $\mathcal{U}\subseteq\mathcal{M}$ an open set and $I$ an interval of the real line.

A (smooth) \emph{curve} on $\mathcal{U}$ is a differentiable function $\gamma\,:\,I\rightarrow\mathcal{\mathcal{U}}$. The tangent vector $\dot{\gamma}(\tau)=u^{\mu}\vpartial_{\mu}$ is called \emph{velocity} of the curve. And the image of $\gamma$ is called \emph{trajectory} or \emph{path}.
\end{defn}
}

\subsection{(Linear) frames and coframes. Anholonomy}

Let $\mathcal{M}$ be a smooth manifold. A basis of $T_{p}\mathcal{M}$ is called a \emph{linear frame} on $p$. This allows us to construct a bundle by considering all of the possible linear frames at each point of the base manifold:

\boxdefinition{
\begin{defn}\label{def:LinFrameBundle} We define the \emph{(linear) frame bundle} as the bundle whose total space is
\begin{equation}
L(\mathcal{M})\coloneqq\bigsqcup_{p\in\mathcal{M}}\{\text{linear frames on}\ p \}\,,
\end{equation}
with the projection $\pi(p, \{\vfre_a\})\coloneqq p$.

Analogously one can define the \emph{(linear) coframe bundle }$L^{*}(\mathcal{M})$ by taking all the basis of the cotangent bundle at each point.
\end{defn}
}

The bundle of linear frames $L(\mathcal{M})$ over a $\dimM$-dimensional manifold $\mathcal{M}$ is a principal fiber bundle with structure group $\mathrm{GL}(\dimM,\mathbb{R})$.

\vspace{-0.2cm}
\boxdefinition{
\begin{defn}
\textbf{((Co)Frame over a manifold)} A \emph{linear (co)frame field} is a section of the (co)frame bundle. It can be global or local.\footnotemark
\end{defn}
}\footnotetext{
If there exists a global frame over $\mathcal{M}$, the manifold is said to be  \emph{parallelizable}. The 2-sphere is an example of a non-parallelizable manifold.}

From now on when we talk about a frame\footnote{
    We will generally omit the word ``field'' (that refers to the fact that it is not a basis just at one point). We will also omit ``linear'', although in the next chapter, in which we will talk about ``affine frames'', we will use it again to avoid confusion.} 
$\{\vfre_a \}$ and a coframe $\{\cofr^a\}$ we are going to assume that they are dual, i.e.,
\begin{equation}
\cofr^{a}(\vfre_{b})=\delta_{b}^a\,.\label{eq: duality frame}
\end{equation}
Let us express them in terms of some reference coordinate basis:\footnote{
    In order to distinguish between coordinate (co)frames and general ones we will use different notations for the indices: Greek ($\mu$, $\nu$, $\rho$, $\lambda$, $\sigma$, $\eta$...) for coordinate frames and Latin ($a$, $b$, $c$, $d$...) for arbitrary frames.}
\begin{equation}
\vfre_{a}=e^\mu{}_a \vpartial_{\mu}\,,\qquad\cofr^{a}=e_{\mu}{}^a\dex x^{\mu}\,,\label{eq: Vielb}
\end{equation}
where the coefficients $e^\mu{}_a$ and $e_{\mu}{}^{a}$ are sometimes called \emph{Vielbein} (or \emph{tetrads} in $\dimM=4$). The duality condition \eqref{eq: duality frame} and the one of $\vpartial_{\mu}$ with $\dex x^{\nu}$ are now expressed as
\[
e_{\mu}{}^{a}e^\mu{}_{b}=\delta_{b}^a\,,\qquad e_{\mu}{}^{a}e^\nu{}_{a}=\delta_{\mu}^{\nu}\,.
\]
According to \eqref{eq: Vielb}, the Vielbein are nothing but the basis transformation matrix between the holonomic frame $\{\vpartial_{\mu}\}$ and the arbitrary frame $\{\vfre_a \}$. Therefore, whenever we have a tensor expressed in one of them, its components in the other frame are obtained by simply multiplying the expression with the appropriate Vielbein, e.g.
\begin{equation}
H_{abc}{}^{d}=H_{\mu\nu\rho}{}^{\lambda}e^\mu{}_{a}e^\nu{}_{b}e^{\rho}{}_{c}e_{\lambda}{}^{d}\qquad\leftrightarrow\qquad H_{\mu\nu\rho}{}^{\lambda}=H_{abc}{}^{d}e_{\mu}{}^{a}e_{\nu}{}^{b}e_{\rho}{}^{c}e^{\lambda}{}_{d}\,.
\end{equation}
Actually, for practical reasons, it is useful to see the Vielbein as an object that ``transforms indices'' $\{\mu,\nu...\}\leftrightarrow\{a,b...\}$.\footnote{
    However, we have to be careful, because the resulting components with different indices $X^{\mu\nu...}\leftrightarrow X^{ab...}$ only represent the components in the other frame if the abstract object $X$ is tensorial. For non-tensorial objects this is just a convenient way to abbreviate the expressions:
    \[
     X^{ab...}\coloneqq X^{\mu\nu}e_{\mu}{}^{a}e_{\nu}{}^b \qquad\text{if }\,X\,\text{is not tensorial.}
    \]
}

It is important to highlight that a general frame is not necessarily associated to coordinates as $\{\vpartial_{\mu}\}$ does. For $\{\vfre_a \}$ to be a coordinate basis there must exist certain functions $y^{a}(x)$ (the new coordinates) such that
\begin{equation}
e^\mu{}_{a}=\frac{\partial x^{\mu}}{\partial y^{a}}\qquad\text{(integrability condition)}\,,
\end{equation}
and this could not happen. For that purpose it is useful to introduce the Lie bracket of vector fields.

\boxdefinition{
\begin{defn}
\textbf{(Lie bracket)} Given two vector fields $\vecV,\,\vecW\in\mathfrak{X}(\mathcal{M})$, their \emph{Lie bracket} is another vector field whose components
are
\begin{equation}
\left[\vecV,\,\vecW\right]^{\mu}\coloneqq V^{\sigma}\partial_{\sigma}W^{\mu}-W^{\sigma}\partial_{\sigma}V^{\mu}\,.
\end{equation}
\end{defn}
}

\boxproposition{
\begin{prop}
The set of smooth vector fields $\mathfrak{X}(\mathcal{M})$ is a Lie algebra with the Lie Bracket, i.e. the Lie bracket is (1) bilinear, (2) antisymmetric and (3) satisfies the Jacobi identity
\begin{equation}
\left[\left[\vecV,\,\vecW\right],\,\vecZ\right]^{\mu}+\left[\left[\vecZ,\,\vecV\right],\,\vecW\right]^{\mu}+\left[\left[\vecW,\,\vecZ\right],\,\vecV\right]^{\mu}=0\qquad\forall\vecV,\,\vecW,\,\vecZ\in\mathfrak{X}(\mathcal{M})\,.
\end{equation}
\end{prop}
}

With this operation in mind, if we have a frame $\{\vfre_a \}$ over our manifold, we can compute the Lie bracket of its vectors
\begin{equation}
[\vfre_{a},\,\vfre_{b}]=-\Omega_{ab}{}^c\vfre_{c}\,,
\end{equation}
where we have introduced the \emph{anholonomy coefficients} 
\begin{equation}
\Omega_{ab}{}^c\coloneqq e^\mu{}_{a}e^\nu{}_b \Omega_{\mu\nu}{}^c\,,\qquad\Omega_{\mu\nu}{}^c\coloneqq2\partial_{[\mu}e_{\nu]}{}^c\,.
\end{equation}
The powerful result that allows us to find out whether a frame has coordinates associated to it or not is the following:

\boxtheorem{
\begin{thm}
The basis $\{\vfre_a \}$ is integrable (i.e., is a coordinate frame) if and only if the anholonomy coefficients vanish ($\Omega_{\mu\nu}{}^{c}=0$).
\end{thm}
}

This basically says that zero anholonomy is connected with the existence of coordinates. That is the reason why we sometimes refer to coordinate frames as \emph{holonomic} frames. In gauge gravity an arbitrary (not necessarily holonomic) frame will play a crucial role and will be even more fundamental than the very metric of General Relativity (which we will introduce in a few sections).

~

Before ending this section we are going to introduce the Levi-Civita pseudotensor,\footnote{More information about anti/symmetrisation of indices in p. \pageref{chap:Conventions}. }
\begin{equation}
\epsilon_{a_{1}...a_{\dimM}}\coloneqq\dimM!\delta_{[a_1}^{1}...\delta_{a_{\dimM}]}^{\dimM}\,,\label{eq: LC pseudo}
\end{equation}
i.e. its components are $+1$ if $(a_1,a_2,...,a_{\dimM})$ is an even permutation of $(1,2,...,\dimM)$, $-1$ if the permutation is odd, and zero if there are repeated indices. This object has, by definition, this same form in any frame (holonomic or not) and, consequently,

\boxproposition{
\begin{prop}$\epsilon_{a_{1}...a_{\dimM}}$ is a pseudotensor density of weight $+1$. \end{prop}
}

\section{Exterior algebra}

\subsection{Exterior product}

\boxdefinition{
\begin{defn}
\textbf{(Exterior or wedge product of two covariant tensor fields)} Given two covariant tensor fields $\tenT$ and $\tenS$ with covariances $k$ and $l$, respectively, we define their wedge product as the following $(k+l)$-covariant tensor: 
\[
(\tenT\wedge\tenS)(\vecV_1,\,...,\,\vecV_{k+l})\coloneqq\frac{1}{k!l!}\sum_{\{\sigma\}}\sign(\sigma)\tenT(\vecV_{\sigma(1)},\,...,\,\vecV_{\sigma(k)})\tenS(\vecV_{\sigma(k+1)},\,...,\,\vecV_{\sigma(k+l)})\,,
\]
where $\vecV_1,\,...,\,\vecV_{k+l}\in\mathfrak{X}(\mathcal{M})$ and the sum covers all permutations $\sigma$. Its components in certain coframe $\{\cofr^a\}$ can be read from
\begin{equation}
\tenT\wedge\tenS=\frac{(k+l)!}{k!l!}T_{[a_{1}...a_{k}}S_{a_{k+1}...a_{k+l}]}\cofr^{a_1}\otimes...\otimes\cofr^{a_{k+l}}\,.
\end{equation}
\end{defn}
}

Note that the result of the wedge product of two covariant tensors (with covariances $k$ and $l$) is a totally antisymmetric covariant tensor, i.e., a $(k+l)$-form.

\boxproposition{
\begin{prop}
\textbf{\textup{(Properties of the wedge product)}} For $\lambda\in\mathbb{R}$ and arbitrary covariant tensors $\ten{A}$, $\ten{B}$, $\ten{B}'$ and $\ten{C}$ with covariance $k$, $l$, $l$ and $m$, respectively, the following properties hold:
\begin{itemize}
\item Antisymmetry
\begin{equation}
\ten{A}\wedge\ten{B}=(-1)^{kl}\ten{B}\wedge\ten{A}\,.
\end{equation}
\item Linearity (actually, bilinearity)
\begin{equation}
\ten{A}\wedge(\ten{B}+\ten{B}')=\ten{A}\wedge\ten{B}+\ten{A}\wedge\ten{B}'\,,\qquad\ten{A}\wedge(\lambda\ten{B})=(\lambda\ten{A})\wedge\ten{B}=\lambda(\ten{A}\wedge\ten{B})\,.
\end{equation}
\item Associative
\begin{equation}
\ten{A}\wedge(\ten{B}\wedge\ten{C})=(\ten{A}\wedge\ten{B})\wedge\ten{C}\,.
\end{equation}
\end{itemize}
\end{prop}
}

\boxexample{
\begin{example}
Exterior product of two $1$-forms
\begin{align}
(\alpha_a \cofr^a)\wedge(\beta_b \cofr^b)= & \frac{2!}{1!1!}\alpha_{[a}\beta_{b]}\cofr^a\otimes\cofr^{b}=\left(\alpha_a \beta_{b}-\alpha_b \beta_a \right)\cofr^a\otimes\cofr^b \,.
\end{align}
\end{example}
\begin{example}
Product of $N$ $1$-forms ($k_{1}=k_{2}=...=k_{N}=1$)
\begin{equation}
\tenT^{(1)}\wedge\tenT^{(2)}\wedge...\wedge\tenT^{(N)}=N!T^{(1)}{}_{[a_1}\,\ldots\,T^{(N)}{}_{a_{N}]}\cofr^{a_1}\otimes...\otimes\cofr^{a_{N}}\,.
\end{equation}
In particular if we do the exterior product of the elements of the
coframe
\begin{equation}
\cofr^{b_1}\wedge...\wedge\cofr^{b_{N}}=N!\delta_{[a_1}^{b_1}\,\ldots\,\delta_{a_{N}]}^{b_{N}}\cofr^{a_1}\otimes...\otimes\cofr^{a_{N}}=N!\cofr^{[b_1}\otimes...\otimes\cofr^{b_{N}]}\,.\label{eq: wedge to otimes basis}
\end{equation}
\end{example}
}

It is not difficult to check that, at each point, the set $\{(\cofr^{b_1}\wedge...\wedge\cofr^{b_{k}})|_{p}\,:\,b_{i}=1,...,\dimM\}$ is a basis of the space of $k$-forms $\Lambda_p^k\mathcal{M}$, and therefore, under $C^{\infty}(\mathcal{M})$-linear combinations,\footnote{
    Linear combinations in which the coefficients can be not only numbers but also smooth functions.} it
generates the full space of smooth $k$-form fields $\Omega^{k}(\mathcal{M})$.

Consider an arbitrary differential form of rank $k$, or $k$-form,
\begin{equation}
\dfal=\alpha_{a_{1}...a_{k}}\cofr^{a_1}\otimes...\otimes\cofr^{a_{k}}\qquad\text{where}\qquad\alpha_{a_{1}...a_{k}}=\alpha_{[a_{1}...a_{k}]}\,.
\end{equation}
 By virtue of \eqref{eq: wedge to otimes basis}, it can be expressed
\begin{equation}
 \boxed{\dfal=\frac{1}{k!}\alpha_{a_{1}...a_{k}}\cofr^{a_{1}...a_{k}}}\,.
\end{equation}
where we have introduced the very convenient abbreviation (that we will maintain through the whole thesis), 
\begin{equation}
\cofr^{a_{1}...a_{k}}\coloneqq\cofr^{a_1}\wedge...\wedge\cofr^{a_{k}}\,.
\end{equation}
Similarly, when dealing with coordinate coframes we will use
\begin{equation}
\dex x^{\mu_{1}...\mu_{k}}\coloneqq\dex x^{\mu_1}\wedge...\wedge\dex x^{\mu_{k}}\,.
\end{equation}

\subsection{Exterior derivative}

\boxdefinition{
\begin{defn}
\textbf{(Exterior derivative)} The exterior derivative is an operator \\ $\dex\,:\,\Omega^{k}(\mathcal{M}) \rightarrow \Omega^{k+1}(\mathcal{M})$ defined through the expression:
\begin{equation}
\dex\dfal=\frac{1}{k!}\partial_{[\mu_1}\alpha_{\mu_{2}...\mu_{k+1}]}\dex  x^{\mu_{1}...\mu_{k+1}}\,.
\end{equation}
with respect to an arbitrary holonomic coframe.
\end{defn}
}

If the coframe is not holonomic, in order to have the $\alpha$ components with Latin indices, one needs to introduce the Vielbeins inside the partial derivative and therefore some anholonomy coefficients arise.\footnote{
    It can also be written as $\dex\dfal=\frac{1}{k!}\bar{\nabla}_{[a_1}\alpha_{a_{2}...a_{k+1}]}\cofr^{a_{1}...a_{k+1}}$ where $\bar{\nabla}$ is any torsion-free connection (in particular the Levi-Civita connection). See Section \ref{sec: Connection} for more information about connections.}

\boxproposition{
\begin{prop}
\textbf{\textup{(Properties of the exterior derivative)}} For $\lambda_1,\,\lambda_2\in\mathbb{R}$, $\dfal\in\Omega^{k}(\mathcal{M})$ and $\dfbe\in\Omega^{l}(\mathcal{M})$, the exterior derivative is linear, nilpotent and verifies a graded Leibniz rule. Respectively,
\begin{equation}
\dex(\lambda_1\dfal+\lambda_2\dfbe) =\lambda_1\dex\dfal+\lambda_2\dex\dfbe\,,\qquad \dex(\dex\dfal) =0\,,\nonumber
\end{equation}
\begin{equation}
\dex(\dfal\wedge\dfbe)=\dex\dfal\wedge\dfbe+(-1)^{k}(\dfal\wedge\dex\dfbe)\,.
\end{equation}
\end{prop}
}

In the language of differential forms, the anholonomy coefficients are encoded in the \emph{anholonomy $2$-form}, defined as the exterior derivative of the coframe:
\begin{equation}
\dex\cofr^{a}=\frac{1}{2}\Omega_{\mu\nu}{}^a\dex x^{\mu}\wedge\dex x^{\nu}\,.\label{eq: def anhol 2forms}
\end{equation}

\subsubsection*{Closed forms, exact forms and de Rham cohomology}

\boxdefinition{
\begin{defn}
\textbf{(Closed and exact differential form)} A differential form $\dfal\in\Omega^{k}(\mathcal{M})$ is said to be exact if $\dfal=\dex\dfbe$ for some $\dfbe\in\Omega^{k-1}(\mathcal{M})$, and closed if $\dex\dfal=0$.

The vector space of exact $k$-forms is denoted $B_{\dex}^{k}(\mathcal{M})$ and the one conformed by all closed $k$-forms is called $Z_{\dex}^{k}(\mathcal{M})$.
\end{defn}
}
\vspace{-0.2cm}

The fact that $\dex$ is nilpotent implies that any exact form is also closed and one can construct the quotient space of closed forms which are equal up to an exact term. The resulting set is the $k$\emph{-th de Rham cohomology group},
\begin{equation}
H_{\dex}^{k}(\mathcal{M})=Z_{\dex}^{k}(\mathcal{M})/B_{\dex}^{k}(\mathcal{M})\,.
\end{equation}
It can be shown that the dimension of these spaces coincide with the \emph{Betti numbers} $b^{k}(\mathcal{M})$, which determine the Euler characteristic. There is indeed an isomorphism $H_{\dex}^{k}(\mathcal{M})\simeq H_{\dex}^{\dimM-k}(\mathcal{M})$ (consequence of the Poincaré duality), which leads us to:

\boxproposition{
\begin{prop}
The Euler characteristic of an odd dimensional manifold vanishes. If the dimension is even, $\dimM=2N$, then:
\begin{equation}
\chi_{\mathrm{Euler}}(\mathcal{M})=(-1)^{N}b^{N}\,.
\end{equation}
\end{prop}
}

\boxexample{
\begin{example}
For example, for the $\dimM$-dimensional sphere, $\mathbb{S}^{\dimM}$, the sequence of Betti numbers is $\{b^{0},...,b^{\dimM}\}=\{1,0,0,...,0,0,1\}$, which implies
\begin{equation}
\chi_{\mathrm{Euler}}(\mathbb{S}^{\dimM})=1+(-1)^{\dimM}\,.
\end{equation}
Here we clearly see how the odd dimensional spheres have zero Euler characteristic. The even-dimensional spheres all have $\chi_{\mathrm{Euler}}(\mathbb{S}^{\dimM\,\text{(even)}})=2$.
\end{example}
\begin{example}
For the torus $\mathbb{S}^{1}\times\mathbb{S}^{1}$, we have $\{b^{0},b^{1},b^{2}\}=\{1,2,1\}$, so \\ $\chi_{\mathrm{Euler}}(\mathbb{S}^{1}\times\mathbb{S}^{1})=0$.
\end{example}
}

\newpage
\subsection{Interior product}

\boxdefinition{
\begin{defn}
\textbf{(Interior product)} Consider a vector field $\vecV$. The \emph{interior product} with respect to $\vecV$ is a map 
\begin{align}
\dint{\vecV}\,:\,\Omega^{k}(\mathcal{M}) & \longrightarrow\Omega^{k-1}(\mathcal{M})\nonumber \\
\dfal & \longmapsto\dint{\vecV}\dfal\,
\end{align}
where the resulting $(k-1)$-form is the one that acts on vector fields $\vecX_{i}\in\mathfrak{X}(\mathcal{M})$ as
\begin{equation}
\left(\dint{\vecV}\dfal\right)\left(\vecX_1,\,...,\,\vecX_{k-1}\right)\coloneqq\dfal\left(\vecV,\,\vecX_1,\,...,\,\vecX_{k}\right)\,.
\end{equation}
\end{defn}
}
In components,
\begin{equation}
\dint{\vecV}\dfal \ =\ V^c\alpha_{cb_{1}...b_{k-1}}\cofr^{b_1}\otimes...\otimes\cofr^{b_{k-1}} \ =\ \frac{1}{(k-1)!}V^c\alpha_{cb_{1}...b_{k-1}}\cofr^{b_{1}...b_{k-1}}\,.
\end{equation}

\boxproposition{
\begin{prop}
\textbf{\textup{(Properties of the interior product)}} Consider the scalars $\lambda_1,\,\lambda_2\in\mathbb{R}$, the vector fields $\vecV,\vecW\in\mathfrak{X}(\mathcal{M})$ and the differential forms $\dfal\in\Omega^{k}(\mathcal{M})$ and $\dfbe\in\Omega^{l}(\mathcal{M})$. Then, the interior product is linear in both variables, i.e.,
\begin{align}
\dint{(\lambda_1\vecV+\lambda_2\vecW)}\dfal & =\lambda_1\,\,\dint{\vecV}\dfal+\lambda_2\,\,\dint{\vecW}\dfal\,,\\
\dint{\vecV}(\lambda_1\dfal+\lambda_2\dfbe) & =\lambda_1\,\,\dint{\vecV}\dfal+\lambda_2\,\,\dint{\vecV}\dfbe\,;
\end{align}
antisymmetric,
\begin{equation}
\dint{\vecV}\left(\dint{\vecW}\dfal\right)=-\dint{\vecW}(\dint{\vecV}\dfal)\,,
\end{equation}
(and hence, nilpotent, $\dint{\vecV}(\dint{\vecV}\dfal)=0$); and verifies the graded Leibniz rule
\begin{equation}
\dint{\vecV}(\dfal\wedge\dfbe)=(\dint{\vecV}\dfal)\wedge\dfbe+(-1)^{k}\dfal\wedge\left(\dint{\vecV}\dfbe\right)\,.
\end{equation}
\end{prop}
}

Let $\{\vfre_a \}$ be a frame with dual coframe $\{\cofr^a\}$ (i.e., $\dint{\vfre_a}\cofr^b =\delta_a^b$), then one can easily prove 
\begin{equation}
\dint{\vecV}\left(\cofr^{a_1...a_k}\right) =kV^{[a_1}\cofr^{a_2...a_k]}\,,\qquad
\dint{\vfre_a}\dfal =\frac{1}{(k-1)!}\alpha_{a b_1...b_{k-1}}\cofr^{b_1...b_{k-1}}\,,
\end{equation}
which imply
\begin{equation}\label{eq: dint prop}
\dint{\vfre_a}\cofr^{b_1...b_k} =k\delta_a^{[b_1}\cofr^{b_2...b_k]}\,.
\end{equation}
Finally we provide a couple of very useful properties valid for arbitrary $\dfal\in\Omega^{k}(\mathcal{M})$, 
\begin{align}
\cofr^a\wedge\big(\dint{\vfre_a}\dfal\big)&=k\dfal\,,\label{eq: ext int contraido}\\
\dint{\vfre_a}\big(\cofr^a\wedge\dfal\big)&=(\dimM-k)\dfal\,.\label{eq: int ext contraido}
\end{align}

\subsection{Volume forms and integration in manifolds}

\boxdefinition{
\begin{defn}
\textbf{(Volume form)} In a $\dimM$-dimensional smooth manifold $\mathcal{M}$, a volume form is a globally non-vanishing $\dimM$-form. 
\end{defn}
}

If we choose a positively oriented frame/coframe\footnote{This means that the volume form $\volf$ verifies $\volf(\vfre_1,\,...,\,\vfre_{\dimM})>0$.} a volume form can be expressed 
\begin{equation}
\volf\coloneqq\mathfrak{f}\cofr^{1}\wedge...\wedge\cofr^{\dimM}=\frac{1}{\dimM!}\mathfrak{f}\epsilon_{a_{1}...a_{D}}\cofr^{a_{1}...a_{D}}\,,\label{eq: volf}
\end{equation}
where $\mathfrak{f}$ is a scalar density of weight $-1$ and $\epsilon_{a_{1}...a_{D}}$ is the Levi-Civita pseudotensor in \eqref{eq: LC pseudo}.

Some important remarks:
\begin{itemize}
\item Observe that under orientation-preserving frame transformations, the whole object transforms tensorially, since the weights of $\mathfrak{f}$ and $\epsilon_{a_{1}...a_{D}}$ cancel out. 
\item If the manifold is not orientable, it is not possible to find a (globally defined) volume form. 
\item Any two volume forms are equal up to a global non-vanishing function (consequence of the fact that the space of $\dimM$-forms is a $1$-dimensional $C^{\infty}$-module).
\end{itemize}
Consider an open set $\mathcal{U}\subset\mathbb{R}^{\dimM}$ and a $\dimM$-form with coordinate expression
\begin{equation}
\dfal=\alpha\dex x^{1}\wedge...\wedge\dex x^{\dimM}\,,\qquad\alpha\in C^{\infty}(\mathcal{U})\,,
\end{equation}
and such that its support is compact and contained in $\mathcal{U}$. Then its integral is defined: 
\begin{equation}
\int_{\mathcal{U}}\dfal\coloneqq\int_{\mathcal{U}}\alpha\,\dex x^{1}...\dex x^{\dimM}\,.\label{eq: integral Rn}
\end{equation}

To define the integration in a manifold we usually proceed in two steps. First we introduce the generalization of the previous concept: 

\boxdefinition{
\begin{defn}
\textbf{(Integration within charts)} Consider an oriented manifold $\mathcal{M}$, an open set $\mathcal{U}\subset\mathcal{M}$ and a $\dimM$-form $\dfal$ whose support is compact and contained in $\mathcal{U}$. Then its integral is defined:
\begin{equation}
\int_{\mathcal{M}}\dfal=\int_{\mathcal{U}}\dfal\coloneqq\int_{\varphi(\mathcal{U})}(\varphi^{-1})^{*}\dfal\,,
\end{equation}
where $(\mathcal{U},\,\varphi)$ is an arbitrary chart over $\mathcal{U}$ that preserves the orientation.
\end{defn}
}

The definition of the pullback ${}^*$ can be found in Appendix \ref{app:pullback}. Note that $(\varphi^{-1})^{*}\dfal$ is a $\dimM$-form in $\mathbb{R}^{\dimM}$, so we can use \eqref{eq: integral Rn} to solve it. 

However, in general, we want to be able to integrate $\dimM$-forms with a support not necessarily included in a chart. Since the differential form has compact support, there must exist a finite cover of it made of positively-oriented charts $\{(\mathcal{U}_{i},\varphi_{i})\}_{i=1}^{m}$. Let $\{\nu_{i}\}_{i=1}^{m}$ be a partition of the unity subordinate to $\{\mathcal{U}_{i}\}{}_{i=1}^{m}$,\footnote{
    By definition it is a family of smooth maps $\nu_{i}\,:\,\mathcal{M}\to[0,1]$, such that the following three conditions are fulfilled: (1) $\mathrm{supp}(\nu_{i})\subset\mathcal{U}_{i}$, (2) $\sum_{i}\nu_{i}(p)=1$ for all $p\in\mathcal{M}$, and (3) any point of $\mathcal{M}$ admits a chart around it that only intersects a finite number of elements of $\{\mathrm{supp}(\nu_{i})\}$.} 
then the differential form $\nu_{i}\dfal$ has compact support included in $\mathcal{U}_{i}$. And now we can apply the previous definition of each of these pieces and add all of them up:
\begin{equation}
\boxed{\int_{\mathcal{M}}\dfal\coloneqq\sum_{i=1}^{n}\int_{\mathcal{M}}(\nu_{i}\dfal)}\,.
\end{equation}
It can be proved that the resulting quantity in independent of the family $\{(\mathcal{U}_{i},\varphi_{i})\}_{i=1}^{m}$ and also independent of the partition of unity $\{\nu_{i}\}_{i=1}^{m}$ chosen (see a more detailed explanation in \cite{SanchezCaja2012}).

\section{Metric structure}

\subsection{Metric tensor}

\boxdefinition{
\begin{defn}
\textbf{(Metric)} Let $\mathcal{M}$ be a smooth manifold. A \emph{metric} is a 2-covariant tensor field $\teng$ over the manifold that is symmetric and non-degenerate. The pair $(\mathcal{M},\teng)$ is called \emph{metric manifold}.
\end{defn}
}

Due to the non-degeneracy, each metric (scalar product in the tangent space), $\teng=g_{\mu\nu}\dex x^{\mu}\otimes\dex x^{\nu}$, determines a unique \emph{inverse metric} (scalar product in the cotangent space), i.e. a 2-contravariant symmetric and non-degenerate tensor field $\teng^{-1}=g^{\mu\nu}\vpartial_{\mu}\otimes\vpartial_{\nu}$, such that:
\begin{equation}
g^{\mu\rho}g_{\rho\nu}=\delta_{\nu}^{\mu}\,.
\end{equation}

The metric determines a natural isomorphism between the tangent space and the cotangent space at each point $p\in\mathcal{M}$, the \emph{musical isomorphism}s:
\begin{align}
\flat\quad:\quad\vecV=V^{\mu}\vpartial_{\mu}|_{p} & \quad\longmapsto\quad\vecV^{\flat}=V^{\nu}g_{\mu\nu}\dex x^{\mu}|_{p}\quad\equiv V_{\mu}\dex x^{\mu}|_{p}\,,\\
\sharp\quad:\quad\dfal=\alpha_{\mu}\dex x^{\mu}|_{p} & \quad\longmapsto\quad\dfal^{\sharp}=\alpha_{\nu}g^{\mu\nu}\vpartial_{\mu}|_{p}\quad\equiv\alpha^{\mu}\vpartial_{\mu}|_{p}\,.
\end{align}
These isomorphisms also relate $(r,\,s)$-tensors with $(r',\,s')$-tensors, provided that $r+s=r'+s'$. In components notation this is indeed the ``raising and lowering indices'' operation
\begin{equation}
T_{\mu\nu\rho\sigma}{}^{\lambda}=T_{\mu\nu\rho}{}^{\tau\lambda}g_{\sigma\tau}=T_{\mu\nu\rho\sigma\tau}g^{\lambda\tau}=...\,.
\end{equation}

\boxdefinition{
\begin{defn}
\textbf{(Classification of vectors)} A vector\footnotemark {} is called \emph{timelike} if $\teng(\vecV,\,\vecV)>0$, \emph{spacelike} if $\teng(\vecV,\,\vecV)<0$, and \emph{lightlike} if $\teng(\vecV,\,\vecV)=0$ and $\vecV\neq0$. 
\end{defn}

\begin{defn}
\textbf{(Signature)} The \emph{signature} of a metric is the pair $(p,n)$ where $p$ and $n$ are, respectively, the number of positive and negative eigenvalues.
\end{defn}

}\footnotetext{A similar definition can be given for $1$-forms by using the inverse metric.}

The metrics with signatures $(0, \dimM)$ (all eigenvectors are spacelike) and $(1, \dimM-1)$ (only one is timelike) are called, respectively, \emph{Riemannian} and \emph{Lorentzian}. A generic metric is sometimes called \emph{pseudo-Riemannian}. 

For physical applications, we will consider that the metric is Lorentzian, i.e. each tangent space has the structure of Minkowski space, and that the manifold is a connected topological space.\footnote{
    One can be even more restrictive and add the requirement of \emph{time orientability}. In a Lorentzian manifold, one can select in each point which of the two lightcones is the one pointing \emph{to the future}, but it is not always possible to make such choice everywhere in the manifold in a continuous way. If it is possible to do that, then the manifold is said to be time orientable (one can then say that there is a ``global'' notion of time flow).} 

\subsection{Canonical volume form and Hodge duality}

The metric also induces a \emph{canonical volume form}:
\begin{equation}
\volfg\coloneqq\frac{1}{\dimM!}\LCten_{a_{1}...a_{\dimM}}\cofr^{a_{1}...a_{D}}\,,\label{eq: volfg def}
\end{equation}
where $\LCten_{a_{1}...a_{\dimM}}$ is the Levi-Civita tensor,\footnote{
    $\epsilon_{a_{1}...a_{\dimM}}$ is a pseudotensor density of weight $w=1$ and $\sqrt{|\det(g_{ab})|}$ a scalar density of weight $w=-1$. Therefore their product, $\LCten_{a_{1}...a_{\dimM}}$, is strictly speaking a pseudotensor (zero weight).} 
given in terms of the Levi-Civita symbol:
\begin{equation}
\LCten_{a_{1}...a_{\dimM}}\coloneqq\sqrt{|\det(g_{ab})|}\epsilon_{a_{1}...a_{\dimM}}\,.\label{eq: LCTensor}
\end{equation}
In other words, a metric defines naturally a volume form \eqref{eq: volf} with $\mathfrak{f}=\sqrt{|\det(g_{ab})|}$. The expression \eqref{eq: volfg def} is valid in arbitrary frames, but in a coordinate one, we find:
\begin{equation}
\volfg=\frac{1}{\dimM!}\sqrt{|\det(g_{\mu\nu})|}\epsilon_{\mu_{1}...\mu_{\dimM}}\dex x^{\mu_1}\wedge...\wedge\dex x^{\mu_{\dimM}}=\sqrt{|\det(g_{\mu\nu})|}\,\underbrace{\dex x^{1}\wedge...\wedge\dex x^{\dimM}}_{\dex^{\dimM}x}\,.
\end{equation}
From now on we will denote $\sqrt{|g|}\coloneqq\sqrt{|\det(g_{\mu\nu})|}$, i.e., whenever we see $g$ we are referring to the determinant in the coordinate basis. One important property of the Levi-Civita tensor is the following (that can be easily proved by using the properties of the Levi-Civita symbol $\epsilon_{a_{1}...a_{\dimM}}$):
\begin{equation}
\LCten_{c_{1}...c_{k}a_{1}...a_{\dimM-k}}\LCten^{c_{1}...c_{k}b_{1}...b_{\dimM-k}}=\sgng\,k!\,(\dimM-k)!\delta_{[a_1}^{b_1}\ldots\delta_{a_{\dimM-k}]}^{b_{\dimM-k}}\,.\label{eq: LCTensorLCTensor}
\end{equation}
which can be used to deduce
\begin{equation}
\cofr^{a_{1}...a_{\dimM}}=\sgng\,\LCten^{a_{1}...a_{\dimM}}\volfg\,.\label{eq: cofr as eps volf}
\end{equation}

In addition, we also get a duality between spaces of differential forms:

\boxproposition{
\begin{prop}
\textbf{\textup{(Hodge duality)}} Consider a manifold $\mathcal{M}$ equipped with a metric. For any $0\leq k\leq\dimM$, there is an isomorphism (\emph{Hodge duality}) of vector spaces between $k$-forms and $(\dimM-k)$-forms given by:
\begin{align}
\star\,:\,\Omega^{k}(\mathcal{M}) & \longrightarrow\Omega^{\dimM-k}(\mathcal{M})\nonumber \\
\dfal\quad & \longmapsto\quad\star\dfal\ \coloneqq\ \frac{1}{(\dimM-k)!k!}\alpha^{b_{1}...b_{k}}\LCten_{b_{1}...b_{k}c_{1}...c_{\dimM-k}}\cofr^{c_{1}...c_{\dimM-k}}\,.
\end{align}
\end{prop}
}

By using \eqref{eq: LCTensorLCTensor}, it can be shown that, for an arbitrary $k$-form $\dfal$,
\begin{equation}
\boxed{\star\star\dfal=(-1)^{k(\dimM-k)}\sgng\dfal}\,.\label{eq: starstar}
\end{equation}
Some interesting particular cases are:
\begin{align}
\star\cofr^{c_{1}...c_{k}} & =\frac{1}{(\dimM-k)!}\LCten^{c_{1}...c_{k}}{}_{a_{1}...a_{\dimM-k}}\cofr^{a_{1}...a_{\dimM-k}}\,,\\
\star\cofr^{c_{1}...c_{\dimM}} & =\LCten^{c_{1}...c_{\dimM}}\,,\label{eq: starcof D}\\
\star\volfg & =\sgng\,,\\
\star1 & =\volfg\,.
\end{align}

The following expressions are extremely useful when we are interested in extracting a Hodge star from an expression involving a $k$-form $\dfal$:
\begin{align}
\dint{\vfre_a}\left(\star\dfal\right)&=\star\left(\dfal\wedge\cofr_a \right)\,,\label{eq: dint star to star wedge}\\
\cofr^a\wedge\star\dfal &=(-1)^{k+1}\star\left(\dint{\vfre^{a}}\dfal\right)\,
\end{align}
(indices have been raised/lowered with $g_{ab}$ as usual). It is interesting to see the particular case $\dfal=\cofr^{b_{1}...b_{k}}$
\begin{align}
\dint{\vfre_a}(\star\cofr^{b_{1}...b_{k}}) & =\star\cofr^{b_{1}...b_{k}}{}_a \,,\\
\cofr^a\wedge\star\cofr^{b_{1}...b_{k}} & =k(-1)^{k+1}g^{a[b_1}\star\cofr^{b_{2}...b_{k}]}\,,
\end{align}
where, in the last one, we made use of \eqref{eq: dint prop}. 

The property
\begin{equation}
\dfal\wedge\star\dfbe=\frac{1}{k!}\alpha^{a_{1}...a_{k}}\beta_{a_{1}...a_{k}}\volfg
\end{equation}
is going to be very useful to built invariants for a Lagrangian in terms of forms.

\section{\label{sec: Connection}Connection}

\subsection{Linear connection}

In principle there is no natural way to compare vectors (and, by extension, tensors) in different points of the manifold, since there is no criterion that tells us how we should move a vector in $p\in\mathcal{M}$ to a different point $q\in\mathcal{M}$, where we have another vector, to compare them. We then need an additional structure called linear connection.\footnote{
    We do not use the name ``affine connection'' because it can be confused with connections in an affine bundle.} 
Suppose that $p$ and $q$ are infinitesimally close, so that they can be assumed to be connected by a straight line with displacement vector $\delta x^{\mu}$. A connection basically defines, for a given vector $\vecv\in T_{p}\mathcal{M}$, a representative $\vecv_{p\to q}$ in $T_{q}\mathcal{M}$. In components in a particular coordinate basis, we have
\begin{equation}
(v_{p\to q})^{\mu}=v^{\mu}-\Gamma_{\lambda\rho}{}^{\mu}\delta x^{\lambda}v^{\rho}\,,\label{eq:partransinf}
\end{equation}
where $\Gamma_{\lambda\rho}{}^{\mu}$ are certain functions. This correspondence allows to define a covariant derivation of vector fields $V^{\mu}$, i.e. a limit
\begin{equation}
\left(\nabla_{\nu}V^{\mu}\right)(q)=\lim_{\delta x^{\nu}\rightarrow0}\frac{V^{\mu}(q)-(V(p){}_{p\to q})^{\mu}}{\delta x^{\nu}}\,.
\end{equation}

Now we give the formal definition:

\boxdefinition{
\begin{defn}
\textbf{(Covariant derivative)} Let $\mathcal{M}$ be a differentiable manifold. A \emph{covariant derivative} over $\mathcal{M}$, $\nabla$, is a map 
\begin{align*}
\nabla\,:\,\mathfrak{X}(\mathcal{M})\times\mathfrak{X}(\mathcal{M}) & \longrightarrow\mathfrak{X}(\mathcal{M})\\
(\vecV,\,\vecW) & \longmapsto\nabla_{\vecV}\vecW
\end{align*}
satisfying the following properties for arbitrary vector fields $\vecU,\,\vecV,\,\vecW\in\mathfrak{X}(\mathcal{M})$,
any arbitrary functions $f,\,g\in C^{\infty}(\mathcal{M})$ and any
real numbers $a,\,b\in\mathbb{R}$:
\end{defn}
\begin{itemize}
\item $C^{\infty}(\mathcal{M})$-linearity in the first variable:
\begin{equation}
\nabla_{f\vecV+g\vecW}\vecU=f\nabla_{\vecV}\vecU+g\nabla_{\vecW}\vecU\,.
\end{equation}
\item $\mathbb{R}$-linearity and Leibniz rule in the second variable:
\begin{align}
\nabla_{\vecU}\left(a\vecV+b\vecW\right) & =a\nabla_{\vecU}\vecV+b\nabla_{\vecU}\vecW\,,\\
\nabla_{\vecU}\left(f\vecV\right) & =\vecU(f)\vecV+f\nabla_{\vecU}\vecV\,.
\end{align}
\end{itemize}
}
$\nabla$ is also known as \emph{linear connection} (or, simply, \emph{connection}), although we will use this name for the associated components:
\begin{equation}
\Gamma_{\mu\nu}{}^{\rho}\coloneqq\dex x^{\rho}\left(\nabla_{\vpartial_{\mu}}\vpartial_{\nu}\right)\,.
\end{equation}

Now we show how this $\nabla$ acts in components:
\begin{align}
\nabla_{\vecU}\vecV=\nabla_{U^{\mu}\vpartial_{\mu}}\left(V^{\nu}\vpartial_{\nu}\right) & =U^{\mu}\left(\vpartial_{\mu}(V^{\nu})\vpartial_{\nu}+V^{\nu}\nabla_{\vpartial_{\mu}}\vpartial_{\nu}\right)\nonumber \\
 & =U^{\mu}\left(\partial_{\mu}V^{\rho}+\Gamma_{\mu\nu}{}^{\rho}V^{\nu}\right)\vpartial_{\rho}\qquad \equiv U^{\mu}\nabla_{\mu}V^{\rho}\vpartial_{\rho}\,,
\end{align}
i.e.,
\begin{equation}
\nabla_{\mu}V^{\rho}=\partial_{\mu}V^{\rho}+\Gamma_{\mu\nu}{}^{\rho}V^{\nu}\,.
\end{equation}
For covectors we have
\begin{equation}
\nabla_{\mu}V_{\rho}=\partial_{\mu}V_{\rho}-\Gamma_{\mu\rho}{}^{\nu}V_{\nu}\,
\end{equation}
and, in general, for an arbitrary tensor density of weight $w$ (for tensors just take $w=0$):
\begin{align}
\nabla_{\mu}\mathfrak{T}^{\nu...}{}_{\rho...} & \coloneqq\partial_{\mu}\mathfrak{T}^{\nu...}{}_{\rho...}+\Gamma_{\mu\sigma}{}^{\nu}\mathfrak{T}^{\sigma...}{}_{\rho...}+\text{...same for all upper indices...}\nonumber \\
 & \qquad\qquad\qquad-\Gamma_{\mu\rho}{}^{\sigma}\mathfrak{T}^{\nu...}{}_{\sigma...}-\text{...same for all lower indices...}\nonumber \\
 & \qquad\qquad\qquad +w\Gamma_{\mu\sigma}{}^{\sigma}\mathfrak{T}^{\nu...}{}_{\rho...}\,.\label{eq: nabla def}
\end{align}

Notice that the object $\Gamma_{\mu\nu}{}^{\rho}$ cannot be associated with the components of a tensor, because under a g.c.t. $x^{\mu}\rightarrow y^{\alpha}$, it transforms as
\begin{equation}
\Gamma'{}_{\alpha\beta}{}^{\gamma}=\frac{\partial x^{\mu}}{\partial y^{\alpha}}\frac{\partial x^{\nu}}{\partial y^{\beta}}\frac{\partial y^{\gamma}}{\partial x^{\rho}}\Gamma_{\mu\nu}{}^{\rho}-\frac{\partial x^{\mu}}{\partial y^{\alpha}}\frac{\partial x^{\nu}}{\partial y^{\beta}}\frac{\partial^{2}y^{\gamma}}{\partial x^{\mu}\partial x^{\nu}}\,.
\end{equation}
The last term cancels a term coming from the partial derivative term in \eqref{eq: nabla def}, making the whole object $\nabla_{\mu}X_{...}$ be always tensorial (or a tensor density depending on $X_{...}$). One can also clearly check that the difference between connections transforms tensorially under g.c.t., and consequently, the quantity $\delta\Gamma_{\mu\nu}{}^{\rho}$ is treated as a tensor when performing variations of connection-dependent actions.

\boxdefinition{
\begin{defn}
\textbf{((Auto-)parallel vector (tensor) field)} Consider a curve $\gamma$ with velocity $u^{\mu}$, a connection $\Gamma_{\mu\nu}{}^{\rho}$ with covariant derivative $\nabla$ and a vector field over the image of the curve $\vecV\in\mathfrak{X}(\text{Im}(\gamma))$. The vector field is said to be \emph{parallel with respect to} $\Gamma_{\mu\nu}{}^{\rho}$ \emph{along} $\gamma$ if $u^{\mu}\nabla_{\mu}V^{\nu}=0$. 

A general vector field (defined in some open set) is called \emph{parallel with respect to }$\Gamma_{\mu\nu}{}^{\rho}$ (without mentioning the curve) if $\nabla_{\mu}V^{\nu}=0$.
\end{defn}
}
These definitions are straightforwardly generalized to arbitrary tensor fields and densities. In some situations, to avoid ambiguity we will use the word ``auto-parallel''.

The connection then defines the notion of parallelism. In particular, the correspondence between tangent spaces that we saw at the begining of this section can be extended to arbitrary points $p$ and $q$ of the manifold connected by a curve $\gamma$:

\boxdefinition{
\begin{defn}
\textbf{(Parallel transport)} For any curve $\gamma$ with velocity $u^{\mu}$ and any point $p\in\text{Im}(\gamma)$, the connection defines uniquely a map $(\tau_{\gamma}^{\nabla})_{p}$, called the \emph{parallel transport}, which is given by
\begin{align}
(\tau_{\gamma}^{\nabla})_{p}\,:\,T_{p}\mathcal{M} & \longrightarrow\mathfrak{X}(\text{Im}(\gamma))\,.\\
v^{\mu} & \longmapsto\left((\tau_{\gamma}^{\nabla})_{p}(\vecv)\right)^{\mu}\equiv W^{\mu}
\end{align}
where $W^{\mu}$ is the unique parallel vector field along $\gamma$ such that $W^{\mu}(p)=v^{\mu}$. 
\end{defn}
}

If we fix the end point we get an isomorphism of tangent spaces
\begin{equation}
\left((\tau_{\gamma}^{\nabla})_{p}(\,.\,)\right)(q)\,:\,T_{p}\mathcal{M}\longrightarrow T_{q}\mathcal{M}\,.
\end{equation}
Indeed, if the path is infinitesimal, it can be approximated by a straight line and we can drop $\gamma$ from the notation. The result is (compare with \eqref{eq:partransinf})
\begin{equation}
\left.\left((\tau_{\gamma}^{\nabla})_{p}(\vecv)\right)^{\mu}(q)\right|_{\text{Infinitesimal }\gamma}\equiv(v_{p\to q})^{\mu}\,.
\end{equation}

\subsection{Curvature and torsion}

\boxdefinition{
\begin{defn}
\textbf{(Curvature and torsion tensors)} Consider a manifold equipped with a connection $\Gamma_{\mu\nu}{}^{\rho}$. The \emph{curvature} and \emph{torsion} tensors (both antisymmetric in the first two indices) are the tensors given in components by
\begin{align}
R_{\mu\nu\rho}{}^{\lambda} & \coloneqq\partial_{\mu}\Gamma_{\nu\lambda}{}^{\rho}-\partial_{\nu}\Gamma_{\mu\lambda}{}^{\rho}+\Gamma_{\mu\sigma}{}^{\rho}\Gamma_{\nu\lambda}{}^{\sigma}-\Gamma_{\nu\sigma}{}^{\rho}\Gamma_{\mu\lambda}{}^{\sigma}\,,\label{eq:defcurvc}\\
T_{\mu\nu}{}^{\rho} & \coloneqq\Gamma_{\mu\nu}{}^{\rho}-\Gamma_{\nu\mu}{}^{\rho}\,.\label{eq:deftorc}
\end{align}
\end{defn}
}

They naturally appear when computing the commutator of covariant derivatives acting on a vector
\begin{equation}
\left[\nabla_{\mu},\,\nabla_{\nu}\right]V^{\lambda}=R_{\mu\nu\rho}{}^{\lambda}V^{\rho}-T_{\mu\nu}{}^{\rho}\nabla_{\rho}V^{\lambda}\,.\label{eq: commutator nabla}
\end{equation}
Indeed, acting on an arbitrary tensor density of weight $w$ we get
\begin{align}
\left[\nabla_{\mu},\,\nabla_{\nu}\right]\mathfrak{T}^{\lambda...}{}_{\sigma...} & =\quad R_{\mu\nu\rho}{}^{\lambda}\mathfrak{T}^{\rho...}{}_{\sigma...}+\text{...same for all upper indices...}\nonumber \\
 & \quad-R_{\mu\nu\sigma}{}^{\rho}\mathfrak{T}^{\lambda...}{}_{\rho...}+\text{...same for all lower indices...}\nonumber \\
 & \quad-T_{\mu\nu}{}^{\rho}\nabla_{\rho}\mathfrak{T}^{\lambda...}{}_{\sigma...}+wR_{\mu\nu\rho}{}^{\rho}\mathfrak{T}^{\lambda...}{}_{\sigma...}\,.
\end{align}

\begin{figure}
\begin{centering}
\includegraphics[width=0.9\textwidth]{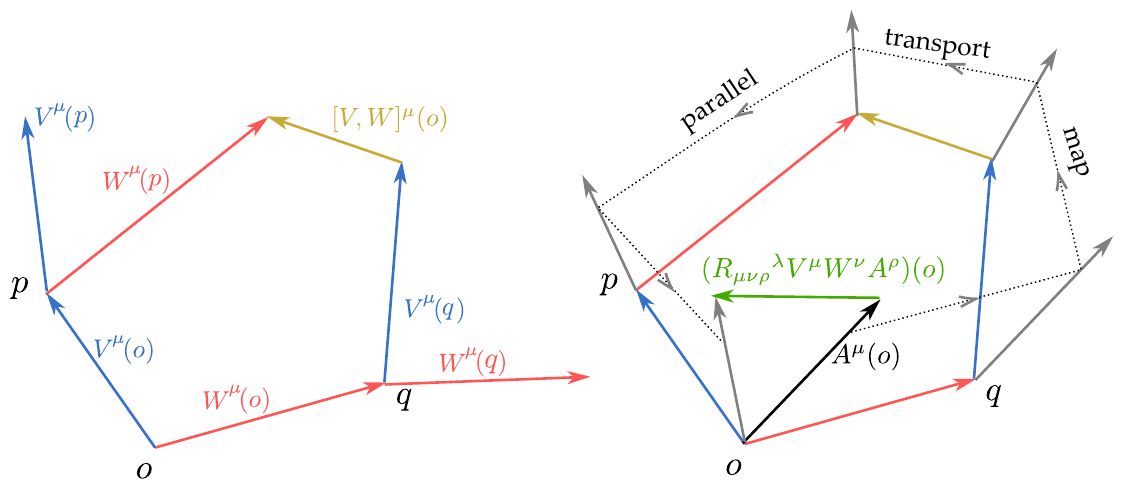}
\par\end{centering}
   \caption{\label{fig:curvinterpret}\textbf{Left}: Schematic view (2 dimensional) of a non-holonomic infinitesimal frame $\{V^\mu, W^\nu\}$ (see them as the elements of an ordinary frame multiplied by some infinitesimal parameter). The non-vanishing Lie bracket indicate that they do not constitute a coordinate basis.  \textbf{ Right}: Geometrical interpretation of the curvature.}
\end{figure}

At this point it is interesting to revise the interpretation of these two tensorial quantities. Suppose that we have a vector in a particular point $p$ of the manifold and consider a closed curve (loop) $\gamma(\tau)$ such that $p\in\text{Im}(\gamma(\tau))$. The idea is to move the vector along the loop by using the parallel transport of $\Gamma_{\mu\nu}{}^{\rho}$. If, when reaching again the initial point, we do not recover the original vector, we say that the curvature is non-vanishing. To be precise, we considered that the infinitesimal loop is an rectangle constructed with two frame-dragged vectors $V^{\mu}$ and $W^{\mu}$ (not parallely transported vectors \cite{Smalley1984}) as in Fig. \ref{fig:curvinterpret} left. Then, if we perform the parallel transport\footnote{See also \cite{MisnerThorneWheeler1973}, although our curvature tensors differ by a sign (and the ordering of the indices).} of some (not necessarily infinitesimal) vector $A^{\mu}$, we find (see Fig. \ref{fig:curvinterpret} right),
\begin{equation}
A^{\lambda}|_{\text{after loop}}-A^{\lambda}=R_{\mu\nu\rho}{}^{\lambda}V^{\mu}W^{\nu}A^{\rho}\,.
\end{equation}

\begin{figure}
\begin{centering}
\includegraphics[width=0.5\textwidth]{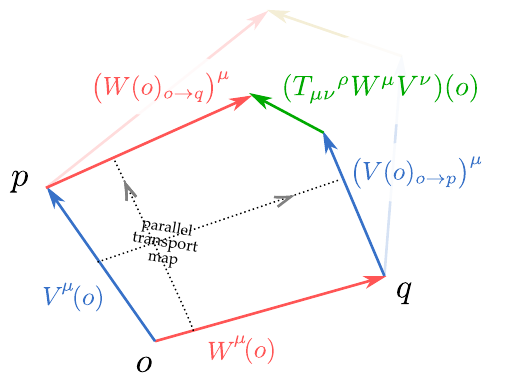}
\par\end{centering}
\caption{\label{fig:torinterpret}Geometrical interpretation of the torsion.}
\end{figure}

On the other hand, the torsion measures the ``closure failure'', when trying to construct a parallelogram: if we take two different vectors at some point and parallely transport each of them along the other, the resulting parallelogram could not close. This indicates the presence of torsion. To be precise, if we do such operation with two infinitesimal vectors $V^{\mu}$ and $W^{\mu}$ (see Fig. \ref{fig:torinterpret}), we see that the vector measuring the ``closure failure'' is given by $T_{\mu\nu}{}^{\rho}V^{\mu}W^{\nu}$.

\boxdefinition{
\begin{defn}
\textbf{(Flat and torsion-free connection)} Let $\mathcal{M}$ be a differentiable manifold equipped with a connection $\Gamma_{\mu\nu}{}^{\rho}$. We say that the latter is \emph{flat} if the curvature vanishes, and \emph{torsion-free} if the torsion vanishes.
\end{defn}
}

We also introduce the following notation for the two independent traces of the curvature\footnote{
    There is another one, but we need a metric to construct it.} 
and for the torsion trace
\begin{align}
R_{\mu\rho}\equiv R^{(1)}{}_{\mu\rho} & \coloneqq R_{\mu\lambda\rho}{}^{\lambda} &  & \text{(Ricci tensor)}\,,\\
R^{(3)}{}_{\mu\nu} & \coloneqq R_{\mu\nu\lambda}{}^{\lambda}\,,\\
T_{\mu} & \coloneqq T_{\mu\lambda}{}^{\lambda}\,.\label{eq: trTor def}
\end{align}
\boxproposition{
\begin{prop}
The curvature tensor satisfies:
\begin{align}
R_{[\mu\nu\rho]}{}^{\lambda} & =\nabla_{[\mu}T_{\nu\rho]}{}^{\lambda}-T_{[\mu\nu}{}^{\sigma}T_{\rho]\sigma}{}^{\lambda}\,,\label{eq:BianchiIdTorcom}\\
\nabla_{[\mu}R_{\nu\rho]\tau}{}^{\lambda} & =T_{[\mu\nu}{}^{\sigma}R_{\rho]\sigma\tau}{}^{\lambda}\,,\label{eq:2BianchiIdGen}\\
-2R_{[\mu\nu]} & =\nabla_{\mu}T_{\nu}-\nabla_{\nu}T_{\mu}+\nabla_{\lambda}T_{\mu\nu}{}^{\lambda}+T_{\mu\nu}{}^{\rho}T_{\rho}+R^{(3)}{}_{\mu\nu}\,.
\end{align}
Therefore, in the torsion-free case:
\begin{equation}
R_{[\mu\nu\rho]}{}^{\lambda}=0\,,\qquad\nabla_{[\mu}R_{\nu\rho]\tau}{}^{\lambda}=0\,,\qquad2R_{[\mu\nu]}=R^{(3)}{}_{\mu\nu}\,.
\end{equation}
\end{prop}
}

Finally one may wonder about the relation between the curvature and the torsion of two connections. The result is the following:

\boxproposition{
\begin{prop}
Consider two connections $\Gamma'{}_{\mu\nu}{}^{\rho}$ and $\Gamma{}_{\mu\nu}{}^{\rho}=\Gamma'{}_{\mu\nu}{}^{\rho}+A_{\mu\nu}{}^{\rho}$,
then their curvatures and torsions are related via:
\begin{align}
R_{\mu\nu\rho}{}^{\lambda} & =R'_{\mu\nu\rho}{}^{\lambda}+2\nabla'{}_{[\mu}A_{\nu]\rho}{}^{\lambda}+T'{}_{\mu\nu}{}^{\sigma}A_{\sigma\rho}{}^{\lambda}+2A_{[\mu|\sigma|}{}^{\lambda}A_{\nu]\rho}{}^{\sigma}\,,\label{eq:RtoRprima}\\
T_{\mu\nu}{}^{\rho} & =T'_{\mu\nu}{}^{\rho}+2A_{[\mu\nu]}{}^{\rho}\,,
\end{align}
where the objects with $'$ are referred to $\Gamma'{}_{\mu\nu}{}^{\rho}$.
\end{prop}
}

\subsection{Nonmetricity}

So far in this section about linear connections we have provided general results that do not require a metric. Now, in the presence of both, metric and connection, we can introduce:

\boxdefinition{
\begin{defn}
\textbf{(Nonmetricity)} Let $(\mathcal{M},\teng)$ be a metric manifold equipped with a connection $\Gamma_{\mu\nu}{}^{\rho}$. The \emph{nonmetricity tensor} is the 3-covariant tensor given in components by 
\begin{equation}
Q_{\mu\nu\rho}\coloneqq-\nabla_{\mu}g_{\nu\rho}\,.
\end{equation}
\end{defn}
}

The geometrical interpretation of the nonmetricity is quite direct: $Q_{\mu\nu\rho}$ measures how far the metric $g_{\mu\nu}$ is from being a parallel tensor. This has to do, of course with the change in the scalar product under parallel transport. Suppose we have two arbitrary vectors at some point $p$, $v^\mu$ and $w^\mu$, and we are interested in understanding the meaning of $Q_{\rho\mu\nu}$ at the point $p$. The idea is: first take the coordinate curve $\gamma$ generated by $\vpartial_\rho$, and construct the parallel transport of $v^\mu$ and $w^\mu$ along that curve. Let us call the resulting auto-parallel fields $V^\mu$ and $W^\mu$, respectively. The scalar product $g_{\mu\nu}V^\mu W^\nu$ is a function and we can calculate its directional derivative in the direction of the curve on the point $p$ as
\begin{align}
  \partial_\rho (g_{\mu\nu}V^\mu W^\nu)(p) = \nabla_\rho (g_{\mu\nu}V^\mu W^\mu)(p) = - Q_{\rho\mu\nu}(p)v^\mu w^\nu\,.
\end{align}
Therefore, zero nonmetricity means that the scalar product of any two auto-parallel vectors is a constant function.

The nonmetricity has two independent traces, that we will represent as:
\begin{align}
Q_{\mu} & \coloneqq Q_{\mu\lambda}{}^{\lambda}\,,\label{eq: trQ1 def}\\
\Qb_{\mu} & \coloneqq Q^{\lambda}{}_{\lambda\mu}=Q^{\lambda}{}_{\mu\lambda}\,.\label{eq: trQ2 def}
\end{align}
The first one is also called the \emph{Weyl vector} (or, more rigorously, \emph{Weyl 1-form}).

Notice that, when a metric is present we can additionally compute a third independent trace of the curvature and the total trace (scalar):
\begin{align}
R^{(2)}{}_{\mu}{}^{\lambda} & \coloneqq R_{\mu\rho}{}^{\rho\lambda}\,,\\
R & \coloneqq R^{(1)}{}_{\rho}{}^{\rho}=-R^{(2)}{}_{\rho}{}^{\rho} &  & \text{(Ricci curvature scalar)}\,.
\end{align}
\boxproposition{
\begin{prop}
The curvature tensor of an affine structure, when a metric is present, satisfies:
\begin{align}
R_{\mu\nu(\rho\lambda)} & =\nabla_{[\mu}Q_{\nu]\rho\lambda}+\frac{1}{2}T_{\mu\nu}{}^{\sigma}Q_{\sigma\rho\lambda}\,,\label{eq:BianchiIdQcom}\\
R^{(3)}{}_{\mu\nu} & =\partial_{[\mu}Q_{\nu]}\,,\\
R^{(2)}{}_{\mu\nu} & =-R_{\mu\nu}+2g^{\lambda\rho}\nabla_{[\mu}Q_{\rho]\nu\lambda}+g^{\lambda\rho}T_{\mu\rho}{}^{\sigma}Q_{\sigma\nu\lambda}\,,
\end{align}
\end{prop}
}

\boxdefinition{
\begin{defn}
\textbf{(Metric-compatible connection)} Let $(\mathcal{M},\teng)$ be a metric manifold equipped with a connection $\Gamma_{\mu\nu}{}^{\rho}$. We say that the latter is \emph{metric-compatible} (or compatible with the metric) if the nonmetricity tensor vanishes.
\end{defn}
}

\boxcorollary{
\begin{cor}
The curvature tensor of a metric-compatible connection satisfies:
\begin{equation}
R_{\mu\nu(\rho\lambda)}=0\,,\qquad R^{(3)}{}_{\mu\nu}=0\,,\qquad R^{(2)}{}_{\mu\nu}=-R_{\mu\nu}\,.
\end{equation}
\end{cor}
}

\subsection{Levi-Civita connection}

\boxdefinition{
\begin{defn}
\textbf{(Levi-Civita connection)} Let $(\mathcal{M},\teng)$ be a metric manifold. The Levi-Civita connection associated to the metric is the one whose components are the Christoffel symbols:
\begin{equation}
\mathring{\Gamma}_{\mu\nu}{}^{\rho}=\left\{ _{\mu\nu}{}^{\rho}\right\} \coloneqq\frac{1}{2}g^{\rho\sigma}\left[\partial_{\mu}g_{\sigma\nu}+\partial_{\nu}g_{\mu\sigma}-\partial_{\sigma}g_{\mu\nu}\right]\,.\label{eq:defLCconn}
\end{equation}
\end{defn}
}

\boxproposition{
\begin{prop}
In a metric manifold $(\mathcal{M},\,\teng)$, the Levi-Civita connection of the metric is the only connection with zero torsion and zero nonmetricity.
\end{prop}
\begin{prop}
The curvature tensor of the Levi-Civita connection has the symmetries
\begin{equation}
\mathring{R}_{\mu\nu\rho\lambda}=-\mathring{R}_{\mu\nu\lambda\rho}\,,\qquad\mathring{R}_{\mu\nu\rho\lambda}=\mathring{R}_{\rho\lambda\mu\nu}\,,
\end{equation}
and fulfills the (first and second, respectively) Bianchi identities:
\begin{equation}
\mathring{R}_{[\mu\nu\rho]}{}^{\lambda}=0\,,\qquad\mathring{\nabla}_{[\sigma}\mathring{R}_{\mu\nu]\rho}{}^{\lambda}=0\,.
\end{equation}
Consequently, there is only one independent trace (Ricci tensor)
\begin{equation}
\mathring{R}_{\mu\nu}=\mathring{R}_{\nu\mu}=-\mathring{R}^{(2)}{}_{\mu\nu}\,,\qquad \mathring{R}^{(3)}{}_{\mu\nu}=0\,.
\end{equation}
\end{prop}
}

\subsection{Linear connection in an arbitrary frame }

As a matter of generality it is better to work with the connection in an arbitrary frame. The transformation reads, in terms of the Vielbeins,
\begin{equation}
e^\nu{}_{a}e_{\lambda}{}^b \Gamma_{\mu\nu}{}^{\lambda}+e_{\sigma}{}^b \partial_{\mu}e^{\sigma}{}_a \eqqcolon\omega_{\mu a}{}^b \,,\label{eq: wconn def}
\end{equation}
which can be easily inverted
\begin{equation}
\omega_{\mu a}{}^{b}=e^\nu{}_{a}e_{\lambda}{}^b \Gamma_{\mu\nu}{}^{\lambda}+e_{\sigma}{}^b \partial_{\mu}e^{\sigma}{}_a \qquad\Leftrightarrow\qquad\Gamma_{\mu\nu}{}^{\lambda}=e_{\nu}{}^{a}e^{\lambda}{}_b \omega_{\mu a}{}^{b}+e^{\lambda}{}_a \partial_{\mu}e_{\nu}{}^a\,.
\end{equation}
Thus, $\omega_{\mu a}{}^{b}$ and $\Gamma_{\mu\nu}{}^{\rho}$ contain the same information, once the Vielbein are known.

The derivative associated to $\omega_{\mu a}{}^b$ acts as follows:
\begin{align}
\mathcal{D}_{\mu}T^{a...}{}_{b...} 
  & \coloneqq\partial_{\mu}T^{a...}{}_{\rho b...}+\omega_{\mu c}{}^a T^{c...}{}_{b...}+\text{...same for all upper Latin indices...}\nonumber \\
  & \qquad\qquad\qquad\,\,-\omega_{\mu b}{}^c T^{a...}{}_{c...}-\text{...same for all lower Latin indices...}\,.\label{eq:defcalD}
\end{align}
Coordinate indices are however transparent to this derivative, i.e., $\mathcal{D}_{\mu}T^\rho = \partial_\mu T^\rho $. It is then interesting to introduce the \emph{total covariant derivative} (the one that affects both kind of indices). We extend the notation $\nabla_{\mu}$ (see \eqref{eq: nabla def}) to represent this total covariant derivative. For a tensor density with weight $w$ under g.c.t., and both types of indices, we have
\begin{align}
\nabla_{\mu}\mathfrak{T}^{\nu a...}{}_{\rho b...} & \coloneqq\partial_{\mu}\mathfrak{T}^{\nu a...}{}_{\rho b...}+\Gamma_{\mu\sigma}{}^{\nu}\mathfrak{T}^{\sigma a...}{}_{\rho b...}+\omega_{\mu c}{}^a\mathfrak{T}^{\nu\rho c...}{}_{b...}+\text{...same for all upper indices...}\nonumber \\
 & \qquad\qquad\qquad\,\,-\Gamma_{\mu\rho}{}^{\sigma}\mathfrak{T}^{\nu a...}{}_{\sigma b...}-\omega_{\mu b}{}^c\mathfrak{T}^{\nu a...}{}_{\rho c...}-\text{...same for all lower indices...}\nonumber \\
 & \qquad\qquad\qquad\,\,+w\Gamma_{\mu\sigma}{}^{\sigma}\mathfrak{T}^{\nu a...}{}_{\rho b...}\,.
\end{align}
This derivative is both tensorial under g.c.t. and under frame transformations. This definition also implies that the Vielbein is covariantly constant
\begin{equation}
\nabla_{\mu}e^\nu{}_{a} = ... =(e_{\sigma}{}^b \partial_{\mu}e^{\sigma}{}_{a} +\Gamma_{\mu\sigma}{}^{\rho}e^{\sigma}{}_{a} e_{\rho}{}^{b} -\omega_{\mu a}{}^b )e^\nu{}_b \overeq{\eqref{eq: wconn def}}0\,.
\end{equation}

From here one can construct the curvature and the torsion associated to $\omega_{\mu a}{}^{b}$ through the analogue of \eqref{eq: commutator nabla},
\begin{equation}
\left[\nabla_{\mu},\,\nabla_{\nu}\right]V^{a}=R_{\mu\nu c}{}^{a}V^{c}+T_{\mu\nu}{}^{b}e^{\rho}{}_b \nabla_{\rho}V^a\,,
\end{equation}
and one gets
\begin{align}
R_{\mu\nu a}{}^{b} & =\partial_{\mu}\omega_{\nu a}{}^{b}-\partial_{\nu}\omega_{\mu a}{}^{b}+\omega_{\mu c}{}^b \omega_{\nu a}{}^{c}-\omega_{\nu c}{}^b \omega_{\mu a}{}^c\,,\\
T_{\mu\nu}{}^{c} & =\partial_{\mu}e_{\nu}{}^{c}-\partial_{\nu}e_{\mu}{}^{c}+\omega_{\mu b}{}^{c}e_{\nu}{}^{b}-\omega_{\nu b}{}^{c}e_{\mu}{}^b \qquad=\mathcal{D}_{\mu}e_{\nu}{}^{c}-\mathcal{D}_{\nu}e_{\mu}{}^c\,.
\end{align}
Observe that the torsion can be rewritten $T_{\mu\nu}{}^{c}=\Omega_{\mu\nu}{}^{c}+2\omega_{[\mu\nu]}{}^c$. Then,
\boxproposition{
\begin{prop}
The torsion components corresponds to the antisymmetric part of the connection if and only if the anholonomy is zero (i.e. if the frame is associated to some coordinates).
\end{prop}
}
\vspace{-0.3cm}

Note that everything is consistent. Since $\omega_{\mu a}{}^{b}$ is nothing but $\Gamma_{\mu\nu}{}^{\lambda}$ expressed in another frame, one can easily check that $R_{\mu\nu a}{}^b$ and $T_{\mu\nu}{}^c$ correspond to $R_{\mu\nu\rho}{}^{\lambda}$ and $T_{\mu\nu}{}^{\rho}$ defined in \eqref{eq:defcurvc} and \eqref{eq:deftorc} under the frame transformation:
\begin{equation}
R_{\mu\nu a}{}^{b}=R_{\mu\nu\rho}{}^{\lambda}e^{\rho}{}_{a}e_{\lambda}{}^b \,,\qquad T_{\mu\nu}{}^{c}=T_{\mu\nu}{}^{\rho}e_{\rho}{}^c\,.
\end{equation}

In the presence of a metric $g_{\mu\nu}=g_{ab}e_{\mu}{}^{a}e_{\nu}{}^{b}$, the components of its Levi-Civita connection in an arbitrary frame are equal to the Christoffel symbols of the anholonomic metric $g_{ab}$ plus an additional term that depends on the anholonomy coefficients:
\begin{align}
e^\mu{}_{c}\mathring{\omega}_{\mu a}{}^{b} & =e^\mu{}_{c}(e^\nu{}_{a}e_{\lambda}{}^b \mathring{\Gamma}_{\mu\nu}{}^{\lambda}+e_{\sigma}{}^b \partial_{\mu}e^{\sigma}{}_a )\\
 & =e^\mu{}_{c}e^\nu{}_{a}e_{\lambda}{}^b \frac{1}{2}g^{\rho\sigma}\left(\partial_{\mu}g_{\sigma\nu}+\partial_{\nu}g_{\mu\sigma}-\partial_{\sigma}g_{\mu\nu}\right)+e^\mu{}_{c}e_{\sigma}{}^b \partial_{\mu}e^{\sigma}{}_a\,,
\end{align}
which, after some abuses of notation can be express\footnote{To be precise $\Omega_{cad}\coloneqq\Omega_{ca}{}^{e}g_{ed}$ and $\mathring{\omega}_{cab}\coloneqq g_{db}e^\mu{}_{c}\mathring{\omega}_{\mu a}{}^{d}$.}
\begin{equation}
\boxed{\mathring{\omega}_{cab}=\frac{1}{2}\left(\partial_{c}g_{ba}+\partial_{a}g_{cb}-\partial_{b}g_{ca}\right)-\frac{1}{2}\left(\Omega_{cab}+\Omega_{bca}-\Omega_{abc}\right)}\,,
\end{equation}
i.e.,
\begin{equation}\label{eq: LCw decom}
\mathring{\omega}_{c(ab)} =\frac{1}{2}\partial_{c}g_{ab}\qquad
\mathring{\omega}_{c[ab]} =\partial_{[a}g_{b]c}-\Omega_{c[ab]}+\frac{1}{2}\Omega_{abc}\,.
\end{equation}
Since we have a metric, one can additionally define the nonmetricity tensor associated to the arbitrary connection $\omega_{\mu a}{}^{b}$,
\begin{equation}
Q_{\mu ab}=-\nabla_{\mu}g_{ab}=-\partial_{\mu}g_{ab}+2\omega_{\mu(a}{}^{c}g_{b)c}=-\partial_{\mu}g_{ab}+2\omega_{\mu(ab)}\qquad=-\mathcal{D}_{\mu}g_{ab}\,,\label{eq: Q for w}
\end{equation}
which, again, corresponds to the nonmetricity $Q_{\mu\nu\rho}$ translated to the general frame:
\begin{equation}
Q_{\mu ab}=Q_{\mu\nu\rho}e^\nu{}_{b}e^{\rho}{}_b \,.
\end{equation}

From \eqref{eq: Q for w} one can easily derive an interesting result:

\boxproposition{
\begin{prop}\label{prop:symconnzeroQ}
Consider a frame in which $g_{ab}$ is constant. A connection is metric-compatible (i.e. the nonmetricity vanishes) if and only if the object $\omega_{\mu ab}=\omega_{\mu a}{}^{c}g_{cb}$ is antisymmetric in the last two indices.
\end{prop}
}

\subsection{General decomposition of a connection}

\boxdefinition{
\begin{defn}
\textbf{(Distorsion)} The \emph{distorsion} tensor is the deviation of a connection with respect to the Levi-Civita connection of the metric, i.e.,
\begin{equation}
\Xi_{\mu a}{}^b \coloneqq\omega_{\mu a}{}^{b}-\mathring{\omega}_{\mu a}{}^b \,.
\end{equation}
\end{defn}
}

The distorsion can be split into two parts, one depending on the torsion (\emph{contorsion} tensor) and another one depending on the nonmetricity (\emph{disformation} tensor):
\begin{equation}
\boxed{\Xi_{cab}=\frac{1}{2}\left(T_{cab}+T_{bca}-T_{abc}\right)+\frac{1}{2}\left(Q_{cab}+Q_{abc}-Q_{bca}\right)}\,.\label{eq: distor decom}
\end{equation}
So, in general, an arbitrary linear connection takes the form
\begin{align}
\omega_{cab} & =\underbrace{\frac{1}{2}\left(\partial_{c}g_{ba}+\partial_{a}g_{cb}-\partial_{b}g_{ca}\right)-\frac{1}{2}\left(\Omega_{cab}+\Omega_{bca}-\Omega_{abc}\right)}_{\text{Levi-Civita connection}}\nonumber \\
 & \quad+\underbrace{\underbrace{\frac{1}{2}\left(T_{cab}+T_{bca}-T_{abc}\right)}_{\text{contorsion}}+\underbrace{\frac{1}{2}\left(Q_{cab}+Q_{abc}-Q_{bca}\right)}_{\text{disformation}}}_{\text{distorsion}}\,,
\end{align}
from which we can read
\begin{align}
\omega_{c(ab)} & =\frac{1}{2}(\partial_{c}g_{ab}+Q_{cab})\,,\\
\omega_{c[ab]} & =\partial_{[a}g_{b]c}+Q_{[ab]c}+(T_{c[ab]}-\Omega_{c[ab]})-\frac{1}{2}(T_{abc}-\Omega_{abc})\,.
\end{align}

\subsection{Connection and associated objects in differential form notation}

In metric-affine gravity it is especially useful to work in differential form notation, and define a set of new variables different from $\{g_{\mu\nu},\,\Gamma_{\mu\nu}{}^{\rho}\}$, although at the end of the day the results are equivalent. The set of fundamental variables that we will consider in the physical framework is made of three objects: the anholonomic metric $g_{ab}$ (which, as we will see, can be fixed by a gauge), the coframe $\cofr^{a}$ and the \emph{connection 1-form} defined as
\begin{equation}
\boxed{\dfom_{a}{}^b \coloneqq\omega_{\mu a}{}^b \dex x^{\mu}}\,.
\end{equation}
In the language of differential forms all of the objects that are going to appear have their coordinate indices (if there are any) hidden, so no Greek indices are going to appear in our expressions. We will only have to deal with Latin indices ($a,b,c,...$) and internal indices for the matter fields. However, let us forget about the latter for simplicity (they will be introduced in the next chapter).

\boxdefinition{
\begin{defn}
\textbf{(Metric-affine geometry)} A tuple $(g_{ab},\,\cofr^a,\,\dfom_a{}^b)$ is called a \emph{metric-affine geometry} over the considered manifold.
\end{defn}
}

If we move on in our description in terms of differential forms, the next step would be to generalize appropriately the notions of covariant derivative, curvature, torsion and nonmetricity. Such generalization is quite simple and can be easily understood. Concerning the covariant derivative, instead of generalizing $\partial_{\mu}$, the idea will be to generalize the exterior derivative $\dex$. We will do it as follows

\boxdefinition{
\begin{defn}
\textbf{(Exterior covariant derivative)} The \emph{exterior covariant derivative} acting on an arbitrary tensor-valued differential form $\dfal_{a...}{}^{b...}$ is defined as the operator:
\begin{align}
\Dex\dfal_{a...}{}^{b...} & \coloneqq\dex\dfal_{a...}{}^{b...}+\dfom_{c}{}^b \wedge\dfal_{a...}{}^{c...}+\text{...same for all upper indices...}\nonumber \\
 & \qquad\qquad-\dfom_{a}{}^c\wedge\dfal_{c...}{}^{b...}-\text{...same for all lower indices...}
\end{align}
\end{defn}
}

If we extract the components:

\boxproposition{
\begin{prop}
For an arbitrary tensor-valued $k$-form $\dfal_{a...}{}^{b...}$,
\begin{align}
\Dex\dfal_{a...}{}^{b...} & =\frac{1}{k!}\mathcal{D}_{[\rho}\alpha_{\mu_{1}...\mu_{k}]a...}{}^{b...}\dex x^{\rho\mu_{1}...\mu_{k}}\\
 & =\frac{1}{k!}\left(\nabla_{[\rho}\alpha_{\mu_{1}...\mu_{k}]a...}{}^{b...}+\frac{k}{2}T_{[\rho\mu_1}{}^{\sigma}\alpha_{|\sigma|\mu_{2}...\mu_{k}]a...}{}^{b...}\right)\dex x^{\rho\mu_{1}...\mu_{k}}\,.
\end{align}
\end{prop}
}

Now consider the covariant objects constructed with the derivatives of the basic fields. For the coframe and the metric, we can directly write $\Dex\cofr^{a}$ and $\Dex g_{ab}$, but since $\dfom_{a}{}^{b}$ is not tensorial in its external indices (i.e., the connection is not a tensor-valued form), one cannot write ``$\Dex\dfom_{a}{}^{b}$''. However it is not difficult to check that the combination $\dex\dfom_{a}{}^{b}+\dfom_{c}{}^b \wedge\dfom_{a}{}^{c}$ is a tensor valued 2-form. With all of this in mind, we introduce the following definitions:

\boxdefinition{
\begin{defn}
\textbf{(Curvature, torsion, nonmetricity and distorsion forms)} We define the \emph{curvature 2-form}, the \emph{torsion 2-form} and the \emph{nonmetricity 1-form} associated to some metric-affine geometry $(g_{ab},\,\cofr^{a},\,\dfom_{a}{}^{b})$, respectively as: 
\begin{align}
\dfR_{a}{}^{b} & \coloneqq\dex\dfom_{a}{}^{b}+\dfom_{c}{}^b \wedge\dfom_{a}{}^c\,,\\
\dfT^{a} & \coloneqq\Dex\cofr^a\,,\\
\dfQ_{ab} & \coloneqq-\Dex g_{ab}\,.
\end{align}
We can also define a \emph{distorsion 1-form} representing the difference with respect to the Levi-Civita connection 1-form
\begin{equation}
\dfXi_{a}{}^b \coloneqq\dfom_{a}{}^{b}-\mathring{\dfom}_{a}{}^b \,.
\end{equation}
\end{defn}
}

\noindent It is not difficult to prove that, once we extract the components of these differential forms, what we get is nothing but the components of the tensors we defined in previous sections:
\begin{equation}
\dfR_{a}{}^{b}=\frac{1}{2}R_{\mu\nu a}{}^b \dex x^{\mu\nu}\,,\quad\dfT^{a}=\frac{1}{2}T_{\mu\nu}{}^a\dex x^{\mu\nu}\,,\quad\dfQ_{ab}=Q_{\mu ab}\dex x^{\mu}\,,\quad\dfXi_{a}{}^{b}=\Xi_{\mu a}{}^b \dex x^{\mu}\,.
\end{equation}
It is specially interesting to decompose the torsion, the curvature and the nonmetricity forms according to the irreducible representations of the group $\mathrm{GL}(\dimM,\mathbb{R})$. But, since we work in the presence of a Lorentzian metric, we can go even further and decompose those parts into other smaller ones with respect to the pseudo-orthogonal group $\mathrm{SO}(1,\dimM-1)$ (basically by extracting the traces). Indeed, these irreducible parts will be key ingredients in many situations within metric-affine gravity and related theories. See Appendix \ref{app:irreds} for their explicit expressions together with some useful properties.

\boxproposition{
\begin{prop}
Consider two connection 1-forms $\dfom'{}_{a}{}^{b}$ and $\dfom_{a}{}^{b}=\dfom'{}_{a}{}^{b}+\dfA_{a}{}^{b}$.
Then, their curvatures are related via
\begin{equation}
\dfR_{a}{}^{b}=\dfR'{}_{a}{}^{b}+\Dex'\dfA_{a}{}^{b}+\dfA_{c}{}^b \wedge\dfA_{a}{}^c\,.
\end{equation}
\end{prop}
}

\noindent This is the equivalent in differential form notation of the expression \eqref{eq:RtoRprima}. In particular, this can be applied to the Levi-Civita connection, which, in terms of the distorsion reads
\begin{equation}
\dfR_{a}{}^{b}=\mathring{\dfR}_{a}{}^{b}+\mathring{\Dex}\dfXi_{a}{}^{b}+\dfXi_{c}{}^b \wedge\dfXi_{a}{}^c\,.\label{eq: R RLC distor}
\end{equation}

\boxproposition{
\begin{prop}
\textbf{\textup{\label{prop:BianchiTQR}(Bianchi identities)}} The curvature, torsion and nonmetricity forms fulfill
\begin{align}
\Dex\dfR_{a}{}^{b} & =0\,,\\
(\Dex\Dex\cofr^{a}=)\qquad\Dex\dfT^{a} & =\dfR_{b}{}^a\wedge\cofr^b \,,\\
(-\Dex\Dex g_{ab}=)\qquad\Dex\dfQ_{ab} & =2\dfR_{(ab)}\,.
\end{align}
\end{prop}
}

\noindent These are nothing but the expressions \eqref{eq:2BianchiIdGen}, \eqref{eq:BianchiIdTorcom} and \eqref{eq:BianchiIdQcom}, respectively, rewritten in another language. Here one can see how useful and compact can be the differential form notation. In addition, applying twice the exterior covariant derivative only generates curvature terms:

\boxproposition{
\begin{prop}
For an arbitrary tensor-valued differential form $\dfal_{a...}{}^{b...}$,
\begin{align}
\Dex\Dex\dfal_{a...}{}^{b...} & =\dfR_{c}{}^b \wedge\dfal_{a...}{}^{c...}+\text{...same for all upper indices...}\nonumber \\
 & \qquad-\dfR_{a}{}^c\wedge\dfal_{c...}{}^{b...}-\text{...same for all lower indices...}\,.
\end{align}
\end{prop}
}

Other useful formulae:
\begin{equation}
\Dex\mathcal{E}_{a_{1}...a_{\dimM}}=-\frac{1}{2}\dfQ_{c}{}^c\mathcal{E}_{a_{1}...a_{\dimM}}\,,\label{eq: Dex del tens LC}
\end{equation}
\begin{equation}
\Dex\star\cofr_{a_{1}...a_{k}}=-\frac{1}{2}\dfQ_{c}{}^c\wedge\star\cofr_{a_{1}...a_{k}}+\dfT^c\wedge\star\cofr_{a_{1}...a_{k}c}\,.\label{eq: Dstarcofr}
\end{equation}

Finally, let us mention that metric-affine geometries are also commonly called in the literature post-Riemannian or non-Riemannian geometries, and depending on the properties of the connection, they receive special names:
\begin{enumerate}
\item \emph{Riemannian geometry}, if the connection is Levi-Civita.\footnote{
    Riemannian/post-Riemannian geometries should not be confused with the concepts of Riemannian/pseudo-Riemannian metrics we previously introduced and that refer to the signature of the metric.}
\item \emph{Riemann-Cartan geometry}, if the connection is metric-compatible (arbitrary torsion).
\item \emph{Weyl-Cartan geometry}, if the nonmetricity is purely Weyl, i.e. $Q_{\rho\mu\nu}=\frac{1}{\dimM}Q_\rho g_{\mu\nu}$.
\item \emph{Torsion-free geometry} (alse called symmetric), if the connection is torsion-free  (the nonmetricity is arbitrary).
\item \emph{Teleparallel geometry}, if the curvature of the connection is identically zero. Subcases of this are:
  \begin{enumerate}
  \item \emph{Symmetric teleparallel geometry}, if, in addition, the torsion is zero ($R=T=0$).
  \item \emph{Weitzenb\"ock geometry}, if, in addition the nonmetricity vanishes ($R=Q=0$).
  \end{enumerate}
\end{enumerate}

\newpage
\section{Miscellany}

\subsection{Autoparallels, geodesics and congruences}

If we have a connection, we can define the notion of autoparallel curves:

\boxdefinition{
\begin{defn}
\textbf{(Pre-autoparallel and autoparallel)} Consider a manifold with a connection $\Gamma_{\mu\nu}{}^{\rho}$. A \emph{pre-autoparallel} is a curve $\gamma(\tau)$ whose velocity $\dot{\gamma}(\tau)=u^{\mu}\vpartial_{\mu}$ is invariant under parallel transport up to a term proportional to the velocity, i.e.,
\begin{equation}
(u^{\nu}\nabla_{\nu}u^{\mu}\equiv)\qquad\frac{\mathrm{d}u^{\mu}}{\mathrm{d}\tau}+\Gamma_{\nu\rho}{}^{\mu}u^{\nu}u^{\rho}=f(\tau)\,u^{\mu}\,.
\end{equation}
A pre-autoparallel with $f=0$ is called simply \emph{autoparallel}.
\end{defn}
}

\noindent Given a pre-autoparallel, there is always an autoparallel with the same image. Essentially they represent the same path over the manifold but with a different parameterization. The parameter for which $f=0$ is called \emph{affine parameter}. 

\boxdefinition{
\begin{defn}
\textbf{(Pre-geodesics and geodesics)} The (pre-)autoparallels of the Levi-Civita connection of some metric $\teng$ are called \emph{(pre)-geodesics} of the metric.
\end{defn}
}

\noindent Indeed, if we have a metric tensor $\teng$ in the manifold, there is a canonical notion of length: for a given curve $\gamma\,:\,[\tau_{i},\tau_{f}]\rightarrow\mathcal{\mathcal{U}}$ with velocity $u^{\mu}$, then
\[
\text{Length}_{\teng}(\gamma)\coloneqq\int_{\tau_{i}}^{\tau_{f}}\sqrt{\left|g_{\mu\nu}(\gamma(\tau))\,u^{\mu}(\tau)\,u^{\nu}(\tau)\right|}\dex\tau\,.
\]
This quantity is independent of the parameterization. One can show that indeed the (pre-)geodesics are those curves of stationary length, i.e., those such that 
\begin{equation}
\delta\text{Length}_{\teng}(\gamma)=0\,.
\end{equation}

Notice that, these two families of paths do not coincide in general:
\begin{equation}
u^{\nu}\nabla_{\nu}u^{\mu}-u^{\nu}\mathring{\nabla}_{\nu}u^{\mu}=\Xi_{(\nu\rho)}{}^{\mu}u^{\nu}u^{\rho}\overeq{\eqref{eq: distor decom}}\left[T^\mu{}_{(\nu\rho)}+Q_{(\nu\rho)}{}^{\mu}-\frac{1}{2}Q^\mu{}_{\nu\rho}\right]u^{\nu}u^{\rho}\,.
\end{equation}

\boxdefinition{
\begin{defn}
\textbf{(Congruence)} A (smooth) \emph{congruence} on $\mathcal{U}$ is a family of (smooth) curves on $\mathcal{U}$ such that each point of $\mathcal{U}$ belongs to the image of one and only one curve of the family.
\end{defn}
}

The set of velocities $u^{\mu}\vpartial_{\mu}$ of a smooth congruence in each point defines a smooth vector field on $\mathcal{U}$. In the presence of a metric, the associated 1-form $\dfu=u_{\mu}\dex x^{\mu}$ can be expressed as an exact form,
\begin{equation}
u_{\mu}=\partial_{\mu}u\qquad\Leftrightarrow\qquad\dfu=\dex u\,,
\end{equation}
i.e. the gradient of a (coordinate) function $u$. This can always be completed to give a chart that covers the entire $\mathcal{U}$. These sets of coordinates are called \emph{adapted coordinates} for the congruence. We will say that a congruence is \emph{(pre-)geodetic} if the curves that constitute the congruence are (pre-)geodesics. Similarly we will say that a congruence is \emph{timelike/lightlike}... if the velocity vector field is timelike/lightlike/... everywhere in $\mathcal{U}$.

\subsection{Lie derivative and covariant Lie derivative}\label{sec:Lieder}

\boxdefinition{

Consider a vector field $\vecV=V^{\mu}\vpartial_{\mu}\in\mathfrak{X}(\mathcal{M})$.
\begin{defn}
\textbf{\label{Def: Lieder}(Lie derivative)} 

The \emph{Lie derivative with respect to} $\vecV$ is an operator $\dLie_{\vecV}\,:\,\mathscr{T}^{(r,\,s)}(\mathcal{M})\rightarrow\mathscr{T}^{(r,\,s)}(\mathcal{M})$ that acting on an arbitrary tensor field $\tenS\in\mathscr{T}^{(r,\,s)}(\mathcal{M})$ with components $S^{\mu_{1}...\mu_{r}}{}_{\nu_{1}...\nu_{s}}$, gives another tensor in the same space with components:
\begin{align}
\left(\dLie_{\vecV}(\tenS)\right)^{\mu_{1}...\mu_{r}}{}_{\nu_{1}...\nu_{s}} & \coloneqq V^{\sigma}\partial_{\sigma}S^{\mu_{1}...\mu_{r}}{}_{\nu_{1}...\nu_{s}}\nonumber \\
 & \qquad-\partial_{\sigma}V^{\mu_1}S^{\sigma...\mu_{r}}{}_{\nu_{1}...\nu_{s}}-\text{...same for all upper indices...}\label{eq: LieDerDef}\\
 & \qquad+\partial_{\nu_1}V^{\sigma}S^{\mu_{1}...\mu_{r}}{}_{\sigma...\nu_{s}}+\text{...same for all lower indices...}\nonumber 
\end{align}
We are going to use the usual notation in the physics literature, $\dLie_{\vecV}S^{\mu_{1}...\mu_{r}}{}_{\nu_{1}...\nu_{s}}$.
\end{defn}
}

The particular cases of a scalar, a vector and the metric are, respectively,
\begin{equation}
\dLie_{\vecV}f =V^{\sigma}\partial_{\sigma}f\,,\quad
\dLie_{\vecV}W^{\mu} =V^{\sigma}\partial_{\sigma}W^{\mu}-\partial_{\sigma}V^{\mu}W^{\sigma}=\left[\vecV,\,\vecW\right]^{\mu}\,,\quad
\dLie_{\vecV}g_{\mu\nu}=2\mathring{\nabla}_{(\mu}V_{\nu)}\,.
\end{equation}
It is important to recall that the Lie derivative is an operator that can be constructed in any smooth manifold (with no further structure), i.e. it is connection-independent.

If we apply the Lie derivative to the elements of $\Omega^{k}(\mathcal{M})$, it can be shown that $\dLie_{\vecV}$ is nothing but a very simple combination of the exterior derivatives and interior products:
\begin{equation}
\dLie_{\vecV}=\dex\circ(\dint{\vecV})+(\dint{\vecV})\circ\dex\,.\label{eq: Lieder df}
\end{equation}
This is sometimes called the \emph{Cartan magic formula}.

This definition for scalar-valued differential forms can be extended to tensor-valued forms in order to be covariant under internal transformations (in the space in which the differential form takes values). We define then:\footnote{
    For an exhaustive study of Lie derivatives and generalizations see the PhD Thesis \cite{Matteucci2003_PhDThesis}.}

\boxdefinition{
\begin{defn}
\textbf{\label{Def: LieDer}(Covariant Lie derivative)} 

The \emph{covariant Lie derivative with respect to the vector field  $\vecV\in\mathfrak{X}(\mathcal{M})$ and the linear connection $\dfom_a{}^b$} is the operator defined over the space of tensor-valued differential forms given by
\begin{equation}
\DLie_{\vecV}=\Dex\circ(\dint{\vecV})+(\dint{\vecV})\circ\Dex\,,
\end{equation}
where $\Dex$ is the exterior covariant derivative of  $\dfom_a{}^b$.
\end{defn}
}
For example, when dealing with $k$-forms with tensor values of the type $\dfal_{a...}{}^{b...}$, the ordinary Lie derivative \eqref{eq: Lieder df} is covariant only under diffeomorphisms, but not under local $\mathrm{GL}(\dimM,\mathbb{R})$ transformations of the frame. However, the covariant version in Definition \ref{Def: LieDer} is well-behaved under both types of transformations.

\chapter{Metric-Affine gauge theories\label{ch:MAGfoundations}}

\boxquote{We are all agreed that your theory is crazy. The question that divides us is whether it is crazy enough to have a chance of being correct.}{Niels Bohr, said to W. Pauli after presenting his (and Heisenberg's) nonlinear field theory of elementary particles at Columbia U. (1958).}

The idea of this chapter is to introduce the gauge approach to metric-affine gravity \cite{Hehl1995,Blagojevic2001,BlagojevicHehl2012,PonomarevObukhov2017}. We will describe the fundamental geometrical construction without entering in subtle mathematical details, as well as the gravitational currents and Noether identities for a generic action. At the end, we will build the general quadratic MAG Lagrangian, its associated objects and have a look at the Einstein-Palatini theory, as a particular example.

\section{Gauge theory: connections in principal bundles}

Geometrically, a \emph{gauge theory} can be described in terms of connections defined over principal bundles and sections of associated bundles to it \cite{Trautman1981}. The aim of this first section is to clarify this statement, specially for the case of internal symmetries (e.g., the usual Yang-Mills theory).

\subsection{Principal connection and gauge fields}

In a general bundle $\mathcal{B}\xrightarrow{\pi}\mathcal{M}$, the directions tangent to the fibers are called ``vertical'' directions. Let us call $V_u\mathcal{B}$ the vertical space at the point $u\in\mathcal{B}$ and the corresponding bundle $V\mathcal{B}\coloneqq\sqcup_u V_u\mathcal{B}$. Nonetheless, there is not such thing as a canonical notion of ``horizontal'' directions in the total space. One can smoothly define a distribution of subspaces $H_u$ (called \emph{horizontal subspaces}) at each point, in such a way that
\begin{equation}
  T_u\mathcal{B} = V_u\mathcal{B} \oplus H_u\qquad \forall u\in\mathcal{B}\,.
\end{equation}
The distribution $\{H_u\}$ is what is called a \emph{connection} over the bundle (sometimes called \emph{Ehresmann connection}).\footnote{
    See \cite{Goldberg2008} for a very pedagogical introduction to the concept of connection.} 
This definition, based on horizontal subspaces, does not look very practical. Interestingly, in particular types of bundles one can look for \emph{connection 1-forms}, which are in one-to-one correspondence to connections. One example, as we will see in the following sections, is the object $\dfom_a{}^b$ we defined in the previous chapter. 

Let us now jump to the particular case of principal bundles, i.e. those that locally look as the base manifold times a Lie group (see Definition \ref{def:princbundle}). In this case, we ask the distribution of horizontal subspaces to respect the right action of the group over the fibers. To be precise, what we mean is that if we translate an horizontal subspace to another point of the same fiber with the right action, the result must be the horizontal space at that point. These connections are called \emph{principal connections} and are in one-to-one correspondence with \emph{principal connection 1-forms} (see Definition \ref{eq:principalcon}). Let us briefly see how this works. 

Let $\mathcal{P}\xrightarrow{\pi}\mathcal{M}$ be a principal bundle with structure group $\mathcal{G}$ and right action $\mathcal{R}_g$ ($g\in\mathcal{G}$). Let us also denote the Lie algebra of the structure group as $\mathfrak{g}$ and its Lie bracket as $[\cdot,\cdot]$. 

First we introduce the concept of fundamental field:
\boxdefinition{
\begin{defn}
\textbf{(Fundamental vector field)} The \emph{fundamental vector field} associated with $\vecT\in\mathfrak{g}$ is the vector field  $\vecT^\#\in\mathfrak{X}(\mathcal{P})$ given at each point $u\in\mathcal{P}$ by
\begin{equation}
  \vecT^\# |_u \coloneqq \left. \frac{{\rm d}}{{\rm d} t}\right|_{t=0} \left(\mathcal{R}_{\exp(t\vecT)}(u)\right)\,,
\end{equation}
where $\exp$ is the usual exponential map from the Lie algebra to the group.\footnotemark
\end{defn}
}\footnotetext{
    For those familiar with Lie group theory, this definition is quite interesting. The elements of the Lie algebra (seen as $T_e\mathcal{G}$), $\{\vecT\}$, are in one-to-one correspondence to the so-called \emph{left-invariant vector fields}, $\{L(\vecT)\}$. Fundamental fields constitute the canonical construction based on left-invariant vector fields and the fact that the fibers are homeomorphic to the group.}
The fundamental fields are vertical fields (tangent to the fibers) and constitute a basis of the $C^\infty$-module of vertical vector fields $\Gamma(V\mathcal{B})$. This concept allows us to define:

\boxdefinition{
\begin{defn}\label{eq:principalcon}
\textbf{(Principal connection 1-form)} A $\mathfrak{g}$-valued 1-form over the total space, $\dfom\in\Omega^1(\mathcal{P};\mathfrak{g})$, is called a \emph{principal connection 1-form} if the following requirements are fulfilled:
\begin{itemize}
\item $\dfom (\vecT^\#) = \vecT$ \quad $\forall\vecT\in\mathfrak{g}$.
\item $\dfom (\vecX) = \mathrm{Ad}_{g} \left(\left(\mathcal{R}_{g}{}^{*}\dfom\right)(\vecX)\right)$ \quad $\forall g\in\mathcal{G}$~~and~~ $\forall\vecX\in\mathfrak{X}(\mathcal{P})$.
\end{itemize}
Here $\mathrm{Ad}$ is the adjoint representation on the Lie algebra.
\end{defn}
}

In principal bundles, the one-to-one correspondence between the description with horizontal subspaces and the one with connection 1-forms goes as follows:
\begin{itemize}
\item ($\Rightarrow$) For a given principal connection $\{H_u\}$, we can define the $\mathfrak{g}$-valued 1-form  $\dfom (\vecX) \coloneqq \vecT $ where $\vecT$ is the only element of $\mathfrak{g}$ such that $\vecT^\#$ is the vertical part of $\vecX$. It can be shown that this $\dfom$ verifies the two conditions in Definition \ref{eq:principalcon}.
\item ($\Leftarrow$) For a given principal connection 1-form, we can introduce the distribution of subspaces $H_u \coloneqq \ker(\dfom|_u)$, which turns out to be a principal connection. 
\end{itemize}
This latter point is remarkable: \emph{a vector is horizontal with respect to a given connection if and only if the associated connection 1-form vanishes on it}. 

Before continuing, let us introduce a notion directly associated with the connection:\footnote{
    Formally, it is more instructive to introduce the notion of exterior covariant derivative in the principal bundle and then prove the equation \eqref{eq:structOmega} (called \emph{structure equation}) as a corollary. However our idea is to avoid details that do not contribute to a general understanding of the structure of gauge theories.}

\vspace{-1mm}
\boxdefinition{
\begin{defn}\label{def:curvatureform}
\textbf{(Curvature 1-form)} The curvature 2-form associated with a principal connection 1-form is the object $\dfOm\in\Omega^2(\mathcal{P};\mathfrak{g})$ defined by
\begin{equation}
   \dfOm \coloneqq \dex \dfom^\mathfrak{a}\otimes{\rm J}_\mathfrak{a} + {\frac 12}(\dfom^\mathfrak{a}\wedge\dfom^\mathfrak{b})\otimes [{\rm J}_\mathfrak{a}, {\rm J}_\mathfrak{b}]\,, \label{eq:structOmega}
\end{equation}
where  $\{{\rm J}_\mathfrak{a}\}$ is some arbitrary basis of the Lie algebra $\mathfrak{g}$.
\end{defn}
}

This surely continues being very abstract. But now we proceed to relate all of this with the very physical concept of \emph{gauge field}.

\boxdefinition{
\begin{defn}
\textbf{(Gauge section. Gauge transformation. Potential and field strength)} A \emph{gauge section} (or, in more physical terms, a \emph{gauge choice}) is a section of the principal bundle $\sigma\in\Gamma(\mathcal{P})$. 

Given a principal connection 1-form $\dfom$, its pullback\footnotemark {} with respect to a gauge section, $\dfA \coloneqq \sigma^* \dfom$, is called the \emph{gauge field} or \emph{gauge potential}. The pullback of the associated curvature, $\dfF \coloneqq \sigma^* \dfOm$, is called the \emph{gauge field strength}.

We define the \emph{gauge transformation} with group element $g:\mathcal{P}\to \mathcal{G}$ to be the change of gauge section $\sigma\to\sigma'$ given by
\begin{equation}
  \sigma(p) \xrightarrow{g}  \sigma'(p) \coloneqq \mathcal{R}_{g(\sigma(p))^{-1}} \big( \sigma(p)\big)\,.
\end{equation}
\end{defn}
}\footnotetext{See the definition of pullback and pushforward in Appendix \ref{app:pullback}.} Notice that the gauge field and the gauge field strength are also $\mathfrak{g}$-valued differential forms but over the base manifold $\mathcal{M}$, so we can expand them in a particular coordinate basis $\dfA = \dex x^\mu \otimes \dfA_\mu$ and $\dfF = \frac{1}{2} (\dex x^\mu\wedge \dex x^\nu)\otimes \dfF_{\mu\nu}$ (where $\dfA_\mu$ and $\dfF_{\mu\nu}$ are $\mathfrak{g}$-valued functions). Indeed, it can be checked from Definition \ref{def:curvatureform} that
\begin{equation}
  \dfF_{\mu\nu}= (\partial_\mu A_\nu{}^\mathfrak{a} - \partial_\nu A_\mu{}^\mathfrak{a} + f_\mathfrak{bc}{}^\mathfrak{a} A_\mu{}^\mathfrak{b} A_\nu{}^\mathfrak{c} )\ {\rm J}_\mathfrak{a} ,\qquad [{\rm J}_\mathfrak{b}, {\rm J}_\mathfrak{c}] \eqqcolon f_\mathfrak{bc}{}^\mathfrak{a} {\rm J}_\mathfrak{a}\,.
\end{equation}

From now on (as it is usual in physics), we are going to assume that the structure group is a \textit{matrix} Lie group, i.e., a closed subgroup of $\mathrm{GL}(n,\mathbb{C})$ for some $n$. In that case, the corresponding Lie algebra is also a matrix space and we can represent $\mathfrak{g}$-valued forms as matrices whose elements are (scalar-valued) differential forms. This simplifies the formal expressions of bundle theory enormously; for instance, the formulas that express how the gauge field and the field strength change under a gauge transformation with group element $g$ are\footnote{
    Let us insist on that this is an abuse of notation. One interested in understanding deeply how these structures work should be able to derive and work at the abstract level. The actual formulae involve the adjoint representation and the Maurer-Cartan form of the structure group. However, again, we are trying to avoid introducing many definitions.}
\begin{equation}\label{eq:transfAF}
   \dfA'{}_\mu =  g\!\cdot\!\dfA_\mu\!\cdot\!g^{-1}  +  g\!\cdot\!\partial_\mu g^{-1}\,,\qquad \dfF'{}_{\mu\nu} =  g\!\cdot\!\dfF_{\mu\nu} \!\cdot\!g^{-1}
\end{equation}
where $\cdot$ is just the matrix multiplication (which we will omit from now on) and $\partial_\mu$ acts on each element of the matrix that follows it. Here we recognize the transformation rules of the usual Yang-Mills connection, where $g\in\mathrm{SU}(n)$.\footnote{
    In physics, some imaginary factors $\iN$ appear, because it is more convenient to work in the complexified algebra. In particular, it is usual to do a substitution of the type ${\rm J}_\mathfrak{a}\to \iN {\rm J}_\mathfrak{a}$, to ensure that the generators ${\rm J}_\mathfrak{a}$ are Hermitian matrices.}

\subsection{Matter fields in standard gauge theory}

We have all of the ingredients of a gauge theory except two: the matter fields and the covariant derivative that acts on them. Let us quickly revise how this is built in bundle theory. 

The matter fields in physics are usually elements of some vector space $W$ associated with a linear representation of the structure group. Formally, the idea is the following: first we take the representation over the vector space where our matter field takes values, i.e. a smooth map
\begin{equation}
  \rho : \mathcal{G} \longrightarrow \mathrm{GL}(W)
\end{equation}
such that $\rho(gh)=\rho(g)\cdot\rho(h)$ for all $g,h\in\mathcal{G}$. Secondly, we introduce the following natural right action $\Phi : g \mapsto \Phi_g$ ($g\in\mathcal{G}$) over the product $\mathcal{P}\times W$:
\begin{equation}
  \Phi_g(u, \vecv) \coloneqq \left(\mathcal{R}_g (u),\  \rho(g^{-1})\,(\vecv)\right)\,.
\end{equation}
Now we identify elements connected by this action and construct the corresponding orbit space (which has a manifold structure):
\begin{equation}
  E\equiv \mathcal{P}\times_\rho W \coloneqq (\mathcal{P}\times W) / \Phi \,.
\end{equation}
By using local trivializations of the principal bundle $\mathcal{P}$, one can construct local trivializations $\mathcal{U}\times W \to E$ (for some open set $\mathcal{U}\subset\mathcal{M}$), i.e., $E$ locally looks as the product $\mathcal{M}\times W$. Furthermore, we can introduce a projection map  $\pi_E ([u, \vecv]) \coloneqq \pi_\mathcal{P} (u) $. We realize that this orbit space has a bundle structure:
\boxdefinition{
\begin{defn}\label{def:assocBmatter}
\textbf{(Associated bundle. Matter fields)} 
The fiber bundle we have constructed above is called \emph{associated bundle} of $\mathcal{P}$ with respect to the representation $\rho$. 

Given an associated bundle, a \emph{matter field} of the gauge theory is a section (local or global) of it $\varPsi\in \Gamma(E)$.
\end{defn}
}

The key point of all of this construction is that the principal connection 1-form $\dfom$ in $\mathcal{P}$ induces \emph{canonically} a derivation of matter fields in any of its associated bundles:
\begin{itemize}
\item On the one hand, remember that the representation $\rho$ is a smooth map between manifolds. Thus, we can take its differential (pushforward) at the identity,
\begin{equation}\label{eq:rhobarrho}
  \bar\rho\coloneqq (\rho_*)_e : \mathfrak{g} \longrightarrow \mathfrak{gl}(W) \cong \{(\dim W)\text{-dimensional real matrices}\}.
\end{equation}
This is indeed a representation of the Lie algebra over the same vector space $W$. We can then use the gauge field $\dfA$ to construct the following object:
\begin{equation}
  \dfA= A_\mu{}^\mathfrak{a} \dex x^\mu \otimes {\rm J}_\mathfrak{a} \qquad \longrightarrow \qquad
  \bar\rho (\dfA) \coloneqq A_\mu{}^\mathfrak{a} \dex x^\mu \otimes \bar\rho({\rm J}_\mathfrak{a})\,.
\end{equation}

\item On the other hand, consider a matter field compatible with the gauge choice $\sigma : \mathcal{U}\to \mathcal{P}$, i.e., a section of the associated bundle of the type:
\begin{equation}
  \varPsi (p) \coloneqq [(\sigma(p), \varPsi^\mathtt{M} \vecv_\mathtt{M})] \equiv \varPsi^\mathtt{M} (p) \vecxi_\mathtt{M}(p) .
\end{equation}
where $\{\vecv_\mathtt{M}\}$ is some basis of $W$, and $\{\vecxi_\mathtt{M}(p) \coloneqq[(\sigma(p),\vecv_\mathtt{M})]\}$ is a frame of matter fields over $\mathcal{U}$ under the gauge choice $\sigma$.
\end{itemize}

\boxdefinition{
\begin{defn}
\textbf{(Covariant derivative of matter fields)} The \emph{covariant derivative} along a certain vector field $\vecX\in\mathfrak{X}(\mathcal{U})$ is the map 
\begin{align}
\mathcal{D}_{\vecX}\ :\ \Gamma(E) & \longrightarrow\Gamma(E)\nonumber\\
\varPsi &\longmapsto \mathcal{D}_{\vecX}\varPsi \coloneqq X^\mu (\mathcal{D}_\mu\varPsi)  \,,
\end{align}
(i.e., the result is another matter field of the same type) where
\begin{equation}
(\mathcal{D}_\mu \varPsi) (p) \coloneqq \Big(\partial_\mu \varPsi^\mathtt{M} + A_\mu{}^\mathfrak{a} (\bar\rho({\rm J}_\mathfrak{a}))^\mathtt{M}{}_\mathtt{N}\,\varPsi^\mathtt{N} \Big) \vecxi_\mathtt{M}(p)\,.
\end{equation}
\end{defn}
}
In differential form notation $(\mathcal{D}_{\vecX} \varPsi) (p) \equiv \dint{\vecX}\Dex \varPsi$,
\begin{equation}
\Dex \varPsi (p) = \Big(\dex \varPsi^\mathtt{M} + A_\mu{}^\mathfrak{a} \dex x^\mu (\bar\rho({\rm J}_\mathfrak{a}))^\mathtt{M}{}_\mathtt{N}\,\varPsi^\mathtt{N} \Big) \vecxi_\mathtt{M}(p)\,.
\end{equation}

These matter fields are $W$-valued functions (sections), but this definition can be extended to $W$-valued differential forms straightforwardly:
\begin{equation}
\Dex \dfvarPsi(p) = \Big(\dex \dfvarPsi^\mathtt{M} + A_\mu{}^\mathfrak{a} \dex x^\mu \wedge \big((\bar\rho({\rm J}_\mathfrak{a}))^\mathtt{M}{}_\mathtt{N}\,\dfvarPsi^\mathtt{N}\big) \Big)\otimes\vecxi_\mathtt{M}(p)\,.
\end{equation}

\section{Gauge approach to metric-affine gravity}

The way the fundamental gravitational fields arise from a gauge approach is not as direct as in Yang-Mills theory. Let us assume that we want to obtain a metric-affine geometry $\{g_{ab},\cofr^a,\dfom_a{}^b\}$ from a gauge construction, i.e., two 1-form fields $\dfom_a{}^b$ and $\cofr^a$, and a Lorentzian metric $g_{ab}$ over the manifold that transform under a $\mathrm{GL}(\dimM,\mathbb{R})$ transformation $\{\vfre_a\} \to \{\vfre'{}_a=\vfre_b (M^{-1})^b{}_a(x)\}$ as follows:
\begin{equation}
  \dfom_a{}^b \to \dfom'{}_a{}^b=  M^b{}_d\, \dfom_c{}^d\, (M^{-1})^c{}_a + M^a{}_c\, \dex (M^{-1})^c{}_b\,,\nonumber
\end{equation}
\begin{equation}
  \cofr^a \to \cofr'{}^a = M^a{}_b\, \cofr^b\,,\qquad g_{ab}\to g'{}_{ab} = (M^{-1})^c{}_a(M^{-1})^d{}_b \, g_{cd} \,.
\end{equation}
and such that $\cofr^a$ fulfills $\det(\dint{\vpartial_\mu}\cofr^a)=\det(e_\mu{}^a)\neq0$ (non-degenerate).

The linear connection shows up in principle quite naturally, as we will see. However, the construction of the metric and the coframe is more involved. In this thesis we will follow the approach in \cite{PonomarevObukhov2017, TresguerresMielke2000} (see also \cite{Gronwald1997}).

\subsection{The principal bundle of MAG: the affine frame bundle}

\subsubsection*{The affine group and its Lie algebra. Some remarks}

\boxdefinition{
\begin{defn}
\textbf{((Real) affine group)} The $\dimM$\emph{-dimensional (real) affine group} is the Lie group given by the semi-direct product
\begin{equation}\label{eq:AffGLTr}
  \mathrm{Aff}(\dimM, \mathbb{R}) \coloneqq \mathrm{Tran}(\mathbb{R}^\dimM) \rtimes\mathrm{GL}(\dimM, \mathbb{R})\,,
\end{equation}
where $\mathrm{Tran}(\mathbb{R}^\dimM)$ is the group of translations of  $\mathbb{R}^\dimM$.
\end{defn}
}

Now we present some remarks about this group and its Lie algebra:
\begin{itemize}
\item Due to the decomposition \eqref{eq:AffGLTr} of the group, we have a corresponding decomposition of Lie algebras,
\begin{equation}\label{eq:affgltr}
  \mathfrak{aff}(\dimM, \mathbb{R}) = \mathfrak{tran}(\mathbb{R}^\dimM) \oplus \mathfrak{gl}(\dimM, \mathbb{R})\,.
\end{equation}

\item Let us revise the explicit Lie algebra structure of $\mathfrak{aff}(\dimM, \mathbb{R})$. For the translational part, since we have the canonical isomorphism $\mathfrak{tran}(\mathbb{R}^\dimM)\cong \mathbb{R}^\dimM$, it is natural to choose for the translational algebra the basis $\{{\rm P}_a\}$ associated with the standard basis of $\mathbb{R}^\dimM$. For  $\mathfrak{gl}(\dimM, \mathbb{R})$, which corresponds to the set of all real $\dimM$-dimensional matrices, we choose the basis $\{{\rm L}^a{}_b\}$, where ${\rm L}^a{}_b$ is the matrix with a 1 in the $(a,b)$ position and zeros elsewhere. The commutation relations in this basis are
\begin{align}
  [{\rm L}^a{}_b,\, {\rm L}^c{}_d] &=\delta_d^a{\rm L}^c{}_b-\delta^c_b{\rm L}^a{}_d\,,\\
  [{\rm L}^a{}_b,\,{\rm P}_c]& = \delta^a_c {\rm P}_b\,,\\
  [{\rm P}_a,\, {\rm P}_b] &= 0\,.
\end{align}

\item The affine group is a matrix Lie group, since it can be embedded in the space of $(\dimM+1)$-dimensional matrices. Here we present the explicit form of this isomorphism and the induced one in the Lie algebra:
\begin{align}
  \mathrm{Aff}(\dimM, \mathbb{R}) &\cong 
  \left\{ \begin{pmatrix}M & b\\ 0 & 1\end{pmatrix}\,
  :\,M\in\mathrm{GL}(\dimM,\mathbb{R}),\,b\in\mathrm{Tran}(\mathbb{R}^{\dimM})\right\}\,,\label{eq:isomAffmatrix}\\
  \mathfrak{aff}(\dimM, \mathbb{R}) &\cong 
  \left\{ \begin{pmatrix}N & d\\ 0 & 0\end{pmatrix}\,
  :\,N\in\mathfrak{gl}(\dimM,\mathbb{R}),\,d\in\mathfrak{tran}(\mathbb{R}^{\dimM})\right\}\,.\label{eq:isomaffmatrix}
\end{align}

\end{itemize}

\subsubsection*{The affine frame bundle}

\boxdefinition{
\begin{defn}
\textbf{(Affine tangent space)} The \emph{affine tangent space} at a point $p\in\mathcal{M}$, denoted as $A_p\mathcal{M}$, is the affine space canonically constructed from the tangent space (the one in which  $T_p\mathcal{M}$ is both the space of points and the vector space of directions).
\end{defn}

\begin{defn}
\textbf{(Affine frame)} An \emph{affine frame} at a point is a pair $(z, \{\vfre_a\})$ where $z$ is a point in $A_p\mathcal{M}$ and $ \{\vfre_a\}$ is a linear frame.
\end{defn}

\begin{defn}
\textbf{(Affine frame bundle)} The \emph{affine frame bundle} is the bundle whose total space is
\begin{equation}
\mathcal{A}(\mathcal{M})\coloneqq\bigsqcup_{p\in\mathcal{M}}\{\text{affine frames on}\ p \}\,.
\end{equation}
with the projection $\pi(p, (z, \{\vfre_a\}))\coloneqq p$.
\end{defn}
}
From now on, we will use the abbreviations $(p, \vfre_a)\equiv(p, \{\vfre_a\})$ and $(p, z, \vfre_a)\equiv(p, (z, \{\vfre_a\}))$.

The affine frame bundle $\mathcal{A}(\mathcal{M})$ is a principal bundle whose structure group is $\mathrm{Aff}(\dimM, \mathbb{R})$. Indeed, the right action of the structure group over the fibers is given by
\begin{equation}
  (p, z, \vfre_a) \to (p, z', \vfre'{}_b)\coloneqq (p, ~~z + b^a\vfre_a, ~~\vfre_a M ^a{}_b)
\end{equation}
where $M\in\mathrm{GL}(\dimM, \mathbb{R})$ and $b\in\mathrm{Tran}(\mathbb{R}^\dimM)$. 

As a consequence of the decomposition \eqref{eq:affgltr}, a connection 1-form in the affine frame bundle $\tilde\dfom$ can always be decomposed into two parts, one living in the general linear algebra and another one in the translational algebra:
\begin{equation}
  \tilde\dfom = \dfomtL + \dfomtT \,.
\end{equation}

In addition, if we use the isomorphism \eqref{eq:isomaffmatrix}, we can write our gauge field as follows (this is an abuse of notation):
\begin{equation}
  \dfA = \sigma^* \tilde\dfom = \begin{pmatrix} \dfAL & \dfAT \\ 0 & 0\end{pmatrix} \,,
\end{equation}
where $\dfAL \coloneqq \sigma^* \dfomtL$ and $\dfAT = \sigma^* \dfomtT$. Under a gauge transformation, i.e., a change of section
\begin{equation}\label{eq:gaugetransf}
  \sigma(p)= (p, z, \vfre_a) \qquad \to \qquad \sigma'(p)= (p, ~~z+b^a(p) \vfre_a, ~~\vfre_a (M^{-1})^a{}_b(p))\,,
\end{equation}
we have that (we used \eqref{eq:transfAF}) 
\begin{align}
  \dfAL \to \dfAL' &= M\ \dfAL\ M^{-1} + M \dex M^{-1}\,, \nonumber\\
  \dfAT \to \dfAT' &= M\ \dfAT\ -\dex b -(M \ \dfAL\ M^{-1}  +  M\dex M^{-1})b \,. \label{eq:transALT}
\end{align}
We observe here that the linear part of the gauge field can be identified naturally with our connection $\dfom_a{}^b$, once we extract the generators. However, the translational part has not a tensorial behavior as the one of the coframe.

\subsection{\label{sec:nonlinrealiz}Non-linear treatment of translational connection}

If we decompose the translational gauge field as $\dfAT=\dfAT^a \otimes {\rm P}_a$ and rewrite it as
\begin{equation} \label{eq:ATcofrDxi}
  \dfAT^a = \cofr^a - \dex\chi^a-\dfAL_b{}^a\chi^b \qquad \equiv \cofr^a - \Dex^\mathrm{L}\chi^a \,,
\end{equation}
for some 0-form $\chi^a\in C^\infty(\mathcal{M})$ transforming $\chi^a\to M^a{}_b\chi^b$ under a gauge transformation \eqref{eq:gaugetransf}, we can prove that the object $\cofr^a$ transforms as a coframe. Let us try to understand what is this $\chi$ and show that $\cofr^a$ is an example of something called \emph{non-linear connection}.

In geometrical terms, a non-linear realization \cite{Coleman1969}\footnote{
    This is related, though not equivalent, to spontaneous symmetry breaking in physics.} 
is the \emph{reduction} of the $\mathcal{G}$-principal bundle of the theory to a principal subbundle with a subgroup $\mathcal{H}\subset\mathcal{G}$ as structure group. It can be shown that such a reduction is possible if and only if there exists a global section of an associated bundle with fiber homeomorphic to the orbit space $\mathcal{G}/\mathcal{H}$ \cite{Sardanashvily2006, PonomarevObukhov2017}. We can then express such a section as $\sigma(p)\coloneqq\exp(\xi^\mathfrak{a}(p)\, {\rm K}_\mathfrak{a})$.\footnote{
    Using a more physical jargon, the field $\sigma(x)$ lives in the ``part'' of the group covered by the exponential of the ``broken'' generators ${\rm K}_\mathfrak{a}$.}
\boxdefinition{
\begin{defn}
{\bf (Non-linear realization)} Consider the object $\varPsi_\mathcal{H}\coloneqq \rho(\sigma^{-1}) \varPsi$ where $\varPsi$ is a certain matter field transforming under the representation $\rho$. Then we say that the pair $(\sigma, \varPsi_\mathcal{H})$ defines a \emph{non-linear realization} if under the action of the group $\mathcal{G}$, it transforms as\footnotemark
\begin{equation}
  (\sigma, \varPsi_\mathcal{H}) \xrightarrow{~~g~~} (\sigma',\, \rho(h)\ \varPsi_\mathcal{H})\,,
\end{equation}
where $h\in\mathcal{H}$ and $\sigma'\in\mathcal{G}/\mathcal{H}$ are related by
\begin{equation}\label{eq:gxixih}
  g\, \sigma = \sigma'\, h\,.
\end{equation}
\end{defn}
}
\footnotetext{This can also be defined at the level of Lie algebra, i.e., in terms of $\{\xi^a, \xi'{}^a\}$, instead of $\{\sigma, \sigma'\}$.}
Here $\varPsi_\mathcal{H}$ is the \emph{non-linear matter field} and the parameters $\xi^a$ are called \emph{Goldstone fields}.
Notice that $h$ depends on $g$ and $\xi$ non-linearly.

Consider a principal connection in $\mathcal{P}$. We omit the construction of the non-linear connection in the bundle and directly formulate it locally (at the level of gauge fields). The \emph{non-linear gauge field} with respect to the Goldstone fields $\xi^a$ is  \cite{PonomarevObukhov2017}:
\begin{equation}
  \dfB \coloneqq \sigma^{-1}\ \dfA\ \sigma + \sigma^{-1}\, \dex \sigma \,,  \label{eq:defBconn}
\end{equation}
with $\dfA$ our gauge field. This object transforms under gauge transformations as
\begin{equation}
  \dfB \quad \xrightarrow{~~g~~}\quad h\ \dfB\ h^{-1} +  h\dex h^{-1}\,, \label{eq:transfB}
\end{equation}
where $h$ is defined in \eqref{eq:gxixih}. Consider now the non-linear covariant derivative of the non-linear field (same as the usual covariant derivative but with $\dfB$ instead of $\dfA$). Interestingly, \eqref{eq:transfB} implies that the result of taking such derivative is also a non-linear field, i.e., it only ``feels'' the $\mathcal{H}$ part of the group:
\begin{align}
  \Dex^{\dfB}\  \varPsi_\mathcal{H} \quad \xrightarrow{~~g~~}\quad \rho(h)\ \Dex^{\dfB}\ \varPsi_\mathcal{H}\,.
\end{align}

~

Now let us apply this to our gravitational gauge theory. For the particular case of $\mathcal{G}=\mathrm{Aff}(\dimM, \mathbb{R})$ and $\mathcal{H}=\mathrm{GL}(\dimM, \mathbb{R})$, we decompose the non-linear connection $\dfB$ according to \eqref{eq:affgltr}, $\dfB=\dfBL+\dfBT$ \cite{PonomarevObukhov2017}. As a consequence, \eqref{eq:transfB} can be split as 
\begin{align}
  \dfBL \to \dfBL' &= h\ \dfBL\ h^{-1}  + h\ \dex h^{-1}\,, \nonumber\\
  \dfBT \to \dfBT' &= h\ \dfBT\ h^{-1} \,.
\end{align}
If we compare this with \eqref{eq:transALT} we notice that the linear part continues transforming inhomogeneously under the $\mathcal{H}$ part,\footnote{
    Notice that $h\neq M$. $M$ is the linear transformation contained in $g$, whereas $h$ is the linear transformation fixed by the non-linear realization (see \eqref{eq:gxixih}).} 
while the translational part transforms homogeneously. In fact one can prove that\footnote{
    One easy way to see this is by using the isomorphisms with $(\dimM+1)$-dimensional matrices  \eqref{eq:isomAffmatrix}-\eqref{eq:isomaffmatrix}. We have $\sigma \to \begin{pmatrix} \identity & \sigma \\ 0 & 1\end{pmatrix}$, so the expression \eqref{eq:defBconn} can be rewritten as:
    \[
      \dfB \to \begin{pmatrix} \identity & \sigma^{-1} \\ 0 & 1\end{pmatrix} 
           \begin{pmatrix} \dfAL & \dfAT \\ 0 & 0\end{pmatrix} \begin{pmatrix} \identity & \sigma \\ 0 & 1\end{pmatrix} + 
           \begin{pmatrix} \identity & \sigma^{-1} \\ 0 & 1\end{pmatrix}\begin{pmatrix} 0 & \dex\sigma \\ 0 & 0\end{pmatrix}
           \quad = \quad 
           \begin{pmatrix} \dfAL & \dfAT+ \dex\sigma + \dfAL\  \sigma \\ 0 & 0\end{pmatrix}\,.
    \]
}
\begin{align}
  \dfBL = \dfAL\,, \qquad \qquad \dfBT = \dfAT + \dex\sigma + \dfAL\ \sigma \quad \equiv \dfAT +\Dex^\mathrm{L}\sigma\,.
\end{align}
So we can finally conclude, just by comparing with \eqref{eq:ATcofrDxi}, that $\chi$ was nothing but the section generated by the Goldstone fields $\xi^a$ and that $\cofr^a$ corresponds to the translational part of the non-linear connection $\dfBT$. Notice that for this particular reduction, the information of  $\dfAL$ is directly encoded in $\dfBL$ with no extra terms.

So far we have found an object that can be identified with the connection 1-form of our metric-affine geometry ($\dfBL = \dfAL \equiv \dfom_a{}^b\otimes{\rm L}^a{}_b$) and another one ($\dfBT\equiv \cofr^a \otimes {\rm P}_a$) with the same transformation rules as the coframe.\footnote{
     Due to the special structure of the affine Lie algebra, the geometry we are describing is a \emph{reductive Cartan geometry} \cite{Wise2010} and $\dfB$ is a \emph{Cartan connection}. Indeed, the affine Lie algebra is even more interesting, because the vector space $\mathfrak{aff}(\dimM,\mathbb{R})/\mathfrak{gl}(\dimM,\mathbb{R})$ of ``broken'' generators has the same dimension as the tangent spaces of the base manifold. The global section $\sigma$ allows to identify the tangent to the fibers at the points $\sigma(p)$ of the abstract $\mathrm{Aff}(\dimM,\mathbb{R})/\mathrm{GL}(\dimM,\mathbb{R})$-bundle with the tangent spaces $T_p\mathcal{M}$. This is the so called \emph{soldering process}, which is a key feature of gauge theories of gravity (see also \cite{Gronwald1997, Gronwald1998} for a different view in terms of a section of origins).} 
Under this reduction, the degrees of freedom of the gauge theory (those corresponding to the fundamental object, $\tilde\dfom$) are rearranged within these two objects. But, where is the metric?

\subsection{The origin of the metric in MAG}

To obtain the entire metric-affine geometry we have to consider a further reduction of the subgroup $\mathrm{GL}(\dimM,\mathbb{R})$ into the Lorentz subgroup $\mathrm{SO}(1, \dimM-1)$ \cite{PonomarevObukhov2017,GiachettiRicci1981, TresguerresMielke2000}. Let us call $r$ the section that takes values in the orbit space $\mathrm{GL}(\dimM,\mathbb{R})/\mathrm{SO}(1, \dimM-1)$ and $\dfGa$ the corresponding non-linear connection (analogues of $\sigma$ and $\dfB$, respectively). Following the steps of the previous section, and after performing a decomposition $\dfGa = \dfGaL + \dfGaT$, one can prove the following expressions \cite{PonomarevObukhov2017, TresguerresMielke2000}:
\begin{align}
  \dfGaL & = r^{-1}\ \dfAL\ r + r^{-1}\, \dex r
  \,,\\
  \dfGaT & = r^{-1} \big(\dfAT + \Dex^\mathrm{L}\sigma\big) = \mathring{\cofr}^b\otimes {\rm P}_b\qquad\qquad \mathring{\cofr}^b\coloneqq (r^{-1})^b{}_a \cofr^a\,.
\end{align}
In this case, $\dfGaL$ is a connection 1-form transforming (inhomogeneously) under the Lorentz group, whereas $\dfGaT$ transforms tensorially under the same group. The important thing now is that the Lorentz group has naturally associated the Minkowski metric $\eta_{ab}$, and this allows to define the following object:
\begin{align}
  g_{ab} \coloneqq (r^{-1})^c{}_a{} (r^{-1})^d{}_b{} \eta_{cd}\,.
\end{align}
Observe that, since $r$'s are invertible matrices, this operation preserves the symmetry and the signature of $\eta_{cd}$. The resulting symmetric and non-degenerate tensor, $g_{ab}$, corresponds to the MAG metric. 

Notice that the invariants of the geometry can be either expressed in terms of Lorentz-non-linear objects or GL-non-linear objects  \cite{TresguerresMielke2000}. For instance, for the line element, we have
\begin{align}
  \dex s^2 = \eta_{ab} \mathring{\cofr}{}^a \otimes \mathring{\cofr}{}^b =  g_{ab} \cofr^a \otimes \cofr^b\,.
\end{align}

Therefore, when we fix the MAG metric to be Minkowski, what we are doing is using the reduction introduced in this section. Therefore the $r$'s, which contain the true degrees of freedom, are hidden in the coframe. Alternatively, one can also choose a description of MAG purely in terms of the metric (this is the usual $(g_{\mu\nu},\Gamma_{\mu\nu}{}^\rho)$ formulation, which does not require the coframe). As we have seen, the metric has a Goldstone nature in MAG \cite{TresguerresMielke2000}, playing the role of a ``generalized Higgs field'' \cite{Trautman1979}.

\subsection{Final comments on the gauge construction of MAG}

Let us briefly sum up what we have done. At the beginning, we had a smooth manifold with no extra structure, i.e., a blank canvas waiting for extra structures that will come from the gauge approach. We considered the affine frame bundle, whose structure group is the affine group, as the basic principal bundle of the gauge procedure. We selected a connection there and then performed reductions into the general linear group and the Lorentz group. The coframe arises as the non-linear translational connection, whereas the metric can be expressed purely in terms of Goldstone fields.

Although we did not derive this in detail, let us mention that the curvature $\dfR_a{}^b$ is nothing but the field strength of the general linear part, whereas the torsion $\dfT^a$ comes from the non-linear translational part. 

If we follow the latter approach (the reduction into the Lorentz group), $\dfR_a{}^b$ and $\dfT^a$ are the \emph{true} field strengths in MAG. The nonmetricity however, is derived from the metric which has a Goldstone nature. In fact, when we fix $g_{ab}$ to be Minkowski, the nonmetricity is just the symmetric part of the connection (see Proposition \ref{prop:symconnzeroQ}). Therefore, when constructing the action, the terms quadratic in the nonmetricity should be seen as mass terms for the connection and not as kinetic terms for the gauge potentials.\footnote{
    In the description entirely in terms of the metric ($g_{\mu\nu},\Gamma_{\mu\nu}{}^\rho$), the torsion is just $2\Gamma_{[\mu\nu]}{}^\rho$ (no derivatives of the coframe) so it is not a \emph{true} field strength. In the meantime, the nonmetricity acquires the term $-\partial_\rho g_{\mu\nu}$, which contains the derivative of the translational degrees of freedom.} 

Finally, it is worth noticing that one can find alternative ways to do gauge gravity or generalizations of structures that give other perspectives. For instance, based on some works of Lord \cite{Lord1987, Lord1978} Tresguerres framed all of this in the language of composite principle bundles \cite{Tresguerres2002} (see also \cite{Tresguerres2012}).

\section{Field theory machinery for a Metric-Affine Gauge action}

In this section we move to the physical (dynamical) part of MAG and establish some general results. The gravitational basic fields are the metric $g_{ab}$, the coframe $\cofr^a$ and the connection $\dfom_a{}^b$.

\subsection{The general MAG action}

Due to the gauge symmetry under linear transformations of the coframe, the Lagrangian must be a function of the metric and the coframe, their exterior covariant derivatives ($\dfT^{a}$, $\dfQ_{ab}$) and the field strength associated with $\dfom_a{}^b$, i.e. $\dfR_a{}^b$, but there cannot be any explicit dependence on $\dfom_a{}^b$.

Let us assume that in this metric-affine spacetime, there are some matter fields described by certain vector-valued differential forms $\{\dfvarPsi^{(i)}\} _{i=1}^{\text{nº matter fields}}$. From now on, we will drop the index $(i)$ and suppose that, when needed, there is an omitted summation over all the matter fields. In general we have
\begin{equation}
\dfvarPsi=\dfvarPsi^\mathtt{M}\otimes\vecxi_\mathtt{M}=\frac{1}{k!}\varPsi_{\mu_{1}...\mu_{k}}{}^\mathtt{M}\underbrace{(\dex x^{\mu_{1}}\wedge...\wedge\dex x^{\mu_{k}})}_{\dex x^{\mu_{1}...\mu_{k}}}\otimes\vecxi_\mathtt{M}\,,
\end{equation}
where $\{\vecxi_\mathtt{M}\}$ is a basis of sections of the corresponding associated bundle. In the usual components/tensor notation, one works with the functions $\varPsi_{\mu_{1}...\mu_{k}}{}^\mathtt{M}\in C^{\infty}(\mathcal{M})$ while, in the differential form notation of MAG, the basic objects are the differential forms $\dfvarPsi^\mathtt{M}\in\Omega^{k}(\mathcal{M})$. 

With all of this in mind, the most general MAG action that we are going to consider is a functional of the metric-affine geometry $\left\{ g_{ab},\,\cofr^a,\,\dfom_a{}^b\right\} $ and some matter fields $\dfvarPsi^\mathtt{M}$ of the type
\begin{equation}
S[g_{ab},\,\cofr^a,\,\dfom_a{}^b,\,\dfvarPsi^\mathtt{M}]=\int\dfL(g_{ab},\,\cofr^a,\,\dfR_a{}^b,\,\dfT^a,\,\dfQ_{ab},\,\dfvarPsi^\mathtt{M},\,\Dex\dfvarPsi^\mathtt{M})\,.\label{eq: Total lagrangian}
\end{equation}
Note that higher gauge derivatives in the matter fields have been dropped since,
\begin{equation}
\Dex\Dex\dfvarPsi^\mathtt{M}=\dfR_a{}^b\,\,\bar\rho^{(\dfvarPsi)}({\rm L}^a{}_{b}){}^\mathtt{M}{}_\mathtt{N}\wedge\dfvarPsi^\mathtt{M}\qquad\sim f(\dfR_a{}^b,\,\dfvarPsi^\mathtt{M})\,,
\end{equation}
where $\bar\rho^{(\dfvarPsi)}$ is Lie algebra representation associated with the Lie group representation $\rho^{(\dfvarPsi)}$ defined in the space where $\dfvarPsi$ takes values (see \eqref{eq:rhobarrho}). Higher derivatives of the gravitational fields can be reduced to field strength higher order terms, thanks to the Bianchi identities (see Proposition \ref{prop:BianchiTQR}).

Obviously all of this admits a corresponding formulation in terms of tensor components. If we call
\begin{equation}
   \dfL \eqqcolon \mathfrak{L}\ \dex^\dimM x \eqqcolon \mathcal{L}\ \volfg\,,
\end{equation}
where $\mathfrak{L}$ is a scalar density of weight $-1$ and $\mathcal{L}$ is a pure scalar, then,\footnote{
    We omitted the indices in the functional dependence of $S$ to abbreviate.}
\begin{equation}
S[g,\,e,\,\omega,\,\varPsi]=\int\mathfrak{L}(g_{ab},\,e_{\mu}{}^a,\,R_{\mu\nu a}{}^b,\,T_{\mu\nu}{}^a,\,Q_{\mu ab},\,\varPsi_{\mu_{1}...\mu_{k}}{}^\mathtt{M},\,\nabla_{[\rho}\varPsi_{\mu_{1}...\mu_{k}]}{}^\mathtt{M})\,\mathrm{d}^{\dimM}x\,.\label{eq: L MAG comp}
\end{equation}
We will give the relevant results in both notations. But first we will provide some expressions to translate between the language of tensor components and the differential form notation.

\subsection{Functional variations in MAG}

\boxsimple{ {\bf Convention}. For variations and partial derivatives in the language of differential forms we choose the convention in which the chain rule is applied to the left so, in particular, the variations $\delta\dfal$ are extracted from the left.
}
For example:
\begin{align}
\frac{\partial\dfL(\dfbe(\dfal))}{\partial\dfal}\eqqcolon\frac{\partial\dfbe}{\partial\dfal}\wedge\frac{\partial\dfL}{\partial\dfbe}\qquad \text{or} \qquad \delta_{\dfal}S[\dfal] \eqqcolon \int\delta\dfal\wedge\frac{\delta S}{\delta\dfal}\,.
\end{align}

Consider a generic action of some tensor-valued $k$-form field $\dfal^\mathtt{M}$, $S[\dfal]$. It is useful to have a dictionary to translate the variations between the components approach in which the fundamental objects are the components $\alpha_{\mu_{1}...\mu_{k}}{}^\mathtt{M}$ and the differential form notation where $\dfal^\mathtt{M}$ is the basic object. If we perform a variation of the action with respect to this field in both ways we get\footnote{
    The notation $\delta_{\mathrm{c}}$ to distinguish both types of variations is normally not necessary except when we vary with respect to a 0-form field because in that case there is no distinction between the differential form and its components. One example is the metric
    \[
      \delta_{\dfal}S[\dfal] =\int\,\delta g_{ab}\frac{\delta S}{\delta g_{ab}}\,,\qquad\delta_{\dfal}S[\dfal]=\int\,\delta g_{ab}\frac{\delta_{\mathrm{c}}S}{\delta g_{ab}}\,\mathrm{d}^{\dimM}x\,.
    \]
    where $\delta S/\delta g_{ab}$ is a $\dimM$-form and $\delta_{\mathrm{c}} S/\delta g_{ab}$ is a 0-form. In Chapter \ref{ch:ECG} and Chapter \ref{ch:QTG} we will drop the $\mathrm{c}$ because we will work entirely in components notation and there will be no confusion.} 
\begin{equation}
\delta_{\dfal}S[\dfal] =\int\,\delta\dfal^\mathtt{M}\wedge\frac{\delta S}{\delta\dfal^\mathtt{M}}\,,\qquad
\delta_{\dfal}S[\dfal]  =\int\,\delta \alpha_{\mu_{1}...\mu_{k}}{}^\mathtt{M} \frac{\delta_{\mathrm{c}}S}{\delta\alpha_{\mu_{1}...\mu_{k}}{}^\mathtt{M}}\,\mathrm{d}^{\dimM}x\,.
\end{equation}
The question is: how are the objects $\frac{\delta S}{\delta\dfal^\mathtt{M}}$ and $\frac{\delta_{\mathrm{c}}S}{ \delta\alpha_{\mu_{1}...\mu_{k}}{}^\mathtt{M}}$ related? One can straightforwardly establish such a relation:
\boxproposition{
\begin{prop}
\label{Prop: var diccionary}The variations in both languages are related through the equations:
\begin{equation}
\frac{1}{\sqrt{\left|g\right|}}\frac{\delta_{\mathrm{c}}S}{\delta\alpha_{\mu_{1}...\mu_{k}}{}^\mathtt{M}}=\frac{1}{k!}\sgng(-1)^{k(\dimM-k)}\dint{\vpartial^{\mu_{k}}}...\dint{\vpartial^{\mu_{1}}}\left(\star\frac{\delta S}{\delta\dfal^\mathtt{M}}\right)\,,
\end{equation}
\begin{equation}
\frac{\delta S}{\delta\dfal^\mathtt{M}}=\frac{1}{\sqrt{\left|g\right|}}\frac{\delta_{\mathrm{c}}S}{\delta\alpha_{\mu_{1}...\mu_{k}}{}^\mathtt{M}}\star\dex x_{\mu_{1}...\mu_{k}}\,.
\end{equation}
where $\vpartial^{\mu}\coloneqq g^{\nu\mu}\vpartial_{\nu}$ and $\dex x_{\mu}\coloneqq g_{\mu\nu}\dex x^{\nu}$.
\end{prop}
}

We continue with the following powerful result that gives the functional derivatives for any tensor-valued $k$- form $\dfal^\mathtt{M}$:

\boxproposition{
\begin{thm}
\label{The: var wrt tensor form}Let $\dfal^\mathtt{M}$ be a tensor-valued $k$-form. Consider the following expression of the covariant derivative acting on $\dfal^\mathtt{M}$,
\begin{equation}
\Dex\dfal^\mathtt{M}=\dex\dfal^\mathtt{M}+\dfA_\mathtt{N}{}^\mathtt{M}\wedge\dfal^\mathtt{N}\,,
\end{equation}
and the following for the associated co-objects (with opposite indices),
\begin{equation}
\Dex\dfbe_\mathtt{M}=\dex\dfbe_\mathtt{M}-\dfA_\mathtt{M}{}^\mathtt{N}\wedge\dfbe_\mathtt{N}\,,
\end{equation}
where $\dfA_\mathtt{N}{}^\mathtt{M}$ is the connection 1-form. Then, for any action
\begin{equation}
S[\dfal,...]=\int\dfL(\dfal^\mathtt{M},\,\Dex\dfal^\mathtt{M},...)\,,
\end{equation}
the following holds:
\begin{equation}
\boxed{\frac{\delta S}{\delta\dfal^\mathtt{M}}=\frac{\partial\dfL}{\partial\dfal^\mathtt{M}} -(-1)^{\mathrm{rank}(\dfal^\mathtt{M})}\Dex\frac{\partial\dfL}{\partial\Dex\dfal^\mathtt{M}}}\,.
\end{equation}
\end{thm}
}

This theorem is valid for any exterior covariant derivative $\Dex$ or, equivalently, for any connection 1-form $\dfA_\mathtt{N}{}^\mathtt{M}$: it can be the gravitational connection, an internal one or a combination of both. For the particular case of pure metric-affine gravity \eqref{eq: Total lagrangian} (in which $\dfA_\mathtt{N}{}^\mathtt{M}=\dfom_a{}^b\bar{\rho}^{(\dfal)}({\rm L}^a{}_{b}){}^\mathtt{M}{}_\mathtt{N}$), one can use this result to derive the functional variation with respect to the metric, the coframe and the matter fields, since all of them are tensor-valued forms. However, notice that $\Dex\dfom_a{}^b$ is not well-defined since $\dfom_a{}^b$ is not a tensor-valued form (it is not tensorial in its Latin indices). For this reason, when we vary with respect to the connection we cannot use the previous theorem, only valid for tensor-valued forms,
\begin{equation}
\delta_{\dfom}S=\int\delta\dfom_a{}^b\wedge\frac{\delta S}{\delta\dfom_a{}^b}\overset{\Danger}{\neq}\int\delta\dfom_a{}^b\wedge\Bigg[\frac{\partial\dfL}{\partial\dfom_a{}^b}+\Dex\frac{\partial\dfL}{\partial\underbrace{\Dex\dfom_a{}^b}_{??}}\Bigg]\,.
\end{equation}
After carefully doing the whole computation with the connection, one can collect all of the variations for the general MAG action \eqref{eq: Total lagrangian}:

\boxproposition{
\begin{prop}
\label{Prop: MAG functional variations} For a MAG action of the type \eqref{eq: Total lagrangian}, the functional variation with respect to each of the fields are given by:
\begin{align}
\frac{\delta S}{\delta\dfvarPsi^\mathtt{M}} & =\frac{\partial\dfL}{\partial\dfvarPsi^\mathtt{M}}-(-1)^{\mathrm{rank}(\dfvarPsi^\mathtt{M})}\Dex\frac{\partial\dfL}{\partial\Dex\dfvarPsi^\mathtt{M}}\,,\\[2mm]
\dfrac{\delta S}{\delta g_{ab}} & =\dfrac{\partial\dfL}{\partial g_{ab}}+\Dex\dfrac{\partial\dfL}{\partial\dfQ_{ab}}\,,\\[2mm]
\dfrac{\delta S}{\delta\cofr^a} & =\dfrac{\partial\dfL}{\partial\cofr^a}+\Dex\dfrac{\partial\dfL}{\partial\dfT^a}\,,\label{eq: varcofr MAG}\\[2mm]
\dfrac{\delta S}{\delta\dfom_a{}^b} & =\big(\bar\rho^{(\dfvarPsi)}({\rm L}^a{}_{b}){}^\mathtt{M}{}_\mathtt{N}\dfvarPsi^\mathtt{N}\big)\wedge\frac{\partial\dfL}{\partial\Dex\dfvarPsi^\mathtt{M}}+2g_{bc}\dfrac{\partial\dfL}{\partial\dfQ_{ac}}+\cofr^a\wedge\dfrac{\partial\dfL}{\partial\dfT^b}+\Dex\dfrac{\partial\dfL}{\partial\dfR_a{}^b}\,.\label{eq: varw MAG}
\end{align}
\end{prop}
}

Now one can derive the corresponding variations in components notation directly from \eqref{eq: L MAG comp} or by using Proposition \ref{Prop: MAG functional variations}. In any case, the result is:

\boxproposition{
\begin{prop}
\label{Prop: MAG functional variations comp} For an action of the type \eqref{eq: L MAG comp}, it follows
\begin{align}
\frac{\delta_{\mathrm{c}}S}{\delta\varPsi_{\mu_{1}...\mu_{k}}{}^\mathtt{M}} & =\frac{\partial\mathfrak{L}}{\partial\varPsi_{\mu_{1}...\mu_{k}}{}^\mathtt{M}}-\left(\nabla_{\rho}+T_\rho\right)\left(\frac{\partial\mathfrak{L}}{\partial\nabla_{[\rho}\varPsi_{\mu_{1}...\mu_{k}]}{}^\mathtt{M}}\right)\,,\label{eq: matt gen var}\\
\frac{\delta_{\mathrm{c}}S}{\delta g_{ab}} & =\dfrac{\partial\mathfrak{L}}{\partial g_{ab}}+\left(\nabla_{\rho}+T_\rho\right)\left(\dfrac{\partial\mathfrak{L}}{\partial Q_{\rho ab}}\right)\\
\frac{\delta_{\mathrm{c}}S}{\delta e_{\mu}{}^a} & =\dfrac{\partial\mathfrak{L}}{\partial e_{\mu}{}^a}-2\left(\nabla_{\rho}+T_\rho\right)\left(\dfrac{\partial\mathfrak{L}}{\partial T_{\rho\mu}{}^a}\right)+T_{\rho\sigma}{}^{\mu}\dfrac{\partial\mathfrak{L}}{\partial T_{\rho\sigma}{}^a}\,,\\
\frac{\delta_{\mathrm{c}}S}{\delta\omega_{\mu a}{}^b} & =\bar\rho^{(\dfvarPsi)}({\rm L}^a{}_{b})^\mathtt{M}{}_\mathtt{N}\frac{\partial\mathfrak{L}}{\partial\nabla_{[\mu}\varPsi_{\nu_{1}...\nu_{k}]}{}^\mathtt{M}}\varPsi_{\nu_{1}...\nu_{k}}{}^\mathtt{N}+2g_{bc}\frac{\partial\mathfrak{L}}{\partial Q_{\mu ac}}+2\frac{\partial\mathfrak{L}}{\partial T_{\mu\rho}{}^b}e_{\rho}{}^a\nonumber \\
 & \qquad-2\left(\nabla_{\rho}+T_\rho\right)\left(\frac{\partial\mathfrak{L}}{\partial R_{\rho\mu a}{}^b}\right)+T_{\rho\sigma}{}^{\mu}\frac{\partial\mathfrak{L}}{\partial R_{\rho\sigma a}{}^b}\,.
\end{align}
\end{prop}
}

Now that we know the dynamical equations for any MAG theory \eqref{eq: Total lagrangian} we derive the Noether identities under the general linear group and the diffeomorphisms. These identities will lead us to a very relevant result: the equation of motion of the metric $g_{ab}$ is \emph{redundant} and we will be able to drop it in our future studies.\footnote{This is of course not surprising since we know from previous sections that the metric is just a Goldstone field. Its degrees of freedom can be completely translated to the coframe.}

\newpage
\subsection{Noether identities}

This section is essentially extracted and adapted from \cite[Sec. 5.2]{Hehl1995}, where more details on these derivations can be found.

\subsubsection*{Noether identity under diffeomorphisms}

Consider a vector $\vecX$ generating a one-parameter subgroup of $\mathrm{Diff}(\mathcal{M})$. The Lagrangian form must be a scalar under $\mathrm{Diff}(\mathcal{M})$ or, in other words, its variation must be a boundary term.\footnote{
    This is clear since $\DLie_{\vecX}\dfL=0+\dex(\dint{\vecX}\dfL)$ so it will only contribute to the expression of $\dfB$ below. But it is irrelevant for the derivation of the Noether identity, whose information is contained in $\dfA$.} 
Since $\dfL$ is a ${\rm GL}(\dimM,\mathbb{R})$-scalar, the operators $\DLie_{\vecV}$ and $\dLie_{\vecV}$ are equivalent (see Section \ref{sec:Lieder}). But it is convenient to choose the covariant version. Therefore, if we act on the general MAG Lagrangian \eqref{eq: Total lagrangian},
\begin{align}
\DLie_{\vecX}\dfL & =(\DLie_{\vecX}g_{ab})\wedge\frac{\partial\dfL}{\partial g_{ab}}+(\DLie_{\vecX}\cofr^a)\wedge\frac{\partial\dfL}{\partial\cofr^a}\nonumber \\
 & \qquad+(\DLie_{\vecX}\dfR_a{}^{b})\wedge\frac{\partial\dfL}{\partial\dfR_a{}^b}+(\DLie_{\vecX}\dfT^a)\wedge\frac{\partial\dfL}{\partial\dfT^a}+(\DLie_{\vecX}\dfQ_{ab})\wedge\frac{\partial\dfL}{\partial\dfQ_{ab}}\nonumber \\
 & \qquad+(\DLie_{\vecX}\dfvarPsi^\mathtt{M})\wedge\frac{\partial\dfL}{\partial\dfvarPsi^\mathtt{M}}+(\DLie_{\vecX}\Dex\dfvarPsi^\mathtt{M})\wedge\frac{\partial\dfL}{\partial\Dex\dfvarPsi^\mathtt{M}}\,.
\end{align}
Taking into account the definition of the covariant Lie derivative one can straightforwardly derive \cite{Hehl1995}:
\boxproposition{
\begin{prop}
For the Lagrangian \eqref{eq: Total lagrangian} the following equation holds
\begin{equation}
0=\dfA+\dex\dfB\,,
\end{equation}
where
\begin{align}
\dfA & \coloneqq-(\dint{\vecX}\dfQ_{ab})\frac{\delta S}{\delta g_{ab}}-(\dint{\vecX}\cofr^a)\Dex\frac{\delta S}{\delta\cofr^a}\nonumber \\
 & \qquad+(\dint{\vecX}\dfT^a)\wedge\frac{\delta S}{\delta\cofr^a}+(\dint{\vecX}\dfR_a{}^{b})\wedge\frac{\delta S}{\delta\dfom_a{}^b}\nonumber \\
 & \qquad+(\dint{\vecX}\Dex\dfvarPsi^\mathtt{M})\wedge\frac{\delta S}{\delta\dfvarPsi^\mathtt{M}}+(-1)^{\mathrm{rank}(\dfvarPsi^\mathtt{M})}(\dint{\vecX}\dfvarPsi^\mathtt{M})\wedge\Dex\frac{\delta S}{\delta\dfvarPsi^\mathtt{M}}\\
\dfB & \coloneqq-\dint{\vecX}\dfL+\Big[(\dint{\vecX}\dfQ_{ab})\frac{\partial\dfL}{\partial\dfQ_{ab}}+(\dint{\vecX}\cofr^a)\frac{\partial\dfL}{\partial\cofr^a}+(\dint{\vecX}\dfT^a)\wedge\frac{\partial\dfL}{\partial\dfT^a}+(\dint{\vecX}\dfR_a{}^{b})\wedge\frac{\partial\dfL}{\partial\dfR_a{}^b}\nonumber \\
 & \qquad\qquad\qquad\qquad\qquad+(\dint{\vecX}\dfvarPsi^\mathtt{M})\wedge\frac{\partial\dfL}{\partial\dfvarPsi^\mathtt{M}}+(\dint{\vecX}\Dex\dfvarPsi^\mathtt{M})\wedge\frac{\partial\dfL}{\partial\Dex\dfvarPsi^\mathtt{M}}\Big]\,.
\end{align}
\end{prop}
}

Notice how beautifully, the partial derivatives of the Lagrangian disappear from the expression of $\dfA$ and combine into variations of the action (the equations of motion). Observe also that $\dfA$ and $\dfB$ are proportional to the components $X^a$, so if we introduce $\dfA\eqqcolon X^a\dfA_{a}$ and $\dfB\eqqcolon X^a\dfB_{a}$
we get
\begin{equation}
0=X^a (\dfA_{a}+\dex\dfB_{a})+\dex X^a\wedge\dfB_a\,.
\end{equation}
Since we can take $X^{a}$ and $\dex X^{a}$ to be independent and arbitrary, we arrive at:
\begin{equation}
\dfA_{a}=\dfB_{a}=0\qquad\Rightarrow\qquad\dfA=\dfB=0\,.
\end{equation}

{\noindent}In particular, $\dfA=0$ leads us to:
\boxtheorem{
\begin{thm}
\label{The: 1NoetherId} Consider a Lagrangian \eqref{eq: Total lagrangian} which is a scalar under diffeomorphisms, then:
\begin{enumerate}
\item Its variations with respect to the gravitational fields and the matter are related via the identity:
\begin{align}
\Dex\frac{\delta S}{\delta\cofr^c} & =-(\dint{\vfre_c}\dfQ_{ab})\frac{\delta S}{\delta g_{ab}}+(\dint{\vfre_c}\dfT^a)\wedge\frac{\delta S}{\delta\cofr^a}+(\dint{\vfre_c}\dfR_a{}^{b})\wedge\frac{\delta S}{\delta\dfom_a{}^b}\nonumber \\
 & \qquad+(\dint{\vfre_c}\Dex\dfvarPsi^\mathtt{M})\wedge\frac{\delta S}{\delta\dfvarPsi^\mathtt{M}}+(-1)^{\mathrm{rank}(\dfvarPsi^\mathtt{M})}(\dint{\vfre_c}\dfvarPsi^\mathtt{M})\wedge\Dex\frac{\delta S}{\delta\dfvarPsi^\mathtt{M}}\,.
\end{align}
\item In particular, if either the matter is on-shell (i.e. $\frac{\delta S}{\delta\dfvarPsi^\mathtt{M}}=0$) or the Lagrangian does not contain any matter fields (only metric, coframe and linear connection), we arrive at the identity\footnotemark
\begin{equation}
\Dex\frac{\delta S}{\delta\cofr^c}\eqonshell{\varPsi}-(\dint{\vfre_c}\dfQ_{ab})\frac{\delta S}{\delta g_{ab}}+(\dint{\vfre_c}\dfT^a)\wedge\frac{\delta S}{\delta\cofr^a}+(\dint{\vfre_c}\dfR_a{}^{b})\wedge\frac{\delta S}{\delta\dfom_a{}^b}\,.
\end{equation}
\end{enumerate}
\end{thm}
}
\footnotetext{The symbol $\eqonshell{f,\,g...}$ means ``equal if the fields $f$, $g$... are on-shell or absent in the Lagrangian''.}
\vspace{-3mm}

{\noindent}The latter looks in components as follows
\begin{equation}
\left(\nabla_{\mu}+T_{\mu\lambda}{}^{\lambda}\right)\left(\frac{\delta_{\mathrm{c}}S}{\delta e_{\mu}{}^c}\right)\eqonshell{\varPsi}-e^{\nu}{}_c\left[Q_{cab}\frac{\delta_{\mathrm{c}}S}{\delta g_{ab}}-T_{\mu\nu}{}^a\frac{\delta_{\mathrm{c}}S}{\delta e_{\mu}{}^a}-R_{\mu\nu a}{}^b\frac{\delta_{\mathrm{c}}S}{\delta\omega_{\mu a}{}^b}\right]\,.
\end{equation}

\subsubsection*{Noether identity under the general linear group}

Now we revise the implications of the invariance under $\mathrm{GL}(\dimM,\mathbb{R})_{\mathrm{local}}$. We start, as usual, with a frame transformation $\{\vfre_a\}\to\{\vfre_b M^b{}_a(x)\}$ ($M^b{}_a(x)\in\mathrm{GL}(\dimM,\mathbb{R})_{\mathrm{local}}$) and assume it to be infinitesimal, i.e.,
\begin{equation}
   M^b{}_a(x)=\delta_a^b+\varpi_a{}^b(x)\,,
\end{equation}
where $\varpi_a{}^b$ is a small parameter. One can easily check that our fields transform as:
\[
\delta_{\varpi}g_{ab}=2\varpi_{(ab)}\,,\qquad\delta_{\varpi}\cofr^{a}=-\varpi_b{}^a\cofr^b\,,\qquad\delta_{\varpi}\dfom_a{}^b =\Dex\varpi_a{}^b\,,
\]
\begin{equation}
\delta_{\varpi}\dfvarPsi^\mathtt{M}=-\varpi_a{}^b\bar\rho^{(\dfvarPsi)}({\rm L}^a{}_{b})^\mathtt{M}{}_\mathtt{N}\dfvarPsi^\mathtt{N}\,,
\end{equation}
or, in components,
\[
\delta_{\varpi}g_{ab}=2\varpi_{(ab)}\,,\qquad\delta_{\varpi}e_{\mu}{}^{a}=-\varpi_b{}^{a}e_{\mu}{}^b\,,\qquad\delta_{\varpi}\omega_{\mu a}{}^b =\mathcal{D}_{\mu}\varpi_a{}^b =\nabla_{\mu}\varpi_a{}^b\,,
\]
\begin{equation}
\delta_{\varpi}\varPsi_{\mu_{1}...\mu_{k}}{}^\mathtt{M}=-\varpi_a{}^b\bar\rho^{(\dfvarPsi)}({\rm L}^a{}_{b})^\mathtt{M}{}_\mathtt{N}\varPsi_{\mu_{1}...\mu_{k}}{}^\mathtt{N}\,.
\end{equation}

And now we get \cite{Hehl1995}
\boxproposition{
\begin{prop}
For the Lagrangian \eqref{eq: Total lagrangian} the following equation holds
\begin{align}
\delta_{\varpi}\dfL 
  & =-\varpi_a{}^b\left[-2g_{bc}\frac{\delta S}{\delta g_{ac}}+\cofr^a\wedge\frac{\delta S}{\delta\cofr^b}+\Dex\frac{\delta S}{\delta\dfom_a{}^b}+\bar\rho^{(\dfvarPsi)}({\rm L}^a{}_{b})^\mathtt{M}{}_\mathtt{N}\dfvarPsi^\mathtt{N}\wedge\frac{\delta S}{\delta\dfvarPsi^\mathtt{M}}\right]\nonumber \\
  & \quad+\dex\bigg\{ \varpi_a{}^b\bigg[\frac{\delta S}{\delta\dfom_a{}^b}-\bar\rho^{(\dfvarPsi)}({\rm L}^a{}_{b})^\mathtt{M}{}_\mathtt{N}\dfvarPsi^\mathtt{N}\wedge\frac{\partial\dfL}{\delta\Dex\dfvarPsi^\mathtt{M}}\nonumber\\
  &\quad\qquad\qquad\qquad -2g_{bc} \frac{\partial\dfL}{\partial\dfQ_{ac}} -\cofr^a\wedge\frac{\partial\dfL}{\partial\dfT^b} -\Dex\frac{\partial\dfL}{\partial\dfR_a{}^b}\bigg]\bigg\} \,.\label{eq: NoethIdGL}
\end{align}
\end{prop}
}

For the same argument we previously used for $\mathrm{Diff}(\mathcal{M})$, the two square brackets in \eqref{eq: NoethIdGL} must vanish. The second one is identically zero due to \eqref{eq: varw MAG}, and from the first one we obtain the corresponding Noether identity:
\boxtheorem{
\begin{thm}
\label{The: 2NoetherId} Consider a Lagrangian \eqref{eq: Total lagrangian} which is a scalar under local transformation in $\mathrm{GL}(\dimM,\mathbb{R})_{\mathrm{local}}$ then:
\begin{enumerate}
\item Its variations with respect to the gravitational fields and the matter are related via the identity:
\begin{align}
\Dex\frac{\delta S}{\delta\dfom_a{}^b} & =2g_{bc}\frac{\delta S}{\delta g_{ac}}-\cofr^a\wedge\frac{\delta S}{\delta\cofr^b}-\bar\rho^{(\dfvarPsi)}({\rm L}^a{}_{b})^\mathtt{M}{}_\mathtt{N}\dfvarPsi^\mathtt{N}\wedge\frac{\delta S}{\delta\dfvarPsi^\mathtt{M}}\,.
\end{align}
\item In particular, if either the matter is on-shell (i.e. $\frac{\delta S}{\delta\dfvarPsi^\mathtt{M}}=0$) or the Lagrangian does not contain any matter fields (only metric, coframe and linear connection), we arrive at the identity
\begin{equation}
\Dex\frac{\delta S}{\delta\dfom_a{}^b}\eqonshell{\varPsi}2g_{bc}\frac{\delta S}{\delta g_{ac}}-\cofr^a\wedge\frac{\delta S}{\delta\cofr^b}\,.
\end{equation}
\end{enumerate}
\end{thm}
}
The latter in components reads
\begin{equation}
\left(\nabla_{\mu}+T_{\mu\lambda}{}^{\lambda}\right)\left(\frac{\delta_{\mathrm{c}}S}{\delta\omega_{\mu a}{}^b}\right)\eqonshell{\varPsi}2g_{bc}\frac{\delta_{\mathrm{c}}S}{\delta g_{ac}}-e_{\mu}{}^a\frac{\delta_{\mathrm{c}}S}{\delta e_{\mu}{}^b}\,.
\end{equation}

This immediately leads us to one of the most powerful results in MAG:
\boxtheorem{
\begin{thm}
For any MAG Lagrangian \eqref{eq: Total lagrangian}:
\begin{enumerate}
\item Either the equation of motion of the metric or the equation of motion of the coframe (Vielbein) is redundant.
\item If the matter and the connection are on-shell (or they do not appear in the Lagrangian) then:
\begin{enumerate}
\item The symmetric part of the Vielbein equation coincides with the equation of the metric:
\begin{equation}
2\frac{\delta S}{\delta g_{ab}}\eqonshell{\varPsi,\,\omega} g^{c(a}\cofr^{b)}\wedge\frac{\delta S}{\delta\cofr^c} \quad \xrightarrow{\text{In components}}  \quad 2\frac{\delta_{\mathrm{c}}S}{\delta g_{ab}}\eqonshell{\varPsi,\,\omega} e_{\mu}{}^{(a}g^{b)c}\frac{\delta_{\mathrm{c}}S}{\delta e_{\mu}{}^c}\,.\label{eq: rel EMtensors}
\end{equation}
\item The antisymmetric part of the Vielbein equation vanishes,
\begin{equation}
   g^{c[a}\cofr^{b]}\wedge\frac{\delta S}{\delta\cofr^c}\eqonshell{\varPsi,\,\omega} 0\quad \xrightarrow{\text{In components}} \quad e_{\mu}{}^{[a}g^{b]c}\frac{\delta_{\mathrm{c}}S}{\delta e_{\mu}{}^c}\eqonshell{\varPsi,\,\omega} 0\,.
\end{equation}
\end{enumerate}
\end{enumerate}
\end{thm}
}
It is important to remark that this results are true for any configuration of the metric and the coframe (even for off-shell configurations).

From now on, we eliminate the equation of the metric from our set of gravitational equations, and concentrate just in the coframe and connection equations. In the next section we are going to define a few objects that will allow as to rewrite the MAG equations of motion in a very nice form, which will also be very useful in order to calculate the dynamics in real situations (for example with a computer program).

\subsection{Momenta and matter currents}

\subsubsection*{Matter currents}

So far, we have been dealing with a totally general action of the type \eqref{eq: Total lagrangian}. Consider now the following splitting of it:
\begin{equation}
S[g_{ab},\,\cofr^a,\,\dfom_a{}^b,\,\dfvarPsi^\mathtt{M}]=S_{\mathrm{Grav}}[g_{ab},\,\cofr^a,\,\dfom_a{}^b]+S_{\mathrm{Matt}}[g_{ab},\,\cofr^a,\,\dfom_a{}^b,\,\dfvarPsi^\mathtt{M}]\,,\label{eq: SGrav SMatt}
\end{equation}
where
\begin{align}
S_{\mathrm{Grav}}[g_{ab},\,\cofr^a,\,\dfom_a{}^b] & \coloneqq \int\dfL_{\mathrm{Grav}}(g_{ab},\,\cofr^a,\,\dfR_a{}^b,\,\dfT^a,\,\dfQ_{ab})\,,\label{eq: LGrav MAG}\\
S_{\mathrm{Matt}}[g_{ab},\,\cofr^a,\,\dfom_a{}^b,\,\dfvarPsi^\mathtt{M}] & \coloneqq \int\dfL_{\mathrm{Matt}}(g_{ab},\,\cofr^a,\,\dfR_a{}^b,\,\dfT^a,\,\dfQ_{ab},\,\dfvarPsi^\mathtt{M},\,\Dex\dfvarPsi^\mathtt{M})\,.
\end{align}
It is worth mentioning that this separation has been done in such a way that all the terms in $\dfL_{\mathrm{Matt}}$ contain the matter fields. In other words, any purely gravitational term has been extracted and placed in $\dfL_{\mathrm{Grav}}$. Now we introduce the matter currents:

\boxdefinition{
\begin{defn}
\textbf{(Matter currents)}. We define respectively the \emph{hypermomentum}, the \emph{canonical energy-momentum} and the \emph{metric energy-momentum} currents:
\begin{equation}
\dfvarDe^a{}_b\coloneqq\dfrac{\delta S_{\mathrm{Matt}}}{\delta\dfom_a{}^b}\,,\qquad\dfSi_a\coloneqq\dfrac{\delta S_{\mathrm{Matt}}}{\delta\cofr^a}\,,\qquad\dfsi^{ab}\coloneqq2\dfrac{\delta S_{\mathrm{Matt}}}{\delta g_{ab}}\,.
\end{equation}
\end{defn}
}

When working in components we will use the following notation\footnote{
    Notice that $\mathcal{T}_{\mu\nu}\equiv e_{\mu}{}^{a}e_{\nu}{}^b\mathcal{T}_{ab}$ is the usual Hilbert energy-momentum tensor used in GR.}
\begin{align}
\sqrt{\left|g\right|}\mathcal{T}^{ab}  & \coloneqq2\frac{\delta_{\mathrm{c}}S_\mathrm{Matt}}{\delta g_{ab}}\,,\label{eq:defEMtensorg}\\
\sqrt{\left|g\right|}\Sigma^{\mu}{}_a  & \coloneqq \frac{\delta_{\mathrm{c}}S_\mathrm{Matt}}{\delta e_{\mu}{}^a}\,,\\
\sqrt{\left|g\right|}\varDelta^{\mu a}{}_b& \coloneqq \frac{\delta_{\mathrm{c}}S_\mathrm{Matt}}{\delta\omega_{\mu a}{}^b}\,.
\end{align}
By virtue of our dictionary (Proposition \ref{Prop: var diccionary}), one can easily check that
\begin{equation}
\dfvarDe^a{}_b=\varDelta^{\mu a}{}_b\star\dex x_{\mu}\,,\qquad\dfSi_a=\Sigma^{\mu}{}_a\star\dex x_{\mu}\,,\qquad\dfsi^{ab}=\mathcal{T}^{ab}\volfg\,.
\end{equation}

Via the Noether identities, one can show that $\dfsi^{ab}$ can be expressed in terms of the other currents. The fundamental matter currents in MAG are then the hypermomentum and the canonical energy-momentum. The latter is associated with the translational part of the gauge group, whereas the hypermomentum corresponds to the general linear group. In particular, the hypermomentum can be decomposed into antisymmetric, trace and traceless symmetric parts. Each of them has a different physical meaning according to the following decomposition of the generators of the algebra $\mathfrak{gl}(\dimM,\mathbb{R})$  (possible in the presence of $g_{ab}$)
\begin{equation}
  g_{ac}{\rm L}^c{}_b = {\rm L}_{[ab]}  + \frac{1}{\dimM} g_{ab}{\rm L}^c{}_c + \rmLNoTr_{ab} \,.
\end{equation}
${\rm L}_{[ab]}$ are the Lorentz generators, the second term generates the dilations, and the remaining term $\rmLNoTr_{ab}\coloneqq {\rm L}_{(ab)}- \frac{1}{\dimM} g_{ab}{\rm L}^c{}_c$ is the shear part. According to these parts, the hypermomentum splits, respectively, into the \emph{spin density current} $\dfvarDe_{[ab]}$, the \emph{dilation current} $\dfvarDe^c{}_c$, and the \emph{shear density current}  $\dfvarDe_{(ab)}-\frac{1}{\dimM}g_{ab}\dfvarDe^c{}_c$. These are, respectively, the sources of contorsion (i.e., torsion), the Weyl 1-form and the traceless part of the nonmetricity 1-form. 

The spin current is one of the main characters in PG, since the connection is antisymmetric and, hence, the hypermomentum coincides with the spin current. For instance, in \cite{HehlKerlick1974} it is shown that, in the context of Einstein-Cartan gravity with a fluid made of neutrons, the torsional gravitational effects (associated with the spin density) would be appreciable for densities of the order of $10^{57}\ {\rm kg}/{\rm m}^3$ (much higher than the density of a neutron star which is $10^{17}\ {\rm kg}/{\rm m}^3$). Interestingly, in earlier epochs of the Universe, scale invariance is expected to arise, and this has a dilation current associated which might be the one that naturally appears in MAG. Therefore, the latter seems to be a quite reasonable extension of PG to be considered. Nevertheless, the shear effects are also expected at higher energies, and their interpretation is subjected to speculation.\footnote{See \cite{NeemanSijacki1979, NeemanSijacki1988, NeemanSijacki1988b,Hehl1995, HehlObukhov1997}.} These currents, which are associated with the microstructure of matter, are the essential quantities that should be used to detect post-Riemannian geometry \cite{NeemanHehl1997, YasskinStoeger1980,PuetzfeldObukhov2007, ObukhovPuetzfeld2015}. 

~

More explicit expressions for the matter currents can be obtained by using the relations given in Proposition \ref{Prop: MAG functional variations}. The presence of $\dfR_a{}^b$, $\dfT^a$ and $\dfQ_{ab}$ in the matter Lagrangian reflects what we are going to call non-minimal couplings:\footnote{
    Interestingly, GR can be recovered within the Poincaré gravity framework under the assumption of an appropriate nonminimal coupling of the matter to the geometry \cite{ObukhovHehl2020}.} 

\boxdefinition{
\begin{defn}
\textbf{(Minimal coupling)}. For a given matter field $\dfvarPsi$ living in some representation of the total gauge group of a theory, we will say that it is \emph{minimally couple to the gauge structure} if the matter-dependent part of the Lagrangian depends exclusively on $\dfvarPsi$ and their total gauge exterior covariant derivative (including not only gravity but also Yang-Mills connections if other gauge structures are involved). Consequently, no curvatures or field strengths appear in it. 
\end{defn}
}

As immediate consequence of Proposition \ref{Prop: MAG functional variations} we find:
\boxproposition{
\begin{prop}
For a matter field $\dfvarPsi^\mathtt{M}$ minimally coupled to the MAG action, the matter currents are given by:
\begin{equation}
\dfvarDe^a{}_{b}  =\big(\bar\rho^{(\dfvarPsi)} ({\rm L}^a{}_{b}){}^\mathtt{M}{}_\mathtt{N}\dfvarPsi^\mathtt{N}\big)\wedge\frac{\partial\dfL_{\mathrm{Matt}}}{\partial\Dex\dfvarPsi^\mathtt{M}}\,,\qquad
\dfSi_{a} =\frac{\partial\dfL_{\mathrm{Matt}}}{\partial\cofr^a}\,,\qquad
\dfsi^{ab} =2\frac{\partial\dfL_{\mathrm{Matt}}}{\partial g_{ab}}\,.
\end{equation}
\end{prop}
}

\subsubsection*{Gravitational momenta and currents}

In the gravitational sector we are going to distinguish between momenta and currents, because they play a different role in the equations of motion:

\boxdefinition{
\begin{defn}
\textbf{\label{Def: GMom Gcurr}(Gravitational momenta and gravitational currents)}. For a general gravitational Lagrangian \eqref{eq: LGrav MAG}, we define the \emph{gravitational momenta} (or \emph{excitations}):
\begin{equation}
\dfH[\omega]{}^a{}_b\coloneqq-\frac{\partial\dfL_{\mathrm{Grav}}}{\partial\dfR_a{}^b}\,,\qquad\dfH[\vartheta]{}_a\coloneqq-\frac{\partial\dfL_{\mathrm{Grav}}}{\partial\dfT^a}\,,\qquad\dfH[g]{}^{ab}\coloneqq-2\frac{\partial\dfL_{\mathrm{Grav}}}{\partial\dfQ_{ab}}\,.\label{eq: def momenta}
\end{equation}

The \emph{gravitational currents} are:
\begin{align}
\dfE[\omega]{}^a{}_{b} & \coloneqq-\cofr^a\wedge\dfH[\vartheta]{}_{b}-g_{bc}\dfH[g]{}^{ac}\,,\label{eq: E(w)}\\
\dfE[\vartheta]{}_{a} & \coloneqq\frac{\partial\dfL_{\mathrm{Grav}}}{\partial\cofr^a}\,,\\
\dfE[g]{}^{ab} & \coloneqq2\frac{\partial\dfL_{\mathrm{Grav}}}{\partial g_{ab}}\,.
\end{align}
\end{defn}
}

They are called momenta because (up to maybe a sign or a constant factor) they coincide with the canonical momenta of the gravitational fields in the Hamiltonian formulation. One interesting consequence of the Noether identities (see eqs. (5.4.11) and (5.4.15) of \cite{Hehl1995}) is that the currents $\dfE[\vartheta]{}_{a}$ and $\dfE[g]{}^{ab}$ are totally determined by the gravitational
momenta through the equations
\begin{align}
\dfE[\vartheta]{}_{a} & =\dint{\vfre_{a}}\dfL_{\mathrm{Grav}}+(\dint{\vfre_{a}}\dfT^{b})\wedge\dfH[\vartheta]{}_{b}+(\dint{\vfre_{a}}\dfR_b{}^{c})\wedge\dfH[\omega]{}^b{}_{c}+\frac{1}{2}(\dint{\vfre_{a}}\dfQ_{bc})\dfH[g]^{bc},\label{eq:E(e)}\\
\dfE[g]{}^{ac}g_{cb} & =\cofr^a\wedge\dfE[\vartheta]{}_{b}+\dfQ_{bc}\wedge\dfH[g]^{ac}-\dfT^a\wedge\dfH[\vartheta]{}_{b}-\dfR_c{}^a\wedge\dfH[\omega]{}^c{}_{b}+\dfR_b{}^c\wedge\dfH[\omega]{}^a{}_c\,.
\end{align}
We see that the Lagrangian and the momenta determine $\dfE[\vartheta]{}_a$. Moreover, $\dfE[\vartheta]{}_{a}$ and the momenta fix the current $\dfE[g]{}^{ab}$.\footnote{
     Indeed, computing the quantity $\dfE[g]{}^{ab}$ is not needed. The reason is that it is only relevant for the metric equation of motion and, as we have seen, this equation is redundant in metric-affine gravity.} 
As a consequence of this result and the equation \eqref{eq: E(w)}, we only need to compute the momenta $\dfH[\omega]{}^a{}_{b}$, $\dfH[\vartheta]{}_{a}$ and $\dfH[g]{}^{ab}$, since the rest of the objects are determined by them (and the Lagrangian).

\subsection{Equations of motion and procedure to explore solutions in MAG}

Taking into account the previous definitions, we can write the equations of motion of MAG in terms of the momenta and the gravitational and matter currents:

\boxtheorem{
\begin{thm}
\textbf{\textup{(Eq. of motion of general MAG)}} \\The equations of motion of \textup{\eqref{eq: SGrav SMatt}} are
\begin{align}
0=\dfrac{\delta S}{\delta\dfom_a{}^b} & \equiv -\Dex\dfH[\omega]{}^a{}_{b}+\dfE[\omega]{}^a{}_{b} + \dfvarDe^a{}_b\,,\\
0=\dfrac{\delta S}{\delta\cofr^a} & \equiv -\Dex\dfH[\vartheta]{}_a+\dfE[\vartheta]{}_a +\dfSi_a\,,\\
0=2\dfrac{\delta S}{\delta g_{ab}} & \equiv -\Dex\dfH[g]{}^{ab}+\dfE[g]{}^{ab} +\dfsi^{ab}\,,\\
0=\frac{\delta S}{\delta\dfvarPsi^\mathtt{M}} & \equiv \frac{\partial\dfL}{\partial\dfvarPsi^\mathtt{M}} - (-1)^{\mathrm{rank}(\dfvarPsi^\mathtt{M})} \Dex\frac{\partial\dfL}{\partial\Dex\dfvarPsi^\mathtt{M}} \,.
\end{align}
\end{thm}
}

But we know that some of the objects involved in these equations can be computed from others and, additionally, that the equation of motion of the metric is on-shell redundant. So a standard procedure to find solutions of a MAG theory is the following:
\begin{enumerate}
\item Compute the gravitational momenta $\dfH[\omega]{}^a{}_{b}$, $\dfH[\vartheta]{}_{a}$
and $\dfH[g]{}^{ab}$ and the matter currents $\dfvarDe^a{}_b$ and $\dfSi_a$.
\item Compute $\dfE[\omega]{}^a{}_{b}$ and $\dfE[\vartheta]{}_{a}$.
\item Evaluate $\dfvarDe^a{}_b$, $\dfSi_a$, $\dfH[\omega]{}^a{}_{b}$, $\dfH[\vartheta]{}_{a}$, $\dfE[\omega]{}^a{}_{b}$
and $\dfE[\vartheta]{}_{a}$ in the specific Ansatz we are interested in. Fix also the gauge by a taking, e.g., a simple form for the metric  (usually Minkowski $g_{ab}\to\eta_{ab}$).
\item Solve the dynamical equations
\begin{align}
\Dex\dfH[\vartheta]{}_{a}-\dfE[\vartheta]{}_{a}  & = \dfSi_a\,,\nonumber \\
\Dex\dfH[\omega]{}^a{}_{b}-\dfE[\omega]{}^a{}_{b}& = \dfvarDe^a{}_b\,.\label{eq: MAGeq vacuum}
\end{align}
\end{enumerate}

This procedure can also be followed in components notation and at the end one has to solve
\begin{align}
\sqrt{\left|g\right|}\Sigma^{\mu}{}_{a} & =2\left(\nabla_{\rho}+T_\rho\right)\left(\dfrac{\partial\mathfrak{L}_{\mathrm{Grav}}}{\partial T_{\rho\mu}{}^a}\right)-T_{\rho\sigma}{}^{\mu}\dfrac{\partial\mathfrak{L}_{\mathrm{Grav}}}{\partial T_{\rho\sigma}{}^a}-\dfrac{\partial\mathfrak{L}_{\mathrm{Grav}}}{\partial e_{\mu}{}^a}\,,\\
\sqrt{\left|g\right|}\Delta^{\mu a}{}_{b} & =2\left(\nabla_{\rho}+T_\rho\right)\left(\frac{\partial\mathfrak{L}_{\mathrm{Grav}}}{\partial R_{\rho\mu a}{}^b}\right)-T_{\rho\sigma}{}^{\mu}\frac{\partial\mathfrak{L}_{\mathrm{Grav}}}{\partial R_{\rho\sigma a}{}^b}-2g_{bc}\frac{\partial\mathfrak{L}_{\mathrm{Grav}}}{\partial Q_{\mu ac}}-2\frac{\partial\mathfrak{L}_{\mathrm{Grav}}}{\partial T_{\mu\rho}{}^b}e_{\rho}{}^a\,.
\end{align}
These expressions are easier to be programmed in xAct for example, so one can avoid dealing with differential forms. However we will exploit the nice algebra of forms e.g. in Chapter \ref{ch:GWsolutions} to find some exact solutions. Differential forms notation can be very useful (specially in gauge theories) due to their algebraic properties, its compactness and because several identities take a very simple form, such as the Bianchi identities. 

\section{The (quadratic) Metric-Affine Gauge action}

Now that we are familiar with the general metric-affine theory, we proceed to introduce the general \emph{quadratic MAG theory}. The corresponding Lagrangian is defined as \emph{the most general linear combination of MAG invariants up to quadratic order in the curvature, torsion and nonmetricity}. In addition, the resulting dynamical equations are going to be \emph{quasilinear} in the basic fields \cite{HehlMacias1999}. Of course this choice is also a matter of simplicity since, as we will see, it is already considerably complex to consider just up to quadratic invariants. The construction of the basis of invariants that we use is explained in more detail in Appendix \ref{app:MAGinvariants}.\footnote{
    If one tries to construct the usual Yang-Mills like Lagrangian, due to the fact that the gauge group is not semi-simple, the resulting action does not lead to the correct results in the case of PG \cite{Tseytlin1982}. Something analogous will happen in MAG.}

\subsection{\label{subsec: qMAG action}The (quadratic) Metric-Affine Gauge action}

The most general quadratic MAG action can be decomposed into an even part (dimension independent) and the odd part corresponding to the chosen dimension $\dimM$. For the even part, we consider the following parameterization\footnote{
    The signs and factors of 2 have been adjusted so that the Lagrangian coincides with the Lagrangian $V$ in \cite{JCAObukhov2021a} (except for the cosmological constant which is not considered in that paper).}
\begin{align}
\dfL_{\mathrm{MAG}}^{\mathrm{even}} & =-\frac{1}{2\kappa^{(\dimM)}}\bigg[ 2\lambda\volfg-a_{0}\dfR^{ab}\wedge\star\cofr_{ab}+\dfT^a\wedge\star\sum_{I=1}^{3}a_{I}\irrdfT{I}{}_{a}+\dfQ_{ab}\wedge\star\sum_{I=1}^{4}b_{I}\irrdfQ{I}{}^{ab}\nonumber \\
 & \qquad\qquad+2b_{5}(\irrdfQ{3}{}_{ac}\wedge\cofr^a) \wedge\star(\irrdfQ{4}{}^{bc}\wedge\cofr_{b})- 2\sum_{I=1}^{3}c_{I}\irrdfQ{I+1}{}_{ab}\wedge\cofr^a\wedge\star\dfT^b\bigg]\nonumber \\
 & \quad-\frac{1}{2\rho^{(\dimM)}} \dfR_{ab}\wedge\star\Bigg[\sum_{I=1}^{6}w_{I} \irrdfW{I}{}^{ab}+v_{1}\cofr^a\wedge\big(\dint{\vfre_c}\irrdfW{5}{}^{cb}\big)\nonumber \\
 & \qquad\qquad\qquad +\sum_{I=1}^{5}z_{I}\irrdfZ{I}{}^{ab}+ v_{2}\cofr_c\wedge\big(\dint{\vfre^a} \irrdfZ{2}{}^{cb}\big)+\sum_{I=3}^{5}v_{I} \cofr^a\wedge\big(\dint{\vfre_c}\irrdfZ{I}{}^{cb}\big)\Bigg]\,.\label{eq: qMAGLeven}
\end{align}
Some remarks about the constants and parameters introduced:
\begin{itemize}
\item The usual cosmological constant (with units of energy density) is given by $\Lambda\coloneqq\frac{\lambda}{\kappa}$.
\item There are 29 parameters: $\lambda$ is dimensionful and the other 28 ($a_{i}$, $b_{i}$, $c_{i}$, $w_{i}$, $z_{i}$, $v_{i}$) are dimensionless.
\item $\kappa^{(\dimM)}$ and $\rho^{(\dimM)}$ are, respectively, the weak and the strong gravitational couplings of the corresponding dimension. In four dimensions we will simply write
\begin{equation}
\kappa^{(4)}\eqqcolon\kappa\,,\qquad\rho^{(4)}\eqqcolon\rho\,.
\end{equation}
The dimensions of these parameters as well as those of the cosmological constant are shown in Table \ref{tab: dims param}. 
\item Among the 29 parameters, only 28 are physical, since one global factor can be extracted from the action. The parameter that will play such a role is $a_{0}$, the one in front of the metric-affine Einstein term. We will write it explicitly so that the derived formulae can be used also in the study of MAG without the Einstein term ($a_{0}=0$). In any other case it will be (virtually) a 1 in order to have the Einstein term correctly normalized.
\end{itemize}
All of this was for the even part in general dimensions. 

\begin{table}
\begin{centering}
\renewcommand\arraystretch{1.5}
\begin{tabular}{|c|c|c|}
\hline 
 & $c=1$ & $c,\hbar=1$\tabularnewline
\hline 
\hline 
$\left[\kappa^{\dimM}\right]=\big[G_{\mathrm{N}}^{(\dimM)}\big]$ & $\mathrm{M}^{-1}\mathrm{L}^{\dimM-3}$ & $\mathrm{M}^{-(\dimM-2)}$\tabularnewline
\hline 
$\left[\rho^{\dimM}\right]$ & $\mathrm{M}^{-1}\mathrm{L}^{\dimM-5}$ & $\mathrm{M}^{-(\dimM-4)}$\tabularnewline
\hline 
$\left[\lambda\right]$ & $\mathrm{L}^{-2}$ & $\mathrm{M}^{2}$\tabularnewline
\hline 
$\left[\Lambda\right]$ & $\mathrm{M}\mathrm{L}^{-(\dimM-1)}$ & $\mathrm{M}^{\dimM}$\tabularnewline
\hline 
\end{tabular}
\renewcommand\arraystretch{1}
\end{centering}
\centering{}\caption{\label{tab: dims param}Dimensionful MAG parameters}
\end{table}

On the other hand, the odd-parity Lagrangian in $\dimM=4$ is given by
\begin{align}
\dfL_{\mathrm{MAG}}^{\mathrm{odd}(4)} & =-\frac{1}{2\kappa}\Big[-\odda_{0}\dfR^{ab}\wedge\cofr_{ab}+\dfT^a\wedge\sum_{I=1}^{3}\odda_{I}\irrdfT{I}{}_a\nonumber \\
 & \qquad+\oddb_5(\irrdfQ{2}{}_{ab}\wedge\cofr^{b})\wedge(\irrdfQ{2}{}^{ac}\wedge\cofr_{c})-2\sum_{I=1}^{3}\oddc_{I}\irrdfQ{I+1}{}_{ab}\wedge\cofr^a\wedge\dfT^b\Big]\nonumber \\
 & \quad-\frac{1}{2\rho}\dfR_{ab}\wedge\Bigg[\sum_{I=1}^{6}\oddw_{I}\irrdfW{I}{}^{ab}+\oddv_{1}\cofr^a\wedge\big(\dint{\vfre_c}\irrdfW{5}{}^{cb}\big)\nonumber \\
 & \qquad\qquad\qquad+\sum_{I=1}^{5}\oddz_{I}\irrdfZ{I}{}^{ab}+\oddv_{2}\cofr_c\wedge\big(\dint{\vfre^a}\irrdfZ{2}{}^{cb}\big)+\sum_{I=3}^{5}\oddv_{I}\cofr^a\wedge\big(\dint{\vfre_c}\irrdfZ{I}{}^{cb}\big)\Bigg]\,,\label{eq: qMAGLodd}
\end{align}
Here we have introduce 24 dimensionless parameters ($\odda_{0}$, $\odda_{i}$, $\oddb_5$, $\oddc_{i}$, $\oddw_{i}$, $\oddz_{i}$, $\oddv_{i}$), but due to the properties \eqref{eq:redundantodd1}, \eqref{eq:redundantodd2} and \eqref{eq:redundantodd3}, only 20 are independent. We will assume the choice 
\begin{equation}\label{eq:oddparcond}
\odda_3=\odda_2,\qquad\oddw_4=\oddw_2,\qquad\oddw_6=\oddw_3,\qquad\oddz_4=\oddz_2\,.
\end{equation}
Indeed (see next chapter), two more parameters can be dropped since there are two topological invariants that can be used to eliminate two of the terms from the odd Lagrangian (see Section \ref{sec:PontrNieh}).

Finally, the metric-affine Lagrangian is the addition of these two parts:
\begin{equation}
\dfL_{\mathrm{MAG}}=\dfL_{\mathrm{MAG}}^{\mathrm{even}}+\dfL_{\mathrm{MAG}}^{\mathrm{odd}(4)}\,.\label{eq: MAG Lag form}
\end{equation}
In Table \ref{tab: param} we show the 45 parameters of the MAG action.
\begin{table}
\begin{centering}
\begin{tabular}{c|c|c}
\hline 
Type of term & Parameters of the even part & Parameters of the 4D odd part\tabularnewline
\hline 
\hline 
$\sim\volfg$ & $\lambda$ & \tabularnewline
\hline 
$\sim R$ & $a_{0}$ & $\odda_{0}$\tabularnewline
\hline 
$\sim TT$ & $a_{1}$, $a_{2}$, $a_{3}$ & $\odda_{1}$, $\odda_{2}$, $(\odda_{3})$\tabularnewline
\hline 
$\sim QQ$ & $b_{1}$, $b_{2}$, $b_{3}$, $b_{4}$, $b_{5}$ & $\oddb_5$\tabularnewline
\hline 
$\sim TQ$ & $c_{1}$, $c_{2}$, $c_{3}$ & $\oddc_{1}$, $\oddc_{2}$, $\oddc_{3}$\tabularnewline
\hline 
$\sim RR$ ($WW$) & $w_{1}$, $w_{2}$, $w_{3}$, $w_{4}$, $w_{5}$, $w_{6}$,  & $\oddw_{1}$, $\oddw_{2}$, $\oddw_{3}$, $(\oddw_{4})$,
$\oddw_{5}$, $(\oddw_{6})$\tabularnewline
\hline 
$\sim RR$ ($ZZ$) & $z_{1}$, $z_{2}$, $z_{3}$, $z_{4}$, $z_{5}$ & $\oddz_{1}$, $\oddz_{2}$, $\oddz_{3}$, $(\oddz_{4})$,
$\oddz_{5}$\tabularnewline
\hline 
$\sim RR$ (mixed) & $v_{1}$, $v_{2}$, $v_{3}$, $v_{4}$, $v_{5}$ & $\oddv_{1}$, $\oddv_{2}$, $\oddv_{3}$, $\oddv_{4}$, $\oddv_{5}$\tabularnewline
\hline 
\end{tabular}
\par\end{centering}
\centering{}\caption{\label{tab: param}Table of independent parameters of the MAG (quadratic) Lagrangian  (29 from the even part, and 20 from the odd part). Parenthesis indicate that they are not independent parameters and can be fixed due to \eqref{eq:redundantodd1}, \eqref{eq:redundantodd2} and \eqref{eq:redundantodd3}.}
\end{table}

\subsubsection*{Momenta for the (quadratic) MAG action}

Recall that the essential objects to construct the gravitational equations of motion are the momenta. If we introduce the following useful decomposition:
\begin{align}
\dfH[g]{}^{ab} & \eqqcolon\frac{2}{\kappa^{(\dimM)}}\dfm^{ab}+\frac{2}{\kappa}\odddfm^{ab}\,,\\
\dfH[\vartheta]{}_{a} & \eqqcolon\frac{1}{\kappa^{(\dimM)}}\dfh_{a}+\frac{1}{\kappa}\odddfh_a\,,\\
\dfH[\omega]{}^a{}_{b} & \eqqcolon\bigg(-\frac{1}{2\kappa^{(\dimM)}}a_{0}\star\cofr^a{}_{b}+\frac{1}{\rho^{(\dimM)}}\dfh^a{}_b\bigg)+\left(-\frac{1}{2\kappa}\odda_{0}\cofr^a{}_{b}+\frac{1}{\rho}\odddfh^a{}_b\right)\,,\label{eq:decomHw}
\end{align}
where the second terms in the r.h.s. of the three equations are those contributions coming from the odd part of the Lagrangian. After a pretty long calculation based on the results in Appendix \ref{app:variations}, it can be shown that (as a matter of generality, we have kept the dimension $\dimM$ arbitrary in the even objects):\footnote{See also \cite{Esser1996_Diploma}.}
\begin{align}
\dfm^{ab} & =\star\Big\{\sum_{I=1}^{4}b_{I}\irrdfQ{I}{}^{ab}-b_{5} \Big[\cofr^{(a}(\dint{\vfre^{b)}}\dfQ) +\frac{1}{\dimM}g^{ab}(\dfvarLa-\dfQ)\Big]\nonumber \\
    & \qquad+c_{1}\dint{\vfre^{(a}}\dfT^{b)} -\frac{\dimM c_{1}-c_{2}-(\dimM-1)c_{3}}{\dimM(\dimM-1)} g^{ab}\dfT+\frac{c_{1}-c_{2}}{\dimM-1} \cofr^{(a}(\dint{\vfre^{b)}}\dfT)\Big\}\,,\label{eq: GMom g}\\
\odddfm^{ab} & =-\oddb_5\cofr^{(a}\star\oddvar{\dfvarLa}{}^{b)}-\oddc_{1}\cofr^{(a}\wedge\dfT^{b)}-\frac{\oddc_{2}-\oddc_{1}}{3}\cofr^{(a}\wedge\dint{\vfre^{b)}}\star\odddfT-\frac{\oddc_{3}-\oddc_{2}}{4}g^{ab}\star\odddfT \,,\\
\dfh_a & =\star\Big\{ \sum_{I=1}^{3}a_{I}\irrdfT{I}{}_{a} +\cofr^b\wedge\sum_{I=2}^{4}c_{I-1} \irrdfQ{I}{}_{ab}\Big\}\,,\label{eq: GMom e}\\
\odddfh_{a} & =\odda_{1}\irrdfT{1}{}_{a} +\odda_{2}(\irrdfT{2}{}_{a} +\irrdfT{3}{}_{a})+\cofr^b\wedge\sum_{I=2}^{4}\oddc_{I-1}\irrdfQ{I}{}_{ab}\,,\\
\dfh_{ab} & = \star \Big\{ \sum_{I=1}^{6}w_{I}\irrdfW{I}{}_{ab}+\sum_{I=1}^{5}z_{I}\irrdfZ{I}{}_{ab}\nonumber \\
    & \qquad+\frac{1}{2}v_{1} \Big[\cofr_a\wedge(\dint{\vfre_c}\irrdfW{5}{}^c{}_{b})+\frac{1}{2}\cofr_{[a}\wedge\dint{\vfre_{b]}}\dfP\Big]\nonumber \\
    & \qquad+\frac{1}{2}v_{2} \Big[\cofr_c\wedge(\dint{\vfre_{(a}}\irrdfW{2}{}^c{}_{b)})+\cofr_c\wedge(\dint{\vfre_{[a}}\irrdfZ{2}{}^c{}_{b]})-2\irrdfZ{2}{}_{ab}\Big]\nonumber \\
    & \qquad+\frac{1}{2}v_{3} \Big[\cofr_a\wedge(\dint{\vfre_c}\irrdfZ{3}{}^c{}_{b})-\frac{1}{2}\cofr_{(a}\wedge(\dint{\vfre_{b)}}\dfP)+\frac{1}{\dimM}g_{ab}\dfP\Big]\nonumber \\
    & \qquad+\frac{1}{2}v_{4} \Big[\cofr_{[a}\wedge(\dint{\vfre_{|c|}}\irrdfZ{4}{}^c{}_{b]})+\cofr_{(a}\wedge(\dint{\vfre_{|c|}}\irrdfW{4}{}^c{}_{b)})+\dimM\irrdfZ{4}{}_{ab}\Big]\nonumber \\
    & \qquad+\frac{1}{2}v_{5} \Big[\cofr_a\wedge(\dint{\vfre_c}\irrdfZ{5}{}^c{}_{b})+\frac{1}{\dimM}g_{ab}\dfP\Big]\Big\}\,,\label{eq: GMom w}\\
\odddfh_{ab} & =\sum_{I=1}^{6} \oddw_{I}\irrdfW{I}{}_{ab} +\sum_{I=1}^{5}\oddz_{I}\irrdfZ{I}{}_{ab}\nonumber \\
    & \quad+\frac{1}{2} \oddv_{1} \Big[\cofr_a\wedge(\dint{\vfre_c} \irrdfW{5}{}^c{}_{b})-\frac{1}{2} \cofr_{[a}\wedge\dint{\vfre_{b]}} \star\odddfP\Big]\nonumber \\
    & \quad+\frac{1}{2}\oddv_{2} \Big[\cofr_c\wedge(\dint{\vfre_{[a}} \irrdfZ{2}{}^c{}_{b]})-\cofr_{(a}\wedge(\dint{\vfre_{|c|}}\irrdfW{4}{}^c{}_{b)}) -\irrdfZ{2}{}_{ab}-\irrdfZ{4}{}_{ab}\Big]\nonumber \\
    & \quad+\frac{1}{2}\oddv_{3} \Big[\cofr_a\wedge(\dint{\vfre_c} \irrdfZ{3}{}^c{}_{b}) -\frac{1}{2}\cofr_{(a} \wedge(\dint{\vfre_{b)}} \star\odddfP) +\frac{1}{4}g_{ab}\star\odddfP\Big]\nonumber \\
    & \quad+\frac{1}{2} \oddv_{4}\Big[\cofr_{[a} \wedge(\dint{\vfre_{|c|}}\irrdfZ{4}{}^c{}_{b]}) -\cofr_c\wedge(\dint{\vfre_{(a}} \irrdfW{2}{}^c{}_{b)}) +2\irrdfZ{2}{}_{ab}+2\irrdfZ{4}{}_{ab}\Big]\nonumber \\
    & \quad+\frac{1}{2} \oddv_{5}\Big[\cofr_a \wedge(\dint{\vfre_c} \irrdfZ{5}{}^c{}_{b}) -\frac{1}{4}g_{ab}\star\odddfP\Big]\,,\label{eq: GMomwod}
\end{align}
where the forms $\dfT$, $\odddfT$, $\dfvarLa$, $\odddfvarLa_a$, $\dfP$ and $\odddfP$ are defined in Appendix \ref{app:irreds}.

\newpage
\subsection{Equations of motion of the quadratic MAG action in $\dimM=4$}

Now we are in position to rewrite the currents and equations of motion in terms of the quantities \eqref{eq: GMom g}-\eqref{eq: GMomwod}. Similarly as we did in \eqref{eq:decomHw}, it is also useful to separate the quadratic part of the Lagrangian from the linear one. In $\dimM=4$,
\begin{equation}
\dfL_{\mathrm{MAG}}=\frac{1}{2\kappa}\Big[\dfR^{ab}\wedge(a_{0}\star\cofr_{ab}+\odda_{0}\cofr_{ab})-2\lambda\volfg\Big]+\frac{1}{\kappa}\dfL_{(2)}\,.
\end{equation}
In addition, let us introduce the following parameter with dimensions of area:
\begin{equation}
  \lrho  \coloneqq \frac{\kappa}{\rho} \,,\label{eq: lr}
\end{equation}
which measures the contribution of the curvature square terms of the Lagrangian.

First we prove some previous results:

\boxlemma{
\begin{lem}\label{lem:currents1}
The gravitational current $\dfE[\vartheta]{}_{a}$ is given by:
\begin{align}
\dfE[\vartheta]{}_{a} & =\frac{1}{2\kappa}\left(a_{0}\dfR_{bc}\wedge\star\cofr^{abc}+2\odda_{0}\dfR_{[ac]}\wedge\cofr^{c}-2\lambda\star\cofr_a\right)+\frac{1}{\kappa}\dfq_a\,,
\end{align}
where 
\begin{equation}
\dfq_{a} \coloneqq\dint{\vfre_{a}}\dfL_{(2)}\!+(\dint{\vfre_{a}}\dfT^{b})\wedge(\dfh_{b}+\odddfh_{b})+(\dint{\vfre_{a}}\dfQ_{bc})(\dfm^{bc}+\odddfm{}^{bc})
 +\lrho(\dint{\vfre_{a}}\dfR_b{}^{c})\wedge(\dfh^b{}_{c}+\odddfh{}^b{}_{c})\,.\label{eq:qdef}
\end{equation}
\end{lem}
}

\boxproof{
\begin{proof}
We compute
\begin{align}
 & (\dint{\vfre_{a}}\dfR_b{}^c) \wedge\dfH[\omega]{}^b{}_c-\frac{1}{\rho}(\dint{\vfre_a}\dfR_b{}^{c})\wedge(\dfh^b{}_c+\odddfh{}^b{}_c)\nonumber \\
 & \quad=-\frac{1}{2\kappa}( \dint{\vfre_{a}}\dfR_b{}^{c})\wedge\left(a_{0}\star\cofr^b{}_c +\odda_{0}\cofr^b{}_c \right)\\
 & \quad=\dint{\vfre_{a}}\left[\frac{1}{\kappa} \dfL_{(2)}-\frac{\lambda}{\kappa} \volfg-\dfL_{\mathrm{MAG}}\right]+\frac{1}{2\kappa}\dfR_{bc} \wedge\dint{\vfre_{a}} \left(a_{0}\star\cofr^{bc}+\odda_{0}\cofr^{bc}\right)\\
 & \quad=\dint{\vfre_{a}}\left[\frac{1}{\kappa} \dfL_{(2)}-\dfL_{\mathrm{MAG}}\right]+\frac{1}{2\kappa}\left(a_{0}\dfR^{bc}\wedge\star \cofr{}_{bca}+2\odda_{0}\dfR_{[ac]} \wedge\cofr^{c}-2\lambda\star\cofr_a\right).
\end{align}
And, finally, we substitute this in \eqref{eq:E(e)}.
\end{proof}
}

\boxlemma{
\begin{lem}
The following relation holds in general for the quadratic MAG action in four dimensions:
\begin{align}
\kappa\Dex\dfH[\omega]{}^a{}_{b} & =-\frac{1}{2}a_{0}\left(\dfT^c\wedge\star\cofr{}^a{}_{bc}+\dfQ^{ac}\wedge\star\cofr{}_{cb}-2\dfQ\wedge\star\cofr{}^a{}_b\right)\\
 & \quad-\frac{1}{2}\odda_{0}\left(\dfT^a\wedge\cofr_{b}-\dfT_b\wedge\cofr^a-\dfQ_{cb}\wedge\cofr^{ac}\right)+\lrho\Dex(\dfh^a{}_{b}+\odddfh^a{}_{b})\,.
\end{align}
\end{lem}
}

\boxproof{
\begin{proof}
To prove this, after the substitution \eqref{eq:decomHw}, we just need to use
\begin{align}
\Dex\star\cofr^a{}_{b}=\Dex\star(g_{bc}\cofr^a\wedge\cofr^{c}) & =\dfQ^{ac}\wedge\star\cofr{}_{cb}-2\dfQ\wedge\star\cofr{}^a{}_{b}+\dfT^c\wedge\star\cofr{}^a{}_{bc}\,,\\
\Dex\cofr^a{}_{b}=\Dex(g_{bc}\cofr^a\wedge\cofr^{c}) & =\dfT^a\wedge\cofr_{b}-\cofr^a\wedge\dfT_{b}-\dfQ_{cb}\wedge\cofr^{ac}\,.
\end{align}
To compute the first one we made use of \eqref{eq: Dstarcofr}.

Then we lower the index
\begin{align}
\kappa\Dex\dfH[\omega]{}_{ab} & =-\frac{1}{2}a_{0}\left(\dfT^c\wedge\star\cofr{}_{abc}-\dfQ_a{}^c\wedge\star\cofr{}_{bc}-2\dfQ\wedge\star\cofr{}_{ab}\right)\\
 & \quad-\frac{1}{2}\odda_{0}\left(2\dfT_{[a}\wedge\cofr_{b]}-\dfQ_{cb}\wedge\cofr{}_a{}^c\right)-\lrho\dfQ{}_{ca}\wedge(\dfh{}^c{}_{b}+\odddfh{}^c{}_{b})+\lrho\Dex(\dfh{}_{ab}+\odddfh{}_{ab})
\end{align}
and from here one trivially obtains the symmetric and antisymmetric parts above.

\end{proof}
}

These previous results make it immediate to derive the following final form for the equations of motion of the 4-dimensional quadratic MAG theory:

\boxtheorem{
\begin{thm}\textbf{\textup{(Eq. of motion of quadratic MAG)}} \\ \label{th:EoMqMAG}
The variations with respect to the coframe and the connection of \textup{$S=\int\dfL_{\mathrm{MAG}}+S_{\mathrm{Matt}}$} in $\dimM=4$, with $\dfL_{\mathrm{MAG}}$ given in \textup{\eqref{eq: MAG Lag form}} and $S_{\mathrm{Matt}}$ being a general matter action, can be written
\begin{align}
\kappa\frac{\delta S}{\delta\cofr^a} & =\frac{a_{0}}{2}\dfR^{bc}\wedge\star\cofr{}_{bca} +\odda_{0}\dfR_{[ac]}\wedge\cofr^{c}-\lambda\star\cofr_{a}+\dfq_{a}\nonumber\\
& \quad-\Dex(\dfh_{a}+ \odddfh_{a})+\kappa\dfSi_a\,, \label{eq:qMAGEqe0}\\
\kappa\frac{\delta S}{\delta\dfom_a{}^b} & =-\cofr^a\wedge(\dfh_{b}+\odddfh_{b}) -2(\dfm^a{}_{b}+\odddfm{}^a{}_{b})\nonumber \\
 & \quad+\frac{a_{0}}{2} \left(\dfT^c\wedge\star\cofr{}^a{}_{bc} +\dfQ^{ac}\wedge\star\cofr{}_{cb}-2\dfQ\wedge\star\cofr{}^a{}_b\right)\nonumber \\
 & \quad+\frac{\odda_{0}}{2} \left(2g^{ac}\dfT_{[c}\wedge\cofr_{b]}- \dfQ_{cb}\wedge\cofr^{ac}\right)-\lrho\Dex(\dfh^a{}_b+\odddfh^a{}_b)+\kappa\dfvarDe^a{}_b \label{eq:qMAGEqw0}\,,
\end{align}
where $\dfq_{a}$ is given by \textup{\eqref{eq:qdef}}.
\end{thm}
}

\subsection{Comments on exact solutions}

The search and study of exact solutions is a crucial step to understand the physical aspects and the implications of any theory. The construction of concrete models is an excellent way to get a reduced framework in which one can compute physical observables. In the context of (quadratic) MAG, this also allows to fix the structure of the Lagrangian, in order to avoid problematic solutions or propagating modes (see Chapter \ref{ch:gravityviability}) and also to ensure certain consistency conditions (e.g. appropriate GR limit  \cite{PonomariovObukhov1982, Gronwald1997, HehlMacias1999}). Moreover, the exact solutions of MAG could also bring new insight on the microstructure of the spacetime. An exhaustive collection of exact solutions for the parity even Lagrangian (and sub-cases of it) can be found in \cite{HehlMacias1999}. We extract some of the references therein and some posterior ones for the present discussion.\footnote{
    It is worth mentioning the so-called \emph{triplet Ansatz} technique used e.g. in \cite{GarciaHehlLaemmerzahl1998, VlachynskyTresguerres1996, ObukhovVlachynskyEsser1996, Puetzfeld2000_Diploma}, which consists in selecting a purely trace torsion and nonmetricity, i.e. $\dfT^a =\irrdfT{2}{}^a$ and $\dfQ_{ab} =\irrdfQ{3}{}_{ab}+\irrdfQ{4}{}_{ab}$ such that (see the definitions of the traces in Appendix \ref{app:irreds}) $\dfQ = k_0 \dfA$, $\dfvarLa = k_1 \dfA$ and $\dfT = k_2 \dfA$ for some 1-form $\dfA$ and some real parameters $\{k_0,k_1,k_2\}$. This has proven to be a very effective method to derive exact solutions in MAG \cite{ObukhovVlachynskyEsser1997}.} 
                                                        
Regarding spherically and axially symmetric solutions (describing some compact gravitational source distribution), one can find in the literature solutions with gravito-electric charge of the Reissner-Nordström type \cite{Tresguerres1995a, Tresguerres1995b,ObukhovVlachynskyEsser1996, HoChernNester1997, TuckerWang1995} and of the Kerr-Newmann type \cite{VlachynskyTresguerres1996}, as well as solutions with both gravito-electric and gravito-magnetic charges \cite{MaciasMielkeSocorro1998,MaciasSocoro1999}. Other electrovacuum solutions have been explored extending a Pleba\'nski-Demia\'nski\footnote{
    This is an important family of Petrov type D solutions of the Einstein–Maxwell equations, which includes as sub-cases the Pleba\'nski-Carter, the Kerr–Newman, and the Kerr solutions.} 
metric structure to the metric-affine framework  \cite{GarciaHehlLaemmerzahl1998, HehlSocorro1998, GarciaMaciasSocorro1999}. In these solutions, in addition to the mass, some of the other MAG currents (dilation, shear and spin currents) are also present. 
Recently in \cite{BahamondeGigante2020} new solutions of the Black-Hole type have been studied for a restricted MAG action.

Metric-affine extensions of pp-waves have also been explored, see \cite{MaciasLammerzahlGarcia2000, GarciaLammerzahlMielke1998, Obukhov2006, Puetzfeld2002, GarciaMaciasPuetzfeld2000, TuckerWang1995, KingVassiliev2001, Vassiliev2002,Vassiliev2005, PasicVassiliev2005, PasicBarakovic2014, PasicBarakovic2017}. Some contributions in this direction are provided in Chapter \ref{ch:GWsolutions}.

At last, but not least, let us also comment a bit on cosmological solutions (see e.g. \cite{Tresguerres1994, ObukhovTresguerres1993, ObukhovVlachynskyEsser1997}. One interesting solution here is the one presented in \cite{Tresguerres1994}, which is a Weyl-Cartan geometry whose Weyl vector (the only nonmetricity of the solution) exponentially decays in time. This is quite interesting since, one would expect deviations with respect to Poincaré gauge gravity only at the very beginning of the Universe, when the dilation invariance (and hence the Weyl vector) plays a role. This is of course a simplified model, in which the arising of shear type excitations of the multispinor matter is not taken into account, although MAG predicts its existence. The development of PG and MAG cosmological models requires to generalize the GR perfect fluid by adding spin current to it  (e.g. the \emph{Weyssenhoff fluid} model \cite{HehlHeydeKerlickNester1976}), as well as shear and dilation currents, something called \emph{hyperfluid} (see \cite{ObukhovTresguerres1993} and the recent developments \cite{Iosifidis2020, Iosifidis2021}). This is also an interesting line of research from which we can learn about matter microstructure, depending on their compatibility with cosmological phenomenology.

Interestingly, the metric-affine framework allows for singularity-free solutions; a great example in the context of PG is the cosmological solution \cite{Kerlick1976} (see also \cite{Minkevich1980, Poplawski2010}). Of course, we also have to be aware of the new ones that could arise in this framework, due to the existence of new matter currents. For instance, torsion singularities have been noticed in the context of Einstein-Cartan gravity\footnote{
    Essentially the Einstein-Palatini action with zero nonmetricity.} 
\cite{NesterIsenberg1977} (see also \cite{ZhangCheng1995}).

\subsection{\label{subsec: Example EP}A particular example: Einstein-Palatini}

The simplest sub-case of the quadratic MAG theory is the metric-affine generalization of the Einstein-Hilbert Lagrangian (\emph{Einstein-Palatini theory}),
\begin{equation}
\dfL_{\mathrm{EP}}=\frac{1}{2\kappa^{(\dimM)}}\left[-2\lambda\volfg+\dfR^{ab}\wedge\star\cofr_{ab}\right]\qquad=\left(-\Lambda+\frac{1}{2\kappa^{(\dimM)}}R\right)\volfg\,,
\end{equation}
We assume also some matter Lagrangian and that the dimension is $\dimM>2$. The gravitational momenta are
\begin{equation}
\dfH[g]{}^{ab}=0,\qquad\dfH[\vartheta]{}_{a}=0,\qquad\dfH[\omega]{}^a{}_b\eqqcolon -\frac{1}{2\kappa^{(\dimM)}}\star\cofr^a{}_b\,,
\end{equation}
which allow to compute the currents
\begin{align}
\dfE[\omega]{}^a{}_{b} & =0\,,\\
\dfE[\vartheta]{}_{a} & =\dint{\vfre_{a}}\dfL_{\mathrm{EP}}+(\dint{\vfre_{a}}\dfR_b{}^{c})\wedge\dfH[\omega]{}^b{}_c=...=\frac{1}{2\kappa^{(\dimM)}}\left[-2\lambda\star\cofr_{a}+\dfR^{bc}\wedge\star\cofr_{bca}\right]\,.
\end{align}
The equations of motion, in this case are (compare with \eqref{eq:qMAGEqe0}-\eqref{eq:qMAGEqw0}):
\begin{align}
2\kappa^{(\dimM)}\frac{\delta S_\mathrm{EP}}{\delta\cofr^a} & =\dfR^{bc}\wedge\star\cofr{}_{bca} -2\lambda\star\cofr_{a}+2\kappa^{(\dimM)}\dfSi_a\,,\label{eq: eq cofr EP} \\
2\kappa^{(\dimM)}\frac{\delta S_\mathrm{EP}}{\delta\dfom_a{}^b} & =\dfQ^{ac}\wedge\star\cofr{}_{cb}-2\dfQ\wedge\star\cofr{}^a{}_b+\dfT^c\wedge\star\cofr{}^a{}_{bc} +2\kappa^{(\dimM)}\dfvarDe^a{}_b \label{eq: eq conn EP}\,,
\end{align}
Let us now extract the components of these equations.

\subsubsection*{Equation of the connection}

Consider a matter Lagrangian free of hypermomentum for simplicity. If we expand the differential forms in the coframe basis, the equation of the connection can be written
\begin{align}
2\kappa^{(\dimM)}\frac{\delta S_\mathrm{EP}}{\delta\dfom_a{}^b} 
 & =\left(Q_{e}{}^{ca}-\frac{1}{2}Q_e g^{ac}\right)\cofr^{e}\wedge\star\cofr_{cb}+\frac{1}{2}T_{ef}{}^{d}g^{ac}\cofr^{ef}\wedge\star\cofr_{cbd}\\
 & =\left(Q_{e}{}^{ca}-\frac{1}{2}Q_e g^{ac}\right)(-1)2\delta_{[c}^{e}\star\cofr_{b]}+\frac{1}{2}T_{ef}{}^{d}g^{ac}(-1)(3\times2)\delta_{[c}^{f}\delta_{b}^{e}\star\cofr_{d]}.
\end{align}
If we simplify the previous expression and lower the indices, we get
\begin{equation}
Q_{bac}-T_{bac}-\left(T_{a}-\frac{1}{2}Q_{a}+\Qb_a\right)g_{bc}+\left(T_{b}-\frac{1}{2}Q_b\right)g_{ac}=0\,.\label{eq: eq conn EP com}
\end{equation}
The general solution of this equation is Levi-Civita up to an arbitrary
mode $\dfA=A_{\mu}\dex x^{\mu}$ \cite{DadhichPons2012, BernalJCA2017},
\begin{equation}
\dfom_a{}^b =\mathring{\dfom}_a{}^{b}+\dfA\delta_a^b\qquad\Leftrightarrow\qquad\omega_{\mu a}{}^b =\mathring{\omega}_{\mu a}{}^{b}+A_{\mu}\delta_a^b\,.\label{eq: EP solw}
\end{equation}

This mode is called projective mode and is a consequence of the \emph{projective symmetry} \cite{Eisenhart1927} that this Lagrangian has (see for instance \cite{JuliaSilva1998}):
\begin{equation}
\dfom_a{}^b\to\dfom_a{}^{b}+\dfA\delta_a^b\qquad\forall\dfA\in\Omega^{1}(\mathcal{M})\,.
\end{equation}
Indeed, the Noether identity under this transformation basically tells that the variation of the EP action with respect to the connection is traceless as can be easily shown:
\begin{equation}
0=\delta_{\text{proj}}S_{\mathrm{EP}}=\int\delta_{\text{proj}}\dfom_a{}^b\wedge\frac{\delta S_{\mathrm{EP}}}{\delta\dfom_a{}^b}=\int\dfA\wedge\delta_a^b\frac{\delta S_{\mathrm{EP}}}{\delta\dfom_a{}^b}\qquad\Rightarrow\qquad\delta_a^b\frac{\delta S_{\mathrm{EP}}}{\delta\dfom_a{}^b}=0\,.
\end{equation}
which can be immediately checked by contracting with $g^{ab}$ in \eqref{eq: eq conn EP com} or by taking the trace of \eqref{eq: eq conn EP} with zero hypermomentum. 

\subsubsection*{Equation of the coframe}

Let us recover the Einstein equations in this formalism. We first compute
\begin{align}
\dfR^{bc}\wedge\star\cofr_{bca} & =\frac{1}{2}R_{mn}{}^{bc}\cofr^{mn}\wedge\star\cofr_{bca}\\
 & =\frac{1}{2}R_{mn}{}^{bc}(-1)(3\times2)\delta_{[b}^{n}\delta_{c}^{m}\star\cofr_{a]}\\
 & =-R_{mn}{}^{bc}\left(\delta_{b}^{[n}\delta_{c}^{m]}\star\cofr_{a}-\delta_a^{[n}\delta_{c}^{m]}\star\cofr_{b}-\delta_{b}^{[n}\delta_a^{m]}\star\cofr_c\right)\\
 & =+R\star\cofr_{a}-R^{(1)}{}_a{}^b\star\cofr_{b}+R^{(2)}{}_a{}^c\star\cofr_{c}
\end{align}
Therefore, the equation of the coframe \eqref{eq: eq cofr EP} can be written
\begin{equation}
\frac{1}{2}\left[R^{(1)}{}_{ab}-R^{(2)}{}_{ab}-Rg_{ab}\right]\star\cofr^b =\kappa^{(\dimM)}(-\Lambda g_{ab}+\Sigma{}_{ba})\star\cofr^b\,.
\end{equation}
We extract the components:
\begin{equation}\label{eq:cofrEP 2}
\frac{1}{2}\left(R^{(1)}{}_{ab}-R^{(2)}{}_{ab}-g_{ab}R\right)=\kappa^{(\dimM)}(-\Lambda g_{ab}+\Sigma{}_{ba})\,.
\end{equation}
If we assume that the matter has zero hypermomentum, the solution of the connection equation is Levi-Civita plus a projective mode. Such connection satisfies 
\begin{equation}
R^{(1)}{}_{ab}-R^{(2)}{}_{ab}=2R^{(1)}{}_{ab}\equiv2 \mathring{R}_{ab}\,.
\end{equation}
Then, our equation \eqref{eq:cofrEP 2} becomes
\begin{equation}
\mathring{R}_{ab}-\frac{1}{2}g_{ab}\mathring{R}=\kappa^{(\dimM)}(\Sigma{}_{ba}-\Lambda g_{ab})\,,
\end{equation}
which can be decomposed into symmetric and antisymmetric parts
\begin{equation}
\mathring{R}_{ab}-\frac{1}{2}g_{ab}\mathring{R}=\kappa^{(\dimM)}(\Sigma{}_{(ab)}-\Lambda g_{ab})\,,\qquad\Sigma{}_{[ab]}=0\,.
\end{equation}
The second condition is a restriction for the matter sector. Assuming that the matter is on-shell, the absence of hypermomentum implies that $\Sigma{}_{(ab)}$ is nothing but $\mathcal{T}_{ab}$ (see eq. \eqref{eq: rel EMtensors}), as a consequence of the gauge symmetry, and we obtain the Einstein equations:
\begin{equation}
\mathring{R}_{ab}-\frac{1}{2}g_{ab}\mathring{R}=\kappa^{(\dimM)}(\mathcal{T}_{ab}-\Lambda g_{ab})\,.
\end{equation}

\chapter{Metric-Affine Lovelock gravity} \label{ch:Lovelock}

\boxquote{Scientific theories can always be improved and are improved. That is one of the glories of science. It is the authoritarian view of the Universe that is frozen in stone and cannot be changed, so that once it is wrong, it is wrong forever.}{Isaac Asimov, ``The Nearest Star'' (1989)}

In the preliminary works \cite{BernalJCA2017, JanssenJCA2019a} the equivalence between metric and Palatini formulation was explored for the Einstein and the Gauss-Bonnet theories. In \cite{JanssenJCA2019a} some families of non-Levi-Civita solutions of the pure Gauss-Bonnet-Palatini theory were found. These works constitute a clear motivation  analyze the topological character of the metric-affine Lovelock terms in general. Are they boundary terms in their corresponding critical dimension? This is the question we are going to address.

\section{Introduction}\label{sec:introLovelock}

\emph{Lovelock gravities} constitute a family of higher-curvature Lagrangian terms that form a natural extension to standard General Relativity. Introduced in the early 1970s by Lovelock \cite{Lovelock1970, Lovelock1971} (though the simplest non-trivial case, Gauss-Bonnet gravity, was already identified by Lanczos in 1938 \cite{Lanczos1938}), they are characterized as \emph{the unique higher-curvature terms for which the equation of motion of the metric (the only field) is second-order}. This result is also known as the \emph{Lovelock theorem}. As we will see along this section, these terms exhibit many special properties. Interestingly, they appear as string corrections to supergravity \cite{CandelasHorowitz1985, GrossWitten1986, GrisaruZanon1986, Tseytlin1986, ParkZanon1987, MetsaevTseytlin1987} and over the years have attracted a lot of attention as alternatives of Dark Matter or Dark Energy, and in order to obtain corrections to black hole, cosmology and some holographic models (see for example \cite{BentoBertilami1996, CveticNojiri2002, NojiriOdintsov2005, BuchelEscobedoMyers2010, deBoerKulaxizi2011, CamanhoEdelstein2014, DadhichDurka2016}).

We start by introducing their analytic form:

\boxdefinition{\begin{defn} \label{def:Lovelock}
\textbf{((Metric) Lovelock invariant)}\\
Let $(\mathcal{M},\teng)$ be a $\dimM$-dimensional manifold with a metric structure. The \emph{$k$-th order (metric) Lovelock invariant} is the $\dimM$-form
\begin{equation}
  \mathring{\dfL}_{k}^{(\dimM)} \coloneqq \mathring{\dfR}^{a_{1}a_{2}}\wedge\ldots\wedge\mathring{\dfR}^{a_{2k-1}a_{2k}}\wedge\star\cofr_{a_{1}...a_{2k}}\,.
\end{equation}
\end{defn}}
After extracting the volume form, $\mathring{\dfL}_{k}^{(\dimM)}\eqqcolon\mathring{\mathcal{L}}_{k}^{(\dimM)}\volfg$, we obtain the scalar Lagrangian
\begin{align}
\mathring{\mathcal{L}}_{k}^{(\dimM)} &=\frac{1}{2^{k}}\begin{vmatrix}
\delta_{\mu_{1}}^{\nu_{1}} & ... & \delta_{\mu_{2k}}^{\nu_{1}}\\
\vdots &  & \vdots\\
\delta_{\mu_{1}}^{\nu_{2k}} & ... & \delta_{\mu_{2k}}^{\nu_{2k}}
\end{vmatrix} \mathring{R}_{\nu_1 \nu_2}{}^{\mu_1 \mu_2}\ldots\mathring{R}_{\nu_{2k-1}\nu_{2k}}{}^{\mu_{2k-1}\mu_{2k}} \\ 
&=\frac{1}{2^{k}}\  (2k)!  \delta_{\mu_1}^{[\nu_1}...\delta_{\mu_{2k}}^{\nu_{2k}]} \ \mathring{R}_{\nu_1 \nu_2}{}^{\mu_1 \mu_2}\ldots\mathring{R}_{\nu_{2k-1}\nu_{2k}}{}^{\mu_{2k-1}\mu_{2k}}\,,
\end{align} 
where the vertical lines represent the determinant.

Some examples of (metric) Lovelock invariants are:
\begin{enumerate}
\item The Einstein term ($k=1$):
\begin{equation}
  \mathring{\dfL}_{1}^{(\dimM)}=  \mathring{\dfR}_{ab} \wedge\star\cofr^{ab} \qquad \Rightarrow\qquad \mathring{\mathcal{L}}_{1}^{(\dimM)} = \mathring{R}\,.
\end{equation}

\item The Gauss-Bonnet term ($k=2$):
\begin{equation}
  \mathring{\dfL}_{2}^{(\dimM)}= \mathring{\dfR}_{ab}\wedge\mathring{\dfR}_{cd}\wedge\star\cofr^{abcd} 
  \qquad \Rightarrow\qquad \mathring{\mathcal{L}}_{2}^{(\dimM)} =  \mathring{R}^2-4\mathring{R}_{\mu\nu}\mathring{R}^{\mu\nu} +\mathring{R}_{\mu\nu\rho\lambda} \mathring{R}^{\mu\nu\rho\lambda} \,.\label{eq:defGB}
\end{equation}
\end{enumerate}

Two interesting properties of the Lovelock invariants are:
\begin{itemize}
  \item $\mathring{\dfL}_{k}^{(\dimM)}\equiv0$ for all $\dimM<2k$. \\ This is easy to check from the definition, since $\cofr_{a_{1}...a_{2k}}\equiv 0$ if $2k>\dimM$ and $\star0=0$. \\ Consequently, in a given dimension $\dimM$ the most general Lagrangian containing Lovelock invariants is ($\dimM$\emph{-dimensional Lovelock theory})
    \begin{equation}
      \mathring{\dfL}_\mathrm{Lov}^{(\dimM)}= \lambda_1 \mathring{\dfL}_{1}^{(\dimM)} + \lambda_2 \mathring{\dfL}_{2}^{(\dimM)} + \ldots \lambda_m \mathring{\dfL}_{m}^{(\dimM)} \, \qquad m\coloneqq \lfloor\tfrac{\dimM}{2}\rfloor \,,
   \end{equation}
  where $\lfloor x\rfloor$ is the floor function and $\lambda_k$ are certain dimensionful parameters.

  \item $\mathring{\dfL}_{k}^{(\dimM)}$ is a boundary term if $\dimM=2k$ (from now on, \emph{critical dimension}) \cite{YalePadmanabhan2011}. Indeed, the integrals of these invariants correspond to the Euler characteristic as can be proved via the generalized Gauss-Bonnet Theorem (see \cite{EguchiGilkeyHanson1980, Nakahara2003} for a pedagogical introduction).
\end{itemize}
In addition to these, other important properties of them, now from a field-theoretical point of view, are the following:
\begin{itemize}
  \item They propagate only the two degrees of freedom of a massless graviton.
  \item As we have already mentioned, the equations of motion for the metric (the only field) are second-order and free of Ostrogradski ghosts\footnote{See section \ref{sec: Ostrog} for more information on Ostrogradski ghosts.} \cite{Zwiebach1985, Zumino1985}. 
\end{itemize}
Due to these properties, Lovelock gravities are singled out with respect to all other higher-curvature extensions, which generically do suffer ghostly propagations. There have been some recent attempts to endow the Gauss-Bonnet term with nontrivial dynamics in $\dimM=4$ \cite{GlavanLin2020}. We will comment some of the problems that arise in the original formulation of that theory in Chapter \ref{ch:gravityviability}.

\quad

Now we generalize these invariants in order to include the curvature of a connection not necessarily equal to the Levi-Civita connection of the metric:
\boxdefinition{\begin{defn}
\textbf{(Metric-affine Lovelock invariant)}\\
Consider a $\dimM$-dimensional manifold equipped with a metric and a connection. The \emph{$k$-th order metric-affine Lovelock invariant} is the $\dimM$-form
\begin{equation}\label{eq:MAGlov}
  \dfL_{k}^{(\dimM)} \coloneqq \dfR^{a_{1}a_{2}}\wedge\ldots\wedge\dfR^{a_{2k-1}a_{2k}}\wedge\star\cofr_{a_{1}...a_{2k}}\,,
\end{equation}
where $\dfR_a{}^b$ is the curvature 2-form associated to the connection.
\end{defn}}
At this point one important observation to take into account is that we cannot construct these invariants without a metric. Note that we are making use of the Hodge star operator and, in addition, we are using $g_{ab}$ to raise/lower some indices. 

If we consider these metric-affine Lovelock invariants isolated (without matter or any other gravitational sector), some results are known about the space of allowed connections. In \cite{Exirifard2008, Borunda2008, DadhichPons2011} it was shown that a general metric-affine Lagrangian $\mathcal{L}(g_{\mu\nu},\,R_{\mu\nu\rho}{}^\lambda)$ allow the Levi-Civita connection as a solution only if the Lagrangian is Lovelock. In this sense, the metric formulation is always a consistent truncation of metric-affine Lovelock theories \cite{DadhichPons2011}. Furthermore, there are indications that Levi-Civita is in general not the only allowed connection. In particular, we already know for the special case $k=1$, i.e. Einstein-Palatini, that an extra projective mode is permitted \cite{DadhichPons2012, BernalJCA2017} (see also Section \ref{subsec: Example EP}).

Of course the property $\dfL_{k}^{(\dimM)}\equiv0$ for all $\dimM<2k$ holds exactly for the same argument we previously used. However the fact that this invariant is a boundary term in the critical dimension is not necessarily true in principle. It has been shown that if we assume the connection to be metric-compatible (i.e. zero nonmetricity), then we get a boundary term \cite{VonderHeyde1975b}. The question we are going to address in this chapter is: what happens with this fact when we switch on the nonmetricity? 

To gain some intuition, we will first analyze the simplest cases $k=1$ and $k=2$ (which are not boundary terms as it is suggested by the results in Section \ref{subsec: Example EP}, as well as \cite{JanssenJCA2019a, HehlMcCreaKopczynski1991}). But before that, let us introduce a very convenient decomposition of the connection.

\section{Useful decomposition of the connection}

The Lovelock term is a metric-affine Lagrangian of the type \eqref{eq: Total lagrangian} with no explicit dependence on the torsion and the nonmetricity. Consequently, the corresponding momenta vanish $\dfH[g]{}^{ab}=0=\dfH[\vartheta]{}_a$, so the variations with respect to the coframe \eqref{eq: varcofr MAG} and the connection \eqref{eq: varw MAG} can be expressed as:
\begin{align}
\dfrac{\delta S_{k}^{(\dimM)}}{\delta\cofr^{a}} & = \dfrac{\partial \dfL_{k}^{(\dimM)}}{\partial\cofr^{a}} \label{eq: varcofr D} \,, \\
\dfrac{\delta S_{k}^{(\dimM)}}{\delta\dfom_a{}^b} &=\Dex\dfrac{\partial\dfL_{k}^{(\dimM)}}{\partial\dfR_a{}^b}\,. \label{eq: varw D}
\end{align}

This would be the dynamics (or the contribution of this term to the dynamics in a more general Lagrangian) described in terms of the coframe and the connection 1-form. Of course, as a field theory, it is completely equivalent to work, for instance, with the coframe, the nonmetricity and the torsion as the independent variables. However we are going to use a slightly different approach.

\boxproposition{\begin{prop}\label{prop:defmcconn}
Consider a general metric-affine geometry $(g_{ab},\cofr^a,\dfom_a{}^b)$ in an arbitrary $\dimM$- dimensional manifold. Then, under the hypothesis $\dex g_{bc}=0$, the object $\mcdfom_{ab} \coloneqq \dfom_{[ab]}$, which can be expressed
\begin{equation}
  \mcdfom_{ab} = \dfom_{ab} - \frac{1}{2} \dfQ_{ab}  \,,
\end{equation}
is a metric-compatible connection with torsion
\begin{equation}\label{eq: T mcT}
   \mcdfT^a = \dfT^a-\frac{1}{2}\dfQ_b{}^a\wedge\cofr^b = \frac{1}{2}\big(T_{cb}{}^a-\frac{1}{2}Q_{[cb]}{}^a\big) \cofr^{cb}  \,,
\end{equation}
where $\dfT^a$ and $\dfQ_{ab}$ are the torsion and the nonmetricity of $\dfom_a{}^b$.
\end{prop}
}
\boxproof{\begin{proof}
  This proof can be easily done in components by using the decomposition \eqref{eq: distor decom}. For the first part we have:
  \begin{align}
    e^\mu{}_a\omega_{\mu bc} \equiv \omega_{abc} &=\mathring{\omega}_{abc}+\frac{1}{2}\left(T_{abc}+T_{cab}-T_{bca}\right)+\frac{1}{2}\left(Q_{abc}+Q_{bca}-Q_{cab}\right) \\
    &=\mathring{\omega}_{abc}+\underbrace{\frac{1}{2}T_{a[bc]}-\frac{1}{2}T_{bca}+Q_{[bc]a}}_{\text{antisym. in}\,bc}
     +\underbrace{\frac{1}{2}Q_{abc}}_{\text{sym. in}\,bc}
  \end{align}
  In order to include the full Levi-Civita part into the antisymmetric part we have to ensure that $\mathring{\omega}_{a(bc)}=0$. From \eqref{eq: LCw decom}, one can easily check that $\mathring{\omega}_{a(bc)}=\frac{1}{2}\partial_a g_{bc}$ so it is enough to impose the metric $g_{ab}$ to be constant. Therefore, if $\dex g_{bc}=0$, indeed, $\omega_{a[bc]}=\omega_{abc}-\frac{1}{2}Q_{abc}$ or, in differential form notation, $\dfom_{[bc]}=\dfom_{bc}-\frac{1}{2}\dfQ_{bc}$. 
  
  The resulting object is $\omega_{abc}$ up to a tensorial part, $\frac{1}{2}Q_{abc}$, therefore $\mcdfom_{ab}$ is a connection. It is metric-compatible by construction because it is antisymmetric and the metric in the chosen frame is constant (see Proposition \ref{prop:symconnzeroQ}). Finally, the torsion can be obtained by direct computation:
  \begin{align}
    \mcdfT^a =\mcDex\cofr^a & =\dex\cofr^a+\mcdfom_b{}^a\wedge\cofr^b \\
                            & =\dex\cofr^a+\dfom_b{}^a\wedge\cofr^b-\frac{1}{2}\dfQ_b{}^a\wedge\cofr^b\\
                            & =\dfT^a-\frac{1}{2}\dfQ_b{}^a\wedge\cofr^b \qquad = \frac{1}{2}\big(T_{cb}{}^a-\frac{1}{2}Q_{[cb]}{}^a\big) \cofr^{cb}\,.
  \end{align}
\end{proof}
}

Notice that this relation is trivially invertible:
\begin{equation}
  \mcdfT^a = \dfT^a-\frac{1}{2}\dfQ_b{}^a\wedge\cofr^b \qquad \Leftrightarrow \qquad \dfT^a = \mcdfT^a+\frac{1}{2}\dfQ_b{}^a\wedge\cofr^b \,.
\end{equation}
Therefore we can redistribute the degrees of freedom contained in the connection as
\begin{equation}
  \dfom_a{}^b \qquad \leftrightarrow \qquad \dfT^a, \dfQ_{ab} \qquad \leftrightarrow \qquad \mcdfT^a, \dfQ_{ab} \qquad \leftrightarrow \qquad \boxed{\mcdfom_a{}^b, \dfQ_{ab}} .
\end{equation}
Indeed we will further split the nonmetricity as the Weyl vector  (i.e., the principal trace $\irrdfQ{4}{}_{ab}=\frac{1}{\dimM}\dfQ_c{}^c g_{ab} = \dfQ g_{ab}$) plus the remaining traceless part, $\dfQ_{ab} = \dfQ g_{ab} + \dfQNoTr_{ab}$.

To sum up, we are going to consider the following splitting of the connection,
\begin{equation} \label{eq: con mccon}
  \boxed{\dfom_{ab} = \mcdfom_{ab} + \frac{1}{2} \dfQ g_{ab} + \frac{1}{2} \dfQNoTr_{ab}}\,,
\end{equation}
and, hence, the set of fundamental variables we will use for our Lovelock theories is
\begin{equation}
  \{g_{ab},\,\cofr^a,\, \mcdfom_a{}^b,\,\dfQ, \, \dfQNoTr_{ab}\}\qquad \text{under}~~\dex g_{ab}=0.
\end{equation}

Let us also show the relation between the curvatures of $\dfom_a{}^b$ and $\mcdfom_a{}^b$: 
\boxproposition{\begin{prop}
Consider a general geometry $(g_{ab},\cofr^a,\dfom_a{}^b)$ for which $\dex g_{ab}=0$. The curvature of $\dfom_a{}^b$  can be expressed in terms of the objects previously defined as
\begin{equation} \label{eq: R mcR}
  \dfR^{ab} =\mcdfR^{ab}  +\frac{1}{4}\dfQNoTr_c{}^b\wedge\dfQNoTr^{ac} +\frac{1}{2}\mcDex\dfQNoTr^{ab}+\frac{1}{2}g^{ab} \dex\dfQ \,,
\end{equation}
where $\mcdfR_a{}^b$ is the curvature 2-form of $\mcdfom_a{}^b$.
\end{prop}
}
\boxproof{\begin{proof}
  We use \eqref{eq: R RLC distor} with $\dfXi_a{}^b=\frac{1}{2}\dfQ_a{}^b$
 \begin{align}
    \dfR_a{}^b & =\mcdfR_a{}^b+\mcDex\left[\frac{1}{2}\dfQ_a{}^b\right]+\left[\frac{1}{2}\dfQ_c{}^b \right]\wedge\left[\frac{1}{2}\dfQ_a{}^c\right]\\
    & =\mcdfR_a{}^b+\frac{1}{2}\mcDex\dfQNoTr_a{}^b+\frac{1}{2}\delta_a^b\dex\dfQ+\frac{1}{4}\dfQNoTr_c{}^b\wedge\dfQNoTr_a{}^c
  \end{align}
Finally, since $\mcDex$ is metric-compatible we can raise the indices here with no worries, and that is the end of the proof.

\end{proof}
}
In the Lovelock terms, only the antisymmetric part of the curvature is relevant. This implies that the last two terms in \eqref{eq: R mcR} completely drop from the Lovelock invariants. This has two main consequences: the first one is that no derivatives of $\dfQNoTr_{ab}$ appear in the Lagrangian (it only enters polynomially), and the second one is that the Weyl 1-form  $\dfQ$ does not play any role in these theories. This last fact is a manifestation of the projective symmetry \cite{Eisenhart1927},
\begin{equation}
  \dfom_a{}^b \to \dfom_a{}^b + \dfA \delta_a^b ,
\end{equation}
that all of these terms exhibit in arbitrary dimensions. Indeed, this transformation in the connection is nothing but a (local) shift of the Weyl 1-form.

To finish this section, it is important to highlight what we have done. We have taken the connection and split its degrees of freedom into two objects (three, but $\dfQ$ disappears) the metric-compatible connection $\mcdfom_{ab}$ and the traceless part of the nonmetricity 1-form $\dfQNoTr_{ab}$. We have managed to re-express the Lovelock terms as Lagrangians of the type
\begin{equation}
  \mcdfL_{k}^{(\dimM)}(g_{ab}, \cofr^a, \mcdfR_a{}^b, \dfQNoTr_{ab}) \coloneqq \dfL_{k}^{(\dimM)}\big(g_{ab}, \cofr^a, \dfR_a{}^b(g_{ab}, \mcdfR_a{}^b, \dfQNoTr_{ab})\big),
\end{equation}
Now instead of the equation of motion of the connection, we have that same information encoded into two variations,
\begin{align}
  \dfrac{\delta \mcS_{k}^{(\dimM)}}{\delta\mcdfom_a{}^b} &=\mcDex\dfrac{\partial\mcdfL_{k}^{(\dimM)}}{\partial\mcdfR_a{}^b}\,,\label{eq: varmcw}\\
  \dfrac{\delta \mcS_{k}^{(\dimM)}}{\delta\dfQNoTr_{ab}} &=\dfrac{\partial\mcdfL_{k}^{(\dimM)}}{\partial\dfQNoTr_{ab}}\,,\label{eq: varQNoTr}
\end{align}
whereas the equation of motion of the coframe remains the same since the change of variables does not involve the coframe.

\section{The Einstein-Palatini action in \texorpdfstring{$\dimM=2$}{dim=2}}\label{sec: D2EP}

\subsection{Solving the theory}

Consider the Einstein-Palatini action in $\dimM=2$ and in the absence of matter. In Section \ref{subsec: Example EP} we compute the equation of the connection in components \eqref{eq: eq conn EP com},
\begin{equation}
Q_{\rho\mu\nu}-T_{\rho\mu\nu}-\left(T_\mu-\frac{1}{2}Q_\mu +\Qb_\mu \right)g_{\rho \nu}+\left(T_\rho -\frac{1}{2}Q_\rho \right)g_{\mu\nu}=0\,.
\end{equation}
For $\dimM>2$ the general solution is Levi-Civita plus an arbitrary projective mode \eqref{eq: EP solw}. Nevertheless, in $\dimM=2$ the situation is degenerate and should be studied separately. First, notice that in dimension 2, the torsion is pure trace, i.e., 
\begin{equation}
T_{\mu\nu}{}^\rho = 2  T_{[\mu} \delta_{\nu]}^\rho \,,
\end{equation}
In other words, $\irrdfT{1}^a=0=\irrdfT{3}^a$. If we plug this into the equation of the connection the trace of the torsion completely drops from the equation (i.e., the torsional degrees of freedom are not constrained by the dynamics):
\begin{equation} \label{eq: eqcon EP D2}
Q_{\rho\mu\nu}+\left(\frac{1}{2}Q_\mu -\Qb_\mu \right)g_{\rho \nu}-\frac{1}{2}Q_\rho g_{\mu\nu}=0
\end{equation}
If we take the trace in $ac$ we arrive at
\begin{equation}
Q_\mu=2\Qb_\mu
\end{equation}
And if we substitute this into our equation \eqref{eq: eqcon EP D2} we get 
\begin{equation} \label{eq: solQ EP}
Q_{\rho\mu\nu}=\frac{1}{2}Q_\rho g_{\mu\nu}\,.
\end{equation}
Hence, the nonmetricity is pure Weyl-trace, i.e., only the irreducible component $\irrdfQ{4}_{ab}$ is non-zero.

It is specially useful to introduce a couple of auxiliary variables and express
\begin{equation}
  T_\mu \eqqcolon A_\mu - B_\mu,\qquad Q_\mu \eqqcolon 4 A_\mu \,,
\end{equation} 
so that the general solution of the equation of the connection is
\begin{equation}
   T_{\mu\nu}{}^\rho= 2 ( A_{[\mu} - B_{[\mu}) \delta_{\nu]}^\rho ,\qquad Q_{\rho\mu\nu}=2 A_\rho g_{\mu\nu},,
\end{equation}
or, equivalently,
\begin{equation} \label{eq: GBPsolconn}
  \boxed{\Gamma_{\mu\nu}{}^\rho= \mathring{\Gamma}_{\mu\nu}{}^\rho + A_\mu \delta_\nu^\rho + B_\nu \delta_\mu^\rho - B^\rho g_{\mu\nu}} \,.
\end{equation}
We thus find that the two-dimensional metric-affine Einstein term leaves the trace of the nonmetricity and the (pure trace) torsion completely undetermined. At this point, it is important to highlight that the pure-trace conditions for the torsion and nonmetricity have a completely different origin. The first one is an intrinsic property of the irreducible decomposition of the torsion in $\dimM=2$. However, the fact that the nonmetricity is equal to its Weyl trace is derived from the dynamical equations of this theory. 

\begin{table}[t]
  \begin{center} \renewcommand{\arraystretch}{1.5}
  {\small
  \begin{tabular}{|c|c|c|c|}
    \hline 
    Tensor & Components in $\dimM$ dim. & Components in 2 dim. & Condition imposed by EoM\tabularnewline
    \hline \hline 
    $T_{\mu\nu}{}^{\rho}$ & $\frac{1}{2}\dimM^{2}(\dimM-1)$ & 2 (pure trace) & None\tabularnewline
    \hline 
    $Q_\mu$ & $\dimM$ & 2 & None \tabularnewline
    \hline 
    $\QNoTr_{\mu\nu\rho}$ & $\frac{1}{2}\dimM(\dimM+2)(\dimM-1)$ & 4 & They are zero\tabularnewline
    \hline 
  \end{tabular}\renewcommand{\arraystretch}{1}
  }
      \caption{\label{tab:tabledof}Splitting of the independent components of the connection in general dimension and in $\dimM=2$. The last column shows the conditions imposed by the equations of motion of the two-dimensional metric-affine Einstein theory. Observe that the indetermination of the trace of the non-metricity holds in arbitrary $\dimM$ due to projective symmetry.}
  \end{center}
  
\end{table}

Once the solution of the connection equation is known, let us look at the other dynamical equation. The curvature tensor constructed from \eqref{eq: GBPsolconn} is
\begin{equation}
  R_{\mu\nu}{}^{\rho\lambda} = \mathring{R}_{\mu\nu}{}^{\rho\lambda} + F_{\mu\nu}(A)g^{\rho\lambda} + 4 \delta_{[\mu}^{[\rho}\mathring{\nabla}_{\nu]}B^{\lambda]} + 4 B^{[\lambda}B_{[\mu}\delta_{\nu]}^{\rho]} + 2 B_\sigma B^\sigma \delta_{[\mu}^\rho \delta_{\nu]}^\lambda .
\end{equation}
where $F_{\mu\nu}(A)\coloneqq 2\partial_{[\mu}A_{\nu]}$. Then, 
\begin{align}
  R^{(1)}{}_{\mu\nu} &= \mathring{R}_{\mu\nu} + F_{\mu\nu}(A) + g_{\mu\nu} \mathring{\nabla}_\lambda B^\lambda\,,\\
  R^{(2)}{}_{\mu\nu} &=-\mathring{R}_{\mu\nu} + F_{\mu\nu}(A) - g_{\mu\nu} \mathring{\nabla}_\lambda B^\lambda\,, \\
                   R &= \mathring{R} + 2\mathring{\nabla}_\lambda B^\lambda\,,\\
\end{align}
and the coframe equation becomes:
\begin{equation}
  \mathring{R}_{\mu\nu} - \frac{1}{2} g_{\mu\nu}\mathring{R} =0 \,,
\end{equation}
i.e., the Einstein equations, which are trivial in 2 dimensions. The reason is that the Einstein tensor vanishes identically in $\dimM=2$. An easy way to see this is to use the result that any 2-dimensional metric is conformally flat, $g_{\mu\nu}=\eN^{2\phi(x)}\eta_{\mu\nu}$. Consequently the expressions of the Ricci tensor and the curvature scalar are,
\begin{equation}
  \mathring{R}_{\mu\nu} =\eta_{\mu\nu} \mathring{\nabla}_\lambda\mathring{\nabla}^\lambda\phi  \, \qquad  \mathring{R} = 2 \eN^{-2\phi(x)} \mathring{\nabla}_\lambda\mathring{\nabla}^\lambda \phi \,
\end{equation}
so
\begin{equation}
  \mathring{G}_{\mu\nu} = \mathring{R}_{\mu\nu} - \frac{1}{2} g_{\mu\nu}\mathring{R} \equiv 0 \,.
\end{equation}

As a result of this analysis, the on-shell geometry depends on three objects,  $\phi$, $B_\mu$ and $A_\mu$, which are related to the only independent components of the metric, the torsion and the trace of the nonmetricity, respectively. All of them remain undetermined. However, observe that, beside the trace (see Table \ref{tab:tabledof}), the nonmetricity in dimension 2 also has a traceless part with 4 independent components, and this part is forced to be zero by the equations of motion. This last result is crucial because it means that a field configuration whose nonmetricity has a non-trivial traceless part, is not a solution of the variational problem. Therefore, this Lagrangian \emph{cannot be a boundary term} as long as the nonmetricity is present. This fact will be explored in the next section by performing the splitting of the connection introduced in the previous section.

\subsection{The \texorpdfstring{$\dimM=2$}{dim=2} Einstein-Palatini term is not a total derivative}

In 2 dimensions, the splitting \eqref{eq: con mccon} (or, for the curvatures \eqref{eq: R mcR}) permits to rewrite the Einstein-Palatini action in differential form notation as\footnote{
    Recall that $S$ and $\mcS$ represent the same action but with different functional dependence:
    \[ \mcS_k^{(2k)}[g,\cofr,\mcdfom,\dfQNoTr] =  S_k^{(2k)}[g,\cofr,\dfom(\mcdfom,\dfQNoTr)]\, . \]
    Analogous notation will be used with the Lagrangian.}
\begin{equation}
S_1^{(2)} =  \frac{1}{2\kappa} \int \LCten^a{}_b \, \dfR_a{}^b(\dfom)
\qquad \Rightarrow \qquad \mcS_1^{(2)} = \frac{1}{2\kappa} \int  \LCten_{ab} \, \Bigl[\mcdfR{}^{ab}(\mcdfom) \ - \  \frac{1}{4}\, \dfQNoTr^{ac}\wedge \dfQNoTr_c{}^b\Bigr],
\label{eq: D2EP}
\end{equation}
where we have used \eqref{eq: starcof D}. Here we see maybe more clearly that the presence of the Levi-Civita tensor $\LCten_{ab}$ is what antisymmetrizes the curvature and, hence, eliminates the Weyl 1-form, in agreement with the projective symmetry \cite{JanssenJCA2019a}.

Considering an orthonormal gauge $g_{ab}=\eta_{ab}$, we find that
  \begin{equation}
    \LCten_{ab}  \mcdfR{}^{ab}  = \LCten_{ab} \, \dex\mcdfom^{ab}  =   \dex (\LCten_{ab}\,\mcdfom^{ab})\,.
  \end{equation}
In the first step we have used that $\LCten_{ab} \ \mcdfom^{ac} \wedge \mcdfom_c{}^b =  0$, due to the antisymmetry of both $\LCten_{ab}$ and $\mcdfom^{ab}$ and the fact that the theory lives in $\dimM=2$ (the indices $a$, $b$ and $c$ have to be all different, but at the same time can only take values in the set $\{1,2\}$). In the second step we used the fact that $\LCten_{ab}=\epsilon_{ab}$ (because the determinant of the anholonomic metric is just a sign) and since $\epsilon_{ab}$ is a constant object,  $\dex\LCten_{ab}=\dex\epsilon_{ab}=0$. The two-dimensional Einstein-Palatini action therefore reduces to
  \begin{equation}\label{dim=2EH-forms-final}
    \mcS_1^{(2)} = \frac{1}{2\kappa} \int \Bigl[  \dex(\LCten_{ab}  \mcdfom^{ab}) - \frac{1}{4} \LCten_{ab}  \dfQNoTr^{ac}\wedge \dfQNoTr_c{}^b\Bigr]\,,
  \end{equation}
which is not a boundary term, unless the connection verifies $\dfQNoTr_{ab}=0$. It is only the (traceless part of) the nonmetricity what spoils the topological character of the theory.

Finally, let us quickly re-derive the results of the previous subsection, but in the language of differential forms. As can be directly seen from $\eqref{eq: D2EP}$, the only dynamical variable is $\dfQNoTr{}_{ab}$. The corresponding equation of motion is:
  \begin{equation}
    0 = \delta_{~\dfQNoTr~} \mcS^{(2)}_1  = \frac{1}{2\kappa} \int \delta \dfQNoTr{}^{ac}\wedge \Big(- \frac{1}{2} \LCten_{ab} \dfQNoTr_c{}^b\Big) \quad  \Rightarrow \quad  \LCten_{b(a} \dfQNoTr_{c)}{}^{b}  =  0\,,
  \end{equation}
whose only solution is the one we found in \eqref{eq: solQ EP},
  \begin{equation} 
    \dfQNoTr_{ab}=0\,. 
  \end{equation}

\section{The Gauss-Bonnet-Palatini action in \texorpdfstring{$\dimM=4$}{dim=4}}

In this section we are going to focus on the next Lovelock term, the Gauss-Bonnet-Palatini Lagrangian, in its critical dimension ($\dimM=4$). Unfortunately, its dynamical equations are already too complicated to be solved in full generality, as we have done for the Einstein-Palatini theory in $\dimM = 2$. However, as we will see, again it is the traceless part of the nonmetricity $\dfQNoTr_{ab}$ what prevents the theory from being a boundary term.

The fact that Gauss-Bonnet-Palatini is not topological was already pointed out in \cite{HehlMcCreaKopczynski1991}. Our alternative proof is in complete agreement with those results. After the present analysis, in Section \ref{sec:kLovelock}, we will see how our proof can be straightforwardly generalized to higher order Lovelock terms. 

\subsection{The \texorpdfstring{$\dimM=4$}{dim=4} Gauss-Bonnet-Palatini term is not a total derivative}

We start by rewriting the action in terms of the new variables via the splitting \eqref{eq: con mccon} (or, for the curvatures \eqref{eq: R mcR}):
\begin{equation}
\mcdfL_{2}^{(4)} \ = \ \LCten_{abcd}{}\, \Big[ \mcdfR{}^{ab}\wedge\mcdfR{}^{cd}
   -  \frac{1}{2} \mcdfR{}^{ab}\wedge\dfQNoTr{}^{cf}\wedge\dfQNoTr_f{}^{d}
   +  \frac{1}{16} \dfQNoTr{}^{ae} \wedge\dfQNoTr_e{}^b \wedge \dfQNoTr{}^{cf} \wedge \dfQNoTr_f{}^d\Big]\,,
\label{eq: D4GB}
\end{equation}
where we have used \eqref{eq: starcof D}. The first term is the four-dimensional Euler characteristic and can easily be written as a total derivative (see for example \cite{YalePadmanabhan2011, Hehl1995}). If we choose again the orthonormal gauge $g_{ab}=\eta_{ab}$, we find that the Lagrangian is of the form
\begin{equation}
\mcdfL_{2}^{(4)} \ = \ \dex \dform{{\cal C}} \ - \ \LCten_{abcd}{}\, 
  \Big[\frac{1}{2}\mcdfR{}^{ab} \wedge \dfQNoTr{}^{cf} \wedge \dfQNoTr_f{}^{d}
  \ - \ \frac{1}{16} \dfQNoTr{}^{ae} \wedge\dfQNoTr_e{}^b \wedge \dfQNoTr{}^{cf} \wedge \dfQNoTr_f{}^d\Big], 
\end{equation}
where
\begin{equation}
\dform{{\cal C}} \coloneqq \LCten^a{}_b{}^c{}_d \,\Big[\, \mcdfR_a{}^b \wedge \mcdfom_c{}^d \ + \ \frac{1}{3}  \mcdfom_a{}^b \wedge \mcdfom_c{}^f \wedge \mcdfom_f{}^d \Big]\,.
\end{equation}
One might think that there could be a way to re-express the last term as a total derivative. It is true that one can extract other exact parts by making use of the derivative contained in the curvature that appears in the first term inside the square bracket. However, it is not enough to cancel all of the nonmetricity terms. 

In order to see that indeed this Lagrangian is not a boundary term we are going to use another strategy. It is well-known that a boundary term has trivial equations of motion ($0=0$), namely, there are no dynamical conditions on the fields. In other words, any possible field  configuration is allowed as a solution. The strategy will actually be to find a configuration that violates at least one of the dynamical equations, since the existence of such configurations implies that some of the fields are constrained. Consequently, the dynamical equations cannot be reduced to $0=0$ (at least not all of them). 

The equations of motion of the auxiliary connection $\mcdfom_a{}^b$ and the traceless part of the nonmetricity $\dfQNoTr_{ab}$ can be obtained from the general equations \eqref{eq: varmcw} and \eqref{eq: varQNoTr}, and the results are, respectively,
\begin{align}
0 & = \mcDex \left[\dfQNoTr_{c}{}^{a}\wedge\dfQNoTr{}^{bc}\right] \label{eq: eommcwGB}\,, \\
0 & = \LCten_{cde}{}^{(a} \dfQNoTr^{b)e} \wedge  \left[\mcdfR{}^{cd} -\tfrac{1}{4}\dfQNoTr_{f}{}^{c} \wedge\dfQNoTr^{df}\right]  \,.
\end{align}
The first one has been contracted by another Levi-Civita tensor to eliminate the one coming from the Lagrangian. This produces an  antisymmetrisation in $\{ab\}$, which can be dropped, because the combination $\dfQNoTr_{c}{}^{a}\wedge\dfQNoTr^{bc}$ is already antisymmetric.

Recall that the goal is to find a field configuration violating at least one of the dynamical equations. For example, consider the following one
  \begin{equation}
    g_{ab}  = \eta_{ab}\,, \qquad 
    \cofr^a = \dex x^a \,, \qquad 
    \mcdfom^{ab}  = \mathring{\dfom}^{ab} + f \dfal^{[a}\delta_{t}^{b]} \,, \qquad 
    \dfQNoTr{}^{ab}= 2 \dfal^{(a}\delta_{t}^{b)}\,,\label{eq: ansatzGB}
  \end{equation}
where $f$ is an arbitrary function and the vector-valued 1-form $\dfal^a$ is defined by the following expression in terms of the  Cartesian coframe $\{\dex t, \dex x, \dex y, \dex z\}(\equiv \cofr^a)$,
  \begin{equation}
    \dfal^a \coloneqq \eN^t \, \left(\delta_y^a \dex y  + \delta_z^a \dex z \right)\,.
  \end{equation}
Note that this Ansatz is consistent with the fact that $\dfQNoTr{}^{ab}$ is traceless, since $\dfal_c\delta_{t}^c=0$. Furthermore, observe also that we can everywhere drop the Levi-Civita connection, since the considered metric is the Minkowski one and the anholonomic (Latin) indices are referred to the Cartesian basis of the space. 

In addition, $\dfal^a$ verifies
\begin{equation}
  \dfQNoTr_{c}{}^{a}\wedge\dfQNoTr{}^{bc} = \dfal^{a} \wedge \dfal^{b} \,.
\end{equation}
With this in mind, it is not difficult to check that the Ansatz \eqref{eq: ansatzGB} violates the equation of motion of $\mcdfom_a{}^b$ \eqref{eq: eommcwGB}:
\begin{equation}
   \mcDex \left[\dfQNoTr_c{}^a\wedge\dfQNoTr^{bc}\right] 
     \ = \ \dex\left[\dfal^a \wedge \dfal^b\right]
     \ = \ 2\eN^{2t} \left(\delta_y^a \delta_z^b  - \delta_y^b \delta_z^a \right)\dex t \wedge\dex y \wedge\dex z  \ \neq \ 0\,.
\end{equation}
It is worth remarking that this last inequality holds in the entire manifold, because the chosen set of coordinates is globally defined. This result proves that the metric-affine generalization of the Gauss-Bonnet term in $\dimM=4$ cannot be written as a total derivative, since only some field configurations are allowed by the equations of motion.

\section{The \texorpdfstring{$k$}{k}-th order metric-affine Lovelock term in \texorpdfstring{$\dimM=2k$}{dim=2k}} \label{sec:kLovelock}

With the previous two cases in mind, it is not difficult to generalize the procedure to the general Lovelock term in its critical dimension.

\subsection{Proving that the theory is not a boundary term}

The starting point is the Lagrangian \eqref{eq:MAGlov} in critical dimension $\dimM=2k$. By using \eqref{eq: starcof D}, we get
\begin{equation}\label{n=2kaction}
  \dfL_{k}^{(2k)} \ = \ \LCten^{a_1}{}_{a_2} ...{}^{a_{2k-1}}{}_{a_{2k}}\dfR_{a_1}{}^{a_2} \wedge\ldots \wedge \dfR_{a_{2k-1}}{}^{a_{2k}} \,.
\end{equation}
Now we substitute the splitting \eqref{eq: con mccon} (or, for the curvatures \eqref{eq: R mcR}) and the Lagrangian can be written as a power series in $\mcdfR$ and $\dfQNoTr \wedge \dfQNoTr$ terms, 
\begin{align}
\mcdfL_{k}^{(2k)} & = \LCten_{a_1 \dots a_{2k}}\ \sum_{m=0}^k  \frac{1}{4^{k-m}} \ \frac{k!}{m! (k-m)!} \ \mcdfR{}^{a_1a_2} \wedge \ldots \wedge  \mcdfR{}^{a_{2m-1}a_{2m}}\wedge  \\
                & \qquad\qquad \wedge \,\dfQNoTr{}^{a_{2m+1}f_1} \wedge \dfQNoTr_{f_1}{}^{a_{2m+2}} \wedge \ldots \wedge\dfQNoTr{}^{a_{2k-1}f_{k-m}} \wedge \dfQNoTr_{f_{k-m}}{}^{a_{2k}} \,, 
\end{align}
The $m=k$ term $\mcdfR_{a_1}{}^{a_2} \wedge \ldots \wedge  \mcdfR_{a_{2k-1}}{}^{a_{2k}}$ is in fact a boundary term \cite{TroncosoZanelli1999, Zanelli2012} (see also \cite{Concha2013, Deruelle2017}). This can be easily seen by using the Bianchi identity $\mcDex\mcdfR_a{}^b=0$ to show that it is a closed form and, hence, locally exact by the Poincar{\'e} lemma. For this reason, we will ignore it in the subsequent computations, as it does not contribute to the functional variations.

The equation of motion of $\mcdfom_a{}^b$ \eqref{eq: varmcw} is in this case
\begin{align} \label{eq: varmcw D2k}
    g_{ca}\frac{\delta \mcS_k^{(2k)}}{\delta \mcdfom_{c}{}^{b}} 
      & = \LCten_{ab a_3 \dots a_{2k}}\ \sum_{m=1}^{k-1} \frac{1}{4^{k-m}} \ \frac{k!}{m! (k-m)!}  \  \mcdfR{}^{a_3a_4} \wedge\ldots \wedge  \mcdfR{}^{a_{2m-1}a_{2m}}\wedge \nonumber \\ 
      & \qquad \qquad \wedge \mcDex \Big[\dfQNoTr^{a_{2m+1}f_1} \wedge \dfQNoTr_{f_1}{}^{a_{2m+2}} \wedge\ldots \wedge \dfQNoTr^{a_{2k-1}f_{k-m}} \wedge \dfQNoTr_{f_{k-m}}{}^{a_{2k}} \Big] \,,
\end{align}
where we have taken into account the Bianchi identity $\mcDex \mcdfR_a{}^b=0$. 

As we did in the Gauss-Bonnet case, we are going to design a counterexample that violates this equation. Consider the Ansatz
\begin{equation}
 g_{ab}  = \eta_{ab}\,, \qquad \cofr^a =\dex x^a\,,\qquad \mcdfom^{ab} = \mathring{\dfom}^{ab} \,, \qquad \dfQNoTr^{ab} = 2 \dfal^{(a}\delta_t^{b)}\,,
\end{equation}
where $x^a$ takes values in the set $\{x^1=t,~x^2,...,~x^{2k}\}$ and we have defined
\begin{equation}
  \dfal^a \coloneqq \eN^t \left(\delta_3^a \dex x^3 + \ldots +  \delta_{2k}^a \dex x^{2k} \right) \,,
\end{equation}
which is consistent with $\dfQNoTr_c{}^c=0$. Similarly as in the Gauss-Bonnet case, this configuration satisfies
\begin{equation} \label{eq: QNoTr aa}
  \dfQNoTr_{c}{}^{a}\wedge\dfQNoTr^{bc} = \dfal^a \wedge \dfal^b \,.
\end{equation}

Note that the connection $\mcdfom^{ab}$ is flat ($\mcdfR^{ab}=\mathring{\dfR}^{ab}=0$) and identically vanishing because the chosen coframe is Cartesian. There are two important consequences of this: first we have that $\mcDex=\dex$, and secondly that only the  $m=1$ term in \eqref{eq: varmcw D2k} survives, as it is the only one that does not contain $\mcdfR_{ab}$. Therefore, \eqref{eq: varmcw D2k} becomes
\begin{align}\label{eq: varmcw D2k 2}
    \eta_{ca}\frac{\delta \mcS_k^{(2k)}}{\delta \mcdfom_{c}{}^{b}} 
       & = \LCten_{ab a_3 \dots a_{2k}} \  \frac{k}{4^{k-1}} \ \dex\Big[\dfQNoTr^{a_{3}f_1} \wedge \dfQNoTr_{f_1}{}^{a_{4}} \wedge \ldots \wedge\dfQNoTr^{a_{2k-1}f_{k-1}} \wedge \dfQNoTr_{f_{k-1}}{}^{a_{2k}} \Big] \,.
\end{align}
If we now use the property \eqref{eq: QNoTr aa}, we get
\begin{align}
 & \frac{-4^{k-1}}{k 2!(2k-2)!} \ \LCten_c{}^{b a_3 \dots a_{2k}} \frac{\delta \hat{S}_k^{(2k)}}{\delta \mcdfom_{c}{}^{b}} \nonumber \\
 & \qquad\qquad = \dex \big(\dfal^{a_3} \wedge \ldots \wedge \dfal^{a_{2k}}\big) \nonumber \\ 
 & \qquad\qquad = 2(k-1)  \eN^{2(k-1)t}  (2k-2)!  \delta_3^{[a_3}\ldots\delta_{2k}^{a_{2k}]} \ \dex t \wedge\dex x^3 \wedge\dex x^4\wedge\ldots\wedge \dex x^{2k}\,.
\end{align}
Again, it is easy to see that this expression is non-zero in the entire manifold, except for $k=1$. Nevertheless, the case $k=1$ was completely solved in Section \ref{sec: D2EP}, so in practice we are only interested in $k>1$. In summary, by finding a field configuration for which the variation with respect to  $\mcdfom_a{}^b$ is not zero, we have extended the argument from the metric-affine Gauss-Bonnet term to the general $k$-th order critical Lovelock term. We have proved that none of the latter is a boundary term in the metric-affine formulation due to the presence of nonmetricity.

\subsection{Exploring non-trivial solutions of the critical case of arbitrary \texorpdfstring{$k$}{k}}\label{subsec:nontrivsolLovk}

According to the previous result, it makes sense to search for non-trivial solutions for the critical metric-affine Lovelock theory of arbitrary order. We are going to come back to the initial description in terms of the basic fields $\{g_{ab},\cofr^a, \dfom_a{}^b\}$. 

Let us stop for a moment and comment a little bit on the coframe equation. Consider that the dimension is arbitrary in principle. Since the torsion does not appear explicitly in the Lovelock term, the variation with respect to the coframe coincides with the partial derivative \eqref{eq: varcofr D}:
  \begin{equation}
    \frac{\delta S_k^{(\dimM)}}{\delta \cofr^a} = \ \frac{\partial \dfL_k^{(\dimM)}}{\partial \cofr^a}\,.
  \end{equation}
If we expand the Hodge star in the definition of the metric-affine Lovelock term, we get
  \begin{align}
    \dfL_{k}^{(\dimM)} 
    & = \ \dfR^{a_1 a_2}\wedge...\wedge\dfR^{a_{2k-1} a_{2k}}\wedge \star \cofr_{a_1... a_{2k}} \\
    & = \ \frac{1}{(\dimM-2k)!} \ \LCten_{a_1...a_{2k} b_1...b_{\dimM-2k}} \, \dfR^{a_1 a_2}\wedge\ldots\wedge\dfR^{a_{2k-1} a_{2k}}\wedge \cofr^{b_1... b_{\dimM-2k}} \,.
  \end{align}
We see that the coframe appears just $\dimM-2k$ times in this expression. Therefore, in the critical dimension $\dimM=2k$, the Lovelock term only depends on the curvature and the Levi-Civita tensor (which is a purely $g_{ab}$-dependent object). In conclusion, the equation of motion of the coframe is trivially satisfied in the critical dimension,
  \begin{equation}
    \frac{\delta S_k^{(2k)}}{\delta \cofr^a} \equiv 0\,.
  \end{equation}

Consequently, we only have to solve the equation of motion of the connection \eqref{eq: varw D}, which for the $k$-th Lovelock term reads
\begin{align}
  0 & = \Dex\LCten^{a_1}{}_{a_2}...{}^{a}{}_{b}\wedge\dfR_{a_1}{}^{a_2}\wedge...\wedge\dfR_{a_{2k-3}}{}^{a_{2k-2}}\\
    & = \Big[ \delta^d_{a_{1}} \LCten_{ca_2 ...a_{2k-2}ab} \ + \ ... \ + \ \delta^d_{a_{2k-3}} \LCten_{a_{1}...a_{2k-4}ca_{2k-2}ab} \nonumber\\
    & \qquad\qquad + \delta^d_a \, \LCten_{a_1 ...a_{2k-2}cb}\Big] \dfQNoTr{}^c{}_d \wedge \dfR^{a_1 a_2}\wedge...\wedge\dfR^{a_{2k-3}a_{2k-2}}\,. \label{eq: eomw D2k}
\end{align}
By observing \eqref{eq: eomw D2k} one can easily obtain various non-trivial connections that are solutions of the theory:
\begin{itemize}
  \item \textbf{Solutions for arbitrary $k$:} In general, the equation is fulfilled by any connection with $\dfQNoTr_{ab}=0$ (i.e., the nonmetricity 1-form coincides with its trace $\irrdfQ{4}_{ab}$). Note that an interesting sub-case is the connection \eqref{eq: GBPsolconn}, that was presented in \cite{JanssenJCA2019a} as a particular non-trivial solution for the $k=2$ case, but conjectured to hold for arbitrary $k$.

  \item \textbf{Solutions for $k>1$:} For the second or higher order Lovelock critical Lagrangian, in  \eqref{eq: eomw D2k} there is at least one curvature multiplying the whole expression, so any teleparallel connection ($\dfR_c{}^d=0$) is a solution. Indeed, we can infer a slightly more general result: any connection fulfilling
  \begin{equation}
    \dfQNoTr_{ab}\wedge \dfR_c{}^d = 0
  \end{equation}
  solves the equation of motion \eqref{eq: eomw D2k}. For instance, the configurations we explore in Chapter \ref{ch:GWsolutions} with $\dfU=0$, belong to this class (see also the original publication \cite{JCAObukhov2021a}). One can also find solutions of this kind by restricting appropriately the geometries in Chapter \ref{ch:GWgen}.

  \item \textbf{Solutions for $k>2$:} In these cases, there are at least two curvatures in the equation of motion \eqref{eq: eomw D2k}. Hence, any connection with curvature of the form
  \begin{equation}
    \dfR_{ab} = \dfal_{ab}\wedge \dfbe\,, \label{eq:Reqalwedgek}
  \end{equation}
  for certain 1-forms $\dfal_{ab}$ and $\dfbe$, is a solution (since $\dfbe\wedge\dfbe=0$). One example in this category is the Ansatz for the connection used in \cite{Obukhov2017} in the context of gravitational waves in Poincar\'e gravity, where $\dfbe$ is the dual form of the wave vector. Other example is given by any metric-affine geometry satisfying the generalized Lichnerowicz criterion (Definition \ref{def: GenLC}) since, as a consequence of Proposition \ref{prop: Lich crit and quad general}, the condition \eqref{eq:Reqalwedgek} is guaranteed under such a criterion, with $\dfbe$ the wave form $\dfk$.
\end{itemize}

\section{Metric-affine extension of the Pontryagin and Nieh-Yan invariants}\label{sec:PontrNieh}

If we now concentrate in $\dimM=4$, one may ask what happens with the other two boundary terms used in Poincar\'e gauge gravity (zero nonmetricity): the Pontryagin invariant and the Nieh-Yan invariant (see some complementary derivations in \cite{HehlMcCreaKopczynski1991}). The idea is to see if, contrary to the metric-affine Gauss-Bonnet term, they are boundary terms and can be used to eliminate terms from the general MAG quadratic action (\eqref{eq: qMAGLeven} and \eqref{eq: qMAGLodd}).

The first one, can be easily generalized from the Riemann-Cartan case to the general metric-affine case as follows:
\begin{equation}
  \dfL_\mathrm{Pontr} \coloneqq \dfR_a{}^b\wedge\dfR_b{}^a \,.
\end{equation}
The same argument used in the Riemann-Cartan case applies in the metric-affine one: the Bianchi identity of the curvature $\dfR_b{}^a $ makes the equation of the connection to be identically satisfied as can be checked by direct computation (we use \eqref{eq: varw MAG}): 
\begin{equation}
  \dfrac{\delta S_\mathrm{Pontr}}{\delta\dfom_{a}{}^{b}} = \Dex\dfrac{\partial\dfL_\mathrm{Pontr}}{\partial\dfR_{a}{}^{b}} = 2 \Dex \dfR_b{}^a=0 \,.
\end{equation}
This Lagrangian can be considered in the absence of metric and coframe, since it is only connection-dependent. However, in the context of metric-affine geometry, the nonmetricity is well defined and we can perform the splitting \eqref{eq: con mccon}. By doing so, the Lagrangian can be easily expressed as a total derivative:
\begin{equation}
  \dfR_a{}^b\wedge\dfR_b{}^a = \dex \Big[ \mcdfR_a{}^b\wedge\mcdfom_b{}^a +\tfrac{1}{3}\mcdfom_a{}^b\wedge\mcdfom_b{}^c\wedge\mcdfom_c{}^a +\tfrac{1}{4} \dfQNoTr^{ab}\wedge \mcDex\dfQNoTr_{ab} +\tfrac{1}{16}\dfQ \wedge \dex\dfQ\,.  \Big] \,.
\end{equation}
This permits to eliminate one term from the odd part of the quadratic MAG Lagrangian in 4 dimensions \eqref{eq: qMAGLodd}.

The Nieh-Yan invariant in Riemann-Cartan geometry (i.e. zero nonmetricity) with connection $\dfom_a{}^b$ is given by
\begin{equation}\label{eq: RC NiehYan}
  \dfL_\mathrm{Nieh-Yan} [\dfom] \coloneqq \dfT^a\wedge\dfT_a+\cofr_{ab}\wedge \dfR^{ab}\qquad = \dex(\cofr_a\wedge\dfT^a) \,.
\end{equation}
This invariant is nothing but the divergence of the torsion axial vector, which is only a vector in $\dimM=4$:
\begin{equation}
  \dfL_\mathrm{Nieh-Yan} [\dfom] = \mathring{\nabla}_\lambda\Big(\frac{1}{2}T_{\mu\nu\rho}\LCten^{\mu\nu\rho\lambda}\Big)\volfg \,.
\end{equation}

Consider the same expression \eqref{eq: RC NiehYan} in the metric-affine formalism (i.e., from now on $\dfom_a{}^b$ will have non-trivial nonmetricity). The result is not a total derivative as can be seen for instance in components, due to the presence of nonmetricity:
\begin{equation}
  \dfT^a\wedge\dfT_a+\cofr_{ab}\wedge \dfR^{ab} = \frac{1}{2}\Big[ \mathring{\nabla}_\lambda\Big( T_{\mu\nu\rho}\LCten^{\mu\nu\rho\lambda}\Big) -Q_{\mu\nu\sigma} T_{\rho\lambda}{}^{\sigma} \LCten^{\mu\nu\rho\lambda}\Big]\volfg \,.
\end{equation}
In order to recover the same boundary term we have to move the nonmetricity term to the l.h.s. of this equation. By direct computation, it is straightforward to check that
\begin{equation}\label{eq:NiehYanbt}
  \dex(\cofr_a\wedge\dfT^a)\  =\ \Dex(\cofr_a\wedge\dfT^a)\ =\ \dfT^a\wedge\dfT_a+\cofr_{ab}\wedge \dfR^{ab} -\dfT^a\wedge\dfQ_{ca}\wedge\cofr^c \,.
\end{equation}
This suggests the following generalization to the Nieh-Yan invariant to the metric-affine framework:
\begin{equation}
  \dfL_\mathrm{Nieh-Yan} [\dfom]\coloneqq \dfT^a\wedge(\dfT_a- \dfQ_{ca}\wedge\cofr^c )+\cofr_{ab}\wedge \dfR^{ab} \,.
\end{equation}
Notice that the contribution we need to add in order to recover a boundary term is linear in the curvature, and quadratic in torsion and nonmetricity. Therefore, we can make use of it to eliminate one term from the odd part of the quadratic MAG Lagrangian in 4 dimensions \eqref{eq: qMAGLodd}. Observe that this does not happen with the Gauss-Bonnet invariant, since the necessary term in that case is {\it quartic} in the nonmetricity.

\section{Conclusions}

In this chapter we have focused on metric-affine Lovelock terms in critical dimensions ($\dimM = 2k$ for the $k$-th order term). It is standard lore that these terms are topological invariants when equipped with the Levi-Civita connection and also for a metric-compatible connection. What we have shown is that this does not hold for general connections. To be precise, it is the traceless part of the non-metricity $\dfQNoTr_{ab} = \dfQ_{ab} - \dfQ g_{ab}$ the responsible for this violation of the topological character.

We showed that the coframe equation is identically satisfied in critical dimension, but the equation of the connection imposes non-trivial restrictions. To prove that we have proceeded in two steps. First, after assuming a constant metric by fixing the $\mathrm{GL}(\dimM,\mathbb{R})$ gauge, we split the connection in a very specific way in terms of two objects: its antisymmetric part (which is a connection in its own) $\mcdfom_a{}^b$ and the traceless part of the nonmetricity. We showed that due to the projective symmetry the trace of the nonmetricity disappears from the Lagrangian, so we can ignore it from a field-theoretical point of view. In the second step we provided an explicit configuration that violates the equation of motion of $\mcdfom_a{}^b$. This implies that there are restrictions in the space of fields (i.e. the space of fields does not coincide with the space of solutions) and, consequently, the associated Lagrangian cannot be a boundary term.

It is worth mentioning that this general result was obtained after a case by case study. In particular, we first analyzed the lowest order case, the Einstein-Palatini action, for which we found the most general solution. The corresponding geometry can be described by just a function $\phi$ (related to the conformal class of the metric $g_{\mu\nu}$) and two 1-forms that parameterize a general 2-dimensional connection with vanishing  $\dfQNoTr_{ab}$. This last constraint, is an indicator that the Lagrangian cannot be a boundary term. The analysis we made for the Gauss-Bonnet-Palatini case is in agreement with our previous work \cite{JanssenJCA2019a} and \cite{HehlMcCreaKopczynski1991}, and was  extended to the general critical metric-affine Lovelock term. Regarding the solutions of the theories with $k>1$, the equations are too complicated to be solved in general. However, from the equation of the connection some non-trivial families of solutions can be obtained for different values of $k$.

In conclusion, in the metric-affine formulation, one should be careful when using Lovelock terms to rewrite curvature invariants in terms of other ones through integration by parts, since additional terms depending on the traceless part of the non-metricity come into play. A similar analysis (that does not appear in the paper \cite{JanssenJCA2019b}) for the metric-affine extensions of the Pontryagin and Nieh-Yan invariants has been presented in this chapter, in agreement with \cite{HehlMcCreaKopczynski1991}. The result of this study is that only the first one is a boundary term, although the generalization of the Nieh-Yan invariant also allows to eliminate a term from the quadratic MAG action.

\subsubsection*{Limitations of this work/future directions}
\begin{itemize}
\item To further split the traceless part of $\dfQ_{ab}$ in its three irreducible components and understand the role of each of them in the destruction of the topological character.
\item  It would be interesting to investigate the full Lovelock theory including all the terms with $k \leq \lfloor \dimM/2 \rfloor $, and look for non-trivial solutions of it. However, the strong restriction that the Einstein-Palatini term imposes, suggests that Levi-Civita plus a projective mode is the only solution allowed in such theories.
\end{itemize}

\part{Gravitational wave exact solutions in MAG \label{part:GW}}

\chapter{Generalizations of GW geometries to metric-affine formulation}\label{ch:GWgen}

\boxquote{General relativity is certainly a very beautiful theory, but how does one judge the elegance of physical theories generally?}{Roger Penrose, ``Fashion, faith, and fantasy \\  in the New Physics of the Universe'', p. 7.}

In this chapter we are going to analyze different criteria used in General Relativity to characterize spacetimes that contain gravitational radiation. We will explore their extrapolation (from a kinematical point of view) to the metric-affine setting and see applications to a particular geometry.

\section{Introduction}

In the last few years, the gravitational wave research has experienced an important development due to the recent detection of the first gravitational wave signals \cite{AbbottLIGO2016,AbbottLIGO2016b}. Interestingly, in purely metric geometry (zero torsion and zero nonmetricity), there is not a \emph{general} covariant definition of what does it mean for a metric to represent a spacetime with gravitational radiation (e.g. in terms of certain property of its curvature). We have instead different criteria and conditions, as well as known geometries such as Kundt spacetimes, where we can find e.g. hypersurfaces playing the role of wave fronts \cite{Kundt1961, KundtTrumper2016}. Many of those criteria are collected in \cite{Zakharov1973} (see also \cite{Puetzfeld2000_Diploma} for a summary of some of them). Moreover, these criteria are constructed in the context of General Relativity, so when going to a generalization we have to ensure that the differential equations satisfied by the metric continue being compatible with the criteria. We are not going to address this aspect here, and just try to make a kinematic generalization (without referring to any particular theory). 

The already mentioned criteria for gravitational wave metrics are usually conditions on the Riemann tensor associated to it or, to be more precise, to its Levi-Civita connection. Therefore, they can be seen as a natural window to explore generalizations of these criteria, simply by considering their application to curvature tensors that come from other connections different from the Levi-Civita one. Indeed, if we consider a broader notion of curvature within a gauge theoretical context, the torsion, which appears related to the field strength of the translational part, can be regarded as a curvature as well and, hence, it can be subjected to these conditions. This way of extending the metric criteria is the idea we are going to explore in more detail throughout the following sections, focusing on one particular criterion with an interesting meaning.

Finally, we would like to recall that exact gravitational wave solutions have been already explored in metric-affine geometry \cite{MaciasLammerzahlGarcia2000, GarciaLammerzahlMielke1998, Obukhov2006, Puetzfeld2002, GarciaMaciasPuetzfeld2000, KingVassiliev2001, Vassiliev2002,Vassiliev2005, PasicVassiliev2005, PasicBarakovic2014, PasicBarakovic2017}. In our study, we will try to generalize the geometries used by Obukhov in \cite{Obukhov2017, Obukhov2006} but, as we previously mentioned, respecting some criteria that can be obtained from the Riemannian ones by making a reasonable generalization.

\section{Gravitational waves in Riemannian geometry} \label{sec: GW Rieman}

\subsection{Transversal space}
Given a lightlike vector field $k^{\mu}$, in order to define the transversal space we need to introduce another lightlike vector $l^{\mu}l_{\mu}=0$, such that $k^{\mu}l_{\mu}\neq0$. Since the normalization for a null vector is arbitrary, let us consider without loss of generality that the field $l^{\mu}$ verifies
\begin{equation}
k^{\mu}l_{\mu}=1\,.
\end{equation}

\boxdefinition{
\begin{defn}
\textbf{(Transversal)}. Given a lightlike congruence with velocity $k^{\mu}$ and another non-colinear lightlike vector $l^{\mu}$ that satisfy $l_{\mu}k^{\mu}=1$, the orthogonal 
\begin{equation}
\left(\mathrm{span}_{\mathbb{R}}\{k^{\mu}\vpartial_{\mu},\,l^{\mu}\vpartial_{\mu}\}\right)^{\bot}\,,
\end{equation}
is called \emph{transversal space} of the congruence (with respect
to $l^{\mu}$). We will say that a tensor $H^{\mu_{1}...\mu_{r}}{}_{\nu_{1}...\nu_{s}}$
is \emph{transversal} if the contraction of any of its indices with
$l^{\mu}$ and $k^{\mu}$ vanishes.
\end{defn}
}

At this point it is useful to introduce the projector onto the transversal spatial slices (see a more detailed explanation in \cite{Blau2011}),
\begin{equation}
\Pi^\mu{}_\nu \coloneqq \delta^\mu_\nu-k^\mu l_\nu-l^\mu k_\nu\,,
\end{equation}
that allows us to extract the transversal part of any tensor, which we will denote as\footnote{
    In general, we will use the underlined notation throughout this thesis to indicate transversality (or something related to it): underlined tensor components means (as we have just defined) total transversal part; underlined tensor-valued differential forms are transversal in the internal indices (not necessarily in the external ones); underlined indices (next chapter) are (spatial) indices raised/lowered with the signature-$(+,+,+)$ part of the transversal metric, etc.}
\begin{equation}
\sobj{H}{}^{\mu \nu ...}{}_{\rho \lambda ...}\coloneqq \Pi^\mu{}_\alpha \Pi^\nu{}_\beta \cdots \ \Pi^\gamma{}_\rho \Pi^\delta{}_\lambda\cdots \  H^{\alpha \beta ...}{}_{\gamma \delta ...}\,.
\end{equation}

\subsection{Null congruences and optical scalars}

Let us now present some quantities that characterize the behavior of a given lightlike congruence with velocity $k^\mu$. Consider the tensor
\begin{equation}
B^{\nu}{}_{\mu}\coloneqq\mathring{\nabla}_{\mu}k^{\nu}\,.\label{eq: def B tensor}
\end{equation}
Its transversal part can be decomposed as\footnote{We absorb the factor $(\dimM-2)^{-2}$ of the trace part into the definition of $\theta$, as it is usual in the literature.}
\begin{equation}
\sobj{B}{}_{\mu\nu} = \omega_{\mu\nu}+\sigma_{\mu\nu}+\Pi_{\mu\nu}\theta\,,
\end{equation}
where we have introduced
\begin{align}
\omega_{\mu\nu} & \coloneqq \sobj{B}{}_{[\mu\nu]}\,,\\
\theta & \coloneqq\frac{1}{\dimM-2}\Pi^{\mu\nu}\sobj{B}{}_{(\mu\nu)}\,,\\
\sigma_{\mu\nu} & \coloneqq \sobj{B}{}_{(\mu\nu)}-\Pi_{\mu\nu}\theta \,.
\end{align}
called, respectively, the \emph{twist tensor}, the \emph{expansion} scalar and the \emph{shear tensor}, and whose expressions in terms of $k^\mu$, $l^\nu$ and the projector $\Pi^\mu{}_\nu$ are collected in the Appendix \ref{app:opticaldecom}. For a given $l^\mu$, this decomposition is unique. Making use of these objects one can   construct\footnote{
    Note that the quantities $\omega_{\mu\nu}\omega^{\mu\nu}$ and $\sigma_{\mu\nu}\sigma^{\mu\nu}$ are non-negative due to the transversality of $\omega_{\mu\nu}$ and $\sigma_{\mu\nu}$. This can be easily seen if we go to the basis that diagonalizes the transversal metric. If we call $-g_i (<0)$ the eigenvalues of the transversal part of the inverse metric, then:
    \[
      \omega_{\mu\nu}\omega^{\mu\nu} =\sum_{i,j=3}^{\dimM}g_i g_j(\omega_{ij})^{2}\geq0\,,\qquad \sigma_{\mu\nu}\sigma^{\mu\nu} =\sum_{i,j=3}^{\dimM}g_i g_j(\sigma_{ij})^{2}\geq0\,.
    \]}
\begin{align}
\omega & \coloneqq\sqrt{\frac{1}{\dimM-2}\omega_{\mu\nu}\omega^{\mu\nu}}\,, \\
|\sigma| & \coloneqq\sqrt{\frac{1}{\dimM-2}\sigma_{\mu\nu}\sigma^{\mu\nu}}\,. 
\end{align}
The objects $\left\{\theta,\,\omega,\,|\sigma|\right\}$ (expansion, twist and shear) are known as the \emph{optical scalars} of the congruence. 

\boxdefinition{
\begin{defn}
\textbf{(Normal congruence)}. A congruence is \emph{normal} if there exists a family of hypersurfaces orthogonal to the curves of the congruence.
\end{defn}
}

A very interesting result is the following \cite{BekaertMorand2013}:

\boxproposition{
\begin{prop}
\label{prop:normalgeod} In a pseudo-Riemannian manifold $(\mathcal{M},\,\teng)$, any normal lightlike congruence is pre-geodetic. Therefore it can be reparameterized to get a geodetic congruence.
\end{prop} 
}
\boxproof{\begin{proof} See Appendix \ref{app:proofnormalgeod}. \end{proof}}

Due to this result, we concentrate on geodetic congruences for which the optical scalars are given by
\begin{align}
\theta & =\frac{1}{\dimM-2}\mathring{\nabla}_{\sigma}k^{\sigma} 
  && =\frac{1}{\dimM-2}\sign(g)\star\dex\star\dfk\,,\,,\\
\omega^2 & =\frac{1}{\dimM-2}\partial_{[\mu}k_{\nu]}\partial{}^{\mu}k^{\nu}
  && =\frac{1}{2(\dimM-2)}\sign(g)\star\left(\dex\dfk\wedge\star\dex\dfk\right)\,,\\
|\sigma|^2 & =\frac{1}{\dimM-2}\mathring{\nabla}_{(\mu}k_{\nu)}\mathring{\nabla}^{\mu}k^{\nu}-\theta^{2}
  && =\frac{1}{\dimM-2}\left[\dint{\vfre_{(a}}\mathring{\Dex}\left(\dint{\vfre_{b)}}\dfk\right)\right]\left[\dint{\vfre^{a}}\mathring{\Dex}\left(\dint{\vfre^{b}}\dfk\right)\right]-\theta^{2}\,.
\end{align}
 
We end this subsection on optical scalars by remarking a very useful property of lightlike congruences that relates the nullity of the twist with the existence of wave fronts (see \cite{Kundt1961, KundtTrumper2016,Poisson2002} and \cite[p.~59]{Witten1962}):

\boxproposition{
\begin{prop}
\label{prop:normalnullw0} A lightlike geodetic congruence is normal if and only if the twist $\omega$ vanishes.
\end{prop}
}
\boxproof{\begin{proof} See Appendix \ref{app:proofnormalnullw0}. \end{proof}}

More information on twist-free solutions of pure radiation can be found in \cite{KundtTrumper2016}.

\section{Kundt and Brinkmann spaces}

Now we focus our interest in the concept of plane-fronted waves. Kundt defined them in \cite{Kundt1961}, a definition that was also presented in \cite[p.~85--86]{Witten1962} together with a theorem that introduces a characterization: \emph{a plane-fronted wave is a vacuum field that admits a normal null congruence with $|\sigma|=\theta=0$}. Observe that, as a consequence of Propositions \ref{prop:normalgeod} and \ref{prop:normalnullw0}, the congruence of a plane-fronted wave is pre-geodetic (so it is geodetic after an appropriate reparameterization) and, additionally, $\omega=0$. 

These definitions correspond to ``vacuum solutions'' of General Relativity, which might not be solutions for other more general theories. Since we are interested in spacetimes defined in a theory-independent way, we start by introducing:\footnote{
    Note that no allusion to vacuum has been made.}

\boxdefinition{
\begin{defn}
\textbf{(Kundt space)}. A \emph{Kundt space}  is a Lorentzian manifold
that admits a geodetic null congruence with $|\sigma|=\theta=\omega=0$.
\end{defn}
}

Every point in a Kundt space admits a coordinate chart $\{x^{\mu}\}=\{u,\,v,\,z^{2},...,z^{\dimM-1}\}$
(a \emph{Kundt chart}) in which the line element is expressed:
\begin{equation}
\mathrm{d} s^{2}=2\dex u\dex v+H(u,\,v,\,z)\dex u^{2}+2W_{i}(u,\,v,\,z)\dex u\dex z^{i}+\sobj{g}{}_{ij}(u,\,z)\dex z^{i}\dex z^{j}\,,
\end{equation}
where $i,j=2,\,...,\,\dimM-1$ and $\sobj{g}{}_{ij}$ is the spatial metric with signature $(-,...,-)$. We have then a local foliation by spacelike surfaces, those with constant $u$. The expressions for the Christoffel symbols, Riemann and Ricci tensors can be found in \cite{Podolsky2009} or \cite[p.~230--231]{BicakLedvinka2014}. Solutions of this kind in different backgrounds (e.g. with and without cosmological constant) are given in \cite[chap.~18]{GriffithsPodolsky2009} and \cite[chap.~31]{Stephani2003}. 

Observe that the coordinate field $\vpartial_{v}\eqqcolon k^\mu \vpartial_{\mu}$, which satisfies
\begin{equation}
k_{\mu}k^{\mu}=0\,,\qquad\qquad k^\rho\mathring{\nabla}_{\rho}k^{\mu}=0\,, \label{eq:wavevector}
\end{equation}
is indeed the velocity field of the congruence that appears in the definition. Moreover, note that $\vpartial_{v}$ is not a covariantly constant field with respect to $\mathring{\nabla}$,
\begin{equation}
\mathring{\nabla}_{\mu}k^\rho=\mathring{\Gamma}_{\mu v}{}^{\rho} =\frac{1}{2}\left(g^{\rho u}\partial_{v}g_{\mu u}+g^{\rho i}\partial_{v}g_{\mu i}\right)\neq 0\,,
\end{equation}
i.e., the tensor $B_{\mu\nu}$ defined in \eqref{eq: def B tensor} is not trivial for Kundt spaces. This expression vanishes if $H$ and $W_{i}$ are independent of the coordinate $v$. This is a well-known particular kind of Kundt spaces called Brinkmann spaces \cite{Brinkmann1925}:

\boxdefinition{
\begin{defn}
\textbf{(Brinkmann space)}. A \emph{Brinkmann space}  is a Lorentzian manifold that admits a non-vanishing vector field $k^\mu \vpartial_\mu$ which is lightlike and covariantly constant with respect to the Levi-Civita conection, namely
\begin{equation}
k_\mu k^\mu=0\,\quad\text {and}\quad\mathring{\nabla}_{\rho}k^{\mu}=0\,.
\end{equation}
\end{defn}
}

If we introduce the associated 1-form $\dfk=k_{\mu}\dex x^{\mu}$, these two conditions can be written in the exterior notation, respectively, as
\begin{equation}
\dfk\wedge\star\dfk=0\,,\qquad\qquad\mathring{\Dex}k^a =0\,.
\end{equation}
In an analogous way as for the Kundt case, there is a coordinate chart we can always find, $\{x^{\mu}\}=\{u,\,v,\,z^{2},...,z^{\dimM-1}\}$ (\emph{Brinkmann chart}), that allows to express the metric (see for example \cite{BlancoSanchezSenovilla2013}):
\begin{equation}
\mathrm{d} s^{2}=2\dex u\dex v+H(u,\,z)\dex u^{2}+2W_{i}(u,\,z)\dex u\dex z^{i}+\sobj{g}{}_{ij}(u,\,z)\dex z^{i}\dex z^{j}\,.\label{eq: Brinkmann general metric}
\end{equation}
Moreover, $H$ or $W_{i}$ (but not both) can always be set to zero with an appropriate redefinition of the spatial coordinates $\{z^{i}\}$ (see for example \cite{Ortin2004}). From now on, when we refer to the Brinkmann metric we will take $W_{i}=0$, so the metric becomes block-diagonal.

\boxdefinition{
\begin{defn}
\textbf{\label{def: ppwave}(pp-wave)}. A \emph{plane-fronted wave with parallel rays }(or\emph{ pp-wave}) is a Brinkmann space admitting a coordinate chart such that
\begin{equation}
\mathrm{d} s^{2}=2\dex u\dex v+H(u,\,z)\dex u^{2}-\delta_{ij}\dex z^{i}\dex z^{j}\,.\label{eq: ppwave}
\end{equation}
\end{defn}
}

By calculating the Einstein tensor of \eqref{eq: ppwave} it is straightforward to prove that this is a vacuum solution of the Einstein equations if and only if $H$ is a harmonic function of the transversal coordinates,
\begin{equation}
\delta^{ij}\partial_{i}\partial_j H(u,\,z)=0\,.
\end{equation}
This condition obviously will no longer be true in more general theories. Indeed, in Chapter \ref{ch:GWsolutions} we will find solutions of quadratic MAG that violate this condition.

\subsection{Criteria for gravitational wave spacetimes}

As we have already mentioned, there are many different attempts in the literature trying to (covariantly) characterize spacetimes with gravitational radiation in General Relativity. These approaches are based on a previous analysis of the Einstein equations and the existence of characteristic submanifolds (wave fronts) and bicharacteristics (rays). Several of these criteria are extensively studied in \cite{Zakharov1973}. Based on that reference and the overview in \cite{Puetzfeld2000_Diploma}, here we present some of them:

\begin{itemize}

\item \noindent \textbf{Pirani criterion}. We will say there are free gravitational waves in an empty region of a spacetime if and only if the curvature there is type $\mathbf{II}$, $\mathbf{III}$ or $\mathbf{N}$ in the Petrov classification.\footnote{See the original formulation in \cite{Pirani1957}.}

\item \noindent \textbf{Lichnerowicz criterion}. For a non-vanishing curvature $\mathring{\dfR}_{a}{}^{b}\neq0$, we will say there is gravitational radiation if and only if there exists a non-vanishing 1-form $\dfk=k_{\mu}\dex x^{\mu}$ satisfying
\begin{align}
\dfk\wedge\star\mathring{\dfR}_{a}{}^{b} & =0 &  & \xrightarrow{\text{components}} & k^{\mu}\mathring{R}_{\mu\nu a}{}^{b} & =0\,,\label{eq: LichneC1}\\
\dfk\wedge\mathring{\dfR}_{a}{}^{b} & =0 &  & \xrightarrow{\text{components}} & k_{[\mu}\mathring{R}_{\nu\rho]a}{}^{b} & =0\,.\label{eq: LichneC2}
\end{align}
Lichnerowicz proved that these two conditions, under the hypothesis $\mathring{\dfR}_{a}{}^{b}\neq0$, imply that $k^{\mu}$ is both lightlike and geodetic \cite{Zakharov1973}. For example, the lightlike condition is immediate contracting \eqref{eq: LichneC2} with $k^{\mu}$ and then substituting \eqref{eq: LichneC1}. In addition, another consequence is that the curvature can be written:
\begin{equation}
\mathring{R}_{\mu\nu\rho\lambda}=b_{\mu\rho}k_{\nu}k_{\lambda}+b_{\nu\lambda}k_{\mu}k_{\rho}-b_{\mu\lambda}k_{\nu}k_{\rho}-b_{\nu\rho}k_{\mu}k_{\lambda}\quad\Leftrightarrow\quad \mathring{R}_{\mu\nu}{}^{\rho\lambda}=4b_{[\mu}{}^{[\rho}k_{\nu]}k^{\lambda]}\,,
\end{equation}
for some symmetric tensor $b_{\mu\nu}=b_{(\mu\nu)}$ with the property
$k^{\mu}b_{\mu\nu}=0$. 

\item \noindent \textbf{Zel'manov criterion}. We will say there is gravitational radiation in a spacetime region if and only if the curvature of this region is not covariantly constant, i.e. $\mathring{\nabla}_{\sigma}\mathring{R}_{\mu\nu a}{}^{b}\neq0$, and verifies the following covariant generalization of the wave equation
\begin{equation}
\mathring{\nabla}^{\sigma}\mathring{\nabla}_{\sigma}\mathring{R}_{\mu\nu a}{}^{b}=0\,.
\end{equation}

This condition is formulated in terms of a particular connection (Levi-Civita). Another criterion very similar to this one but formulated independently of any connection is the Maldybaeva criterion, which is based on a special (metric dependent) operator that acts on differential forms:

\item \noindent \textbf{Maldybaeva criterion}. We will say there is gravitational radiation in a spacetime region if and only if the (non-trivial) curvature 2-form satisfies the wave equation
\begin{equation}
 \Delta^{\teng}_{\rm LdR} \mathring{\dfR}_{a}{}^{b}=0\,,
\end{equation}
where $\Delta^{\teng}_{\rm LdR}\coloneqq\dex\delta_\teng+\delta_\teng\dex$ is the Laplace-de Rham operator.\footnote{Here, $\delta_\teng$ is the codifferential of $\dex$ with respect to the Hodge star operator of the metric $\teng$. It is generally defined (up to signs that depend on the convention) as $\delta_\teng\coloneqq (\text{sign})\star \dex \star$.}

\end{itemize}

In the context of General Relativity one can easily find the following relations between these criteria for the particular case of Einstein spaces:

\boxproposition{
\begin{prop}
\textbf{\textup{\label{prop: relation criteria}Relations between criteria for Einstein spaces.}}

Let $(\mathcal{M},\teng)$ be an Einstein space, i.e., one satisfying $R_{\mu\rho\nu}{}^\rho=cg_{\mu\nu}$ for some real constant $c$. Then, the following statements hold:
\begin{itemize}
\item In vacuum $(c=0)$: Lichnerowicz criterion $\Leftrightarrow$ Petrov type $\mathbf{N}$.
\item Maldybaeva criterion $\Leftrightarrow$ vacuum $(c=0)$ and Petrov type $\mathbf{N}$.
\item Zel'manov criterion $\Rightarrow$ vacuum $(c=0)$ and Petrov type $\mathbf{N}$. \\
The converse (vacuum $+\mathbf{N}\Rightarrow$ Zel'manov) is also true with only the metrics\textup{ \cite[eq. (7.12)]{Zakharov1973}} as exceptions.
\end{itemize}
\end{prop}
}

\noindent The result is the following diagram for the criteria we have seen in the case of Einstein spaces:
\begin{equation}
\begin{array}{ccccc}
\text{Zel'manov} & \begin{array}{c}
\Longrightarrow\\
\Longleftarrow_{*}
\end{array} & \text{Pirani (N)}+\text{vacuum} & \Leftrightarrow & \text{Maldybaeva}\\
 &  & \Updownarrow\\
 &  & \text{Lichnerowicz}\\
 &  & +\text{vacuum}
\end{array}
\end{equation}
where $*$ denotes that there are two exceptions. For non-Einstein spaces, the relations become more obscure. 

Finally, it is worth mentioning that there are other criteria, such as the ones by Debever, Bel, etc. More information about them can be found in \cite{Zakharov1973}. 

\newpage

\section{Extension to metric-affine geometries} \label{sec: GW non-Rieman}

In the previous section we have seen that there are different criteria to classify or categorize metrics in a gravitational wave type. However, in a metric-affine framework we have an additional field, a linear connection $\Gamma_{\mu\nu}{}^\rho$ (or equivalently $\omega_{\mu a}{}^b$), and the idea now is to analyze how can we restrict in a reasonable way an arbitrary connection to explore gravitational wave scenarios in these theories.

We start by recalling that Petrov types are based on the classification of the principal null directions of the Weyl tensor. Therefore, a generalization of the Pirani criterion could be possible by understanding the behavior of the irreducible part $\irrdfW{1}{}_{ab}$ of the new curvature, which is the one that reduces to the Weyl tensor in a Riemannian geometry \cite{Hehl1995}.

For the Maldybaeva and Zel'manov criteria one could generalize the differential operator, the curvature or both. Actually, when working either in MAG or in PG gravity, the field strengths are the curvature $\dfR_a{}^b$ and the torsion $\dfT^a$ of the connection, so another possibility might be to apply the criterion to both objects.

Finally, we examine the Lichnerowicz criterion for gravitational waves. If one checks in detail electromagnetic wave configurations in classical Maxwell theory, it is easy to see that the curvature form  $\dfF= \frac{1}{2}F_{\mu\nu} \dex x^\mu\wedge\dex x^\nu$ associated to $A_\mu$ satisfies the analogous conditions
\begin{equation}
\dfk\wedge\dfF=0\,,\qquad \dfk\wedge\star\dfF=0\,.
\end{equation}
If we now look at these equations in components,
\begin{equation}
k_{[\mu}F_{\nu\rho]}=0\,,\qquad  k^{\mu}F_{\mu\nu}=0\,,
\end{equation}
we realize that they essentially encode the well-known radiation conditions for the electromagnetic field,
\begin{equation}
\delta_{ij}k^{i}E^{j}=\delta_{ij}k^{i}B^{j}=0\,,\qquad\epsilon_{ijk}\frac{k^{i}}{k^{0}}E^{j}=\delta_{ik}B^{i}\,.
\end{equation}
After this motivation, and inspired
  by\footnote{Although, in \cite{BlagojevicCvetkovic2017} the authors use a different generalization: they contract $k$ with the internal indices:
  \[
  k^a R_{\mu\nu ab}=0\,,\qquad R_{\mu\nu[ab}k_{c]}=0\,,\qquad k^a T_{\mu\nu a}=0\,,\qquad T_{\mu\nu[a}k_{c]}=0\,.
  \]}
\cite{Obukhov2017} and \cite{BlagojevicCvetkovic2017} we are going to focus on the Lichnerowicz criterion and its generalization to a metric-affine geometry imposing the corresponding conditions over our curvatures. Considering we are working in a MAG framework, we define the following generalization involving both fieldstrengths:

\boxdefinition{
\begin{defn} {\bf (Generalized) Lichnerowicz Criteria \label{def: GenLC}}
\begin{align}
\text{1LCR}&&  \dfk\wedge\dfR_{a}{}^{b} & =0 &  & \xrightarrow{\text{components}} & k_{[\mu}R_{\nu\rho]a}{}^{b} & =0\,,\nonumber \\
\text{2LCR}&& \dfk\wedge\star\dfR_{a}{}^{b} & =0 &  & \xrightarrow{\text{components}} & k^{\mu}R_{\mu\nu a}{}^{b} & =0\,,\label{eq: General Lich R}\\
\text{1LCT}&& \dfk\wedge\dfT{}^{a} & =0 &  & \xrightarrow{\text{components}} & k_{[\mu}T_{\nu\rho]}{}^{a} & =0\,,\nonumber \\
\text{2LCT}&& \dfk\wedge\star\dfT{}^{a} & =0 &  & \xrightarrow{\text{components}} & k^{\mu}T_{\mu\nu}{}^{a} & =0\,.\label{eq: General Lich T}
\end{align}
\end{defn}
}

~

~

Now we provide an important general result which allows to express the curvature and the torsion form under these conditions in a very special way.\footnote{
    From now on \textit{transversal} means transversal to the congruence generated by $\dfk$ with respect to $\dfl$. Notice also that the differential forms that appear in the following proposition are not transversal in their Latin indices, but in their internal (coordinate) indices, which are hidden in this notation.}
\boxproposition{
\begin{prop}
\label{prop: Lich crit and quad general}Consider a lightlike congruence with velocity $k^\mu$. Let $\dfk=k_\mu \dex x^\mu$ be the associated 1-form and $\dfl$ another lightlike 1-form such that $l^{\mu}k_{\mu}=1$. For an arbitrary tensor-valued 2-form $\dfal_{a...}{}^{b...}$, the following results hold:
\begin{enumerate}
\item If  $\dfk\wedge\dfal_{a...}{}^{b...}=0$, we can express
\begin{equation}
\dfal_{a...}{}^{b...}=\dfk\wedge(s_{a...}{}^{b...}\, \dfl +\sobj{\dfbe}{}_{a...}{}^{b...})\,,
\end{equation}
where $s_{a...}{}^{b...}$ is a tensor-valued 0-form and $\sobj{\dfbe}{}_{a...}{}^{b...}$ is a tensor-valued transversal 1-form.

\item If  $\dfk\wedge\star\dfal_{a...}{}^{b...}=0$, then
\begin{equation}
\dfal_{a...}{}^{b...}=\dfk\wedge\sobj{\dfbe}{}_{a...}{}^{b...}+\sobj{\dfga}_{a...}{}^{b...} \,,
\end{equation}
for certain tensor-valued transversal forms $\sobj{\dfbe}{}_{a...}{}^{b...}$ and $\sobj{\dfga}_{a...}{}^{b...}$.

\item If the two conditions of the previous points are fulfilled, the 2-form reduces to
\begin{equation}
\dfal_{a...}{}^{b...}=\dfk\wedge\sobj{\dfbe}{}_{a...}{}^{b...}\,,
\end{equation}
and the following quadratic condition is
  satisfied,\footnotemark
\begin{equation}
\dfal_{a...}{}^{b...}\wedge\star\dfal_{c...}{}^{d...}=0\, \qquad\big(\alpha_{\mu\nu a...}{}^{b...} \ \alpha^{\mu\nu}{}_{c...}{}^{d...}=0\big) \,.
\end{equation}
\end{enumerate}
\end{prop}
}

\footnotetext{Applied to electromagnetic waves in Maxwell theory, this quadratic condition lead us to $\dfF\wedge\star\dfF=0$, namely $F_{\mu\nu} F^{\mu\nu}=0$, that corresponds to the equality $  \delta_{ij}E^iE^j =\delta_{ij}B^iB^j$.}

\boxproof{
\begin{proof}
We drop the external indices for simplicity. An arbitrary two form can be decomposed:
\[\dfal= s \ \dfk\wedge\dfl+ \dfk\wedge\sobj{\dfbe}+\dfl\wedge\sobj{\dfbe}' +\sobj{\dfga}\,.\]
where $\sobj{\dfbe}$ and $\sobj{\dfbe}'$ are transversal 1-forms and $\sobj{\dfga}$ is a transversal 2-form. 
\begin{enumerate}
\item The condition tells 
$0  =\dfk\wedge\dfal =\dfk\wedge\dfl\wedge\sobj{\dfbe}'+\dfk\wedge\sobj{\dfga}$ so, by linear independence, $\sobj{\dfbe}'=\sobj{\dfga}=0$.
\item $0 = \dfk\wedge\star\dfal=-\star(-s\ \dfk+\sobj{\dfbe}')$, which is true if and only if  $s=\sobj{\dfbe}'=0$.
\item They are immediate consequences of the previous results.
\end{enumerate}
\vspace{-4mm}
\end{proof}
}
Consequently, a metric-affine geometry satisfying the generalized Lichnerowicz criteria (Definition \ref{def: GenLC}) fulfills
\begin{equation}
  \dfR_{ab} = \dfk \wedge \sobj{\dfal}_{ab} , \qquad \dfT^a = \dfk \wedge \sobj{\dfbe}^{a},
\end{equation}
for some tensor-valued transversal 1-forms $\sobj{\dfal}_{ab}$ and $\sobj{\dfbe}^{a}$. Notice that the condition for the curvature guarantees that this is a solution of the equations of motion of the metric-affine Lovelock term with $k>2$ in its critical dimension (see Section \ref{subsec:nontrivsolLovk}).

\section{A particular case: analysis}
\subsection{Generalities}

\subsubsection*{Metric structure}

Due to its simplicity, we are interested in a spacetime metric of the Brinkmann type,\footnote{From now on we choose the orientation $\mathcal{E}_{u,v,z^2,...,z^{\dimM-1}}= \sqrt{|g|}$.}
\begin{equation}
\mathrm{d} s^{2}=2\dex u\dex v+H(u,\,z)\dex u^{2}+\sobj{g}{}_{ij}(u,\,z)\dex z^{i}\dex z^{j}\,.\label{eq: Brink metric}
\end{equation}
As we have seen previously, from this metric we can obtain two relevant (dual) objects. The first one is the \emph{wave vector} $k^{\mu}\vpartial_{\mu}\coloneqq \vpartial_{v}$, that points towards the direction of propagation of the wave. It is autoparallel with respect to $\mathring{\nabla}$ and lightlike by definition of the $v$ coordinate. The other one is the exact form 
\begin{equation}
  \dfk\coloneqq k_\mu\dex x^{\mu} \qquad = \dex u
\end{equation}
which we will call the \emph{wave form}.

In addition, consider the 1-form 
\begin{equation}
\dfl=l_{\mu}\dex x^{\mu}\coloneqq\frac{1}{2}H(u,\,z)\dex u+\dex v\,.
\end{equation} 
It is not difficult to see that it is lightlike and verifies $l_{\mu}k^{\mu}=1$. Clearly, the transversal space of the congruence generated by $k^\mu$ with respect to $l^\mu$ is the one generated by the coordinate vectors in $\{z^{i}\}$ directions,
\begin{equation}
\sobj{T_{p}\mathcal{M}}\coloneqq\mathrm{span}_{\mathbb{R}} \left\{ \vpartial_{i}|_{p} \right\}_{i=2}^{\dimM-1} \ \subset \ T_{p}\mathcal{M} \,.
\end{equation}

\subsubsection*{Coframe}

In a theory with  $\mathrm{GL}(\dimM,\,\mathbb{R})$ freedom, normally fixing the anholonomic metric $g_{ab}$ to be Minkowski is quite convenient. This choice is indeed the one we will make in the next chapter. However, when dealing with gravitational waves, there is another gauge that can be also quite convenient: a \emph{lightcone gauge} for the first two directions and an \emph{orthonormal gauge} for the transversal space. In this chapter we will make use of this one. Therefore (we use $A,B...$ for the transversal anholonomic indices),
\begin{equation}
\teng\coloneqq\cofr^{0}\otimes\cofr^{1}+\cofr^{1}\otimes\cofr^{0}-\delta_{AB}\cofr^{A}\otimes\cofr^{B}\quad=g_{ab}\cofr^a \otimes\cofr^b \,, \label{eq: anhol metric}
\end{equation}
There are several coframes compatible with this gauge. But one special coframe (we will call \emph{gauge basis}) that makes this possible for the Ansatz we have taken for the metric \eqref{eq: Brink metric} is:
\begin{equation}
\{\cofr^a \}=\left\{ \begin{array}{rlr}
\cofr^{0} & \coloneqq\dfk=\dex u \,,\\
\cofr^{1} & \coloneqq\dfl=\frac{1}{2}H\dex u+\dex v\,,\\
\cofr^{A} & \coloneqq e_{i}{}^{A}\dex z^{i}\,
\end{array}\right.
\end{equation}
with dual frame
\begin{equation}
\{\vfre_{a}\}=\left\{ \begin{array}{rlr}
\vfre_{0} & =-\frac{1}{2}H\vpartial_{v}+\vpartial_{u}\,,\\
\vfre_{1} & =\vpartial_{v}\,,\\
\vfre_{A} & =e^{i}{}_{A}\vpartial_{i}\,
\end{array}\right.
\end{equation}
where the Vielbeins $e_{i}{}^{A}(u,\,z)$ satisfy
\begin{equation}
-\delta_{AB}e_{i}{}^{A}e_{j}{}^{B}=\sobj{g}{}_{ij}\,\qquad
e^{i}{}_{B}e_{i}{}^{A}=\delta_{B}^{A}\,\qquad e^{i}{}_{A}e_{j}{}^{A}=\delta_{j}^{i}\,.
\end{equation}
In this chapter we will try to work always with $\dimM$-dimensional indices and avoid using the indices $A,B...$. For this purpose, we define the transversal coframe,
\begin{equation}
\spatcofr^{a}\coloneqq\delta_{A}^{a}\cofr^{A}\,.
\end{equation}
Observe that, as the name suggests, these objects only cover the transversal part of the cotangent space (orthogonal to the 1-forms $\dfk$ and $\dfl$). Using this, an arbitrary element of the coframe can be expressed
\begin{equation}
\cofr^a =l^{a}\dfk+k^a \dfl+\spatcofr^{a}\,,
\end{equation}
The anholonomy two form associated to this coframe and the Levi-Civita connection 1-form of the metric are given 
  by,\footnote{Note that the last term, $-\frac{1}{2}(\sobj{\Omega}_{cab} +\sobj{\Omega}_{bca}-\sobj{\Omega}_{abc})
$, is indeed the Levi-Civita connection 1-form $\mathring{\sobj{\omega}}_{cab}$ associated to the metric $\sobj{g}{}_{ij}$ of the transversal sections.}
\begin{align}
\dex\cofr^a & =\left(-\frac{1}{2}k^a \partial_{b}H+\sobj{\Omega}_{b}{}^{a}\right)\dfk\wedge\spatcofr^{b}+\frac{1}{2}\sobj{\Omega}_{bc}{}^{a}\spatcofr^b\wedge\spatcofr^c \,,\\
\mathring{\dfom}_{ab}& =\left(\spartial_{[a}Hk_{b]}-\sobj{\Omega}_{[ab]}\right)\dfk-2\sobj{\Omega}_{(cd)}\delta_{[a}^{d}k_{b]}\spatcofr^{c}-\frac{1}{2}\left(\sobj{\Omega}_{cab}+\sobj{\Omega}_{bca}-\sobj{\Omega}_{abc}\right)\spatcofr^{c}\,\label{eq: LC general}
\end{align}
where we have defined the transversal objects
\begin{align}
\spartial_{a} & \coloneqq\delta_{a}^{A}\partial_{A}\qquad =(\delta_{a}^{b}-l^b k_a -k^b l_{a})\partial_{b}\,,\label{eq: transv partial}\\
\sobj{\Omega}_{a}{}^{b} & \coloneqq\delta_{a}^{B}e^{i}{}_{B}\delta_{A}^{b}\Omega_{ui}{}^{A}\,, \label{eq: ntriv anhol 1}\\
\sobj{\Omega}_{ab}{}^{c}&\coloneqq\delta_{a}^{B}\delta_{b}^{C}\delta_{A}^{c}\Omega_{BC}{}^{A} \label{eq: ntriv anhol 2}\,.
\end{align}

\subsubsection*{Connection}

Based on the connections used in \cite{Obukhov2017, Obukhov2006}, we start by analyzing the following generalization
\begin{equation}
\dfom_{ab}=\mathring{\dfom}_{ab}+(\mathcal{C}_{ab}k_c +\mathcal{P}_{cab})\cofr^c +k_a k_b \dfA+g_{ab}\dfB\,, \label{eq: gen conn}
\end{equation}
where $\dfA=A_a\cofr^a$ and $\dfB=B_a\cofr^a$ are in principle general 1-forms, and $\mathcal{C}_{ab}$ and $\mathcal{P}_{cab}$ are arbitrary tensors satisfying
\begin{equation}
\mathcal{C}_{ab}=\mathcal{C}_{[ab]}\,,\qquad\mathcal{P}_{cab}=\mathcal{P}_{c[ab]}\,,\qquad k^{c}\mathcal{P}_{cab}=0=l^{c}\mathcal{P}_{cab}\,.
\end{equation}

$\dfom_{ab}$, as a whole, is completely independent of the metric. However, in this chapter we are going to decompose it into Levi-Civita plus distorsion in order to work in terms of the tensorial objects $\mathcal{C}_{ab}$, $\mathcal{P}_{cab}$, $A_{a}$ and $ B_{a}$. Of course, the objects $\mathcal{C}_{ab}$ and $\mathcal{P}_{cab}$ are somehow metric-dependent because of the extraction of the Levi-Civita part. The actual explicitly metric-independent expression of the connection is
\begin{equation}
\dfom_{ab}=(\mathsf{C}_{ab}k_c +\mathsf{P}_{cab})\cofr^c +k_a k_b \dfA+g_{ab}\dfB\,,
\end{equation}
where
\begin{align}
\mathsf{C}_{ab} & \coloneqq\mathcal{C}_{ab}+\spartial_{[a}Hk_{b]}-\sobj{\Omega}_{[ab]}\,,\\
\mathsf{P}_{cab} & \coloneqq \mathcal{P}_{cab}-2\sobj{\Omega}_{(cd)} \delta_{[a}^{d}k_{b]}-\frac{1}{2}\left(\sobj{\Omega}_{cab} +\sobj{\Omega}_{bca}-\sobj{\Omega}_{abc}\right)\,.
\end{align}
In a strict gauge setting (e.g. in the next chapter in which we are going to explore solutions), it is convenient to work in terms of $\mathsf{C}_{a}{}^b$ and $\mathsf{P}_{\mu a}{}^b$ because they contain purely information about the connection (and not the metric or the coframe). However they do not transform tensorially, and, for the present study, we found it convenient to work in terms of the tensors.\footnote{At the end of the day, the equations of motion of a covariant metric-affine Lagrangian can be written in terms of the curvature, the torsion and the nonmetricity that can only depend on the combinations $\mathcal{C}_{ab}$ and $\mathcal{P}_{cab}$, due to their tensorial nature.}

For future purposes we introduce the following decompositions
\begin{align}
A_{a}            &=\sA_{a}+\kA l_{a}+Ak_a \,,\\
B_{a}            &=\sB_{a}+\kB l_{a}+Bk_a \,,\\
\mathcal{C}_{ab} & = \sC_{ab}+2\kC_{[a}l_{b]}+2C_{[a}k_{b]}+2Ck_{[a}l_{b]}\,,\\
\mathcal{P}_{cab}& =\sP_{cab}+2\kP_{c[a}l_{b]}+2P_{c[a}k_{b]}+ 2P_{c}k_{[a}l_{b]}\,,
\end{align}
where the tensors $\sA_a$, $\sB_a$, $\kC_a$, $C_a$, $\sC_{ab}$, $P_c$, $P_{ca}$, $\kP_{ca}$ and $\sP_{cab}$ are totally transversal. The curvature, torsion and nonmetricity as well as the irreducible decomposition of the last two are collected in Appendix \ref{app: RTQ general conn}.

\subsubsection*{Summary and Lichnerowicz criteria}

Putting together all of the structures described in this section we have the following geometry
\begin{align}
\teng & =\cofr^{0}\otimes\cofr^{1}+\cofr^{1}\otimes \cofr^0-\delta_{AB}\cofr^{A}\otimes\cofr^{B}\,\nonumber\\
\cofr^a &= \Big\{ \cofr^0=\dfk=\dex u, \quad \cofr^1 = \dfl=\frac{1}{2}H(u,\,z)\dex u+\dex v, \quad \cofr^{A} = e_i{}^{A}(u,\,z)\dex z^i \Big\} \,\nonumber\\
\dfom_{ab}&=\mathring{\dfom}_{ab}+(\mathcal{C}_{ab}k_c +\mathcal{P}_{cab})\cofr^c +k_a k_b \dfA+g_{ab}\dfB\,. \label{eq: typeI}
\end{align}

\boxtheorem{
\begin{thm} \label{thm: LichC typeI}
For a geometry of the type \eqref{eq: typeI}, the condition $\mathrm{1LCT}$ is equivalent to
\begin{equation}
0=P_c=\kP_{ca}=P_{[cd]}=\sP_{[cd]}{}^a= \kB = \sB_a \,;
\end{equation}
the condition $\mathrm{2LCT}$ is equivalent to
\begin{align}
0 & =\kC_a=\kP_{ca}= \kB\,,\nonumber\\
0 & = \sB_c - P_c\,, \nonumber \\
0 & = \kA + C - B \,;
\end{align}
the condition $\mathrm{1LCR}$ is equivalent to
\begin{align}
0&=\dfk\wedge\dex\dfB\,, \nonumber\\
0&=e^i{}_a e^j{}_b \partial_{[i}\sA_{j]}+2\sA_{[a}P_{b]}\,, \nonumber\\
0&=\partial_v \sA_a-\spartial_a \kA+2 \kA P_a\,, \nonumber\\
0&=\kP_{ab} \kA\,, \nonumber\\
0&=\sA_{[a}\bar{P}_{b]c}\,,\nonumber\\
0&=\partial_{v}\mathcal{P}_{cab} \ (=k^d \mathring{\nabla}_{d} \mathcal{P}_{cab})\,, \nonumber\\
0&=\spatcofr^d\wedge\spatcofr^c \left(\mathring{\nabla}_{[d}\mathcal{P}_{c]ab}+\mathcal{P}_{[d|eb}\mathcal{P}_{|c]a}{}^{e}\right) \,;
\end{align}
and, finally, the condition $\mathrm{2LCR}$ is equivalent to
\begin{align}
0 & =\partial_v\mathcal{C}_{ab}\ (=k^d\mathring{\nabla}_d\mathcal{C}_{ab})\,,\nonumber\\
0 & =\partial_v\mathcal{P}_{cab}\ (=k^{d}\mathring{\nabla}_{d}\mathcal{P}_{cab})\,,\nonumber\\
0 & = 2 \partial_{[v}A_{u]} + 2C\kA = \partial_v \Big(A-\frac{1}{2}H \kA\Big) -\partial_u\kA + 2C\kA\,, \nonumber\\
0 & = \partial_v \sA_a-\spartial_a \kA+2 \kA P_a\,,  \nonumber\\
0 & =\kC_{a}\kA\,,\nonumber\\
0 & =\bar{P}_{ab}\kA\,,\nonumber\\
0 & =\partial_v \sB_i - \partial_i \kB \,,\nonumber\\
0 & =2\partial_{[v} B_{u]} =\partial_v \Big(B-\frac{1}{2}H\kB\Big) - \partial_u \kB .
\end{align}
\end{thm}
}

\boxproof{
\begin{proof}
It follows straightforwardly from the application of Proposition \ref{prop: Lich crit and quad general} to our particular torsion and curvature 2-forms (see Appendix \ref{app: RTQ general conn}).
\end{proof}
}

Moreover, combining the four conditions we easily arrive at
\boxcorollary{
\begin{cor} \label{cor: LichC typeI together}
For a geometry of the type \eqref{eq: typeI}, the generalized Lichnerowicz criteria for torsion and curvature (Definition \textup{\ref{def: GenLC}}) is verified if and only if
\begin{align}
0&=P_c=\kP_{ca}=P_{[cd]}=\sP_{[cd]}{}^a= \kB = \sB_a =\kC_a \nonumber\\
0& = \kA + C - B \,,\nonumber\\
0& =\partial_v B =\partial_v\mathcal{C}_{ab}=\partial_v\mathcal{P}_{cab} \,,\nonumber\\
0&=\partial_v \sA_a-\spartial_a \kA\,, \nonumber\\
0&= \partial_v \Big(A-\frac{1}{2}H \kA\Big) -\partial_u\kA + 2C\kA\,,\nonumber\\
0&=\partial_{[i}\sA_{j]}\,, \nonumber\\
0&=\spatcofr^d\wedge\spatcofr^c\left(\mathring{\nabla}_{[d}\sP_{c]}{}^{ab} +2\mathring{\nabla}_{[d}P_{c]}{}^{[a}k^{b]}+ 2\sP_{[d}{}^{e[a}k^{b]}P_{c]e}+\sP_{[d}{}^{ea}\sP_{c]e}{}^{b}\right) \,.
\end{align}
\end{cor}
}

As can be proved, the connections chosen in \cite{Obukhov2017, Obukhov2006} are LCR but not LCT, since the condition $0= \kA + C - B$ is not satisfied. For future convenience and due to we are interested in doing part of our calculations generalizing the geometries in those papers, we introduce the abbreviation LCT* for those connections satisfying all of the conditions except that one. In addition, to make the violation of the LCT as explicit as possible we introduce the scalar function
\begin{equation}
\mathcal{Y}\coloneqq \kA+C-B\,. \label{eq: def Y LCT*}
\end{equation}
At this point, it is not difficult to see that under LCT* the violation of the quadratic condition for the torsion is indeed proportional to the square of it,
\begin{equation}
\dfT{}^a\wedge\star\dfT^b=-\mathcal{Y}^2 k^a k^b\ \volfg\,.
\end{equation}
Recall that for our Ansatz, $\volfg\coloneqq\star 1= \sqrt{|g|}\dex u\wedge\dex v\wedge\dex z^2 ...\wedge z^{\dimM-1}$.

\subsection{Subcase: pp-waves and other simplifications}

We are going to restrict further the theory by imposing two simplifications, one in the coframe (equivalently in the holonomic metric $g_{\mu\nu}$) and the other in the connection.

\subsubsection*{Metric and coframe}

First we take the pp-wave case (see Definition \ref{def: ppwave}), i.e., we will assume the transversal space to be only $u$-dependent $\sobj{g}{}_{ij}(u,\,z)=\sobj{g}{}_{ij}(u)$, consequently there should be a redefinition of the transversal coordinates $z^i$ such that $\sobj{g}{}_{ij}$ becomes diagonal. Because of this, consider the case
\begin{equation}
\mathrm{d} s^2 =2\dex u\dex v+H(u,\,z)\dex u^2-\delta_{ij}\dex z^i\dex z^j\,.
\end{equation}
In the metric-affine context where we work in terms of the coframe, this is equivalent to
\begin{equation}
e_i{}^{A} =\delta_i^{A}\quad\Leftrightarrow\quad\cofr^{A}=\delta_i^{A}\dex z^i\,.
\end{equation}
Now, the anholonomy gets simplified and only $\Omega_{ui}{}^{1}=-\frac{1}{2}\partial_i H$ survives, i.e.
\begin{equation}
\dex\cofr^a =-\frac{1}{2} k^a\spartial_b H\dfk\wedge\cofr^b\,,
\end{equation}
so
\begin{equation}
\sobj{\Omega}_a{}^b=\sobj{\Omega}_{ab}{}^c=0\,. \label{eq: typeII zero anhol}
\end{equation}

\boxproposition{
\begin{prop}
In the gauge coframe, the Levi-Civita connection 1-form of the pp-wave metric is:
\begin{equation}
\mathring{\dfom}_{ab}=\spartial_{[a}Hk_{b]}\dfk=\partial_{[a}Hk_{b]}\dfk\qquad\Leftrightarrow\qquad\mathring{\omega}_{\mu ab}=k_{\mu}\partial_{[a}Hk_{b]}\,.
\end{equation}
\end{prop}
}

\boxproof{
\begin{proof}
Obtaining $\mathring{\dfom}_{ab}=\spartial_{[a}Hk_{b]}\dfk$ is immediate starting from \eqref{eq: LC general}. To see the rest we need to expand the transversal derivative,
\[
\spartial_{[a}Hk_{b]}=(\delta_{[a|}^{c}-l^{c}k_{[a|}-k^{c}l_{[a|})\partial_c Hk_{|b]}=\partial_{[a}Hk_{b]}\,.
\]
\end{proof}
}

When working with indices in the gauge base (because the connection is a non-covariant object under frame transformations), we have
\begin{align}
k^{c}\mathring{\omega}_{cab} & =0\,,\\
l^{c}\mathring{\omega}_{cab} & =\spartial_{[a}Hk_{b]}\,,\\
\mathring{\omega}_{ca}{}^{c}\equiv g^{cb}\mathring{\omega}_{cab} & =0\,.
\end{align}
Consequently for any totally transversal tensor   $\sobj{S}_{ab...}{}^{c...}$,
\begin{equation}
\spartial^a\sobj{S}_{ab...}{}^{c...}=\partial^a\sobj{S}_{ab...}{}^{c...}=\mathring{\nabla}^{a}\sobj{S}_{ab...}{}^{c...}=\mathring{\nabla}_{a}\sobj{S}^a{}_{b...}{}^{c...}=\partial_{a}\sobj{S}^a{}_{b...}{}^{c...}=\spartial_a\sobj{S}^a{}_{b...}{}^{c...}\,.
\end{equation}
Let us insist on that these equations are only true in the gauge basis because $\partial_c g_{ab}=0$ in that particular frame and, as a consequence of our basis choice, $k_a $, $l_a$, $k^a $ and $l^a$ are also constant. If we change the frame, the new anholonomy coefficients would enter the game.

Another consequence is that the covariant derivative of $l^a$ can be written in the following covariant way (valid in any frame, not only in the gauge basis),
\begin{equation}
\mathring{\nabla}_c l^a = \frac{1}{2} k_c (k^a l^b- g^{ab})\partial_bH\,
\end{equation}
which implies 
\begin{equation}
\mathring{\nabla}_c l^c = 0\,.
\end{equation}

Finally we present the Levi-Civita curvature and its irreducible parts in the gauge basis which gives
\begin{align}
\mathring{\dfR}_{ab} & =\spartial_{c}\spartial_{[a}Hk_{b]}\spatcofr^{c}\wedge\dfk\,,\\
\irrdfLCW{1}{}_{ab} & =\left(\spartial_{c}\spartial_{[a}Hk_{b]}-\frac{1}{\dimM-2}\spartial^{e}\spartial_{e}Hk_{[b}g_{a]c}\right)\spatcofr^{c}\wedge\dfk\,,\\
\irrdfLCW{4}{}_{ab} & =\frac{1}{\dimM-2}\spartial^{e}\spartial_{e}Hk_{[b}g_{a]c}\spatcofr^{c}\wedge\dfk\,,\\
\irrdfLCW{6}{}_{ab} & =0\,.
\end{align}
The non-trivial ones can be covariantized as follows
\begin{align}
\mathring{\dfR}_{ab} & =\mathring{\nabla}_{c}\partial_{[a}Hk_{b]}\cofr^c \wedge\dfk\,,\\
\irrdfLCW{1}{}_{ab} & =\left(\mathring{\nabla}_{c}\partial_{[a}Hk_{b]}-\frac{1}{\dimM-2}\mathring{\nabla}^{2}Hk_{[b}g_{a]c}\right)\cofr^c \wedge\dfk\,,\\
\irrdfLCW{4}{}_{ab} & =\frac{1}{\dimM-2}\mathring{\nabla}^{2}Hk_{[b}g_{a]c}\cofr^c \wedge\dfk\,,
\end{align}
(now the Latin indices refer to any basis). These are totally $\mathrm{GL}(\dimM,\,\mathbb{R})$-covariant equations. Note that 
\begin{equation}
k^a\mathring{\dfR}_{ab}=k^a\irrdfLCW{1}{}_{ab}=k^a\irrdfLCW{4}{}_{ab}=0\,.
\end{equation}

\subsubsection*{Connection}

In addition to the pp-wave condition we also impose on the connection the following metric-independent restriction in the gauge basis,
\begin{equation}
\mathsf{P}_{cab}=0\,.
\end{equation}
Note that we specify the basis because $\mathsf{P}_{cab}$ is not a tensor and this condition only holds in very particular frames. This, together with \eqref{eq: typeII zero anhol}, imply that $\mathcal{P}_{cab}=0$ and, therefore,
\begin{equation}
\dfom_{ab}=\mathring{\dfom}_{ab}+\mathcal{C}_{ab}\dfk+k_a k_b \dfA+g_{ab}\dfB\,, \label{eq: conn 2}
\end{equation}
whose torsion and curvature get simplified
\begin{align}
\dfT^a & =\mathcal{C}_c{}^a\dfk\wedge\cofr^c+k^a\dfA\wedge\dfk+\dfB\wedge\cofr^a\,,\nonumber\\
 & =\left[-Ck^a -\kC^{a}-\kA k^a +B k^a -\kB l^{a}\right]\dfk\wedge\dfl\nonumber \\
 & \quad+\left[\kC_c l^{a}+C_{c}k^a +\sC_{c}{}^{a}-k^a \sA_{c}+B\delta_{c}^{a}-\sB_{c}l^{a}\right]\dfk\wedge\spatcofr^c\nonumber \label{eq: T for type II}\\
 & \quad+\left[\kB \delta_{c}^{a}-\sB_{c}k^a \right]\dfl\wedge\spatcofr^c +\sB_c\spatcofr^c\wedge\spatcofr^a\,.\\
\dfR_{a}{}^{b} & =\mathring{\dfR}_{a}{}^{b}+\Dex\mathcal{C}_a{}^b\wedge\dfk+k_a k^b\dex\dfA+\delta_{a}^{b}\dex\dfB \nonumber\\
 & =\mathring{\dfR}_a{}^b+\mathring{\Dex}\mathcal{C}_a{}^b\wedge\dfk+k_a k^b \dex\dfA+\delta_a^b\dex\dfB- k_c (k_a \mathcal{C}^{bc}+k^b \mathcal{C}_a{}^c)\dfk\wedge\dfA \label{eq: R for type II} \,,
\end{align}
while no changes in the nonmetricity have been made with respect to that of the connection \eqref{eq: gen conn} (see Appendix \ref{app: RTQ general conn}). More details on them and their irreducible decomposition are presented for completeness in Appendix \ref{app: RTQ conn 2}.

\subsubsection*{Summary and Lichnerowicz criteria}

Again we summarize the geometry we have considered in this section,
\begin{align}
\teng & =\cofr^{0}\otimes\cofr^{1}+\cofr^{1}\otimes \cofr^0-\delta_{AB}\cofr^{A}\otimes\cofr^{B}\,\nonumber\\
\cofr^a &= \Big\{ \cofr^0=\dfk=\dex u, \quad \cofr^1 = \dfl=\frac{1}{2}H(u,\,z)\dex u+\dex v, \quad \cofr^{A} = \delta_i^{A}\dex z^i \Big\} \,\nonumber\\
\dfom_{ab}&=\mathring{\dfom}_{ab}+\mathcal{C}_{ab}k_c \cofr^c +k_a k_b \dfA+g_{ab}\dfB\,. \label{eq: typeII}
\end{align}
In this case, the generalized Lichnerowicz criteria give essentially the same as in Theorem \ref{thm: LichC typeI} and Corollary \ref{cor: LichC typeI together} but setting $P_c=\kP_{ab}=P_{ab}=\sP_{abc}=\mathcal{P}_{abc}=0$.

\subsection{Further restrictions}

Finally, we consider an additional restriction of the subcase treated in the last section. The metric and the coframe continue being the same, but we consider the connection to be subjected to the following constraints 
\begin{align}\label{eq: conditions final conn}
C & =C(u) \,, &\dfB & = B(u)\dfk \,,\nonumber\\
\kC_a & =0 \,,             &0&=\partial_v \sA_a-\spartial_a \kA\,, \nonumber\\
C_a & =C_a(u,\,z) \,, &0&=\partial_v \Big(A-\frac{1}{2}H \kA\Big) -\partial_u\kA + 2C\kA \,, \nonumber\\
\sC_{ab} & =\sC_{ab}(u)\,, &0&=\partial_{[i}\sA_{j]}\,.
\end{align}

A few remarks:
\begin{itemize}
\item We have as an immediate corollary $\dex\dfB=0$.
\item The Levi-Civita part remains the same as in \eqref{eq: typeII}. So the Riemannian
curvature is purely Weyl ($\irrdfLCW{1}{}_{ab}$) and Ricci
($\irrdfLCW{4}{}_{ab}$), while the curvature scalar vanishes.
\item This configuration together with the conditions
\begin{equation}
C_a=\sC_{ab}=C=B=0
\end{equation}
reproduces the Ansatz for the connection in \cite{Obukhov2006}. If, instead, we impose 
\begin{equation}
\sA_a=A=\kA=\sC_{ab}=C=B=0\,,
\end{equation}
we obtain the one in \cite{Obukhov2017}.
\end{itemize}

\boxtheorem{
\begin{thm}
The geometry \eqref{eq: typeII} together with \eqref{eq: conditions final conn} satisfies both \textup{LCR} and \textup{LCT*}. If, additionally, $\mathcal{Y}=0$, then the full generalized Lichnerowicz criterion is fulfilled.
\end{thm}
}

Let us now focus on the basic tensors associated to the connection and the properties they acquire under \eqref{eq: conditions final conn}. From now on we will use \eqref{eq: def Y LCT*} to eliminate $\kA$ from all the equations.

\subsubsection*{Torsion and its properties}

Now we present the torsion and its irreducible components,
\begin{align}
\dfT^a & =-\mathcal{Y}k^a \dfk\wedge\dfl+\big[\sC_{c}{}^{a}+(C_{c}- \sA_{c})k^a + B\delta_{c}^{a}\big]\dfk\wedge\spatcofr^c \\
\irrdfT{2}{}^{a} & =\frac{1}{\dimM-1}\left[(\dimM-2) B-\mathcal{Y}\right]\dfk\wedge\cofr^a \,,\\
\irrdfT{3}{}^{a} & =\frac{1}{3}\sC_{bc}\left(k^a \spatcofr^b\wedge\spatcofr^c+2g^{ac}\dfk\wedge\spatcofr^{b}\right)\\
\irrdfT{1}{}^{a} & =\left[\frac{1}{3}\sC_{b}{}^a +(C_b- \sA_{b})k^a +\frac{1}{\dimM-1}(\mathcal{Y}+ B)\delta_b^a\right]\dfk\wedge\spatcofr^b\\
 & \quad-\frac{\dimM-2}{\dimM-1}(\mathcal{Y}+ B)k^a \dfk\wedge\dfl-\frac{1}{3}\sC_{bc}k^a \spatcofr^b\wedge\spatcofr^c
\end{align}
It is worth remarking that the totally antisymmetric component is directly connected with the antisymmetric transversal tensor $\sC_{ab}$, and that the trace of the torsion,
\begin{equation}
-\dint{\vfre_a}\dint{\vfre_b}\irrdfT{2}{}^b=T_{ab}{}^b=\left[(\dimM-2)B-\mathcal{Y}\right]k_a \,,
\end{equation}
 is proportional to $B$ for geometries LCT (i.e. with $\mathcal{Y}=0$).

\begin{itemize}
\item The operator $\dfk\wedge$ gives zero also when acting on $\irrdfT{2}{}^{a}$ since it is proportional to $\dfk$, but for the rest of the irreducible components we have
\begin{equation}
\dfk\wedge\irrdfT{3}{}^{a}=-\dfk\wedge\irrdfT{1}{}^{a}=\frac{1}{3}\sC_{bc}k^a \dfk\wedge\spatcofr^b\wedge\spatcofr^c\,.
\end{equation}

\item However the operator $\dfk\wedge\star$ gives zero for $\irrdfT{3}{}^{a}$ because it only has components in the directions $\dfk$ and $\spatcofr^{c}$. For the rest,
\begin{align}
\dfk\wedge\star\irrdfT{2}{}^{a} & =\frac{1}{\dimM-1}\left[(\dimM-2) B-\mathcal{Y}\right]k^a \star\dfk\,,\\
\dfk\wedge\star\irrdfT{1}{}^{a} & =-\frac{\dimM-2}{\dimM-1}(\mathcal{Y}+ B)k^a \star\dfk\,.
\end{align}
\item The contractions of the torsion 2-form and its irreducible components with $k^a $ all vanish,
\begin{equation}
k_a \irrdfT{3}{}^a=k_a \irrdfT{2}{}^a=k_a \irrdfT{1}{}^{a}=0\qquad\Rightarrow\qquad k_a \dfT{}^a =0\,.
\end{equation}
This is not true for contractions in the first two indices of the torsion tensor $T_{bc}{}^a$ and their irreducible components, since there are coefficients in the direction of $\dfl$, which give 1 instead of 0 when contracting with $k^a$.

\end{itemize}

\subsubsection*{Nonmetricity and its properties}

Now we have
\begin{equation}
\dfQ_{ab}=2 k_a k_b \dfA+2g_{ab} B\dfk\,.
\end{equation}
Therefore the traces are
\begin{align}
\dfQ_{c}{}^{c} =Q_{ac}{}^c \cofr^a & = 2\dimM B\dfk\,,\\
\dint{\vfre^{c}}\dfQ_{ca}=Q_{ca}{}^c &= 2(\mathcal{Y}+2 B-C)k_a \,
\end{align}
The irreducible components of $\dfQ_{ab}$ do not experience any changes with respect to those in Appendix \ref{app: RTQ general conn}, apart from the substitution $\dfB=B\dfk$. 

Let us show some properties of this nonmetricity:

\begin{itemize}
\item Contractions of the nonmetricity with $k^a$
\begin{align}
k^a \dfQ{}_{ab}=k^a \irrdfQ{4}{}_{ab} & =2k_b  B\dfk\,,\\
k^a \irrdfQ{3}{}_{ab} & =\frac{2(\dimM-2)}{(\dimM-1)(\dimM+2)}(\mathcal{Y}+ B-C)k_b \dfk\,,\\
k^a \irrdfQ{1}{}_{ab} & =\frac{2(\dimM-2)}{3(\dimM+2)}(\mathcal{Y}+ B-C)k_b \dfk\,,\\
k^a \irrdfQ{2}{}_{ab} & =-\frac{2(\dimM-2)}{3(\dimM-1)}(\mathcal{Y}+ B-C)k_b \dfk\,.
\end{align}
An immediate consequence is
\begin{equation}
k^a k^b \irrdfQ{I}{}_{ab}=0\qquad\qquad\forall I\,.
\end{equation}
\item Contracted wedge with the coframe
\begin{align}
\irrdfQ{4}{}_{ab}\wedge\cofr^a  & =-2 B\cofr_b \wedge\dfk\,,\label{eq: Q wedge cofr 1}\\
\irrdfQ{3}{}_{ab}\wedge\cofr^a  & =\frac{2}{\dimM-1}(\mathcal{Y}+ B-C)\cofr_b \wedge\dfk\,,\\
\irrdfQ{2}{}_{ab}\wedge\cofr^a  & =2k_b \dfA\wedge\dfk-\frac{2}{\dimM-1}(\mathcal{Y}+ B-C)\cofr_b \wedge\dfk\,,\\
\irrdfQ{1}{}_{ab}\wedge\cofr^a  & =0\,.\label{eq: Q wedge cofr 4}
\end{align}
\item Finally we apply the operators $\dfk\wedge $ and $\dfk\wedge\star $ on the nonmetricity form,
\begin{align}
\dfk\wedge\dfQ_{ab} & =2k_a k_b (\mathcal{Y}+ B-C)\dfk\wedge\dfl+2k_a k_b  A_{c}\dfk\wedge\spatcofr^{c}\,,\\
\dfk\wedge\star\dfQ_{ab} & =2k_a k_b (\mathcal{Y}+ B-C)\volfg\,.
\end{align}
\end{itemize}

\subsubsection*{Curvature and its properties}

Only the following parts of the curvature survive to the conditions \eqref{eq: conditions final conn}
\begin{align}
\irrdfZ{1}{}_{ab} & =k_a k_b(2C \sA_c-l^d \calF_{dc})\spatcofr^c\wedge\dfk\,,\\
\irrdfW{1}{}_{ab} & =\left[2\VdHC_{(cd)}-\frac{2}{\dimM-2}g_{cd}\VdHC\right]\delta_{[a}^{d}k_{b]}\spatcofr^c\wedge\dfk\,,\\
\irrdfW{2}{}_{ab} & =2\VdHC_{[cd]}\delta_{[a}^{d}k_{b]}\spatcofr^c \wedge\dfk\,,\\
\irrdfW{4}{}_{ab} & =\frac{2}{\dimM-2}\VdHC g_{c[a}k_{b]}\spatcofr^c\wedge\dfk\,,
\end{align}
where we have introduced the following transversal tensors
\begin{align}
\VdHC_{a} & \coloneqq\frac{1}{2}\spartial_a H+C_a\,,\label{eq: objects R type III 1}\\
\VdHC_{ab} & \coloneqq\spartial_a\VdHC_b=\mathring{\nabla}_b\VdHC_a-k_b l^c \mathring{\nabla}_c\VdHC_a\,,\\
\VdHC & \coloneqq\VdHC_c{}^c =\spartial_c \VdHC^c =\mathring{\nabla}_c\VdHC^c\,,
\end{align}
which are going to dominate the antisymmetric part of the curvature $\dfW{}_{ab}$ and its parts, and the antisymmetric tensor
\begin{equation}
\calF_{dc}\coloneqq2\mathring{\nabla}_{[d} A_{c]}=2e^{\mu}{}_{d}e^{\nu}{}_{c}\partial_{[\mu} A_{\nu]}\,.\label{eq: objects R type III 2}
\end{equation}
These new objects fulfill the relations
\begin{equation}
\spartial^{c}\VdHC_{cd}=\spartial^{2}\VdHC_{d}\,,\qquad\spartial^{c}\VdHC_{dc}=\spartial_{d}\VdHC^{c}{}_{c}=\spartial_{d}\VdHC\,,
\end{equation}
\begin{equation}
l^d \calF_{dc} \spatcofr^c = (\partial_u\sA_c - \spartial_c A) \spatcofr^c \,.
\end{equation}

As a consequence, the final expression for the total curvature is
\begin{equation}
\dfR_{ab}=k_a k_b (2C \sA_c-l^d\calF_{dc})\spatcofr^c\wedge\dfk+2\VdHC_{c[a}k_{b]}\spatcofr^c\wedge\dfk\,. 
\end{equation}
Observe that from this expression we can immediately read the symmetric ($\dfZ{}_{ab}$) and antisymmetric ($\dfW{}_{ab}$) parts of the curvature. Notice also that the antisymmetric one is totally controlled by the tensor $\VdHC_{ca}$, while the symmetric part only depends on $A_a$ and its derivatives. Finally we provide some nice properties of this curvature:

\begin{itemize}
\item The Lichnerowicz conditions also hold independently for $\irrdfZ{1}{}_{ab}$, $\irrdfW{1}{}_{ab}$, $\irrdfW{2}{}_{ab}$ and $\irrdfW{4}{}_{ab}$ since all of them are linear combinations of $\spatcofr^c\wedge\dfk$.
\item In addition to the Lichnerowicz condition, we also have
\begin{equation}
k^a \dfZ_{ab}=k^a \irrdfW{1}{}_{ab}=k^a \irrdfW{2}{}_{ab}=k^a \irrdfW{4}{}_{ab}=0\,.
\end{equation}
thanks to the fact that $k^aC_a=0=k_c\spatcofr^c$. Consequently
\begin{equation}
k^a \dfR_a{}^b{}=0 \,,\qquad k_b\dfR_a{}^b=0\,.
\end{equation}
This result together with the fact that $\dfR_{ab}$ goes in the direction of $\spatcofr^c\wedge\dfk$ tells that any contraction of the curvature tensor $R_{abcd}$ (and hence of any of its irreducible components) with the wave vector $k^a$ vanishes.

\item The traces of the curvature are
\begin{align}
\dint{\vfre^{b}}\dfR_{ba} & =\dint{\vfre^{b}}\irrdfW{4}{}_{ba}=\VdHC k_a \dfk\,,\\
\dint{\vfre^{b}}\dfR_{ab} & =-\dint{\vfre^{b}}\irrdfW{4}{}_{ba}=-\VdHC k_a \dfk\,,\\
\dfR_a{}^a & = 0\,,\\
\dint{\vfre^{a}}\dint{\vfre^{b}}\dfR_{ba} & =0\,.
\end{align}
or, equivalently in components,
\begin{align}
R_{acb}{}^c = - R_{ac}{}^{c}{}_b & = \VdHC k_a k_b  \,,\\
R_{abc}{}^c &= 0  \,,\\
R_{ab}{}^{ab} & = 0\,.
\end{align}
\end{itemize}

\subsubsection*{Evaluated MAG Lagrangian}

Finally we would like to end this section with a useful result when working in Metric-Affine Gauge gravity. Consider the (even) MAG Lagrangian in arbitrary dimensions, \eqref{eq: qMAGLeven} and the 4-dimensional odd parity extension \eqref{eq: qMAGLodd},\footnote{
    Here we extend the result in \cite{JCA2020}. In the paper we did not consider the most general odd parity (quadratic) extension, as we are doing in this thesis (see \eqref{eq: qMAGLodd}).} 
\boxtheorem{
\begin{thm}
Let $\mathcal{G}=(g_{ab},\,\cofr^a ,\,\dfom_{a}{}^{b})$ be a geometry of the type treated this section, i.e. \eqref{eq: typeII} under the restrictions \eqref{eq: conditions final conn}. Then, in arbitrary dimensions, any even-parity linear or quadratic invariant involving exclusively the curvature, the torsion and the nonmetricity of the connection, and no derivatives of them, is identically zero. Furthermore, the 4-dimensional odd-parity invariants that satisfy the previous requirements also vanish.
\end{thm}
}

\boxproof{
\begin{proof}
First we use that the irreducible components $\irrdfZ{I}$ with $I=2,3,4,5$ and $\irrdfW{I}$ with $I=3,5,6$ are zero. Then using the properties in the previous subsections, it is almost immediate to check that all of the terms that appear in \eqref{eq: qMAGLeven} vanish independently. Since they form a basis of all possible (linear and quadratic) invariants involving the curvature, the torsion and the nonmetricity, then all possible invariants are zero at this order. Something similar happens with the basis of odd invariants in four dimensions built with the terms appearing in \eqref{eq: qMAGLodd}.
\end{proof}
}

In particular, when looking for solutions of this type for the complete 4-dimensional MAG action $\dfL_{\mathrm{MAG}}$ (defined in \eqref{eq: MAG Lag form}), only the cosmological constant term contributes to the evaluated Lagrangian,
\begin{equation}
\left.\dfL_{\mathrm{MAG}}\right|_{\mathcal{G}}= -\frac{\lambda}{\kappa}\volfg\,.
\end{equation}
This result simplifies considerably the equation of motion of the coframe. To be precise, the term with the interior derivative of the Lagrangian reduces to
\begin{equation}
\dint{\vfre_a}\big(\left.\dfL_{\mathrm{MAG}}\right|_{\mathcal{G}}\big)=-\frac{\lambda}{\kappa}\star\cofr_a \,.
\end{equation}

\section{Final comments and conclusions}

In this chapter we have revised several criteria that can be found in the literature to discern whether a metric spacetime or a region of it belongs to a gravitational wave category, i.e. it contains gravitational radiation. We also recalled that, in the context of General Relativity, some of them are equivalent in vacuum for very simple kinds of metrics. Then we discussed some possibilities for them to be extended to a metric-affine geometry and focused on one of them, the Lichnerowicz criteria. The main motivation for this choice is that this criterion reflects some common features between electromagnetic radiation and gravitational waves. We therefore proposed a generalization of it and showed its implications for a particular geometry. For the metric (or, equivalently the pair formed by the anholonomic metric and the coframe) we considered a Brinkmann space, whereas the linear connection was chosen as a generalization of those studied in the works \cite{Obukhov2006, Obukhov2017}. We then collected the conditions this connection should satisfy in order to respect the proposed generalization of the Lichnerowicz criteria. Finally, we analyzed some particular cases providing several properties of the associated curvature, torsion and nonmetricity.

\subsubsection*{Limitations of this work/future directions}
\begin{itemize}
\item We have concentrated here on generalizing the criteria used in Riemannian geometry (i.e. with the Levi-Civita connection), but there are other conditions to be taken into account, for instance, the symmetries of the metric (\emph{isometries}). In the Brinkmann case, the wave vector $\vpartial_v=k^\mu \vpartial_\mu$ is indeed a Killing vector, which can be seen in the fact that none of the metric components in the Brinkmann chart depends on the $v$ coordinate. Encouraged by this fact, one may also require the linear connection to have zero Lie derivative in the direction of $k^\mu$. Since this is true for the Levi-Civita part, it will be guaranteed whenever the distorsion tensor has zero Lie derivative. For instance, for our configuration \eqref{eq: typeII} expressed in the Brinkmann chart, this condition gives essentially
\begin{equation}
0= \partial_v\mathcal{C}_{\mu\nu}k^{\rho}+k_{\mu}k_{\nu}\partial_vA^\rho + g_{\mu\nu}\partial_v B^\rho\,.
\end{equation}
Contracting appropriately this equation one obtains that all of the tensors that the connection depends on must be $v$-independent. For $\mathcal{C}_{\mu\nu}$ this is true under the generalized Lichnerowicz criteria, but for $A^\mu$ and $B^\mu$ we get new conditions to be considered, which will obviously simplify further our geometries.

\item It is also worth mentioning the role of the metric in theories beyond General Relativity. The criteria explained in \cite{Zakharov1973} are defined in the context of the differential equations of motion of General Relativity. So in order for our metric Ansatz to be associated to gravitational radiation (in the sense of Lichnerowicz) it should be guaranteed that the equations of motion of the theory for the metric sector are of the same type. In the MAG case, this is true e.g. if the parameters of the action are such that the Riemannian (Levi-Civita) quadratic part in the curvature gives the Gauss-Bonnet invariant. In that case, the four dimensional theory becomes simply General Relativity plus additional fields (torsion, nonmetricity and their derivatives). The compatibility of the criteria with other theories that do not respect these requirements should be carefully studied.  In addition, the precise physical meaning of the generalized Lichnerowicz criterion (Definition \ref{def: GenLC}) in relation to the dynamical equations for the connection in each particular theory is another important question to address. These points and their implications in the MAG theory are left for future research.
\end{itemize}


\chapter{Exact GW solutions in quadratic MAG}\label{ch:GWsolutions}

\boxquote{If I had an hour to solve a problem I'd spend 55 minutes thinking about the problem and 5 minutes thinking about solutions.}{Albert Einstein? (Actually, it is not clear whether\\ Einstein said this, but it is a nice quote)}

As we have already commented in the previous chapter, the plane-fronted gravitational waves generalize the basic properties of electromagnetic waves in flat spacetime to the case of curved spacetime geometry. In the framework of GR, the theoretical study of the gravitational waves has a long and rich history \cite{Barnett2014, Bondi1957, BondiPirani1959, Brinkmann1923, Brinkmann1923b, Brinkmann1925, ChenNester2017, ColeyMcNutt2012,CroppVisser2010, CroppVisser2011, EinsteinRosen1937, EhlersKundt1962, FlanaganHughes2005, Griffiths1991, JordanEhlersKundt2009, JordanEhlersSachs2013, McNuttMilson2013, Penrose1965, Peres1959, RosenVirbhadra1993, Torre2006,Rosen1937,Kundt1961,Zakharov1973,Stephani2003}. A wide variety of exact gravitational plane wave solutions have been obtained in Poincar\'e gauge gravity \cite{Adamowicz1980, ChinChern1983, Sippel1986, Zhytnikov1994, SinghGriffiths1990, BabourovaFrolov1999, BlagojevicCvetkovic2014, BlagojevicCvetkovic2014b, BlagojevicCvetkovic2015, BlagojevicCvetkovic2015b, BlagojevicCvetkovic2017, Obukhov2017, BlagojevicCvetkovic2017b}, in teleparallel gravity \cite{MuellerHoissen1983, ConroyKoivisto2018,Hohmann2018,HohmannPfeifer2019,CaiCapozziello2016, CapozzielloCapriolo2020}, in a number of modified gravity theories \cite{Baykal2016, GleiserDotti2005, GursesHalilsoy1978, Mohseni2012}, as well as in supergravity \cite{AichelburgDereli1978, DereliTucker1980, Hull1984, Urrutia1981} and in superstring theories \cite{GimonHashimoto2003,GimonHashimoto2003b, ChamblinGibbons2000, HorowitzTseytlin1995, MarolfRoss2002, Michelson2002, Tseytlin1995}. The higher-dimensional generalizations of the gravitational wave solutions were discussed in \cite{Sokolowski1991, ColeyMilson2003, Hervik2003,Obukhov2004}.

The earlier studies \cite{MaciasLammerzahlGarcia2000, GarciaLammerzahlMielke1998, Obukhov2006, Puetzfeld2002, GarciaMaciasPuetzfeld2000, KingVassiliev2001, Vassiliev2002,Vassiliev2005, PasicVassiliev2005, PasicBarakovic2014, PasicBarakovic2017} had demonstrated the existence of solutions of that kind in the metric-affine theory of gravity with propagating torsion and nonmetricity fields. For instance, in some of these studies the solutions were found for a specific set of parameters of the MAG Lagrangian \eqref{eq: qMAGLeven} and by using the triplet technique \cite{ObukhovVlachynskyEsser1997,HehlMacias1999}. In the present study, we will consider the full even parity Lagrangian together with a particular Ansatz containing both torsion and nonmetricity.

The goal of this chapter, is to study plane gravitational waves for the general quadratic MAG Lagrangian. Analyzing wave exact solutions is in fact a good approach to the understanding of the particle spectrum of these MAG models, extending the earlier results \cite{Karananas2015, BlagojevicCvetkovic2018, BaikovHayashi1992, PercacciSezgin2020, LinHobson2019, LinHobson2020, LinHobson2020b}. One good example was given in \cite{BlagojevicCvetkovic2017b}, where some parameters coming from the wave analysis were found to be related to the masses of the spin-$2^\pm$ torsional modes.

\section{Lagrangian and Ansatz}

\subsection{Metric-affine setting}

We are going to explore four-dimensional solutions for the gravitational action \eqref{eq: qMAGLeven}, i.e., the most general even-parity quadratic metric-affine action with zero cosmological constant. To simplify the notation we will simply denote the Lagrangian as $\dfL$ throughout this chapter. We have removed the odd sector just for simplicity. 

The Ansatz we chose is a plane wave configuration that we will describe in this section in terms of the basic gravitational fields $(g_{ab}, \cofr^a, \dfom_a{}^b)$. Essentially, we will extend the approach \cite{Obukhov2006,Obukhov2017, BlagojevicCvetkovic2017b} in which the gravitational waves are patterned by the electromagnetic waves on a curved spacetime. 

We fix the $\mathrm{GL}(\dimM,\mathbb{R})$ gauge by choosing the metric to be the one of Minkowski space in Cartesian coordinates:
\begin{equation}\label{eq: gab}
g_{ab} = {\rm diag}(+1,-1,-1,-1) \,. 
\end{equation}

Now we specify the Ansatz for the coframe 1-form that is going to be\footnote{
    We have introduce a factor of 2 in front of $\dex v$ that does not appear in the article \cite{JCAObukhov2021a}. This and the changes in the names of the free function $H$ and the null coordinates are done just in order to use the same notation for the Brinkmann metric as in the previous chapter. The extra factor of 2 is completely irrelevant for our computations.}
\begin{align}
  \cofr^0 &= \frac{1}{2}(H + 1)\dex u + \dex v,\label{eq: cof0}\\ 
  \cofr^1 &= \frac{1}{2}(H - 1)\dex u + \dex v,\label{eq: cof1}\\
  \cofr^A &= \dex x^A,\qquad A = 2,3,\label{eq: cof23}
\end{align}
where $H = H(u, x^A)$. Here we continue using the family of indices $A,B,C,D,...$ for the transversal directions of either the coframe or the coordinate basis (since they both coincide in the transversal subspace according to \eqref{eq: cof23}). We additionally introduce the family of indices $\mm, \nn,... = 0,1$ for the non-transversal anholonomic indices. To sum up, we have:
\begin{align}
  \mu &=\big\{u,\,v,\,\underbrace{2,\,3}_A \big\}         &\rightsquigarrow x^\mu   &= (x^u\equiv u, \, x^v\equiv v,\, x^A) \,,\\ 
    a &=\big\{\underbrace{0,\,1}_\mm,\,\underbrace{2,\,3}_A \big\} &\rightsquigarrow \cofr^a &=(\cofr^\mm,\cofr^A)  \,.
\end{align}

This choice for the coframe implies that the line element (i.e., the holonomic components of the metric $g_{\mu\nu}$) corresponds to a Brinkmann metric
\begin{equation}\label{ds_2}
  \der s^2 = g_{ab}\cofr^a\otimes\cofr^b = 2\dex u \dex v + H \dex u^2 - \delta_{AB}\dex x^A \dex x^B \,.
\end{equation}

Similarly as we did in the previous chapter, we introduce the wave 1-form $\dfk$,
\begin{equation}
  \dfk \coloneqq \dex u \qquad  (=\cofr^0 - \cofr^1) \,,\label{eq: kdef}
\end{equation}
which, as we saw, verifies $\dfk\wedge \star \dfk = 0$. This is just reflecting that $\dint{\vfre_a}\dfk=k_a = (1, -1, 0, 0)$ is a null field, i.e., $k_a k^a = 0$. 

For the connection 1-form, we assume the form
\begin{equation}\label{eq: conW}
  \dfom_a{}^b = -\dfk \left(k_a V^b + k^b W_a\right) + k_a k^b \dfU\,, \qquad\qquad \dfU \coloneqq U_a\cofr^a ,
\end{equation}
where three new vector variables are introduced: $W_a$, $V_a$ and $U_a$. These variables are considered to be functions of the coordinates $(u, x^A)$ only.

The 1-form $\dfU$, as well as $W^a$ and $V^a$, are assumed to be orthogonal to the wave covector, i.e.
\begin{equation}
  \dfk\wedge \star \dfU = 0\qquad (\Leftrightarrow \quad k^a U_a = 0)\,,
\end{equation}
\begin{equation}
  k_a W^a = 0,\qquad k_a V^a = 0.\label{eq: kW0}
\end{equation}
These conditions are guaranteed if we choose these objects to be purely transversal (the only non-vanishing components are those in the $x^2$ and $x^3$ directions). In other words, they have the following form:
\begin{equation}\label{eq: WVUa0}
  W^a= \delta^a_A \, W^A(u, x^B)\,,\qquad  V^a= \delta^a_A\, V^A(u, x^B)\,,\qquad  U_a= \delta_a^A\, U_A(u, x^B)\,. 
\end{equation}
In total, our metric-affine Ansatz is described by 7 variables: $H = H(u, x^B)$, $W^A = W^A(u, x^B)$, $V^A = V^A(u, x^B)$ and $U_A = U_A(u, x^B)$. These functions determine the wave profile and their explicit form should be found from the equations of motion.

One immediately verifies that the wave 1-form is closed, and the wave covector is both, constant and covariantly constant:
\begin{equation}
\dex \dfk = 0,\qquad \dex k_a = 0,\qquad \Dex k_a = 0.\label{dk0}
\end{equation}
With this in mind, we compute the nonmetricity 1-form and the torsion and curvature 2-forms:
\begin{align}
  \dfQ_{ab} &=-2\dfk k_{(a} \pW_{b)} + 2k_a k_b\dfU \,,\label{eq: nonW}\\
  \dfT^a    &=- \dfk\wedge k^a\,\dfThe ,\label{eq: torW}\\
  \dfR_a{}^b&=  \dfk\wedge\left(k_a \sdex V^b + k^b \sdex W_a\right) + k_a k^b \dex \dfU \,,\label{eq: curW}
\end{align}
where we introduced
\begin{align}
  \dfThe     &\coloneqq \frac{1}{2}\sdex H + W_a\cofr^a + \dfU \quad = \left(\frac{1}{2}\partial_A H - \delta_{AB} W^B + U_A \right) \cofr^A ,\label{eq: THW}\\
  \pmW_a &\coloneqq W_a \pm V_a \,,\label{eq: pmW}
\end{align}
and the transversal exterior derivative $\sdex$, which only acts in the 2-space spanned by $x^A = (x^2, x^3)$:
\begin{equation}
 \sdex \coloneqq \cofr^A \dint{\vfre_A} \dex = \dex x^A\partial_A,\qquad A = 2,3.\label{eq: ud}
\end{equation}

From \eqref{eq: curW} we can immediately deduce
\begin{align}
  (\dfR^{[ab]} \eqqcolon)\!\!\quad \dfW^{ab} &= \dfk \wedge \mOm^{[a}k^{b]}\,, \label{eq: WcurvW}\\
  (\dfR^{(ab)} \eqqcolon)    \quad \dfZ^{ab} &= \dfk \wedge \pOm^{(a}k^{b)} + k^a k^b \dex \dfU \,,
\end{align}
where
\begin{align}
   \pmOm^a \coloneqq \sdex \pmW^a \qquad =(\delta_A^a \partial_B \pmW^A) \cofr^B \,. \label{eq: OMW}
\end{align}
The new objects \eqref{eq: THW} and \eqref{eq: OMW} verify the properties:
\begin{equation}
  \dfk \wedge\star \dfThe = 0,\quad \dfk\wedge\star \pmOm^a = 0\,,   \quad   k_a\pmOm^a = 0\,.\label{eq: kTOM}
\end{equation}
Applying the transversal differential to \eqref{eq: THW}, and making use of \eqref{eq: OMW}, we find
\begin{equation}
  \sdex \dfThe = \frac{1}{2}(\pOm_a + \mOm_a)\wedge\cofr^a + \sdex \dfU.\label{eq: dTHW}
\end{equation}
This expression is essentially equivalent to the Bianchi identity $\Dex \dfT^a = \dfR_b{}^a\wedge\cofr^b$, which can be immediately checked by applying $\Dex$ to \eqref{eq: torW} and using \eqref{eq: curW}. It is worthwhile to notice that
\begin{equation}\label{du}
  \dex\dfU = \dfk\wedge\dot{\dfU} + \sdex \dfU,\qquad \dot{\dfU} = (\partial_u U_a)\,\cofr^a\,,
\end{equation}
where $\dot{\dfU}$ is transversal. This immediately implies
\begin{equation}
  k^b\dint{\vfre_b}\dex\dfU=0
\end{equation} 
or, equivalently,
\begin{equation}
  \dfk\wedge\star\dex\dfU=0\,.
\end{equation}

\subsubsection*{Comments on the wave-like structure of R, T and Q}

It is worthwhile to notice that the 2-forms of the gravitational gauge field strengths \eqref{eq: torW} and \eqref{eq: curW} have the same structure as the electromagnetic field strength of a plane wave, when $\dfU = 0$. Indeed, we have (we also include the nonmetricity)
\begin{equation}
\dfQ_{ab} = \dfk\,q_{ab},\quad \dfT^a = \dfk\wedge \dfa^a,\quad \dfR_a{}^b = \dfk\wedge \dfa_a{}^b,
\end{equation}
where the objects $q_{ab} \coloneqq -(k_a \pW_b + k_b \pW_a)$, $\dfa^a \coloneqq -\, k^a\dfThe$ and $\dfa_a{}^b \coloneqq k_a \sdex \,V^b + k^b \sdex \,W_a$ play the role of the gravitational ``polarization'' forms, in complete analogy to the polarization 1-form $\dfA$ in the electromagnetic plane wave field $\dfF = \dfk\wedge \dfA$, which is orthogonal to the wave covector $\dfk\wedge \star \dfA = 0$ (see Section \ref{sec: GW non-Rieman}).

By using Proposition \ref{prop: Lich crit and quad general}, it can be easily checked that the generalized Lichnerowicz criteria is fulfilled (indeed, also the nonmetricity verifies an analogous criterion):
\begin{align}
\dfk\wedge\star \dfQ_{ab} = 0,\qquad \dfk\wedge\star\dfT^a = 0,\qquad \dfk\wedge\star\dfR_a{}^b = 0,\\
\dfk\wedge \dfQ_{ab} = 0,\qquad \dfk\wedge \dfT^a = 0,\qquad \dfk\wedge \dfR_a{}^b = 0, \\
\dfQ_{ab}\wedge\star \dfQ_{cd} = 0,\qquad \dfT^a\wedge\star \dfT^b = 0,\qquad \dfR_a{}^b\wedge\star \dfR_c{}^d = 0,
\end{align}
in complete analogy to the electromagnetic plane wave (see Section \ref{sec: GW non-Rieman}). This can be also seen as a consequence of the following properties that the gravitational ``polarization'' forms verify
\begin{align}
\dfk\wedge\star  \dfa^a = 0,\qquad \dfk\wedge\star  \dfa_a{}^b = 0,\\
k^a q_{ab} = 0,\qquad k_a \dfa^a = 0,\qquad k_b \dfa_a{}^b = 0,\qquad k^a \dfa_a{}^b = 0.
\end{align}

In addition to this discussion, notice that the following conditions are satisfied in general (also for $\dfU\neq0$):
\begin{equation}
k^a \dfQ_{ab} = 0,\quad k_a \dfT^a = 0,\quad k_b \dfR_a{}^b = 0,\quad k^a \dfR_a{}^b = 0,\label{eq:kTRW}.
\end{equation}

\subsection{Transversal geometry. Useful properties} 
Although the geometry of the transversal 2-space spanned by $x^A = (x^2, x^3)$ is fairly simple, it is convenient to describe it explicitly. This space is a flat Euclidean space with metric $\delta_{AB}$. As any other metric, it defines a canonical volume form and a Hodge star operator on the transversal space:
\begin{align}
  \svolf         &\coloneqq \frac{1}{2}\LCten_{AB} \cofr^A\wedge\cofr^B = \dex x^2\wedge \dex x^3 \,\\
  \sstar \cofr_A &\coloneqq \LCten_{AB} \cofr^B .
\end{align}
where $\LCten_{23} \equiv \epsilon_{23} = 1$. Notice that this identification between the Levi-Civita tensor and the Levi-Civita symbol can be done since $\sqrt{|\delta|}=1$.

It is not difficult to check that the canonical volume of the entire spacetime (associated to $g_{\mu\nu}$) can be rewritten as
\begin{equation}
  \cofr^0 \wedge \cofr^1 \wedge\cofr^2 \wedge\cofr^3 \equiv\qquad  \volfg = \dfk\wedge \dex v\wedge\svolf
\end{equation}

As a consequence of this, one can deduce
\begin{equation}
  \star \dfk = - \dfk \wedge \svolf \,,\label{eq: dualk}
\end{equation}
from which the following properties can be derived:
\begin{align}
  \star (\dfk\wedge\dfal) &= \dfk\wedge\sstar\dfal \,,\label{eq: kalpha}\\
  \star (\dfk\wedge\dfbe) &= (k^a\beta_a)\svolf \,, \\
  \star \dfal &= \cofr^0 \wedge\cofr^1 \wedge \sstar \dfal\,,\label{eq: star al}
\end{align}
where $\dfal= \alpha_A\cofr^A$ is an arbitrary transversal 1-form and $\dfbe= \beta_0\cofr^0 +\beta_1\cofr^1$ is an arbitrary non-transversal 1-form. If, in addition, the components of the transversal 1-form $\dfal$ only depend on the coordinates $u$ and $x^A$, the following relation holds:
\begin{equation}
  \dex(\dfk\wedge\dfal) = -\dfk\wedge\sdex\dfal \label{eq: dkalpha}\,.
\end{equation}

It also worth remarking that the dual of the transversal volume form is closed, i.e., $\dex \star \svolf = 0$. This is an immediate consequence of
\begin{equation}
  \star \svolf = \cofr^0\wedge \cofr^1 = \dex u \wedge \dex v\,.
\end{equation}

\subsection{Auxiliary decomposition of Omega}

For the irreducible decomposition of the curvature, it will be very useful to perform the following splitting of $\pmOm^a$: 
\begin{equation}\label{eq: Omm}
  \pmOm^a = \irrpmOm{1}^a + \irrpmOm{2}^a +\irrpmOm{4}^a ,
\end{equation}
which read explicitly: 
\begin{align}
\irrpmOm{1}^a &\coloneqq \frac{1}{2}\Big(\pmOm^a + \cofr^b \dint{\vfre^a} \pmOm_b - \cofr^a \dint{\vfre_b} \pmOm^b \Big),\label{eq: OM1}\\
\irrpmOm{2}^a &\coloneqq \frac{1}{2}\Big(\pmOm^a - \cofr^b \dint{\vfre^a} \pmOm_b \Big),\label{eq: OM2}\\
\irrpmOm{4}^a &\coloneqq \frac{1}{2}\,\cofr^a \dint{\vfre_b}\pmOm^b \,.\label{eq: OM4}
\end{align}

Let us now analyze in more detail the components of these 1-forms:
\begin{itemize}
  \item The objects $\irrpmOm{I}^A$ are purely transversal, i.e., they can be expanded as $\irrpmOm{I}^A = \irrpmOmc{I}^A{}_B \cofr^B$. The matrices $\irrpmOmc{I}^A{}_B $ correspond to, respectively, the symmetric traceless part, the skew-symmetric part and the trace of the $2\times 2$ matrix $\partial_B \pmW^A$:
  \begin{align}
    \irrpmOmc{1}^A{}_B &= \frac{1}{2} \Big(\partial_B\pmW^A + \partial^A \pmW_B - \delta^A_B \partial_C \pmW^C \Big), \label{eq: OMAB1}\\
    \irrpmOmc{2}^A{}_B &= \frac{1}{2} \Big(\partial_B\pmW^A - \partial^A \pmW_B \Big)\,,\label{eq: OMAB2}\\
    \irrpmOmc{4}^A{}_B &= \frac{1}{2} \delta^A_B \partial_C \pmW^C \,.\label{eq: OMAB4}
  \end{align}

  \item On the contrary, $\irrpmOm{I}^\mm$ are purely non-transversal 1-forms and are given by
  \begin{align}
    \irrpmOm{1}^\mm = - \, \irrpmOm{4}^\mm = -\frac{1}{2}\cofr^\mm \dint{\vfre_b}\pmOm^b \,, \qquad
    \irrpmOm{2}^\mm = 0 \,.  
  \end{align}
  Notice that $\irrpmOm{1}^\mm+\irrpmOm{2}^\mm+\irrpmOm{4}^\mm=0$, as it should be because $\pmOm^\mm=0$.

\end{itemize}

With all of this information, one can demonstrate the following properties of these 1-forms:
\begin{itemize}
  \item Antisymmetric part\hfill{}($\cofr_a\wedge\dfal^a= \alpha_{[ab]}\cofr^{ab}$)
  \begin{equation}
    \cofr_a\wedge\irrpmOm{1}^a=0 \,,\qquad \cofr_a\wedge\irrpmOm{2}^a=\cofr_a\wedge\pmOm^a \,,\qquad \cofr_a\wedge\irrpmOm{4}^a=0\,.
  \end{equation}

  \item Trace\hfill{}($\dint{\vfre_a}\dfal^a= \alpha_a{}^a$)
  \begin{equation}
    \dint{\vfre_a}\irrpmOm{1}^a=-\dint{\vfre_a}\pmOm^a \,,\qquad \dint{\vfre_a}\irrpmOm{2}^a=0 \,,\qquad \dint{\vfre_a}\irrpmOm{4}^a=2\dint{\vfre_a}\pmOm^a\,.
  \end{equation}

  \item Contraction of the external index  with $k_a$
  \begin{equation}\label{eq: kOM}
    k_a\irrpmOm{1}^a = - k_a \irrpmOm{4}^a = -\frac{1}{2} (\dint{\vfre_a}\pmOm^a) \dfk \,, \qquad k_a\irrpmOm{2}^a =0 \,.
  \end{equation}

  \item Contraction of the internal index with $k^b$\hfill{}($\dfk\wedge\star\dfal^a = \star(k^b\alpha_b{}^a)$)
  \begin{equation}
    \dfk\wedge\star\irrpmOm{1}^a = - \dfk\wedge\star\irrpmOm{4}^a = -\frac{1}{2} k^a \cofr_b\wedge\star\pmOm^b \,, \qquad \dfk\wedge\star\irrpmOm{2}^a =0 \,.
  \end{equation}

\end{itemize}

\subsection{Irreducible decompositions of T, Q and R}\label{sec: irreps ansatz}

Taking into account \eqref{eq: kW0} one can check, directly from \eqref{eq: torW}, that the trace and the totally symmetric part of the torsion vanish, i.e.,
\begin{equation}
  \dint{\vfre_a} \dfT^a = 0 \, \qquad \cofr_a\wedge \dfT^a=0\,.
\end{equation}
This implies that, respectively, the second and third irreducible parts of the torsion are zero. In other words, the torsion is purely tensorial, so its irreducible decomposition reads:
\begin{align}
  \irrdfT{1}^a &= \dfT^a = -\dfk \wedge k^a\,\dfThe \,,\\
  \irrdfT{2}^a &= 0 \\
  \irrdfT{3}^a &= 0 \,.
\end{align}

Similarly it can be easily proven that the two traces of the nonmetricity \eqref{eq: nonW} are zero
\begin{equation}
  \dint{\vfre^a} \dfQNoTr_{ab} = 0 \,, \qquad \dfQ_a{}^a=0\,.
\end{equation}
Hence, only the first and the second irreducible components survive. The decomposition now is:
\begin{align}
  \irrdfQ{1}_{ab} &= -\frac{4}{3}\dfk k_{(a} \pW_{b)} -\frac{2}{3}k_a k_b\pW_c\cofr^c +\frac{4}{3}\dfk k_{(a}U_{b)} + \frac{2}{3}k_a k_b\dfU, \,,\\
  \irrdfQ{2}_{ab} &= -\frac{2}{3}\dfk k_{(a} \pW_{b)} +\frac{2}{3}k_a k_b\pW_c\cofr^c -\frac{4}{3}\dfk k_{(a}U_{b)} + \frac{4}{3}k_a k_b\dfU. \\
  \irrdfQ{3}_{ab} &= 0 \\
  \irrdfQ{4}_{ab} &= 0 \,.
\end{align}

Finally, concerning the structure of the curvature $\dfR^{ab} = \dfW^{ab} + \dfZ^{ab}$, five irreducible pieces are trivial,
\begin{equation}
  \irrdfW{3}^{ab}=\irrdfW{5}^{ab}=\irrdfW{6}^{ab}=0\qquad \irrdfZ{3}^{ab}=\irrdfZ{5}^{ab}=0\,,
\end{equation}
whereas the rest can be expressed in terms of $\irrpmOm{I}^a$ and $\dfU$:
\begin{align}
  \irrdfW{1}^{ab} &= \dfk\wedge \irrmOm{1}^{[a}k^{b]}\,,\label{eq: curW1}\\
  \irrdfW{2}^{ab} &= \dfk\wedge \irrmOm{2}^{[a}k^{b]}\,,\label{eq: curW2}\\
  \irrdfW{4}^{ab} &= \dfk\wedge \irrmOm{4}^{[a}k^{b]}\,,\label{eq: curW4}\\
  \irrdfZ{1}^{ab} &= \frac{1}{2}\dfk\wedge\irrpOm{1}^{(a}k^{b)}+ \frac{1}{4}k^a k^b\, \cofr_c\wedge\pOm^c  +\frac{1}{2}\dfk\wedge k^{(a}\dint{e^{b)}} \dex \dfU + \frac{1}{2}k^a k^b \dex \dfU +\frac{1}{2}\dfk\wedge \pOm^{(a}k^{b)}\,,\\
  \irrdfZ{2}^{ab} &= \frac{1}{2}\dfk\wedge \irrpOm{2}^{(a}k^{b)} -  \frac{1}{4}k^a k^b \cofr_c\wedge\pOm^c-\frac{1}{2}\dfk\wedge k^{(a} \dint{e^{b)}} \dex \dfU + \frac{1}{2}k^a k^b \dex \dfU\,,\\
  \irrdfZ{4}^{ab} &= \frac{1}{2}\dfk\wedge \irrpOm{4}^{(a}k^{b)}\label{eq: ZW4}\,.
\end{align}

Notice that although some of the irreducible parts $\irrpmOm{I}^a$ are not orthogonal to the wave covector (see \eqref{eq: kOM}), all irreducible parts of the curvature, nonmetricity and torsion satisfy
\begin{equation}
k_a \irrdfW{I}^{ab} = 0 \,,\qquad k_a \irrdfZ{I}^{ab} = 0 \,,\qquad
k^a \irrdfQ{I}_{ab} = 0 \,, \qquad k_a\irrdfT{I}^a = 0\,,\label{eq: kWZQ}
\end{equation}
in complete agreement with \eqref{eq:kTRW}.

\section{Equations of motion}

\subsection{Gravitational momenta}

According to what we presented in Section \ref{subsec: qMAG action}, the gravitational momenta can be written (in the absence of the odd sector):
\begin{equation}
  \dfH[g]{}^{ab}=\frac{2}{\kappa}\dfm{}^{ab}\,, \qquad \dfH[\vartheta]{}_a = \frac{1}{\kappa}\dfh{}_{a}\,, \qquad \dfH[\omega]{}^a{}_b= \frac{1}{\kappa}\left(-\frac{1}{2}a_0\star\cofr^{a}{}_{b}+\lrho\dfh{}^a{}_b\right)\,,\label{eq: GMom expan}
\end{equation}
where the  $\dfm{}^{ab}$, $\dfh{}_a$ and $\dfh{}_{ab}$ are given by \eqref{eq: GMom g}, \eqref{eq: GMom e} and \eqref{eq: GMom w}, respectively. If we impose the nullity of the trivial irreducible components of our Ansatz, i.e.,
\begin{equation}
  \irrdfW{3}^{ab}=\irrdfW{5}^{ab}=\irrdfW{6}^{ab} = 0 \,,\qquad \irrdfZ{3}^{ab} = \irrdfZ{5}^{ab} =0 \,, \nonumber
\end{equation}
\begin{equation}
  \irrdfQ{3}_{ab} = \irrdfQ{4}_{ab} =0 \,, \qquad \irrdfT{2}^a =\irrdfT{4}^a = 0\,,
\end{equation}
we get
\begin{align}
  \dfm{}^{ab}  &=\star\Big[ b_1\irrdfQ{1}{}^{ab}+b_2\irrdfQ{2}{}^{ab} +c_1 \dint{\vfre^{(a}}\dfT^{b)}\Big], \label{eq: GMom g GW}\\[2mm]
  \dfh{}_a     &=\star\Big[a_1\dfT_{a}-c_1 \irrdfQ{2}{}_{ab}\wedge\cofr^b\Big] \,, \label{eq: GMom e GW}\\[2mm]
  \dfh{}_{(ab)}&=\star\Big[z_1\irrdfZ{1}{}_{ab}+(z_2-v_2)\irrdfZ{2}{}_{ab}+(z_4+2v_4)\irrdfZ{4}{}_{ab} \nonumber \\
             & \quad\qquad +\frac{v_2}{2}\cofr_c \wedge(\dint{\vfre_{(a}}\irrdfW{2}{}^{c}{}_{b)})+ \frac{v_4}{2}  \cofr_{(a|} \wedge(\dint{\vfre_{|c|}}\irrdfW{4}{}^c{}_{|b)})\Big], \label{eq: GMom w1 GW}\\[2mm]
  \dfh{}_{[ab]}&=\star\Big[w_1\irrdfW{1}{}_{ab}+w_2 \irrdfW{2}{}_{ab}+w_4\irrdfW{4}{}_{ab} \nonumber\\
             & \quad\qquad +\frac{v_2}{2} \cofr_c\wedge(\dint{\vfre_{[a}}\irrdfZ{2}{}^c{}_{b]})+ \frac{v_4}{2} \cofr_{[a}\wedge (\dint{\vfre_{|c|}}\irrdfZ{4}{}^c{}_{b]})\Big]. \label{eq: GMom w2 GW}
\end{align}
In view of \eqref{eq: kWZQ}, we verify the orthogonality properties for the duals:
\begin{equation}
  k_a \dfm^{ab} = 0\,,\qquad k^a \dfh_a = 0\,, \qquad k^a \dfh_{[ab]} = 0\,, \qquad k^a \dfh_{(ab)} = 0\,.\label{eq: kmhh}
\end{equation}

\subsubsection*{Coframe momentum}
From \eqref{eq: GMom e GW} it is straightforward to prove that
\begin{equation}
  \boxed{\dfh_a  = -\star \big[ k_a \dfk\wedge\dfXi\big] } \,,\label{eq: GWHa1}
\end{equation}
where we have introduced the following transversal 1-form
\begin{equation}
  \dfXi \coloneqq a_1\,\dfThe - c_1 \pW_A \cofr^A - 2c_1 \dfU \,.
\end{equation}

\subsubsection*{Metric momentum}
If we introduce
\begin{align}
  \mu_a   &\coloneqq c_1 \dint{\vfre_a}\dfThe - \frac{4b_1 + 2b_2}{3} \pW_a + \frac{4(b_1 - b_2)}{3} U_a \,,\label{eq: mua}\\
  \check\dfmu &\coloneqq c_1 \dfThe + \frac{2b_1 - 2b_2}{3} \pW_b\cofr^b - \frac{2b_1 + 4b_2}{3} \dfU \,,\label{eq: mu}
\end{align}
the equation \eqref{eq: GMom g GW} can be expressed
\begin{equation}
  \boxed{\dfm^{ab} = \star \big[ \dfk \wedge\mu^{(a} k^{b)} - k^a k^b \check\dfmu \big]} \,.\label{eq: GWMab1}\\
\end{equation}

\subsubsection*{Connection momentum}
Finally, by substituting the irreducible components of the curvature for our Ansatz \eqref{eq: curW2}-\eqref{eq: ZW4} in \eqref{eq: GMom w1 GW} and \eqref{eq: GMom w1 GW}, we obtain, respectively, 
\begin{equation}
  \boxed{\dfh_{[ab]}= \star \big[ \dfk\wedge\wUps_{[a}k_{b]}\big]\,,\qquad\qquad \dfh_{(ab)}= \star\big[ \dfk\wedge\zUps_{(a}k_{b)} + \frac{1}{4} k_a k_b \dfvarUps\big] } \,,
\end{equation}
where we have introduce some auxiliary objects
\begin{align}
  \wUps_a &\coloneqq w_1 \irrmOm{1}_a + w_2 \irrmOm{2}_a + w_4 \irrmOm{4}_a  - \frac{v_2}{2}\irrpOm{2}_a + \frac{v_4}{2}\irrpOm{4}_a + \frac{v_2}{2}\dint{\vfre_a}\sdex \dfU,\label{eq: wups}\\[2mm]
  \zUps_a &\coloneqq z_1 \irrpOm{1}_a + \frac{z_1 + z_2 - v_2}{2}\irrpOm{2}_a + \frac{z_1 + z_4 + 2v_4}{2}\irrpOm{4}_a \nonumber\\
          & \quad\qquad -\frac{v_2}{2}\irrmOm{2}_a + \frac{v_4}{2}\irrmOm{4}_a + \frac{z_1 - z_2 + v_2}{2} \dint{\vfre_a}\sdex \dfU,\label{eq: zups}\\[2mm]
\dfvarUps &\coloneqq \cofr^c\wedge \big[(z_1 - z_2 + v_2)\pOm_c + v_2\mOm_c\big] + 4z_1\dfk\wedge\dot{\dfU} + 2(z_1 + z_2 - v_2)\sdex \dfU.\label{eq: ups}
\end{align}
These objects have the following properties:
\begin{itemize}
\item  The 1-forms $\wUps_A$ and $\zUps_A$ are purely transversal. 
\item  The 1-forms $\wUps_\mm$ and $\zUps_\mm$ are purely non-transversal and can be expressed as:
\begin{equation}
  \wUps_\mm = \frac{1}{4}\wphi \cofr_\mm \,\qquad  \zUps_\mm = \frac{1}{4}\zphi \cofr_\mm \label{Lwza}
\end{equation}
where we have introduced the scalar functions
\begin{align}
  \wphi &\coloneqq 2(w_4 - w_1)\,\dint{\vfre_b} \mOm^b + v_4 \dint{\vfre_b} \pOm^b \,,\label{eq: phiw}\\
  \zphi &\coloneqq (z_4 - z_1 + 2v_4)\,\dint{\vfre_b} \pOm^b + v_4\,\dint{\vfre_b}\mOm^b \,.\label{eq: phiz}
\end{align}
\end{itemize}

If we take into account these properties it can be shown that
\begin{align}
  \dfh{}_{[AB]}&=0\,, &\dfh{}_{[\mm\nn]}&=0\,, &\dfh{}_{[A\mm]}  &=\frac{1}{2}\dfk\wedge\sstar\wUps_A k_\mm\,, \label{eq: starcompH1}\\
   \dfh{}_{(AB)}&=0\,, &\dfh{}_{(\mm\nn)}&=\frac{1}{4}k_\mm k_\nn \left(\star\oUps-\zphi\svolf\right)\,,&\dfh{}_{(A\mm)}  &=\frac{1}{2}\dfk\wedge\sstar\zUps_A k_\mm \,. \label{eq: starcompH2}
\end{align}

\subsection{Dynamical equations and reduction of the derivatives}

If we use Theorem \ref{th:EoMqMAG}, the gravitational equations of motion can be recast as
\begin{align}
\kappa\frac{\delta S}{\delta\cofr^{a}} & =\frac{a_{0}}{2}\dfR^{bc}\wedge\star\cofr{}_{bca} -\Dex\dfh_{a}\,, \\
\kappa\frac{\delta S}{\delta\dfom_a{}^{b}} & =-\cofr^a\wedge\dfh_{b} -2\dfm^a{}_{b} +\frac{a_{0}}{2} \left(\dfT^c\wedge\star\cofr{}^a{}_{bc} +\dfQ^{ac}\wedge\star\cofr{}_{cb}\right)-\lrho \ \Dex\dfh^a{}_b \,.
\end{align}
Notice that we have already dropped $\dfq_a$ from the first equation and the term proportional to $\dfQ$ from the second one, since they vanish for our particular Ansatz.

For the derivative term of the coframe equation we use the structure of $\dfh_a$ in \eqref{eq: GWHa1} to show that
\begin{equation}
  \dfom_a{}^b\wedge \dfh_b = \dfom_a{}^b k_b \wedge \star(...) =0,
\end{equation}
which implies
\begin{equation}
  \Dex\dfh_a = \dex \dfh_a \,.
\end{equation}

Finally, we have to deal with $\Dex\dfh^a{}_b$. We start by lowering the index $a$ in the whole equation, so the derivative term becomes
\begin{equation}
  g_{ca}\Dex\dfh^c{}_b = \Dex\dfh_{ab} - \dfQ_{ca}\wedge\dfh^c{}_b\,.
\end{equation}
The last term vanishes if we take into account the orthogonality $k_c\dfh^c{}_b=0$ and the decomposition \eqref{eq: starcompH1}-\eqref{eq: starcompH2}. Similarly to the case of $\dfh_a$ it can be shown that
\begin{equation}
  \dfom_a{}^b\wedge \dfh_{[bc]} = 0,\qquad \dfom_a{}^b\wedge \dfh_{(bc)} = 0 \,,
\end{equation}
so,
\begin{align}
  g_{ca}\Dex\dfh^c{}_b = \dex\dfh_{ab}\,.
\end{align}

These three results completely eliminate the exterior covariant derivative from the equations of motion and reduce it to the ordinary exterior derivative in some cases. The equations of motion after these substitutions read

\boxtheorem{
\begin{align}
\kappa\frac{\delta S}{\delta\cofr^{a}} & =\frac{a_{0}}{2}\dfR^{bc}\wedge\star\cofr{}_{bca} -\dex\dfh_{a}\,,\label{eq: EoM1 F} \\
\kappa g_{ac}\frac{\delta S}{\delta\dfom_c{}^b} & =-\cofr_a\wedge\dfh_b -2\dfm_{ab} +\frac{a_{0}}{2} (\star\cofr_{abc}\wedge\dfT^c +\dfQ_a{}^c\wedge\star\cofr{}_{cb})-\lrho \ \dex\dfh_{ab} \,.\label{eq: EoM2 F}
\end{align}
~
}

\newpage
\subsection{Coframe equation}

Let us now focus on the coframe equation \eqref{eq: EoM1 F}. The term with the curvature can be expressed
\begin{equation}
  \dfR_{[bc]}\wedge\star\cofr^{abc} \overeq{\eqref{eq: WcurvW}} \star\cofr_{abc}\wedge \dfk\wedge \mOm^b k^c =...=  k_a \star \dfk \, \dint{\vfre_b} \mOm^b \overeq{\eqref{eq: dualk}} -k_a \dfk \wedge \svolf\, \dint{\vfre_b} \mOm^b \,.
\end{equation}
In addition, according to \eqref{eq: GWHa1},
\begin{equation}
  \dfh_a = -k_a \star (\dfk \wedge\dfXi) \overeq{\eqref{eq: kalpha}} -k_a \dfk \wedge\sstar\dfXi\,,  
\end{equation}
and we can evaluate the exterior differential 
\begin{equation}
  \dex\dfh_a = -k_a \dex (\dfk \wedge\sstar\dfXi) \overeq{\eqref{eq: dkalpha}} k_a  \dfk \wedge\sdex\,\sstar\dfXi  \,.
\end{equation}

In conclusion, the equation of the coframe \eqref{eq: EoM1 F} can be recast as
\begin{equation}
- k_a \dfk \wedge \Big[ \frac{a_0}{2} \svolf\,\dint{\vfre_b} \mOm^b + \sdex\, \sstar \dfXi\Big] = 0 \,.
\end{equation}
Notice that the free index $a$ appears in a global $k_a$, so it can be removed (only the component ``$\mm$'' of the equation is non vanishing).

\subsection{Connection equation}\label{mag2}

Now we are going to simplify the terms in \eqref{eq: EoM2 F}. By evaluating the torsion \eqref{eq: torW} we can easily rewrite the first term as
\begin{equation}
  \frac{a_0}{2} \star\cofr_{abc}\wedge \dfT^c \ = \ a_0 \star \dfk (\dint{\vfre_{[a}}\dfThe) k_{b]} \overeq{\eqref{eq: dualk}} - a_0 \dfk\wedge \svolf \, (\dint{\vfre_{[a}}\dfThe) k_{b]},
\end{equation}
For the following two terms we substitute the nonmetricity \eqref{eq: nonW} and use $\cofr_a\wedge\dfh_b = - \star (\dint{\vfre_a} \dfh_b)$
\begin{align}
  \frac{a_0}{2} \dfQ_a{}^c\wedge\star\cofr_{cb} - \cofr_a\wedge \dfh_b
  &~~~=~~~ \star\dfk \nu_a k_b - k_a k_b \star \check\nu + a_0 \star \dfk k_a U_b, \\
  &\overeq{\eqref{eq: dualk}} - \dfk\wedge\svolf (\nu_a k_b + a_0 k_a U_b) - k_a k_b \star \check\dfnu \,,
\end{align}
where
\begin{align}
  \nu_a       &\coloneqq a_1\dint{\vfre_a}\dfThe - (\frac{a_0}{2} + c_1)\pW_a - 2c_1 U_a \,, \\
  \check\dfnu &\coloneqq a_1\dfThe  + (\frac{a_0}{2} - c_1)\pW_a \cofr^a - 2c_1 \dfU \,.
\end{align}

With all of this and after substituting \eqref{eq: GWMab1}, the equation of motion of the connection \eqref{eq: EoM2 F} becomes
\begin{align}
\dfk\wedge \svolf \Big[- a_0  \, (\dint{\vfre_{[a}}\dfThe) k_{b]}- (\nu_a k_b + a_0 k_a U_b) + 2\mu_{(a} k_{b)} \Big] 
 +k_a k_b \star( - \check\dfnu+ 2 \check\dfmu)
 - \lrho \,\dex\dfh_{ab} =0 \,.
\end{align}
From now on we are going to work separately with the symmetric and antisymmetric parts of this equation which are, respectively,
\begin{align}
\dfk\wedge \svolf \Big[- (\nu_{(a} k_{b)} + a_0 k_{(a} U_{b)}) + 2\mu_{(a} k_{b)} \Big] 
 +k_a k_b \star( - \check\dfnu+ 2 \check\dfmu)
 - \lrho \,\dex\dfh_{(ab)} &=0 \label{eq: EoM2s F}\,,\\
\dfk\wedge \svolf \Big[- a_0  \, (\dint{\vfre_{[a}}\dfThe) k_{b]}- (\nu_{[a} k_{b]} + a_0 k_{[a} U_{b]})  \Big]  - \lrho \,\dex\dfh_{[ab]} &=0 \label{eq: EoM2a F}\,.
\end{align}

Finally, the derivative terms read
\begin{align}
\dex \dfh_{[ab]} &= -\dfk \wedge \sdex\,\sstar \wUps_{[a}k_{b]},\label{eq: dha}\\
\dex \dfh_{(ab)} &= \dex \star (\dfk\wedge \zUps_{(a}) k_{b)} + \frac{1}{4} k_a k_b \dex \star \oUps \,.\label{eq: dhS}
\end{align}
Notice that in \eqref{eq: dhS} we have the full differential instead of the transversal one. Its evaluation is somewhat nontrivial. 

Now that we have all of the ingredients, we are going to specialize to the subsets of indices: $a = (\mm, A)$, where $\mm = 0,1$ and $A = 2,3$. For this purpose, it is important to keep in mind \eqref{eq: starcompH1} and \eqref{eq: starcompH2}.  Regarding the antisymmetric equation \eqref{eq: EoM2a F}, only the component ``$[A\mm]$'' is not trivial:
\begin{equation}
\dfk\wedge\Big[\svolf (a_0 \dint{\vfre_A} \dfThe + \nu_A - a_0 U_A ) - \lrho \ \sdex\,\sstar \wUps_A\Big] k_\mm = 0.
\end{equation}
However, the symmetric part \eqref{eq: EoM2s F} encompasses two nontrivial components, ``$(\mm\nn)$'' and ``$(A\mm)$'', respectively,
\begin{align}
k_\mm k_\nn \Big[ \star(2\check\dfmu - \check\dfnu) - \frac{\lrho}{4} \ \dex (\star\oUps - \zphi \svolf)\Big] & = 0, \label{eq: EoM2s1 F2x}\\
\dfk\wedge\Big[\svolf (\nu_A - 2\mu_A + a_0 U_A) - \lrho \ \sdex\,\sstar\zUps_A\Big] k_\mm &= 0.
\end{align}

Finally, let us rewrite a little bit the equation \eqref{eq: EoM2s1 F2x}. Since from \eqref{eq: ups} we have
\begin{equation}\label{eq: ups decom}
  \oUps = \chi\,\svolf + \dfk\wedge\dfxi \,,
\end{equation}
with
\begin{align}
  \dfxi&\coloneqq 4z_1\dot{\dfU}\,,\\
  \chi&\coloneqq \LCten^{AB}\partial_A \big[(z_1 - z_2 + 2v_2)\delta_{CB}W^C + (z_1 - z_2)\delta_{CB}V^C + 2(z_1 + z_2 - v_2) u_B\big] \,. \label{eq: upschi}
\end{align}
By using \eqref{eq: star al} and \eqref{eq: kalpha}, 
\begin{equation}
  \star\oUps = \chi \, \cofr^0\wedge\cofr^1 + \dfk\wedge\sstar\dfxi \,,
\end{equation}
which implies
\begin{equation}\label{eq: dups decom}
  \dex \star \oUps = \cofr^0\wedge\cofr^1\wedge\sdex\chi - \dfk\wedge \sdex \, \sstar \dfxi.
\end{equation}
As a result, with the help of \eqref{eq: star al} and \eqref{eq: dups decom} we recast \eqref{eq: EoM2s1 F2x} into
\begin{equation}
  \cofr^0\wedge\cofr^1\wedge\Big[\sstar(2\check\dfmu - \check\dfnu) - \frac{\lrho}{4}\sdex \chi\Big] + \frac{\lrho}{4}\dfk\wedge\Big[\svolf\partial_u \zphi + \sdex\, \sstar \dfxi\Big] = 0.
\end{equation}

\newpage
\subsection{Final set of equations in components}
From the last two subsections we have obtained the following system of equations:

\boxtheorem{
\begin{align}
  &(\text{EoM}~\cofr^a)_\mm  & 
-  \dfk \wedge \Big[ \frac{a_0}{2} \svolf\,\dint{\vfre_b} \mOm^b + \sdex\,\sstar \Xi\Big] &=0 \label{eq: EoM1 F2} \,,\\
  &(\text{EoM}~\dfom_a{}^b)_{[A\mm]} & \dfk\wedge\Big[\svolf (a_0 \dint{\vfre_A} \dfThe + \nu_A - a_0 U_A ) - \lrho \ \sdex\, \sstar \wUps_A\Big] &=0 \,,\label{eq: EoM2a F2}\\
  &(\text{EoM}~\dfom_a{}^b)_{(\mm\nn)} &  \cofr^0\wedge\cofr^1\wedge\Big[\sstar(2\check\dfmu - \check\dfnu) - \frac{\lrho}{4}\sdex \chi\Big] + \frac{\lrho}{4}\dfk\wedge\Big[\svolf\partial_u \zphi + \sdex\, \sstar \dfxi\Big]&=0 \,,\label{eq: EoM2s1 F2}\\
  &(\text{EoM}~\dfom_a{}^b)_{(A\mm)} & \dfk\wedge\Big[\svolf (\nu_A - 2\mu_A + a_0 U_A) - \lrho \ \sdex\sstar\zUps_A\Big] &=0 \,.\label{eq: EoM2s2 F2}
\end{align}
}

The final steps before the exploration of the space of solutions are: firstly, expanding all of the objects in these equations (undoing all of the definitions) in terms of our basic variables, which we will fix to be
\begin{equation}
  \{H, W^A, V^A, U_A\}
\end{equation}
(with the indices in that position); and, secondly, extracting the components of these equations, i.e., eliminating the differential forms and expressing them as scalar differential equations.

Since the metric $g_{ab}$ is going to disappear from the equations and will be split into $H$ and $\delta_{AB}$, it is useful to introduce the following notation for the basic vectorial variables with the indices raised/lowered by the transversal flat metric (to be precise, its part with signature $(+,+,+)$)\footnote{Let us insist on this detail to avoid confusion. In the previous chapter, we have been calling ``transversal metric'' to the transversal part of the metric, i.e. $\sobj{g}{}_{AB}$. But this is equal to $-\delta_{AB}$.}
\begin{equation}
  \uW{A} \coloneqq \delta_{AB}W^B\,,\qquad \uV{A} \coloneqq \delta_{AB}V^B\,,
  \qquad \uU{A} \coloneqq \delta^{AB}U_B\, ,
\end{equation}
as well as the differential operator
\begin{equation}
  \upartial{A} \coloneqq \delta^{AB}\partial_B\,.
\end{equation}
This convention is extremely important when we recast the 4-dimensional expressions in the formulas \eqref{eq: EoM1 F2}-\eqref{eq: EoM2s2 F2} into the 2-dimensional transversal ones. In particular, one should be always careful with $\partial^a$, $W_a$, $V_a$, and $U^a$, when we specialize to $a = A$, since then
\begin{equation}
  \partial^A = - \upartial{A}\,,\qquad W_A = -\uW{A}\,,\qquad V_A = -\uV{A}\,,\qquad U^A = -\uU{A}\,.
\end{equation}
It is also convenient to introduce the transversal Laplacian
\begin{equation}
  \sDelta \coloneqq \delta^{AB}\partial_A\partial_B\qquad\qquad (=\partial_A\upartial{A}=-\partial_A\partial^A )\,.
\end{equation}

In components, the equations \eqref{eq: EoM1 F2}, \eqref{eq: EoM2a F2} and \eqref{eq: EoM2s2 F2} read
\begin{align}
  0 &= -\frac{a_1}{2} \sDelta H + \Big(\frac{a_0}{2} - c_1 +   a_1\Big)\partial_A W^A  - \Big(\frac{a_0}{2} + c_1\Big)\partial_AV^A - (a_1 - 2c_1)\partial_A\uU{A} \,,\label{eq: EoM comp1} \\[4mm]
  0 &= \frac{a_0 + a_1}{2}\partial_AH + \Big(c_1 - \frac{a_0 + 2a_1}{2}\Big)\uW{A} + \Big(\frac{a_0}{2} + c_1\Big)\uV{A}  + (a_1 - 2c_1) U_A\nonumber\\
    &\qquad-\frac{\lrho}{4}\Big[ 2w_1\sDelta\,\uW{A} - 2w_1\sDelta\,\uV{A} + (2w_4 + v_4)\partial_A\partial_BW^B + (-2w_4 + v_4)\partial_A\partial_BV^B \Big]\nonumber\\
    &\qquad-\frac{\lrho}{4}\LCten_{AB} \upartial{B} \Big\{ \LCten^{CD}  \Big[(- 2w_2 + v_2)\partial_C\uW{D} + (2w_2 + v_2) \partial_C\uV{D} - 2v_2 \partial_C U_D \Big]\Big\} \,,\label{eq: EoM comp2} \\[4mm]
  0 &= \frac{a_1 - 2c_1}{2}\partial_AH + \Big(\frac{a_0}{2}-a_1+3c_1-\frac{8b_1 + 4b_2}{3} \Big)\uW{A} \nonumber\\
    & \qquad\qquad+ \Big( \frac{a_0}{2}+c_1-\frac{8b_1 + 4b_2}{3}\Big) \uV{A}  + \Big(a_0+a_1-4c_1-\frac{8b_1-8b_2}{3}\Big) U_A\nonumber\\
    &\qquad - \frac{\lrho}{4}\Big[2z_1\sDelta\,\uW{A} + 2z_1\sDelta\,\uV{A} + (z_1 + z_4 + 3v_4)\partial_A\partial_BW^B + (z_1 + z_4 + v_4)\partial_A\partial_BV^B \Big]\nonumber\\
    &\qquad-\frac{\lrho}{4}\LCten_{AB} \upartial{B}\Big\{ \LCten^{CD}  \Big[(- z_1 - z_2 + 2v_2)\partial_C\uW{D} - (z_1 + z_2)\partial_C\uV{D}
- 2(z_1 - z_2 + v_2)\partial_C U_D \Big]\Big\} \,,\label{eq: EoM comp3}
\end{align}
whereas, the equation \eqref{eq: EoM2s1 F2} yields two equations:
\begin{align}
  0 &= \frac{2c_1 - a_1}{2}\partial_AH + \Big(\frac{a_0 + 2a_1}{2}-3c_1-\frac{4b_1 - 4b_2}{3}\Big)\uW{A} + \nonumber\\
    &\qquad\qquad+ \Big(\frac{a_0}{2}-c_1-\frac{4b_1 - 4b_2}{3} \Big) \uV{A}  + \Big(4c_1-a_1-\frac{4b_1 + 8b_2}{3}\Big) U_A\nonumber\\
    &\qquad+ \frac{\lrho}{4}\LCten_{AB} \upartial{B}\Big\{ \LCten^{CD}  \Big[(z_1 - z_2 + 2v_2)\partial_C\uW{D} + (z_1 - z_2)\partial_C\uV{D} + 2(z_1 + z_2 - v_2) \partial_C U_D \Big]\Big\} \,, \label{eq: EoM comp4}\\ 
  0 &= \partial_u \Big[(z_4 - z_1 + 3v_4)\partial_AW^A + (z_4 - z_1 + v_4)\partial_AV^A - 4z_1\partial_A\uU{A}\Big] \,.\label{eq: EoM comp5}
\end{align}
The total number of equations \eqref{eq: EoM comp1}-\eqref{eq: EoM comp4} is 7, which coincides with the number of unknown variables $H, W^A, V^A, U_A$; so it is expected that one can find the latter as functions of transversal coordinates $x^A$. The additional equation \eqref{eq: EoM comp5} does not make the system over-determined, since it merely fixes the dependence on $u$. 

\section{Particular solutions}

We are now in a position to solve the field equations \eqref{eq: EoM comp1}-\eqref{eq: EoM comp5}. Before studying the general case, we are going to check some particular ones.

\subsection{Riemannian gravitational waves}

The nonmetricity \eqref{eq: nonW} and the torsion \eqref{eq: torW} vanish when $\dfU = 0$, $\pW^a = 0$, and $\dfThe = 0$ which is realized for 
\begin{equation}
W^A = -\,V^A = \frac{1}{2}\delta^{AB}\partial_BH.\label{eq: notorW}
\end{equation}
Substituting this into our equations, we find that \eqref{eq: EoM comp4} is identically satisfied, the equation of motion of the coframe \eqref{eq: EoM comp1} becomes
\begin{equation}
  a_0 \sDelta H = 0\, ,\label{eq: notW1}
\end{equation}
whereas the rest reduce to 
\begin{align}
  v_4 \partial_u \sDelta H &= 0\,,\\
  \lrho(w_1 + w_4)\partial_A \sDelta H &= 0\,,\\
  \lrho v_4 \partial_A \sDelta H &= 0.
\end{align}

An immediate conclusion is that the well-known plane wave solution of GR, with the function $H$ satisfying the Laplace equation in the transversal space, is an exact solution of the general quadratic MAG model. Notice that this is consistent with earlier results on the torsion-free solutions in Poincar\'e gauge theory \cite{Obukhov2006b, ObukhovPonomarev1989, Obukhov2019}.

In fact, there is a very strong result that one can derive from here. According to the results in \cite{ObukhovVlachynskyEsser1997}, the Riemannian wave \eqref{eq: notorW}-\eqref{eq: notW1} is not just a particular solution but the only possible solution for the purely torsion + nonmetricity quadratic class of MAG models ($w_I = 0$, $z_I = 0$, $v_I = 0$), except for one special choice of the coupling constants:
\begin{equation}
a_0= -\,a_1 = {\frac {a_2}{2}} = 2a_3 = 4b_1 = -2b_2 = - 8b_3 = {\frac {8b_4}{3}} = 2b_5 = -2c_1 = c_2 = c_3 \,.
\end{equation}
This is indeed the choice that reduces the Lagrangian to just the Einstein-Hilbert term (see \eqref{eq:RtoROmega2}). Since the Lagrangian in that case is only metric dependent, the connection remains unfixed by the dynamics and, hence, Levi-Civita is not the only solution.

\subsection{(General) Teleparallel solutions}

In this subsection we consider the general teleparallel solutions, i.e., configurations satisfying the condition of zero curvature, $\dfR_a{}^b = 0$. In the framework of MAG, the torsion and the nonmetricity are nontrivial and can propagate, as long as the curvature is zero.

The curvature \eqref{eq: curW} vanishes if and only if
\begin{equation}
  \sdex W^a = 0\,,\qquad  \sdex V^a = 0 \,,\qquad \dex \dfU =0\,.
\end{equation}
These conditions tell that $W^A$ and $V^A$ are independent of the transversal coordinates, whereas $U_A$ is pure gradient and independent of $u$:
\begin{equation}
  W^a = W^a(u)\,,\qquad V^a = V^a(u)\,,\qquad U_A = \frac{1}{2} \partial_A \, \cU(x^B)\,. \label{eq: telehyp}
\end{equation}

Under these conditions, \eqref{eq: EoM comp5} is identically fulfilled, while the three equations \eqref{eq: EoM comp4}, \eqref{eq: EoM comp2} and \eqref{eq: EoM comp3} after a lengthy but straightforward derivation can be recast into the algebraic system
\begin{align}
  (a_0 - 4b_1)\,\varPhi_A &= 0 \,,\label{eq: telesys1}\\
  3(a_0 + 2c_1)\Theta_A + 2(a_0 + 2b_2)\,\varPsi_A &= 0 \,,\\
  2(a_0 + a_1)\,\Theta_A + (a_0 + 2c_1)\,\varPsi_A &= 0 \,,\label{eq: telesys3}
\end{align}
where we introduced the abbreviations
\begin{align}
  \Theta_A \equiv \dint{\vfre_A}\dfThe &= \frac{1}{2}\partial_AH - \uW{A} + U_A \,, \label{eq: auxvartele1}\\
  \varPhi_A &\coloneqq \uW{A} + \uV{A} + U_A \,,\\
  \varPsi_A &\coloneqq \uW{A} + \uV{A} - 2U_A \,.\label{eq: auxvartele3}
\end{align}
Besides, the equation of the coframe \eqref{eq: EoM comp1} is reduced to
\begin{equation}
  a_1 \sDelta H + (a_1 - 2c_1) \sDelta  \cU = 0.\label{eq: teleEoMe}
\end{equation}

If we write the previous system \eqref{eq: telesys1}-\eqref{eq: telesys3} in matrix form,
\begin{equation}
  \left(\begin{array}{ccc} 
    0 & (a_0 - 4b_1) & 0 \\
    2(a_0 + a_1) & 0 & (a_0 + 2c_1) \\
    3(a_0 + 2c_1) & 0 & 2(a_0 + 2b_2) \end{array}\right)
  \left(\begin{array}{c}\Theta_A \\ \varPhi_A \\ \varPsi_A\end{array}\right) =  0,\label{matrix}
\end{equation}
we conclude that a nontrivial solution exists whenever the determinant vanishes, i.e.
\begin{equation}
  (a_0 - 4b_1)\left[3(a_0 + 2c_1)^2 - 4(a_0 + a_1)(a_0 + 2b_2)\right] = 0 \,. \label{det}
\end{equation}
This must be seen as a restriction on the coupling constants of the general Lagrangian \eqref{eq: qMAGLeven}. Only the MAG models for which this condition holds, admit (nontrivial) teleparallel gravitational waves. 

Let us clarify what happens if the determinant is not zero (i.e. what we mean by ``trivial'' solution). In principle, the only possible solution is $\Theta_A = \varPhi_A = \varPsi_A = 0$. It is not difficult to check that this implies that the nonmetricity and the torsion are zero. Then, the connection is Levi-Civita. Therefore the curvature (which is zero by hypothesis) coincides with the one of the metric. The only geometry compatible with this condition is the Minkowski space. 

\subsection{Weitzenb{\"o}ck solutions (standard teleparallelism)}

The nonmetricity \eqref{eq: nonW} is zero if and only if
\begin{equation}
  \pW^a = 0\quad\Big(\Leftrightarrow~~W^a(u,x^B) = - V^a(u,x^B)\Big)\,,\qquad \dfU=0\,.
\end{equation}
Then, for the standard teleparallel setting we have
\begin{equation}
  - V^A=W^A=W^A(u)\,,\qquad \dfU=0
\end{equation}
which, in terms of the auxiliary variables \eqref{eq: auxvartele1}-\eqref{eq: auxvartele3}, corresponds to
\begin{equation}
  \Theta_A = \frac{1}{2}\partial_AH - \uW{A}\,,\qquad\varPhi_A = \varPsi_A = 0\,.
\end{equation}

The system \eqref{eq: telesys1}-\eqref{eq: telesys3} is considerably simplified
\begin{equation}
  (a_0 + 2c_1)\Theta_A = 0 \,,\qquad  (a_0 + a_1)\,\Theta_A =0\,. 
\end{equation}
The torsion is controlled by the value of $\Theta_A$, so a nontrivial solution only exists in a class of quadratic models restricted by the conditions
\begin{equation}
a_0 + a_1 = 0,\qquad 2 c_1 + a_0 = 0 \, \label{eq: WeiSolcond}.
\end{equation}

In addition, the equation of the coframe \eqref{eq: teleEoMe} becomes
\begin{equation}
  a_1 \sDelta\,H = 0 \qquad \xrightarrow{\eqref{eq: WeiSolcond}} \qquad a_0 \sDelta\,H = 0 \,.
\end{equation}
Then, the metric structure coincides with the solution of GR, whereas the torsion will be determined by the non-vanishing $\Theta_A$.

\subsection{Symmetric teleparallel solutions}

\emph{Symmetric teleparallel geometry} is characterized by the vanishing curvature and torsion, along with a nontrivial nonmetricity \cite{NesterYo1998, Adak2006, AdakKalay2006, BeltranHeisenbergKoivisto2018a, ConroyKoivisto2018, Hohmann2018, HohmannPfeifer2019}. The torsion \eqref{eq: torW} is zero under the conditions
\begin{equation}
  \dfThe=0 \quad \Big(\Leftrightarrow~~\tfrac{1}{2}\partial_AH(u,x^B) - \uW{A}(u,x^B) + U_A(u,x^B)=0 \Big)\,.
\end{equation}
This, together with the conditions \eqref{eq: telehyp} for the nullity of the curvature gives
\begin{equation}
  W^a = W^a(u)\,,\quad V^a = V^a(u)\,,\quad U_A = \frac{1}{2} \partial_A \, \cU(x^B)\,,\quad
  \partial_A (H+\cU) - 2\uW{A}=0  \,.
\end{equation}
If we take the derivative $\upartial{A}$ of the last condition, since $uW_A$ is independent of $x^B$, gives
\begin{equation}
  \sDelta(H+\cU)=0\,.
\end{equation}
This allows to rewrite the equation of the coframe \eqref{eq: teleEoMe} as
\begin{equation}
  c_1 \sDelta H = 0\,.
\end{equation}

The algebraic system \eqref{eq: telesys1}-\eqref{eq: telesys3} is simplified to 
\begin{equation}
  (a_0 - 4b_1)\varPhi_A = 0\,,\qquad (a_0 + 2b_2)\varPsi_A = 0\,,\qquad (a_0 + 2c_1)\,\varPsi_A = 0\,.
\end{equation}

Consequently, nontrivial symmetric teleparallel wave solutions exist when $\varPhi_A\neq0$ or $\varPsi_A\neq0$. Otherwise, solutions reduce to the flat Minkowski spacetime. Moreover, the general solutions (with nontrivial $\varPhi_A$ and $\varPsi_A$) are only allowed in those MAG models fulfilling
\begin{equation}
a_0 - 4b_1 = 0\,, \qquad a_0 + 2b_2 = 0\,,\qquad c_1 + \frac{a_0}{2} = 0\,.
\end{equation}

\section{General solutions}

\subsection{Potential-copotential decomposition and splitting of the equations}

The first step to find the general solution will be to perform an appropriate splitting of the two independent components contained in each of our vector variables: $W^A$, $V^A$ and $U_A$. Since they are defined in a flat 2-dimensional space, we can consider a decomposition of the type:
\begin{align}
  W^A &= \frac{1}{2}\left(\upartial{A}\cW  + \LCten^{AB}\partial_B\cWb \right)\,,\label{eq:Wpotdec}\\
  V^A &= \frac{1}{2}\left(\upartial{A}\cV  + \LCten^{AB}\partial_B\cVb \right)\,,\\
  U_A &= \frac{1}{2}\left(\partial_A\cU  + \LCten_{AB}\,\upartial{B}\,\cUb \right)\,.\label{eq:Upotdec}
\end{align}
Physically, the six new variables $\cW, \cV, \cU$ (\emph{potentials}) and $\cWb, \cVb, \cUb$ (\emph{copotentials}) are analogues of the well-known Hertz potentials in classical electrodynamics. The overline denotes the three parity-odd variables $\cWb, \cVb, \cUb$  to distinguish them from the parity-even variables $U, \cW, \cV, \cU$. These decompositions are quite practical because if we have
\begin{equation}
  F^A = \frac{1}{2}\left(\upartial{A}\mathcal{F} + \LCten^{AB}\partial_B\oddvar{\mathcal{F}} \right)\,,\label{eq: pot copot F}\\
\end{equation}
then one immediately shows that
\begin{equation}
  \partial_A F^A = \frac{1}{2} \sDelta \mathcal{F}\qquad\text{and}\qquad \LCten_{AB}\upartial{B} F^A = \frac{1}{2} \sDelta \oddvar{\mathcal{F}}\,.
\end{equation}

In our system of dynamical equations  \eqref{eq: EoM comp1}- \eqref{eq: EoM comp5}, there are two types of equations depending on the number of free indices. The equations \eqref{eq: EoM comp1} and \eqref{eq: EoM comp5} do not have free indices and depend only on the divergences of $W^A$, $V^A$ and $U_A$. Therefore, only the potentials appear on them. These equations can be recast, respectively, as
\begin{align}
  0&=-a_1\sDelta H + \Big(\frac{a_0}{2} - c_1 + a_1\Big)\sDelta\cW - \Big(\frac{a_0}{2} + c_1\Big)\sDelta\cV - (a_1 - 2c_1)\sDelta\cU \,,\label{eq8} \\[2mm]
  0&=\partial_u \Big[(z_4 - z_1 + 3v_4)\,\sDelta\,\cW  + (z_4 - z_1 + v_4)\,\sDelta\,\cV - 4z_1\,\sDelta\,\cU \Big] .\label{eq3}
\end{align}

Let us now focus on the other three equations \eqref{eq: EoM comp2}- \eqref{eq: EoM comp4}, which have a free transversal index. The idea that we are going to apply to each of them is the following: the entire equation can be seen as some transversal vector equal to zero, and such a vector admits a decomposition of the type \eqref{eq: pot copot F}. A vector decomposed in this way is zero if and only if its potential and copotential vanish up to a harmonic function. Then, the three equations are split into six (with one less derivative) with an arbitrary harmonic function as an inhomogeneous term. However, we have six variables in \eqref{eq:Wpotdec}-\eqref{eq:Upotdec} and the equations are linear, so we have enough freedom in general to absorb those harmonic functions by redefining the variables $\{\cW, \cV, \cU, \cWb, \cVb, \cUb\}$. After applying this procedure to \eqref{eq: EoM comp2} (component ``$[A,\mm]$'' of the equation of the connection) we get,
\begin{align}
  0 &= (a_0 + a_1) H + \Big(c_1 - \frac{a_0+2a_1}{2}\Big)\cW + \Big(\frac{a_0}{2}+c_1\Big)\cV +(a_1-2c_1)\cU \nonumber\\ 
    &\quad -\frac{\lrho}{4}\Big[(2w_1+2w_4+v_4)\sDelta\cW- (2w_1+2w_4-v_4)\sDelta\cV \Big] \,,\label{eq4}\\[4mm]
  0 &= \Big(c_1-\frac{a_0+2a_1}{2}\Big)\cWb + \Big(\frac{a_0}{2} + c_1\Big)\cVb + (a_1 - 2c_1)\cUb \nonumber\\ 
    &\quad -\frac{\lrho}{4} \Big[(2w_1+2w_2-v_2)\sDelta\cWb -(2w_1+2w_2+v_2)\sDelta\cVb +2v_2\sDelta\cUb \Big]\,.\label{eq5}
\end{align}
If we do the same for \eqref{eq: EoM comp3} (component ``$(A,\mm)$'' of the equation of the connection) the result is
\begin{align}
  0 &= (a_1 - 2c_1) H + \Big(\frac{a_0}{2}-a_1+3c_1-\frac{8b_1+4b_2}{3}\Big)\cW + \Big(\frac{a_0}{2}+c_1- \frac{8b_1+4b_2}{3}\Big)\cV \nonumber\\
    &~~ + \Big(a_0 + a_1 - 4c_1 - {\frac {8b_1 - 8b_2}{3}}\Big)\cU - \frac{\lrho}{4} \Big[(3z_1+z_4+3v_4)\sDelta\cW + (3z_1+z_4+v_4)\sDelta\cV \Big],\label{eq6}\\[4mm]
  0 &= \Big(\frac{a_0}{2}-a_1+3c_1-\frac{8b_1+4b_2}{3}\Big)\cWb +\! \Big(\frac{a_0}{2}+c_1- \frac{8b_1+4b_2}{3}\Big)\cVb +\! \Big(a_0+a_1-4c_1- {\frac {8b_1 - 8b_2}{3}}\Big)\cUb \nonumber\\
    &\quad -\frac{\lrho}{4}\Big[(3z_1+z_2-2v_2)\sDelta\cWb +(3z_1+z_2)\sDelta\cVb +2(z_1-z_2+v_2)\sDelta\cUb\Big]\,.\label{eq7}
\end{align}
And, finally, for \eqref{eq: EoM comp4} (part of the component ``$(\mm,\nn)$'' of the equation of the connection) we obtain the following two conditions:
\begin{align}
  0 &= (2c_1-a_1) H + \Big(\frac{a_0+2a_1}{2}-3c_1- \frac{4b_1-4b_2}{3}\Big)\cW \nonumber\\
    &\quad +\Big(\frac{a_0}{2}-c_1- \frac{4b_1-4b_2}{3} \Big) \cV +\Big(4c_1-a_1- \frac{4b_1 + 8b_2}{3}\Big)\cU\,,\label{eq1}\\[4mm]
  0 &= \Big(\frac{a_0+2a_1}{2}-3c_1- \frac{4b_1-4b_2}{3}\Big)\cWb+ \Big(\frac{a_0}{2}-c_1- \frac{4b_1-4b_2}{3} \Big)\cVb+ \Big(4c_1-a_1- \frac{4b_1 + 8b_2}{3}\Big)\cUb \nonumber\\
    &\quad -\frac{\lrho}{4} \Big[(z_1-z_2+2v_2)\sDelta\cWb+ (z_1-z_2)\sDelta\cVb+ 2(z_1+z_2-v_2)\sDelta\cUb\Big]\,.\label{eq2}
\end{align}
It is worth noticing that after this splitting, the even-parity variables and the odd-parity ones do not mix in the same equation. In other words, we have obtained two uncoupled systems of equations: one, with four equations, for the potentials and $H$, and another system of three equations for the copotentials. This is a consequence of the fact that the Lagrangian only contains even parity invariants. If we had included the odd parity sector, this separation does not generically happen. Good examples of the failure of this decoupling in the presence of odd-parity invariants can be found in \cite{Obukhov2017, BlagojevicCvetkovic2017b}, in the context of Poincar\'e gauge gravity.

The analysis of this system of equations is considerably simplified by a convenient choice of variables. The key to this is discovered when we substitute \eqref{eq:Wpotdec}-\eqref{eq:Upotdec} into (\ref{eq: auxvartele1})-(\ref{eq: auxvartele3}), which yields
\begin{align}
  \Theta_A &= \frac{1}{2}\left(\partial_A\cX_1 + \LCten_{AB}\upartial{B} \cXb_1\right),\label{thA1}\\
  \varPhi_A &= \frac{1}{2}\left(\partial_A\cX_2 + \LCten_{AB}\upartial{B} \cXb_2\right),\label{psA1}\\
  \varPsi_A &= \frac{1}{2}\left(\partial_A\cX_3 + \LCten_{AB}\upartial{B} \cXb_3\right),\label{phiA1}
\end{align}
where
\begin{align}
  \cX_1 &= H-\cW+\cU   \,,& \cXb_1 &= -\cWb+\cUb \,,     \nonumber\\
  \cX_2 &= \cW+\cV+\cU \,,& \cXb_2 &= \cWb +\cVb+\cUb \,,\nonumber\\
  \cX_3 &= \cW+\cV-2\cU\,,& \cXb_3 &= \cWb +\cVb-2\cUb\,.\label{eq: defXs}
\end{align}
We choose these, together with
\begin{equation}
  \cX_0 = \cW -\cV \label{eq: defX0}
\end{equation}
as the new set of variables. The inverse relations are
\begin{align}
  H   &= \frac{1}{2}\cX_0 + \cX_1 + \frac{1}{2}\cX_3\,,&& \nonumber\\
  \cW &= \frac{1}{2}\cX_0 + {\frac 13}\cX_2 + {\frac 16}\cX_3\,,&
  \cWb&= -\,\cXb_1 + {\frac 13}\cXb_2 - {\frac 13}\cXb_3\,,\nonumber\\
  \cV &= - \,\frac{1}{2}\cX_0 + {\frac 13}\cX_2 + {\frac 16}\cX_3\,,&
  \cVb&= \cXb_1 + {\frac 13}\cXb_2 + \frac{2}{3}\cXb_3\,,\nonumber\\
  \cU &= {\frac 13}\cX_2 - {\frac 13}\cX_3\,,&
  \cUb&= {\frac 13}\cXb_2 - {\frac 13}\cXb_3\,. \label{eq: invdefXs}
\end{align}
In the following two subsections we will provide the final form of our dynamical equations in terms of these new variables and analyze each sector separately.

\subsection{Analysis of the even-parity sector}
Substituting \eqref{eq: invdefXs} into the field equations, we recast \eqref{eq1}, \eqref{eq4}, \eqref{eq6}, \eqref{eq8}, \eqref{eq3}, respectively, into

\boxtheorem{
\begin{align}
  0 &= (2c_1 - a_1)\cX_1 +\frac{1}{3}(a_0 - 4b_1)\cX_2 + \Big[ -\frac{a_0}{2}-c_1+\frac{2}{3}(a_0 + 2b_2)\Big]\cX_3 \,,\label{eq: Xeq1}\\
  0 &= (a_0 + a_1)\cX_1 + \Big(\frac{a_0}{2}+c_1\Big)\cX_3 -\frac{\lrho}{4}\Big[2(w_1+w_4)\sDelta\cX_0 + \frac{2}{3}v_4 \sDelta\cX_2 + \frac{1}{3}v_4\sDelta\cX_3 \Big] \,,\label{eq: Xeq4} \\
  0 &= (a_1 - 2c_1)\cX_1 + \frac{2}{3}(a_0 - 4b_1)\cX_2 + \Big[\frac{a_0}{2}+c_1-\frac{2}{3}(a_0 +2b_2)\Big]\cX_3 \nonumber\\
    &\qquad- \frac{\lrho}{4}\Big[ v_4\sDelta\cX_0 + \frac{2}{3}(3z_1+z_4+2v_4)\sDelta\cX_2 +\frac{1}{3}(3z_1+z_4+2v_4) \sDelta\cX_3\Big]\,,\label{eq: Xeq6}\\
  0 &= \frac{a_0}{2}\sDelta\cX_0 -a_1\sDelta\cX_1 -c_1\sDelta\cX_3 \,,\label{eq: Xeq8}\\
  0 &= \partial_u\Big[v_4\sDelta\cX_0 + \frac{2}{3}(z_4-3z_1+2v_4)\sDelta\cX_2+ \frac{1}{3}(z_4+3z_1+2v_4)\sDelta\cX{}_3\Big]\,.\label{eq: Xeq3}
\end{align}
}

It is important to notice the existence of a very special solution of this system. If we take $\cX_2 = 0$ and $\cX_3  = 0$ or, equivalently
\begin{equation}
  \cU  = 0\,,\qquad \cW  = -\cV\,,
\end{equation}
the remaining system for $\cX_0$ and $\cX_1$ can be recast into
\begin{equation}
  v_4 \sDelta \cX_0 = 0,\quad (2c_1 - a_1)\cX_1 = 0, \quad 
-\frac{a_0}{2}\sDelta\cX_0 +a_1 \sDelta\cX_1 = 0,\quad
(a_0+a_1)\cX_1 - \frac{\lrho}{2}(w_1+w_4)\sDelta\cX_0 = 0.\label{eq: X2X3zero}
\end{equation}
Assuming $v_4\neq 0$ and $a_1\neq 2c_1$, we find $\sDelta\cX_0 = 0 =\cX_1$ and the four equations \eqref{eq: X2X3zero} are automatically satisfied. In summary, we have found the solution for which the even-parity variables fulfill
\begin{equation}
\sDelta H = 0,\qquad \cW  = -\,\cV  = H,\qquad \cU  = 0\,,
\end{equation}
and this describes the massless graviton mode.

In general, \eqref{eq: Xeq1}-\eqref{eq: Xeq3} is a second order differential systems with constant coefficients, for which solutions are sought in the form
\begin{equation}
  \cX_I = \cX_I^{(0)}(u) \ \eN^{\iN\,q_A\,x^A},\quad I = 0,1,2,3\,.\label{eq: XXeven}
\end{equation}
Under this Ansatz, the first four equations \eqref{eq: Xeq1}-\eqref{eq: Xeq8} become an algebraic system for the amplitudes $\cX_I^{(0)}$ which, in matrix form, reads
\renewcommand\arraystretch{1.5}
\begin{equation}
  \left(\begin{array}{cccc} 
    -\frac{a_0}{2} & a_1 & 0 & c_1 \\
    0 & 2c_1 - a_1 & {\frac 13}(a_0 - 4b_1) & - \frac{a_0}{2} - c_1 + \frac{2}{3}(a_0 + 2b_2) \\
    2(w_1 + w_4){\mathcal Q}^2 & a_0 + a_1 & \frac{2}{3}v_4{\mathcal Q}^2 & \frac{a_0}{2} + c_1 + {\frac 13}v_4{\mathcal Q}^2 \\
    {\mathcal Q}^2v_4 & 0 & a_0 - 4b_1 + \frac{2}{3}{\mathcal Q}^2\lambda_0 & {\frac 13}{\mathcal Q}^2\lambda_0 
  \end{array}\right)
  \left(\begin{array}{c}
    \cX_0^{(0)} \\ \cX_1^{(0)} \\ \cX_2^{(0)} \\ \cX_3^{(0)}
  \end{array}\right) = 0.\label{eq: algX}
\end{equation}
\renewcommand\arraystretch{1}
Here we are denoting
\begin{equation}
  \mathcal{Q}^2 \coloneqq \frac{\lrho}{4}q_Aq_B\delta^{AB}\,,\qquad \lambda_0 \coloneqq 3z_1 + z_4 + 2v_4\,.
\end{equation}
In order to have non-trivial solutions, the determinant of the $4\times 4$ matrix in \eqref{eq: algX} must vanish. It is interesting to notice that, although the matrix is $4\times 4$, the quantity  ${\mathcal Q}^2$ only appears in two rows of it. Then, the condition of vanishing determinant is a quadratic equation in the variable  ${\mathcal Q}^2$. In \cite{BlagojevicCvetkovic2017b}, it was proven that the solutions of such quadratic equation in PG are related to the masses of the spin-$2^\pm$ torsional modes of PG around Minkowski spacetime. Since we do not currently have a detailed analysis of the full spectrum of MAG,\footnote{Some preliminary steps are shown in Chapter \ref{ch:MAGspectrum}.} we cannot make a strong claim here. Nevertheless, we conjecture that something similar will happen in our MAG model.

Finally, the remaining field equation \eqref{eq: Xeq3} constrains the $u$-dependence of the even-parity amplitudes imposing the relation
\begin{equation}\label{Xeq30}
  v_4\,\partial_u\cX_0^{(0)} + \frac{2}{3}(z_4-3z_1+2v_4) \partial_u \cX_2^{(0)} + \frac{1}{3}(z_4+ 3z_1+2v_4)\,\partial_u\cX_3^{(0)} = 0\,. 
\end{equation}

\subsection{Analysis of the odd sector}

Similarly as we did for the even sector, the substitution of \eqref{eq: invdefXs} into the field equations allows to recast \eqref{eq2}, \eqref{eq5} and \eqref{eq6}, respectively, into
\begin{align}
  0 &= (2c_1-a_1)\cXb_1 +\frac{1}{3}(a_0-4b_1)\cXb_2 +\Big[-\frac{a_0}{2}-c_1+\frac{2}{3}(a_0+2b_2)\Big]\cXb_3 \nonumber\\
    &\qquad-\frac{\lrho}{4}\Big[-2v_2\sDelta\cXb_1 +\frac{4}{3}z_1\sDelta\cXb_2 -\frac{1}{3}(z_1+3z_2)\sDelta\cXb_3\Big] \,,\label{eq: Xeq2}\\ 
  0 &= (a_0 + a_1)\cXb_1 +\Big(\frac{a_0}{2}+c_1\Big)\cXb_3 -\frac{\lrho}{4}\Big[-4(w_1+w_2)\sDelta\cXb_1 -(2w_1+2w_2+v_2) \sDelta\cXb_3\Big]\,,\label{eq: Xeq5}\\
  0 &= (a_1-2c_1)\cXb_1 +\frac{2}{3}(a_0-4b_1)\cXb_2 +\Big[\frac{a_0}{2} +c_1-\frac{2}{3}(a_0+2b_2)\Big]\cXb_3 \nonumber\\
    &\qquad-\frac{\lrho}{4}\Big[2v_2\sDelta\cXb_1 +\frac{8}{3}z_1 \sDelta\cXb_2 + \frac{1}{3}(z_1 + 3z_2)\sDelta\cXb_3\Big]\,.\label{eq: Xeq7}
\end{align}

A peculiar property of this system is that the variable $\cXb_2$ is decoupled from the other two. To see this, let us construct an equivalent system of equations. We take the sum \eqref{eq: Xeq2} and \eqref{eq: Xeq7}, which yields an equation exclusively for $\cXb_2$. Then, we add \eqref{eq: Xeq2} and \eqref{eq: Xeq5} and use the latter to eliminate $\cXb_2$. The resulting equations, together with \eqref{eq: Xeq5}, read

\boxtheorem{
\begin{align}
  0 &= (a_0-4b_1)\cXb_2 - \lrho z_1 \sDelta \cXb_2\,, \label{eq: DX2}\\ 
  0 &= (a_0+2c_1)\cXb_1 + \frac{2}{3}(a_0 + 2b_2)\cXb_3\nonumber \\
    &\quad - \frac{\lrho}{4} \Biggl\{-2\left[2w_1+2w_2+v_2\right]\sDelta \cXb_1 - \Big[2w_1 + 2w_2+v_2+\frac{1}{3}(z_1+3z_2)\Big] \sDelta\cXb_3\Biggr\}\,,\label{eq: Xeq2a}\\
  0 &= (a_0+a_1)\cXb_1 +\Big(\frac{a_0}{2}+c_1\Big)\cXb_3 \nonumber\\
    &\quad -\frac{\lrho}{4}\Big[-4(w_1+w_2)\sDelta\cXb_1 -(2w_1+2w_2+v_2) \sDelta\cXb_3\Big]\,.\label{eq: Xeq52}
\end{align}
}

Again, we consider solutions of the following form:
\begin{equation}
\cXb_J = \cXb_J^{(0)}(u) \ \eN^{\iN \oddvar{q}_Ax^A},\quad J = 1,2,3, \label{eq: XXodd}
\end{equation}
where $\oddvar{q}_A$ is of course not necessarily equal to the $q_A$ introduced in \eqref{eq: XXeven}.

For the odd sector, the resulting algebraic system for the amplitudes $\cXb{}_I^{(0)}$ has the following matrix form:
\renewcommand\arraystretch{1.5}
\begin{equation}
  \left(\begin{array}{ccc} 
    0 & a_0 - 4b_1 + 4z_1\overline{\mathcal Q}{}^2 & 0 \\
    a_0 + 2c_1 - 2\overline{\mathcal Q}{}^2\lambda_2 & 0 & \frac{2}{3}(a_0 + 2b_2) - \overline{\mathcal Q}{}^2(\lambda_2 + \lambda_3)\\
    a_0 + a_1 - \overline{\mathcal Q}{}^2\lambda_1 & 0 & \frac{a_0}{2} + c_1 - \overline{\mathcal Q}{}^2\lambda_2
  \end{array}\right)
  \left(\begin{array}{c}
    \cXb{}_1^{(0)} \\ \cXb{}_2^{(0)} \\ \cXb{}_3^{(0)}
  \end{array}\right) =  0.\label{algXo}
\end{equation}
\renewcommand\arraystretch{1}
Here we introduced
\begin{equation}
  \oddvar{\mathcal Q}{}^2 \coloneqq \frac{\lrho}{4}\oddvar{q}_A\oddvar{q}_B\delta^{AB}\,,
\end{equation}
and the abbreviations
\begin{equation}
  \lambda_1 \coloneqq 4(w_1+w_2)\,, \qquad \lambda_2 \coloneqq 2(w_1+w_2)+v_2\,, \qquad\lambda_3 \coloneqq \frac{1}{3}(z_1+3z_2)\,.
\end{equation}

The system \eqref{algXo} shows that there are three propagating parity-odd modes which are determined by
\begin{equation}
  a_0-4b_1+4z_1\oddvar{\mathcal Q}{}^2 = 0,\qquad \mathcal{A}\oddvar{\mathcal Q}{}^4+ {\mathcal B}\oddvar{\mathcal Q}{}^2 + {\mathcal C}= 0\,,
\end{equation}
where we denoted the combinations of the coupling constants
\begin{align}
  \mathcal{A} &\coloneqq 2\lambda_2^2 +\lambda_1(\lambda_2+\lambda_3)\,,\\
  \mathcal{B} &\coloneqq -4\left(\frac{a_0}{2}+c_1\right)\lambda_2 +(a_0+a_1)(\lambda_2+\lambda_3) -\frac{2}{3}(a_0+2b_2)\lambda_1 \,,\\
  \mathcal{C} &\coloneqq 2\left(\frac{a_0}{2}+c_1\right)^2-\frac{2}{3}(a_0+2b_2)(a_0+a_1)\,.
\end{align}

Since the equation \eqref{eq3} (the only one that constrains the $u$-dependence) is independent of the copotentials, the parity-odd amplitudes $\cXb{}_I^{(0)} = \cXb{}_I^{(0)}(u)$,  $J = 1,2,3$, are arbitrary functions of $u$.

\subsection{``Pseudo-instanton'' solutions}

In the article \cite{Vassiliev2002}, Vassiliev considered the particular class of MAG models containing only the $(\irrW{I}^{ab})^2$ and $(\irrZ{I}^{ab})^2$ invariants. In other words, the considered Lagrangian is \eqref{eq: qMAGLeven} with $a_1 = a_2 = a_3 = 0$, $b_1 = \dots = b_5 = 0$, $c_1 = c_2 = c_3 = 0$ and also $v_I=0$.  In this context, Vassiliev defined ``pseudo-instantons'' as those solutions of the vacuum MAG field equations with {\it metric-compatible connection and purely irreducible curvature}. The idea of this section is to explore if solutions of this kind are allowed under our particular Ansatz for the gravitational variables \eqref{eq: cof0}-\eqref{eq: cof23} and \eqref{eq: conW}.

For our wave Ansatz, trivial nonmetricity $\dfQ_{ab} = 0$ means that $\dfU = 0$ and $W^A = - V^A$. In terms of potentials and copotentials, this condition is given by
\begin{equation}
  \cU  = \cUb  = 0\,, \qquad \cW  = -\cV \,, \qquad \cWb  = -\cVb \,,\label{eq: VasnoQ1}
\end{equation}
or equivalently
\begin{align}
  \cX_0 = 2\cW \,,\quad\cX_1 = H-\cW \,,\quad \cX_2 = \cX_3 = 0\,,\\ 
                       \cXb_1=- \cWb \,,\quad\cXb_2 =\cXb_3 = 0\,.\label{eq: VasnoQ2}
\end{align}

Consider the general curvature quadratic model, i.e., all of the parameters (including $a_0$) vanish except $w_I$, $z_J$, and $v_K$. Under the conditions \eqref{eq: VasnoQ2}, our MAG field equations reduce to
\begin{equation}
\sDelta\cW  = 0,\qquad \sDelta\cWb  = 0.\label{DWW}
\end{equation}
By virtue of the decomposition \eqref{eq:Wpotdec}, we find
\begin{equation}
  \partial_A W^A  = 0,\qquad \LCten^{AB}\partial_A\uW{B} = 0\,
\end{equation}
and, therefore, from \eqref{eq: OMAB1}-\eqref{eq: OMAB4} and \eqref{eq: VasnoQ1} we conclude that 
\begin{equation}\label{nOm}
\irrmOm{2}^a = \irrmOm{4}^a = 0,\qquad \pOm^a = 0  \quad (\text{i.e.,} \irrpOm{I}^a = 0 ~~\forall I).
\end{equation}
The resulting curvature is
\begin{equation}
  \dfR_a{}^b = \irrdfW{1}{}_a{}^b \,,
\end{equation}
which is purely irreducible and, hence, a ``pseudo-instanton'' solution in the sense of \cite{Vassiliev2002}. We have found that all metric-compatible solutions respecting our Ansatz are Weyl-pseudo-instantons (i.e. the curvature is equal to $\irrdfW{1}{}_a{}^b$ which is the generalization of the Weyl tensor to metric-affine geometry).

\section{Conclusions}

In this chapter we have considered the Lagrangian \eqref{eq: qMAGLeven}, which contains all possible parity-even linear and quadratic invariants in curvature, torsion and nonmetricity. The cosmological constant has been dropped for simplicity and the work has been focused on vacuum solutions (i.e., zero energy-momentum and  hypermomentum content). We assumed a pp-wave Ansatz for the coframe \eqref{eq: cof0}-\eqref{eq: cof23} and the linear connection \eqref{eq: conW}. The resulting curvature, torsion and nonmetricity exhibit very special structure (see Section \ref{sec: irreps ansatz}).

We have shown that the pp-wave solutions of GR and teleparallel gravity arise as special cases of our solutions.

We have extended the method used in \cite{Obukhov2017} to MAG under our particular Ansatz. We first performed a Helmholtz-like decomposition of each of the three transversal 2-vectors $W^A$, $V^A$ and $U_A$ into its scalar potential and copotential. The MAG dynamical equations, when written in terms of these six variables, together with the function $H$ from the pp-wave metric Ansatz, constitute a system of eight equations \eqref{eq8}-\eqref{eq2}. The equation \eqref{eq: EoM comp5} (equivalently \eqref{eq3} in terms of the potentials), constrains the dependence on the null coordinate $u$, whereas the remaining seven equations are of the Helmholtz or screened Laplace type and determine the seven unknown variables as functions of the transversal coordinates $x^A$.

The structure of the field equations is very special and allows to decouple the parity-even (the three potentials and $H$) and the parity-odd (the three copotentials) variables into two separate sets of equations. As we mentioned above, this decoupling is a consequence of the absence of parity-odd invariants in the action. We considered the standard exponential substitution \eqref{eq: XXeven} and \eqref{eq: XXodd} for the seven scalar variables, extracting the transversal dependence in a plane-wave-like factor. After substituting this into our dynamical equations, they can be rewritten as an algebraic system of equations for the wave amplitudes.

We took advantage of the previous result and focused on the type of models in which the Lagrangian is constructed only from the curvature invariants, whereas the quadratic in torsion and nonmetricity terms are set to zero. In this context we checked that ``pseudo-instanton'' solutions (in the sense of \cite{Vassiliev2002}) are allowed under our Ansatz and we found the general solution, which are of the Weyl-pseudo-instanton type.

It is important to stress that most of our results were obtained without or under very mild restrictions imposed on the parameters (coupling constants) of the action. Therefore, the resulting geometries are exact solutions for large families of MAG models of the type \eqref{eq: qMAGLeven}, and not for specific sets of parameters.

\subsubsection*{Limitations of this work/future directions}
Finally, we comment on the possible future directions arising from the limitations (hypothesis) of this work. One possibility is of course to consider another Ansatz for the geometry (a Kundt metric Ansatz for $g_{\mu\nu}$, connections with other nonvanishing irreducible parts, etc.). The other option is, instead, modifying the Lagrangian. One could then search for:

\begin{itemize}
\item {\bf Solutions with non-vanishing cosmological constant}. This case would require important modifications in the Ansatz for the coframe (along the lines of \cite{Obukhov2004} and \cite{BlagojevicCvetkovic2017b}).
\item {\bf Solutions with a nontrivial odd-parity gravitational sector}. This direction could be very relevant since a possible violation of parity is widely discussed in the current literature \cite{ChenHoNester2009, HoNester2012, HoNester2011, HoNester2012, HoChenNester2015, Diakonov2011, IosifidisRavera2020, BaeklerHehl2011, BaeklerHehlNester2011, Obukhov2020}.
\item {\bf Non-vacuum solutions}. It is also interesting to consider solutions in the presence of some realistic (but simple) matter distributions such as in-falling dust or collapsing spheres of relativistic particles.
\end{itemize}

\part{Viability of MAG and other modified theories of gravity \label{part:viability}}

\chapter{On the stability of gravitational theories}\label{ch:gravityviability}

\newcommand\jose[1]{\revisar{**}{\color{red} #1}\revisar{**}}

\boxquote{People who want to improve should take their defeats as lessons, and endeavor to learn what to avoid in the future. You must also have the courage of your convictions. If you think your move is good, make it.}{José Raúl Capablanca}

When we extend a well-behaved theory by introducing new degrees of freedom, we often pay attention just to the contributions to those physical observables we are interested in. However, it is important to adopt a broader perspective, because some fields or couplings can render unstable behaviors, for instance. In some cases, the instability appears just in some backgrounds, invalidating them as reliable models, but in others, its presence can spoil the general dynamics of the theory. In this chapter we are going to focus on revising some basic types of instabilities that usually appear in modified theories of gravity.

\section{Introduction: stability in field theory}

The intuitive notion of stability corresponds to the idea that if we slightly perturb a system, the resulting evolution should not be so different with respect to the unperturbed one, for instance, oscillating around it. In particular, if there are dissipative effects, one would expect the system to evolve towards the unperturbed solution. These can be formalized in the framework of dynamical systems: that perturbative excitations of the original trajectory with sufficiently small amplitude will remain in a certain neighborhood of the original solution in phase space. Usually, in theoretical physics, we require our models (by this we mean a solution of a particular theory) to be perturbatively well-behaved. Therefore it is crucial to know and identify the instabilities that can invalidate them.

In principle, we can separate the instabilities into two groups: those that manifest in particular backgrounds (background-dependent) and those that are intrinsic to the theory. The first ones indicate that the chosen model is problematic and, generally, should not be treated as a reliable physical model (e.g. in the context of cosmology, it would not describe the real Universe). The second ones are more severe since the propagating unstable modes will generically be present around any solution, implying that the entire theory is pathological. In this chapter we are going to focus on four types of instabilities, gradient ones, tachyons, ghosts and strong coupling.\footnote{
    Another relevant one, which is worth remarking, is the stability under quantum corrections. This is an extra test that classically stable theories have to pass.} 
For the description of the pathologies, we base our discussion on the references \cite{Rubakov2014, JoyceJain2015, Elder2017PhD}.

Modified theories of gravity are usually intricate theories with several degrees of freedom coupled in very complicated ways. If a potentially problematic term is noticed, it must be checked whether the instability is actually present or if some miracle could prevent the dynamics from the unstable behavior \cite{Woodard2007} (in Section \ref{sec:stabMTG} we will see the case of $\mathring{R}^2$ gravity, which is an example of such a miracle). 

\newpage
\subsection{Background stability I: Gradient instabilities, tachyons and ghosts}

Consider a field theory in Minkowski space for certain scalar $\Phi$. Suppose that we perform a perturbative expansion $\Phi=\Phi_0+\phi$ around some background solution $\Phi_0$. The action for the perturbation (at the lowest order) will be something like
\begin{equation}
  \mathcal{L}= F_1(\Phi_0) \dot{\phi}^2  - F_2(\Phi_0) |\anabla\phi|^2 - F_3(\Phi_0) \phi^2\,.
\end{equation}
Let us assume now a slow-varying background, i.e. that $\frac{\partial \Phi_0}{\Phi_0}\ll\frac{\partial \phi}{\phi}$.\footnote{
    This is valid for perturbations $\phi$ of high frequency.} 
In that case, the previous Lagrangian can be recast as
\begin{equation}\label{eq:Lagslowbg}
  \mathcal{L}= {\frac 1 {2a}} \dot{\phi}^2  - {\frac 12} b |\anabla\phi|^2 - {\frac 12} m^2 \phi^2\qquad\text{where}\quad b, m^2\in\mathbb{R},\quad a\in\mathbb{R}\setminus\{0\}\,.
\end{equation}
Here the dot represents the time derivative $\partial_0$ of the field, $\anabla$ is the gradient operator in $\mathbb{R}^3$ and $|.|$ is the module taken with the Euclidean metric of $\mathbb{R}^3$.  The equations of motion are
\begin{equation}
  \ddot{\phi}= a(b\anabla^2\phi  -m^2\phi) \,.
\end{equation}
If we now take the Fourier transform of the field 
\begin{equation}
  \hat\phi(k)\coloneqq \frac{1}{\sqrt{2\pi}} \int_{\mathbb{R}^3} \eN^{-\iN k_\mu x^\mu} \phi(x)\ {\rm d}t{\rm d}^3x\,,\qquad k_\mu = (\omega, \vec{k}),
\end{equation}
we get the following dispersion relation
\begin{equation}
  \omega^2 = a(b|\vec{k}|^2+m^2)\,.
\end{equation}
Notice that depending on the values of $\{a,b,m^2\}$ we might be allowing modes with imaginary frequencies, i.e. exponentially growing/decaying modes.

The corresponding Hamiltonian functional is\footnote{
    In the context of the Hamiltonian formulation of field theory, it is useful to work in terms of the Hamiltonian functional (instead of the Hamiltonian density). This is constructed not in terms of the Lagrangian density $\mathcal{L}(\phi,\partial_\mu\phi,...)$, which is a function of the fields and their derivatives, but in terms of the Lagrangian functional. The latter is a functional of the fields and their time derivatives (seen as independent variables), $L[\phi,\dot{\phi},\ddot{\phi}...]\coloneqq \int {\rm d}^3x \mathcal{L}$. The spatial derivatives are then considered within the functional dependence on the fields.}
\begin{equation}
  H[\phi,\pi] = \int {\rm d}^3x \  \frac{1}{2}\left(a \pi^2  +  b |\anabla\phi|^2 +  m^2 \phi^2\right)\,, \qquad \pi\coloneqq \frac{\delta\mathcal{L}}{\delta \dot{\phi}}= \frac{1}{a} \dot{\phi}.
\end{equation}
Here we see the problems from another perspective. For instance, observe that certain values of $\{a,b,m^2\}$ prevent the Hamiltonian from being bounded from below.

It is not difficult to see that the case $a,b,m^2>0$ is free of these exponentially growing modes (since all of the frequencies are real). Therefore, the Hamiltonian is non-negative and, consequently, bounded from below. Among these cases, we can find the Klein-Gordon Lagrangian
\begin{equation}
  \mathcal{L}_\mathrm{KG}= {\frac 12} \partial_\mu\phi \partial^\mu\phi- {\frac 12} m^2 \phi^2\,,
\end{equation}
which corresponds to $a=b=1$ and $m^2\geq0$.

\newpage
\begin{figure}[H]
\begin{center}
\includegraphics[width=\textwidth]{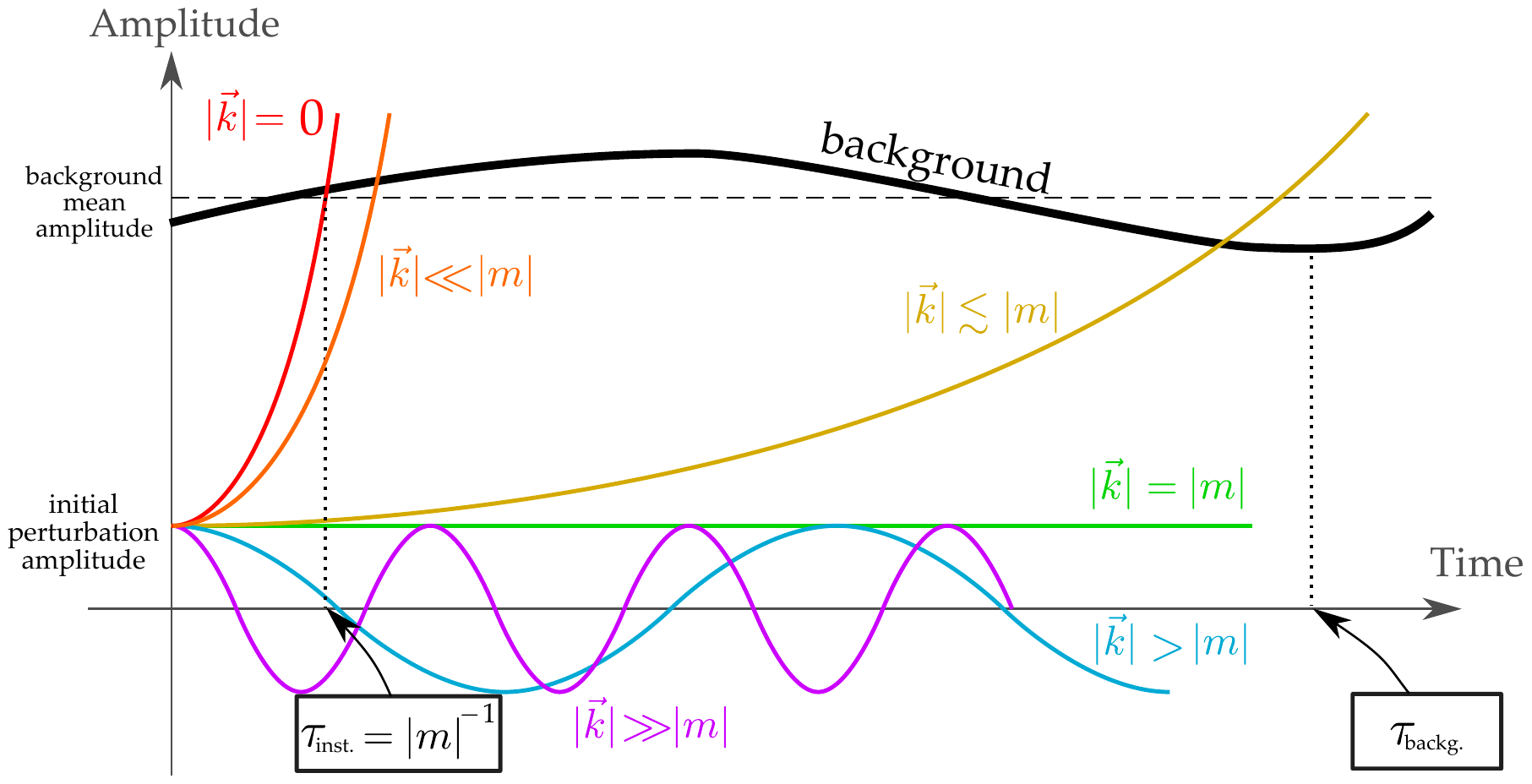}
\end{center}
\caption{\label{Fig:tachyon}
Here we schematically show the evolution of the modes for different $|\vec{k}|$ in the presence of a  tachyonic instability. For simplicity we took $|b|=1$. Observe that the instability time scale $\tau_\mathrm{inst.}$ is momentum independent and defined by the mass $|m|$ of the tachyon. For times much smaller than $\tau_\mathrm{inst.}$ the instability is controlled.}
\end{figure}

\subsubsection*{Tachyons}

We say that a field is a \emph{tachyon}, if the mass term has the wrong sign or, equivalently, if the mass parameter $m$ is purely imaginary. Usually tachyons indicate that the background is not the true vacuum of the theory.\footnote{In fact, tachyon instabilities play a role in cosmology. See e.g. \cite{LiCheung2014, Jassal2004}.}

In our example, if we assume $a>0$ (since it is a global normalization we choose $a=1$), the tachyonic instability appears whenever $b\geq0$ and $m^2<0$. Then, if we look at the dispersion relation 
\begin{equation}
  \omega^2 = b|\vec{k}|^2+m^2,
\end{equation}
we see that for sufficiently small 3-momentum ($b|\vec{k}|^2<-m^2$), the r.h.s. becomes negative and the associated modes have exponential behaviors \cite{Rubakov2014, JoyceJain2015},
\begin{equation}
  \phi(x) \propto \eN^{\pm |\omega|t} = \eN^{\pm \sqrt{ |m|^2 - b|\vec{k}|^2}t} .
\end{equation}
The presence of growing modes indicate that the theory is unstable. Observe that we do not have modes growing arbitrarily fast. The limiting case (highest $|\omega|$) corresponds to $\vec{k}=0$. This mode defines the scale of the instability
\begin{equation}
  \tau_\mathrm{inst.} = |m|^{-1}\,.
\end{equation}
At this time, the unstable modes will reach amplitudes comparable to the background (see Fig \ref{Fig:tachyon}). Notice that for larger masses of the tachyon, $\tau_\mathrm{inst.}$ decreases (for a fixed initial amplitude) and the instability is even more severe. In the opposite case, the dynamics is reliable at short times, whereas at large times one cannot say anything about stability. The reason for this is that the slow-varying hypothesis we assumed is violated, since $\tau_\mathrm{inst.}$ is comparable with the characteristic time scale of the background $\tau_\mathrm{backg.}$. The analysis of stability in that case cannot be done with the Lagrangian \eqref{eq:Lagslowbg}.

\newpage 
\begin{figure} 
\begin{center}
\includegraphics[width=\textwidth]{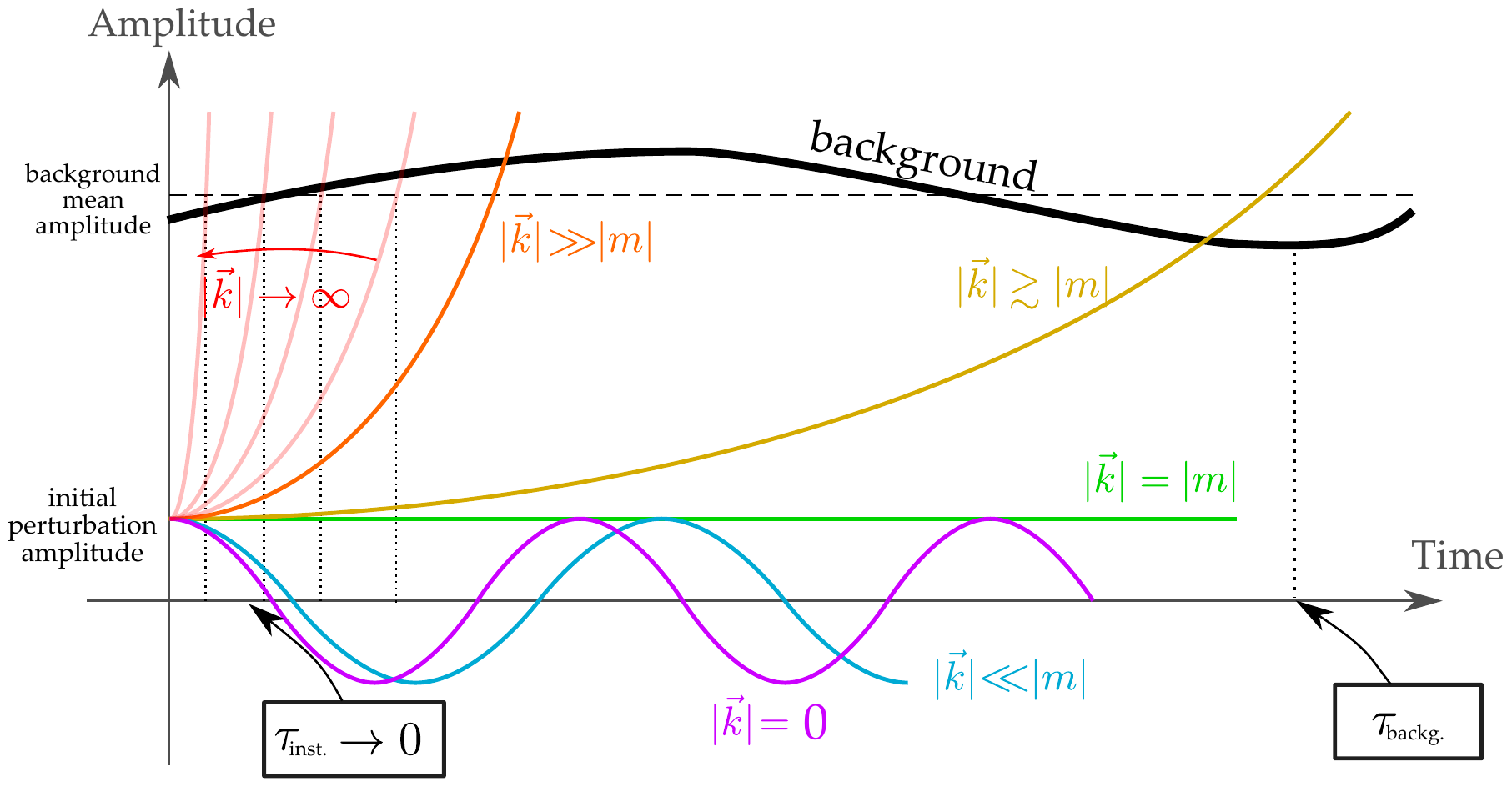}
\end{center}
\caption{\label{Fig:gradientinst}
Here we schematically show the evolution of the modes for different $|\vec{k}|$ in the presence of a  gradient/Laplacian instability. For simplicity we took $|b|=1$. Observe that the instability time scale $\tau_\mathrm{inst.}$ is essentially zero (ignoring a possible cutoff) since the exponential growing is faster for higher momenta.}
\end{figure}

\subsubsection*{Gradient instabilities}

\emph{Gradient instabilities} (also called Laplacian) are those characterized by imaginary frequencies at high momenta. This is a consequence of the wrong sing in front of the gradient term in the Lagrangian, leading to imaginary sound speed parameter ($c_s^2<0$).

Let us take our example with the scalar field with the normalization $a=1$ and a ``good''  mass term with $m^2\geq0$. The dispersion relation for $b<0$,  
\begin{equation}
  \omega^2 = b|\vec{k}|^2+m^2,
\end{equation}
shows, as we announced at the beginning, that for high momenta ($|\vec{k}|^2>m^2/|b|$) the frequencies are imaginary. In other words, there will be exponentially growing modes. When the mass is negligible with respect to the momentum we have \cite{Rubakov2014, JoyceJain2015}
\begin{equation}
  \phi(x) \propto \eN^{|\omega|t} \approx \eN^{|b|^{1/2}|\vec{k}|t}.
\end{equation}
Contrary to the tachyonic case, we see that in a system with a gradient instability, modes with higher momenta grow faster and faster. Consequently, the characteristic time of the instability goes to 0, i.e., the instability develops arbitrarily fast. (see Fig. \ref{Fig:gradientinst}).

One might think that this situation can be alleviated and we can do phenomenology if we put some cutoff $\Lambda$ to the momenta, but the problem continues. First of all notice that the time scale characteristic of the background ($\tau_\mathrm{backg.}$) should be larger than the cutoff time scale $\Lambda^{-1}$ (if not, we would be working beyond the limits of the Effective Field Theory). In addition, there are exponentially growing modes developing the instability up to the cutoff scale \cite{Rubakov2014}. Then, for $|m|<|\vec{k}|<\Lambda$ the background is ruined by the instability, similarly as in the tachyonic case for $0<|\vec{k}|<|m|$. On the other hand, the region of momentum space $|\vec{k}|>\Lambda$ is directly outside the regime of validity of the Effective Field Theory. Therefore, a gradient instability completely invalidates the theory and makes it non-predictive \cite{JoyceJain2015}.

\newpage

\begin{figure} 
\begin{center}
\includegraphics[width=0.55\textwidth]{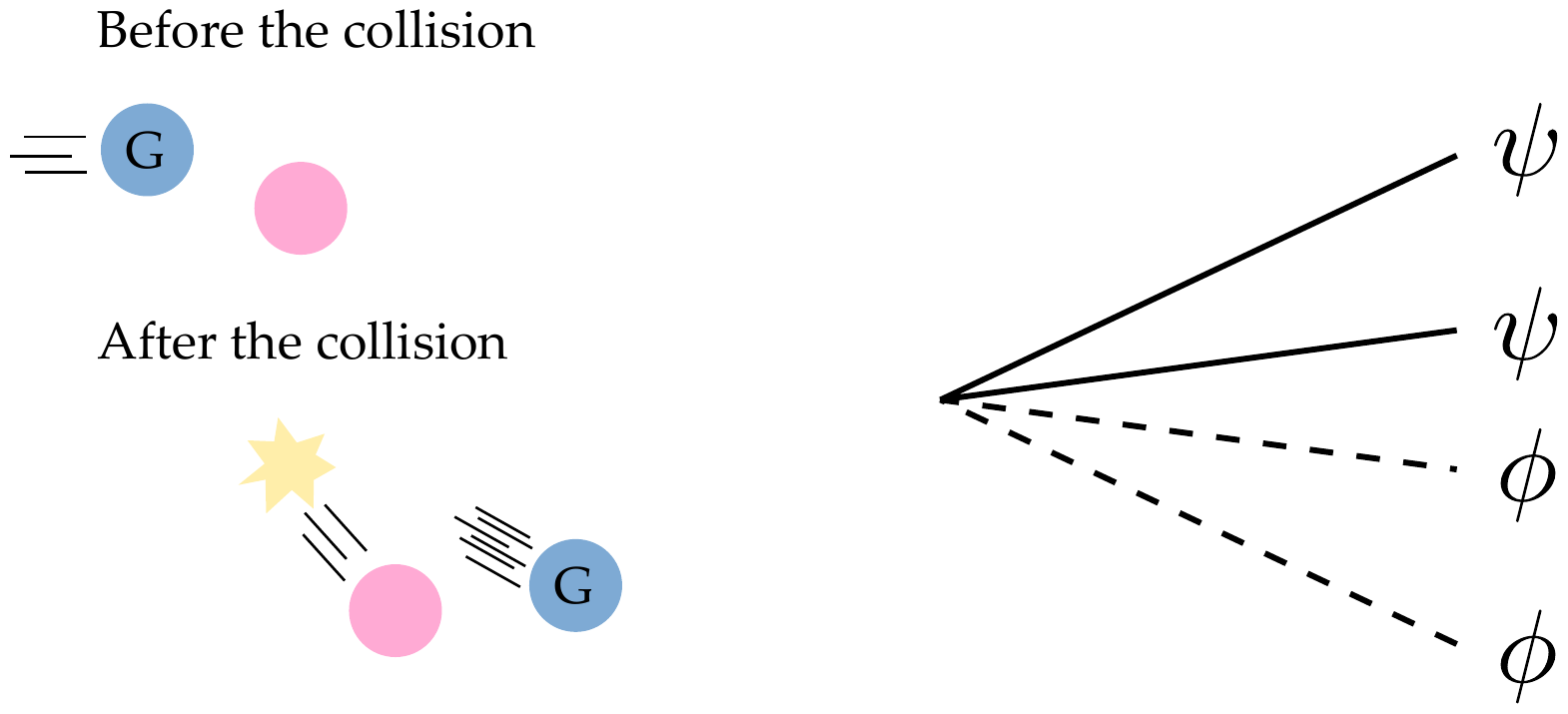}
\end{center}
\caption{\label{Fig:ghostinst}
{\bf Left}: A classical ghostly particle moving towards an ordinary particle at rest. As a result of an elastic collision, the latter acquire some kinetic energy, which also implies a boost on the velocity of the ghost. Since its kinetic energy is negative, this process does not violate energy conservation. Notice that also the conservation of the momentum is different in the presence of a ghost.
{\bf Right}: Example of the typical pathological quantum process allowed in theories with ghosts. Here we see the vacuum decaying into two healthy particles and two ghosts (based on the example \eqref{eq:ghostex}). This process would quickly happen everywhere, making the vacuum unstable.}
\end{figure}
\subsubsection*{Ghosts}

In classical theory, a \emph{ghost} is a field with negative kinetic energy. From a quantum perspective, the associated quanta have either negative energy or negative norm.

In a theory with just one field, the corresponding kinetic term can always be canonically normalized. The problem arises when there are couplings to other fields whose kinetic terms have the correct sign. Consider our previous example with $a=b=-1$ and $m^2>0$, coupled to another scalar field $\psi$ with the canonical kinetic term and the same mass for simplicity:
\begin{equation}\label{eq:ghostex}
  \mathcal{L}= {\frac 12} \partial_\mu\psi \partial^\mu\psi -{\frac 12} \partial_\mu\phi \partial^\mu\phi- {\frac 12} m^2 (\phi^2+\psi^2) + \alpha \phi^2\psi^2 \,.
\end{equation}
If we compute the Hamiltonian we can clearly see the problem:
\begin{align}
  H[\phi,\pi_\phi,\psi,\pi_\psi] = \int {\rm d}^3x \left[\frac{1}{2} (\pi_\psi^2-\pi_\phi^2)  + {\frac 12}(|\anabla\phi|^2-|\anabla\psi|^2) + {\frac 12} m^2 (\phi^2+\psi^2) - \alpha \phi^2\psi^2 \right]\,.
\end{align}
Indeed, the term $-\pi_\phi^2$ prevents the Hamiltonian from being bounded from below. These contributions have catastrophic consequences even in the classical regime (see Fig. \ref{Fig:ghostinst} left). However, the problem is even worse in the quantum theory since there is no ground state. In that case the problem is not that the energy becomes lower and lower, since the energy (assuming the system is isolated) is conserved, but that certain sectors become highly excited, triggering the unstable behavior \cite{Woodard2007, CarrollHoffman2003, ClineJeon2004}. The vacuum of the theory will copiously decay into ghosts and non-ghostly particles (with arbitrary positive energy) without violating energy conservation.\footnote{
    For a scalar ghost, the propagator can be expressed in any of these two forms \cite{ClineJeon2004},
    \[ \frac{-\iN}{p^2-m^2+\iN \epsilon},\qquad \frac{-\iN}{p^2-m^2-\iN \epsilon}\,.\]
    The first one gives negative norms, and hence violation of unitarity. However, the second one gives a unitary theory with negative energy states that can produce these uncontrolled decays.} 
In our particular example, this would happen through the process $0\to \phi+\phi+\psi+\psi$ (see Fig. \ref{Fig:ghostinst} right).\footnote{
    In quantum-field-theoretical terms, the available phase space of the decay is typically large so the time scale of instability is short \cite{Rubakov2014}. Of course if we work at scales much lower than the mass of the ghost, the instability is not active.}

Some authors have been working in the direction of clarifying that having a Hamiltonian with no ground state does not necessarily imply a pathological theory (see e.g. \cite{BenderMannheim2008, ChenFasiello2013, Smilga2017}). These solutions usually involve non-standard approaches and their classical limit is not clear, so we are not going to enter into details since the complexity of the field is remarkable.\footnote{In some cases, the ghosts can be eliminated via Gupta-Bleuler formalism, as in Quantum Electrodynamics, by consistently restricting the Hilbert space.} For the purposes of this thesis, we will adopt the most extreme position and try to totally eliminate the ghosts. Nevertheless, each particular theory should be carefully studied to reveal whether the instability is actually present. In any case, the alarm should turn on whenever these terms or negative energy densities are present.

\subsection{Background stability II: Strongly coupled backgrounds}\label{sec:strongcoupling}

In this section we are going to briefly comment on \emph{strong coupling}. This is another type of background-dependent instability that is very common in modified theories of gravity and that cannot be directly seen at linear order in perturbations.

The general idea is the following. Suppose we have a theory that we know that contains $n$ propagating modes in its general spectrum (or $n$ degrees of freedom). Now we perform perturbations around certain solution and check that only $m$ ($<n$) propagate at linear order but at certain higher order, some of the remaining ones become active. This discontinuity in the number of propagating modes at different orders in the perturbation expansion indicates an instability of the background called strong coupling.

One way to see the problem is that, after canonical normalization, the vanishing coefficient of the kinetic term will affect all the interaction terms and consequently the effective couplings in the interactions will become infinite (this motivates the denomination ``strong coupling'').

From the point of view of dynamical system theory, strongly coupled solutions typically lie in a phase space surface that presents some kind of singular behavior. In Chapter \ref{ch:ECG}, we will see that indeed the strongly coupled background is a singular point of the principal part of the dynamical system of differential equations. To be precise, what happens there is that the matrix of coefficients for the highest derivative sector does not have maximum rank at that point. In other words, the order of the system is abruptly reduced in that solution. It is worth highlighting that this singular point is not a point in spacetime (whose presence could allow regular solutions), but a singularity in phase space, which is something much more severe.

One important consequence of having a strongly coupled background is that solutions around it cannot be perturbatively computed in the standard way, i.e., part of the information about them could not be capture by the perturbative method (see Appendix \ref{app:toyexample}).

\subsection{Background-independent instabilities: Ostrogradski ghosts}\label{sec: Ostrog}

\subsubsection*{Ostrogradski theorem}

The previously mentioned instabilities usually appear in field theory when performing perturbations around certain backgrounds and can be absent in other solutions. Unfortunately, there could be instabilities totally inherent to the very theory, which are not artefacts of the specific background under study. A quite notable example, which is often present in modified gravity, are those ghosts predicted by the Ostrogradski theorem (\emph{Ostrogradski ghosts}). We are going to start by revising such result, which is indeed one of the most powerful results about stability in field theory.

In 1850, Ostrogradski formulated an extension of the ordinary Hamiltonian formalism to higher derivative theories \cite{Ostrogradski1850}. Here we reproduce the theorem as was stated in \cite{AokiMotohashi2020}:\footnote{See also \cite{Pons1989, GanzNoui2021}.}
\boxtheorem{
\begin{thm} 
\textbf{\textup{(Ostrogradski Theorem)}} \\ Let a Lagrangian involve $n$-th order finite time derivatives of variables. If $n\geq2$ and the Lagrangian is non-degenerate with respect to the highest-order derivatives, the Hamiltonian of this system linearly depends on a canonical momentum.
\end{thm}
}
Here \emph{non-degenerate Lagrangian} means that the Hessian has maximum rank. In this context, the Hessian should be understood as the matrix of second variations of $L$ with respect to the highest time derivatives of each field. See e.g. \eqref{eq:checkdegLBfi}.

For completeness, we reproduce here the Ostrogradski construction with second time derivatives in the context of point particles, just to fix ideas \cite{Woodard2015}. Consider the Lagrangian (we use $L$ and $H$ for the Lagrangian and the Hamiltonian in point particle dynamics) $L = L(x,\dot{x},\ddot{x})$ and assume that it is non-degenerate, i.e.,
\begin{equation}
  \frac{\partial^2 L}{\partial \ddot{x}^2}\neq0\,.
\end{equation}
Notice that this condition permits the Euler-Lagrange equation,
\begin{equation}\label{eq:ELeqOstro}
  \frac{\partial L}{\partial x}-\frac{\dex}{\dex t}\frac{\partial L}{\partial \dot{x}}+\frac{\dex^2}{\dex t^2}\frac{\partial L}{\partial \ddot{x}}=0\,
\end{equation}
to be re-expressed as \cite{Woodard2015}
\begin{equation}
   x(t)=f(t,x_0,\dot{x}_0,\ddot{x}_0,\dddot{x}_0)\,.
\end{equation}
Thus, instead of two, four initial values are needed. This reflects the fact that there are $4/2=2$ degrees of freedom and not 1, even when we have just one unknown function $x(t)$. This ``hidden'' degree of freedom is a consequence of having higher-derivatives. 

Let us move on to the Hamiltonian view. For this particular case, the Ostrogradski procedure requires to introduce two generalized coordinates and two generalized momenta as follows
\begin{equation}\label{eq:qpiOstro}
  q_1 \coloneqq x,\qquad q_2 \coloneqq \dot{x}\,,\qquad \pi_1\coloneqq \frac{\partial L}{\partial \dot{x}}-\frac{\dex}{\dex t}\frac{\partial L}{\partial \ddot{x}}\,,\qquad \pi_2\coloneqq \frac{\partial L}{\partial \ddot{x}}\,.
\end{equation}
The non-degeneracy condition implies that one can invert these definitions to write
\begin{equation}
  \ddot{x} = F(q_1, q_2, \pi_2)\,,
\end{equation}
such that
\begin{equation}
  \pi_2 = \frac{\partial L}{\partial \ddot{x}} \Big|_{x=q_1, \, \dot{x}=q_2,\, \ddot{x}=F}\,.
\end{equation}

At this point, we introduce the Ostrogradski Hamiltonian, defined in the usual way as the Legendre transformation of $L$ with respect to the velocities (derivatives of $q_i$):
\begin{align}
 H &= \dot{q}_1 \pi_1 + \dot{q}_2\pi_2 - L(x,\dot{x},\ddot{x}) \\
   &= q_2 \pi_1 +  F(q_1, q_2, \pi_2)\pi_2 - L(q_1,q_2, F(q_1, q_2, \pi_2)) \,.
\end{align}
We clearly see here that the momentum $\pi_1$ only appears once in the Hamiltonian and linearly, as we wanted to prove. Let us finally show that this Hamiltonian is not just an artificial construction but  indeed reproduces the same dynamics as the Lagrangian we started with. Three of the Hamilton equations are nothing but the definitions of $q_2$, $F$ and $\pi_1$, respectively,
\begin{equation}
  \dot{q}_1= \frac{\partial H}{\partial \pi_1} = q_2\,, \qquad
  \dot{q}_2= \frac{\partial H}{\partial \pi_2} = F(q_1, q_2, \pi_2)\,,\qquad
  \dot{\pi}_2=-\frac{\partial H}{\partial q_2} = -\pi_1 + \frac{\partial L}{\partial \dot{x}}\,,
\end{equation}
whereas the remaining one,
\begin{equation}
  \dot{\pi}_1=-\frac{\partial H}{\partial q_1} = \frac{\partial L}{\partial x}\,,
\end{equation}
is the Euler-Lagrange equation \eqref{eq:ELeqOstro} (after the substitutions \eqref{eq:qpiOstro}). It is important to note that, also in this formulation, $H$ continues being the energy of the system, i.e., the Noether current for the time translation symmetry \cite{Woodard2015}.

~

These notions can be generalized to field theory. Let us see a particular example (see Appendix D of \cite{JoyceJain2015}):
\boxexample{
\begin{example}\label{example:Ostro}
Consider the following Lagrangian for a scalar field
\begin{equation}
  \mathcal{L}= {\frac 12} \partial_\mu\phi\partial^\mu\phi + {\frac \lambda 2}(\square\phi)^2 - V(\phi),\qquad \lambda \neq0.
\end{equation}
where $\square\coloneqq \eta^{\mu\nu} \partial_\mu\partial_\nu$. We construct the Lagrangian functional
\begin{equation}
  L[\phi, \dot{\phi}, \ddot{\phi}] = \int {\rm d}^3x \mathcal{L} =  \int {\rm d}^3x \Big[ {\frac 12} (\dot{\phi}^2 - |\anabla\phi|^2) + {\frac \lambda 2}(\ddot{\phi} - \anabla^2\phi)^2 - V(\phi)\big]\,,
\end{equation}
which is indeed non-degenerate,
\begin{equation}
  \det \big[\mathrm{Hess}(L)\big]=\frac{\delta^2 L}{\delta \ddot{\phi}\delta \ddot{\phi}} = \lambda  \quad\neq0\,.
\end{equation}
We introduce canonical variables according to the Ostrogradski procedure:
\begin{align}
  q_1 & \coloneqq \phi\,,
  &\pi_1 &\coloneqq \frac{\delta L}{\delta \dot{\phi}}-\frac{\dex}{\dex t}\frac{\delta L}{\delta \ddot{\phi}}= \dot{\phi} - \lambda \dddot{\phi} + \lambda\anabla^2\dot{\phi} \,,\nonumber\\
  q_2 & \coloneqq \dot{\phi}\,,
  &\pi_2 &\coloneqq \frac{\delta L}{\delta \ddot{\phi}} =\lambda(\ddot{\phi} - \anabla^2\phi)\,.
\end{align}
From the expression of $\pi_2$ we can obtain the function $F$,
\begin{equation}
  \ddot{\phi} = F(q_1,q_2,\pi_2) = {\frac 1 \lambda} \pi_2 + \anabla^2 q_1\,.
\end{equation}
The resulting Hamiltonian functional is indeed linear in $\pi_1$:
\begin{align}
   & H[q_1,q_2,\pi_1, \pi_2] \nonumber\\
   &\qquad = \int {\rm d}^3x \big[\dot{q}_1 \pi_1 + \dot{q}_2\pi_2 - \mathcal{L} \big] \nonumber\\
   &\qquad =  \int {\rm d}^3x \Big[q_2 \pi_1 + \Big(\frac{1}{2\lambda}\pi_2 + \anabla^2 q_1\Big)\pi_2 - {\frac 12} \Big((q_2)^2 - |\anabla q_1|^2\Big) + V(q_1) \Big] \,.
\end{align}
\end{example}
}

The appearance of the ghostly mode can be also seen directly in Lagrangian formalism by introducing auxiliary variables (at the end of the day this is just a ``dof redefinition'' that allows to extract the pathological part). Let us revisit the previous example:
\boxexample{
\begin{example}\label{example:Ostro2}
Consider the following Lagrangian density depending on two fields $\phi$ and $\chi$
\begin{equation}\label{eq:LagExampleOstr}
  \mathcal{L}= {\frac 12} \partial_\mu\phi \partial^\mu\phi + \chi \square\phi - \frac{1}{2\lambda}\chi^2 - V(\phi)\,.
\end{equation}
This theory is on-shell equivalent to the one of the Example \ref{example:Ostro}. Indeed $\chi$ is an \emph{auxiliary field}, since no derivatives of it appear in the action and it does not appear linearly (in such case it would not be an auxiliary field but a Lagrange multiplier). Its equation of motion is algebraic and can be solved as
\begin{equation}
  0=\frac{\partial \mathcal{L}}{\partial \chi}= \square\phi - \frac{1}{\lambda} \chi \qquad\Rightarrow \qquad \chi = \lambda \square\phi\,.
\end{equation}
If the equation of motion for certain field can be solved exactly and we plug it back in the Lagrangian, we do not alter the dynamics of the rest of the fields (i.e. it is consistent to do that). If we plug this solution we immediately get the Lagrangian of the Example \ref{example:Ostro}. The thing is now to realize (see \cite{JoyceJain2015}) that we can perform a field redefinition $\phi \to \chi - \alpha$ (which is completely regular) and the Lagrangian \eqref{eq:LagExampleOstr} becomes
\begin{equation}
  \mathcal{L}= {\frac 12} \partial_\mu\alpha \partial^\mu\alpha - {\frac 12} \partial_\mu\chi \partial^\mu\chi -  \frac{1}{2\lambda}\chi^2 - V(\chi-\alpha)\,.
\end{equation}
We can see that the kinetic term of $\chi$ has the wrong sign and thus is responsible for the Hamiltonian not to be bounded from below. In this final rearrangement of the fields, $\chi$ is the one that we can identify with the Ostrogradski ghost.
\end{example}
}

\subsubsection*{Avoiding Ostrogradski ghosts and gauge symmetries: the massive spin 1 particle}

The most general kinetic term one can construct for a relativistic field theory of a vector field $A_{\mu}$ is a combination of the invariants 
\begin{equation}
\partial_{\mu}A_{\nu}\partial^{\mu}A^{\nu}\,,\qquad\partial_{\mu}A_{\nu}\partial^{\nu}A^{\mu}\,,\qquad(\partial_{\mu}A^{\mu})^{2}\,.
\end{equation}
Up to boundary terms, only two are independent, so let us consider the following Lagrangian
\begin{equation}
\mathcal{L}_A=-\frac{1}{4}aF_{\mu\nu}(A)F^{\mu\nu}(A)-\frac{1}{2}b(\partial_{\mu}A^{\mu})^{2}+\frac{1}{2}m^{2}A_{\mu}A^{\mu}\,,
\end{equation}
where we have introduced a mass term for the field and also the usual notation $F_{\mu\nu}(X)\coloneqq2\partial_{[\mu}X_{\nu]}$. This Lagrangian is assumed to be part of a larger one with other fields that are canonically normalized and such that (for simplicity) the only derivatives of $A_{\mu}$ are those of $\mathcal{L}_A$. 

If we separate spatial and time components $A_{\mu}\equiv(A,A_{i})$, we obtain
\begin{align}
\mathcal{L}_A & =\frac{1}{2}a\delta^{ij}\dot{A}_{i}\dot{A}_{j}-\frac{1}{2}b\dot{A}^{2}+a\dot{A}_{i}\partial^{i}A-b\dot{A}\partial_{i}A^{i}\nonumber \\
 & \qquad-\frac{1}{2}a\partial_{i}A\partial^{i}A-\frac{1}{4}aF_{ij}(A)F^{ij}(A)-\frac{1}{2}b(\partial_{i}A^{i})^{2}+\frac{1}{2}m^{2}(A^{2}+A_{i}A^{i})\,.
\end{align}
Here we see that either $a<0$ or $b>0$ (or both) implies the presence of a ghost. If we jump to the Hamiltonian formulation by introducing the momenta
\begin{align}
\pi\coloneqq\frac{\delta L_A}{\delta\dot{A}} & =-b(\dot{A}+\partial_{i}A^{i})\,,\qquad\pi^{i}\coloneqq\frac{\delta L_A}{\delta\dot{A}_{i}}=-a(\dot{A}^{i}-\partial^{i}A)\,,
\end{align}
we can identify the ghosts in the following kinetic contributions to the Hamiltonian functional (assuming both $a,b\neq0$)
\begin{equation}
H_A[A,\pi,A_{i},\pi^{i}]=\int\mathrm{d}^{3}x\left(-\frac{1}{b}\pi^{2}+\frac{1}{2a}\delta_{ij}\pi^{i}\pi^{j}+...\right)\,.
\end{equation}

Interestingly, the term with $b$ contains a ghost even when $b<0$. Indeed, this mode can be transformed into an Ostrogradski one. To see this, let us perform, instead of a $3+1$ decomposition of the field, a splitting into longitudinal $\phi$ and transversal part $B_{\mu}$ as follows\footnote{This decomposition can always be done. See \cite[sec. 8.7.1]{Schwartz2014}.}
\begin{equation}
A_{\mu}=\partial_{\mu}\phi+B_{\mu}\qquad\text{with}\qquad\partial_{\mu}B^{\mu}=0\,,
\end{equation}
which implies $\partial_{\mu}A^{\mu}=\square\phi$ and $F_{\mu\nu}(A)=F_{\mu\nu}(B)$.
The Lagrangian can be then recast as
\begin{equation}
\mathcal{L}_{B\phi}=-\frac{1}{4}aF_{\mu\nu}(B)F^{\mu\nu}(B)-\frac{1}{2}b(\square\phi)^{2}+\frac{1}{2}m^{2}(\partial_{\mu}\phi+B_{\mu})(\partial^{\mu}\phi+B^{\mu}).
\end{equation}
The Lagrangian functional $L_{B\phi}[B_\mu, \dot{B}_\mu, \phi, \dot\phi, \ddot\phi]$ is non-degenerate for nontrivial $a$ and $b$:
\setlength{\arraycolsep}{2pt}
\begin{equation}\label{eq:checkdegLBfi}
  \renewcommand\arraystretch{2}
  \det \big[ \mathrm{Hess}(L_{B\phi})\big]=\det\begin{pmatrix}
   \dfrac{\delta^2 L_{B\phi}}{\delta\dot{B}_{\mu}\delta\dot{B}_{\nu}}&
   \dfrac{\delta^2 L_{B\phi}}{\delta\dot{B}_{\mu}\delta\ddot{\phi}}  \\
   \dfrac{\delta^2 L_{B\phi}}{\delta\ddot{\phi}\delta\dot{B}_{\nu}}  & 
   \dfrac{\delta^2 L_{B\phi}}{\delta\ddot{\phi}\delta\ddot{\phi}}
   \end{pmatrix}
   \renewcommand\arraystretch{1.5}
   =\det\begin{pmatrix}-a\eta^{\mu\nu} & 0 \\ 0 & -b\square\phi \end{pmatrix}
  \renewcommand\arraystretch{1}
   =-ab\square\phi\,.
\end{equation}
\setlength{\arraycolsep}{6pt}
One can now check that there are two degrees of freedom associated to $\phi$ instead of one. To see this, we rewrite the Lagrangian density as
\begin{equation}
\mathcal{L}_{B\phi} =-\frac{1}{4}aF_{\mu\nu}(B)F^{\mu\nu}(B)+\frac{1}{2}m^{2}B_{\mu}B^{\mu}+\frac{1}{2}\phi\left(-b\square^{2}-m^{2}\square\right)\phi+\text{total der.}
\end{equation}
and compute the propagator in momentum space (we follow the steps of \cite{Schwartz2014})
\begin{equation}\label{eq:ghostprop}
\Pi_{\phi}=\frac{1}{bk^{4}-m^{2}k^{2}}=\frac{1}{m^{2}}\left[\frac{1}{k^{2}}-\frac{b}{bk^{2}-m^{2}}\right]\,.
\end{equation}
Note that the second term has negative norm for generic $b$. To avoid the Ostrogradski ghost we need to impose $b=0$. In addition, in order to get the correct normalization for the kinetic term of $B_{\mu}$ we need $a=1$ (we avoid the other ghostly modes since $a>0$). The resulting Lagrangian (\emph{St\"uckelberg Lagrangian}),\footnote{
    The elimination of the ghost is a degenerate case of the theory in terms of $A$ and $A_{i}$, because a 0 is introduced in the (diagonal) kinetic matrix
    \[
    \mathcal{L}_A|_{a=1,b=0}=\frac{1}{2}(\dot{A},\,\dot{A}_{i})\begin{pmatrix}0 & 0\\
    0 & \delta^{ij}
    \end{pmatrix}\begin{pmatrix}\dot{A}\\
    \dot{A}_{j} \end{pmatrix}+...\,.
    \]
    From the Hamiltonian perspective, it is not possible to solve for the velocities in terms of the momenta (we get $\pi=0$). The correct way to continue with the Hamiltonian approach is the Dirac procedure for constrained systems \cite{Dirac_LecturesQM, Wipf1993, Tavakoli2014Lec1} (see also \cite{Date2010} and the explanations in \cite{Blagojevic2001, PonomarevObukhov2017}).}
\begin{equation}
\mathcal{L}_\text{St\"uck}=-\frac{1}{4}aF_{\mu\nu}(B)F^{\mu\nu}(B)+\frac{1}{2}m^{2}(\partial_{\mu}\phi+B_{\mu})(\partial^{\mu}\phi+B^{\mu})\,,
\end{equation}
is gauge invariant under 
\begin{equation}
B_{\mu}\to B_{\mu}+\partial_{\mu}g ~\,,\qquad\phi\to\phi-g\,.
\end{equation}
If we fix the gauge that makes $\phi=0$ we recover the Proca Lagrangian, which describes the three degrees of freedom of a massive spin-1 particle.

The introduction of gauge symmetries (redundancies) allows to eliminate degrees of freedom. This can be used to cure theories from ghostly behaviors. In the previous example under the longitudinal+transversal decomposition, the symmetry makes the Lagrangian degenerate so that the Ostrogradski theorem cannot be applied (we will explicitly use this in several examples in Chapter \ref{ch:QTG}). In general, one should be careful and ensure that the introduced gauge symmetry has indeed eliminated pathological modes (as in this case) and not healthy ones.

Similarly, the $\mathrm{U}(1)$ gauge symmetry $A_{\mu}\to A_{\mu}+\partial_{\mu}g$ prevents the Maxwell Lagrangian from being pathological, since it allows to eliminate the ghost associated to the longitudinal mode. Another example of this is the Fierz-Pauli Lagrangian for a massless spin-2 field, where the ghosts can be removed thanks to the gauge symmetry under $h_{\mu\nu}\to h_{\mu\nu}+\partial_{\mu}\xi_{\nu}+\partial_{\nu}\xi_{\mu}$. 

\section{Stability problems in different modified theories of gravity}\label{sec:stabMTG}

If we look at (massless) Fierz-Pauli as the linearization of GR around Minkowski background, we can check that the gauge transformation $h_{\mu\nu}\to h_{\mu\nu}+\partial_{\mu}\xi_{\nu}+\partial_{\nu}\xi_{\mu}$ is just the linearization of a general diffeomorphism. The already mentioned healthiness of the Fierz-Pauli Lagrangian is then a crucial result that reflects the linear stability of GR around Minkowski. A more careful Hamiltonian analysis reveals that the theory is indeed healthy at the full non-linear level.\footnote{
    In this thesis we do not enter in Hamiltonian analysis, so we leave the details for the reader.} 

If we construct a modified theory of gravity at the non-linear level (combining curvatures with other tensor quantities for instance), instabilities of different kinds are expected. Here we present some examples of modified theories of gravity and currently known pathologies they exhibit:
\begin{itemize}
\item Except Lovelock terms, higher order curvature theories of gravity (in the metric formulation) propagate two extra fields in addition to the massless graviton: a ghostly spin-2 field with nontrivial mass and a scalar. In general, these theories will have the ghostly modes propagating. One example is (Cosmological) Einsteinian Cubic gravity (see Chapter \ref{ch:ECG}), which contains a propagating ghost and whose cosmological backgrounds (at least with flat spatial slices \cite{BeltranJCA2021}) are strongly coupled. \\

However, there are some very special cases such as $f(\mathring{R})$, in which a miracle happens. Of course, if we go to the Einstein frame, we are done (because GR is healthy). But it is interesting to see more closely why the ghost disappears in more physical terms. To see this, consider the particular example of $\mathring{R}^2+a \mathring{R}_{\mu\nu} \mathring{R}^{\mu\nu}$. In the limit $a\rightarrow0$, one can check that the mass of the extra spin-2 field goes to infinity (in appropriate variables) without generating any pathologies \cite{BeltranJCA2021}. This leads us to the well-known result that the term $\mathring{R}^2$ is healthy. This result can be extended to all $f(\mathring{R})$ \cite{Woodard2007}.

\item The case of an arbitrary function of the Gauss-Bonnet invariant, called $f(G)$-gravity (see e.g. \cite{NojiriOdintsov2005, CognolaElizalde2006,DeFeliceTsujikawa2009}), is similar to $f(\mathring{R})$: the ghostly spin-2 is absent and only the massless graviton and the scalar propagate \cite{Kobayashi2011}.

\item Ricci based gravity, as commented in Section \ref{sec:landscapeMTG}, can be mapped into an Einstein frame. The resulting theory is just GR, so it propagates only a healthy graviton. However, the inclusion of the antisymmetric part of the Ricci in the action completely destroys the well behavior of these theories transforming the projective mode in a propagating ghost \cite{BeltranDelhom2019}. See also the extension of this analysis in \cite{BeltranDelhom2020}, where they argue that these problems will be present in more general metric-affine theories.

\item The non-linear extensions of the teleparallel equivalents, $f(T)$ \cite{GolovnevKoivisto2018,BengocheaFerraro2008, LiSotiriouBarrow2010} and $f(Q)$ \cite{BeltranHeisenbergKoivisto2018a, BeltranHeisenbergPekar2019} (we will introduced them in Chapter \ref{ch:QTG}) also suffer from instabilities. The reason is that the nice symmetries (which allow for instance the elimination of ghosts) of the teleparallel equivalents $T$ and $Q$ are in fact symmetries up to a boundary term. These boundary terms cannot be extracted from the function $f$, so the symmetries are lost in the non-linear extensions. In addition, the cosmological backgrounds are generically strongly coupled in $f(T)$ \cite{GolovnevKoivisto2018}, while $f(Q)$ alleviates the situation, suffering the problem only on maximally symmetric backgrounds \cite{BeltranHeisenbergPekar2019} (in particular, in Minkowski spacetime).

\item The quadratic PG gravity Lagrangian contains ghosts and tachyons for generic values of the parameters (see for instance the analysis of the particle spectrum in \cite{BlagojevicCvetkovic2018}). Similar problems are expected in MAG.

\end{itemize}

Of course, it is possible to find sets of parameters in some of these theories that lead to healthy propagations. Relevant examples of this are the already mentioned (see the Section \ref{sec:landscapeMTG}) Horndeski \cite{Horndeski1974, DeffayetDeser2009, DeffayetEsposito2009, NicolisRattazzi2009,CliftonFerreira2012, Kobayashi2011}, beyond Horndeski \cite{GleyzesLanglois2015, GleyzesLanglois2015b, LangloisNoui2016, MotohashiNoui2016, MotohashiSuyama2018, BenAchour2016, KleinRoest2016, MotohashiSuyama2015, Zumalacarregui2013} and generalized Proca \cite{Heisenberg2014,Tasinato2014,BeltranHeisenberg2016, BeltranHeisenbergKoivisto2016}. They remove Ostrogradski ghosts by construction via ensuring either the degeneracy of the Lagrangian or the presence of, at most, 2nd-order derivatives in the dynamical equations. Indeed, in theories involving scalars, vectors, etc., such as MAG, the corresponding terms must belong to one of the known classes of healthy theories.\footnote{Some further recommended literature is \cite{Heisenberg2018} for an extensive review of cosmology in different modified gravity theories and \cite{DeffayetRombouts2005} on massive gravity.}

\section{Inconsistencies in 4DEGB}

In addition to the possible instabilities, the very definition of a theory might contain inconsistencies. Although it has been refined a posteriori, the initial formulation of  \emph{4-dimensional Einstein-Gauss-Bonnet gravity} (abbreviated as 4DEGB) is a good example of this. This theory was recently proposed \cite{GlavanLin2020} as a candidate that incorporates dynamical contributions coming from the Gauss-Bonnet (metric) term in dimension 4 (which, as we know, is just a boundary term). After the publication, there have been many contributions regarding the nature and/or the well-definiteness of 4DEGB (some examples are \cite{Lu2020,Bonifacio2020, Tian2020,Hennigar2020, Kobayashi2020,Tian2020,Mahapatra2020, Fernandes2020,Ai2020, Shu2020}). Here we are going to discuss some of these problematic aspects in the formulation \cite{GlavanLin2020}. The full detailed analysis of inconsistencies was performed in \cite{ArrecheaDelhomJCA2021}\footnote{
    This work contains an additional section about spherically symmetric solutions and the singularity problem (which is supposedly solved in this 4DEGB theory) that we have omitted in this thesis.} 
(see also the comment \cite{ArrecheaDelhomJCA2020}).

\newpage
Essentially, \cite{GlavanLin2020} proposed the following method to bypass the Lovelock theorem (see Section \ref{sec:introLovelock}): we consider Einstein-Gauss-Bonnet (EGB) theory in an arbitrary dimension $\dimM$ with a coupling constant for the Gauss-Bonnet term $\mathring{\mathcal{L}}_{2}^{(\dimM)}$ (given in \eqref{eq:defGB}) re-scaled by a factor of $1/(\dimM-4)$,\footnote{Here and in the following two chapters we use units $\kappa = \mpl^{-2}$, where $\mpl$ is the Planck mass.}
\begin{equation}\label{action}
    S[g_{\mu\nu}]=\int \mathrm{d}^{\dimM}x \sqrt{|g|}\left[-\Lambda + \frac{\mpl^2}{2}\mathring{R}+\frac{\alpha}{\dimM-4}\mathring{\mathcal{L}}_{2}^{(\dimM)}\right],
\end{equation}
and then we vary the dynamical equations and take the limit $\dimM\to 4$. Now we are going to dive into the problems of this procedure.

\subsection{Problem 1. The $(\dimM-4)$-factor and the definition of the equations}

The central point of the approach of \cite{GlavanLin2020} is the statement that the contribution of the Gauss-Bonnet term to the equations of motion is proportional to a factor of $(\dimM-4)$, which is cancelled with the one in the coupling constant, allowing for a well defined $\dimM\to4$ \textit{limit} at the level of the field equations. The problem here is that there are contributions to the equation of motion which are not proportional to $(\dimM-4)$ (this was also noticed in \cite{Gurses2020}). 

Consider the $k$-th order Lovelock term in an arbitrary dimension $\dimM$. When varying the action with respect to the coframe $\cofr^a$, we find
\begin{equation}\label{eq:varLovk}
\star\frac{\delta\mathring{S}_{k}^{(\dimM)}}{\delta \cofr^a} = (\dimM-2k)(\dimM-2k-1)!\ J^{(k)}_{ac}\ \cofr^c\,,
\end{equation}
where $J^{(k)}_{ac}$ is a regular tensor built from combinations of the Riemann tensor that differ for each $k$. The second factor comes from the contraction of two Levi-Civita symbols. Therefore, it is of combinatorial nature. Note that the involved counting is not a continuous process in which the number of indices being counted (or equivalently the dimension) can take any value; rather, the value must be an integer. Indeed, for \eqref{eq:varLovk} to be valid, $\dimM$ must be greater than $2k$ because a $(-1)!$ cannot arise from counting possible permutations. Since \eqref{eq:varLovk} is not valid for $\dimM=2k$, it cannot be stated that the factor $(\dimM-2k)$ is the responsible for the vanishing of \eqref{eq:varLovk} in $\dimM=2k$.
Instead, the nullity of this variation in $\dimM=2k$ is due to two reasons: (1) the curvatures in the Lagrangian (via Levi-Civita connection) do not contribute to the coframe variation (see  \cite{MardonesZanelli1991}) and (2) the explicit dependence on the coframes dissappears from the Lagrangian in $\dimM=2k$, since
\begin{equation}\label{eq:starcofrcritD}
\star (\cofr_{a_1... a_{2k}}) \overeq{\dimM=2k} F^{(k)} \LCten_{a_1 ... a_{2k}}\,,
\end{equation}
where $\LCten_{a_1 ... a_{2k}}$ is the Levi-Civita tensor associated to the anholonomic Minkowski metric (i.e., it is a constant object) and $F^{(k)}$ is a non-zero constant for each $k$. It is important to remark that the property (2) is very special of the $\dimM=2k$ case.

If we work in components, we find that the equation of motion can be rearranged as
\begin{equation}
\frac{1}{\sqrt{|g|}}\frac{\delta_{\mathrm{c}} \mathring{S}_{k}^{(\dimM)}}{\delta g^{\mu\nu}} = (\dimM-2k) A_{\mu\nu} + W_{\mu\nu}, \label{eq: AW decomposition}
\end{equation}
where no $\dimM-2k$ factor can be extracted from $W_{\mu\nu}$. For instance, the first-order Lovelock term (the Einstein-Hilbert action) leads to $A^\text{EH}_{\mu\nu}=0$ and $W^\text{EH}_{\mu\nu}=\mathring{G}_{\mu\nu}$, which vanishes in $\dimM=2$. If we split the Riemann into its irreducible parts, the Gauss-Bonnet term leads to (see also \cite{Gurses2020})
\begin{align}
   A^\text{GB}_{\mu\nu}  &=  \frac{\dimM-3}{(\dimM-2)^2} \Big[\frac{2\dimM}{\dimM-1}\mathring{R}_{\mu\nu} \mathring{R} -\frac{4(\dimM-2)}{\dimM-3}\mathring{R}^{\rho\lambda}\mathring{C}_{\mu\rho\nu\lambda}\nonumber\\
   &\qquad\qquad   -4\mathring{R}_{\mu}{}^\rho \mathring{R}_{\nu\rho}+ 2g_{\mu\nu}\mathring{R}_{\rho\lambda}\mathring{R}^{\rho\lambda} -\frac{\dimM+2}{2(\dimM-1)}g_{\mu\nu} \mathring{R}{}^2 \Big] \,,\label{eq: Amn GaussBonnet}  \\
   W^\text{GB}_{\mu\nu} &=  2 \left[ \mathring{C}_{\mu}{}^{\rho\lambda\sigma}\mathring{C}_{\nu\rho\lambda\sigma} -\frac{1}{4} g_{\mu\nu} \mathring{C}_{\tau\rho\lambda\sigma}\mathring{C}^{\tau\rho\lambda\sigma}\right] \,,
\end{align}
where we have introduced the Weyl tensor $\mathring{C}_{\mu\nu\rho\lambda}$. Taking this into account, the field equations given by \eqref{action} in arbitrary dimension are
\begin{equation}\label{fieldeqs}
\mathring{G}_{\mu\nu}+\frac{1}{\mpl^2}\Lambda g_{\mu\nu}+\frac{2\alpha}{\mpl^2}\left(A^\text{GB}_{\mu\nu}+\frac{W^\text{GB}_{\mu\nu}}{\dimM-4}\right)=0\,.
\end{equation}
We see here that the \textit{regularization} made in \cite{GlavanLin2020} (i.e. evaluating $\dimM=4$ after calculating the equations of motion in arbitrary $\dimM$) works fine for the $A^\text{GB}_{\mu\nu}$ term, giving a nontrivial contribution. However, the  $W^\text{GB}_{\mu\nu}$ term is generically ill-defined. Indeed, the 4-dimensional identity $W^\text{GB}_{\mu\nu}=0$ is due to the loss of independent components of the curvature as the dimension approaches (discretely) $\dimM=4$. It is an algebraic identity, analogous to $\mathring{G}_{\mu\nu}=0$ in $\dimM=2$. Therefore, it is not due to any proportionality to $(\dimM-4)$.\footnote{
    Examples of zero variation due to algebraic reasons are Galileon theory and some interacting massive vector theories. There, it can be seen that due to the Cayley-Hamilton theorem, the interaction Lagrangian of a given order $k$ identically vanishes for dimensions higher than the critical dimension associated to $k$ \cite{BeltranHeisenberg2016}.}

This is of course very different from what is done, e.g. in dimensional regularization. In the latter the divergent integrals in $\dimM=4$ are extended analytically to complex $\dimM$, and then the limit $\dimM\to 4$ is taken, so that the divergent and finite contributions are separated. A key aspect of dimensional regularization is that such analytic continuation is performed over scalar functions,\footnote{
    Typically the tensorial structures within the integrals are extracted from them by employing Lorentz-covariance arguments, and therefore the integral to regularize is always a scalar function.} 
whose algebraic structure is not sensitive to the value of $\dimM$. However, the quantity $W^\text{GB}_{\mu\nu}$ is tensorial, so it is not clear how to extend it to arbitrary (real or complex) dimensions (what does it mean to have $\sqrt{2}+\iN$ number of components?). Of course this problem can be solved if we can generalize appropriately the notion of limit (see, e.g., \cite{Mazur2001})

\subsection{Problem 2. Beyond linear perturbations around maximally symmetric backgrounds}

In order to avoid the previous problem we could simply restrict ourselves to metrics for which $\mathring{C}_{\mu\nu\rho\lambda}=0$ in arbitrary $\dimM$ (implying $W^\text{GB}_{\mu\nu}=0$), i.e., conformally flat geometries. One particular case are maximally symmetric spacetimes, i.e. those for which
\begin{equation}\label{eq: max symm}
  \mathring{\dfR}^{ab}= \frac{\KMaxSym}{\mpl^2(\dimM-1)}\cofr^{ab} \qquad\text{or, equivalently,}\qquad 
  \mathring{R}_{\mu\nu}{}^{\rho\sigma}=\frac{\KMaxSym}{\mpl^2(\dimM-1)}\left(\delta^\rho_\mu \delta^\sigma_\nu-\delta^\sigma_\mu \delta^\rho_\nu\right),
\end{equation}
where $\KMaxSym$ is a real constant. In this case, the (restricted) variation of the Gauss-Bonnet term is indeed proportional to $(\dimM-4)$, and the variation gives
\begin{equation}\label{fieldeqssym}
   \mathring{G}_{\mu\nu}+\frac{1}{\mpl^2}\Lambda g_{\mu\nu}+\frac{2\alpha}{\mpl^2}A^\text{GB}_{\mu\nu}=0 \,.
\end{equation}
Although this is true for conformally flat geometries, one should bear in mind that arbitrary perturbations around these backgrounds are sensitive to the ill-defined contributions that come from the $W^\text{GB}_{\mu\nu}$ dependence of the full 4DEGB field equations \eqref{fieldeqs}.

It is worth noticing, though, that the ill-defined corrections that enter the equations of motion through the $\alpha W^\text{GB}_{\mu\nu}/(\dimM-4)$ term do not contribute to linear order in perturbation theory around a maximally symmetric background. Presumably, this is the reason why these problematic contributions were unnoticed in \cite{GlavanLin2020}, where only linear perturbations were considered. Nonetheless, the ill-defined terms related to $W^\text{GB}_{\mu\nu}$ will enter the perturbations at second-order.
 
To show this, let us consider a general perturbation around a maximally symmetric background by splitting the full metric as
\begin{equation}
    g_{\mu\nu}=\bgg_{\mu\nu}+\epsilon h_{\mu\nu}
\end{equation}
where $\bgg_{\mu\nu}$ is a maximally symmetric solution of \eqref{fieldeqs}. Therefore, the l.h.s. of \eqref{fieldeqs} can be written as a perturbative series in $\epsilon$:
\begin{equation}
     E^{(0)}{}_{\mu\nu}+\epsilon E^{(1)}{}_{\mu\nu}+\epsilon^2 E^{(2)}{}_{\mu\nu}\ldots\,.
\end{equation}
Here $E^{(0)}{}_{\mu\nu}=0$ are the background field equations, $E^{(1)}{}_{\mu\nu}=0$ are the equations for linear perturbations, and so on. Using the zeroth-order equation, the linear perturbations in $\dimM$ dimensions and around a maximally symmetric background are described by
\begin{align}
0=&\left(1+\frac{4(\dimM-3)}{\dimM-1}\frac{\alpha\KMaxSym}{\mpl^4}\right)\times \bigg[\bgnabla^{\rho}\bgnabla^{\mu}h_{\nu\rho}+\bgnabla_{\rho}\bgnabla_{\nu}h^{\mu\rho}-\bgnabla^{\rho}\bgnabla_{\rho} h_{\mu\nu}-\bgnabla^{\mu}\bgnabla_{\nu}h \nonumber\\
& \qquad\qquad\qquad\qquad\qquad
+\delta^{\mu}{}_{\nu}(\bgnabla^{\sigma}\bgnabla_{\sigma}h -\bgnabla_{\rho}\bgnabla_{\sigma}h^{\rho\sigma})-\frac{\KMaxSym}{\mpl^2} (\delta^\mu_\nu h-2h^\mu{}_\nu)\bigg]\,,\label{linearpert}
\end{align}
where $h\coloneqq h^\sigma{}_\sigma$ and the indices have been raised with $\bgg^{\mu\nu}$. This equation is regular in $\dimM=4$ and, as noted in \cite{GlavanLin2020}, coincides with the GR result (up to an overall factor). Nevertheless, the problem arises at second order in perturbations. If we restrict to the simplest case $\KMaxSym=0$ (Minkowski background), the second-order equations $E^{(2)}{}_{\mu\nu}=0$ can be recast as
\begin{align}\label{quadraticpert}
0 & =[\text{GR terms of }\mathcal{O}(h^{2})]_{\mu\nu} +\frac{\alpha}{\mpl^{2}(\dimM-4)} \times \nonumber \\
 & \quad\Big[
 -2\bgnabla_{\gamma}\bgnabla_{\alpha}h_{\nu\beta}\bgnabla^{\gamma}\bgnabla^{\beta}h_{\mu}{}^{\alpha}
 +2\bgnabla_{\gamma}\bgnabla_{\beta}h_{\nu\alpha}\bgnabla^{\gamma}\bgnabla^{\beta}h_{\mu}{}^{\alpha}
 -2\bgnabla_{\mu}\bgnabla^{\gamma}h^{\alpha\beta}\bgnabla_{\nu}\bgnabla_{\beta}h_{\alpha\gamma}\nonumber \\
 &\qquad +4\bgnabla^{\gamma}\bgnabla^{\beta}h_{(\mu}{}^{\alpha}\bgnabla_{\nu)}\bgnabla_{\alpha}h_{\beta\gamma}
 +2\bgnabla_{\mu}\bgnabla^{\gamma}h^{\alpha\beta}\bgnabla_{\nu}\bgnabla_{\gamma}h_{\alpha\beta}
 -4\bgnabla^{\gamma}\bgnabla^{\beta}h_{(\mu}{}^{\alpha}\bgnabla_{\nu)}\bgnabla_{\beta}h_{\alpha\gamma} \nonumber \\
 & \qquad +\bgg_{\mu\nu}\big( 2\bgnabla_{\delta}\bgnabla_{\beta}h_{\alpha\gamma} -\bgnabla_{\delta}\bgnabla_{\gamma}h_{\alpha\beta} -\bgnabla_{\beta}\bgnabla_{\alpha}h_{\gamma\delta}\big)\bgnabla^{\delta}\bgnabla^{\gamma}h^{\alpha\beta} \Big]\,,
\end{align}
where we have substituted the 0th- and 1st-order equations. Notice that, given that the numerator of the $1/(\dimM-4)$ term comes entirely from $W^\text{GB}_{\mu\nu}$, it vanishes identically in $\dimM=4$, rendering an indeterminate $0/0$ after the {\it limit} outlined in \cite{GlavanLin2020} is taken. The situation is similar if $\KMaxSym\neq0$, and all of this seems to be in the line of \cite{Bonifacio2020}, where it was shown that the amplitudes of Gauss-Bonnet in the $\dimM\to4$ \textit{limit} correspond to those of a scalar-tensor theory. Moreover, the scalar was found to be infinitely strongly coupled, suggesting that a new (hidden at linear order) pathological degree of freedom will show up beyond linear order perturbations.

\subsection{Problem 3. An action for the regularized equations?}\label{sec:action}

At this point one may wonder: if the problem is the term $W^\text{GB}_{\mu\nu}$, why not trying to find a diffeomorphism-invariant action whose field equations in $\dimM\geq4$ are of the form \eqref{fieldeqssym} (i.e., only with the $A^\text{GB}_{\mu\nu}$ contribution)? To find such an action, one could try to subtract a scalar from the EGB action so that the contribution of $W^\text{GB}_{\mu\nu}$ disappears after taking the variation with respect to the metric, without losing the diffeomorphism symmetry of the EGB action. However we found the following no-go theorem:

\boxtheorem{
\begin{thm}
There is no Lagrangian, exclusively metric-dependent and invariant under diffeomorphisms, whose equations of motion are \eqref{fieldeqssym} in arbitrary dimensions.
\end{thm}
}

\boxproof{
\begin{proof}
For metric-dependent theory, the Noether identity associated to $\mathrm{Diff}(\mathcal{M})$ in components notation tells that the variation with respect to the metric is identically divergenceless (true off-shell for any configuration). Therefore, the divergence of \eqref{fieldeqssym} should be identically zero. After substituting $A^\text{GB}_{\mu\nu}$ from \eqref{eq: Amn GaussBonnet}, we get
\begin{align}
    \mathring{\nabla}{}^\mu\left[\frac{\mpl^2}{2}\mathring{G}_{\mu\nu}+\frac{1}{2}\Lambda g_{\mu\nu}+\alpha A^\text{GB}_{\mu\nu}\right] = \alpha\mathring{\nabla}{}^\mu A^\text{GB}_{\mu\nu} =  \frac{4\alpha}{\dimM-2}\mathring{C}_{\nu\rho\lambda\mu}\nabla^\mu \mathring{R}^{\rho\lambda}\,.
\end{align}
Now we only have to check that the r.h.s. is not identically zero in an arbitrary dimension. To see that, consider the following counterexample in five dimensions:
\begin{equation}
    {\rm d}s^2={\rm d}t^2 - {\rm e}^{2t}{\rm d}x^2- {\rm e}^{4t}({\rm d}y^2+{\rm d}z^2+{\rm d}w^2)\,,
\end{equation}
for which
\begin{equation}
    \mathring{\nabla}{}^\mu A^\text{GB}_{\mu\nu} = 4 \delta^t_\nu\neq0\,.
\end{equation}
\end{proof}
}

This result means that the $W^{\text{GB}}_{\mu\nu}$ term does not come from a scalar Lagrangian under diffeomorphisms. One might think of breaking this symmetry, but even in that case it is not possible reproduce the equations \eqref{fieldeqssym} \cite{HohmannPfeiferVoicu2021}. The other possibility is to add new fields. In particular, some authors have proposed regularizations of the action involving a scalar field of the  Horndeski type \cite{Lu2020,Fernandes2020, Bonifacio2020, Hennigar2020}.

~

To conclude this section about 4DEGB, it is important to recall that the idea of extracting well-defined contributions from topological terms by considering a divergent coupling constant is an appealing issue with an immense range of applicability. Similar ideas has been seen to lead to well-defined theories in the context of Weyl geometry \cite{BeltranKoivisto2014, BeltranKoivisto2016, BeltranHeisenbergKoivisto2016}. However one should define a consistent way of performing such procedure.

\section{Some general ideas to remember}\label{sec:stabconcl}

This chapter summarizes some of the problems we have to deal with when formulating a modification of GR (or, in general, any field theory). Ghosts, gradient instabilities, tachyons and strong coupling are just some examples of potentially dangerous issues, that are very present in these theories. These pathologies  are normally discussed around particular backgrounds, although ghosts will appear generically in theories with higher-order time derivatives by virtue of the Ostrogradski theorem. However, as we have seen, dodging the consequences of this theorem is possible if we violate the non-degenerancy hypothesis; and one way to do this is by introducing redundancies (gauge symmetries) in the theory.

We have also enumerated some modified theories of gravity with their associated problems. At this point, it is interesting to come back to our gauge metric-affine theories and make an important remark. It is actually quite common the idea that these theories allow to avoid Ostrogradski ghosts since the Lagrangian depend at most on first derivatives of the gravitational fields (within the curvature, the torsion and the nonmetricity). However, some families of metric-affine theories have been studied and they are also generally plagued by ghosts \cite{BeltranDelhom2019}. What happens here (and in extensions of these theories) can be seen from several perspectives. Let us mention two of these views:
\begin{itemize}
\item In those theories in which the connection plays the role of an auxiliary field (e.g. Einstein-Palatini) even if there are no derivatives of  metric in the action, once we solve the equation of the connection and plug the solution back into the action we may generate a Lagrangian containing Ostrogradski ghosts. This same idea can be applied to other metric-affine theories, and it is in this substitution process when the higher derivatives (that may render ghost propagations) are introduced  (see Example \ref{example:Ostro2}).
\item Once we have a Lagrangian, the way we rearrange the degrees of freedom is irrelevant. In any metric-affine theory with, e.g. higher order curvature terms, we can perform a Levi-Civita + distorsion expansion. Then, second derivatives of the metric will appear and  could lead to a pathological propagation for the graviton, for instance.
\end{itemize}
It is also worth remarking that in most of the modified theories of gravity constructed as $f(K)$, with $K$ some special gravitational invariant, new degrees of freedom are introduced due to the loss of symmetries and they are usually problematic. In Chapter \ref{ch:QTG} we will see some examples.

We insist one more time in that these problems should be carefully analyzed. In this thesis we will explore different theories and reveal some of their problems. In some of the cases the catastrophic consequences will be evident; in some others a deeper analysis is needed to finally state whether the theories are healthy. In particular, in the context of strong coupling, it may happen that the full structure of the interactions at all orders is such that this problem never appears \cite{BeltranJCA2021}. 

Finally, we have also revised the formulation of 4DEGB in \cite{GlavanLin2020}, which involves a delicate limit, as an example of theory with inconsistencies. We have highlighted the ill-definiteness of the equations \cite{Gurses2020}, the undetermined terms that enter at 2nd order in perturbations (that seem to be related with strongly coupled modes \cite{Bonifacio2020}), and the fact that one cannot derive the theory from a ``regular'' action \cite{HohmannPfeiferVoicu2021}.

\subsubsection*{About the following chapters}

In the following chapters we are going to apply these ideas to some particular examples. In the next chapter (Chapter \ref{ch:ECG}) we will see a great example of the consequences of perturbing a strongly coupled background. In Chapter \ref{ch:QTG}, among other things, we will use the introduction of gauge symmetries to avoid the presence of pathological modes in the teleparallel restriction of the even MAG Lagrangian \eqref{eq: qMAGLeven}. Finally, in Chapter \ref{ch:MAGspectrum} we will start exploring the spectrum of the full quadratic MAG.


\chapter{Strong coupling of cosmological models in Einsteinian Cubic Gravity and beyond}\label{ch:ECG}

\chaptermark{Strong coupling of cosmological models in ECG and beyond}

\boxquote{
The history of science is full of cases where previously accepted theories and hypotheses have been entirely overthrown, to be replaced by new ideas that more adequately explain the data. (...) This self-questioning and error-correcting aspect of the scientific method is its most striking property, and sets it off from many other areas of human endeavor where credulity is the rule.}{Carl Sagan, ``Broca's Brain'' (1979), p. 96.}

We start the chapter by introducing the theories we are going to deal with. \emph{Einsteinian Cubic Gravity} \cite{BuenoCano2016b} is a higher curvature theory of gravity defined to possess the same linear spectrum as General Relativity, i.e., a massless spin-2 field, around maximally symmetric spacetimes in arbitrary dimension \cite{OlivaRay2010, MyersRobinson2010, BuenoCano2016b}. In our analysis we are going to consider the extension of Einsteinian Cubic Gravity that was introduced in \cite{ArciniegaEdelstein2018} (ECG).\footnote{
    Notice that throughout this chapter, by ECG we do not mean the original Einsteinian Cubic Gravity but its cosmological generalization.}  
The latter has the same property as the original Einsteinian Cubic Gravity but around arbitrary cosmological scenarios (Friedmann-Lema\^itre-Robertson-Walker (FLRW)), not only the maximally symmetric ones. Another way of generalizing the original Einsteinian Cubic Gravity is by relaxing the definition. In particular one can consider the higher curvature theory with the GR linear spectrum around maximally symmetric backgrounds in a given dimension (not necessarily in arbitrary dimensions). The resulting theories are the so-called \emph{Generalized Quasi-Topological Gravity theories} (GQTG) \cite{BuenoCanoHennigar2019b}. 

The main goal of this chapter is to analyze the presence of instabilities in the cosmological solutions of ECG (see e.g. \cite{ArciniegaEdelstein2018, CisternaGrandi2020,  ArciniegaBuenoCano2018}), in which the symmetries allow a more straightforward analysis of the pathologies. However, as we will discuss, other backgrounds are also prone to these problems \cite{HennigarMann2017, BuenoCano2016, HennigarKubiznak2017, BuenoCano2017, Ahmed2017, BuenoCano2017b, Feng2017, Hennigar2017, HennigarPoshteh2018, BuenoCanoHennigar2019, BuenoCanoHennigar2018, PoshtehMann2019, MirHennigar2019,MirMann2019,Mehdizadeh2019,Erices2019}. At the end of the chapter, we will study GQTG corrections and show that cosmological solutions are even more pathological than in the ECG case.

\section{Previous indications of potential problems}

As it is well-known, Lovelock terms (see Definition \ref{def:Lovelock}) are the only higher curvature theories that share the field content of GR around any background or, in other words, at the full non-linear order \cite{Lovelock1970,Lovelock1971}. Therefore, ECG necessarily contains additional dofs which in turn are associated to the higher than second-order nature of the field equations.\footnote{
    More precisely, one should say that these theories are neither any of the Lovelock Lagrangians nor are they related via a regular field redefinition to them. See \cite{BuenoCanoMoreno2019} for an interesting discussion about the role of field redefinitions within the framework of GQTG where it is shown that they constitute a complete basis for the Lagrangians of the type $\mathcal{L}(g_{\mu\nu},\mathring{R}_{\mu\nu\rho\lambda})$.}
Among these additional dofs there will be ghostly modes associated to an Ostrogradski instability. In an attempt to avoid this pathology, ECG selects a particular cubic polynomial of the Riemann tensor so that the field equations become second order around specific backgrounds. This construction amounts to requiring that those backgrounds correspond to surfaces in phase space, or in the space of solutions, where the principal part of the system of equations is singular. This has many associated problems. For instance, standard perturbation theory will not be well-defined around such a singular surface or, at least, not in a standard form. Notice that the principal part becomes singular not at a given point or surface in the spacetime (in which case regular solutions might still exist thanks to the mildness of the singular behavior of the principal part in that situation), but on a surface in the space of solutions, which is a fundamentally more pathological situation. Thus, the very construction of the theory suggests that the backgrounds with the same linear spectrum as GR will be strongly coupled (see Section \ref{sec:strongcoupling}) and, therefore, they cannot correspond to stable trajectories in phase space, i.e., no physical curves in phase space will smoothly evolve towards those solutions. Notice that this pathology is independent from the existing Ostrogradski instability in arbitrary higher order curvature theories and it would exist even if the full theory did not contain any ghosts. What this discussion suggests is that the strongly coupled modes will in turn be associated to the ghosts present in the theory. 

Recently, it has been shown the presence of instabilities in these theories for inflationary solutions in \cite{PookkillathDeFelice2020}. We we will confirm this findings and complement them with extra evidence for the generic pathological character of these solutions. 

~

\boxsimple{
\noindent {\bf Note on conventions:} In this chapter we are using the mostly plus convention for the metric (signature $(-,+,+,+)$) and the Riemann, the Ricci tensor and the curvature scalar differ from our original definition in a global sign. The rest remains the same.
}

\section{(Cosmological) Einsteinian Cubic Gravity}

The action of ECG is
\begin{equation} \label{eq:ECGaction}
    S = \int \dex^4x \sqrt{|g|} \left(- \Lambda + \frac{\mpl^2}{2} \mathring{R} + \frac{\pECG}{\mpl^2} \mathcal{R}_{(3)} \right)\,,
\end{equation}
where the first two terms simply reproduce the pure GR sector with cosmological constant, and $\mathcal{R}_{(3)}$ is a cubic polynomial of the Riemann tensor given by
\begin{align}
    \mathcal{R}_{(3)} & = - \frac{1}{8} \Big(
    12 \mathring{R}_\mu{}^\rho{}_\nu{}^\sigma \mathring{R}_\rho{}^\tau{}_\sigma{}^\eta \mathring{R}_\tau{}^\mu{}_\eta{}^\nu  
    +\mathring{R}_{\mu\nu}{}^{\rho\sigma} \mathring{R}_{\rho\sigma}{}^{\tau\eta} \mathring{R}_{\tau\eta}{}^{\mu\nu}
    +2 \mathring{R} \mathring{R}_{\mu\nu\rho\sigma}\mathring{R}^{\mu\nu\rho\sigma}\nonumber \\
    & \qquad \qquad
    -8 \mathring{R}^{\mu\nu}\mathring{R}_\mu{}^{\rho\sigma \tau}\mathring{R}_{\nu\rho\sigma\tau}
    +4 \mathring{R}^{\mu\nu}\mathring{R}^{\rho\sigma}\mathring{R}_{\mu\rho\nu\sigma}
    -4 \mathring{R} \mathring{R}_{\mu\nu}\mathring{R}^{\mu\nu}
    +8 \mathring{R}_\mu{}^\nu \mathring{R}_\nu{}^\rho \mathring{R}_\rho{}^\mu
    \Big)\,.
\end{align}
The relative coefficients are carefully selected to guarantee that the linear spectrum of the theory around maximally symmetric and cosmological backgrounds is the same as that of GR, i.e., only the usual two polarizations of the gravitational waves propagate.

For a FLRW universe of the type
\begin{equation}
  \dex s^2=-\dex t^2+a^2(t) \delta_{ij} \dex x^i\dex x^j\,, \label{eq:FLRW}
\end{equation}
the gravitational Friedmann equation reads
\begin{equation}\label{eq: Isotropic condition}
3\mpl^2\HubH^2(t)-6\frac{\pECG}{\mpl^2} \HubH^6(t)=\rho+\Lambda\,,
\end{equation}
where $\rho$ is the energy-density of the matter sector and $\HubH(t)\coloneqq \frac{\dot a}{a}$ is the Hubble parameter. In the absence of any matter $\rho=0$, we can see that we have (at most) three branches of expanding de Sitter solutions. Among these de Sitter branches, the stability of the gravitational waves will impose some stability conditions. Indeed, if we consider metric perturbations $g_{ij}=a^2(\delta_{ij}+h_{ij})$ with $h_{ij}$ transverse and traceless, the corresponding quadratic action for the tensor perturbations around the de Sitter solutions is given by (see e.g. \cite{PookkillathDeFelice2020})
\begin{equation} \label{Eq:actiontensormodes}
S^{(2)}= \frac{\mpl^4-6 \HubH^4_0\pECG}{8 \mpl^2} \sum_{\lambda} \int \dex t \, \dex^3x \, a^3 \left[ \dot{h}^2_\lambda- \frac{1}{a^2} (\partial_i h_\lambda)^2\right],
\end{equation}
where the sum extends to the two polarizations of the gravitational waves and $\HubH_0$ is the considered de Sitter branch. From this expression it follows that we need to have $\mpl^4 - 6 \HubH^4_0\pECG>0$ in order to avoid ghostly gravitational waves. As a consequence of this condition, the only allowed de Sitter solutions satisfy $\Lambda > 2 \HubH^2_0>0$ \cite{PookkillathDeFelice2020}.

This quadratic action for the tensor modes explicitly shows the property of these theories that only the usual polarizations of the GWs propagate on a cosmological background at linear order. However, being a higher order curvature theory outside the Lovelock class, the full theory is expected to contain extra degrees of freedom, that will enter at higher orders in perturbation theory.

\section{Approaching FLRW from Bianchi I in ECG}

To show the singular nature of the FLRW solutions of ECG, the idea will be to break the isotropy of the spacetime while keeping homogeneity, and study the evolution of the universe near the isotropic case.\footnote{
    A gravitational wave with a sufficiently long wavelength mimics the shear of a Bianchi I universe, so our study will give the non-linear evolution of gravitational waves, but restricted to the infrared sector.}
In our analysis, which partially overlaps with the results presented in \cite{PookkillathDeFelice2020}, we will discuss some subtleties and give complementary arguments that will support the pathological character of these solutions.

\subsection{Bianchi I solutions. Theory-independent generalities}
Consider a general gravitational action $S[g_{\mu\nu}]$, as well as an arbitrary Bianchi I spacetime. The latter is described by the line element
\begin{equation}
\dex s^2=-\mathcal{N}^{\,2}(t)\dex t^2+a^2(t) \dex x^2+b^2(t) \dex y^2+c^2(t) \dex z^2 \qquad \eqqcolon g^\text{I}{}_{\mu\nu} \,, \label{eq: BianchiI}
\end{equation}
where $\mathcal{N}(t)$ is the lapse function and $a(t)$, $b(t)$ and $c(t)$ stand for the scale factors along the three coordinate axis. We will work in cosmic time, i.e. $\mathcal{N}(t)=1$, but we need to keep it general to correctly derive the dynamical equations. To make a more
direct contact with the isotropic FLRW solutions, it is convenient to introduce the isotropic scale factor ${\bar a}\coloneqq (abc)^{1/3}$ with the corresponding expansion rate
\begin{equation}\label{eq:defH}
\HubH(t)\coloneqq \frac{\dot{\bar a}}{\bar a}=\frac{1}{3}\bigg(\frac{\dot{a}}{a}+\frac{\dot{b}}{b}+\frac{\dot{c}}{c}\bigg)\,.
\end{equation}
In addition, we will encode the anisotropic part in two functions, $\sigma_1(t)$ and $\sigma_2(t)$, defined implicitly by 
\begin{equation} \label{eq: def sigmas}
\frac{\dot{a}}{a}= \HubH + \eanis (2\sigma_1-\sigma_2)\,, \qquad 
\frac{\dot{b}}{b}= \HubH - \eanis (\sigma_1-2\sigma_2)\,, \qquad 
\frac{\dot{c}}{c}= \HubH - \eanis (\sigma_1+\sigma_2)\,,
\end{equation}
where $\eanis$ is certain (not necessarily small) fixed parameter representing the deviation with respect to the isotropic case ($\eanis=0$). Notice that these definitions are consistent with \eqref{eq:defH}. 

Since the metrics of the type \eqref{eq: BianchiI} fulfill the requirements of the \emph{Palais' principle of symmetric criticality} \cite{Palais1979} (see also \cite{FelsTorre2002,DeserTekin2003,DeserFranklinTekin2004}), one can use the minisuperspace approach\footnote{
    This is also known as Weyl method, because Weyl made use of it to derive the Schwarzschild solution of the Einstein field equations in \cite{Weyl1922}.}
 and substitute the Ansatz \eqref{eq: BianchiI} in the action before taking the variation. We denote the resulting action as
\begin{equation}
  \bar{S} [\mathcal{N}, a, b, c] \coloneqq S[g^\text{I}{}_{\mu\nu}(\mathcal{N}, a, b, c)]\,.
\end{equation}
It will be also convenient to introduce the following notation for some combinations of the equations of motion
\begin{align} 
  {\rm E}_{ab} \coloneqq ~~ 0 & = \frac{\delta \bar{S}}{\delta a}a - \frac{\delta \bar{S}}{\delta b}b\,, &
  {\rm E}_{abc}\coloneqq ~~ 0 & =\frac{1}{3} \left(\frac{\delta\bar{S}}{\delta a}a+\frac{\delta\bar{S}}{\delta b}b+\frac{\delta\bar{S}}{\delta c}c\right)\,,\nonumber \\
  {\rm E}_{cb} \coloneqq ~~ 0 & = \frac{\delta \bar{S}}{\delta c}c - \frac{\delta \bar{S}}{\delta b}b\,,  &
  {\rm E}_\mathcal{N} \coloneqq ~~ 0 & =\frac{\delta \bar{S}}{\delta \mathcal{N}}\,,\nonumber \\
  {\rm E}_{ca} \coloneqq ~~ 0 & = \frac{\delta\bar{S}}{\delta c}c-\frac{\delta\bar{S}}{\delta a}a\,.
\end{align}
Notice that not all of them are independent equations. The set $\{{\rm E}_{abc}, {\rm E}_{ab}, {\rm E}_{cb},  {\rm E}_{ca} \}$ is indeed linearly dependent, but there is an additional constraint we should take into account: the Noether identity associated to diffeomorphisms,
\begin{equation}
  \frac{\dex}{\dex t}\left(\frac{\delta \bar{S}}{\delta \mathcal{N}}\right) + 
  \left(\frac{\dot{\mathcal{N}}}{\mathcal{N}}+ \frac{\dot{a}}{a}+\frac{\dot{b}}{b}+\frac{\dot{c}}{c} \right)\frac{\delta \bar{S}}{\delta \mathcal{N}} - \frac{1}{\mathcal{N}}
  \left(\dot{a}\frac{\delta \bar{S}}{\delta a} +\dot{b}\frac{\delta \bar{S}}{\delta b} + \dot{c}\frac{\delta \bar{S}}{\delta c}\right)=0\,.
\end{equation}
The parameterization \eqref{eq: def sigmas} for the shears $\sigma_1$ and $\sigma_2$ has been chosen so that it is convenient to work with $\{{\rm E}_\mathcal{N},  {\rm E}_{cb},  {\rm E}_{ca}\}$. However, in order to obtain a more direct generalization of the results obtained in \cite{PookkillathDeFelice2020}, only in Section \ref{subsec: perturbative sol} we will make the (equivalent) choice $\{{\rm E}_\mathcal{N},  {\rm E}_{ab},  {\rm E}_{ca}\}$, which agrees with the one made by the authors of that paper.

\subsection{Perturbative solution around de Sitter spacetime} \label{subsec: perturbative sol}
In  \cite{PookkillathDeFelice2020} it was argued that de Sitter is a stable perturbative solution of \eqref{eq:ECGaction} in the case of an axisymmetric Bianchi I (the case $c(t)=b(t)$). The idea will be to reproduce the same analysis for the general Bianchi I case in order to clarify some subtle shortcomings of the solutions generated perturbatively. 

From \eqref{eq: Isotropic condition} one can trivially obtain that the de Sitter spacetime given by
\begin{equation}
  \mathcal{N}(t)=1\quad\text{and}\quad a(t)=b(t)=c(t)=\eN^{\HubH_0t}
\end{equation}
is a solution of $\{{\rm E}_\mathcal{N},  {\rm E}_{ab},  {\rm E}_{ca}\}$ if and only if the cosmological constant and the Hubble
parameter fulfil
\begin{equation}
  \Lambda=3\frac{\HubH_0^2}{\mpl^2} (\mpl^4-2\HubH_0^4\pECG)\,.
\end{equation}
Now we are going to take this isotropic configuration as our background and perform a perturbative expansion around it,
\begin{equation}
  a(t) =a^{(0)}(t)+ \sum_{k=1}^\infty \epsilon^k a^{(k)}(t)\,, \quad
  b(t) =a^{(0)}(t)+ \sum_{k=1}^\infty \epsilon^k b^{(k)}(t)\,, \quad
  c(t) =a^{(0)}(t)+ \sum_{k=1}^\infty \epsilon^k c^{(k)}(t)\,,
  \label{Eq:perturbativeAnsatz}
\end{equation}
where $a^{(0)}(t)\coloneqq\eN^{\HubH_0 t}$ and $\epsilon$ is the (small) perturbation parameter, which should not to be confused with the anisotropic parameter $\eanis$ introduced above. 

Since the background is a solution, if we substitute the perturbative expansion \eqref{Eq:perturbativeAnsatz} into our set of dynamical equations $\{{\rm E}_\mathcal{N},  {\rm E}_{ab},  {\rm E}_{ca}\}$, we expect the first non-trivial contribution to appear at first order. Indeed, we get the following system of second-order differential equations
\begin{align}
  0 & = \dot{a}^{(1)} + \dot{b}^{(1)}+ \dot{c}^{(1)}- \HubH_0 \big(a^{(1)}+b^{(1)}+c^{(1)}\big)  \,,\nonumber \\
  0 & =  \ddot{b}^{(1)} - \ddot{a}^{(1)}+ \HubH_0 \big(\dot{b}^{(1)} - \dot{a}^{(1)}\big) - 2\HubH_0^2 \big(b^{(1)} -a^{(1)}\big) \,,\nonumber \\
  0 & =\ddot{c}^{(1)} - \ddot{a}^{(1)}+ \HubH_0 \big(\dot{c}^{(1)} - \dot{a}^{(1)}\big) - 2\HubH_0^2 \big(c^{(1)} -a^{(1)}\big)\,,
  \label{eq:firstorder}
\end{align}
provided $\mpl^4 \neq 6 \HubH_0^4\pECG$. As shown in \eqref{Eq:actiontensormodes}, this is in turn a necessary condition to avoid a pathological behavior of the tensor modes, which requires $\mpl^4 - 6 \HubH_0^4\pECG>0$ (see also \cite{PookkillathDeFelice2020}). The general solution for this first order contribution is
\begin{align}
  a^{(1)} & =C_1\eN^{-2\HubH_0t}+C_3\eN^{\HubH_0t}\,,\nonumber \\
  b^{(1)} & =C_2\eN^{-2\HubH_0t}+C_4\eN^{\HubH_0t}\,,\nonumber \\
  c^{(1)} & =-(C_1+C_2)\eN^{-2\HubH_0t}+C_5 \eN^{\HubH_0t}\,,
\end{align}
for some integration constants $C_i$ ($i=1,...,5$). 
We can proceed analogously to obtain the solution at second order that is found to be
\begin{align}
a^{(2)} & = 
  \frac{C_1^2-C_1C_2-C_2^2}{4}\frac{\mpl^4-258\HubH_0^4\pECG}{\mpl^4-6\HubH_0^4\pECG}\eN^{-5\HubH_0t}+
  \frac{1}{3}\big[\bar{C}+D_1\big]\eN^{-2\HubH_0t}+
  D_3\eN^{\HubH_0t}\,,\nonumber \\
b^{(2)} & =
  \frac{C_2^2-C_1C_2-C_1^2}{4}\frac{\mpl^4-258\HubH_0^4\pECG}{\mpl^4-6\HubH_0^4\pECG}\eN^{-5\HubH_0t}+
  \frac{1}{3}\big[\bar{C}+D_2\big]\eN^{-2\HubH_0t}+
  D_4\eN^{\HubH_0t}\,,\nonumber \\
c^{(2)} & =
  \frac{C_1^2+3C_1C_2+C_2^2}{4}\frac{\mpl^4-258\HubH_0^4\pECG}{\mpl^4-6\HubH_0^4\pECG}\eN^{-5\HubH_0t}+
  \frac{1}{3}\big[\bar{C}-D_1-D_2\big]\eN^{-2\HubH_0t}+
  D_5\eN^{\HubH_0t}\,,
\end{align}
where $D_{i}$ ($i=1,...,5$) are new integration constants and $\bar C\coloneqq C_1C_3+C_2C_4-(C_1+C_2)C_5 $. 

The resulting perturbative expansion reproduces the exactly isotropic de Sitter solution at all orders, since the anisotropic contributions decay exponentially making the whole series converge to it. Notice that the perturbative contributions proportional to $\eN^{\HubH_0 t}$ can be absorbed into the background solution. According to \cite{PookkillathDeFelice2020}, this behavior guarantees the existence of FLRW solutions. However, some care must be taken to correctly interpret this perturbative solution since the zeroth order corresponds to a singular surface in phase space where dynamical dofs disappear, as we will see in the next section. Consequently, a standard perturbative expansion around this surface can be problematic and the conclusions drawn from it can be flawed. 

In the present case, there is something crucial to notice: each order is obtained by solving second order equations, while the full equations are known to be fourth order. This can be seen in the equations at first order \eqref{eq:firstorder}, and it is easy to understand that this will be the case at all orders in perturbation theory. The reason is that the coefficients of the terms with third and fourth derivatives of the scale factors at $n$-th order must be evaluated on the purely isotropic zeroth order solution. But we know that the isotropic case, by definition of the theory, gives second order differential equations, so the coefficients of these terms must vanish necessarily. The importance of this observation is that it implies that we are necessarily missing perturbative modes along specific directions whose stability is not under control and, certainly, they are not captured by the perturbatively-generated solution. From this analysis, we cannot conclude that the de Sitter solution is a good background solution. In Appendix \ref{app:toyexample} we illustrate these issues with a simple one-dimensional mechanical toy example.

\subsection{Shear equations}

In the theory \eqref{eq:ECGaction}, one finds that the highest order derivatives of the anisotropy functions $\sigma_1$ and $\sigma_2$ appear in the evolution equations (shear equations), ${\rm E}_{cb}$ and ${\rm E}_{ca}$. Obviously, these equations trivialize in the isotropic case because they describe the evolution of the shear $\sigma_1$ and $\sigma_2$, which means that there is an overall factor $\eanis$ (that goes to zero in the isotropic limit). In the absence of any anisotropic stress, as we are considering, and away from the isotropic case, the shear evolution equations can be taken to be $\{{\rm E}_{cb}/\eanis, {\rm E}_{ca}/\eanis\}$ (we remove the global $\eanis$ factor), which can be written in the following schematic form:
\renewcommand\arraystretch{1.3}
\begin{equation}\label{eq: principal part}
  \frac{\pECG}{\mpl^2} \left[ \eanis
    {\bf M_1}
    \begin{pmatrix}\dddot{\sigma}_2 \\\dddot{\sigma}_1 \end{pmatrix}
    + \eanis 
    {\bf M_2} \begin{pmatrix}\ddot{\sigma}_2 \\\ddot{\sigma}_1 \end{pmatrix}+  {\bf V} \right] 
    +3 \mpl^2 \begin{pmatrix}3\HubH\sigma_2+\dot{\sigma}_2 \\ 3\HubH\sigma_1+\dot{\sigma}_1 \end{pmatrix} =0 \,.
\end{equation}
\renewcommand\arraystretch{1}
where the matrices ${\bf M}_{1}$ and  ${\bf M}_{2}$, and the column vector ${\bf V}$ start at zeroth order in $\eanis$.\footnote{
    To be more specific, the components of ${\bf M}_1$ depend polynomially on $\sigma_1$, $\sigma_2$ and $\HubH$, whereas those of ${\bf M}_2$ and ${\bf V}$ also depend on $\dot{\sigma}_1$, $\dot{\sigma}_2$ and the derivatives of $\HubH$.}
This equation shows how the higher order terms containing second and third derivatives of the shear trivialize in the isotropic limit $\eanis\rightarrow0$. Consequently, in this limit, the order of the corresponding differential equations is reduced. This does not imply that the shear evolution is not modified by the ECG term in the action, since the usual GR evolution (described by the last term in the l.h.s. of \eqref{eq: principal part}) receives corrections from ${\bf V}$.

Now that we have obtained the non-linear equation for the anisotropic homogeneous modes, we can corroborate that the isotropic solution lies on a singular surface of phase space. This means that solutions near the singular isotropic surface can never end in the isotropic solution. At best, a given trajectory could approach the isotropic solution, but its intrinsically singular nature prevents the possibility of making any reliable claim. In particular, this is the reason why the perturbative expansion of Section \ref{subsec: perturbative sol} fails to capture the full perturbative spectrum around the isotropic solution.

\subsection{Complete dynamical analysis}

In order to go deep into the pathological character of the isotropic solutions we can consider the full system of equations, i.e., the shear equations \eqref{eq: principal part} together with the lapse equation ${\rm E}_\mathcal{N}$, in terms of $\sigma_1$, $\sigma_2$ and $\HubH$ (we take $\eanis=1$ from now on). The highest derivatives of the isotropic Hubble expansion rate $\HubH$ and the shears $\sigma_1$ and $\sigma_2$ that appear in each of them are:
\begin{equation}\label{eq:highestDeqs}
  {\rm E}_\mathcal{N}~:~( \ddot\HubH,  \ddot\sigma_1,  \ddot\sigma_2) \,,\qquad
  {\rm E}_{cb}       ~:~(\dddot\HubH, \dddot\sigma_1, \dddot\sigma_2)\,,\qquad
  {\rm E}_{ca}       ~:~(\dddot\HubH, \dddot\sigma_1, \dddot\sigma_2)\,.
\end{equation}
Although the shear equations contain third order derivatives of $\HubH$, they can be eliminated by taking successive time derivatives of ${\rm E}_\mathcal{N}$. This procedure results in additional corrections to the coefficients of $\dddot\sigma_1$ and $\dddot\sigma_2$. We are interested in obtaining the matrix of the principal part of the equations once they are written in the discussed normal form, i.e., with derivatives only up to the orders $(\dddot\sigma_1, \dddot\sigma_1, \ddot\HubH)$.

At this point, it is convenient to factor the isotropic expansion out by introducing the variables
\begin{equation} X(t)\coloneqq \frac{\sigma_1(t)}{\HubH(t)} \qquad \text{and} \qquad Y(t) \coloneqq\frac{\sigma_2(t)}{\HubH(t)}\,, \end{equation}
and work with the number of e-folds
\begin{equation}  \dex N=\HubH(t)\dex t \label{eq: N efolds}\end{equation}
as time variable (we will use a prime to represent the derivative with respect to $N$). After these manipulations, the full system of equations can be written as
\begin{equation} \label{eq: systemeqs}
\mathcal{H}^i{}_j (\xi^j)' + F^i(X,Y,X',Y',X'',Y'',\HubH,\HubH')=0\,,
\end{equation}
where $\vec{\xi}\coloneqq(\HubH',X'',Y'')$, $\vec{F}$ is a vector that depends on the displayed dynamical variables and $\mathcal{H}^i{}_j$ is the desired matrix of coefficients for the principal part. The determinant of this matrix is given by
\begin{align}\label{eq: detHess}
     \det \mathcal{H}^i{}_j &= - \frac{1458 \pECG^3 \HubH^{15}}{\mpl^{12}} \big(X^2-2Y^2+2Y-X+2XY\big)\times\nonumber\\
     &\qquad\times\big(Y^2-2X^2+2X-Y+2XY\big)\big(Y^2+X^2-4XY-X-Y\big)\,.
\end{align}

Besides the singular curves given by $\det \mathcal{H}^i{}_j=0$, there is an additional separatrix associated to a null eigenvalue along the $\HubH''$ direction given by
\begin{equation}\label{eq: extrasep}
    2\big(X^3+Y^3\big)- 3\big(XY^2+X^2Y\big) -2\big(X^2+Y^2\big)+2XY=0\,.
\end{equation}
This separatrix does not appear from the vanishing of the determinant because the other two eigenvalues diverge on this curve in such a way that the determinant remains finite. Thus, we also have to consider this separatrix in our analysis.
Notice that the equations of the singular separatrices in \eqref{eq: detHess} and \eqref{eq: extrasep} are invariant under  $X\leftrightarrow Y$ (i.e., $\sigma_1\leftrightarrow\sigma_2$) and only depend algebraically on $X$, $Y$. Consequently, these separatrices correspond to singular hypersurfaces in phase space that are orthogonal to the plane $(X,Y)$ or, equivalently, parallel to the other directions.\footnote{
    Let us be more explicit on this and notice that the considered phase space is spanned by the coordinates $(\HubH,X,Y,\HubH',X',Y',X'',Y'')$. The singular surfaces then exhibit a symmetry under translations in the subspace $(\HubH,\HubH',X',Y',X'',Y'')$.}

\begin{figure}[H]
  \begin{center}
    \includegraphics[width=0.4\textwidth]{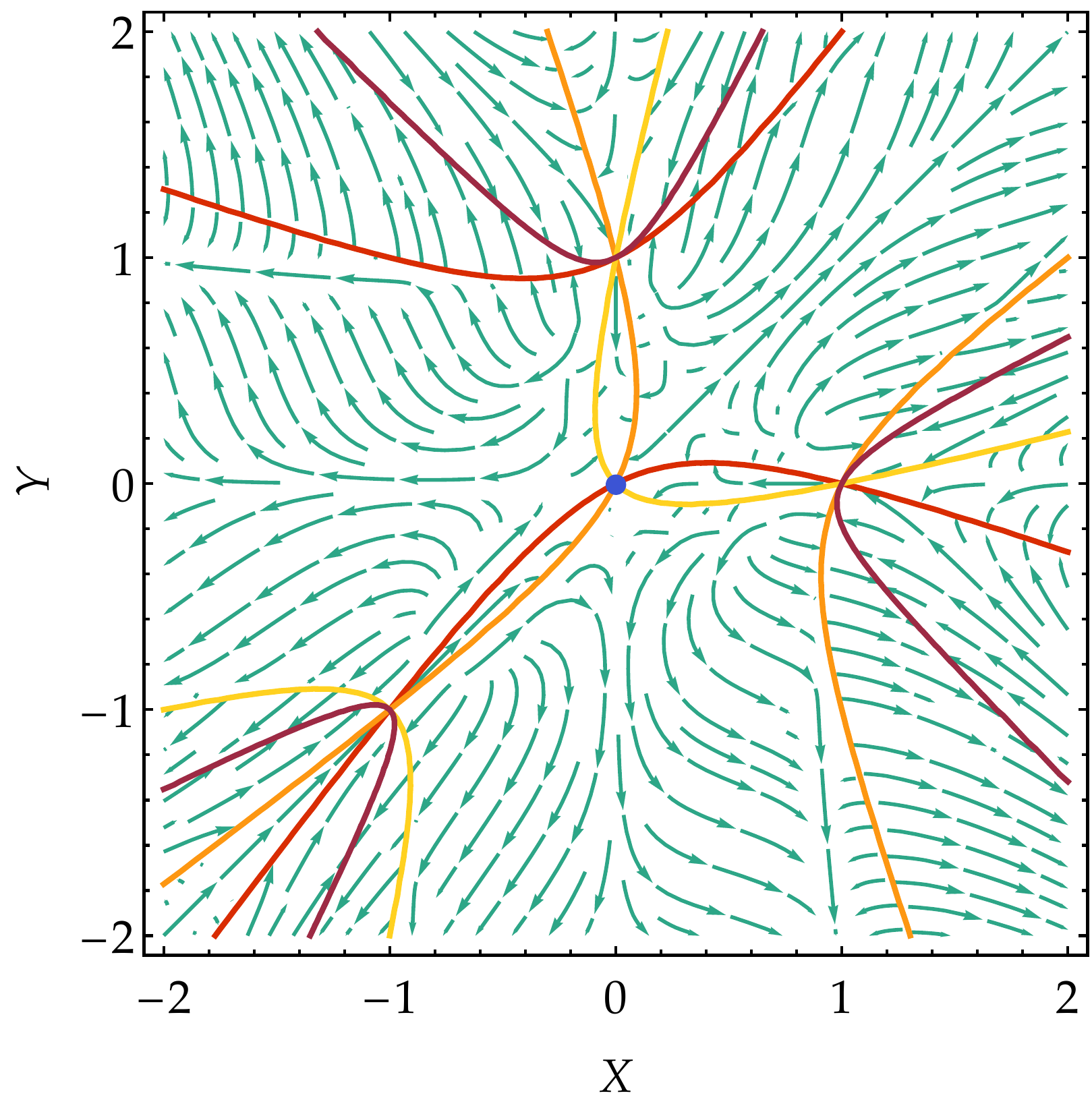} \qquad
    \includegraphics[width=0.42\textwidth]{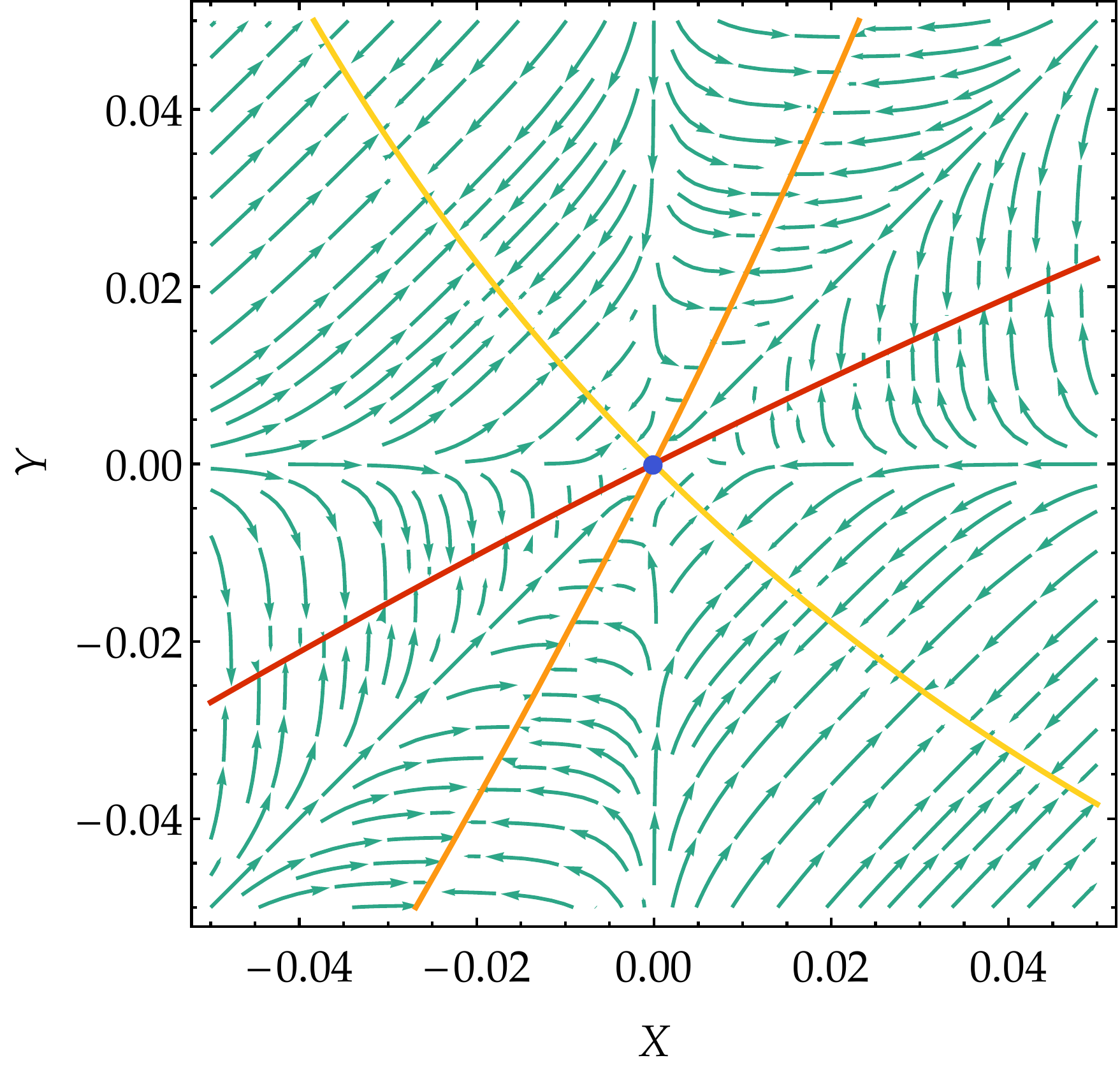}
    \quad
    \includegraphics[width=0.4\textwidth]{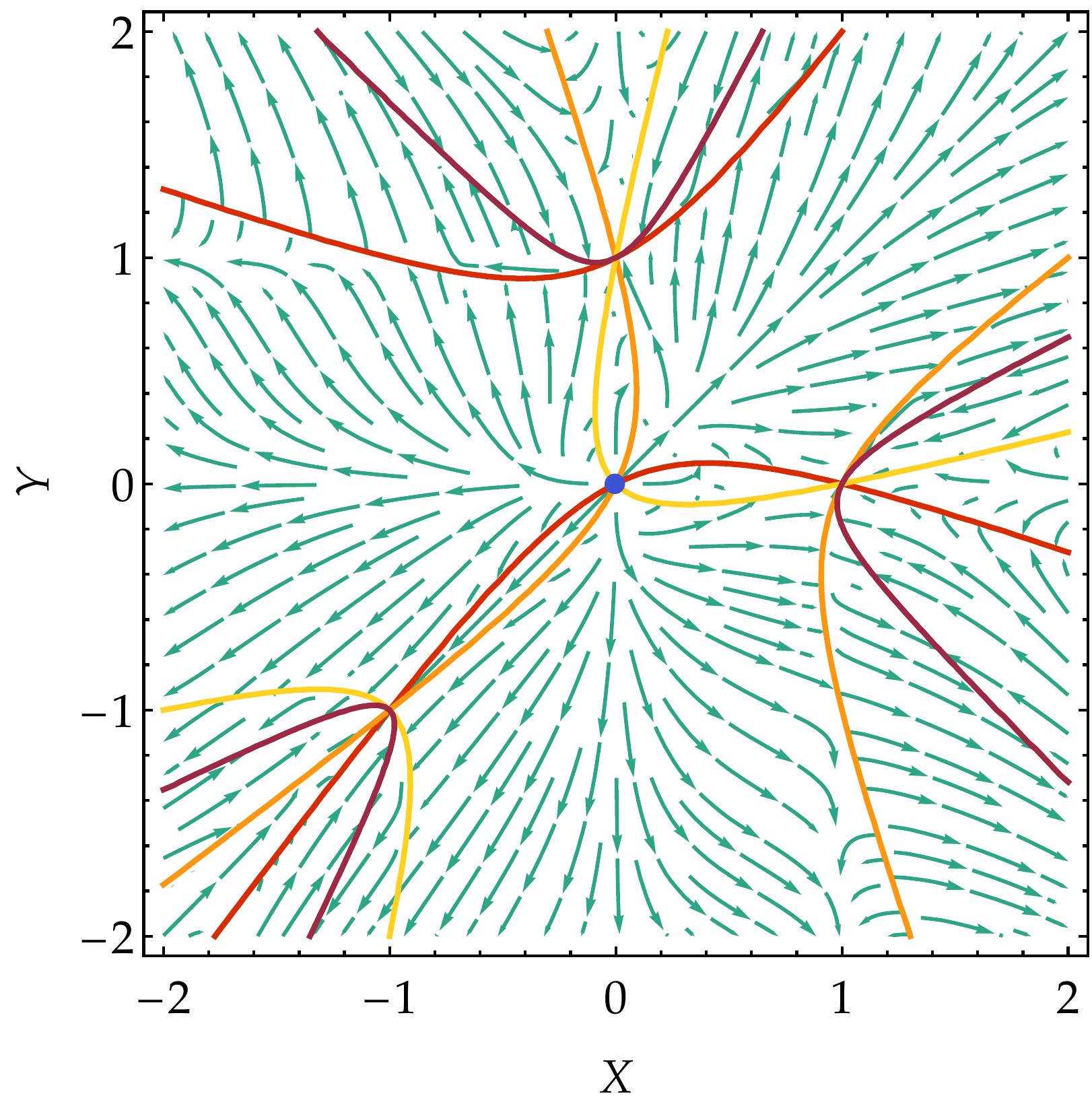}\qquad
    \includegraphics[width=0.42\textwidth]{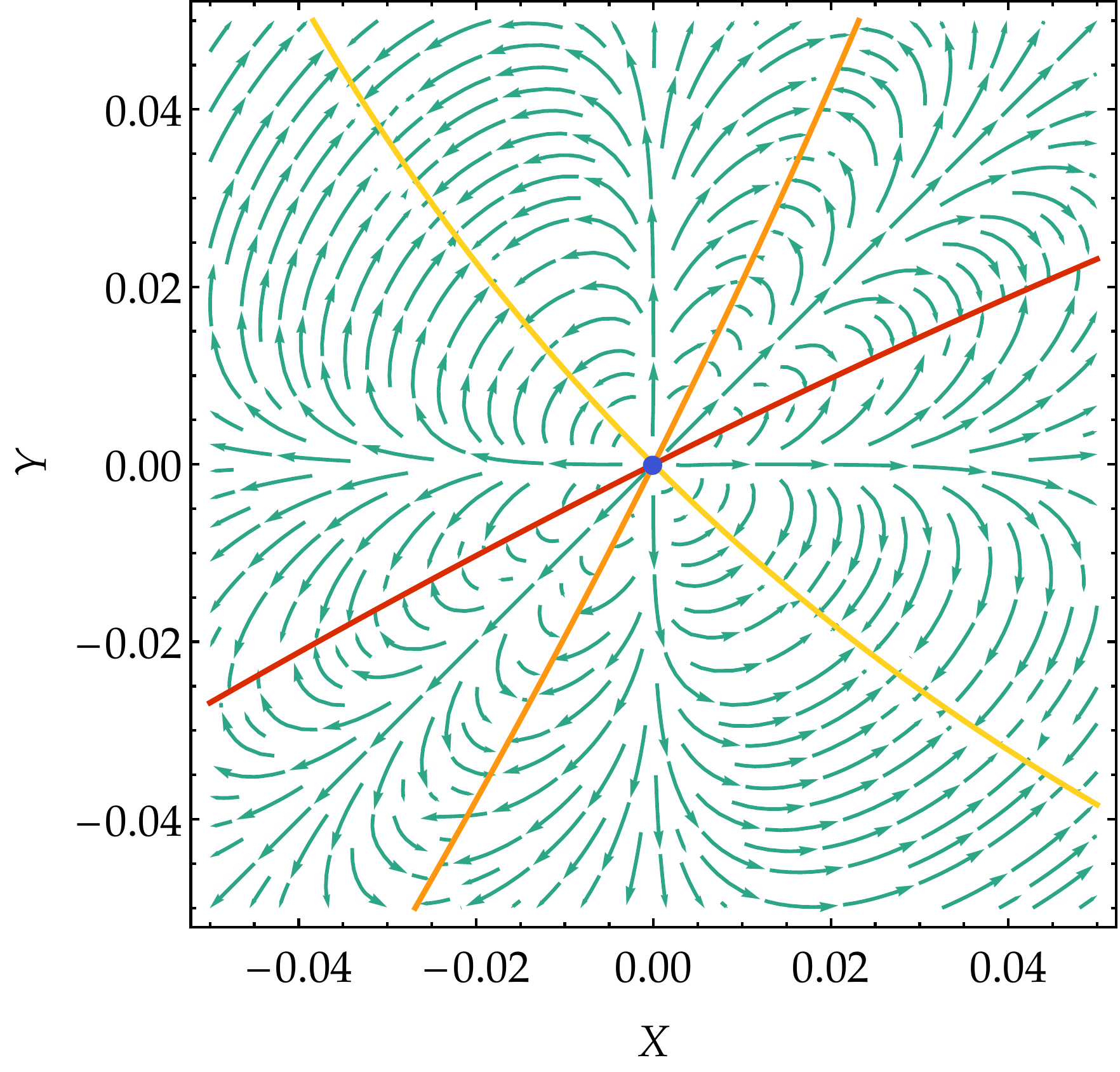}
  \end{center}
  \caption{\label{fig:phasemapECG}
  These plots show how the field $(\xi^2,\,\xi^3)'=(X''',Y''')$ behaves with respect to $X$ and $Y$ under the conditions $\HubH'=X'=Y'=X''=Y''=0$ and $\HubH=\HubH_\text{dS}$. The curves in red, orange and yellow are, respectively, the three branches of singular separatrices as they appear in \eqref{eq: detHess}, while the dark red one is the additional separatrix that does not appear in the Hessian determinant. The isotropic solution corresponds to the blue point at the origin. We can also see other three distinctive singular points that correspond to the physical solutions where only one of the directions expands (see Table \ref{tab: intersections}). The first two plots correspond to the value $\pECG \mpl^4/\HubH^4_\text{dS} = 0.1$ and the last ones to $\pECG \mpl^4/\HubH^4_\text{dS} = 0.001$. 
  }
\end{figure}

In Fig. \ref{fig:phasemapECG}, the curves displayed in yellow, orange and red are the critical curves where the determinant \eqref{eq: detHess} vanishes, while the dark red ones represent the additional separatrix given by  \eqref{eq: extrasep}.
Notice that the isotropic point ($X=Y=0$) is crossed by the first three, whereas it is just an isolated solution of \eqref{eq: extrasep}. In total, the separatrices have only four intersection points as can be seen in Fig. \ref{fig:phasemapECG}. These special points are collected in Table \ref{tab: intersections}. These critical points correspond, in addition to the isotropic solution, to the three FLRW universes with flat spatial slices in which only one of the spacelike directions is dynamical.

The analysis of the full dynamical system and its phase map is quite cumbersome. However, we do not need to perform such analysis for our purposes, since we are only interested in explicitly showing the pathological character of the isotropic solutions. This should already be clear from the fact that such a solution in fact belongs to the discussed separatrices. However, we can be more explicit by considering the following restricted situation: 

\newpage
\begin{table}
  \begin{center}
  \renewcommand\arraystretch{1.6}
  \begin{tabular}{|c|c|c||l|}
    \hline 
    $(X,\,Y)$ & $(\sigma_{1},\,\sigma_{2})$ & $\left(\frac{\dot{a}}{a},\,\frac{\dot{b}}{b},\,\frac{\dot{c}}{c}\right)$ & Description \tabularnewline
    \hline \hline 
    $(0,\,0)$ & $(0,\,0)$ & $\left(\HubH,\,\HubH,\,\HubH\right)$ & Isotropic point \tabularnewline
    \hline 
    $(1,\,0)$ & $(\HubH,\,0)$ & $\left(3\HubH,\,0,\,0\right)$ & $b, c$ constant functions \tabularnewline
    \hline 
    $(0,\,1)$ & $(0,\,\HubH)$ & $\left(0,\,3\HubH,\,0\right)$& $a, c$ constant functions \tabularnewline
    \hline 
    $(-1,\,-1)$ & $(-\HubH,\,-\HubH)$ & $\left(0,\,0,\,3\HubH\right)$& $a, b$ constant functions \tabularnewline
    \hline 
  \end{tabular}
  \renewcommand\arraystretch{1}
  \end{center}
  
  \caption{\label{tab: intersections}
    In this table we summarise the four special points in the plane $XY$ where the different singular branches intersect. These special solutions correspond to universes with isotropic evolution and with evolution along one of the directions while the transverse ones remain static.}
  
\end{table}

\newpage
\begin{enumerate}
  \item First, since the critical surfaces only depend on $X$ and $Y$, we will focus on this plane of the phase space. 

  \item Then we will consider the flow of the trajectories with $\HubH'=X'=Y'=X''=Y''=0$ and $\HubH=\HubH_\text{dS}$ that correspond to trajectories that are anisotropically displaced from the de Sitter solution and left {\it at rest}.

  \item We can then plot the flow of the vector field $(\xi^2,\,\xi^3)'=(X''',Y''')$ in the $(X,Y)$ plane as we show in Fig. \ref{fig:phasemapECG}. 
\end{enumerate}
Let us clarify the procedure we are following: we first take the phase space flow on the hypersurface $\HubH'=X'=Y'=X''=Y''=0$ and $\HubH=\HubH_\text{dS}$. Then we project it onto the plane $(\xi^2,\,\xi^3)=(X'',Y'')$. And, finally, we plot how this projected flow varies with the coordinates $(X,Y)$ (they can be seen as external parameters for the resulting vector field). 

A cautionary word might be in order here. The plots in Fig. \ref{fig:phasemapECG} involve an identification of the $(X,Y)$ axis with the directions of $(\xi^2,\,\xi^3)$. Thus, although these diagrams provide limited information on the physical trajectories, they can be used to clearly see the separatrices as well as the crucial consequence that no physical solutions can smoothly approach them. In particular, we can see that the isotropic point (the origin of the plot) is an unstable point. In the following subsection we will provide some numerical examples to clearly illustrate these arguments.

\subsection{Numerical analysis for a $\Lambda$-dominated era}
We will now examine numerical solutions for the full set of equations in the Bianchi I spacetime. We choose again the independent set of differential equations $\{{\rm E}_\mathcal{N}, {\rm E}_{cb}, {\rm E}_{ca} \}$. As we saw in \eqref{eq:highestDeqs}, in order to reduce the order in $\HubH$, we replace $\ddot{\HubH}$ and $\dddot{\HubH}$ in the shear equations by the expressions obtained by taking successive time derivatives of ${\rm E}_\mathcal{N}$. 

\begin{figure} 
\begin{center}
\includegraphics[width=0.4\textwidth]{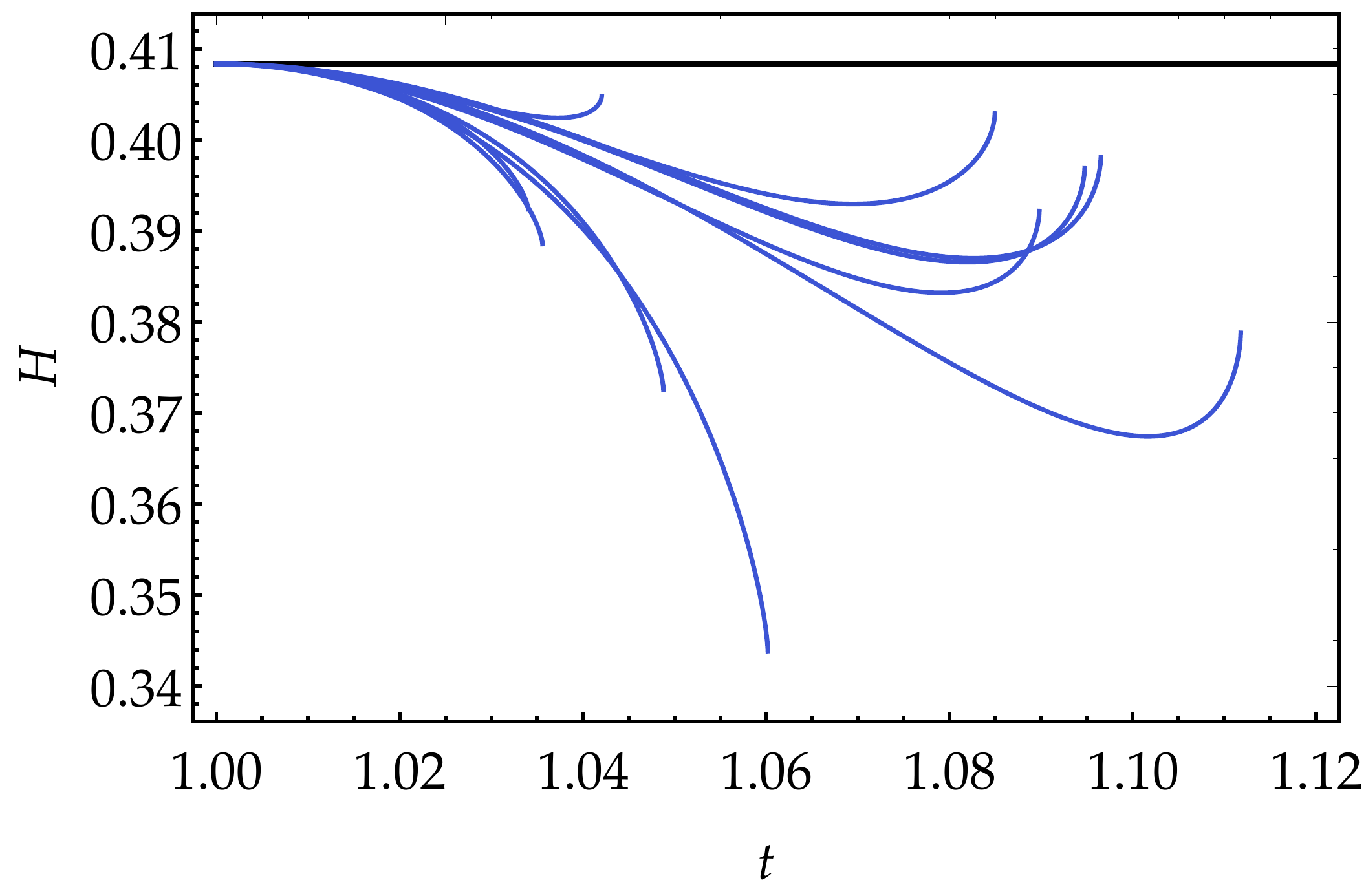} \qquad
\includegraphics[width=0.42\textwidth]{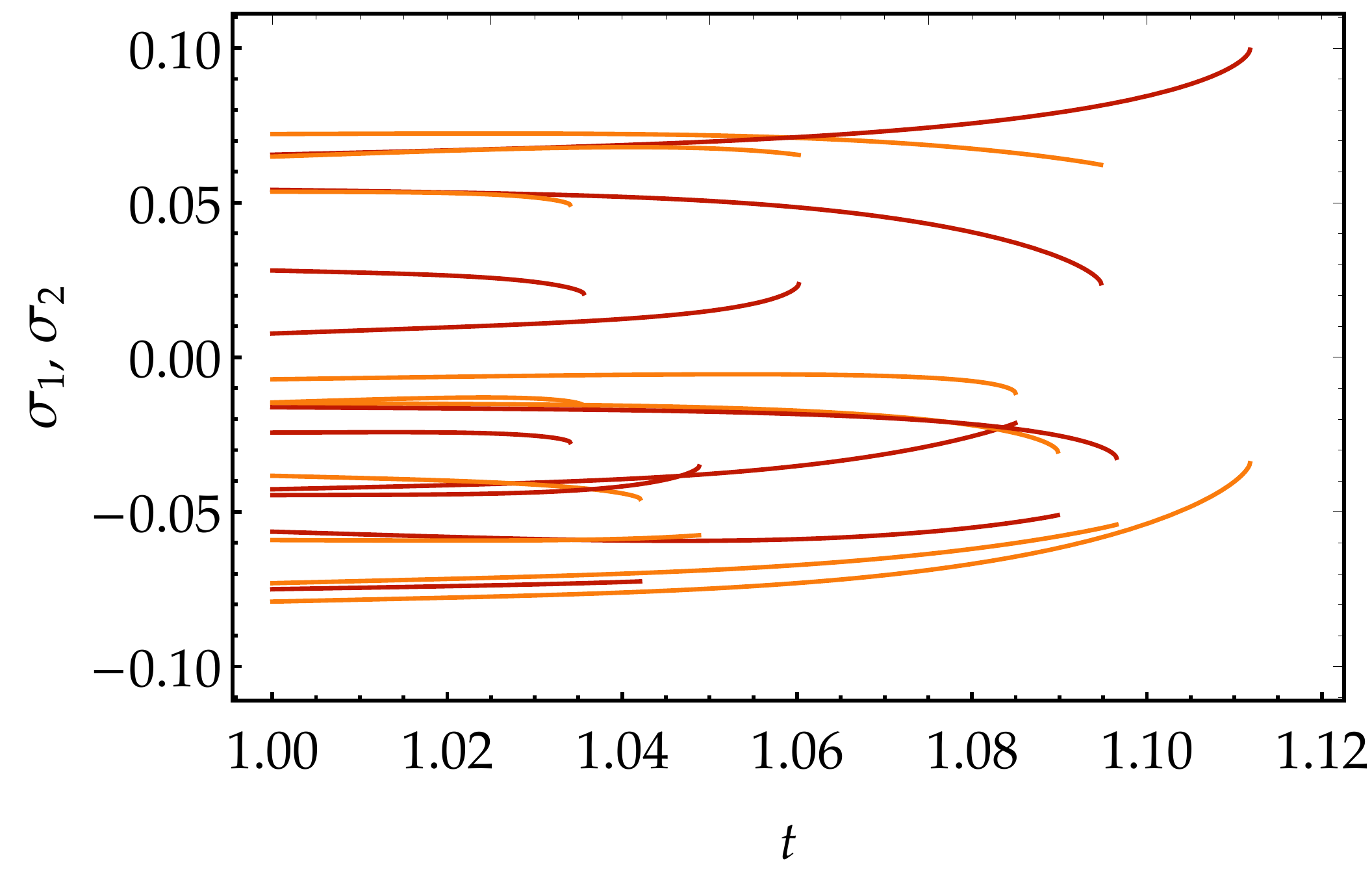}
\end{center}
\caption{
In the first plot we represent the isotropic solution for $\HubH(t)$ (black line) and the numerical ones (blue lines) starting with the same initial conditions for $\HubH$, but for randomly generated initial conditions for $\sigma_1(t)$ and $\sigma_2(t)$. The second plot shows the result of the integration for $\sigma_1(t)$ and $\sigma_2(t)$. For this numerical result we have chosen $\Lambda=0.5$, $\mpl=1$, $\pECG=0.01$ and the initial value $\HubH(t=1)=0.408362$ (the only solution of \eqref{eq: Isotropic condition} compatible with the stability condition that forces $\Lambda>2\HubH^2$  \cite{PookkillathDeFelice2020}).}
\label{Fig:plotdeSitter}
\end{figure}

Our goal is to scan the phase space around the de Sitter solution. With this space as our baseline, we give randomly generated initial condition with a small amplitude to the initial shears and their derivatives. We show the obtained numerical solutions in Fig. \ref{Fig:plotdeSitter}, which confirms our discussion above. The left panel shows the evolution for $\HubH$ together with the exact de Sitter solution. We see that in all of the numerical solutions, which are perturbatively closed to de Sitter at the beginning, quickly deviate from the isotropic one. In addition, the solutions that eventually turn and approach the isotropic solution encounter a point beyond which the evolution ceases. This behavior clearly reflects the fact that the solution reaches a singular point. For illustrative purposes we only plot ten solutions, but we have checked that this is the general tendency.

\subsection{Numerical analysis for a radiation-dominated era}

In this section we will perform a similar numerical integration but in the presence of a matter sector $S_{\rm Matt}$ describing radiation, i.e., one whose energy-momentum tensor (as defined in \eqref{eq:defEMtensorg}) has the form
\begin{equation}
  \mathcal{T}^{\mu\nu} =(\rho_{\rm r}+P_{\rm r})u^\mu u^\nu+P_{\rm r}g^{\mu\nu}\,,
\end{equation}
where $P_{\rm r}=\frac{1}{3}\rho_{\rm r}$ and $u^\mu$ is the fluid 4-velocity. Therefore, among $\{{\rm E}_\mathcal{N},  {\rm E}_{cb},  {\rm E}_{ca}\}$, only the equation of the lapse is modified, according to $ {\rm E}_{\mathcal{N}} \to {\rm E}_{\mathcal{N}}-\rho_{\rm r}(t)$, with respect of the equations of the previous subsection.

\begin{figure}
\begin{center}
\includegraphics[width=0.4\textwidth]{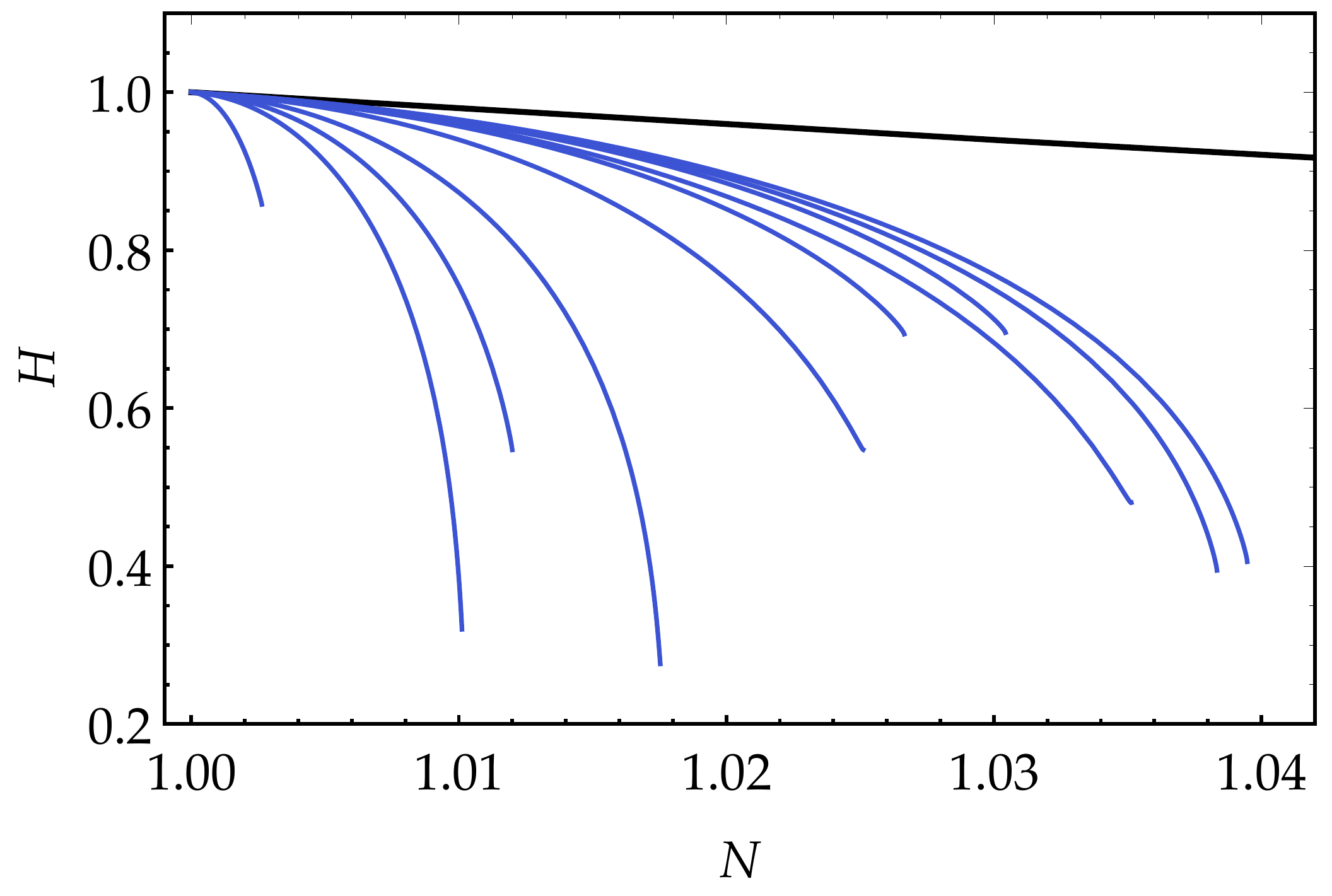} \qquad
\includegraphics[width=0.42\textwidth]{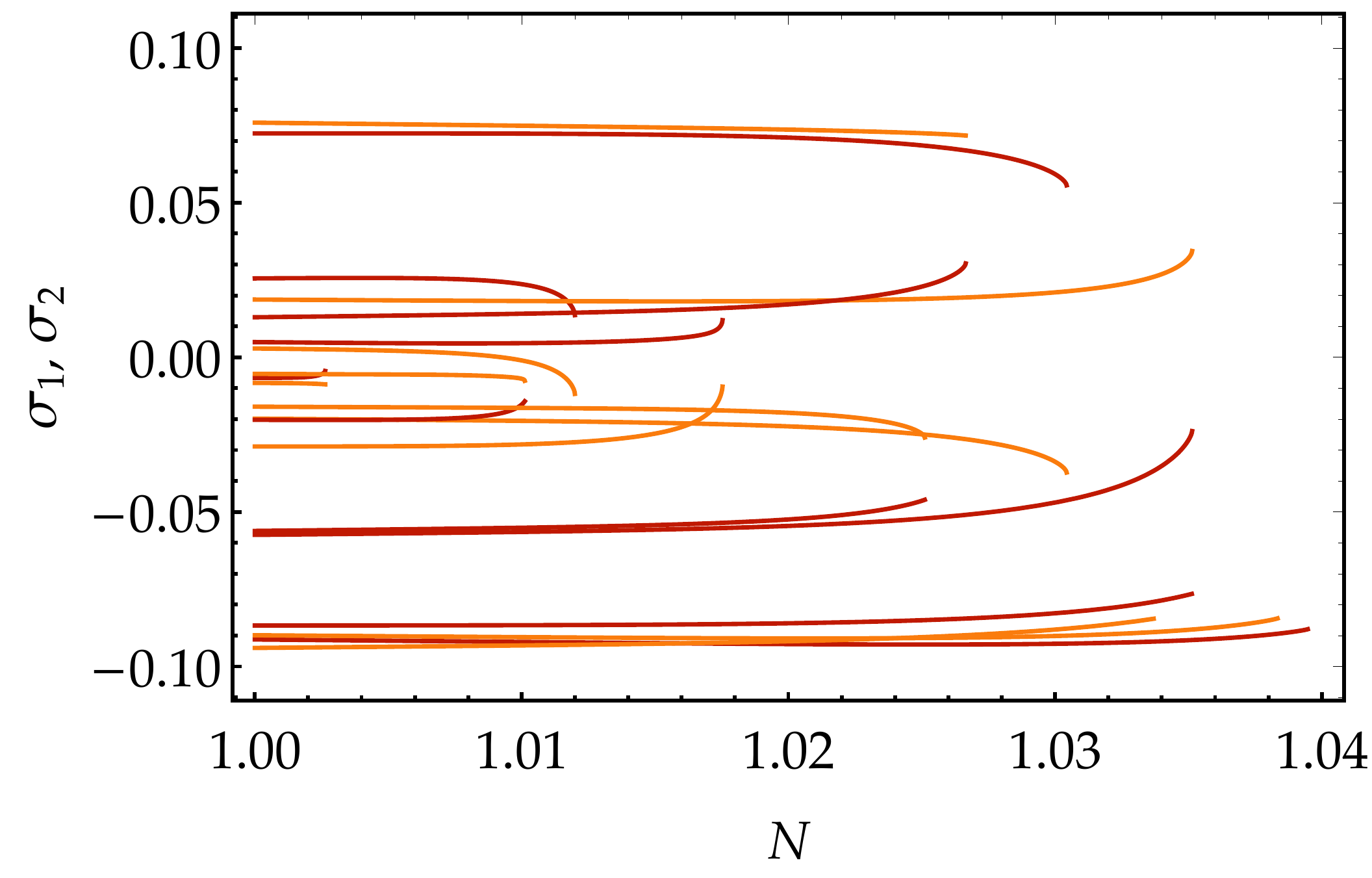}
\end{center}
\caption{In these plots we represent the evolution of the perturbed solution around an isotropic radiation-dominated background similarly as in Fig. \ref{Fig:plotdeSitter}, but now expressing the evolution in terms of the number of e-folds $N$ instead of the cosmic time $t$. For this numerical result we have chosen $\Lambda=0.5$, $\mpl=1$, $\pECG=0.01$ and the initial value $\HubH(N=1)=1$ (which implies $\rho_0=160.519$ due to \eqref{eq: Isotropic condition}).} \label{Fig:plotRadiation}
\end{figure}

In the presence of radiation it is convenient to work in terms of the number of e-folds $N$, defined in \eqref{eq: N efolds}. Then, the Bianchi identity associated to diffeomorphisms for the matter action,
\begin{equation}
  0=\nabla_{\mu}T^{\mu\nu}\qquad\Rightarrow\qquad\frac{\dot{\rho}_{\rm r}(t)}{\rho_{\rm r}(t)}=-4\HubH(t)\,,
\end{equation}
can be immediately integrated:
\begin{equation} \rho_{\rm r}(N)=\rho_0 \eN^{-4N}\,.\end{equation}
With all of this in mind, we proceed in a similar way as in the previous section. For the numerical computation we use the initial value of the Hubble constant, $\HubH(N_\text{ini})$, as an input and employ it to determine $\rho_0$ through the isotropic equation \eqref{eq: Isotropic condition}.\footnote{When solving for $\rho_0$ for the given value of $\HubH(N_\text{ini})$, there are more than one branch of solutions in general. Actually, for the set of parameters employed in Fig. \ref{Fig:plotRadiation}, there is another real branch where the isotropic solution for $\HubH$ is an increasing function. For that case, the same conclusions can be reached.}
If we focus on a radiation-dominated era, we can neglect the cosmological constant term (initially, $\rho_0 \eN^{-4 N_\text{ini}} \gg \Lambda$). In Fig. \ref{Fig:plotRadiation} we show the evolution for ten sets of randomly generated initial values for the shears $\sigma_1$ and $\sigma_2$. As in the case discussed in the previous section, the numerical solutions exhibit an important deviation with respect to the isotropic background. It is worth noticing that the blue curves show no tendency to return to the isotropic curve. Again, we have checked that these ten curves are representative of the general behavior. 

\section{Pathologies in higher order generalized quasi-topological theories} 

Pathologies as the ones we have discussed in the previous section for ECG, are in principle expected in any of the Generalized Quasi-Topological Gravity theories introduced in \cite{ArciniegaBuenoCano2018}. Interestingly, we will find that the cosmological solutions based on these extended GQTG are even more prone to problems than in the ECG case, in a sense that we explain in the following. 

The theory we will consider is ECG \eqref{eq:ECGaction} plus a series of higher order terms in the curvature of the type GQTG. For our purposes here it will be sufficient to restrict our analysis to the first three higher order terms. Consider then the action \eqref{eq:ECGaction} with the correction
\begin{equation} \label{eq: DeltaS quasitop}
  \Delta S=  \int \dex^4x \sqrt{|g|} \left( \frac{\beta_4}{\mpl^4} \mathcal{R}_{(4)} +\frac{\beta_5}{\mpl^6} \mathcal{R}_{(5)} +\frac{\beta_6}{\mpl^8} \mathcal{R}_{(6)} \right)\,,
\end{equation}
where $\beta_i$ ($i=4,5,6$) are dimensionless parameters and $\mathcal{R}_{(i)}$ ($i=4,5,6$) are the curvature invariants given in \cite{ArciniegaBuenoCano2018} and that we reproduce in Appendix \ref{app:GQTG} for completeness. Following the same procedure as in the ECG theory, we can obtain the evolution equations for the shear functions $\sigma_1$ and $\sigma_2$, which now have the form (compare to \eqref{eq: principal part}) 
\renewcommand\arraystretch{1.3}
\begin{align}\label{eq: pp quasitop}
\left[ \dot{\HubH}^2 v_1 \ \mathds{1}_{2\times 2} + \eanis \left(\frac{\pECG}{\mpl^2}{\bf M}_1 +  {\bf N}_1 \right) \right] 
    \begin{pmatrix}\dddot{\sigma}_2 \\\dddot{\sigma}_1 \end{pmatrix}+
    \left[ \dot{\HubH} v_2 \ \mathds{1}_{2\times 2}  + \eanis \left(\frac{\pECG}{\mpl^2}{\bf M}_2 +{\bf N}_2 \right) \right] 
    \begin{pmatrix}\ddot{\sigma}_2 \\\ddot{\sigma}_1 \end{pmatrix}  & \nonumber\\
    + \frac{\pECG}{\mpl^2}{\bf V} + {\bf W}
    +3 \mpl^2 \begin{pmatrix}3\HubH\sigma_2+\dot{\sigma}_2 \\ 3\HubH\sigma_1+\dot{\sigma}_1 \end{pmatrix} &= 0 \,,
\end{align}
\renewcommand\arraystretch{1}
where
\begin{align}
    v_1 &\coloneqq -6 \frac{\beta_4}{\mpl^4} 
    - \frac{2}{5} \left(2\, \HubH^2 + 3\, \dot{\HubH} \right) \frac{\beta_5}{\mpl^6}
    + \frac{3}{1040} \left(16848\, \HubH^4 + 90357\,\HubH^2\dot{\HubH} + 136175\, \dot{\HubH}^2 \right) \frac{\beta_6}{\mpl^8} \,,\\
    v_2 &\coloneqq -12\,  (3\HubH\dot{\HubH}+2\ddot{\HubH})\frac{\beta_4}{\mpl^4}
    - \frac{4}{5} \big(6\,\HubH^3\dot{\HubH}+13\,\HubH\dot{\HubH}^2+4\,\HubH^2\ddot{\HubH} +9\dot{\HubH}\ddot{\HubH} \big) \frac{\beta_5}{\mpl^6} \nonumber \\
    & \quad + \frac{3}{520} \big(50544\, \HubH^5\dot{\HubH} 
    +338463\,\HubH^3\dot{\HubH}^2 + 589239\,\HubH\dot{\HubH}^3  \nonumber \\
    & \quad\qquad\qquad +33696\,\HubH^4\ddot{\HubH}+271071\, \HubH^2\dot{\HubH}\ddot{\HubH}+544700\, \dot{\HubH}^2\ddot{\HubH}\big) \frac{\beta_6}{\mpl^8}\,,
\end{align}
and the rest of the contributions coming from $\Delta S$ are encoded in the matrices ${\bf N}_1$ and ${\bf N}_2$, and in the vector ${\bf W}$. These three objects start at zeroth order in $\eanis$.

As in the ECG case we see that the isotropic de Sitter solution ($\dot{\HubH}=0=\eanis$) corresponds to a singular surface in phase space, thus giving rise to the same type of strong coupling problems due to the disappearance of some dofs. However, we can see that the higher order terms do not trivialize in the case of an arbitrary cosmological background ($\eanis=0$ but $\dot{\HubH}\neq0$). The reason is in the very definition of these theories: they have the linear spectrum of GR only around maximally symmetric backgrounds. On a general isotropic cosmological background the additional modes associated to the higher order nature of the field equations are fully active. Though this prevents any strong coupling issue, it indicates that the ghostly degrees of freedom will propagate on general cosmological backgrounds, thus making them unstable.

A possible improvement of this pathological behavior can be obtained by noticing that there are several inequivalent terms at each order in curvature that lead to second order gravitational equations for an isotropic Ansatz, but differ beyond the isotropic solutions. In that respect, we have taken the particular combination given in \eqref{eq: DeltaS quasitop} to illustrate the present pathologies, but this is not unique nor the most general choice. Thus, it is conceivable that these terms can be combined in such a way that the equations remain of second order around arbitrary FLRW spacetimes and not only for the maximally symmetric ones, as suggested in \cite{ArciniegaBuenoCano2018}. In other words, there could exist combinations so that the higher order contributions completely vanish in the limit $\eanis\rightarrow 0$, similarly to what happens for the ECG. This currently remains as an open question.\footnote{
    We thank Pablo A. Cano for pointing out this possibility to us.}

\section{Discussion and conclusions}

In this chapter we have discussed the pathologies present in ECG and GQTG that arise as a direct consequence of their defining prescription. These theories contain a massless spin 2, a massive spin-2 and a massive scalar fields, but only the former propagates on some specific backgrounds (same linear spectrum as GR on those spacetimes). This condition is imposed in order to get rid of the ghostly modes associated to the higher order nature of the theories. The evanescence of dofs on these backgrounds however can be interpreted as an indication for the presence of strongly coupled modes. Since this cannot be seen at linear order, we have instead studied the full non-linear equations of slightly deformed backgrounds with fewer symmetries than the spacetimes used to define the theories. We have mainly focused on cosmological solutions and the ECG action. In this setting, we have shown that the isotropic solution indeed corresponds to a singular surface in phase space, preventing the solution trajectories from smoothly evolving towards it. Furthermore, we have analyze how standard perturbation theory around the isotropic solution fails to reproduce the full landscape of perturbations and, consequently, the conclusions drawn from a perturbative analysis cannot be fully trusted. We have also discussed these problems for the extended class of GQTG. Here we have found that, not only the same strong coupling problems around the maximally symmetric backgrounds that define the theories persist, but the ghostly degrees of freedom are actually active around general cosmological backgrounds.

Although we have focused on cosmological backgrounds, there is nothing really special about them (other than the simplicity introduced by the additional symmetries) and it is easy to envision that the same class of pathologies will be present e.g. around static and spherically symmetric backgrounds. Likewise, similar problems are expected to arise in extension involving additional fields. Recently, a new class of quasi-topological electromagnetic theories has been introduced in \cite{CanoMurcia2020} where theories featuring non-minimal couplings of a ${\rm U}(1)$ gauge field to gravity are explored along the lines of GQTG. In this respect, it is known that the so-called Horndeski vector-tensor interaction (see e.g. \cite{Horndeski1976}) is the only gauge-invariant non-minimal coupling that gives rise to second order field equations (i.e. the analogue of Lovelock terms). Thus, the Lagrangians obtained in \cite{CanoMurcia2020} without additional modes on spherically symmetric backgrounds will again be prone to the same type of pathologies discussed here.  That would not be the case if those Lagrangians were related to the Horndeski vector-tensor interaction via a field redefinition for instance. In this respect, similar conclusions would apply to GQTG including a scalar field featuring derivative non-minimal couplings and constructed so that the scalar only propagates one additional dof around some specific backgrounds, thus lying outside the class of Horndeski theories or any of the known healthy scalar-tensor theories (see e.g. \cite{Zumalacarregui2013, GleyzesLanglois2015b,LangloisNoui2016}).

Undoubtedly, the general class of GQTG exhibit a series of remarkable properties that make them very interesting and worth investigating. However, it is crucial to bear in mind that their very defining property is intimately related to the presence of pathologies that need to be properly tackled to guarantee the physical viability of models based on these theories (e.g. inflationary scenarios). Of course, this should not preclude exploiting their exceptional properties to draw physically sensible and useful results from them\footnote{
    This happens in other theories. For instance, in massive gravity, open FLRW solutions (the only ones allowed) are plagued by strong coupling issues and non-linear ghost-like instabilities (see e.g. \cite{GumrukcuogluLin2011, DeFeliceGumrukcuoglu2012}). Nevertheless, it is important to remark that in GQTG the full theory generically exhibit the pathology (because it is of the Ostrogradski type).}
(see for instance \cite{BuenoCanoHennigar2019, BuenoanoHennigar2020}, where analytical results for the thermodynamical properties of Taub-NUT solutions were derived). Therefore, the main conclusion from our analysis is that certain backgrounds cannot be regarded, at least a priori, as viable physical models.

\subsubsection*{Other comments and future directions}

\begin{itemize}
\item {\bf Relation to cuscuton model}. It is interesting to notice a certain resemblance of what happens in the theories under consideration in this chapter and the \emph{cuscuton model}, first introduced in \cite{AfshordiChung2006} (see \cite{IyonagaTakahashi2018} for an extended version). This model describes a scalar field whose propagation speed becomes infinite around homogeneous configurations so it does not propagate. This feature was analyzed from a full Hamiltonian approach in \cite{GomesGuariento2017}, where it was shown that indeed the homogeneous configuration corresponds to a singular surface in phase space. The authors argued that the cuscuton could be defined in a sensible manner only if the homogeneity of the field is imposed {\it a priori}. It would be interesting to study if a similar interpretation could be employed for the ECG and GQTG by constraining the space of allowed metrics.

\item {\bf EFT considerations}. Theories with a reduced spectrum around some backgrounds are sometimes interpreted within the realm of Effective Field Theories (EFT). In this respect, we find it convenient to stress that, with that philosophy in mind, one should include all operators complying with the symmetries (diffeomorphims in this case) and field content (the metric and, possibly, a matter sector). In particular, there is no reason not to include the quadratic terms in the curvatures, which also introduce ghostly dofs and would become dominant at a lower scale, and higher order curvature terms that would then be order one whenever the ECG operator becomes non-perturbative. If we do not include these terms, then the question is: how stable is the resulting tuning of the coefficients under quantum corrections (graviton and matter loops)?\footnote{
    It is interesting to note that all higher order curvature terms with up to two covariant derivatives acting on the Riemann can be related to the GQTG Lagrangians via field redefinitions \cite{BuenoCanoMoreno2019}. This seems to suggest, that the GQTG could serve as a basis for the gravitational EFT operators (see \cite{RuhdorferSerra2019}), at least partially for operators not involving higher than second derivatives of the curvatures.}
\end{itemize}

%
\chapter{The General Teleparallel Quadratic theory}\label{ch:QTG}

\boxquote{An expert is a man who has made all the mistakes which can be made in a very narrow field.}{Niels Bohr, as quoted by Edward Teller (10 October 1972)}

Analyzing the stability of the full quadratic MAG Lagrangian is quite a challenging task. In this chapter we are going to focus on the teleparallel restriction of the Lagrangian \eqref{eq: qMAGLeven} in dimension 4, i.e., with vanishing curvature. Imposing a local symmetry under the general linear group turns the Lagrangian into the general teleparallel equivalent of General Relativity carrying both torsion and nonmetricity. We will show how the GR equivalents with either zero torsion or zero nonmetricity can be seen as gauge-fixed versions of the general equivalent. We will also study the linear theory around Minkowski spacetime to prove that the presence of extra gauge symmetries is required for the theory to be potentially viable.

\section{Introduction to general teleparallel theories}

The metric-affine geometries with zero curvature have a well-defined notion of distant parallelism (\emph{teleparallelism}). In this framework it is possible to formulate other fully equivalent descriptions of GR. On the one hand we have the \emph{Teleparallel Equivalent of GR} (TEGR) \cite{AldrovandiPereira2013} formulated in the absence of nonmetricity (Weitzenb\"ock geometries), where gravity is identified with the torsion. On the other hand, flat spacetimes with zero torsion can accomodate the GR effects in the nonmetricity giving rise to the \emph{Symmetric Teleparallel Equivalent of GR} (STEGR) \cite{NesterYo1998,  BeltranHeisenbergKoivisto2018a}. Generalizations of these theories (of course, inequivalent to GR) have been considered by the modified gravity community such as $f(T)$-gravity  \cite{BengocheaFerraro2008, LiSotiriouBarrow2010,GolovnevKoivisto2018} and $f(Q)$-gravity \cite{BeltranHeisenbergKoivisto2018a, BeltranHeisenbergPekar2019}, whose Lagrangians are arbitrary functions of the TEGR and STEGR invariants, respectively.

Consider a teleparallel framework in which we allow both torsion and nonmetricity. The only constraint is then the teleparallel condition,
\begin{equation} \label{eq: zeroR}
  R_{\mu\nu\rho}{}^\lambda=0.
\end{equation}
This can be integrated and the resulting connection (also known as \emph{inertial connection}) can be expressed
\begin{equation}\label{eq:telepconn}
  \Gamma_{\mu\nu}{}^\rho =(\Lambda^{-1})^\rho{}_\sigma\partial_\mu\Lambda^\sigma{}_\nu,
\end{equation}
where $\Lambda^\alpha{}_\beta$ is an arbitrary (constant) invertible matrix (i.e., an element of $\mathrm{GL}(4,\mathbb{R})$). Notice that the connection exhibits a global symmetry $\Lambda^\mu{}_\nu\rightarrow  M^\mu{}_\rho \Lambda^\rho{}_\nu$ for a constant $M^\mu{}_\rho \in\mathrm{GL}(4,\mathbb{R})$ that will be present in the teleparallel theories.\footnote{Notice that $\Lambda^\mu{}_\nu$ is not a tensor under diffeomorphisms, as can be easily seen from the transformation of $\Gamma_{\mu\nu}{}^\rho$ as a connection.} The torsion and the nonmetricity for this geometry can be straightforwardly computed:
\begin{align}
T_{\mu\nu}{}^\rho&=2(\Lambda^{-1})^\rho{}_\sigma\partial_{[\mu}\Lambda^\sigma{}_{\nu]}, \label{eq:Teletorsion}\\
Q_{\rho\mu\nu}   &=-\partial_\rho g_{\mu\nu} + 2(\Lambda^{-1})^\lambda{}_\sigma \partial_\rho\Lambda^\sigma{}_{(\mu} g_{\nu)\lambda}.
\end{align}
These are the two fundamental pieces to construct a general teleparallel theory since, by covariance, the connection will only appear within tensorial quantities (and the only available options are torsion and/or nonmetricity).

Consider a general metric-affine action $S^*[g,\,\Gamma]$ (the ${}^*$ is conventional). We are not interested in its teleparallel solutions (i.e., varying first and then imposing the teleparallel condition), but in studying its teleparallel restriction. The natural way to proceed would be to impose the restriction dynamically through a Lagrange multiplier, although there are other possibilities (see \cite{Hohmann2021}),
\begin{equation}
  S_{\mathrm{Tot}}[g,\,\Gamma,\,l] = S^*[g,\,\Gamma]+\int l^{\mu\nu\rho}{}_{\lambda}R_{\mu\nu\rho}{}^{\lambda}\sqrt{|g|}\mathrm{d}^{\dimM}x\,,
\end{equation}
where the second term contains a Lagrange multiplier $l^{\mu\nu\rho}{}_{\lambda}=l^{[\mu\nu]\rho}{}_{\lambda}$ that forces the theory to be teleparallel. The equations of motion for the Lagrange multiplier, the connection and the metric are, respectively,
\begin{align}
  0 =\frac{\delta S_{\mathrm{Tot}}}{\delta l^{\mu\nu\rho}{}_\lambda} & = R_{\mu\nu\rho}{}^\lambda\,,\label{eq:eomltelep}\\
  0 =\frac{\delta S_{\mathrm{Tot}}}{\delta\Gamma_{\mu\nu}{}^\rho} &=\frac{\delta S^*}{\delta\Gamma_{\mu\nu}{}^\rho}-2(\nabla_\lambda+T_\lambda)\mathfrak{l}^{\lambda\mu\nu}{}_\rho+T_{\lambda\sigma}{}^\mu\mathfrak{l}^{\lambda\sigma\nu}{}_\rho\,, \label{eq:eomGtelep}\\
  0 =\frac{\delta S_{\mathrm{Tot}}}{\delta g_{\mu\nu}} &=\frac{\delta S^*}{\delta g_{\mu\nu}}+\frac{1}{2}g^{\mu\nu}R_{\sigma\tau\rho}{}^{\lambda}\mathfrak{l}^{\sigma\tau\rho}{}_\lambda\,,\label{eq:eomgtelep}
\end{align}
where $\mathfrak{l}^{\mu\nu\rho}{}_{\lambda}\coloneqq\sqrt{|g|} l^{\mu\nu\rho}{}_\lambda$. Notice that the last term in \eqref{eq:eomgtelep} disappears whenever the Lagrange multiplier is on-shell.

Although the Lagrange multiplier appears in the equation of the connection and one has to deal with it, fortunately, it is possible to eliminate it by taking an appropriate derivative \cite{Hohmann2021}:
\boxproposition{
\begin{prop} The equations \eqref{eq:eomltelep} and \eqref{eq:eomGtelep} imply
  \begin{equation}
    (\nabla_\mu+T_\mu)\frac{\delta S^*}{\delta\Gamma_{\mu\nu}{}^\rho}=0\,. \label{eq:eomGtelepRed}
  \end{equation}
\end{prop}
}
\boxproof{
\begin{proof}
  First, let us write the equation \eqref{eq:eomGtelep} as follows:
  \begin{equation}
    \frac{\delta S^*}{\delta\Gamma_{\mu\nu}{}^\rho}= \mathfrak{K}^{\mu\nu}{}_\rho\,\qquad\text{where}\qquad\mathfrak{K}^{\mu\nu}{}_\rho\coloneqq 2(\nabla_\lambda+T_\lambda)\mathfrak{l}^{\lambda\mu\nu}{}_\rho -T_{\lambda\sigma}{}^\mu\mathfrak{l}^{\lambda\sigma\nu}{}_\rho\,.
  \end{equation}
To prove \eqref{eq:eomGtelepRed}, we simply act with the operator $\nabla_\mu+T_\mu$ on both sides of this equation and take into account that
\begin{align}
  (\nabla_{\mu}+T_{\mu})\mathfrak{K}^{\mu\nu}{}_\rho  
&=-[\nabla_{\mu},\,\nabla_{\lambda}]\mathfrak{l}^{\mu\lambda\nu}{}_\rho  +2\nabla_{\mu}T_{\lambda}\mathfrak{l}^{\lambda\mu\nu}{}_\rho + 2T_{\lambda}\nabla_{\mu} \mathfrak{l}^{\lambda\mu\nu}{}_\rho  +2T_{\mu}\nabla_{\lambda}\mathfrak{l}^{\lambda\mu\nu}{}_\rho \nonumber\\
&\quad-\nabla_{\mu}(T_{\lambda\sigma}{}^{\mu} \mathfrak{l}^{\lambda\sigma\nu}{}_\rho )- T_{\mu}T_{\lambda\sigma}{}^{\mu}\mathfrak{l}^{\lambda\sigma\nu}{}_\rho \nonumber\\
&=T_{\mu\lambda}{}^{\sigma}\nabla_{\sigma}\mathfrak{l}^{\mu\lambda\nu}{}_\rho + 2\nabla_{\mu}T_{\lambda}\mathfrak{l}^{\lambda\mu\nu}{}_\rho -\nabla_{\sigma}(T_{\lambda\mu}{}^{\sigma}\mathfrak{l}^{\lambda\mu\nu}{}_\rho )- T_{\mu}T_{\lambda\sigma}{}^{\mu}\mathfrak{l}^{\lambda\sigma\nu}{}_\rho \nonumber\\
&=-(2\nabla_{[\lambda}T_{\mu]}+\nabla_{\sigma}T_{\lambda\mu}{}^{\sigma}+T_{\sigma}T_{\lambda\mu}{}^{\sigma})\mathfrak{l}^{\lambda\mu\nu}{}_\rho = 0\,,
\end{align}
where we have substituted the identity
\begin{equation}
  2R_{[\mu|\sigma|\lambda]}{}^{\sigma}+R_{\lambda\mu\sigma}{}^\sigma =2\nabla_{[\lambda}T_{\mu]}+\nabla_{\rho}T_{\lambda\mu}{}^{\rho}+T_{\lambda\mu}{}^\sigma T_\sigma\,,
\end{equation}
in the last step, and we have used several times the teleparallel condition.
\end{proof}
}

On the other hand, we can consider the action already evaluated in the teleparallel connection \eqref{eq:telepconn}, i.e., directly formulated in terms of the fundamental fields $g_{\mu\nu}$ and $\Lambda^\mu{}_\nu$:
\begin{equation}
  S \ [g_{\mu\nu},\,\Lambda^\mu{}_\nu] \coloneqq S^* \ [g_{\mu\nu},\,\Gamma_{\sigma\lambda}{}^\rho (\Lambda^\mu{}_\nu)].
\end{equation}
Then one can easily check:
\boxproposition{
\begin{prop} The variations of $S$ with respect to $g_{\mu\nu}$ and $\Lambda^\mu{}_\nu$ are given by:
  \begin{equation}
    \frac{\delta S}{\delta g_{\mu\nu}}=\frac{\delta S^*}{\delta g_{\mu\nu}}\,,\qquad\qquad  \frac{\delta S}{\delta\Lambda^\sigma{}_\nu}=  (\Lambda^{-1})^\rho{}_\sigma (\nabla_\mu + T_\mu)\frac{\delta S^* }{\delta\Gamma_{\mu\nu}{}^\rho}\,.
  \end{equation}
\end{prop}
}
\boxproof{
\begin{proof}
  The equation of the metric is straightforward. To compute the field equations of $\Lambda^\mu{}_\nu$, we use the following identity that can be easily checked by direct computation of both sides:
\begin{equation}\label{eq:deltaGamma} 
  \delta\Gamma_{\mu\nu}{}^\rho=\nabla_\mu\big[(\Lambda^{-1})^\rho{}_\sigma\delta\Lambda^\sigma{}_\nu\big].
\end{equation}
And, with this in mind,
\begin{align}
  \int \dex^\dimM x \frac{\delta S}{\delta\Lambda^\mu{}_\nu}\delta\Lambda^\mu{}_\nu = \delta_\Lambda S 
    & =\int \dex^\dimM x \ \frac{\delta S^*}{\delta\Gamma_{\mu\nu}{}^\rho} \delta_\Lambda\Gamma_{\mu\nu}{}^\rho \\
    {}^\eqref{eq:deltaGamma} & =\int \dex^\dimM x \  \frac{\delta S^*}{\delta\Gamma_{\mu\nu}{}^\rho} \nabla_\mu\big[(\Lambda^{-1})^\rho{}_\sigma\delta\Lambda^\sigma{}_\nu\big] \\
    {}^\text{int. by parts}& =\int \dex^\dimM x \ (\nabla_\mu + T_\mu)\frac{\delta S^*}{\delta\Gamma_{\mu\nu}{}^\rho} (\Lambda^{-1})^\rho{}_\sigma\delta\Lambda^\sigma{}_\nu\,.
\end{align}
\end{proof}
}

We see that by assuming the teleparallel condition at the level of the action we recover the same equation for the metric and the equation \eqref{eq:eomGtelepRed}. The remaining one, which is the rest of the equation \eqref{eq:eomGtelep} that is not fixed by \eqref{eq:eomGtelepRed} can be seen as an equation for the Lagrange multiplier that we will ignore. Therefore, from now on we will work in terms of $g_{\mu\nu}$ and $\Lambda^\mu{}_\nu$ at the level of the action.

\section{Teleparallel Quadratic gravity}

The most general (parity-preserving) teleparallel action quadratic in torsion and nonmetricity is
\begin{equation}
  S_\parallel \ [g_{\mu\nu},\,\Lambda^\mu{}_\nu] = \frac{1}{2}\mpl^2 \int \dex^4 x\sqrt{|g|} \ \mathbb{G},
\end{equation}
with\footnote{Since we are using different conventions than \cite{BeltranHeisenbergJCA2020}, some signs have been introduced in the action to ensure the following relations between our parameters and those of that article: $\alpha_i\leftrightarrow a_i$, $\beta_i\leftrightarrow b_i$, $\gamma_i\leftrightarrow c_i$. In addition, we corrected here the misprint in the term $\gamma_1\leftrightarrow c_1$ that appears in \cite{BeltranHeisenbergJCA2020}.}
\begin{align}
  \mathbb{G} &\coloneqq  \alpha_{1}T_{\mu\nu\rho}T^{\mu\nu\rho} +\alpha_{2}T_{\mu\nu\rho}T^{\mu\rho\nu} +\alpha_{3}T_{\mu}T^{\mu} -\beta_{1}Q_{\mu\nu\rho}T^{\mu\nu\rho} -\beta_{2}Q_{\mu}T^{\mu} -\beta_{3}\Qb_{\mu}T^{\mu}    \nonumber\\
             &\qquad  +\gamma_{1}Q_{\rho\mu\nu}Q^{\rho\mu\nu} + \gamma_{2}Q_{\rho\mu\nu}Q^{\mu\nu\rho} + \gamma_{3}Q_{\mu}Q^{\mu}+ \gamma_{4}\Qb_{\mu}\Qb^{\mu}+ \gamma_{5}Q_{\mu}\Qb^{\mu},
\end{align}
where the traces $T^{\mu}$, $Q_{\mu}$ and $\Qb_{\mu}$ agree with the definitions \eqref{eq: trTor def}, \eqref{eq: trQ1 def} and \eqref{eq: trQ2 def}, respectively. This teleparallel action reduces to \emph{New GR} \cite{HayashiShirafuji1979} for a metric-compatible connection, and to \emph{Newer GR} \cite{BeltranHeisenbergKoivisto2018a} for a torsion-free connection.

Let us now present the equations of motion of this theory. Consider the theory $S_{\parallel}+S_\mathrm{Matt}$ for some matter action that does not depend on the connection (i.e., it has zero hypermomentum). Then, according to the previous section,  the field equations for the metric and $\Lambda^\mu{}_\nu$ are, respectively,
\begin{align}
  \mathcal{G}^{\mu\nu}= \frac{1}{\mpl^2} \mathcal{T}^{\mu\nu}\,, \qquad\qquad  (\nabla_\mu+T_\mu )\mathcal{P}^{\mu\nu}{}_\rho=0\,. 
\end{align}
where we have introduced the abbreviations\footnote{
    The notation for the energy-momentum tensor is consistent with \eqref{eq:defEMtensorg}.}
\begin{equation}
  \mathcal{G}^{\mu\nu} \coloneqq\frac{2}{\mpl^2}\frac{1}{\sqrt{|g|}}\frac{\delta S_{\parallel} }{\delta g_{\mu\nu}}\,\qquad
  \mathcal{T}^{\mu\nu} \coloneqq\frac{2}{\sqrt{|g|}}\frac{\delta S_\mathrm{Matt} }{\delta g_{\mu\nu}}\,\qquad
  \mathcal{P}^{\mu\nu}{}_\rho\coloneqq \frac{\delta S^*_{\parallel} }{\delta\Gamma_{\mu\nu}{}^\rho} \label{eq:defGTP} \,,
\end{equation}
and the explicit expressions of $\mathcal{G}^{\mu\nu}$ and $\mathcal{P}^{\mu\nu}{}_\rho$ are
\begin{align}
\mathcal{G}^{\mu\nu} 
&=\mathcal{L}_{\parallel}g^{\mu\nu}+\alpha_1 \big[T^{\sigma\rho\mu} T_{\sigma\rho}{}^{\nu} -2T^{\mu}{}_{\sigma\rho}T^{\nu\sigma\rho}\big]- \alpha_2 T^{\mu}{}_{\sigma\rho}T^{\nu\rho\sigma}-\alpha_3 T^\mu T^\nu  \nonumber\\
&\quad-\beta_1 (Q^{(\mu}{}_{\sigma\rho}-Q_{\sigma\rho}{}^{(\mu})T^{\nu)\sigma\rho}- \beta_2 (Q^{\rho\mu\nu}T_\rho  +Q^{(\mu}T^{\nu)})- \beta_3 (Q^{(\mu\nu)\rho}T_\rho  +\Qb^{(\mu}T^{\nu)})\nonumber\\
&\quad-\gamma_1 (Q^{\mu}{}_{\sigma\rho}Q^{\nu\sigma\rho}+ 2Q^{\sigma\rho\mu}Q_{\sigma\rho}{}^\nu)-\gamma_2 (2Q_{\sigma\rho}{}^{(\mu}Q^{\nu)\sigma\rho}+ Q^{\rho\sigma\mu}Q_{\sigma\rho}{}^{\nu})\nonumber\\
&\quad-\gamma_3 (Q^\mu Q^\nu +2Q_\rho  Q^{\rho\mu\nu})-\gamma_4 (\Qb^\mu \Qb^\nu + 2\Qb_\rho  Q^{(\mu\nu)\rho})\nonumber\\
&\quad -\gamma_5 (Q^{(\mu}\Qb^{\nu)}+\Qb_\rho  Q^{\rho\mu\nu}+Q_\rho  Q^{(\mu\nu)\rho})\nonumber\\
&\quad+\Big(\nabla_\rho  -\frac{1}{2}Q_\rho  + T_\rho  \Big)\Big[2\gamma_1 Q^{\rho\mu\nu}+ 2\gamma_2 Q^{(\mu\nu)\rho}+2\gamma_3 Q^\rho  g^{\mu\nu}+2\gamma_4 g^{\rho(\mu}\Qb^{\nu)}\nonumber\\
&\qquad\qquad\qquad+\gamma_5 (\Qb^\rho  g^{\mu\nu}+ g^{\rho(\mu}Q^{\nu)})-\beta_1 T^{\rho(\mu\nu)}-\beta_2 T^\rho  g^{\mu\nu}-\beta_3 g^{\rho(\mu}T^{\nu)}\Big]\,,\\[2mm]
\mathcal{P}^{\mu\nu}{}_\rho
&=2\alpha_1 T^{\mu\nu}{}_\rho  -2\alpha_2 T_\rho  {}^{[\mu\nu]}+2\alpha_3 T^{[\mu}\delta_\rho^{\nu]}\nonumber\\
&\quad-\beta_1 (Q^{[\mu\nu]}{}_\rho  +T^{\mu(\nu\lambda)}g_{\lambda\rho})-\beta_2 (Q^{[\mu}\delta_\rho^{\nu]}+T^\mu \delta_\rho^\nu )-\beta_3 (\Qb^{[\mu}\delta_\rho^{\nu]}+g^{\mu(\nu}T^{\lambda)}g_{\lambda\rho})\nonumber\\
&\quad+2\gamma_1 Q^{\mu\nu}{}_\rho  +2\gamma_2 Q^{(\nu\lambda)\mu}g_{\lambda\rho}+2\gamma_3 Q^\mu \delta_\rho^\nu +2\gamma_4 g^{\mu(\nu}\Qb^{\lambda)}g_{\lambda\rho}+\gamma_5 (\Qb^\mu \delta_\rho^\nu +g^{\mu(\nu}Q^{\lambda)}g_{\lambda\rho})\,.
\end{align}
In principle we have $10 (g) + 16 (\Lambda)=26$ independent components, but the invariance under diffeomorphisms reduce them to a maximum of 18 propagating fields that can be associated to the 16 components of $\Lambda^\mu{}_\nu$ plus the two polarizations of the graviton contained in $g_{\mu\nu}$. As we will see, further restrictions in the parameters are needed to avoid ghosts. 

\section{On the equivalents of GR}
Here, we are going to elaborate a bit more on the equivalence between GR and their teleparallel equivalents. The starting point is the post-Riemannian expansion of the Ricci scalar of the general connection around the Levi-Civita of the spacetime metric:
\begin{equation}
  R= \mathring{R} - \mathcal{L}_\mathrm{GTEGR} - \mathring{\nabla}_{\mu}\left(\Qb^{\mu}-Q^{\mu}+2 T^{\mu}\right)\,, \label{eq:RtoROmega2}
\end{equation}
where we have defined
\begin{align}\label{eq:GTEGRdef}
  \mathcal{L}_\mathrm{GTEGR} &\coloneqq\frac{1}{4} T_{\mu \nu \rho} T^{\mu \nu \rho}+\frac{1}{2} T_{\mu \nu \rho} T^{\mu \rho \nu}-T_{\mu} T^{\mu}+Q_{\mu \nu \rho} T^{\mu \nu\rho}-Q_{\mu} T^{\mu}+\Qb_{\mu} T^{\mu}\nonumber\\
  &+\frac{1}{4} Q_{\mu \nu \rho} Q^{\mu \nu \rho}-\frac{1}{2} Q_{\mu \nu \rho} Q^{\nu \mu \rho}-\frac{1}{4} Q_{\mu} Q^{\mu}+\frac{1}{2} Q_{\mu} \Qb^{\mu},
\end{align}
which is obtained from $\mathbb{G}$ upon the parameters choice\footnote{
    Here we corrected another misprint in \cite{BeltranHeisenbergJCA2020} concerning the value of $\beta_2 \leftrightarrow b_2$.}
\begin{equation}
  (\alpha_1,\alpha_2,\alpha_3~~|~~ \beta_1, \beta_2, \beta_3 ~~|~~ \gamma_1, \gamma_2, \gamma_3, \gamma_4, \gamma_5)=\left(\tfrac14,\tfrac12,-1 \right| 1, -1, 1 \left| \tfrac14,-\tfrac12,-\tfrac14,0,\tfrac12\right). 
\end{equation}
The general relation \eqref{eq:RtoROmega2} between the Ricci scalars, up to an irrelevant total derivative, is the root for the equivalents of GR in teleparallel geometries. If we evaluate \eqref{eq:RtoROmega2} in the teleparallel case (i.e., $R=0$), we get that the Einstein-Hilbert term $\mathring{R}$ is dynamically equivalent to $\mathcal{L}_\mathrm{GTEGR}$,
\begin{equation}
  \mathring{R} = \mathcal{L}_\mathrm{GTEGR} +  \mathring{\nabla}_{\mu} \left(\Qb^{\mu}-Q^{\mu}+2 T^{\mu}\right). 
\end{equation}
For this reason, $\mathcal{L}_\mathrm{GTEGR}$ is called \emph{General Teleparallel Equivalent of GR} (GTEGR) (general, because both torsion and nonmetricity are non-vanishing).

Now we are in position to give a nice interpretation of TEGR and STEGR as different gauge-fixed versions of the general equivalent:

\begin{itemize}
\item {\bf TEGR gauge}

In the TEGR, the connection is further restricted to be metric-compatible, i.e.,
\begin{equation}
2(\Lambda^{-1})^\lambda{}_\sigma\partial_\rho\Lambda^\sigma{}_{(\mu} g_{\nu)\lambda} =\partial_\rho g_{\mu\nu}\,. 
\label{eq:Lambdag}
\end{equation}
This relates the metric and $\Lambda^\mu{}_\nu$. This gauge does not fix the full $\mathrm{GL}(4,\mathbb{R})$ symmetry. To see this, notice that \eqref{eq:Lambdag} is solved if
\begin{equation}
g_{\mu\nu}=\Lambda^\rho{}_\mu\Lambda^\sigma{}_\nu c_{\rho\sigma}
\end{equation}
for an arbitrary constant $c_{\rho\sigma}$. The gauge \eqref{eq:Lambdag} indeed leaves undetermined the orthogonal subgroup with respect to the metric $c_{\rho\sigma}$. Since we are interested in Lorentzian metrics, it is natural to choose $c_{\rho\sigma}=\eta_{\rho\sigma}$, and the residual symmetry is nothing but a local Lorentz invariance, which is the well-known symmetry of TEGR.

\item {\bf STEGR gauge}

The STEGR on the other hand is obtained by imposing  $T_{\mu\nu}{}^\rho=0$. This forces the condition $\Lambda^\mu{}_\nu=\partial_\nu \xi^\mu$ for some arbitray $\xi^\mu$'s that can be identified with a coordinate transformation.  In fact, the parameters $\xi^\mu$ can be  interpreted as St\"uckelberg fields introduced to restore covariance of the Einstein Lagrangian, giving rise to the Einstein-Hilbert one (see \eqref{eq:LagEINvsEH}).
\end{itemize}

These two formulations of GR purely in terms of nonmetricity and torsion (respectively), together with the usual Einstein-Hilbert formulation in terms of the Levi-Civita Ricci scalar have been dubbed the \emph{geometrical trinity} \cite{BeltranHeisenbergKoivisto2019}. Note that these are just two very specific gauges and one can consider choices that interpolate between them. This opens up the possibility for a whole plethora of modifications of gravity  based on non-linear extensions of the corresponding partially gauge-fixed version of the GR equivalent analogous to the $f(T)$ and $f(Q)$ theories based on the TEGR and STEGR gauges. It is important to emphasize that most of these extensions will be prone to suffer from pathologies due to the loss of symmetries, as we explained in Section \ref{sec:stabMTG}. In particular, a potentially interesting non-linear extension could be that without any partial gauge-fixing, i.e., $f(\mathcal{L}_\mathrm{GTEGR})$ where the full $\Lambda^\alpha{}_\beta$ is allowed to contribute.

\section{Perturbative spectrum on Minkowski}

\subsection{Quadratic Lagrangian and symmetries}
Let us focus in the linear theory on a Minkowski background. Consider the following first-order expansions of the fields:
\begin{equation}
g_{\mu\nu}=\eta_{\mu\nu}+h_{\mu\nu},\qquad \Lambda^\mu{}_\nu=\delta^\mu{}_\nu+\lambda^\mu{}_\nu.
\end{equation}
If we define $H_{\mu\nu}\coloneqq2\lambda_{(\mu\nu)}$ and $B_{\mu\nu}\coloneqq2\lambda_{[\mu\nu]}$, the torsion and the nonmetricity are given by:
\begin{equation}
T_{\mu\nu}{}^\rho=\partial_{[\mu}(H^\rho{}_{\nu]}+B^\rho{}_{\nu]})\,,\qquad Q_{\rho\mu\nu}= -\partial_\rho \big(h_{\mu\nu}-H_{\mu\nu}\big)\,. \label{eq:pertQ}
\end{equation}
If we express the Lagrangian in $S_{\parallel} \eqqcolon\int \dex^4x \sqrt{|g|} \mathcal{L}_{\parallel}$ in terms of the perturbations, at leading (second) order we get, up to boundary terms,
\begin{align}
\frac{1}{\mpl^2}\mathcal{L}_{\parallel}^{(2)}=&\quad\frac{\gamma_1}{2}\partial_\rho h_{\mu\nu} \partial^\rho h^{\mu\nu}+\frac{\gamma_{24}}{2}\partial_\mu h^{\mu\rho}\partial^\nu h_{\nu\rho}+\frac{\gamma_5}{2}\partial_\mu h\partial_\nu h^{\mu\nu}+\frac{\gamma_3}{2} \partial_\mu h \partial^\mu h \nonumber\\
&+\frac{\cc_1}{8}\partial_\rho H_{\mu\nu} \partial^\rho H^{\mu\nu}+\frac{\cc_2}{8}\partial_\mu H^{\mu\rho}\partial^\nu H_{\nu\rho}+\frac{\cc_3}{4}\partial_\mu H\partial_\nu H^{\mu\nu}+\frac{\cc_4}{8} \partial_\mu H \partial^\mu H \nonumber\\
&+\frac{2\alpha_1-\alpha_2}{8}\partial_\mu B_{\rho\beta}\partial^\mu B^{\rho\beta}+\frac{2\alpha_1-3\alpha_2-\alpha_3}{8}\partial_\mu B^{\mu\rho}\partial^\nu B_{\nu\rho}\nonumber\\
&+\frac{\beta_2-4\gamma_3}{4} \partial_\mu h\partial^\mu H+\frac{2\alpha_1+\alpha_2+\alpha_3-\beta_1+\beta_3}{4}\partial_\mu B^{\mu\rho}\partial^\nu H_{\nu\rho}+\frac{\beta_1-\beta_3}{4}\partial_\mu B^{\mu\rho}\partial^\nu h_{\nu\rho}\nonumber\\
&+\frac{\beta_1-4\gamma_1}{4}\partial_\rho h_{\mu\nu} \partial^\rho H^{\mu\nu}-\frac{\beta_1+\beta_3+4\gamma_{24}}{4}\partial_\mu h^{\mu\rho}\partial^\nu H_{\nu\rho}\nonumber\\
&+\frac{\beta_3-3\gamma_5}{4}\partial_\mu H\partial_\nu h^{\mu\nu}-\frac{\beta_2+2\gamma_5}{4}\partial_\mu h\partial_\nu H^{\mu\nu},
\end{align}
where $h\coloneqq h_\lambda{}^\lambda$, $H\coloneqq H_\lambda{}^\lambda$ and we have defined
\begin{align}
\gamma_{24}&\coloneqq\gamma_2+\gamma_4,\quad\cc_1\coloneqq2\alpha_1+\alpha_2-2\beta_1+4\gamma_1,\quad\cc_2\coloneqq-2\alpha_1-\alpha_2+\alpha_3+2(\beta_1+\beta_3)+4\gamma_{24},\nonumber\\
\cc_3 &\coloneqq-\alpha_3+\beta_2-\beta_3+2\gamma_5,\quad\cc_4\coloneqq\alpha_3-2\beta_2+4\gamma_3.
\end{align}
Let us notice that the parameters $\gamma_2$ and $\gamma_4$ only enter through the combination $\gamma_{24}$ at this order. This degeneracy disappears at the non-linear level, due to the interactions. 
For arbitrary parameters, this quadratic Lagrangian contains the 2-symmetric rank-2 fields $h_{\mu\nu}$ and $H_{\mu\nu}$ plus the antisymmetric field $B_{\mu\nu}$. 

~

Before the study of the field content of the linearized theory, let us revise how the different symmetries of the theory are realized in the perturbations of the metric and the connection:
\begin{itemize}
  \item {\bf Diffeomorphisms} (local). At linear order we have
    \begin{equation}
      \delta_\zeta h_{\mu\nu}=-2\partial_{(\mu}\zeta_{\nu)},\quad\quad
      \delta_\zeta \lambda^\alpha{}_\beta=-\partial_\beta\zeta^\alpha. \label{eq:Diff1}
    \end{equation}
    The latter translates into 
    \begin{equation}
      \delta_\zeta H_{\mu\nu}=-2\partial_{(\mu}\zeta_{\nu)}\quad \text{and}\quad \delta_\zeta B_{\mu\nu}=2\partial_{[\mu}\zeta_{\nu]}. \label{eq:Diff2}
    \end{equation}
    As a consequence of the Stewart-Walker lemma \cite{StewartWalker1974}, since the background torsion and nonmetricity vanish, their perturbations are gauge-invariant under linearized diffeomorphisms.

  \item {\bf General linear transformations} (global). This global symmetry of the connection is realized as
    \begin{equation}
      \delta_\varpi\lambda^\mu{}_\nu=\varpi^\mu{}_\nu \qquad \text{or} \qquad \delta_\varpi H_{\mu\nu}=2\varpi_{(\mu\nu)},\quad \delta_\varpi B_{\mu\nu}=2\varpi_{[\mu\nu]},
    \end{equation}
    where $\varpi^\mu{}_\nu\in\mathfrak{gl}(4,\mathbb{R})$ is constant. The absence of any masses for $B_{\mu\nu}$ and $H_{\mu\nu}$ is guaranteed by this global symmetry. Notice that the field $B_{\mu\nu}$ transforms according to the $\mathfrak{so}(1,3)$ subalgebra, i.e., the Lorentz part. On the other hand, $H_{\mu\nu}$ transforms with the generators of the complementary part of the algebra. One interesting consequence of this is that the realization of a local Lorentz symmetry in the quadratic Lagrangian will be connected with the absence of $B_{\mu\nu}$.\footnote{The reason for this is that the Noether identity under a local shift symmetry of a field says that the corresponding equation of motion must be fulfilled off-shell. And this is only true if all the contributions of that field to the Lagrangian are either zero or a boundary term.}
\end{itemize}

\subsection{Minimal field content: GTEGR}
Here we consider the general equivalent of GR \eqref{eq:GTEGRdef}. At second order the Lagrangian reads
\begin{equation}
  \frac{1}{\mpl^2}\mathcal{L}_{\rm GTEGR}^{(2)}=\frac18\partial_\alpha h_{\mu\nu} \partial^\alpha h^{\mu\nu}-\frac14\partial_\mu h^{\mu\alpha}\partial^\nu h_{\nu\alpha}+\frac14\partial_\mu h\partial_\nu h^{\mu\nu}-\frac18 (\partial h)^2.
\end{equation}
This is precisely the Fierz-Pauli Lagrangian for the metric perturbations $h_{\mu\nu}$. Therefore, only a (healthy) spin-2 mode propagates, whereas the dofs associated to the connection enter as a total derivative. This reflects that, at linear level, the local gauge symmetry $\mathrm{GL}(4,\mathbb{R})$ is realized up to the total derivative term that we have dropped.

It is interesting to check that requiring the local $\mathrm{GL}$ symmetry, even at linear order, is a very strong condition that fixes the theory to be GTEGR (i.e. GR) at that order, except for the degeneracy between $\gamma_2$ and $\gamma_4$, so there is only one free parameter in the full theory.\footnote{
    Observe that here we are imposing local $\mathrm{GL}$ symmetry at linear order. If at higher orders such symmetry is violated, this would diagnose a strong coupling problem in Minkowski, because there is a discontinuity in the number of dofs.} 
This can be directly obtained by imposing $H_{\mu\nu}$ and $B_{\mu\nu}$ to have trivial linear field equations.

This is the minimal field content in order to describe gravity. Therefore theories with other parameters, and with gravity as a required sector, will propagate more dofs.

\subsection{Maximal field content}

The general quadratic Lagrangian contains the fields $h_{\mu\nu}$, $H_{\mu\nu}$ and $B_{\mu\nu}$, which are prone to propagate ghost-like modes. To avoid it, in addition to the Diff symmetry, we are going to introduce appropriate extra gauge symmetries. The maximum number of physical dofs that we can have without incurring in ghostly instabilities will correspond to having 2 massless spin-2 fields (2$\times$2) plus a massless Kalb-Ramond field (1), making a total of 5 dofs. Indeed any choice of parameters propagating more than 5 dofs will have ghosts around a Minkowski background. It will then be useful to study those theories that precisely propagate this number of dofs.

\newpage

\noindent\textit{Conditions on $B_{\mu\nu}$}

We are going to consider that the transformation $\delta B_{\mu\nu}= 2\partial_{[\mu}\theta_{\nu]}$ (usual transformation of a massless 2-form) is a gauge symmetry of $\delta S^{(2)}_\parallel$.  The associated Noether identity is
\begin{align}
  0=\partial_\mu\Bigg(\frac{\delta S^{(2)}_\parallel}{\delta B_{\mu\nu}}\Bigg)
     &\propto(2\alpha_1 +\alpha_2 +\alpha_3 +\beta_3-\beta_1)\left(\square\partial_{\sigma}H^{\nu\sigma}-\partial^{\nu}\partial_{\mu}\partial_{\sigma}H^{\mu\sigma}\right)\nonumber\\
     &\qquad-(\beta_3-\beta_1)\left(\square\partial_{\sigma}h^{\nu\sigma}-\partial^{\nu}\partial_{\mu}\partial_{\sigma}h^{\mu\sigma}\right)+(2\alpha_1 +\alpha_2 +\alpha_3 )\square\partial_{\mu}B^{\mu\nu}\,,
\end{align}
where $\square\coloneqq \eta^{\mu\nu} \partial_\mu\partial_\nu$. This condition imposes 
\begin{equation}
\text{Gauge 2-form:}\quad\quad\quad2\alpha_1+ \alpha_2+\alpha_3=0 \quad \text{and}\quad \beta_3=\beta_1. 
\label{eq:gaugeB}
\end{equation}
Since $Q_{\rho\mu\nu}$ is independent of $B_{\mu\nu}$, the pure nonmetricity sector remains completely free. The first condition is the same that one obtains in the context of New GR to ensure the decoupling between the Kalb-Ramond field and the graviton \cite{Ortin2004}. On the other hand, the second condition comes from the mixed sector that combines torsion and nonmetricty, which is absent in New GR and also in Newer GR. Note that imposing this symmetry decouples $B_{\mu\nu}$ from the symmetric sector $h_{\mu\nu}$ and $H_{\mu\nu}$. 

~

\noindent\textit{Conditions on the symmetric sector. Alternative I}

Consistency of the symmetric sector requires additional gauge symmetries. In particular, we will impose invariance under another copy of linearized diffeomorphisms (independent from the one that the full theory exhibits).\footnote{\label{fn:diagHh}
    We have chosen to fully realize the extra copy of Diff with $h_{\mu\nu}$, but it could also be realized with $H_{\mu\nu}$, giving the same results. Indeed, the general transformations $\delta h_{\mu\nu}=\alpha_1\partial_{(\mu}\zeta^1_{\nu)}+\alpha_2\partial_{(\mu}\zeta^2_{\nu)}$ and $\delta H_{\mu\nu}=\beta_1\partial_{(\mu}\zeta^1_{\nu)}+\beta_2\partial_{(\mu}\zeta^2_{\nu)}$ can be trivially diagonalized with a redefinition of the gauge parameters.}
The corresponding Noether identity is
\begin{align}
  0=\partial_\mu\Bigg(\frac{\delta S^{(2)}_\parallel}{\delta h_{\mu\nu}}\Bigg)
     &\propto(\beta_1-\beta_3-8\gamma_1-4\gamma_{24})\square\partial_\mu H^{\mu\nu} -(\beta_1+2\beta_2+\beta_3+4\gamma_{24} +4\gamma_5)\partial^\nu \partial_\mu\partial_\sigma H^{\mu\sigma}\nonumber\\
     &\quad+2(\beta_2+\beta_3-2\gamma_5-4\gamma_3) \square\partial^{\nu}H -(\beta_1-\beta_3)\square\partial_\rho B^{\nu\rho}\nonumber\\
     &\quad+4(2\gamma_1+\gamma_{24})\square\partial_\mu h^{\mu\nu} +4(\gamma_{24}+\gamma_5)\partial^\nu\partial_{\mu}\partial_\sigma h^{\mu\sigma}+4(2\gamma_3+\gamma_5)\square\partial^\nu h\,.
\end{align}
Assuming \eqref{eq:gaugeB}, this identity is identically satisfied if 
\begin{equation}
  \text{Diff}\times\text{Diff:}\quad\quad\quad \gamma_5=2\gamma_1,\quad \gamma_3=-\gamma_1,\quad \gamma_{24}=-2\gamma_1,\quad\text{and}\quad \beta_2=-\beta_1\,.
\end{equation}

Finally, we impose the decoupling of $h_{\mu\nu}$ and $H_{\mu\nu}$, which requires the additional condition $\beta_1=4\gamma_1$. This guarantees that the Newtonian limit is appropriately recovered.\footnote{
  Here we are ensuring that only the metric perturbations $h_{\mu\nu}$ couples to the matter.}

Under all of these conditions the quadratic Lagrangian reduces to
\begin{equation}
  \frac{1}{\mpl^2}\mathcal{L}_{\parallel}^{(2)} =-\gamma_1h_{\rho\lambda}E^{\rho\lambda\mu\nu}h_{\mu\nu}-\frac{2 \alpha_1 + \alpha_2 - 4 \gamma_1}{4}H_{\rho\lambda}E^{\rho\lambda\mu\nu}H_{\mu\nu}+\frac{2\alpha_1-\alpha_2}{24}F_{\mu\nu\rho}F^{\mu\nu\rho},
\end{equation}
where we have introduced the  Kalb-Ramond field strength $F_{\mu\nu\rho}\coloneqq 3\partial_{[\mu} B_{\nu\rho]}$ and the Lichnerowicz operator in Minkowski space,\footnote{
    This operator basically allows to rewrite the Fierz-Pauli Lagrangian in a compact way as follows
    \begin{align}
      \mathcal{L}_\mathrm{FP} & =\tfrac{1}{2}\partial_{\mu}h_{\nu\rho}\partial^{\mu}h^{\nu\rho}-\partial_{\rho}h{}^{\rho\mu}\partial_{\sigma}h^{\sigma}{}_{\mu}+\partial_{\sigma}h^{\sigma}{}_{\mu}\partial^{\mu}h-\tfrac{1}{2}\partial_{\mu}h\partial^{\mu}h \nonumber\\
      &= -h^{\mu\nu}E_{\mu\nu}{}^{\rho\sigma}h_{\rho\sigma}+\text{boundary term}\,.\nonumber
   \end{align}
    }
\begin{equation}
  E_{\mu\nu}{}^{\rho\sigma}\coloneqq\frac{1}{2}\left(\delta_{\mu}^{\rho}\delta_{\nu}^{\sigma}\square-2\delta_{(\mu}^{\rho}\partial_{\nu)}\partial^{\sigma}+\eta^{\rho\sigma}\partial_{\mu}\partial_{\nu}+\eta_{\mu\nu}\left(\partial^{\rho}\partial^{\sigma}-\eta^{\rho\sigma}\square\right)\right)\,.
\end{equation}

One interesting final remark is that the three gauge symmetries that we have obtained give rise to a full decoupling of the original Diffs for the three fields, i.e., the transformations \eqref{eq:Diff1} and $\eqref{eq:Diff2}$ become symmetries with independent parameters for $h_{\mu\nu}$, $H_{\mu\nu}$ and $B_{\mu\nu}$.

~

\noindent\textit{Conditions on the symmetric sector. Alternative II}

An alternative to the previous extra Diff symmetry is imposing an additional Weyl Transverse Diffeomorphism (WTDiff). This also guarantees the propagation of 2 massless spin-2 fields. Firstly, Transverse Diffeomorphism (TDiffs), correspond to the invariance under a diffeomorphism with $\partial_\mu\zeta^\mu=0$. This can also be realized either with $h_{\mu\nu}$ or with $H_{\mu\nu}$ independently.\footnote{
    This can be seen by using an similar argument as in Footnote \ref{fn:diagHh}, thanks to the already existing Diff symmetry and to the fact that TDiff is a subgroup of Diff.} 
In both cases we get
\begin{equation}
\text{TDiff:}\quad\quad\quad2\gamma_1+\gamma_{24}=0.
\end{equation} 
The idea now, instead of completing this symmetry to full Diffs as before, is to add invariance under Weyl transformations. We consider the transformation $\delta h_{\mu\nu}=w_h\varphi\eta_{\mu\nu}$ and $\delta H_{\mu\nu}=w_H\varphi\eta_{\mu\nu}$, where $w_h$ and $w_H$ are not-necessarily equal weights. The corresponding Noether identity is
\begin{equation} 
\eta_{\mu\nu}\left[w_h\frac{\delta}{\delta h_{\mu\nu}}+w_H\frac{\delta }{\delta H_{\mu\nu}}\right]S^{(2)}_\parallel=0,
\end{equation}
which is fulfilled under the following conditions:
\begin{align}
  0 &= 2 (2 \alpha_1 + \alpha_2) w_H - 2 (\gamma_{24} - 8 \gamma_3  - \gamma_5) (w_h - w_H) -\beta_1 (2 w_h - 3 w_H) - \beta_2 (4 w_h - 7 w_H)\,,\nonumber\\
  0 &= \beta_1 w_h + 2 (\beta_2 + \gamma_{24}  + 2 \gamma_5) (w_h - w_H) - (2 \alpha_1 + \alpha_2) w_H \,,\nonumber\\
  0 &= 2 (\gamma_{24} - 8 \gamma_3 - \gamma_5) (w_h - w_H) - (\beta_1 + 3 \beta_2) w_H\,,\nonumber\\
  0 &=2(\gamma_{24} + 2 \gamma_5) (w_h - w_H) - \beta_1 w_H \,.
\end{align}
Two interesting realizations of the additional Weyl symmetry are:\footnote{
    See \cite{IosifidisKoivisto2018} for a detailed analysis of the different realizations of conformal/scale/Weyl transformations within the metric-affine framework.}
\begin{itemize}
\item $w_H=0$: The Weyl symmetry is fully realized on $h_{\mu\nu}$ while the connection does not transform. The solution is then $\beta_1+2\beta_2=3\gamma_{24}-16\gamma_3=8\gamma_3+3\gamma_5=0$.
\item $w_H=w_h$: The Weyl symmetry is covariantly realized with the change in the metric perturbation accompanied by the corresponding change in the connection. In this case we find $2\alpha_1+\alpha_2=\beta_1=\beta_2=0$.
\end{itemize}
Let us finally mention that the parameters $\alpha_i=\beta_i=0$ and $\gamma_5=-\frac83 \gamma_3=-\frac12 \gamma_{24}= \gamma_1 \neq0$ gives the following Lagrangian 
\begin{equation}
  \frac{1}{\mpl^2}\mathcal{L}_{\parallel}^{(2)} = \gamma_{1} \left[     \frac{1}{2}\partial_\mu\tilde{h}_{\nu\rho}\partial^\mu\tilde{h}^{\nu\rho} -\partial_\rho\tilde{h}{}^{\rho\mu}\partial_\sigma\tilde{h}^\sigma{}_\mu +\frac{1}{2}\partial_\sigma\tilde{h}^\sigma{}_\mu\partial^\mu\tilde{h} -\frac{3}{16}\partial_\mu \tilde{h}\partial^\mu\tilde{h} \right] \,,
\end{equation}
for $\tilde{h}_{\mu\nu}\coloneqq h_{\mu\nu}- H_{\mu\nu} $ and $\tilde{h}\coloneqq \tilde{h}_\sigma{}^\sigma$. This is invariant under WTDiff transformations for arbitrary $w_H$ and $w_h$. Indeed, the object in square brackets is the 4-dimensional evaluation of the so-called \emph{WTDiff-theory} in flat space  \cite{Bonifacio2015}:
\begin{equation}\label{eq:WTDiffDdim}
  \mathcal{L}_{\text{WTDiff},\dimM}(\tilde{h}_{\mu\nu}) =      \frac{1}{2}\partial_\mu \tilde{h}_{\nu\rho}\,\partial^\mu \tilde{h}^{\nu\rho} -\partial_\rho \tilde{h}{}^{\rho\mu}\,\partial_\sigma \tilde{h}^\sigma{}_\mu +\frac{2}{\dimM}\partial_\sigma \tilde{h}^\sigma{}_\mu\,\partial^\mu \tilde{h} -\frac{\dimM+2}{2 \dimM^2}\partial_\mu \tilde{h}\,\partial^\mu \tilde{h}  \,.
\end{equation}
This Lagrangian also describes a massless spin-2 field (see e.g. \cite{AlvarezBlasGarriga2006}). The only difference with linearized GR is the appearance of a cosmological constant as an integration constant. In fact, \eqref{eq:WTDiffDdim} corresponds to the linearization of \emph{unimodular gravity} (see \cite{AlvarezFaedo2007} and references therein).

\subsection{Theories with local Lorentz invariance}
An interesting class of theories with enhanced  symmetries are those with a local Lorentz invariance. As we explained above, this is achieved at linear order by imposing the disappearance of $B_{\mu\nu}$ from the Lagrangian or, equivalently, by obtaining the parameters that trivialize its equation of motion. Notice that only the terms involving the torsion are relevant, since the nonmetricity is exclusively $H_{\mu\nu}$-dependent at this order. In particular, this means that the coefficients $\gamma_{i}$ are not constrained by this requirement. Therefore, requiring this local Lorentz symmetry implies $\beta_3=\beta_1$, $\alpha_3=-4\alpha_1$ and $\alpha_2=2\alpha_1$, i.e., the pure torsion sector must reduce to the TEGR Lagrangian, while $\beta_2$ remains unfixed and $\beta_3=\beta_1$. This of course contains the GTEGR case discussed above.

\section{Conclusions}

In this chapter, we have shown how TEGR and STEGR can be seen as particular gauge-fixings of  the general equivalent GTEGR. Indeed, the singular nature of GTEGR has been revealed to be a complete gauging of the global $\mathrm{GL}(4,\mathbb{R})$ symmetry enjoyed by the general inertial connection. In addition to this, we have shown that requiring such a gauging of the global symmetry in the linear theory around Minkowski only leaves one free parameter.

Then we have obtained the 2nd-order perturbative Lagrangian around the Minkowski spacetime with Levi-Civita connection, and discussed the need for additional symmetries in order to avoid ghosts. We provided two alternatives. The first one consists in imposing the usual gauge symmetry in the 2-form field $B_{\mu\nu}$ and an extra copy of diffeomorphisms in the symmetric sector $\{H_{\mu\nu}, h_{\mu\nu}\}$. The other one is identical to the latter concerning the antisymmetric sector; however, in the symmetric sector we introduced an extra WTDiff symmetry, which also permit the propagation of an additional healthy graviton.

Finally we comment on a somewhat extended {\it folk} argument in favor of teleparallel theories stating that they provide a better starting point for modifications of gravity. The alluded reason is that the action only contains first order derivatives of the fields and, consequently, the corresponding extensions are less prone to introducing Ostrogradski instabilities than the curvature based theories that contain second derivatives of the metric. For reasons we explained in Section \ref{sec:stabconcl}, this is not necessarily true. The equivalents of GR crucially realize some symmetries up to total derivatives so one must be very careful when considering either extensions or non-standard matter couplings in order not to introduce additional potentially unstable dofs.

\subsubsection*{Limitations of this work/future directions}
Our enhancement of the symmetries to avoid  pathologies, however, is not the end of the story. There are some limitations/details one should be aware of:
\begin{itemize}
  \item Imposing the required gauge symmetries at linear order is a necessary but not a sufficient condition. A good example of this subtlety is New GR. In this theory it has been observed that the gauge symmetry that renders the 2-form stable at linear order cannot be maintained at the non-linear order \cite{BeltranDialektopoulos2020}.

  \item Even if the additional massless spin-2 field can be made to enjoy the necessary symmetries at linear order, it is expected that at full non-linear order the two spin-2 fields will interact. But it is known that a theory with massless spin-2 fields in its spectrum only admits one single species of this type. Another possibility one could envision is that one of the spin-2 fields becomes massive with healthy interactions. However, also in that situation the two spin-2 fields would present derivative interactions which are prone to the re-introduction of pathological modes \cite{deRhamMatasTolley2013, deRhamMatasTolley2015}. The difficulty to realize in general the safe self-interactions obtained in \cite{Hinterbichler2013} suggests that GTEGR is probably the only consistent general quadratic teleparallel theory that includes gravity.
\end{itemize}


\chapter{Particle spectrum of quadratic MAG}\label{ch:MAGspectrum}

\boxquote{I can live with doubt, and uncertainty, and not knowing. I think it's much more interesting to live not knowing than to have answers which might be wrong. I have approximate answers, and possible beliefs, and different degrees of certainty about different things, but I'm not absolutely sure of anything.}{Richard P. Feynman, ``The Pleasure of Finding Things Out'' (1981)}

In this chapter we present some preliminary results on the particle spectrum of quadratic metric-affine gravity. We consider the perturbation expansion in four dimensions, around Minkowski space and including both the even and the odd parity sectors in \eqref{eq: MAG Lag form}. We will focus just on the spin-0 sector of the theory; the rest is still {\it work in progress} \cite{JCAObukhov2021b}.

\section{Introduction}

The complete stability analysis of the (quadratic) MAG theory \eqref{eq: MAG Lag form} is a really challenging task. Among the different approaches that can be followed to understand the behavior of its dofs, probably the most natural one to begin with is the study of the particle spectrum in flat space. Basically,  the idea will be to perform a perturbation expansion of the fields in our metric-affine geometry $(g_{ab},\cofr^a,\dfom_a{}^b)$ using the flat metric and its Levi-Civita connection as background values.\footnote{
    We say ``flat'' and not Minkowski intentionally, since by ``Minkowski'' metric we usually mean the flat metric {\it in Cartesian coordinates}.} 
At the level of the action (and in terms of field content), (quadratic) MAG theory \eqref{eq: MAG Lag form} can be seen as a direct generalization of (quadratic) PG. Therefore, it is interesting to have a look at the latter first, to see what kind of stability problems are expected in MAG. In \cite{YoNester1999,YoNester2002}, the linear spectrum of (quadratic) PG is analyzed, and it is shown that only an extra scalar or an extra pseudoscalar (besides the graviton) can propagate in order to have a healthy theory. Moreover, in \cite{BeltranMaldonado2020}, the authors follow a different approach. First they performed a Levi-Civita + distorsion splitting to work with the metric, the torsion and the nonmetricity as basic variables (the latter being zero in PG). Then they impose several strong constraints on the vector sector of the theory, such as an appropriate Riemannian limit for the quadratic curvature sector (to avoid ghostly propagations coming from the metric), and the vanishing of certain types of terms to avoid ghosts and strong coupling issues. Interestingly, they reached similar conclusions as \cite{YoNester1999,YoNester2002}. This strong result, one more time, reflects how difficult is to consistently modify GR without incurring in pathologies.

(Quadratic) MAG is even more complex than PG, so all these problems are expected to occur.  Maybe a complete Hamiltonian analysis could be the most powerful tool to find out the number of true degrees of freedom of the theory by studying the associated algebra of constraints \cite{Dirac_LecturesQM, Wipf1993, Tavakoli2014Lec1,Date2010,Blagojevic2001, PonomarevObukhov2017}.\footnote{See for instance \cite{DAmbrosioGarg2020, DAmbrosioHeisenberg2020} as examples of applications of the Hamiltonian analysis to theories with nonmetricity, or \cite{BlixtGuzman2020} (and references therein) for a review on Hamiltonian analysis in teleparallel theories.} However, it is better to start with more simple approaches to learn first about the new subtleties associated to the presence of nonmetricity (which includes a spin-3 mode). Here, we will address the study of the particle spectrum of (quadratic) MAG in flat space.

For the spectrum analysis there are also different ways to proceed. One of the classical ones is by the method of spin projectors.\footnote{
    See \cite{Karananas2015} for the application of this technique to the analysis of the PG spectrum. Other examples are \cite{LinHobson2019, LinHobson2020, LinHobson2020b}.} 
Some work has been done in MAG in this direction \cite{PercacciSezgin2020}, although they focus on the two particular cases with either zero torsion or zero nonmetricity. Our idea here will be not to restrict the fields, at least in a {\it ad hoc} way and, instead of using this spin-projector technique, we will follow the steps of \cite{BlagojevicCvetkovic2018}.

To finish this introduction we present in Table \ref{tab:MAGmodes} the number of irreducible spin-modes that can arise from the connection in MAG via Young scheme \cite{BaikovHayashi1992, PercacciSezgin2020, AlvarezAnero2018}. Notice that this could not necessarily coincide with the number of connection degrees of freedom in MAG, because there could be special constraints fixing some dofs in terms of others or extra modes that were not a priori due to the presence of ghosts (higher derivatives in the longitudinal modes of some fields, see e.g. \eqref{eq:ghostprop}).

\begin{table}
\begin{centering}
\renewcommand\arraystretch{1.5}
\begin{tabular}{|c|c|c|c|c|}
\hline 
         & Spin-0 modes & Spin-1 modes & Spin-2 modes & Spin-3 modes\tabularnewline
\hline 
\hline 
$\irrdfT{1}$ & 0 & 2 & 2 & 0 \tabularnewline
\hline 
$\irrdfT{2}$ & 1 & 1 & 0 & 0 \tabularnewline
\hline 
$\irrdfT{3}$ & 1 & 1 & 0 & 0 \tabularnewline
\hline 
$\irrdfQ{1}$ & 1 & 1 & 1 & 1 \tabularnewline
\hline 
$\irrdfQ{2}$ & 0 & 2 & 2 & 0 \tabularnewline
\hline 
$\irrdfQ{3}$ & 1 & 1 & 0 & 0 \tabularnewline
\hline 
$\irrdfQ{4}$ & 1 & 1 & 0 & 0 \tabularnewline
\hline 
\hline 
Total        & 5 & 9 & 5 & 1 \tabularnewline
\hline 
\end{tabular}
\renewcommand\arraystretch{1}
\end{centering}
\centering{}\caption{\label{tab:MAGmodes}Spin content (number of spin modes) of the irreducible components of the torsion and the nonmetricity.}
\end{table}

\section{Perturbative expansion}

Let  $(\bgg_{ab}, \bgcofr^a, \bgdfcon_a{}^b)$ be our background metric-affine geometry. In principle, we consider it to be a Riemannian geometry (i.e., zero torsion and zero nonmetricity) with constant curvature:
\begin{align}
  \bgDex\,\bgg_{ab}&= \dex\bgg_{ab} - \bgdfcon{}_a{}^c\,\bgg_{cb} - \bgdfcon_b{}^c\,\bgg_{ac} = 0,\\
  \bgDex\bgcofr^a  &= \dex\bgcofr^a + \bgdfcon_b{}^a\wedge\bgcofr^b = 0,\\
  \bgdfR_a{}^b     &= \dex\bgdfcon_a{}^b + \bgdfcon_c{}^b \wedge\bgdfcon_a{}^c = \KMaxSym\,\bgg_{ac} \bgcofr^c\wedge\bgcofr^b, \label{eq:bgR}  
\end{align}
where $\KMaxSym$ is some real constant.

Now we perform a perturbation of these variables around the chosen background,
\begin{align}
  g_{ab}     &= \bgg_{ab} + \perg_{ab}, \\ 
  \cofr^a    &= \bgcofr^a + \percofr^a, \label{eq:cofper}\\
  \dfom_a{}^b&= \bgdfcon_a{}^b + \perdfcon_a{}^b, 
\end{align}
which give the following (exact) expansions for the torsion, the curvature and the nonmetricity:
\begin{align}
  \dfQ_{ab} &= -\bgDex\perg_{ab} + 2\perdfcon_{(a}{}^c \, \bgg_{b)c} + 2\perdfcon_{(a}{}^c\,\perg_{b)c}, \label{eq:Qper} \\ 
  \dfT^a    &= \bgDex\percofr^a + \perdfcon_b{}^a\wedge\bgcofr^b + \perdfcon_b{}^a\wedge\percofr^b,\label{eq:Tper}\\
  \dfR_a{}^b&= \KMaxSym\,\bgg_{ac}\bgcofr^c\wedge \bgcofr^b + \bgDex\perdfcon_a{}^b + \perdfcon_c{}^b\wedge \perdfcon_a{}^c.\label{eq:Rper}  
\end{align}
From now on we will use the metric $\bgg_{ab}$ to raise/lower the indices and write $\star$ instead of $\bgobject{\star}$ in order to alleviate the notation.\footnote{
    Indeed, this can be seen as a direct substitution $\star=\bgobject{\star}$ (and not just as a matter of notation). This is justified because the star will always be in expressions of the type $\star\dfal$ or $\dfal\wedge \star \bgcofr_{a...b}$ where $\dfal$ represents a first-order object. Therefore, the difference between $\star$ and $\bgobject{\star}$ is second-order and can be ignored in our computations, since we will work at linear order.} 
In addition, we will assume small perturbations so we will drop the quadratic contributions represented by the last terms in these expressions. 

It is worthwhile to notice that by combining \eqref{eq:Qper} and \eqref{eq:Tper}, we can express the connection perturbation in terms of the perturbed nonmetricity and the torsion:
\begin{align}
\perdfcon_{ab} =& -\,\dint{\vfre_{[a}} \dfT_{b]}  +\,\dint{\vfre_{[a}} \bgDex \percofr_{b]} + {\frac 12}\,\bgcofr^c \dint{\vfre_a} \dint{\vfre_b} \dfT_c - {\frac 12}\,\bgcofr^c
\dint{\vfre_a} \dint{\vfre_b} \bgDex\percofr_c \nonumber\\
&  + \bgcofr^c \dint{\vfre_{[a}} \dfQ_{b]c}+ \bgcofr^c \dint{\vfre_{[a}} \bgDex \perg_{b]c} +{\frac 12}\,\dfQ_{ab} + {\frac 12}\,\bgDex\perg_{ab} \, .\label{eq:gamchimu}
\end{align}
By using this in \eqref{eq:Rper}, we find 
\begin{equation}\label{eq:Rnu}
\star\bgcofr_{ab}\wedge\dfR^{ab} = 12\KMaxSym\svolf
+ \dex\star (2\dfT - \dfvarLa + 3\dfQ) + \dex\star \Big[-\,\dint{\vfre_a}\bgDex(2\percofr^a + \perg^a{}_b \bgcofr^b)\Big]+\mathcal{O}(2).
\end{equation}
Hence, the perturbed curvature scalar $X\svolf = -\,\star\bgcofr_{ab}\wedge\dfR^{ab}$ is not just a function of the perturbed torsion and nonmetricity, but it also picks up a contribution from the metric and coframe perturbations,
\begin{equation}\label{eq:XdTnu}
(-R\equiv)\quad X = -12\KMaxSym + \star\dex\star (2\dfT - \dfvarLa + 3\dfQ) + \star\dex\star \Big[-\dint{\vfre_a}\bgDex(2\percofr^a + \perg^a{}_b\bgcofr^b)\Big]+\mathcal{O}(2).
\end{equation}

\subsection{Linearized MAG field equations}

Following along the lines of \cite{BlagojevicCvetkovic2018}, we take flat spacetime as background ($\KMaxSym=0$ in \eqref{eq:bgR}). This choice is only consistent if the cosmological constant vanishes. Under these assumptions and after removing the matter sector, the linearization of the general MAG equations presented in Theorem \ref{th:EoMqMAG} yields
\begin{align}
\mathbf{E}_a\coloneqq\kappa\frac{\delta S}{\delta\cofr^a} & =\frac{a_0}{2}\dfR^{bc}\wedge\star\bgcofr_{abc} +\odda_0\dfR_{[ac]}\wedge\bgcofr^c -\bgDex(\dfh_a+ \overline{\dfh}_a) + \mathcal{O}(2) \,, \label{eq:qMAGEqe0lin}\\
\mathbf{C}^a{}_b\coloneqq\kappa\frac{\delta S}{\delta\dfom_a{}^b} & =-\bgcofr^a\wedge(\dfh_{b}+\overline{\dfh}_{b}) -2(\dfm^a{}_{b}+\overline{\dfm}{}^a{}_b)\nonumber \\
 & \quad+\frac{a_0}{2} \big(\dfT^c\wedge\star\bgcofr{}^a{}_{bc} +\dfQ^{ac}\wedge\star\bgcofr_{cb}-2\dfQ\wedge\star\bgcofr{}^a{}_b \big)\nonumber \\
 & \quad+\frac{\odda_0}{2} \big(2g^{ac}\dfT_{[c}\wedge\bgcofr_{b]}- \dfQ_{cb}\wedge\bgcofr^{ac}\big)-\lrho\bgDex(\dfh^a{}_b+\overline{\dfh}^a{}_b)  + \mathcal{O}(2) \label{eq:qMAGEqw0lin}\,.
\end{align}
Notice that the term $\dfq_{a}$ disappeared since it is purely second-order. In these equations the objects $\dfQ_{ab}, \dfT^{a}, \dfR_a{}^b$, and $\dfm^{ab}, \dfh_{a}, \dfh^{a}{}_{b}$, $\odddfm^{ab}, \odddfh_{a}, \odddfh^{a}{}_{b}$ represent the variables up to {\it first order} in perturbation theory (they differ from the original ones by second-order terms). From now on, the symbol $\mathcal{O}(2)$ will be dropped, and all of the expressions will be understood as valid up to second-order terms.

\subsection{Linearized Bianchi identities}

Now we are going to derive some useful expressions from the linearized Bianchi identities (see Proposition \ref{prop:BianchiTQR}),
\begin{align}
\Dex\dfQ_{ab} &= 2\dfR_{(a}{}^c\,g_{b)c}  
   & \Rightarrow&& \bgDex\dfQ_{ab} &=2 \dfZ_{ab}\,,\label{eq:BiaL0}\\
\Dex\dfT^a &= \dfR_b{}^a\wedge\cofr^b
   & \Rightarrow&& \bgDex \dfT^a  &= \dfR_b{}^a\wedge\bgcofr^b\,,\label{eq:Bia1L}\\
\Dex\dfR_a{}^b &= 0 
   & \Rightarrow&& \bgDex \dfW^{ab} &= 0,\qquad  \bgDex \dfZ^{ab}=0\,.\label{eq:BiaL2} 
\end{align}

From the trace of \eqref{eq:BiaL0}, we find
\begin{equation}
\dfZ = 2\dex \dfQ, \qquad \dex\dfZ = 0. \label{eq:ZdQ}
\end{equation}
In fact, the second equation in \eqref{eq:BiaL2} is redundant, being a direct consequence of \eqref{eq:BiaL0}. From the latter we derive, by taking into account \eqref{eq:ZdQ}, 
\begin{equation}
\dfZNoTr_{ab} = {\frac 12}\,\bgDex\dfQNoTr_{ab}.\label{eq:aZdQ}
\end{equation}
In addition, from the linearized Bianchi identities of the curvature \eqref{eq:BiaL2}, we can derive some useful properties for the objects $\dfPsi^a$, $\odddfPsi^a$, $\dfPhi^a$ and $\odddfPhi^a$, which determine some of the irreducible components of the curvature (see Appendix \ref{app:irreds}). If we introduce 
\begin{align}
(\vartheta X) &\coloneqq \bgcofr^a\wedge \dfX_a\,, \\
(\vartheta Y) &\coloneqq \bgcofr^a\wedge \dfY_a\,, \\
\dfPhi &\coloneqq (\dint{\vfre^b}\star \bgDex\star \bgDex\dfQNoTr_{ab})\,\bgcofr^a\,,\label{eq:phi}
\end{align}
then one can show:
\begin{align} 
\bgDex\,\star \odddfPsi^a &=\bgcofr^a\wedge\Bigl[-{\frac 12} \dex(\vartheta X) - {\frac 14}\,\star \dex\oddX\Bigr],\label{eq:DoPsi}\\
\bgDex\,\star \odddfPhi^a &=\bgcofr^a\wedge\Bigl[-{\frac 12}\dex(\vartheta Y) \Bigr],\label{eq:DoPhi}\\
\bgDex\,\star \dfPsi^a      &=\bgcofr^a\wedge\Bigl[{\frac 12}\dex\star (\vartheta X) + {\frac 14}\,\star \dex X\Bigr]\,,\label{eq:DPsi}\\
\bgDex\,\star \dfPhi^a      &=\bgcofr^a\wedge\Bigl[-{\frac 12}\,\dex\star (\vartheta Y) + {\frac 12}\,\star \dfPhi\Bigr]\,.\label{eq:DPhi}
\end{align}
The first three equations are obtained either by multiplying each of equations in \eqref{eq:BiaL2} with $\cofr_a\wedge$, or by taking interior products $\dint{\vfre_a} \dint{\vfre_b}$. For the last one, we used the definition of $\dfPhi_a$ and then \eqref{eq:aZdQ}.

The 1-form $\dfPhi$ we have introduced in \eqref{eq:phi}, is characterized by the crucial property
\begin{equation}
\dex\star \dfPhi = \dex\star \dfDe - {\frac 23}\dex\star\dex\star \dex\star \dfvarLa,\label{eq:dphi} 
\end{equation}
where we defined the 1-form
\begin{align}
\dfDe := (\dint{\vfre^b}\star \bgDex\star \bgDex\,\irrdfQ{1}_{ab})
\,\bgcofr^a.\label{eq:del}
\end{align}

Finally, we linearize the expression \eqref{eq:NiehYanbt} we derived in Chapter \ref{ch:Lovelock} for the metric-affine generalization of the Nieh-Yan invariant. The result is:
\begin{align}\label{eq:XdT}
\oddX = \star\dex\star \odddfT.
\end{align}

\section{Analysis of the particle spectrum: Spin-0 sector}

The spin-0 sector is described by the scalar variables which can be constructed from the irreducible parts of the gravitational fields. In \cite{BlagojevicCvetkovic2018}, the only scalars correspond to the divergences of the trace and the axial part of the torsion. Indeed, for a given 1-form $\dfal=\alpha_a\cofr^a$ if we call $\partial\alpha\equiv\bgnabla_a \alpha^a$, it is not difficult to check that $\partial\alpha=-\,\star \dex\star \dfal$. In MAG, four scalars can be constructed from the connection in this way,
\begin{equation}
\partial T = -\,\star \dex\star \dfT,\qquad \partial \oddT = -\,\star \dex\star \odddfT,\label{eq:dTT}
\end{equation}
\begin{equation}
\partial Q = -\,\star \dex\star \dfQ,\qquad \partial\varLambda = -\,\star \dex\star \dfvarLa.\label{eq:dQL}
\end{equation}
However, from the analysis of the Young diagrams (see Table \ref{tab:MAGmodes}), we expect 5 scalar modes coming from the connection. Indeed, in addition to the four variables above, from \eqref{eq:del} we construct the fifth one
\begin{equation}
\partial\Delta = -\,\star\dex\star \dfDe \qquad (= \bgnabla_a\bgnabla_b\bgnabla_c{}^{(1)}\!Q^{abc}).\label{eq:ddel}
\end{equation}

Recall that we work under the conditions \eqref{eq:oddparcond}. However, we will not take into account the two quadratic metric-affine topological invariants (see Chapter \ref{ch:Lovelock}) to drop two extra parameters from the odd sector. The idea is to work in general and use this freedom after performing the full analysis to simplify the equations appropriately. 

\subsection{Previous results}

Here we are going to derive expressions for the curvature scalar $X$ ($=-R$) and the curvature pseudoscalar $\oddX$ in terms of the variables $\{ \partial T, \partial \oddT, \partial Q, \partial \varLambda\}$, as well as a constraint that will allow to eliminate one of these four scalars. 

The expression of $\oddX$ can be immediately read from \eqref{eq:XdT},
\begin{equation}
  \oddX = -\partial\oddT\,. \label{eq:XdT2}
\end{equation}

The other two equations are obtained from $\bgcofr^a\wedge\mathbf{E}_a = 0$ and $\mathbf{C}^a{}_a = 0$, which, when computed in tensor notation, are nothing but the trace of the equation of the coframe and the trace of the equation of the connection in the last two indices. These equations are explicitly given by 
\begin{align}
a_0\,\star\bgcofr_{ab}\wedge\dfR^{ab} + \odda_0 \,\dfR_{ab}\wedge\bgcofr^{ab} + \dex(\bgcofr^a\wedge\dfh_a + \bgcofr^a\wedge \odddfh_a) &= 0,\label{eq:tr1}\\
\bgcofr^a\wedge\dfh_a + \bgcofr^a\wedge\odddfh_a + 2(\dfm^a{}_a + \odddfm^a{}_a) + \,\ell_\rho^2\, \dex(\dfh^a{}_a + \odddfh^a{}_a) &= 0,\label{eq:tr2}
\end{align}
respectively. In order to simplify them we use the following properties, that can be derived from \eqref{eq: GMom g}-\eqref{eq: GMomwod},
\begin{align}
\bgcofr^a\wedge\dfh_a + \bgcofr^a\wedge\odddfh_a &= \star(-a_2 \dfT + \odda_2\odddfT + c_2\dfvarLa - 3c_3 \dfQ),\label{eq:vh}\\
\dfm^a{}_a + \odddfm^a{}_a  &= \star( c_3 \dfT - \oddc_3 \odddfT - b_5\dfvarLa + 4b_4 \dfQ),\label{eq:maa}\\
\dfh^a{}_a &= {\frac 12}\star[(2z_5 + v_5)\dfZ - (v_1 + v_5)(\vartheta X) + (v_3 + v_5)(\vartheta Y)],\label{eq:heaa}\\
\odddfh^a{}_a &= {\frac 12}[(2\oddz_5 + \oddv_5)\dfZ - (\oddv_1 + \oddv_5)(\vartheta X) + (\oddv_3 - \oddv_5) (\vartheta Y)].\label{eq:hoaa}
\end{align}

If we substitute these equations in \eqref{eq:tr1}, and making use of \eqref{eq:XdT}, we can express the curvature scalar as a linear combination of the divergences of our vector variables:
\begin{equation}
\boxed{a_0 X = -\,a_2\partial T + (\odda_0 + \odda_2)\partial\oddT + c_2\partial\varLambda - 3c_3\partial Q}. \label{eq:scalar}
\end{equation}

Now we take the exterior derivative of \eqref{eq:tr2}, and find
\begin{equation}
\boxed{-(a_2 - 2c_3)\partial T + (\odda_2 - 2\oddc_3)\partial\oddT + (c_2 - 2b_5)\partial\varLambda - (3c_3 - 8b_4)\partial Q = 0}.\label{eq:comb}
\end{equation}
This property means that not all four variables are independent in a generic case. So we only need four equations of motion for our five scalar (spin-0) variables.

\subsection{Deriving the first two equations for the spin-0 sector}

Now we are going to consider the other two independent traces of the equation of the connection: $\bgcofr_a\wedge\mathbf{C}^a{}_b$ (trace in the first two indices) and $\bgcofr^b\wedge\mathbf{C}^a{}_b$ (trace in the first and the third indices). From \eqref{eq:qMAGEqw0lin} we can derive explicitly these expressions, which read respectively 
\begin{align}
{\frac 12}\,\bgcofr_b\wedge\left[-\,a_0\,\star \!(2 \dfT + \dfvarLa + 3\dfQ) - \odda_0 \,\star \odddfT\right] \qquad &\nonumber\\
 - 2\,\bgcofr_a\wedge(\dfm^a{}_b + \odddfm^a {}_b)+ \,\ell_\rho^2\,\bgDex\,(\bgcofr_a\wedge\dfh^a{}_b + \bgcofr_a\wedge \odddfh^a{}_b) &= 0,\label{eq:2ndLa}\\
{\frac 12}\,\bgcofr^a\wedge\left[\,a_0\,\star \!(2\dfT - 3\dfvarLa + 3\dfQ) + \odda_0\,\star \odddfT\right] + \bgcofr^a\wedge (\bgcofr^b\wedge\dfh_b + \bgcofr^b\wedge\odddfh_b )\qquad& \nonumber\\
- \,2\,\bgcofr^b\wedge(\dfm^a{}_b + \odddfm^a{}_b)+ \,\ell_\rho^2\,\bgDex\,(\bgcofr^b\wedge\dfh^a{}_b +\bgcofr^b\wedge \odddfh^a{}_b) &= 0.\label{eq:2ndLb}
\end{align}

Again, from \eqref{eq: GMom g}-\eqref{eq: GMomwod}, we derive the following properties:
\begin{align}
\bgcofr_a\wedge \dfm^a{}_b 
    &= {\frac 14}\bgcofr_b\wedge \left[(4b_3 - b_5)\star\! \dfvarLa + (4b_4 - 9b_5)\star \!\dfQ + (- 3c_2 + c_3)\star \!\dfT\right],\label{eq:vtm1}\\[4mm]
\bgcofr_a\wedge \odddfm^a{}_b 
    &= {\frac 14}\bgcofr_b \wedge (3\oddc_2 - \oddc_3)\star \odddfT,\label{eq:vtm2}\\[4mm]
\bgcofr_a\wedge\dfh^a{}_b 
    &=\bgcofr_b\wedge \star \Bigl[ -{\frac {w_5 + 2v_1}{2}}(\vartheta X) + {\frac {2z_3 + v_1 + 3v_3}{4}}(\vartheta Y) + {\frac {2z_5 + v_1 + 3v_5}{8}}\dfZ \nonumber\\
    &\qquad+ {\frac {v_5 - 3v_3}{8}}\dfP  \Bigr]+(w_4 + v_4)\star\! \dfPsi_b - {\frac {2z_4 + 5v_4}2} \star\! \dfPhi_b + {\frac {w_6}{4}}X\star\!\bgcofr_b ,\label{eq:vth1}\\[4mm]
\bgcofr_a\wedge \odddfh^a{}_b 
    &= \bgcofr_b\wedge \Bigl[ -{\frac {\oddw_5}{2}}(\vartheta X)  - {\frac {\oddz_3}{2}}(\vartheta Y) + {\frac {\oddz_5}{4}}\dfZ + {\frac {-2\oddv_1 + 3\oddv_3 - \oddv_5}{8}}\star\! \odddfP\Bigr] \nonumber\\
    &\qquad+(\oddw_2 + \oddv_4) \star\! \odddfPsi{}_b + (\oddz_2 - \oddv_2 + \oddv_4)\star\! \odddfPhi{}_b + {\frac {\oddw_3}{4}}\oddX\star\!\bgcofr_b ,\label{eq:vtho1}\\[4mm]
\bgcofr^b\wedge\dfh^a{}_b 
    &=\bgcofr^a\wedge \star\Bigl[{\frac {w_5}{2}}(\vartheta X) + {\frac {2z_3 + v_3  - v_1}{4}}(\vartheta Y) + {\frac {2z_5 - v_1 + v_5}{8}}\dfZ + {\frac {v_5 - 3v_3}{8}}\dfP  \Bigr] \nonumber\\
    &\qquad+(- w_4 + v_4)\star\! \dfPsi^a - {\frac {2z_4 + 3v_4}2}\star\! \dfPhi^a - {\frac {w_6}{4}}X\star\!\bgcofr^a ,\label{eq:vth2}\\[4mm]
\bgcofr^b\wedge \odddfh^a{}_b 
    &= \bgcofr^a\wedge \Bigl[ {\frac {\oddw_5 + \oddv_1}{2}}(\vartheta X) - {\frac {\oddz_3 + \oddv_3}{2}}(\vartheta Y) + {\frac{\oddz_5 - \oddv_5}{4}}\dfZ + {\frac {2\oddv_1 + 3\oddv_3- \oddv_5}{8}}\star\! \odddfP\Bigr] \nonumber\\
    &\qquad+(-\oddw_2 + \oddv_4) \star\! \odddfPsi^a + (\oddz_2 + \oddv_4)\star\! \odddfPhi^a - {\frac {\oddw_3}{4}}\oddX\star\!\bgcofr^a .\label{eq:vtho2}
\end{align}

These expressions together with \eqref{eq:PPXYZ} allow to re-express our two equations \eqref{eq:2ndLa} and \eqref{eq:2ndLb} as
\begin{equation}\label{eq:vtF}
\boxed{{\frac 12}\,\bgcofr_b\wedge (\dform{\mathcal F}_1 + \ell_\rho^2\,\dex\dform{\mathcal B}_1) = 0},\qquad
\boxed{{\frac 12}\,\bgcofr^a\wedge (\dform{\mathcal F}_2 + \ell_\rho^2\,\dex\dform{\mathcal B}_2) = 0}\,,
\end{equation}
where
\begin{align}
\dform{\mathcal F}_1 
    &\coloneqq-\,{\frac {\ell_\rho^2(2z_4 + 5v_4)}{2}}\,\star \dfPhi + {\frac {\ell_\rho^2(w_4 + w_6 + v_4)}{2}}\,\star \dex X + {\frac {\ell_\rho^2(\oddw_3 - \oddw_2 - \oddv_4)}{2}}\star \dex\oddX\nonumber\\
    &\qquad + \,(3c_2 - c_3 - 2a_0)\,\star \dfT - (3\oddc_2 - \oddc_3 + \odda_0)\,\star \odddfT \nonumber\\
    &\qquad -(a_0 + 4b_3 - b_5)\,\star \dfvarLa - (3a_0 + 4b_4 - 9b_5)\,\star \dfQ,\label{eq:F1}\\[4mm]
\dform{\mathcal F}_2 
    &\coloneqq-\,{\frac {\ell_\rho^2(2z_4 + 3v_4)}{2}}\,\star \dfPhi -{\frac {\ell_\rho^2 (w_4 + w_6 - v_4)}{2}}\,\star \dex X + {\frac {\ell_\rho^2(\oddw_2 - \oddw_3 - \oddv_4)}{2}}\,\star \dex\oddX\nonumber\\
    &\qquad + \,(2a_0 - 2a_2 + 3c_2 - c_3)\,\star \dfT + (\odda_0 + 2\odda_2 - 3\oddc_2+ \oddc_3)\,\star \odddfT\nonumber\\ 
    &\qquad + \,(- 3a_0 + 2c_2 - 4b_3 + b_5)\,\star \dfvarLa+ (3a_0 -  6c_3 - 4b_4 + 9b_5)\,\star \dfQ,\label{eq:F2}
\end{align}
and
\begin{align}
\dform{\mathcal B}_1 
    &\coloneqq (w_4 + w_5 + v_4 + 2v_1)\,\star (\vartheta X) - {\frac {2z_3 - 2z_4 + v_1+ 3v_3 - 5v_4}{2}}\,\star (\vartheta Y)  \nonumber\\
    &\qquad + \,(\oddw_5 - \oddw_2 - \oddv_4)\,(\vartheta X) + (\oddz_3- \oddz_2 + \oddv_2 - \oddv_4)\,(\vartheta Y)  \nonumber\\
    &\qquad- {\frac {2z_5 + v_1 + 3v_5}{4}}\,\star \dfZ+ {\frac {3v_3 - v_5}{4}}\,\star \dfP- {\frac {\oddz_5}2}\,\dfZ+ {\frac {2\oddv_1 - 3\oddv_3 + \oddv_5}{4}}\,\star \odddfP\,,
\label{eq:B1}\\[4mm]
\dform{\mathcal B}_2 
    &\coloneqq -\,(w_4 + w_5 - v_4)\,\star (\vartheta X)  + {\frac {2z_4 - 2z_3 + v_1 - v_3 + 3v_4}{2}}\,\star (\vartheta Y)  \nonumber\\
    &\qquad + \,(\oddw_2 - \oddw_5 - \oddv_1 - \oddv_4)\,(\vartheta X) + (\oddz_3 - \oddz_2 + \oddv_3 - \oddv_4)\,(\vartheta Y)\nonumber\\
    &\qquad - {\frac {2z_5 - v_1 + v_5}{4}}\,\star \dfZ + {\frac {3v_3 - v_5}{4}}\,\star \dfP - {\frac {\oddz_5 - \oddv_5}2}\,\dfZ - {\frac {2\oddv_1 + 3\oddv_3 - \oddv_5}{4}}\,\star \odddfP.\label{eq:B2}
\end{align}

After applying the interior product $\dint{\vfre^b}$ and $\dint{\vfre_a}$ to \eqref{eq:vtF}, we obtain
\begin{equation}\label{eq:vtF1}
\dform{\mathcal F}_1 + \ell_\rho^2\,\dex\dform{\mathcal B}_1 = 0,\qquad \dform{\mathcal F}_2 + \ell_\rho^2\,\dex\dform{\mathcal B}_2 = 0.
\end{equation}
And, finally, if we take the exterior derivative, we end up with $\dex\dform{\mathcal F}_1 = 0$ and $\dex\dform{\mathcal F}_2 = 0$, which are the first two dynamical equations for the scalar modes. If we make use of \eqref{eq:dphi}, insert the curvature scalar $X$ from \eqref{eq:scalar} and the pseudoscalar $\oddX$ from \eqref{eq:XdT}, as well as the d'Alembertian operator $\bgnabla{}^2 \coloneqq - \star\dex\star\dex$, they can be recast as:
\vspace{-1mm}
\boxtheorem{
\begin{align}
-\,{\frac {\ell_\rho^2(w_4 + w_6 +v_4)}{2a_0}}\left[a_2\bgnabla{}^2\partial T- (\odda_0 +\odda_2)\bgnabla{}^2\partial\oddT - c_2\bgnabla{}^2\partial\varLambda+ 3c_3\bgnabla{}^2\partial Q\right] & \nonumber\\
    - \,{\frac {\ell_\rho^2(\oddw_3 - \oddw_2 - \oddv_4)}{2}}\,\bgnabla{}^2\partial\oddT - {\frac {\ell_\rho^2(2z_4 + 5v_4)}3}\,\bgnabla{}^2\,\partial\varLambda - {\frac {\ell_\rho^2(2z_4 + 5v_4)}{2}}\,\partial\Delta & \nonumber\\
    + \,(3c_2 - c_3 - 2a_0)\,\partial T - (3\oddc_2 - \oddc_3 + \odda_0)\,\partial\oddT \nonumber\\
    - (a_0 + 4b_3 -b_5)\,\partial\varLambda - (3a_0 + 4b_4 - 9b_5)\,\partial Q &= 0,\label{eq:sc1}\\
{\frac {\ell_\rho^2(w_4 + w_6 - v_4)}{2a_0}}\left[a_2\bgnabla{}^2\partial T - (\odda_0+ \odda_2)\bgnabla{}^2\partial\oddT - c_2\bgnabla{}^2\partial\varLambda+ 3c_3\bgnabla{}^2\partial Q\right] & \nonumber\\
    -\,{\frac {\ell_\rho^2(\oddw_2 - \oddw_3 - \oddv_4)}{2}}\,\bgnabla{}^2\partial\oddT - {\frac {\ell_\rho^2(2z_4 + 3v_4)}3}\,\bgnabla{}^2\,\partial\varLambda - {\frac {\ell_\rho^2(2z_4 + 3v_4)}{2}}\,\partial\Delta & \nonumber\\
    + \,(2a_0 - 2a_2 + 3c_2 - c_3)\,\partial T + (\odda_0 + 2\odda_2 - 3\oddc_2+ \oddc_3)\,\partial\oddT & \nonumber\\
    + \,(-3a_0 + 2c_2 - 4b_3 + b_5)\,\partial\varLambda + (3a_0 -  6c_3 - 4b_4 + 9b_5)\,\partial Q &= 0.\label{eq:sc2}
\end{align}
}

\subsection{Deriving the third equation for the spin-0 sector}

The following independent dynamical equation for the scalar modes can be derived, in tensor notation, from the totally antisymmetric part of the equation of the connection. In terms of differential forms, such equation can be encoded in $\star \mathbf{C}^a{}_b\wedge\bgcofr_a{}^b = 0$. 

First we find that
\begin{align}
\frac{a_0}{2} \big(\dfT^c\wedge\star\bgcofr{}^a{}_{bc}+\dfQ^{ac}\wedge\star\bgcofr{}_{cb}-2\dfQ\wedge\star\bgcofr{}^a{}_b \big)\wedge\bgcofr_a{}^b
    &=a_0\,\star \odddfT,\label{eq:sT}\\
\frac{\odda_0}{2} \big(2g^{ac}\dfT_{[c}\wedge\bgcofr_{b]}- \dfQ_{cb}\wedge\bgcofr^{ac}\big)\wedge\bgcofr_a{}^b
    &=-\odda_0\,\star (2\dfT - \dfvarLa + 3\dfQ),\label{eq:soT}\\ 
\star \!\left[\bgcofr^a\wedge (\dfh_b + \odddfh_b)\right]\wedge\bgcofr_a{}^b 
    &= 2\,\star (\dint{\vfre_a}\dfh^a + \dint{\vfre_a}\odddfh^a) \nonumber\\
    &= 2\,\star (a_3\,\odddfT + \odda_2\,\dfT - \oddc_2\,\dfvarLa + 3\oddc_3\,\dfQ).\label{eq:svh}  
\end{align}
Obviously $\star (m^a{}_b + \odddfm^a{}_b)\wedge\bgcofr_a\wedge\bgcofr^b = 0$ in view of the symmetry. 

The last thing we need is to compute the contribution of the derivative terms in \eqref{eq:qMAGEqw0lin}. They can be expressed
\begin{align}\label{eq:sDh}
\star (\bgDex \dfh^a{}_b)\wedge\bgcofr_a{}^b &= \star (\bgDex \dfh^{[ab]})\wedge\bgcofr_{ab} 
    =2\dint{\vfre_a}\bgDex\left[\star \dfA^a + \cofr^a\wedge \star (\vartheta A) - {\frac 12}A \star\cofr^a\right],\\
\star (\bgDex\odddfh^a{}_b)\wedge\bgcofr_a{}^b & = \star (\bgDex\odddfh^{[ab]})\wedge\bgcofr{}_{ab} 
    = 2\dint{\vfre_a}\bgDex\left[\star \odddfA{}^a + \cofr^a\wedge \star (\vartheta \oddA) - {\frac 12}\,\oddA\star\cofr^a\right],\label{eq:sDoh}
\end{align}
where we have introduced 
\begin{equation}
\dfA^a \coloneqq \dint{\vfre_b} \dfh^{[ab]},\qquad 
\odddfA^a \coloneqq \dint{\vfre_b}\odddfh^{[ab]},
\end{equation}
which explicitly read
\begin{align}
\dfA^a &= -\,w_2\,\odddfPsi^a + {\frac {v_2}{2}}\,\odddfPhi^a - {\frac {w_3}{4}}\,\oddX\,\bgcofr^a  \nonumber \\
    &\qquad+\,\dint{\vfre^a}\,\star \!\left[{\frac {w_5}{2}}\,(\vartheta X) - {\frac {v_1}{4}}\,(\dfP - (\vartheta X)) - {\frac {v_3}{4}}\,(\vartheta Y) - {\frac {v_5}{8}}\,\dfZ\right],\label{eq:Aea}\\
\odddfA{}^a &= \oddw_2\,\dfPsi^a - {\frac {\oddv_4}{2}}\,\dfPhi^a +{\frac {\oddw_3}{4}}\,X\,\bgcofr^a  \nonumber \\
    &\qquad+\,\dint{\vfre^a}\left[{\frac {\oddw_5}{2}}\,(\vartheta X) + {\frac {\oddv_1}{4}} \,(\star \odddfP + (\vartheta X)) - {\frac {\oddv_3}{4}}\,(\vartheta Y) - {\frac {\oddv_5}{8}}\,\dfZ\right]\label{eq:Aoa}\,.
\end{align}
and their traces and antisymmetric parts:
\begin{equation}
  A \coloneqq \dint{\vfre_a} \dfA^a,\qquad \oddA \coloneqq \dint{\vfre_a} \odddfA{}^a,\qquad
  (\vartheta A) \coloneqq \bgcofr^a\wedge\dfA_a ,\qquad (\vartheta \oddA) \coloneqq \bgcofr^a\wedge\odddfA_a\,.
\end{equation} 

Now that we have all of the ingredients we can write our equation as
\begin{equation}
\star \mathbf{C}^a{}_b\wedge\bgcofr_a\wedge\bgcofr^b = 
\dform{\mathcal F}_3 + \ell_\rho^2\,\dex \dform{\mathcal B}_3 = 0,\label{eq:vtF3}
\end{equation}
where 
\begin{align}
\dform{\mathcal F}_3 &= {\frac {\ell_\rho^2\oddv_4}{2}}\,\star \dfPhi + {\frac {\ell_\rho^2 (\oddw_3 - \oddw_2)}{2}}\,\star \dex X - {\frac {\ell_\rho^2(w_2 + w_3)}{2}} \,\star \dex\oddX\nonumber\\
    &\qquad -2(\odda_0 + \odda_2)\,\star \dfT + (a_0 - 2 a_3)\,\star \odddfT + (\odda_0+2\oddc_2)\star \dfvarLa - 3(\odda_0+2\oddc_3)\,\star \dfQ,\label{eq:F3}\\
\dform{\mathcal B}_3 &=  -\,{\frac {2\oddw_2 - 2\oddw_5 - \oddv_1}{2}}\,\star (\vartheta X) - {\frac {\oddv_3 + \oddv_4}{2}}\,\star (\vartheta Y) - {\frac {\oddv_5}4}\,\star \dfZ - {\frac {\oddv_1}{2}}\,\odddfP \nonumber\\
    &\qquad -\,{\frac {2w_2 + 2w_5 + v_1}{2}}\,(\vartheta X)  + {\frac {v_2 + v_3}{2}}\,(\vartheta Y) + {\frac {v_5}{4}}\,\dfZ + {\frac {v_1}{2}}\,\dfP.\label{eq:B3}
\end{align}
As we did in the previous section, we take the exterior derivative and obtain $\dex\dform{\mathcal F}_3 = 0$, which is the third equation for the spin-0 sector. Explicitly, this equation reads:

\boxtheorem{
\begin{align}
{\frac {\ell_\rho^2(\oddw_3 - \oddw_2)}{2a_0}}\left[a_2\bgnabla{}^2\partial T
- (\odda_0 +\odda_2)\bgnabla{}^2\partial\oddT - c_2\bgnabla{}^2\partial\varLambda
+ 3c_3\bgnabla{}^2\partial Q\right] && \nonumber\\
- \,{\frac {\ell_\rho^2(w_2 + w_3)}{2}}\,\bgnabla{}^2\partial\oddT
- {\frac {\ell_\rho^2\oddv_4}3}\,\bgnabla{}^2\,\partial\varLambda 
- {\frac {\ell_\rho^2\oddv_4}{2}}\,\partial\Delta && \nonumber\\
+ 2(\odda_0 + \odda_2)\,\partial T - (a_0 - 2a_3)\,\partial\oddT
- (\odda_0+2\oddc_2)\,\partial\varLambda + 3(\odda_0+2\oddc_3)
\,\partial Q &= 0.\label{eq:sc3}
\end{align}
}

\subsection{Results up to the present moment}

By using the constraint \eqref{eq:scalar}, we can eliminate $\partial Q$ from our set of equations. If we introduce the following 3-component object
\begin{equation}
{\bf U} := \begin{pmatrix}\partial T \\ \partial\oddT \\ \partial\varLambda\end{pmatrix},
\end{equation}
we can then recast our three dynamical equations \eqref{eq:sc1}, \eqref{eq:sc2} and \eqref{eq:sc3} in matrix notation as\footnote{
    Here we have focused on scalar modes coming from the torsion and the nonmetricity, but it is important to remark that there could also be spin-0 modes coming from the metric/coframe sector. They are currently under study.}
\begin{equation}
{\bf K}\,\bgnabla{}^2\,{\bf U} + {\bf M}\,{\bf U} + {\bf N}\,\partial\Delta = 0.\label{eq:scM}
\end{equation}
The explicit form of the $3\times 3$ matrices ${\bf K}$ and ${\bf M}$, and the 3-column ${\bf N}$ can be read from \eqref{eq:sc1}, \eqref{eq:sc2} and \eqref{eq:sc3}.

Observe that there is no dynamical term for $\partial\Delta$ in \eqref{eq:scM}. So far it is not clear how to derive this extra equation (a difficulty that was not present in the analysis of PG \cite{BlagojevicCvetkovic2018}). Interestingly, the scalar variable $\partial\Delta$ is defined in terms of the irreducible component $\irrdfQ{1}_{ab}$, whose leading spin-order is 3 (see Table \ref{tab:MAGmodes}). Motivated by this, we conjecture that the derivation of the (still) missing equation for this scalar mode is very related to the analysis of the spin-3 sector of the MAG equation, which remains to be done.

\part{Final comments \label{part:conclusions}}

\chapter{Final comments}\label{ch:conclusions}

\boxquote{Bang...}{Spike Spiegel (last episode of Cowboy Bebop)}

To finish this thesis, we proceed to discuss and collect some general ideas about the results we have obtained.

From the first part of the thesis, we can highlight the enormous interest both mathematical and physical of MAG. In contrast to the ordinary gauge theories of internal symmetries, MAG requires an additional process (soldering) to appropriately generate the coframe, which appears as a non-linear translational connection after the reduction ${\rm Aff}(\dimM, \mathbb{R})\to{\rm GL}(\dimM, \mathbb{R})$. Interestingly, the MAG metric can also be motivated in a similar way by an additional reduction into the Lorentz subgroup of ${\rm GL}(\dimM, \mathbb{R})$. There are however different ways to formulate and motivate this structure, besides the one we presented here. This is a consequence of several subtleties and features that are very characteristic of gauge theories of gravity and that are far from being completely solved. Formulations with the same observational consequences provide different understandings of how gravity and, consequently, the spacetime emerges. Therefore the investigation along this line is well justified, at least from a fundamental perspective.

It is interesting to notice the number of special difficulties of MAG  in contrast to PG. The addition of nonmetricity and, in particular, its traceless part gives rise to many of these theoretical complications. Known examples are the interpretation of the shear current, or the fact that the equivalent to spinor representations in MAG correspond to spaces of infinite dimension. In this thesis we have checked another peculiarity  in this regard which is the violation of the topological character of critical Lovelock terms in this frame, precisely due to terms depending on the traceless part of the nonmetricity. To prove this, we used the fact that the equations of motion of a boundary term are trivial, and explicitly constructed a particular geometry for which one of them is violated. It is worth remarking that for the metric-affine Gauss-Bonnet theory, the extra term needed to get a boundary term is quartic in the nonmetricity, so it cannot be used to simplify the 4-dimensional quadratic MAG Lagrangian. 
Despite this, we have found that the Nieh-Yan invariant admits a direct generalization by adding a quadratic term from the mixed sector with both $Q$ and $T$. Moreover, the Pontryagin invariant is automatically a boundary term in metric-affine formulation. These results tell that two invariants from the odd sector can be dropped from the general quadratic Lagrangian.

Concerning particular MAG geometries, we have revised the GW criteria in GR and proposed some possible metric-affine generalizations that keep a similar algebraic structure in terms of field strengths, although the physical interpretation of them are theory-dependent. We have focused on the Lichnerowicz criteria which are a direct application of the radiation conditions in Maxwell theory to the Levi-Civita curvature tensor. In our extension, we applied the same conditions but over the full curvature and the torsion of the metric-affine geometry. Then, we selected a particular family of geometries and revised the conditions that those criteria impose on the different variables.

As we have already mentioned, the exploration of exact solutions provides very valuable information about the non-linear regime of a theory. In an independent work, we used a particular Ansatz to search for vacuum solutions of the general quadratic MAG (even-parity) theory. Riemannian solutions as well as teleparallel ones and of the pseudo-instanton type have been found. Moreover, the conditions for having nontrivial general solutions (without restricting to any type of geometry) have also been explicitly computed. It is worth noticing that the method used in the general case, based on a potential-copotential decomposition of the 2-vector variables, allowed to write the complicated MAG equations as a set of Helmholtz equations for general values of the even parameters. In the future, these results will be extended to include the odd parity sector, or maybe by adding matter sources.

Physically speaking, a viable theory is in particular characterized by the well-behavior of their degrees of freedom. Furthermore, one should also be aware of possible inconsistencies derived from the very construction of the theory. In this regard, we have analyzed the particular case of 4DEGB, and revised some of its inconsistencies, which are essentially due to a ``0/0'' term that makes the equations of motion ill-defined and that cannot be regularized, because each of the zeros has a completely different origin.

Regarding the stability of the degrees of freedom, the different analysis we have explored in that direction constitute a fascinating field of study in between the purely theoretical and observational sides. As we have seen for the particular case of ECG, even if the linear spectrum of the theory is healthy, strongly coupled modes can render catastrophic unstable behaviors around specific backgrounds. Indeed, in the context of ECG, the isotropic solution (FRLW with flat spatial slices) lies in the intersection of  singular surfaces in phase space. As we saw in general, and also in some numerical simulations, any little initial (anisotropic) deviation from such background push the solution far from the initial configuration, showing that these backgrounds cannot be seen as viable physical models in these theories. We have also added the first three GQTG corrections and checked that the problem is even more severe, since then the ghosts are fully active around these cosmological backgrounds.

In addition to the previous analysis, we have also learned how important is to kill those problematic modes that are intrinsic to the theory. We have seen several examples of this in the context of general quadratic teleparallel gravity. In particular, we ended up with two possibilities that ensure (at linear level) a well-behaved propagation for its maximal field content. In one case the symmetric sector (made of two spin-2 fields) has Diff$\times$Diff symmetry, whereas in the other one they present local invariance under Diff$\times$WTDiff. In both cases, there is an additional Kalb-Ramond field equipped with the standard gauge transformation. To reach these results, we introduced additional symmetries that allow to ``bypass'' the Ostrogradski theorem by violating the non-degeneracy hypothesis. Such technique has been shown to be very useful to remove dangerous ghosts in those theories.

These analyses, together  with what we know about PG, clearly indicate that these problems will be present in MAG. In fact, one would expect that a safe theory free of ghosts within MAG will likely end up being GR with extra scalars or vectors belonging to one of the known types of well-behaved theories. From the analysis of the spin-0 sector, we noticed that it is not obvious to get a dynamical equation for one of the five spin-0 modes of the theory associated with the connection, in particular, the one that is related to the spin-3 sector. The rest of the spectrum is currently under study: the possible additional spin-0 modes coming from the metric/coframe, the spin-1 and spin-2 modes, the latter being usually problematic, and the spin-3 mode, which is prone to generate even more problems. Future results on the analysis of the linear spectrum around flat spacetime will help to restrict the theory to safer subsets of parameters. Other complementary (and necessary) works such as the Hamiltonian analysis, and the explicit stability analysis of certain backgrounds are also needed to further establish the limitations of MAG. All of these theoretical developments are crucial, and must be performed before doing any strong claim about the observational consequences of MAG.


\part{Appendix \label{part:appendix}}
\appendix

\chapter{\label{app:pullback}Pullback and pushforward}

\boxdefinition{

Let $\phi\,:\,\mathcal{M}\to\mathcal{N}$ be a smooth map between manifolds.
\begin{defn}
\textbf{(Pullback of a function)} The \emph{pullback of a smooth function} $f\,:\,\mathcal{N}\to\mathbb{R}$ is the function $\phi^{*}f\,:\,\mathcal{N}\to\mathbb{R}$ given by
\begin{equation}
\phi{}^{*}f\coloneqq f\circ\phi\,.
\end{equation}
\end{defn}
\begin{defn}
\textbf{(Pushforward of a vector)} The \emph{pushforward of a vector} $\vecv\in T_{p}\mathcal{M}$ is the vector $\left(\phi{}_{*}\vecv\right)\in T_{\phi(p)}\mathcal{N}$ that acts on functions as
\begin{equation}
(\phi{}_{*}\vecv)\left(f\right)\coloneqq\vecv\left(\phi{}^{*}f\right)=v^{\mu}\partial_{\mu}|_{p}\left(\phi{}^{*}f\right)\,.
\end{equation}
\end{defn}
\begin{defn}
\textbf{(Pullback of a differential form)} The \emph{pullback of a $k$-form} at $q$, $\dfal\in\Lambda_{q}^{k}\mathcal{N}$, to the point $p\in\mathcal{M}$ such that $\phi(p)=q$, is a $k$-form in $p$, represented as $(\phi{}^{*}\dfal)\in \Lambda_{p}^{k}\mathcal{M}$, which acts as:
\begin{equation}
(\phi{}^{*}\dfal)\left(\vecv_1,\,...,\,\vecv_{k}\right)\coloneqq\dfal\left(\phi{}_{*}\vecv_1,\,...,\,\phi{}_{*}\vecv_{k}\right)\,,
\end{equation}
where $\vecv_1,\,...,\,\vecv_{k}\in T_{p}\mathcal{M}$.
\end{defn}
}

Notice that the first one is always well defined (it is just the composition). Nevertheless, generalising the last two to vector fields or $k$-form fields is not trivial, because if $\phi$ is not surjective, the result will be, at most a field over the image or pre-image of $\phi$ (and there is no natural way to extend it to the entire manifold). Moreover, the pushforward has an additional problem:
\begin{itemize}
\item If $\phi$ is not injective, we do not even get a vector field over ${\rm Im}(\phi)$. The reason is the following: imagine two points $p_1,p_2\in\mathcal{M}$ with the same image $\phi(p_1)=\phi(p_2)\equiv q$, there is an ambiguity on how to chose the vector at $q$ (is it the one coming from $\vecX|_{p_1}$ or the one coming from $\vecX|_{p_2}$ by the pushforward?).
\end{itemize}
Observe how, for the particular case in which $\phi$ is a diffeomorphism, these definitions can be extended to fields. Indeed, since the inverse of the diffeomorphism is also a diffeomorphism, $\phi$ establishes an identification between a field and its pullback / pushforward, making it possible to the define the operations that go in the opposite direction. In other words, if $\phi$ is a diffeomorphism, we can actually speak about the ``pushforward of a $k$-form'' and the ``pullback of a vector field''.

\chapter{\label{app:irreds}Irreducible decompositions}

\section{Irreducible decomposition of the torsion}

\subsection*{Irreducible parts}

Under the pseudo-orthogonal group, the torsion 2-form can be decomposed into three parts:
\begin{equation}
  \dfT^a=\irrdfT{1}{}^a+\irrdfT{2}{}^a+\irrdfT{3}{}^a\,,
\end{equation}
where
\begin{align}
  \irrdfT{2}{}^a & \coloneqq\frac{1}{\dimM-1}\cofr^a\wedge(\dint{\vfre_b}\dfT^b)\,,\\
  \irrdfT{3}{}^a & \coloneqq\frac{1}{3}\dint{\vfre^a}(\dfT^b\wedge\cofr_b)\,,\\
  \irrdfT{1}{}^a & \coloneqq\dfT^a-\irrdfT{2}{}^a -\irrdfT{3}{}^a.
\end{align}
which correspond, respectively, to the trace, the totally antisymmetric part and the remaining tensorial part. The first two can be rewritten as
\begin{equation}
  \irrdfT{2}{}^a=\frac{1}{\dimM-1}\cofr^a\wedge\dfT\,,\qquad\irrdfT{3}{}^a\coloneqq\frac{1}{3}\sgng(-1)^{\dimM-3}\dint{\vfre^a}\star\odddfT\,,
\end{equation}
where we have introduced the torsion trace form (1-form) and the axial torsion ($(\dimM-3)$-form):
\begin{align}
  \dfT & \coloneqq\dint{\vfre_a}\dfT^a &  & =-T_{\mu\rho}{}^{\rho}\dex x^{\mu}\,,\\
  \odddfT & \coloneqq\star(\dfT^b \wedge\cofr_b) &  & =\frac{1}{2(\dimM-3)!}(T_{[\mu\nu\rho]}\LCten^{\mu\nu\rho}{}_{\lambda_{1}...\lambda_{\dimM-3}})\dex x^{\lambda_{1}}\wedge...\wedge\dex x^{\lambda_{\dimM-3}}\,.
\end{align}

\subsection*{Components of the irreducible parts of the torsion}
If we expand the torsion parts $\irrdfT{I}{}^a=\frac{1}{2}\irrT{I}{}_{bc}{}^a\cofr^{bc}$ we have
\begin{align}
  \irrT{2}{}_{bc}{}^a & =\frac{2}{\dimM-1}T_{[b}\delta_{c]}^a\,,\\
  \irrT{3}{}_{bc}{}^a & =T_{[bcd]}g^{ad}\,,\\
  \irrT{1}{}_{bc}{}^a & =T_{bc}{}^a-\irrT{2}{}_{bc}{}^a-\irrT{3}{}_{bc}{}^a\,.
\end{align}

\subsection*{Properties of the irreducible parts of the torsion}
\begin{itemize}
\item Totally antisymmetric parts
\begin{align}
  \irrdfT{I}{}_a\wedge\cofr^a & =0                      & \Leftrightarrow\quad\irrT{I}{}_{[abc]} & =0\qquad(\Leftrightarrow\,\,\irrT{I}{}_{abc}=-2\irrT{I}{}_{c[ab]}) & I & =1,\,2\,,\\
  \irrdfT{3}{}_a\wedge\cofr^a & =\dfT{}_a\wedge\cofr^a  & \Leftrightarrow\quad\irrT{3}{}_{abc}   & =T_{[abc]}\,. & &
\end{align}
\item Traces
\begin{align}
  \dint{\vfre_a}\irrdfT{I}{}^a & =0                        & \Leftrightarrow\quad\irrT{I}{}_{ac}{}^{c} & =0           & I & =1,\,3\,,\\
  \dint{\vfre_a}\irrdfT{2}{}^a & =\dint{\vfre_a}\dfT{}^a & \Leftrightarrow\quad\irrT{2}{}_{ac}{}^{c} & =T_{ac}{}^c\,.& &
\end{align}
\end{itemize}

\begin{table} 
\begin{centering}
\renewcommand{\arraystretch}{1.5}
\begin{tabular}{|c|c|c|}
  \hline                     & Arbitrary dimension $\dimM$     & $\dimM=4$\tabularnewline
  \hline 
  \hline Total               & $\frac{1}{2}\dimM^{2}(\dimM-1)$ & 24\tabularnewline
  \hline 
  \hline $\irrdfT{1}{}^a$    & $\frac{1}{3}\dimM(\dimM+2)(\dimM-2)$ & 16\tabularnewline
  \hline $\irrdfT{2}{}^a$    & $\dimM$ & 4\tabularnewline
  \hline $\irrdfT{3}{}^a$    & $\frac{1}{6}\dimM(\dimM-1)(\dimM-2)$ & 4\tabularnewline
  \hline \multicolumn{1}{c}{}& \multicolumn{1}{c}{}            & \multicolumn{1}{c}{}\tabularnewline
\end{tabular}~~%
\begin{tabular}{|c|c|c|}
  \hline                     & Arbitrary dimension $\dimM$     & $\dimM=4$\tabularnewline
  \hline 
  \hline Total               & $\frac{1}{2}\dimM^{2}(\dimM+1)$ & 40\tabularnewline
  \hline 
  \hline $\irrdfQ{1}{}_{ab}$ & $\frac{1}{6}\dimM(\dimM-1)(\dimM+4)$ & 16\tabularnewline
  \hline $\irrdfQ{2}{}_{ab}$ & $\frac{1}{3}\dimM(\dimM^{2}-4)$      & 16\tabularnewline
  \hline $\irrdfQ{3}{}_{ab}$ & $\dimM$  & 4\tabularnewline
  \hline $\irrdfQ{4}{}_{ab}$ & $\dimM$  & 4\tabularnewline
  \hline 
\end{tabular}
\renewcommand{\arraystretch}{1}
\end{centering}
\caption{Distribution of the independent components of the torsion (left) and the nonmetricity (right).}
\end{table}

\section{Irreducible decomposition of the nonmetricity}
\subsection*{Irreducible parts}
We start by introducing the traceless part (in the external indices) of the nonmetricity,
\begin{equation}
  \dfQNoTr{}_{ab}\coloneqq\dfQ_{ab}-\frac{1}{\dimM}g_{ab}\dfQ_{c}{}^c\,,\label{eq:defQtrless}
\end{equation}
and the following independent 1-forms that contain the two traces of the nonmetricity tensor $Q_{\mu\nu\rho}$,
\begin{align}
  \dfQ  & \coloneqq\frac{1}{\dimM}\dfQ_{c}{}^{c} &  & =\frac{1}{\dimM}Q_{\mu}\dex x^{\mu}\,,\qquad\text{(Weyl 1-form)}\\
  \dfvarLa & \coloneqq(\dint{\vfre^c}\dfQNoTr{}_{cb})\cofr^b &  & =\left[\Qb{}_{\mu}-\frac{1}{\dimM}Q_{\mu}\right]\dex x^{\mu}\,.
\end{align}

If we work in terms of the nonmetricity tensor $Q_{\mu\nu\rho}$ there is no canonical way to separate the two traces (each of them is an irreducible component). In differential form notation there is a canonical choice: first we extract the trace in the last two (second term in the r.h.s. of \eqref{eq:defQtrless}) and then we further split the traceless part, which contains three additional irreducible parts. Therefore, the whole nonmetricity 1-form can be split as
\begin{equation}
  \dfQ_{ab}=\underbrace{\irrdfQ{1}{}_{ab}+\irrdfQ{2}{}_{ab}+\irrdfQ{3}{}_{ab}}_{\dfQNoTr{}_{ab}}+\irrdfQ{4}{}_{ab}\,.
\end{equation}
The last one is the trace in the external indices (contains the Weyl 1-form) and $\irrdfQ{3}{}_{ab}$ is the remaining trace. The other two have totally traceless components: the components of $\irrdfQ{1}{}_{ab}$ are totally symmetric, whereas those of $\irrdfQ{2}{}_{ab}$ constitute the remaining tensorial part. The explicit definitions are:
\begin{align}
  \irrdfQ{4}{}_{ab} & \coloneqq g_{ab}\dfQ\\
  \irrdfQ{3}{}_{ab} & \coloneqq\frac{2\dimM}{(\dimM-1)(\dimM+2)}\left(\cofr_{(a}\dint{\vfre_{b)}}\dfvarLa-\frac{1}{\dimM}g_{ab}\dfvarLa\right)\\
  \irrdfQ{2}{}_{ab} & \coloneqq-\frac{2}{3}\sgng\star\left(\odddfvarLa_{(a}\wedge\cofr_{b)}\right)\\
  \irrdfQ{1}{}_{ab} & \coloneqq\dfQ_{ab}-\irrdfQ{2}{}_{ab}-\irrdfQ{3}{}_{ab}-\irrdfQ{4}{}_{ab}
\end{align}
where we have introduced the auxiliary $(\dimM-2)$-form
\begin{equation}
  \odddfvarLa_a \coloneqq\star\left[\dfQNoTr_{ac}\land\cofr^{c}-\frac{1}{\dimM-1}\cofr_a\wedge\dfvarLa\right]\,.
\end{equation}
By expanding this last object one can prove:
\begin{equation}
  \irrdfQ{2}{}_{ab}=\frac{2}{3}\dfQNoTr{}_{ab}-\frac{2}{3}(\dint{\vfre_{(a}}\dfQNoTr_{b)d})\cofr^{d}+\frac{2}{3}\frac{1}{\dimM-1}(g_{ab}\dfvarLa-\cofr_{(a}\dint{\vfre_{b)}}\dfvarLa)\,.
\end{equation}

\subsection*{Components of the irreducible parts of the nonmetricity}
The nonmetricity parts $\irrdfQ{I}{}_{ab}=\irrQ{I}{}_{cab}\cofr^{c}$ can be expressed
\begin{align}
  \irrQ{3}{}_{cab} & =\frac{2\dimM}{(\dimM-1)(\dimM+2)} \left[\left(\Qb{}_{(a}-\frac{1}{\dimM}Q_{(a}\right)g_{b)c} -\frac{1}{\dimM}g_{ab} \left(\Qb{}_{c}-\frac{1}{\dimM} Q_c\right)\right]\,,\\
  \irrQ{4}{}_{cab} & =\frac{1}{\dimM}Q_{c}g_{ab}\,,\\
  \irrQ{1}{}_{cab} & =Q_{(cab)}-\frac{1}{\dimM+2}g_{(ab}\left(Q_{c)}+2\Qb{}_{c)}\right)\,,\\
  \irrQ{2}{}_{cab} & =Q_{cab}-\irrQ{1}{}_{cab}-\irrQ{3}{}_{cab}-\irrQ{4}{}_{cab}\,,
\end{align}

\subsection*{Properties of the irreducible parts of the nonmetricity}
\begin{itemize}
  \item Antisymmetric parts (of the components) in the first two indices
    \begin{align}
      \irrdfQ{1}{}_{ab}\wedge\cofr^b & =0\qquad\Leftrightarrow & \irrQ{1}{}_{[ab]c} & =0\,,
    \end{align}
    Only the following are non-trivial
    \begin{align}
      \irrdfQ{2}{}_{ab}\wedge\cofr^b & =\sgng\star\odddfvarLa_a \quad\left[=\dfQ{}_{ab}\wedge\cofr^{b}+\big(\tfrac{1}{\dimM-1}\dfvarLa-\dfQ\big)\wedge\cofr_a\right]\,,\\
      \irrdfQ{3}{}_{ab}\wedge\cofr^b & =\frac{1}{\dimM-1}\cofr_a\wedge\dfvarLa\,,\\
      \irrdfQ{4}{}_{ab}\wedge\cofr^b & =\dfQ\wedge\cofr_a\,.
    \end{align}
  \item Traces
    \begin{align}
      \irrdfQ{I}{}_c{}^c            & =0\qquad\Leftrightarrow & \irrQ{I}{}_{ac}{}^{c} & =0 & I & =1,\,2,\,3\\
      \dint{\vfre^a}\irrdfQ{I}{}_{ab} & =0\qquad\Leftrightarrow & \irrQ{I}{}^c{}_{ca} & =0 & I & =1,\,2
    \end{align}
    Only the following are non-trivial
    \begin{align}
      \irrdfQ{4}{}_{c}{}^{c}            & =\dfQ_{c}{}^c\qquad(=\dimM\dfQ)\,,\\
      \dint{\vfre^a}\irrdfQ{3}{}_{ab} & =\dint{\vfre_a}\dfvarLa\,,\\
      \dint{\vfre^a}\irrdfQ{4}{}_{ab} & =\dint{\vfre_a}\dfQ\,.
    \end{align}
  \item Totally symmetric parts:
    \begin{align}
      \dint{\vfre_{(a}}\irrdfQ{2}{}_{bc)} & =0\qquad\Leftrightarrow & \irrQ{2}{}_{[abc]} & =0 & I & =1,\,2
    \end{align}
    Only the following are non-trivial
    \begin{align}
      \dint{\vfre_{(a}}\irrdfQ{3}{}_{bc)} & =\frac{2}{\dimM+2}g_{(ab}\dint{\vfre_{c)}}\dfvarLa\,,\dint{\vfre_{(a}}\irrdfQ{4}{}_{bc)}=g_{(ab}\dint{\vfre_{c)}}\dfQ\,,\\
      \dint{\vfre_{(a}}\irrdfQ{1}{}_{bc)} & =\dint{\vfre_{(a}}\dfQ_{bc)}-g_{(ab}\dint{\vfre_{c)}}\left[\dfQ-\frac{2}{\dimM+2}\dfvarLa\right]\,.
    \end{align}
\end{itemize}

\section{Irreducible decomposition of the curvature}
\subsection*{Irreducible parts}
For the curvature 2-form we first split it into symmetric and antisymmetric parts (in the external indices):
\begin{align}
  \dfR_{ab} & =\dfW_{ab}+\dfZ_{ab}\\
            & =\dfW_{ab}+\dfZNoTr_{ab}+\frac{1}{\dimM}g_{ab}\dfZ
\end{align}
where $\dfW_{ab}\coloneqq\dfR_{[ab]}$, $\dfZ_{ab}\coloneqq\dfR_{(ab)}$, $\dfZ\coloneqq\dfR_{c}{}^{c}$ and $\dfZNoTr_{ab}$ is the traceless part of $\dfZ_{ab}$. 

Under the pseudo-orthogonal group, the antisymmetric part can be separated into six irreducible parts $\dfW_{ab}=\sum_{I=1}^{6}\irrdfW{I}{}_{ab}$, given by
\begin{align}
  \irrdfW{2}{}_{ab} & \coloneqq\sgng\star\left(\cofr_{[a}\wedge\odddfPsi{}_{b]}\right)\\
  \irrdfW{3}{}_{ab} & \coloneqq\sgng\frac{1}{12}\star\left(\odddfX\wedge\cofr_{ab}\right)\\
  \irrdfW{4}{}_{ab} & \coloneqq-\frac{2}{\dimM-2}\cofr_{[a}\wedge\dfPsi_{b]}\\
  \irrdfW{5}{}_{ab} & \coloneqq-\frac{1}{\dimM-2}\cofr_{[a}\wedge\dint{\vfre_{b]}}\left(\cofr^c\wedge\dfX_c\right)\\
  \irrdfW{6}{}_{ab} & \coloneqq-\frac{1}{\dimM(\dimM-1)}X\cofr_{ab}\\
  \irrdfW{1}{}_{ab} & \coloneqq\dfW_{ab}-\sum_{\mathrm{I}=2}^{6}\irrdfW{I}{}_{ab}
\end{align}
where we have introduced the following auxiliary objects (the number at the left of each quantity represents its rank as differential forms)
\begin{align}
 & [1]      \hspace{-12mm} & \dfX^a              & \coloneqq\dint{\vfre_b}\dfW^{ab}\,, &  
 & [\dimM-3]\hspace{-12mm} & \odddfX{}^a   & \coloneqq\star(\dfW^{ba}\wedge\cofr_b)\,, \\
 & [0]      \hspace{-12mm} & X                   & \coloneqq\dint{\vfre_a}\dfX^a &  
 & [\dimM-4]\hspace{-12mm} & \odddfX       & \coloneqq\dint{\vfre_a}\odddfX{}^a\,,
\end{align}
\begin{align}
   & [1]      \hspace{-8mm} & \dfPsi{}_a          & \coloneqq\dfX_a -\frac{1}{\dimM}X\cofr_a -\frac{1}{2}\dint{\vfre_a}(\cofr^b \land\dfX_b)\,,\\
   & [\dimM-3]\hspace{-8mm} & \odddfPsi{}_a & \coloneqq\odddfX{}_a-\frac{1}{4}\cofr_a \land\odddfX -\frac{1}{\dimM-2}\dint{\vfre_a}(\cofr^b \land\odddfX{}_b) \,.
\end{align}

Some of these pieces have a straightforward interpretation: $\irrdfW{3}{}_{ab}$ corresponds to the totally antisymmetric part of the curvature ($\sim R_{[\mu\nu\rho\lambda]}$) and $\irrdfW{6}{}_{ab}$ is the Ricci scalar (since $X= R_{ba}{}^{ab}=-R$). For a metric compatible connection $\dfZ_{ab}\equiv0$ and, in particular, for the Levi-Civita curvatute, only the Ricci scalar, the symmetric part of the Ricci tensor and the Weyl curvature tensor survive, and they are encoded into $\irrdfW{6}{}_{ab}$, $\irrdfW{4}{}_{ab}$ and $\irrdfW{1}{}_{ab}$, respectively.

\begin{table}[H]
\begin{centering}
\renewcommand{\arraystretch}{1.5}
{\small
\begin{tabular}{|c|c|c|}
  \hline                     & Arbitrary dimension $\dimM$         & $\dimM=4$\tabularnewline
  \hline 
  \hline Total $\dfW_{ab}$   & $\frac{1}{4}\dimM^{2}(\dimM-1)^{2}$ & 36\tabularnewline
  \hline 
  \hline $\irrdfW{1}{}_{ab}$ & $\frac{1}{12}(\dimM+2)(\dimM+1)\dimM(\dimM-3)$ & 10\tabularnewline
  \hline $\irrdfW{2}{}_{ab}$ & $\frac{1}{8}(\dimM+2)\dimM(\dimM-1)(\dimM-3)$  & 9\tabularnewline
  \hline $\irrdfW{3}{}_{ab}$ & $\frac{1}{24}\dimM(\dimM-1)(\dimM-2)(\dimM-3)$ & 1\tabularnewline
  \hline $\irrdfW{4}{}_{ab}$ & $\frac{1}{2}(\dimM+2)(\dimM-1)$     & 9\tabularnewline
  \hline $\irrdfW{5}{}_{ab}$ & $\frac{1}{2}\dimM(\dimM-1)$         & 6\tabularnewline
  \hline $\irrdfW{6}{}_{ab}$ & 1                                   & 1\tabularnewline
  \hline 
\end{tabular}~~%
\begin{tabular}{|c|c|c|}
  \hline                     & Arbitrary dimension $\dimM$         & $\dimM=4$\tabularnewline
  \hline 
  \hline Total $\dfZ_{ab}$   & $\frac{1}{4}\dimM^{2}(\dimM^{2}-1)$ & 60\tabularnewline
  \hline 
  \hline $\irrdfZ{1}{}_{ab}$ & $\frac{1}{8}(\dimM-2)(\dimM+4)(\dimM^{2}-1)$  & 30\tabularnewline
  \hline $\irrdfZ{2}{}_{ab}$ & $\frac{1}{8}(\dimM+2)\dimM(\dimM-1)(\dimM-3)$ & 9\tabularnewline
  \hline $\irrdfZ{3}{}_{ab}$ & $\frac{1}{2}\dimM(\dimM-1)$         & 6\tabularnewline
  \hline $\irrdfZ{4}{}_{ab}$ & $\frac{1}{2}(\dimM+2)(\dimM-1)$     & 9\tabularnewline
  \hline $\irrdfZ{5}{}_{ab}$ & $\frac{1}{2}\dimM(\dimM-1)$         & 6\tabularnewline
  \hline \multicolumn{1}{c}{}& \multicolumn{1}{c}{}                & \multicolumn{1}{c}{}\tabularnewline
\end{tabular}
}
\renewcommand{\arraystretch}{1}
\par\end{centering}
\caption{Distribution of the independent components of the curvature (a total
of \\ $\frac{1}{2}\dimM^{3}(\dimM-1)$, which correspond to 96 in $\dimM=4$).}
\end{table}

Furthermore, the symmetric part contains five irreducible parts $\dfZ_{ab}=\sum_{I=1}^{5}\irrdfZ{I}{}_{ab}$ defined as follows\footnote{
    All of the conventions for the irreducible parts of the torsion, nonmetricity and curvature are in agreement with the ones used in \cite{JCAObukhov2021a}. In fact, they correspond to those of \cite{Hehl1995, McCrea1992}, except for $\irrdfZ{4}{}_{ab}$ and $\irrdfZ{5}{}_{ab}$ that have been exchanged.}
\begin{align}
  \irrdfZ{2}{}_{ab} & \coloneqq\frac{1}{2}\sgng\star\left(\cofr_{(a}\wedge\odddfPhi{}_{b)}\right)\\
  \irrdfZ{3}{}_{ab} & \coloneqq\frac{1}{\dimM^{2}-4}\left[\dimM\cofr_{(a}\wedge\dint{\vfre_{b)}}(\cofr^c\wedge\dfY_c)-2g_{ab}(\cofr^c\wedge\dfY_c)\right]\\
  \irrdfZ{4}{}_{ab} & \coloneqq\frac{2}{\dimM}\cofr_{(a}\wedge\dfPhi{}_{b)}\\
  \irrdfZ{5}{}_{ab} & \coloneqq\frac{1}{\dimM}g_{ab}\dfZ\\
  \irrdfZ{1}{}_{ab} & \coloneqq\dfZ_{ab}-\sum_{\mathrm{I}=2}^{5}\irrdfZ{I}{}_{ab}
\end{align}
where we have introduced (again, the number at the left of each quantity represents its rank as differential forms)
\begin{align}
 & [1]      \!\!\!\! & \dfY_a             & \coloneqq\dint{\vfre^b}\dfZNoTr{}_{ab}\,, &  
 & [\dimM-3]\!\!\!\! & \odddfY{}_a  & \coloneqq\star(\dfZNoTr{}_{ba}\wedge\cofr^b)\,, \\
 & [1]      \!\!\!\! & \dfPhi{}_a         & \coloneqq\dfY_a-\frac{1}{2}\dint{\vfre_a}(\cofr^b\wedge\dfY_b)\,, &  
 & [\dimM-3]\!\!\!\! & \odddfPhi{}_a& \coloneqq\odddfY{}_a-\frac{1}{\dimM-2}\dint{\vfre_a}(\cofr^b\wedge\odddfY{}_b)\,.
\end{align}

\subsection*{Components of the irreducible parts of the curvature}
The irreducible parts of the antisymmetric part of the curvature ($W_{cdab}=R_{cd[ab]}$) \\ $\irrdfW{I}{}_{ab}=\frac{1}{2}\irrW{I}{}_{cdab}\cofr^{cd}$ have the following components
\begin{align}
\irrW{2}{}_{cdab} & =-\frac{1}{2}\left(W_{abcd}-W_{cdab}\right)- \frac{2}{\dimM-2}\left(R^{(1)}{}_{[ef]}-R^{(2)}{}_{[ef]}\right) \delta_{[a}^{e}g_{b][c}\delta_{d]}^{f}\,,\\
\irrW{3}{}_{cdab} & =W_{[cdab]}\qquad=R_{[cdab]}\,,\\
\irrW{4}{}_{cdab} & =-\frac{2}{\dimM-2}\left(R^{(1)}{}_{(ef)} -R^{(2)}{}_{(ef)}\right)\delta_{[a}^{e}g_{b][c}\delta_{d]}^{f} -\frac{4}{\dimM(\dimM-2)}g_{c[a}g_{b]d}R\,,\\
\irrW{5}{}_{cdab} & =\quad\frac{2}{\dimM-2} \left(R^{(1)}{}_{[ef]}-R^{(2)}{}_{[ef]}\right) \delta_{[a}^{e}g_{b][c}\delta_{d]}^{f}\,,\\
\irrW{6}{}_{cdab} & =\frac{2}{\dimM(\dimM-1)}g_{c[a}g_{b]d}R\,,\\
\irrW{1}{}_{cdab} & =W_{cdab}-\sum_{\mathrm{I}=2}^{6}\irrW{I}{}_{cdab}\,.
\end{align}
and the five parts of the symmetric curvature ($Z_{cdab}=R_{cd(ab)}$) $\irrdfZ{I}{}_{ab}=\frac{1}{2}\irrZ{I}{}_{cdab}\cofr^{cd}$ correspond to
\begin{align}
\irrZ{2}{}_{cdab} & =\frac{1}{2}\left(Z_{cdab}-Z_{c(ab)d}+Z_{d(ab)c}\right)\nonumber \\
 & \qquad-\frac{1}{2(\dimM-2)}\left(R^{(1)}{}_{[ef]}+ R^{(2)}{}_{[ef]}-R^{(3)}{}_{ef}\right)\left(2\delta_{(a}^{e}g_{b)[c}\delta_{d]}^{f} -g_{ab}\delta_{[c}^{e}\delta_{d]}^{f}\right)\,,\\
\irrZ{3}{}_{cdab} & =\frac{2}{\dimM^{2}-4} \left(R^{(1)}{}_{[ef]}+R^{(2)}{}_{[ef]}-\frac{2}{\dimM}R^{(3)}{}_{ef}\right) (\dimM\delta_{(a}^{e}g_{b)[c}\delta_{d]}^{f}-g_{ab}\delta_{[c}^{e}\delta_{d]}^{f})\,,\\
\irrZ{4}{}_{cdab} & =-\frac{2}{\dimM}\left(R^{(1)}{}_{(ef)}+R^{(2)}{}_{(ef)}\right)\delta_{(a}^{e}g_{b)[c}\delta_{d]}^{f}\,,\\
\irrZ{5}{}_{cdab} & =\frac{1}{\dimM}g_{ab}R^{(3)}{}_{cd}\,,\\
\irrZ{1}{}_{cdab} & =Z_{cdab}-\sum_{\mathrm{I}=2}^{5}\irrZ{I}{}_{cdab}\,.
\end{align}

\subsection*{Properties of the irreducible parts of the curvature}

\begin{itemize}
\item Properties of the auxiliary objects
  \begin{equation}
    \star(\cofr^c\wedge\dfX_c)=-\cofr^c\wedge\odddfX{}_c\,,\qquad\star(\cofr^c\wedge\dfY_c)=-\cofr^c\wedge\odddfY{}_c\,,
    \end{equation}
  \begin{equation}
    \dint{\vfre^a}\dfY_a=\dint{\vfre^a}\odddfY{}_a=0\,,
  \end{equation}
  \begin{equation}
    \dint{\vfre^a}\dfPsi{}_a=\dint{\vfre^a}\odddfPsi{}_a=\dint{\vfre^a}\dfPhi{}_a=\dint{\vfre^a}\odddfPhi{}_a=0\,,
  \end{equation}
  \begin{equation}
    \cofr^a\wedge\dfPsi{}_a=\cofr^a\wedge\odddfPsi{}_a=\cofr^a\wedge\dfPhi{}_a=\cofr^a\wedge\odddfPhi{}_a=0\,.
  \end{equation}
\item Trivial traces and contractions with the coframe
  \begin{align}
    \dint{\vfre^b}\irrdfW{I}{}_{ab} & =0\qquad\Leftrightarrow & \irrW{I}{}_{cda}{}^{c} & =0 & I & =1,\,2,\,3\\
    \dint{\vfre^a}\dint{\vfre^b}\irrdfW{I}{}_{ab} & =0\qquad\Leftrightarrow & \irrW{I}{}_{cd}{}^{dc} & =0 & I & =1,\,2,\,3,\,4,\,5\\
    \cofr^a\wedge\big(\dint{\vfre^b}\irrdfW{I}{}_{ab}\big) & =0\qquad\Leftrightarrow & \irrW{I}{}_{c[da]}{}^{c} & =0 & I & =1,\,2,\,3,\,4,\,\hphantom{5,}\,6\\
    \cofr^a\wedge\irrdfW{I}{}_{ab} & =0\qquad\Leftrightarrow & \irrW{I}{}_{[cda]b} & =0 & I & =1,\,\hphantom{2,\,3,}\,4,\,\hphantom{5,}\,6\\
    \cofr^a\wedge\cofr^b\wedge\irrdfW{I}{}_{ab} & =0\qquad\Leftrightarrow & \irrW{I}{}_{[cdab]} & =0 & I & =1,\,2,\hphantom{\,3,}\,4,\,5,\,6 \\[4mm]
    \irrdfZ{I}{}_{c}{}^{c} & =0\qquad\Leftrightarrow & \irrZ{I}{}_{dac}{}^{c} & =0 & I & =1,\,2,\,3,\,4\\
    \dint{\vfre^b}\irrdfZ{I}{}_{ab} & =0\qquad\Leftrightarrow & \irrZ{I}{}_{cda}{}^{c} & =0 & I & =1,\,2\\
    \cofr^a\wedge\irrdfZ{I}{}_{ab} & =0\qquad\Leftrightarrow & \irrZ{I}{}_{[cda]b} & =0 & I & =1,\,\hphantom{2,\,3,}\,4
  \end{align}
\item Non-trivial traces
  \begin{align}
    \dint{\vfre^b}\irrdfW{4}{}_{ab} & =\dfPsi_a\,, & \dint{\vfre^b}\irrdfZ{3}{}_{ab} & =\frac{1}{2}\dint{\vfre_a}(\cofr^c\wedge\dfY_c)\,,\nonumber \\
    \dint{\vfre^b}\irrdfW{5}{}_{ab} & =\frac{1}{2}\dint{\vfre_a}(\cofr^c\wedge\dfX_c)\,, & \dint{\vfre^b}\irrdfZ{4}{}_{ab} & =\dfPhi_a\,,\nonumber \\
    \dint{\vfre^b}\irrdfW{6}{}_{ab} & =\frac{1}{\dimM}X\cofr_a\,, & \dint{\vfre^b}\irrdfZ{5}{}_{ab} & =\frac{1}{\dimM}\dint{\vfre_a}\dfZ\,.
  \end{align}
\item Non-trivial contractions with the coframe
  \begin{align}
    \cofr^a\wedge\irrdfW{2}{}_{ab} & =\sgng(-1)^{\dimM-1}\star\odddfPsi{}_b\,, & \cofr^a\wedge\irrdfZ{2}{}_{ab} & =\sgng(-1)^{\dimM-1}\star\odddfPhi{}_b\,,\nonumber \\
    \cofr^a\wedge\irrdfW{3}{}_{ab} & =-\frac{1}{4}\sgng\dint{\vfre_b}\star\odddfX\,, & \cofr^a\wedge\irrdfZ{3}{}_{ab} & =-\frac{1}{\dimM-2}\cofr_{b} \wedge\cofr^c\wedge\dfY_c\,,\nonumber \\
    \cofr^a\wedge\irrdfW{5}{}_{ab} & =-\frac{1}{\dimM-2}\cofr_b\wedge\cofr^c\wedge\dfX_c\,, & \cofr^a\wedge\irrdfZ{5}{}_{ab} & =\frac{1}{\dimM}\cofr_{b} \wedge\dfZ\,.
  \end{align}
\item Other special properties
  \begin{equation}
    \irrdfW{5}{}_{ab}=\frac{2}{\dimM-2}\cofr_{[a}\wedge\dint{\vfre_{|c|}}\irrdfW{5}{}^c{}_{b]} \qquad\Leftrightarrow \qquad\irrW{5}{}_{cd}{}^{ab}=-\frac{2}{\dimM-2} \irrW{5}{}_{e[c}{}^{e[a}\delta_{d]}^{b]}\,.
  \end{equation}
  \begin{equation}
    \irrdfZ{2}{}_{ab} =\frac{1}{2}\dint{\vfre_{(a}}(\cofr^c\wedge\irrdfZ{2}{}_{b)c})\qquad \Leftrightarrow\qquad \irrZ{2}{}_{cd}{}^{ab}=-2\irrZ{2}{}_{[c}{}^{(ab)}{}_{d]}\,.
  \end{equation}
  \begin{equation}
    \irrdfZ{5}{}_{ab}=\frac{2}{\dimM}\cofr_{(a}\wedge(\dint{\vfre^c}\irrdfZ{5}{}_{b)c}) \qquad \Leftrightarrow\qquad\irrZ{5}{}^{cd}{}_{ab}= -\frac{4}{\dimM}\irrZ{5}{}^{e[c}{}_{e(a}\delta_{b)}^{d]}\,.
  \end{equation}
\end{itemize}

\newpage
\section{Decomposition of the auxiliary P-objects}

\subsection*{Definitions}

It is useful to introduce the following objects (the numbers at the left of each quantity represents its rank as differential forms):
\begin{align}
  & [2]      \hspace{-8mm} & \dfP_{ab} & \coloneqq\cofr_a\wedge(\dint{\vfre^c}\dfR_{cb})\,, &  
  & [2]      \hspace{-8mm} & \odddfP{}_{ab} & \coloneqq\star\dint{\vfre_a}(\cofr^c\wedge\dfR_{cb})\,,\\
  & [1]      \hspace{-8mm} & \dfP_a & \coloneqq\dint{\vfre^c}\dfR_{cb}\,, &  
  & [\dimM-3]\hspace{-8mm} & \odddfP{}_a & \coloneqq\star(\cofr^c\wedge\dfR_{cb})\,,\\
  & [2]      \hspace{-8mm} & \dfP & \coloneqq\cofr^a\wedge\dfP{}_a\,, &  
  & [\dimM-2]\hspace{-8mm} & \odddfP & \coloneqq\cofr^a\wedge\odddfP{}_a\,. \label{eq:defPs}
\end{align}
Then, one can prove that
\begin{align}
  \dfP_{ab}            & =\cofr_a\wedge\dfP_b\,, & 
  \odddfP{}_{ab} & =\cofr_a\wedge\odddfP{}_b\,,\\
  \dfP_c{}^c           & =\dfP\,, & 
  \odddfP{}_c{}^c& =\odddfP\,.
\end{align}

\subsection*{Decomposition of $\dfP_{a}$ and $\odddfP{}_{a}$. Useful expressions}

Explicitly in terms of the objects used in the irreducible decomposition of the curvature, we get
\begin{align}
  \dfP_a            & =\dfY_a-\dfX_a+\frac{1}{\dimM}\dint{\vfre_a}\dfZ\,,\\
  \odddfP{}_a & =\odddfY{}_a+\odddfX{}_a+\frac{1}{\dimM}\dint{\vfre_a}\star\dfZ\,.
\end{align}

Similarly as the decomposition of the torsion one can extract the trace and the totally antisymmetric parts of $\dfP_a$ and $\odddfP{}_a$ as follows 
\begin{align}
  \dfP_a            & =\dfPNoTr{}_a +\frac{1}{\dimM} \cofr_a(\dint{\vfre_c}\dfP{}^c) +\frac{1}{2}\dint{\vfre_a}\underbrace{(\cofr^c\wedge\dfP{}_c)}_{\dfP}\\
  \odddfP{}_a & =\oddvar{\dfPNoTr}{}_a+ \frac{1}{4}\cofr_a(\dint{\vfre_c}\odddfP{}^c)+ \frac{1}{\dimM-2}\dint{\vfre_a} \underbrace{(\cofr^c\wedge \odddfP{}_c)}_{\odddfP}
\end{align}
where the traces and the tensorial parts are:
\begin{align}
  \dint{\vfre_c}\dfP{}^c  & =-X\,,                    &  \dint{\vfre_c}\odddfP{}^c & =\odddfX\,,\\
  \dfPNoTr{}_a            & =\dfPhi{}_a-\dfPsi{}_a\,, &  \oddvar{\dfPNoTr}{}_a           & =\odddfPhi{}_a+\odddfPsi{}_a\,.
\end{align}
Notice that, by construction, the tensorial parts verify
\begin{equation}
  \dint{\vfre^a}\dfPNoTr{}_a = \dint{\vfre^a}\oddvar{\dfPNoTr}{}_a=0\,,\qquad\qquad
  \cofr^a \wedge\dfPNoTr{}_a = \cofr^a \wedge\oddvar{\dfPNoTr}{}_a=0\,.
\end{equation}
We also derive
\begin{equation}
  \dfP = -\cofr^a\wedge \dfX_a + \cofr^a\wedge \dfY_a + {\frac 12}\dfZ,\qquad 
  \star\odddfP = \cofr^a\wedge \dfX_a + \cofr^a\wedge \dfY_a - {\frac 12}\dfZ\,. \label{eq:PPXYZ}
\end{equation}

\chapter{MAG invariants and variations}

\section{The construction of the quadratic MAG action\label{app:MAGinvariants}}

\subsection{Even parity MAG invariants (up to second order)}

\subsubsection*{MAG zero order and linear invariants (even)}

At zero order in curvature, tensor and nonmetricity the only possibility is the canonical metric volume form:
\begin{equation}
\dfI_{0}\coloneqq\volfg\,.
\end{equation}
This term, which can be multiplied by some dimensionful constant is indeed the \emph{cosmological constant term}.

At linear order in torsion or nonmetricity there is no possible invariant. The reason is that there is no covariant scalar (total trace) of them, since they have an odd number of indices ($T_{\mu\nu}{}^{\lambda}$ and $Q_{\mu\nu\lambda}$). However, from the curvature one can construct in any dimension a linear invariant proportional to the curvature scalar. In differential form notation this can be written as
\begin{equation}
\dfI_{R}^{+}\coloneqq\dfR^{ab}\wedge\star\cofr_{ab}\qquad=R_{\mu\nu}{}^{\mu\nu}\volfg=R\,\volfg\,,
\end{equation}

\subsubsection*{MAG quadratic invariants in torsion and nonmetricity (even)}

One can distinguish three sectors: $TT$, $QQ$ and $TQ$:
\begin{itemize}
\item \textbf{$TT$ sector}\\
The part quadratic in torsion can be generated by three invariants: each irreducible component contracted with itself. In exterior notation:
\begin{align}
\dfI_{TT(1)}^{+} & \coloneqq\irrdfT{1}^a\wedge\star\irrdfT{1}{}_{a} &  & =\tfrac{1}{2}\irrT{1}{}^{\mu\nu\rho}\irrT{1}{}_{\mu\nu\rho}\volfg\,,\\
\dfI_{TT(2)}^{+} & \coloneqq\irrdfT{2}^a\wedge\star\irrdfT{2}{}_{a} &  & =\tfrac{1}{2}\irrT{2}{}^{\mu\nu\rho}\irrT{2}{}_{\mu\nu\rho}\volfg\,,\\
\dfI_{TT(3)}^{+} & \coloneqq\irrdfT{3}^a\wedge\star\irrdfT{3}{}_{a} &  & =\tfrac{1}{2}\irrT{3}{}^{\mu\nu\rho}\irrT{3}{}_{\mu\nu\rho}\volfg\,.
\end{align}
Due to the orthogonality of the irreducible components of $\dfT^{a}$ under the product $\wedge\star$, we can rewrite:
\begin{equation}
\dfI_{TT(I)}^{+}\coloneqq\dfT^a\wedge\star\irrdfT{I}{}_a\,\qquad\text{for }I=1,2,3\,.
\end{equation}

\item \textbf{$QQ$ sector}\\
The part quadratic in nonmetricity admits a basis of five invariants: each irreducible component contracted with itself (4) plus the contraction of both traces (1), i.e. 
\begin{align}
\dfI_{QQ(1)}^{+} & \coloneqq\irrdfQ{1}{}^{ab}\wedge\star\irrdfQ{1}{}_{ab} &  & =\irrQ{1}{}^{\mu\nu\rho}\irrQ{1}{}_{\mu\nu\rho}\volfg\,,\\
\dfI_{QQ(2)}^{+} & \coloneqq\irrdfQ{2}{}^{ab}\wedge\star\irrdfQ{2}{}_{ab} &  & =\irrQ{2}{}^{\mu\nu\rho}\irrQ{2}{}_{\mu\nu\rho}\volfg\,,\\
\dfI_{QQ(3)}^{+} & \coloneqq\irrdfQ{3}{}^{ab}\wedge\star\irrdfQ{3}{}_{ab} &  & =\irrQ{3}{}^{\mu\nu\rho}\irrQ{3}{}_{\mu\nu\rho}\volfg\nonumber \\
 &  &  & \qquad=\tfrac{2\dimM}{(\dimM-1)(\dimM+2)}\left[\tfrac{1}{\dimM}Q_{\mu}-\Qb_{\mu}\right]^{2}\volfg\,.\\
\dfI_{QQ(4)}^{+} & \coloneqq\irrdfQ{4}{}^{ab}\wedge\star\irrdfQ{4}{}_{ab} &  & =\irrQ{4}{}^{\mu\nu\rho}\irrQ{4}{}_{\mu\nu\rho}\volfg=\tfrac{1}{\dimM}Q_{\mu}\Qb{}^{\mu}\volfg\,.\\
(\dfI_{QQ(5)}^{+})' & \coloneqq(\dfQ^{ab}\wedge\star\cofr_{a})(\dint{\vfre_b}\dfQ_{c}{}^{c}) &  & =Q_{\mu}\Qb{}^{\mu}\volfg
\end{align}
however we prefer to work with another one instead of the last one:
\begin{equation}
\dfI_{QQ(5)}^{+}\coloneqq(\irrdfQ{3}{}_{ac}\wedge\cofr^{a})\wedge\star(\irrdfQ{4}{}^{bc}\wedge\cofr_b)\qquad=\tfrac{1}{\dimM}Q_{\mu}\left(\tfrac{1}{\dimM}Q^\mu -\Qb^\mu \right)\volfg\,,
\end{equation}
and they are related by
\begin{equation}
\dfI_{QQ(5)}^{+}=\tfrac{1}{\dimM}\left(\dfI_{QQ(4)}^{+}-(\dfI_{QQ(5)}^{+})'\right)\,.
\end{equation}
Again, due to the orthogonality of the irreducible components of $\dfQ_{ab}$ with the product $\wedge\star$, we can rewrite the first four with a total nonmetricity at one side,
\begin{equation}
\dfI_{QQ(I)}^{+}\coloneqq\dfQ^{ab}\wedge\star\irrdfQ{I}{}_{ab}\,\qquad\text{for }I=1,2,3,4\,.
\end{equation}
\item \textbf{$TQ$ sector}\\
The mixed sector only has three independent invariants: the only nontrivial contraction of the full torsion with the full nonmetricity (1) plus the trace of the torsion contracted with both traces of the nonmetricity (2). Respectively,
\begin{align}
\dfI_{TQ(1)}^{+} & \coloneqq\irrdfQ{2}{}_{ab}\wedge\cofr^a\wedge\star\dfT^b &  & =T^{\mu\nu\rho}\irrQ{2}{}_{\mu\nu\rho}\volfg\,,\\
\dfI_{TQ(2)}^{+} & \coloneqq\irrdfQ{3}{}_{ab}\wedge\cofr^a\wedge\star\dfT^b &  & =T^{\mu\nu\rho}\irrQ{3}{}_{\mu\nu\rho}\volfg\nonumber \\
 &  &  & \qquad=\tfrac{1}{\dimM-1}\left[\tfrac{1}{\dimM}Q_{\mu}-\Qb_{\mu}\right]T^{\mu}\volfg\,,\\
\dfI_{TQ(3)}^{+} & \coloneqq\irrdfQ{4}{}_{ab}\wedge\cofr^a\wedge\star\dfT^b &  & =T^{\mu\nu\rho}\irrQ{4}{}_{\mu\nu\rho}\volfg=\tfrac{1}{\dimM}Q_{\mu}T^{\mu}\volfg\,.
\end{align}
\end{itemize}

\subsubsection*{MAG quadratic invariants in curvature (even)}

In components it is not so difficult to see that there are a total of 16 independent invariants:
\begin{itemize}
\item Six involving the full curvature:
\begin{equation}
R_{abcd}R{}^{abcd}\,,\quad R_{abdc}R{}^{abcd}\,,\quad R_{acbd}R{}^{abcd}\,,\quad R_{acdb}R{}^{abcd}\,,\quad R_{cdab}R{}^{abcd}\,,\quad R_{adcb}R{}^{abcd}\,.
\end{equation}
\item Nine involving the traces:
\begin{align}
 & R^{(1)}{}_{ab}R^{(1)}{}^{ab}\,,\quad R^{(1)}{}_{ab}R^{(1)}{}^{ba}\,,\quad R^{(1)}{}_{ab}R^{(2)}{}^{ab}\,,\quad R^{(2)}{}_{ab}R^{(2)}{}^{ab}\,,\quad R^{(1)}{}_{ab}R^{(2)}{}^{ba}\,,\nonumber \\
 & R^{(2)}{}_{ab}R^{(2)}{}^{ba}\,,\quad R^{(1)}{}_{ab}R^{(3)}{}^{ab}\,,\quad R^{(2)}{}_{ab}R^{(3)}{}^{ab}\,,\quad R^{(3)}{}_{ab}R^{(3)}{}^{ab}\,.
\end{align}
\item One involving the total trace:
\begin{equation}
R^{2}\,.
\end{equation}
\end{itemize}
In differential form notation we are going to make use again of the irreducible components of the curvature 2-form and we separate the 16 independent invariants into three sectors:
\begin{itemize}
\item The sector containing the squares of the irreducible components of the antisymmetric part $\dfW_{ab}=\dfR_{[ab]}$. For $I=1,...,6$ we have
\begin{equation}
\dfI_{RR(I)}^{+}\coloneqq\irrdfW{I}_{ab}\wedge\star\irrdfW{I}{}^{ab}\qquad\qquad=\tfrac{1}{2}\irrW{I}{}^{\mu\nu\rho\lambda}\irrW{I}{}_{\mu\nu\rho\lambda}\volfg\,.
\end{equation}
\item The sector containing the squares of the irreducible components of the symmetric part $\dfZ_{ab}=\dfR_{(ab)}$. For $I=1,...,5$ we have
\begin{equation}
\dfI_{RR(6+I)}^{+}\coloneqq\irrdfZ{I}_{ab}\wedge\star\irrdfZ{I}{}^{ab}\qquad\qquad=\tfrac{1}{2}\irrZ{I}{}^{\mu\nu\rho\lambda}\irrZ{I}{}_{\mu\nu\rho\lambda}\volfg\,.
\end{equation}
\item An additional sector containing five additional independent invariants:
\begin{align}
\dfI_{RR(12)}^{+} & \coloneqq\dfR_{ab}\wedge\star\left[\cofr^a\wedge\big(\dint{\vfre_c}\irrdfW{5}{}^{cb}\big)\right]\qquad\qquad=R^{\mu}{}_{\lambda}{}^{\lambda\nu}\irrW{5}{}_{\mu\sigma}{}^{\sigma}{}_{\nu}\volfg\,.\\
\dfI_{RR(13)}^{+} & \coloneqq\dfR_{ab}\wedge\star\left[\cofr_c\wedge\big(\dint{\vfre^{a}}\irrdfZ{2}{}^{cb}\big)\right]\qquad\qquad=R^{\mu\nu\rho\lambda}\irrZ{2}{}_{\rho[\nu\mu]\lambda}\volfg\,.\\
\dfI_{RR(14)}^{+} & \coloneqq\dfR_{ab}\wedge\star\left[\cofr^a\wedge\big(\dint{\vfre_c}\irrdfZ{3}{}^{cb}\big)\right]\qquad\qquad=R^{\mu}{}_{\lambda}{}^{\lambda\nu}\irrZ{3}{}_{\mu\sigma}{}^{\sigma}{}_{\nu}\volfg\,.\\
\dfI_{RR(15)}^{+} & \coloneqq\dfR_{ab}\wedge\star\left[\cofr^a\wedge\big(\dint{\vfre_c}\irrdfZ{4}{}^{cb}\big)\right]\qquad\qquad=R^{\mu}{}_{\lambda}{}^{\lambda\nu}\irrZ{4}{}_{\mu\sigma}{}^{\sigma}{}_{\nu}\volfg\,.\\
\dfI_{RR(16)}^{+} & \coloneqq\dfR_{ab}\wedge\star\left[\cofr^a\wedge\big(\dint{\vfre_c}\irrdfZ{5}{}^{cb}\big)\right]\qquad\qquad=R^{\mu}{}_{\lambda}{}^{\lambda\nu}\irrZ{5}{}_{\mu\sigma}{}^{\sigma}{}_{\nu}\volfg\,.
\end{align}
\end{itemize}
As in the previous section, the orthogonality under $\wedge\star$, allows to rewrite
\begin{equation}
\dfI_{RR(I)}^{+}\coloneqq\dfR_{ab}\wedge\star\irrdfW{I}{}^{ab}\quad(I=1,...,6)\qquad\text{and}\qquad\dfI_{RR(6+J)}^{+}\coloneqq\dfR_{ab}\wedge\star\irrdfZ{J}{}^{ab}\quad(J=1,...,5)
\end{equation}
so that a global $\dfR_{ab}\wedge\star$ can be extracted in the action.

\subsection{MAG odd parity invariants in four dimensions}

All of the invariants constructed in the previous section exist in any dimension since they all have the form (ignoring the external indices) $\dfal\wedge\star\dfbe$ with $\dfal,\dfbe\in\Omega^{k}(\mathcal{M})$ for some $k$. Then, due to the Hodge star, the combination $\dfal\wedge\star\dfbe$ is a differential form of rank $k+(\dimM-k)=\dimM$, i.e. a perfectly valid term to be considered in the Lagrangian. In principle, combinations that do not involve the Hodge star can be or not $\dimM$-forms depending on the dimension. For instance, the term
\[
\dfQ_{ab}\wedge\dfQ^{ab}\in\Omega^{2}(\mathcal{M})
\]
is a good metric-affine term for the action only in 2 dimensions. These kind of terms in a Lagrangian are called odd-parity since they contain a Levi-Civita tensor inside. For instance, for the previous example in $\dimM=2$,
\[
\dfQ_{ab}\wedge\dfQ^{ab}=Q_{cab}Q_{d}{}^{ab}\cofr^{cd}=-Q_{\mu ab}Q_{\nu}{}^{ab}\LCten^{\mu\nu}\volfg
\]
where we have used the equation \eqref{eq: cofr as eps volf} and that for Lorentzian 2-dimensional metrics $\sgng=-1$. 

Since the most interesting situation is the 4-dimensional case, we are going to provide in this section all of the possible odd parity invariants (linear and quadratic in torsion, nonmetricity and curvature) in that dimension.

\subsubsection*{MAG zero order and linear invariants (odd in 4 dimensions)}

There are no zero order odd parity invariants. Similarly as in the even parity case there are also no linear invariants in torsion or nonmetricity. With the curvature there is one possibility, constructed with its totally antisymmetric part. In differential form notation this can be written as
\begin{equation}
\dfI_{R}^{-}\coloneqq\dfR^{ab}\wedge\cofr_{ab}\qquad=\frac{1}{2}\sgng R_{[\mu\nu\rho\lambda]}\LCten^{\mu\nu\rho\lambda}\volfg\,.
\end{equation}

\subsubsection*{MAG quadratic invariants in torsion and nonmetricity (odd in 4 dimensions)}

One can distinguish three sectors: $TT$, $QQ$ and $TQ$:
\begin{itemize}
\item \textbf{$TT$ sector}\\
The part quadratic in torsion can be generated by two invariants:
\begin{align}
\dfI_{TT(1)}^{-} & \coloneqq\dfT^a\wedge\irrdfT{1}{}_{a}=\irrdfT{1}{}^a\wedge\irrdfT{1}{}_{a} &  & =\frac{1}{4}\sgng\irrT{1}{}_{\mu\nu}{}^{\sigma}\irrT{1}{}_{\rho\lambda\sigma}\LCten^{\mu\nu\rho\lambda}\volfg\,,\\
\dfI_{TT(2)}^{-} & \coloneqq\dfT^a\wedge\irrdfT{2}{}_{a}=\irrdfT{3}{}^a\wedge\irrdfT{2}{}_{a} &  & =\frac{1}{4}\sgng\irrT{3}{}_{\mu\nu}{}^{\sigma}\irrT{2}{}_{\rho\lambda\sigma}\LCten^{\mu\nu\rho\lambda}\volfg\,,
\end{align}
because
\begin{equation}
\dfT^a\wedge\irrdfT{3}{}_{a}=\dfT^a\wedge\irrdfT{2}{}_a\,.\label{eq:redundantodd1}
\end{equation}
In components one can choose $T_{\mu}T_{[\nu\rho\lambda]}\LCten^{\mu\nu\rho\lambda}$
and $T_{\mu\nu}{}^{\sigma}T_{\rho\lambda\sigma}\LCten^{\mu\nu\rho\lambda}$.
\item \textbf{$QQ$ sector}\\
Interestingly, there is only one possible independent invariant. The traces cannot be involved since there are not enough indices with two of these vectors and the Levi-Civita tensor. The traceless totally symmetric part cannot be involved either, so the only possibility is
\begin{align}
\dfI_{QQ}^{-}\coloneqq(\irrdfQ{2}{}_{ac}\wedge\cofr^{a})\wedge(\irrdfQ{2}{}^{bc}\wedge\cofr_b)
  &\qquad =\sgng\irrQ{2}{}_{\mu\nu}{}^{\sigma}\irrQ{2}{}_{\rho\lambda\sigma}\LCten^{\mu\nu\rho\lambda}\\ 
  &\qquad =\sgng Q_{\mu\nu}{}^{\sigma}Q_{\rho\lambda\sigma}\LCten^{\mu\nu\rho\lambda}\,.
\end{align}
\item \textbf{$TQ$ sector}\\
In this sector we have again three possibilities. A basis of invariants in tensorial notation can be $T_{\mu\nu\rho}Q_{\lambda}\LCten^{\mu\nu\rho\lambda}$,  $T_{\mu\nu\rho}\Qb_{\lambda}\LCten^{\mu\nu\rho\lambda}$ and $T_{\mu\nu}{}^{\sigma}Q_{\rho\lambda\sigma} \LCten^{\mu\nu\rho\lambda}$. In terms of the irreducible components and in differential form notation we will take the basis:
\begin{align}
\dfI_{TQ(1)}^{-} & \coloneqq\irrdfQ{2}{}_{ab}\wedge\cofr^a\wedge\dfT^b &  & =\frac{1}{2}\sgng\irrQ{2}{}_{\mu\nu\sigma}T_{\rho\lambda}{}^{\sigma}\LCten^{\mu\nu\rho\lambda}\volfg\,,\\
\dfI_{TQ(2)}^{-} & \coloneqq\irrdfQ{3}{}_{ab}\wedge\cofr^a\wedge\dfT^b &  & =\frac{1}{2}\sgng\irrQ{3}{}_{\mu\nu\sigma}T_{\rho\lambda}{}^{\sigma}\LCten^{\mu\nu\rho\lambda}\volfg\nonumber \\
 &  &  & \qquad=\tfrac{1}{6}\sgng\left[\tfrac{1}{4}Q_{\mu}-\Qb_{\mu}\right]T_{\nu\rho\lambda}\LCten^{\mu\nu\rho\lambda}\volfg\,,\\
\dfI_{TQ(3)}^{-} & \coloneqq\irrdfQ{4}{}_{ab}\wedge\cofr^a\wedge\dfT^b &  & =\frac{1}{2}\sgng\irrQ{4}{}_{\mu\nu\sigma}T_{\rho\lambda}{}^{\sigma}\LCten^{\mu\nu\rho\lambda}\volfg\nonumber \\
 &  &  & =\frac{1}{8}\sgng Q_{\mu}T_{\nu\rho\lambda}\LCten^{\mu\nu\rho\lambda}\volfg\,.
\end{align}
\end{itemize}

\subsubsection*{MAG quadratic invariants in curvature (odd in 4 dimensions)}

We again separate the invariants into three sectors:
\begin{itemize}
\item The sector containing the squares of the irreducible components of the antisymmetric part $\dfW_{ab}=\dfR_{[ab]}$. We construct for $I=1,...,6$ the quantities
\begin{equation}
\dfW_{ab}\wedge\irrdfW{I}{}^{ab}\qquad\qquad=\frac{1}{4}\sgng R_{\mu\nu}{}^{\sigma\tau}\irrW{I}{}_{\rho\lambda\sigma\tau}\LCten^{\mu\nu\rho\lambda}\volfg\,.
\end{equation}
However by virtue of the properties 
\begin{align}
\dfW_{ab}\wedge\irrdfW{2}{}^{ab} & =\dfW_{ab}\wedge\irrdfW{4}{}^{ab}\qquad\irrdfW{2}{}_{ab}\wedge\irrdfW{4}{}^{ab}\,,\nonumber \\
\dfW_{ab}\wedge\irrdfW{3}{}^{ab} & =\dfW_{ab}\wedge\irrdfW{6}{}^{ab}\qquad\irrdfW{3}{}_{ab}\wedge\irrdfW{6}{}^{ab}\,,\label{eq:redundantodd2}
\end{align}
only four of them are independent:
\begin{equation}
\dfI_{RR(1)}^{-}\coloneqq\dfW_{ab}\wedge\irrdfW{1}{}^{ab}\,,\qquad\dfI_{RR(2)}^{-}\coloneqq\dfW_{ab}\wedge\irrdfW{2}{}^{ab}\,,\nonumber
\end{equation}
\begin{equation}
\dfI_{RR(3)}^{-}\coloneqq\dfW_{ab}\wedge\irrdfW{3}{}^{ab}\,,\qquad\dfI_{RR(4)}^{-}\coloneqq\dfW_{ab}\wedge\irrdfW{5}{}^{ab}\,.
\end{equation}
\item The sector containing the squares of the irreducible components of the symmetric part $\dfZ_{ab}=\dfR_{(ab)}$. For $I=1,...,5$ we have
\begin{equation}
\dfZ_{ab}\wedge\irrdfZ{I}{}^{ab}\qquad\qquad=\frac{1}{4}\sgng R_{\mu\nu}{}^{\sigma\tau}\irrZ{I}{}_{\rho\lambda\sigma\tau}\LCten^{\mu\nu\rho\lambda}\volfg\,.
\end{equation}
Now due to
\begin{equation}
\dfZ_{ab}\wedge\irrdfZ{2}{}^{ab}=\dfZ_{ab}\wedge\irrdfZ{4}{}^{ab}\qquad\irrdfZ{2}{}_{ab}\wedge\irrdfZ{4}{}^{ab}\,,\label{eq:redundantodd3}
\end{equation}
we choose the following set of independent invariants:
\begin{equation}
\dfI_{RR(5)}^{-}\coloneqq\dfZ_{ab}\wedge\irrdfZ{1}{}^{ab}\,,\qquad\dfI_{RR(6)}^{-}\coloneqq\dfZ_{ab}\wedge\irrdfZ{2}{}^{ab}\,,\nonumber
\end{equation}
\begin{equation}
\dfI_{RR(7)}^{-}\coloneqq\dfZ_{ab}\wedge\irrdfZ{3}{}^{ab}\,,\qquad\dfI_{RR(8)}^{-}\coloneqq\dfZ_{ab}\wedge\irrdfZ{5}{}^{ab}\,.
\end{equation}
\item An additional sector containing five additional independent invariants:
\begin{align}
\dfI_{RR(9)}^{-} & \coloneqq\dfR_{ab}\wedge\left[\cofr^a\wedge\big(\dint{\vfre_c}\irrdfW{5}{}^{cb}\big)\right]\!\!\qquad=\frac{1}{2}\sgng R_{\mu\nu\rho\tau}\irrW{5}{}_{\lambda\sigma}{}^{\sigma\tau}\LCten^{\mu\nu\rho\lambda}\volfg\,,\\
\dfI_{RR(10)}^{-} & \coloneqq\dfR_{ab}\wedge\left[\cofr_c\wedge\big(\dint{\vfre^{a}}\irrdfZ{2}{}^{cb}\big)\right]\qquad=\frac{1}{2}\sgng R_{\mu\nu\sigma\tau}\irrZ{2}{}^{\sigma}{}_{\lambda\rho}{}^{\tau}\LCten^{\mu\nu\rho\lambda}\volfg\,,\\
\dfI_{RR(11)}^{-} & \coloneqq\dfR_{ab}\wedge\left[\cofr^a\wedge\big(\dint{\vfre_c}\irrdfZ{3}{}^{cb}\big)\right]\qquad=\frac{1}{2}\sgng R_{\mu\nu\rho\tau}\irrZ{3}{}_{\lambda\sigma}{}^{\sigma\tau}\LCten^{\mu\nu\rho\lambda}\volfg\,,\\
\dfI_{RR(12)}^{-} & \coloneqq\dfR_{ab}\wedge\left[\cofr^a\wedge\big(\dint{\vfre_c}\irrdfZ{4}{}^{cb}\big)\right]\qquad=\frac{1}{2}\sgng R_{\mu\nu\rho\tau}\irrZ{4}{}_{\lambda\sigma}{}^{\sigma\tau}\LCten^{\mu\nu\rho\lambda}\volfg\,,\\
\dfI_{RR(13)}^{-} & \coloneqq\dfR_{ab}\wedge\left[\cofr^a\wedge\big(\dint{\vfre_c}\irrdfZ{5}{}^{cb}\big)\right]\qquad=\frac{1}{2}\sgng R_{\mu\nu\rho\tau}\irrZ{5}{}_{\lambda\sigma}{}^{\sigma\tau}\LCten^{\mu\nu\rho\lambda}\volfg\,.
\end{align}
\end{itemize}
\newpage

\section{Summary of MAG invariants up to quadratic order in T, Q, R}

\begin{equation}
  \dimM\text{-dimensional even parity invariants} \qquad\qquad\qquad   \text{4-dimensional odd parity invariants}\qquad\qquad\nonumber
\end{equation}
\begin{align*}
\dfI_{0} & \coloneqq\volfg\,,\\
\dfI_{R}^{+} & \coloneqq\dfR^{ab}\wedge\star\cofr_{ab}\,, & 
\dfI_{R}^{-} & \coloneqq\dfR^{ab}\wedge\cofr_{ab}\,,\\
\dfI_{TT(1)}^{+} & \coloneqq\dfT^a\wedge\star\irrdfT{1}{}_a\,, & 
\dfI_{TT(1)}^{-} & \coloneqq\dfT^a\wedge\irrdfT{1}{}_a\,,\\
\dfI_{TT(2)}^{+} & \coloneqq\dfT^a\wedge\star\irrdfT{2}{}_a\,, & 
\dfI_{TT(2)}^{-} & \coloneqq\dfT^a\wedge\irrdfT{2}{}_a\,,\\
\dfI_{TT(3)}^{+} & \coloneqq\dfT^a\wedge\star\irrdfT{3}{}_a\,,\\
\dfI_{QQ(1)}^{+} & \coloneqq\dfQ^{ab}\wedge\star\irrdfQ{1}{}_{ab}\,, & 
\dfI_{QQ}^{-} & \coloneqq(\irrdfQ{2}{}_{ac}\wedge\cofr^{a})\wedge(\irrdfQ{2}{}^{bc}\wedge\cofr_b)\,,\\
\dfI_{QQ(2)}^{+} & \coloneqq\dfQ^{ab}\wedge\star\irrdfQ{2}{}_{ab}\,,\\
\dfI_{QQ(3)}^{+} & \coloneqq\dfQ^{ab}\wedge\star\irrdfQ{3}{}_{ab}\,,\\
\dfI_{QQ(4)}^{+} & \coloneqq\dfQ^{ab}\wedge\star\irrdfQ{4}{}_{ab}\,,\\
\dfI_{QQ(5)}^{+} & \coloneqq(\irrdfQ{3}{}_{ac}\wedge\cofr^{a})\wedge\star(\irrdfQ{4}{}^{bc}\wedge\cofr_b)\,,\\
\dfI_{TQ(1)}^{+} & \coloneqq\irrdfQ{2}{}_{ab}\wedge\cofr^a\wedge\star\dfT^b\,, & 
\dfI_{TQ(1)}^{-} & \coloneqq\irrdfQ{2}{}_{ab}\wedge\cofr^a\wedge\dfT^b\,,\\
\dfI_{TQ(2)}^{+} & \coloneqq\irrdfQ{3}{}_{ab}\wedge\cofr^a\wedge\star\dfT^b\,, & 
\dfI_{TQ(2)}^{-} & \coloneqq\irrdfQ{3}{}_{ab}\wedge\cofr^a\wedge\dfT^b\,,\\
\dfI_{TQ(3)}^{+} & \coloneqq\irrdfQ{4}{}_{ab}\wedge\cofr^a\wedge\star\dfT^b\,, & 
\dfI_{TQ(3)}^{-} & \coloneqq\irrdfQ{4}{}_{ab}\wedge\cofr^a\wedge\dfT^b\,,\\
\dfI_{RR(1)}^{+} & \coloneqq\dfR_{ab}\wedge\star\irrdfW{1}{}^{ab}\,, & 
\dfI_{RR(1)}^{-} & \coloneqq\dfR_{ab}\wedge\irrdfW{1}{}^{ab}\,,\\
\dfI_{RR(2)}^{+} & \coloneqq\dfR_{ab}\wedge\star\irrdfW{2}{}^{ab}\,, & 
\dfI_{RR(2)}^{-} & \coloneqq\dfR_{ab}\wedge\irrdfW{2}{}^{ab}\,,\\
\dfI_{RR(3)}^{+} & \coloneqq\dfR_{ab}\wedge\star\irrdfW{3}{}^{ab}\,, & 
\dfI_{RR(3)}^{-} & \coloneqq\dfR_{ab}\wedge\irrdfW{3}{}^{ab}\,,\\
\dfI_{RR(4)}^{+} & \coloneqq\dfR_{ab}\wedge\star\irrdfW{4}{}^{ab}\,,\\
\dfI_{RR(5)}^{+} & \coloneqq\dfR_{ab}\wedge\star\irrdfW{5}{}^{ab}\,, & 
\dfI_{RR(4)}^{-} & \coloneqq\dfR_{ab}\wedge\irrdfW{5}{}^{ab}\,,\\
\dfI_{RR(6)}^{+} & \coloneqq\dfR_{ab}\wedge\star\irrdfW{6}{}^{ab}\,,\\
\dfI_{RR(7)}^{+} & \coloneqq\dfR_{ab}\wedge\star\irrdfZ{1}{}^{ab}\,, & 
\dfI_{RR(5)}^{-} & \coloneqq\dfR_{ab}\wedge\irrdfZ{1}{}^{ab}\,,\\
\dfI_{RR(8)}^{+} & \coloneqq\dfR_{ab}\wedge\star\irrdfZ{2}{}^{ab}\,, & 
\dfI_{RR(6)}^{-} & \coloneqq\dfR_{ab}\wedge\irrdfZ{2}{}^{ab}\,,\\
\dfI_{RR(9)}^{+} & \coloneqq\dfR_{ab}\wedge\star\irrdfZ{3}{}^{ab}\,, & 
\dfI_{RR(7)}^{-} & \coloneqq\dfR_{ab}\wedge\irrdfZ{3}{}^{ab}\,,\\
\dfI_{RR(10)}^{+} & \coloneqq\dfR_{ab}\wedge\star\irrdfZ{4}{}^{ab}\,,\\
\dfI_{RR(11)}^{+} & \coloneqq\dfR_{ab}\wedge\star\irrdfZ{5}{}^{ab}\,, & 
\dfI_{RR(8)}^{-} & \coloneqq\dfR_{ab}\wedge\irrdfZ{5}{}^{ab}\,,\\
\dfI_{RR(12)}^{+} & \coloneqq\dfR_{ab}\wedge\star\left[\cofr^a\wedge\big(\dint{\vfre_c}\irrdfW{5}{}^{cb}\big)\right]\,, & 
\dfI_{RR(9)}^{-} & \coloneqq\dfR_{ab}\wedge\left[\cofr^a\wedge\big(\dint{\vfre_c}\irrdfW{5}{}^{cb}\big)\right]\,,\\
\dfI_{RR(13)}^{+} & \coloneqq\dfR_{ab}\wedge\star\left[\cofr_c\wedge\big(\dint{\vfre^{a}}\irrdfZ{2}{}^{cb}\big)\right]\,, & 
\dfI_{RR(10)}^{-} & \coloneqq\dfR_{ab}\wedge\left[\cofr_c\wedge\big(\dint{\vfre^{a}}\irrdfZ{2}{}^{cb}\big)\right]\,,\\
\dfI_{RR(14)}^{+} & \coloneqq\dfR_{ab}\wedge\star\left[\cofr^a\wedge\big(\dint{\vfre_c}\irrdfZ{3}{}^{cb}\big)\right]\,, & 
\dfI_{RR(11)}^{-} & \coloneqq\dfR_{ab}\wedge\left[\cofr^a\wedge\big(\dint{\vfre_c}\irrdfZ{3}{}^{cb}\big)\right]\,,\\
\dfI_{RR(15)}^{+} & \coloneqq\dfR_{ab}\wedge\star\left[\cofr^a\wedge\big(\dint{\vfre_c}\irrdfZ{4}{}^{cb}\big)\right]\,, & 
\dfI_{RR(12)}^{-} & \coloneqq\dfR_{ab}\wedge\left[\cofr^a\wedge\big(\dint{\vfre_c}\irrdfZ{4}{}^{cb}\big)\right]\,,\\
\dfI_{RR(16)}^{+} & \coloneqq\dfR_{ab}\wedge\star\left[\cofr^a\wedge\big(\dint{\vfre_c}\irrdfZ{5}{}^{cb}\big)\right]\,, & 
\dfI_{RR(13)}^{-} & \coloneqq\dfR_{ab}\wedge\left[\cofr^a\wedge\big(\dint{\vfre_c}\irrdfZ{5}{}^{cb}\big)\right]\,.
\end{align*}

\section{Useful variations\label{app:variations}}

Here we collect the variations of the metric-affine invariant that conform the Lagrangians \eqref{eq: qMAGLeven} and \eqref{eq: qMAGLodd} (the most general ones with linear and quadratic terms in nonmetricity, torsion and curvature). The operator $\delta$ that we use below represents an arbitrary variation with respect to $\dfQ_{ab}$, $\dfT^a$ and $\dfR{}^{ab}$ (as independent fields).

\subsection{Even sector in arbitrary dimension}

For arbitrary metric signature in dimension $\dimM$ we have:
\begin{align}
\delta\dfI_{0} & =0\,,\\
\delta\dfI_{R}^{+} & =\delta\dfR^{ab}\wedge\star\Big[\cofr_{ab}\Big]\,,\\
\delta\dfI_{TT(I)}^{+} & =\delta\dfT^a\wedge\star\Big[2\irrdfT{I}{}_a\Big]\qquad\text{for }I=1,2,3\,,\\
\delta\dfI_{QQ(I)}^{+} & =\delta\dfQ_{ab}\wedge\star \left[2\irrdfQ{I}{}^{ab}\right] \qquad\!\!\!\!\!\text{for }I=1,2,3,4\,,\\
\delta\dfI_{QQ(5)}^{+} & =\delta\dfQ_{ab}\wedge \star\left[-\cofr^{(a}(\dint{\vfre^{b)}} \dfQ)-\frac{1}{\dimM}g^{ab}(\dfvarLa-\dfQ)\right]\,,\\
\delta\dfI_{TQ(1)}^{+} & =\delta\dfT^a\wedge\star\left[ \irrdfQ{2}{}_{ab}\wedge\cofr^b\right]+ \delta\dfQ_{ab}\wedge\star\left[\frac{-1}{\dimM-1}(\cofr^{(a}\dint{\vfre^{b)}}-g^{ab})\dfT -\dint{\vfre^{(a}}\dfT^{b)}\right]\,,\\
\delta\dfI_{TQ(2)}^{+} & =\delta\dfT^a\wedge\star\left[\irrdfQ{3}{}_{ab}\wedge\cofr^b\right] +\delta\dfQ_{ab} \wedge\star\left[\frac{1}{\dimM-1}\Big(\cofr^{(a}\dint{\vfre^{b)}} -\frac{1}{\dimM}g^{ab}\Big)\dfT\right]\,,\\
\delta\dfI_{TQ(3)}^{+} & =\delta\dfT^a\wedge\star \left[\irrdfQ{4}{}_{ab}\wedge\cofr^b\right] +\delta\dfQ_{ab}\wedge\star\left[ -\frac{1}{\dimM}g^{ab}\dfT\right]\,,\\
\delta\dfI_{RR(I)}^{+} & =\delta\dfR^{ab}\wedge\star\left[2\irrdfW{I}_{ab}\right]\qquad\!\!\text{for }I=1,2,3,4,5,6\,,\\
\delta\dfI_{RR(6+I)}^{+} & =\delta\dfR^{ab}\wedge\star\left[2\irrdfZ{I}_{ab}\right]\qquad\text{for }I=1,2,3,4,5\,,\\
\delta\dfI_{RR(12)}^{+} & =\delta\dfR^{ab}\wedge\star\left[ \cofr_a\wedge\dint{\vfre_c}\irrdfW{5}{}^c{}_{b} +\frac{1}{2}\cofr_{[a}\wedge\dint{\vfre_{b]}} \dfP\right]\,,\\
\delta\dfI_{RR(13)}^{+} & =\delta\dfR^{ab}\wedge\star\left[\cofr_c\wedge( \dint{\vfre_{(a}}\irrdfW{2}{}^c{}_{b)})+ \cofr_c\wedge(\dint{\vfre_{[a}}\irrdfZ{2}{}^c{}_{b]}) -2\irrdfZ{2}{}_{ab}\right]\,,\\
\delta\dfI_{RR(14)}^{+} & =\delta\dfR^{ab}\wedge\star\left[ \cofr_a\wedge\dint{\vfre_c}\irrdfZ{3}{}^c{}_{b}- \frac{1}{2}\cofr_{(a}\wedge\dint{\vfre_{b)}}\dfP +\frac{1}{\dimM}g_{ab}\dfP\right]\,,\\
\delta\dfI_{RR(15)}^{+} & =\delta\dfR^{ab}\wedge\star\left[\cofr_{[a}\wedge (\dint{\vfre_{|c|}}\irrdfZ{4}{}^c{}_{b]}) +\cofr_{(a}\wedge(\dint{\vfre_{|c|}}\irrdfW{4}{}^c{}_{b)}) +\dimM\irrdfZ{4}{}_{ab}\right]\,,\\
\delta\dfI_{RR(16)}^{+} & =\delta\dfR^{ab}\wedge\star\left[\cofr_a \wedge\dint{\vfre_c}\irrdfZ{5}{}^c{}_{b}+ \frac{1}{\dimM}g_{ab}\dfP\right]\,.
\end{align}

\newpage
\subsection{Odd sector in four dimension}

For arbitrary metric signature in four dimensions we have:
\begin{align}
\delta\dfI_{R}^{-} & =\delta\dfR^{ab}\wedge\Big[\cofr_{ab}\Big]\,,\\
\delta\dfI_{TT(1)}^{-} & =\delta\dfT^a\wedge\Big[2\irrdfT{1}{}_a\Big]\\
\delta\dfI_{TT(2)}^{-} & =\delta\dfT^a\wedge\Big[\irrdfT{2}{}_{a}+\irrdfT{3}{}_a\Big]\\
\delta\dfI_{QQ}^{-} & =\delta\dfQ_{ab}\wedge\left[-\cofr^{(a}\wedge\star\odddfvarLa^{b)}\right]\\
\delta\dfI_{TQ(1)}^{-} & =\delta\dfT^a\wedge\left[\irrdfQ{2}{}_{ab}\wedge \cofr^b\right]+\delta\dfQ_{ab}\wedge\left[\cofr^{(a} \wedge\dfT^{b)}-\frac{1}{3}\cofr^{(a}\wedge\dint{\vfre^{b)}}(\cofr_c\wedge\dfT^c)\right]\,,\\
\delta\dfI_{TQ(2)}^{-} & =\delta\dfT^a\wedge\left[\irrdfQ{3}{}_{ab}\wedge\cofr^b \right]+\delta\dfQ_{ab} \wedge\left[\frac{1}{3}\cofr^{(a}\wedge\dint{\vfre^{b)}}(\cofr_c\wedge\dfT^c)-\frac{1}{4}g^{ab}(\cofr_c\wedge\dfT^c)\right]\,,\\
\delta\dfI_{TQ(3)}^{-} & =\delta\dfT^a\wedge\left[ \irrdfQ{4}{}_{ab}\wedge\cofr^b\right] +\delta\dfQ_{ab} \wedge\left[\frac{1}{4} g^{ab}\cofr_c\wedge\dfT^c\right]\,,\\
\delta\dfI_{RR(1)}^{-} & =\delta\dfR^{ab}\wedge\left[2\irrdfW{1}_{ab}\right]\,,\\
\delta\dfI_{RR(2)}^{-} & =\delta\dfR^{ab}\wedge\left[\irrdfW{2}_{ab}+\irrdfW{4}_{ab}\right]\,,\\
\delta\dfI_{RR(3)}^{-} & =\delta\dfR^{ab}\wedge\left[\irrdfW{3}_{ab}+\irrdfW{6}_{ab}\right]\,,\\
\delta\dfI_{RR(4)}^{-} & =\delta\dfR^{ab}\wedge\left[2\irrdfW{5}_{ab}\right]\,,\\
\delta\dfI_{RR(5)}^{-} & =\delta\dfR^{ab}\wedge\left[2\irrdfZ{1}_{ab}\right]\,,\\
\delta\dfI_{RR(6)}^{-} & =\delta\dfR^{ab}\wedge\left[\irrdfZ{2}_{ab}+\irrdfZ{4}_{ab}\right]\,,\\
\delta\dfI_{RR(7)}^{-} & =\delta\dfR^{ab}\wedge\left[2\irrdfZ{3}_{ab}\right]\,,\\
\delta\dfI_{RR(8)}^{-} & =\delta\dfR^{ab}\wedge\left[2\irrdfZ{5}_{ab}\right]\,,\\
\delta\dfI_{RR(9)}^{-} & =\delta\dfR^{ab}\wedge \left[\cofr_a\wedge\dint{\vfre_c}\irrdfW{5}{}^c{}_{b} +\frac{1}{2}\sgng\cofr_{[a}\wedge \dint{\vfre_{b]}}\star\odddfP\right]\,,\\
\delta\dfI_{RR(10)}^{-} & =\delta\dfR^{ab}\wedge\left[ \cofr_c\wedge(\dint{\vfre_{[a}}\irrdfZ{2}{}^c{}_{b]}) -\cofr_{(a}\wedge(\dint{\vfre_{|c|}} \irrdfW{4}{}^c{}_{b)})-\irrdfZ{2}{}_{ab}-\irrdfZ{4}{}_{ab}\right]\,,\\
\delta\dfI_{RR(11)}^{-} & =\delta\dfR^{ab}\wedge\left[ \cofr_a\wedge \dint{\vfre_c}\irrdfZ{3}{}^c{}_{b}+ \frac{1-2\sgng}{3}\Big(\!\!-\!\frac{1}{2}\cofr_{(a} \wedge\dint{\vfre_{b)}}\star\!\odddfP+ \frac{1}{4}g_{ab}\star\!\odddfP\Big)\right]\,,\\
\delta\dfI_{RR(12)}^{-} & =\delta\dfR^{ab}\wedge\left[\cofr_{[a} \wedge(\dint{\vfre_{|c|}}\irrdfZ{4}{}^c{}_{b]}) -\cofr_c\wedge(\dint{\vfre_{(a}}\irrdfW{2}{}^c{}_{b)}) +2\irrdfZ{2}{}_{ab}+2\irrdfZ{4}{}_{ab}\right]\,,\\
\delta\dfI_{RR(13)}^{-} & =\delta\dfR^{ab}\wedge\left[\cofr_a\wedge\dint{\vfre_c} \irrdfZ{5}{}^c{}_{b} -\frac{1}{4}g_{ab}\star\odddfP\right]\,.
\end{align}
In our computations we performed the substitution $\sgng=-1$, since we are always dealing with a Lorentzian metric in mostly minus convention.

\chapter{On null congruences and optical decomposition} 

\section{Distributions and Frobenius theorem}
Let $\mathcal{M}$ be a smooth manifold and $\mathcal{U}\subseteq\mathcal{M}$ an open set.

\boxdefinition{
\begin{defn}
\textbf{($k$-distribution)}. In each point $p\in\mathcal{U}$ we
define a $k$-dimensional subspace of the tangent space $D_{p}^{(k)}$
satisfying the following: there exists a neighborhood of $p$ and
a set of $k$ linearly independent smooth vector fields $\{\vecX_{1},...,\,\vecX_{k}\}$
such that at $p$ they generate $D_{p}^{(k)}$:
\begin{equation}
\mathrm{span}_{\mathbb{R}}\{(\vecX_{1})_{p}\,,...,\,(\vecX_{k})_{p}\}=D_{p}^{(k)}\,.
\end{equation}
The collection $D^{(k)}=\bigcup_{p\in\mathcal{U}}D_{p}^{(k)}$ is
called a \emph{$k$-distribution} over $\mathcal{U}$.
\end{defn}
\begin{defn}
\textbf{(Involutive distribution)}. A distribution is said to be \emph{involutive}
if for every $p\in\mathcal{U}$, there exists a basis $\{\vecE_{i}\}_{i=1}^{k}$
of the distribution such that in a neighborhood of $p$:
\begin{equation}
[\vecE_{i},\,\vecE_{j}]\in\mathrm{span}_{\mathbb{R}}\{\vecE_{1},...,\,\vecE_{k}\}\,.
\end{equation}
\end{defn}
}

It is interesting to know when a given $k$-distribution can be seen
as ``tangent'' to certain hypersurfaces of our manifold. The following
theorem gives us the recipe:

\boxtheorem{
\begin{thm}
\textbf{\textup{(Frobenius)}} A $k$-distribution $D^{(k)}$ induces
integral submanifolds iff it is involutive.
\end{thm}
}

Indeed, for our purposes there is a much more useful formulation of
this theorem. The idea is to take the $k$-distribution at each point
and consider the $(\dimM-k)$-dimensional annihilator subspace for
the distribution:
\begin{equation}
\mathrm{Ann}D_{p}^{(k)}\coloneqq\{\dfal\in T_{p}^{*}\mathcal{M}\quad|\quad\dfal(\vecv)=0\quad\forall\vecv\in D_{p}^{(k)}\}\,.
\end{equation}

\boxtheorem{
\begin{thm}
\textbf{\textup{\label{Thm: Frobenius}(Frobenius (reformulated))}} A $k$-distribution
$D^{(k)}$ induces integral submanifolds iff every $\dfal\in\mathrm{Ann}D_{p}^{(k)}$
satisfies:
\begin{equation}
\dex\dfal=\dfal\wedge\dfbe\qquad\xrightarrow{\mathrm{components}}\qquad\partial_{[\mu}\alpha_{\nu]}=\alpha_{[\mu}\beta_{\nu]}\label{eq: da eq a wedge b}
\end{equation}
for a certain $\dfbe\in T_{p}^{*}\mathcal{M}$.
\end{thm}
}

\newpage
\boxproposition{
\begin{prop}
Equivalent conditions to \textup{\eqref{eq: da eq a wedge b} }are
the following
\begin{enumerate}
\item Every $\dfal\in\mathrm{Ann}D_{p}^{(k)}$ satisfies:
\begin{equation}
\dfal\wedge\dex\dfal=0\qquad\xrightarrow{\mathrm{components}}\qquad\alpha_{[\mu}\partial_{\nu}\alpha_{\rho]}=0\,.\label{eq: a wedge d a  is zero}
\end{equation}
\item If the hyperplanes of the distribution have codimension 1, every $\dfal\in\mathrm{Ann}D_{p}^{(\dimM-1)}$\textup{ }can be locally expressed\textup{:
\begin{equation}
\dfal=f\dex u\qquad\xrightarrow{\mathrm{components}}\qquad\alpha_{\mu}=f\partial_{\mu}u\,.
\end{equation}
}
for some real functions $f$ and $u$.
\end{enumerate}
\end{prop}
}

\boxproof{
\begin{proof}
First we start with the equivalence between \eqref{eq: da eq a wedge b}
$\Leftrightarrow$ 1.
\begin{itemize}
\item $\Rightarrow$) It is trivial multiplying by $\dfal\wedge$ both sides
of \eqref{eq: da eq a wedge b}. 
\item $\Leftarrow$) We write $\dex\dfal$ in components with respect to
a basis given by $\{\cofr^{1}=\dfal,\,\cofr^{2},...,\,\cofr^{\dimM}\}$.
We will use the indices $a,\,b=1,...,\,\dimM$ and $i,\,j=2,...,\,\dimM$:
\begin{equation}
\dex\dfal=\frac{1}{2}c_{ab}\cofr^a\wedge\cofr^{b}=0+c_{1i}\dfal\wedge\cofr^{i}+\frac{1}{2}c_{ij}\cofr^{i}\wedge\cofr^{j}
\end{equation}
And now:
\begin{equation}
\dfal\wedge\dex\dfal=0+\frac{1}{2}c_{ij}\dfal\wedge\cofr^{i}\wedge\cofr^{j}
\end{equation}
Due to the vector space structure $c_{ij}=0$. Consequently $\dex\dfal=\dfal\wedge\dfbe$
with $\dfbe\equiv c_{1i}\cofr^{i}$.
\end{itemize}
Now we check that 2 $\Rightarrow$ 1: 
\begin{itemize}
\item This is immediate by direct computation: $\dfal\wedge\dex\dfal=f\dex u\wedge\dex\left(f\dex u\right)=0$.
\end{itemize}
Therefore, we have 2 $\Rightarrow$ 1 $\Leftrightarrow$ \eqref{eq: da eq a wedge b}. \\

To finish the proof it is enough to check that \eqref{eq: da eq a wedge b} $\Rightarrow$ 2 or, in other words (thanks to Theorem \ref{Thm: Frobenius}), that the existence of integral submanifolds implies 2:
\begin{itemize}
\item Consider a certain point $p$ in the manifold. If there exist integral submanifolds, then it is always possible to find a chart around $p$ with associated coordinates $\{\partial_{u},\,\partial_{x^{2}},\,...,\partial_{x^{\dimM}}\}$
where $\{\partial_{x^{2}},\,...,\partial_{x^{\dimM}}\}$ generate the distribution, i.e.,
\begin{equation}
\dfal_{p}(\partial_{u})=f(p)\,,\qquad\text{and}\qquad\dfal_{p}(\partial_{x^{i}})=0\qquad\forall i=2,...,\,\dimM\,,
\end{equation}
such that the function $f(p)$ does not vanish (if not, $\partial_{\lambda}$ would be linearly dependent with the rest of the basis). And the only 1-form that satisfies these conditions is
\begin{equation}
\frac{1}{f(p)}\dfal=\dex u\,,
\end{equation}
and this gives the condition 2.
\end{itemize}
\end{proof}
}

\newpage

\section{Optical decomposition}\label{app:opticaldecom}

For a lightlike congruence with velocity $k^\mu$ and for any lightlike vector $l^\mu$ such that $k^\mu l_\mu=\epsilon=\pm1$, the general covariant expressions for the twist tensor, the expansion and the shear tensor are
\begin{align}
\theta & =\tfrac{1}{\dimM-2}\left(\mathring{\nabla}_{\sigma}k^{\sigma}-\epsilon l_{\sigma}\dot{k}^{\sigma}\right)\,,\\
\omega_{\mu\nu} & =\mathring{\nabla}_{[\nu}k_{\mu]}-\epsilon(l^{\sigma}\mathring{\nabla}_{\sigma}k_{[\mu})k_{\nu]}+\epsilon l_{[\mu}\dot{k}_{\nu]}-\epsilon k_{[\mu}(l_{|\sigma|}\mathring{\nabla}_{\nu]}k^{\sigma})+k_{[\mu}l_{\nu]}l_{\sigma}\dot{k}^{\sigma}\,,\\
\sigma_{\mu\nu} & =\left[\mathring{\nabla}_{(\mu}k_{\nu)}-\tfrac{1}{\dimM-2}\Pi_{\mu\nu}\mathring{\nabla}_{\sigma}k^{\sigma}\right]-\epsilon(l^{\sigma}\mathring{\nabla}_{\sigma}k_{(\mu})k_{\nu)}-\epsilon k_{(\mu}(l_{|\sigma|}\mathring{\nabla}_{\nu)}k^{\sigma})\nonumber \\
 & \qquad-\epsilon\left[l_{(\mu}\dot{k}_{\nu)}-\tfrac{1}{\dimM-2}\Pi_{\mu\nu}l_{\sigma}\dot{k}^{\sigma}\right]+k_{(\mu}l_{\nu)}l_{\sigma}\dot{k}^{\sigma}+k_{\mu}k_{\nu}(l_{\lambda}l^{\sigma}\mathring{\nabla}_{\sigma}k^{\lambda})
\end{align}
where $\dot{k}^{\sigma}\coloneqq k^\mu \mathring{\nabla}_\mu k^\sigma$, which vanishes in the geodetic case.

\section{Proof of Proposition \ref{prop:normalgeod}}\label{app:proofnormalgeod}

By the Theorem \ref{Thm: Frobenius}, the associated 1-form $\dfk=k_{\mu}\dex x^{\mu}$
satisfies:
\begin{equation}
\dex\dfk=\dfk\wedge\dfbe\qquad\Leftrightarrow\qquad\partial_{[\mu}k_{\nu]}=k_{[\mu}\beta_{\nu]}\,,
\end{equation}
for a certain 1-form $\dfbe=\beta_{\mu}\dex x^{\mu}$. Then, using $k^{\nu}k_{\nu}=0$,
\begin{align}
k^{\nu}\mathring{\nabla}_{\nu}k^{\mu}=g^{\mu\rho}k^{\nu}\mathring{\nabla}_{\nu}k_{\rho} & =g^{\mu\rho}\Big(2k^{\nu}\partial_{[\nu}k_{\rho]}+k^{\nu}\mathring{\nabla}_{\rho}k_{\nu}\Big)\nonumber\\
 & =g^{\mu\rho}\underbrace{(k^{\nu}k_{\nu})}_{0}\beta_{\rho}-g^{\mu\rho}k^{\nu}k_{\rho}\beta_{\nu}+2g^{\mu\rho}\mathring{\nabla}_{\rho}\underbrace{(k_{\nu}k^{\nu})}_{0}\nonumber\\
 & =(-k^{\nu}\beta_{\nu})\,k^{\mu}\qquad=f(\tau)\,k^{\mu}
\end{align}
so it is a pre-geodesic. So we can reparameterize the curve $k^{\mu}\rightarrow fk^{\mu}$,
where $f$ is some monotonic function of the parameter. The new curve continues being lightlike and the normality condition holds:
\begin{equation}
0=\dfk\wedge\dex\dfk\quad\rightarrow\quad0=f^{2}\dfk\wedge\dex\dfk-f\underbrace{\dfk\wedge\dfk}_{0}\wedge\dex f\quad\Leftrightarrow\quad0=\dfk\wedge\dex\dfk\,.
\end{equation}

\newpage
\section{Proof of Proposition \ref{prop:normalnullw0}}\label{app:proofnormalnullw0}

(We follow \cite[p.~38--39]{Poisson2002}). Consider a lightlike vector $l^\mu$ verifying $k^\mu l_\mu=\epsilon=\pm 1$ and call 
\begin{equation}
H_{\rho\mu\nu}=3k_{[\mu}\partial_{\rho}k_{\nu]}=3k_{[\mu}B_{\nu\rho]}\quad=\quad k_{\mu}B_{[\nu\rho]}+k_{\rho}B_{[\mu\nu]}+k_{\nu}B_{[\rho\mu]}\,.
\end{equation}
By contracting with $\epsilon l^{\rho}$:
\begin{equation}\label{eq:appHBproof}
\epsilon l^{\rho}H_{\rho\mu\nu}=B_{[\mu\nu]}-\epsilon\left[k_{\mu}B_{[\rho\nu]}+k_{\nu}B_{[\mu\rho]}\right]l^{\rho}=B_{[\mu\nu]}-\epsilon\left[k_{[\nu}B_{\mu]\rho}+B_{\rho[\nu}k_{\mu]}\right]l^{\rho}\,.
\end{equation}

In addition to this, it can be shown that the following equation holds for a geodesic:
\begin{equation}
\sobj{B}{}_{\mu\nu} = B_{\mu\nu}-\epsilon l^{\sigma}k_{\nu}B_{\mu\sigma}-\epsilon l^{\sigma}B_{\sigma\nu}k_{\mu}+k_{\mu}k_{\nu}l^{\lambda}l^{\sigma}B_{\lambda\sigma}\,,
\end{equation}
and, finally, taking the antisymmetric part:
\begin{equation}
\omega_{\mu\nu}=B_{[\mu\nu]}-\epsilon\left[k_{[\nu}B_{\mu]\sigma}+B_{\sigma[\nu}k_{\mu]}\right]l^{\sigma}\overeq{\eqref{eq:appHBproof}}\epsilon l^{\rho}H_{\rho\mu\nu}\,.
\end{equation}

$\Rightarrow$) If the congruence is normal, then by the Frobenius
theorem $0=k_{[\mu}\partial_{\rho}k_{\nu]}=\tfrac{1}{3}H_{\rho\mu\nu}$,
so $\omega_{\mu\nu}=0$.

$\Leftarrow$) It can be proved that if $\omega_{\mu\nu}$ vanishes
for a particular $l^{\rho}$, then it is zero for any $l^{\rho}$ and, therefore, for any lightlike vector field non-colinear with $k^{\mu}$. We know $k^{\rho}H_{\rho\mu\nu}=0$,
so
\begin{equation}
n^{\rho}H_{\rho\mu\nu}=0\qquad\forall n^{\rho}\quad\text{lightlike}.
\end{equation}
And since, any vector in a Lorentzian space can be expressed as the
sum of two lightlike vectors, we obtain in each $p\in\mathcal{M}$:
\begin{equation}
w^{\rho}H_{\rho\mu\nu}=0\quad\forall w^{\rho}\in T_{p}\mathcal{M}\qquad\qquad\Rightarrow\qquad\qquad0=H_{\rho\mu\nu}\,,
\end{equation}
so, by Frobenius theorem, the congruence is normal.

\chapter{Curvature, torsion and nonmetricity for the geometries in Chapter \ref{ch:GWgen}}

\section{Curvature, torsion and non-metricity for the  connection \eqref{eq: gen conn}}\label{app: RTQ general conn}

The curvature form \eqref{eq: gen conn} is given by,
\begin{align}
\dfR_{ab} &\nonumber =\mathring{\dfR}_{ab}+\mathring{\Dex}\mathcal{C}_{ab}\wedge\dfk +\mathring{\Dex}\mathcal{P}_{cab}\wedge\spatcofr^{c}+k_a k_b \dex\dfA+g_{ab}\dex\dfB\\
 & \quad+2\left(\mathcal{C}_{c(a}\dfk+\mathcal{P}_{dc(a}\spatcofr^{d}\right)k_{b)}k^{c}\wedge\dfA +\mathcal{P}_{dcb}\mathcal{P}_{ea}{}^{c}\spatcofr^d\wedge\spatcofr^e-2\mathcal{P}_{d[a}{}^{c}\mathcal{C}_{b]c}\dfk\wedge\spatcofr^{d}\,,
\end{align}
and the torsion by 
\begin{align}
\dfT^a & =\mathcal{C}_{c}{}^{a}\dfk\wedge\cofr^c +\mathcal{P}_{cd}{}^{a}\cofr^{cd}+ k^a \dfA\wedge\dfk+\dfB\wedge\cofr^a \,,\\
 & =\left[-Ck^a -\kC^{a}-\kA k^a +B k^a -\kB l^{a}\right]\dfk\wedge\dfl\nonumber \\
 & \quad+\left[\kC_c l^a+C_{c}k^a +\sC_{c}{}^{a}-P_{c}l^{a}+P_{c}{}^{a}-k^a  \sA_c+B\delta_c^a- \sB_c l^a\right]\dfk\wedge\spatcofr^c \nonumber \\
 & \quad+\left[P_{c}k^a +\kP_{c}{}^{a}+\kB\delta_{c}^{a}- \sB_c k^a\right]\dfl\wedge\spatcofr^c \nonumber \\
 & \quad+\left[\kP_{cd}l^{a}+P_{cd}k^a + \sP_{cd}{}^{a}+ \sB_c\delta_d^a\right] \spatcofr^c\wedge\spatcofr^d\,,
\end{align}
with trace and antisymmetric components
\begin{align}
\irrdfT{2}{}^a & =\tfrac{1}{\dimM-1}(C+\kA-P_c{}^c)\cofr^a \wedge\dfk -\tfrac{1}{\dimM-1}\bar{P}_{b}{}^{b}\cofr^a \wedge\dfl\nonumber \\
 & \qquad+\tfrac{1}{\dimM-1}(\kC_c + \sP_{bc}{}^b)\cofr^a\wedge\cofr^c -\cofr^a \wedge\dfB\,,\\
\irrdfT{3}{}^{a} & =g^{ab}\left[2(P_{[b}-\bar{C}_{[b})k_c l_{d]}+ (\sC_{[cd}+2P_{[cd})k_{b]}+2\bar{P}_{[cd} l_{b]}+ \sP_{[bcd]}\right]\cofr^c\wedge\cofr^d\,,
\end{align}
while the other one can be calculated simply by $\irrdfT{1}{}^{a} =\dfT^a-\irrdfT{2}{}^{a}-\irrdfT{3}{}^{a}$. 

For the non-metricity we have the following expression
\begin{align}
\dfQ_{ab} & =2\dfom_{(ab)}=2k_a k_b \dfA+2g_{ab}\dfB\,.
\end{align}
Therefore the traces are
\begin{align}
\dfQ_{c}{}^{c} & =2\dimM\dfB\,,\\
\dint{\vfre^{c}}\dfQ_{cb} & =2\kA k_b +2B_{b}\,,
\end{align}
and its irreducible decomposition,
\begin{align}
\irrdfQ{4}{}_{ab} & =2g_{ab}\dfB\,,\\
\irrdfQ{3}{}_{ab} & =\tfrac{4\dimM}{(\dimM-1)(\dimM+2)}\kA\left[k_{(a}\cofr_{b)}- \tfrac{1}{\dimM}g_{ab}\dfk\right]\,,\\
\irrdfQ{1}{}_{ab} & = 2\big[k_{(a}k_b  A_{c)}- \tfrac{2}{\dimM+2}\kA k_{(a}g_{bc)}\big]\cofr^c\,,\\
\irrdfQ{2}{}_{ab} & = 2k_a k_b \dfA - \irrdfQ{3}{}_{ab} -\irrdfQ{1}{}_{ab} \,.
\end{align}

\section{Irreducible decomposition of the curvature and the torsion for \eqref{eq: conn 2}}\label{app: RTQ conn 2}

The irreducible components of the torsion \eqref{eq: T for type II} are
\begin{align}
\irrdfT{2}{}^{a} & =\tfrac{1}{\dimM-1}(C+k_c  A^c )\cofr^a \wedge\dfk+\tfrac{1}{\dimM-1}\kC_c \cofr^a \wedge\spatcofr^c +\dfB\wedge\cofr^a \,,\nonumber \\
\irrdfT{3}{}^{a} & =g^{ab}\left[-2\kC_{[b}k_c l_{d]}+C_{[cd}k_{b]}\right]\cofr^c\wedge\cofr^d\,,\\
\irrdfT{1}{}^{a} & =-\left[\tfrac{1}{3}\kC^{a}+\tfrac{\dimM-2}{\dimM-1}(C+k_c  A^c )k^a \right]\dfk\wedge\dfl+\tfrac{2\dimM-5}{3(\dimM-1)}\kC_c k^a \dfl\wedge\spatcofr^c \nonumber \\
 & \quad-\left(\tfrac{1}{\dimM-1}\kC_{d}\delta_{c}^{a}+\tfrac{1}{3}C_{cd}k^a \right)\spatcofr{}^{cd}\\
 & \quad+\left[\tfrac{1}{3}C_{c}{}^{a}+(C_{c}-\tilde{ A}_{c})k^a +\tfrac{\dimM-4}{3(\dimM-1)}\kC_c l^{a}+\tfrac{1}{\dimM-1}(C+k_d  A^{d})\delta_{c}^{a}\right]\dfk\wedge\spatcofr^c\,.
\end{align}
For the curvature \eqref{eq: R for type II} we first separate into antisymmetric and symmetric parts
\begin{align}
(\dfR_{[ab]}\equiv) \quad \dfW_{ab}  & =\mathring{\dfR}_{ab}+\mathring{\Dex}\mathcal{C}_{ab}\wedge\dfk\,,\\
(\dfR_{(ab)}\equiv) \quad \dfZ_{ab} & =k_a k_b \dex\dfA+g_{ab}\dex\dfB-2k_c k_{(a}\mathcal{C}_{b)}{}^{c}\dfk\wedge\dfA\,.
\end{align}
Taking this into account, it can be shown that the irreducible components are
\begin{align}
\irrdfW{3}{}_{ab} & =\Big(\partial_{v}\sC_{[ab}l_c k_{d]}+\spartial_{[c}\sC_{ab}k_{d]}\Big)\cofr^c\wedge\cofr^d\,,\\
\irrdfW{4}{}_{ab} & =\irrdfLCW{4}{}_{ab}-2\tfrac{\dimM-1}{\dimM-2}\irrdfW{6}{}_{ab}\nonumber \\
   & \quad+\tfrac{1}{\dimM-2}\Big[\partial_{v}(2Cl_{[a}-C_{[a}) +\spartial_{c}(2C^{c}k_{[a}+\kC^{c}l_{[a})-\spartial_{c}(\sC_{[a}{}^{c}- C\delta_{[a}^{c})\Big]\dfk\wedge\cofr{}_{b]}\nonumber \\
   & \quad+\tfrac{1}{\dimM-2}\Big[\partial_{v}(2Ck_{[a}+\kC_{[a})+ \spartial_{c} \kC^{c}k_{[a}\Big]\dfl\wedge\cofr{}_{b]}\nonumber \\
   & \quad+\tfrac{1}{\dimM-2}\Big[-\partial_{v}(C_{c}k_{[a}- \kC_c l_{[a})+ 2\spartial_{(c}\kC_{d)}\delta_{[a}^{d}- \spartial_{d}(\sC_{c}{}^{d}- C\delta_{c}^{d})k_{[a}\Big]\spatcofr^{c}\wedge\cofr{}_{b]}\,,\\
\irrdfW{5}{}_{ab} & =\quad\tfrac{1}{\dimM-2}\Big[-\partial_{v}C_{[a}+ \spartial_{c} \kC^{c}l_{[a}- \spartial_{c}(\sC_{[a}{}^{c}+ C\delta_{[a}^{c})\Big]\dfk\wedge\cofr{}_{b]}\nonumber \\
   & \quad+\tfrac{1}{\dimM-2}\Big[\partial_{v}\kC_{[a}- \spartial_{c} \kC^{c}k_{[a}\Big]\dfl\wedge\cofr{}_{b]}\nonumber \\
   & \quad+\tfrac{1}{\dimM-2}\Big[\partial_{v}(C_{c}k_{[a}- \kC_c l_{[a})+2\spartial_{[c}\kC_{d]}\delta_{[a}^{d}+ \spartial_{d}(\sC_{c}{}^{d}+C\delta_{c}^{d})k_{[a}\Big]\spatcofr^{c}\wedge\cofr{}_{b]}\,,\\
\irrdfW{6}{}_{ab} & =\tfrac{2}{\dimM(\dimM-1)}(\partial_{v}C +\spartial_{c} \kC^{c})\cofr_a\wedge\cofr_b\,,\\
\irrdfZ{2}{}_{ab} & =\tfrac{1}{2(\dimM-2)}Z_c^{-}\dint{\vfre_{(a|}}\Big\{\dfk\wedge\spatcofr^{c}\wedge[\cofr_{|b)}-(\dimM-2)k_{|b)}\dfl]\Big\}\nonumber \\
   & \quad+\tfrac{1}{2}(e^{i}{}_{c}e^{j}{}_{d}\partial_{[i} A_{j]}+\kC_c \sA_{d})k_{(a}\dint{\vfre_{b)}}\big(\dfk\wedge\spatcofr^c\wedge\spatcofr^d\big)\,,\\
\irrdfZ{3}{}_{ab}  & =\tfrac{\dimM}{\dimM^{2}-4}Z_c^{-}\left[k_{(a}\cofr_{b)}\wedge\spatcofr^c- \delta_{(a}^{c}\cofr_{b)}\wedge\dfk-\tfrac{2}{\dimM}g_{ab}\dfk\wedge\spatcofr^c\right]\,,\\
\irrdfZ{4}{}_{ab} & =\tfrac{1}{\dimM}Z_c^{+}k_{(a}\cofr_{b)} \wedge\spatcofr^{c} +\tfrac{1}{\dimM}\left[2\big(2(C\kA-\partial_{[u} A_{v]}) +\kC_c \sA^c\big)k_{(a}+ Z_{(a}^{+}\right]\cofr_{b)}\wedge\dfk\,,\\
\irrdfZ{5}{}_{ab} & = g_{ab}\dex\dfB\,,
\end{align}
where we have introduced the abbreviation $Z_a^\pm\coloneqq 2e^{i}{}_{a}\partial_{[v} A_{i]}\pm \kC_{a} \kA$, and the other three have been omitted because they can be calculated by the ones above by using
\begin{align}
\irrdfW{2}{}_{ab} & =\tfrac{1}{2}\dfW_{ab}+\tfrac{1}{4}\left(\dint{\vfre_{a}}\dint{\vfre_{b}}\dfW_{dc}\right)\cofr^d\wedge\cofr^c-\irrdfW{5}{}_{ab}\,,\\
\irrdfW{1}{}_{ab} & =\dfW_{ab}-\irrdfW{2}{}_{ab}-\irrdfW{3}{}_{ab}-\irrdfW{4}{}_{ab}-\irrdfW{5}{}_{ab}-\irrdfW{6}{}_{ab}\,,\\
\irrdfZ{1}{}_{ab} & =\dfZ_{ab}-\irrdfZ{2}{}_{ab}-\irrdfZ{3}{}_{ab}-\irrdfZ{5}{}_{ab}-\irrdfZ{4}{}_{ab}\,.
\end{align}

\subsection{Other expressions derived from the connection \eqref{eq: conn 2}}\label{app: other properties conn 2}

The general derivatives of $\dfk$ and $\dfl$ are
\begin{align}
\nabla_{c}k^a  & =-(Ck^a +\kC^{a})k_c +k^a  B_{c}\,,\\
(\nabla_{c}-\mathring{\nabla}_c)l^{a} & =(Cl^{a}-C^{a})k_c +k^a  A_{c}+l^{a} B_{c}\,.
\end{align}
With these equations and the following properties of the distorsion tensor (defined as the difference between the connection and the Levi-Civita one)
\begin{align}
g^{ca}(\dfom_{ca}{}^b-\mathring{\dfom}_{ca}{}^b) & =(\kA-C)k^b+ B^b-\kC^b\,,\\
k^c(\dfom_{ca}{}^b-\mathring{\dfom}_{ca}{}^b) & =\kA k_a k^b+\kB\delta_a^b\,,\\
l^c(\dfom_{ca}{}^b-\mathring{\dfom}_{ca}{}^b) & =\mathcal{C}_a{}^b+A k_a k^b+B\delta_a^b\,.
\end{align}
one can prove for transversal tensors of arbitrary number of indices
\begin{align}
k^{c}\nabla_{c}\sobj{S}_{a...}{}^{b...} & =\underbrace{k^{c}\mathring{\nabla}_{c}}_{\partial_{v}}\sobj{S}_{a...}{}^{b...}+(n^{\text{up}}-n_{\text{down}})\kB \sobj{S}_{a...}{}^{b...}\,,\\
l^{c}\nabla_{c}\sobj{S}_{a...}{}^{b...} & =l^{c}\mathring{\nabla}_{c}\sobj{S}_{a...}{}^{b...}+(n^{\text{up}}-n_{\text{down}})B\sobj{S}_{a...}{}^{b...}\nonumber \\
 & \quad-(\sC_{a}{}^{d}-l_{a}\kC^d-k_a C^d) \sobj{S}_{d...}{}^{b...}- ...+(\sC_{d}{}^{b}+l^b \kC_{d}+ k^b C_{d})\sobj{S}_{a...}{}^{d...}+...\,,\\
k^a \nabla_{c}\sobj{S}_{ab...}{}^{d...} & =\kC^{a}k_c S_{ab...}{}^{d...}\,,\\
l^{a}\nabla_{c}\sobj{S}_{ab...}{}^{d...} & =(C^{a}k_c -\mathring{\nabla}_{c}l^{a})\sobj{S}_{ab...}{}^{d...}\,,\\
\nabla^{a}\sobj{S}_{ab...}{}^{c...} & =\mathring{\nabla}^{a}\sobj{S}_{ab...}{}^{c...}+\left[\kC^{a}+(n^{\text{up}}-n_{\text{down}}-1) \sB^{a}\right]\sobj{S}_{ab...}{}^{c...}\,,
\end{align}
where $n^{\text{up}}$ and $n_{\text{down}}$ are respectively the
number of indices up (contravariance) and down (covariance) of the
tensor $\sobj{S}_{a...}{}^{b...}$. These properties are extremely useful in order to eliminate or reduce derivatives that appear in the equations of motion of metric-affine theories.

\chapter{A toy example} \label{app:toyexample}
In order to illustrate the problematic nature of dwelling on a surface where the principal part of the equations is singular, we will analyze a simple mechanical example. This will also allow us to illustrate the problem with generating perturbative solutions around a singular surface in phase space. Thus, let us consider a system with one degree of freedom $q(t)$ that evolves according to the following equation:\footnote{
    This is not a Hamiltonian system, but this property is not relevant for our purposes here where we want to illustrate the problematic nature of solutions where the principal part vanishes.}
\begin{equation}
 q(t) q''(t)+\Big(1-q'(t)\Big)q'(t)=0.
\end{equation}
 This equation has a singular surface given by $q=0$ where the principal part vanishes, which in turn is an exact solution. The general solution can be written as
\begin{equation}
 q(t)=C_1+C_2 \eN^{-t/C_1}
 \label{eq:Generalq}
\end{equation}
as can be checked by direct substitution. We can see that the family with $C_2=0$ reproduces the obvious constant solutions. However, among those constant solutions, the trivial one $q=0$ would require $C_1=0$, which, as we can see from the general solution, corresponds to the paradigmatic example of an essential singularity. This clearly shows that the trivial solution actually dwells on a singular surface of the space of solutions so one can expect to find difficulties to obtain perturbative solutions around it. To clearly see this, let us try to perturbative solve around $q=0$ so we expand
\begin{equation}
q(t)=q^{(1)}(t)+q^{(2)}(t)+q^{(3)}(t)+\cdots
\end{equation}
where $q^{(n)}(t)$ is assumed to be of order $n$ in some expansion parameter. It is not difficult to see that the term with second derivatives always contributes at order $(n+1)$ so that it plays no role in determining $q^{(n)}(t)$. This is analogous to what happens in the perturbative expansion around FLRW, where the terms with 4-th order derivatives never appear at $n-$th order in perturbations. This is clearly an indication that the perturbative expansion will fail in exploring the whole space of solutions around the trivial one. In our simple example, it is immediate to check that the solution for $q^{(n)}(t)$ is always a constant mode so the full perturbative solution is
\begin{equation}
q=c^{(1)}+c^{(2)}+c^{(3)}+\cdots
\end{equation}
i.e., only the constant mode of the general solution is generated. On the other hand, if we expand around a constant but non-trivial solution $q(t)=q_0$, the perturbative solution can be obtained to be
\begin{align}
q(t)&=q_0+c_2^{(1)}+c_2^{(2)}+c_2^{(3)}+\cdots\nonumber\\
&\quad +\Big[q_0 \Big(c_1^{(1)}+c_1^{(2)}+c_1^{(3)}\Big)+(c_1^{(1)}c_2^{(1)}+c_2^{(1)}c_1^{(2)}+c_1^{(1)}c_2^{(2)})\left(1+\frac{t}{q_0}\right) \nonumber\\
&\quad\qquad+c_1^{(1)}\big(c_2^{(1)}\big)^2\frac{t^2}{2q_0^3}+\cdots\Big]\eN^{-t/q_0}
\end{align}
which reproduces the expansion of the general solution \eqref{eq:Generalq} around $C_1=q_0$ and $C_2=0$, as it should. Notice that this perturbative solution is singular for $q_0=0$, thus showing once again the singular character of the trivial solution.

\begin{figure}
\begin{center}
\includegraphics[width=0.55\linewidth]{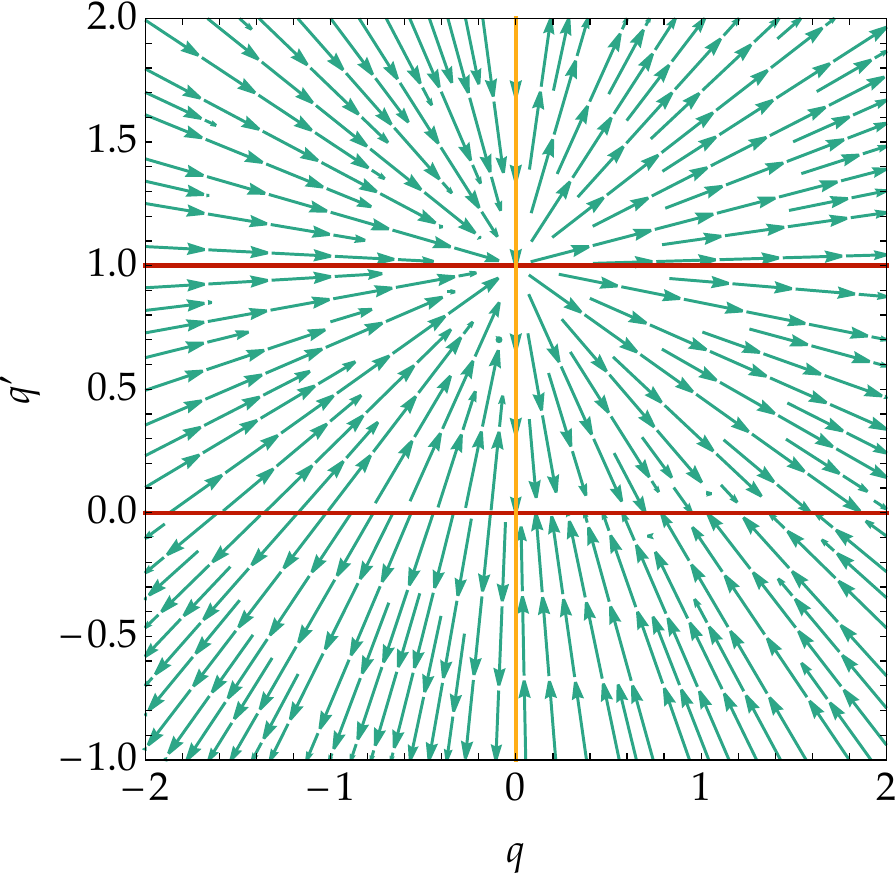}
\end{center}
\caption{In this Figure we show the phase map of the toy example given in \eqref{eq:Generalq}. We can clearly see how the surface $q=0$ corresponds to a separatrix in phase space as expected. }
\label{Fig:phasemapq}
\end{figure}

The second order equation  \eqref{eq:Generalq} can be written in an autonomous first order system as
\begin{equation}
q'=p,\quad\quad
p'=\frac{p-1}{q}p.
\end{equation}
Again it is obvious that $q=0$ represents a singular surface in phase space that describes a separatrix (see Fig. \ref{Fig:phasemapq}). Furthermore, there is one critical trajectory given by $p=0$. It is not difficult to see that the trajectories are straight lines of the form $p=1+cq$. From the phase map it is apparent that the separatrix is not a good physical solution for the system. Among the problems reported above, we can see that the separatrix can never be exactly reached from any point in phase space (unless it already belongs to the separatrix) and it does not correspond to an attractor region so the system will hardly evolve towards there.

\chapter{Curvature invariants in GQTG}\label{app:GQTG}
In this appendix we reproduce the terms given in \cite{ArciniegaBuenoCano2018} for completeness:\footnote{It is worth recalling that here we are using a different convention for the Riemann, Ricci and curvature scalar. All of them have an extra global sign with respect to our original definition.}
\begin{align}
\mathcal{R}_{(4)}
&\coloneqq-\frac{1}{192}\Big[5 \mathring{R}^{4}-60 \mathring{R}^{2} Q_{1}+30 \mathring{R}^{2} Q_{2}-160 \mathring{R} C_{1}+32 \mathring{R} C_{2}-104 \mathring{R} C_{3}\nonumber\\
&\qquad\qquad+272 Q_{1}^{2}-256 Q_{1} Q_{2} +336 A_{10}+48 A_{14}\Big]\,\\[4mm]
\mathcal{R}_{(5)}
&\coloneqq-\frac{1}{5760}\Big[15 \mathring{R}^{5}-36 \mathring{R}^{3} Q_{1}-224 \mathring{R}^{3} Q_{2}-336 \mathring{R}^{2} C_{1}-140 \mathring{R}^{2} C_{2}+528 \mathring{R}^{2} C_{3}\nonumber\\
&\quad\qquad\qquad -592 \mathring{R} Q_{1}^{2}+1000 \mathring{R} Q_{1} Q_{2} +301 \mathring{R} Q_{2}^{2}-912 \mathring{R} A_{2}-928 \mathring{R} A_{10}\nonumber\\
&\quad\qquad\qquad +1680 \mathring{R} A_{14}+1152 Q_{1} C_{1}+264 Q_{1} C_{2}+312 Q_{2} C_{2}-64 Q_{1} C_{3} \nonumber\\
&\quad\qquad\qquad -2080 Q_{2} C_{3}+4992 I_{1}\Big]\,\\[4mm]
\mathcal{R}_{(6)}
&\coloneqq \frac{1}{3594240}\Big[ 56813 \mathring{R}^{6}-523188 \mathring{R}^{4} Q_{1}+6234 \mathring{R}^{4} Q_{2}+798849 \mathring{R}^{3} C_{2}-558622 \mathring{R}^{3} C_{3}\nonumber\\
&\qquad\qquad\qquad +1235848 \mathring{R}^{2} Q_{1}^{2} -163250 \mathring{R}^{2} Q_{1} Q_{2}+42084 \mathring{R}^{2} Q_{2}^{2}-707808 \mathring{R}^{2} A_{2}\nonumber\\
&\qquad\qquad\qquad +231048 \mathring{R}^{2} A_{10}+439920 \mathring{R}^{2} A_{14}-5265366 \mathring{R} Q_{1} C_{2} +23208 \mathring{R} Q_{2} C_{2}\nonumber\\
&\qquad\qquad\qquad +4902132 \mathring{R} Q_{1} C_{3}+44880 \mathring{R} Q_{2} C_{3}-704400 Q_{1}^{3}+289200 Q_{1} Q_{2}^{2}\nonumber\\
&\qquad\qquad\qquad -62400 Q_{2}^{3} +1168128 \mathring{R} I_{1}+792000 Q_{1} A_{2}+374400 Q_{2} A_{2} \nonumber\\
&\qquad\qquad\qquad -723600 Q_{2} A_{10}-676800 C_{1}^{2}+7903368 C_{1} C_{2}-8581680 C_{1} C_{3}\nonumber\\
&\qquad\qquad\qquad -3782484 C_{2}^{2} +15454692 C_{2} C_{3}-12753720 C_{3}^{2}\Big]\,,
\end{align}
where we are using the abbreviations 
\begin{align}
Q_1 &\coloneqq \mathring{R}_{\mu\nu}\mathring{R}^{\mu\nu} \,,\\
Q_2 &\coloneqq \mathring{R}_{\mu\nu\rho\sigma}\mathring{R}^{\mu\nu\rho\sigma} \,,\\
C_1 &\coloneqq \mathring{R}_\mu{}^\rho{}_\nu{}^\sigma \mathring{R}_\rho{}^\tau{}_\sigma{}^\eta \mathring{R}_\tau{}^\mu{}_\eta{}^\nu \,,\\
C_2 &\coloneqq \mathring{R}_{\mu\nu}{}^{\rho\sigma} \mathring{R}_{\rho\sigma}{}^{\tau\eta} \mathring{R}_{\tau\eta}{}^{\mu\nu} \,, \\
C_3 &\coloneqq \mathring{R}_{\mu\nu\rho\lambda} \mathring{R}^{\mu\nu\rho}{}_\sigma \mathring{R}^{\lambda \sigma} \,, \\
A_2 &\coloneqq \mathring{R}_{\mu}{}^{\sigma}{}_{\rho}{}^{\tau} \mathring{R}^{\mu\nu\rho\lambda} \mathring{R}_{\nu \alpha\lambda \beta} \mathring{R}_{\sigma}{}^{\alpha}{}_{\tau}{}^{\beta} \,,\\
A_{10} &\coloneqq  \mathring{R}^{\mu\nu} \mathring{R}_{\mu}{}^{\rho}{}_{\nu}{}^{\lambda} \mathring{R}_{\sigma\tau \alpha\rho} \mathring{R}^{\sigma\tau \alpha}{}_{\lambda} \,,\\
A_{14} &\coloneqq  \mathring{R}^{\mu\nu} \mathring{R}^{\rho\lambda} \mathring{R}_{\sigma\rho\tau\lambda} \mathring{R}^{\sigma}{}_{\mu}{}^{\tau}{}_{\nu} \,,\\
I_1 &\coloneqq \mathring{R}_{\rho\sigma}{}^{\mu\nu} \mathring{R}_{\mu\tau}{}^{\rho\lambda} \mathring{R}_{\alpha \gamma}{}^{\sigma\tau} \mathring{R}_{\nu\delta}{}^{\alpha \beta}\mathring{R}_{\lambda \beta}{}^{\gamma\delta} \,.
\end{align}


\begingroup
  \renewcommand*{\bibfont}{\small}
  \renewcommand{\cleardoublepage}{}
  \renewcommand{\clearpage}{}
  \printbibliography
\endgroup

\mbox{}
\newpage
\thispagestyle{empty}
\setcounter{footnote}{0}

\begin{center}{\LARGE {\bf Erratum}}\end{center}

\vspace{5mm}

After submitting the thesis I found several typos.\footnote{I would like to thank Thomas Z{\l}osnik for pointing out some of them.} Most of them are irrelevant misprints in the notation or English errors, so I will not comment on them. The relevant changes I have performed are collected in the following list:

\begin{itemize}

\item Page 23. I rewrote the last of the three remarks about the volume form. The proportionality function does not need to be positive (as in the previous versions); that depends on the chosen orientations.

\item Page 23 I added the important fact that the support must be compact in order to define the integration.

\item Page 31. I dropped equation (2.4.40) from the Corollary 2.50, since it is already present in Proposition 2.53.

\item Page 45. In the previous version, before Definition 3.11., the generators of the translations appeared in the expression of $\sigma(p)$. Since, at this point, I am doing a general discussion without specifying the involved Lie groups, I changed the expression of $\sigma$ by introducing arbitrary ``broken'' generators $\mathrm{K}_\mathfrak{a}$.

\item Page 49. I added a last line in the footnote, clarifying why in Chapters 8 and 9 I do not use the ``c'' notation (just for simplicity since there is no confusion).

\item Page 51-52. There were some errors here. In eq. (3.3.24) and (3.3.27) the Lie derivative of the Lagrangian was already included in $\dfB$, so I have dropped it. In (3.3.26) there was a global minus sign missing (which also does not appear in \cite{Hehl1995}). I also rewrote some paragraphs of these pages, correcting some details.

\item Page 52-53. The last term of (3.3.35) is identically vanishing. I kept it in the equation but mentioned it right after the Proposition 3.18.

\item Page 56. I added a 2 in the l.h.s. of (3.3.59).

\item Page 59. The parameter $\overline{c}_4$ has been removed from Table 3.4.2. In addition, I corrected the number of odd parameters in the caption of that table (20 instead of the previous 16).

\item Page 61. Some missing $\frac{1}{\kappa}$ have been added in (3.4.15), (3.4.20) and (3.4.21).

\item Page 74 and 75. Clarifications added after (4.4.1) and before (4.5.1) regarding the appearance of the Levi-Civita tensor.

\item Page 90. The footnote 8 has been moved to avoid a LaTeX bug in the previous version (in which there were two ``footnote 9'').

\item Page 116. I corrected an error right before and within equation (6.3.6) and added a little paragraph before in order to clarify the point. In the new (6.3.6), I corrected some sign errors that appear in the paper \cite{JCAObukhov2021a}.

\item Page 141. After (7.3.1) the limit $\alpha \to 0$ has been substituted by the correct one, which is $\dimM\to 4$.

\newpage
\thispagestyle{empty}

\item Page 149. The Subsection 8.2.1 of the previous version is actually the Section 8.3 of the thesis. This has been corrected.

\item Page 155. The old Subsection 8.2.6 (now 8.3.5) has been renamed. I added to the title ``for a $\Lambda$-dominated era'' to be more specific.

\item Page 163. Some (irrelevant) parts of the proof of Proposition 9.2 have been removed.

\item Page 208. The last part of the proof of Proposition D.5 has been re-arranged to make it more understandable.

\item Page 209-210. I have reordered the Sections D.2, D.3 and D.4 to be more consistent with the order in which they are referenced in the main text.

\item Page 209 (now 210). At the beginning of the proof, I have introduced the $\epsilon$ that appears below.

\item I corrected the notation for the Levi-Civita Riemann tensor $R\to\mathring{R}$ at the end of Chapter 3, at the beginning of Chapter 8 and in Appendix G.

\end{itemize}

\noindent {\bf Message for the reader}: please, let me know if any other errors or misprints are found. For possibly updated versions of this text, please check my personal webpage.

\vspace{2cm}

\hfill --- The author

\end{document}